\documentclass[12pt,openany]{book}
\usepackage{amsmath,amsfonts,amssymb,amsthm,a4wide,graphics}
\usepackage{makeidx}
\usepackage{nameref}
\usepackage[hyperindex]{hyperref}

\numberwithin{equation}{section}

\makeindex

\begin{document}

\title{{\bf\Huge Classical Systems in Quantum Mechanics}}
 \author{Pavel B\'ona, \\ e-mails:\ pavel.bona@gmail.com \\
 bona1@uniba.sk \\
 Home web-page: http://www.st.fmph.uniba.sk/~bona1/\\
Department of Theoretical Physics, Comenius University \\ SK-842
48 Bratislava, Slovakia}

\maketitle
\newcommand{\nl}{\hfill\break}
\newcommand{\lhs}{left hand side\ }
\newcommand{\rhs}{right hand side\ }



\pagebreak


\vspace*{\fill}



\hyphenation{co-ad-joint re-pre-sen-ta-ti-on Bra-ti-sla-va iso-morph-ism}

\def\refer#1#2#3#4#5{{\rm #1}:\ {\rm #2}\ {\bf #3}\ {(#4)}\ #5;\ }

\def\={\hbox{-}}

\def\acc{$\!\!$\'{}$\!$}
\def\dti{\!\cdot\!}
\def\rref#1~{~(\ref{#1})}
\def\dref#1~{~Definition~\ref{df;#1}}
\def\pgr#1~{~\pageref{eq;#1}}
\def\pgd#1~{~\pageref{df;#1}}
\def\pg#1~{~\pageref{#1}}
\def\bs#1{$\boldsymbol{#1}$}
\def\mbs#1{\boldsymbol{#1}}

\def\Ran{{\rm Ran}}
\def\Ker#1{${\rm Ker}(#1)$}
\def\mKer#1{{\rm Ker}(#1)}

\def\supp{{\rm supp}\ }
\def\rank{{\rm rank}}

\def\deg#1{{\rm deg}($#1$)}
\def\mdeg#1{{\rm deg}(#1)}

\def\mT#1{\mcl T_{#1}}
\def\T#1{$\mcl T_{#1}$}
\def\cw{\curlywedge}

\def\LM#1{$\mbs{\Lambda^{#1}}(M)$}
\def\mLM#1{\mbs{\Lambda^{#1}}(M)}
\def\LN#1{$\mbs{\Lambda^{#1}}(N)$}
\def\mLN#1{\mbs{\Lambda^{#1}}(N)}

\def\mSs{{\cal S}_*}
\def\Ss{${\cal S}_*$}
\def\cS{${\cal S}$}
\def\mS{{\cal S}}
\def\Ej{$E_{\{j\}}$}
\def\mEj{E_{\{j\}}}
\def\mrj{\mrh_{\{j\}}}
\def\rj{$\mrh_{\{j\}}$}
\def\psitr{\psi_t^\mrh}


\def\pika{\ $\diamondsuit$}
\def\bpika{\ $\blacklozenge$}
\def\zel{\ $\spadesuit$}
\def\zal{\ $\clubsuit$}
\def\dovi{\ $\heartsuit$}

\def\pI{${\mbs p}_I$}
\def\pII{${\mbs p}_{II}$}
\def\mpI{{\mbs p}_I}
\def\mpII{{\mbs p}_{II}}

\def\bT{$\Bbb T$}
\def\mbT{{\Bbb T}}
\def\bA{$\Bbb A$}
\def\mbA{{\Bbb A}}
\def\bK{$\Bbb K$}
\def\mbK{{\Bbb K}}
\def\bC{$\Bbb C$}
\def\mbC{{\Bbb C}}
\def\bR{$\Bbb R$}
\def\mbR{{\Bbb R}}
\def\bN{$\Bbb N$}
\def\mbN{{\Bbb N}}
\def\bZ{$\Bbb Z$}
\def\mbZ{{\Bbb Z}}
\def\bI{$\Bbb I$}
\def\mbI{{\Bbb I}}
\def\bF{$\Bbb F$}
\def\bG{$\Bbb G$}
\def\mbF{{\Bbb F}}
\def\mbG{{\Bbb G}}
\def\bbf{${\bf f}$}
\def\mbbf{{\bf f}}
\def\cbF{${\bf F}$}
\def\mcbF{{\bf F}}
\def\cbU{${\bf U}$}
\def\mcbU{{\bf U}}
\def\bD{{\bf D}}
\def\bx{{\bf x}}
\def\by{{\bf y}}
\def\bz{{\bf z}}
\def\bv{{\bf v}}
\def\bw{{\bf w}}

\def\fg{{\mfk g}}
\def\fgs{{\mfk g^*}}

\def\bbs#1{$\boldsymbol{#1}$}
\def\mbbs#1{\boldsymbol{#1}}

\def\ia{{\it a}}
\def\ib{{\it b}}
\def\ix{{\it x}}
\def\iy{{\it y}}
\def\iz{{\it z}}
\def\iu{{\it u}}

\def\fpsxe{$\rf^\psi_{[\xi,\eta]}$}
\def\mfpsxe{\rf^\psi_{[\xi,\eta]}}
\def\fpsx{$\rf^\psi_\xi$}
\def\mfpsx{\rf^\psi_\xi}
\def\fpse{$\rf^\psi_\eta$}
\def\mfpse{\rf^\psi_\eta}

\def\rq{{\rm q}}
\def\rf{{\rm f}}
\def\rrh{{\rm h}}
\def\rx{{\rm x}}
\def\ry{{\rm y}}
\def\rz{{\rm z}}
\def\ra{{\rm a}}
\def\rrb#1{{\rm b}_{#1}}
\def\rc{{\rm c}}
\def\rd{{\rm d}}
\def\rQ{{\rm Q}}
\def\ru{{\rm u}}
\def\rv{{\rm v}}
\def\rn{{\rm n}}
\def\mrho{{\rm h}_0}
\def\mrhoo{{\rm h}^0}

\def\CH{${{\frak C}({\cal H})}$}
\def\mCH{{{\frak C}({\cal H})}}
\def\TH{${{\frak T}({\cal H})}$}
\def\UH{${\cal U(H)}$}
\def\mTH{{{\frak T}({\cal H})}}
\def\mUH{{\cal U(H)}}

\def\Lqn{$L^2(\mbR^n)$}
\def\mLqn{L^2(\mbR^n)}
\def\LHs{${\cal L(H)}_s$}
\def\mLHs{{\cal L(H)}_s}
\def\mH{{\cal H}}
\def\mHp{{\cal H}_\Pi}
\def\mK{{\cal K}}
\def\K{${\cal K}$}
\def\M{${\cal M}$}
\def\mM{{\cal M}}
\def\Ms{${\cal M_*}$}
\def\mMs{{\cal M_*}}
\def\N{${\cal N}$}
\def\mN{{\cal N}}
\def\Ns{${\cal N_*}$}
\def\mNs{{\cal N_*}}
\def\mLH{{\cal L(H)}}
\def\LH{${\cal L(H)}$}
\def\LHo{${\cal L(H)_\mome}$}
\def\mLHo{{\cal L(H)_\mome}}
\def\LHp{$\cal L(H)_\Pi$}
\def\mLHp{\cal L(H)_\Pi}
\def\H{${\cal H}$}
\def\Hp{${\cal H}_\Pi$}

\newcommand{\PH}{$P({\cal H})$}
\def\mPH{P({\cal H})}
\newcommand{\A}{${\mathcal{A}}$}
\def\mA{{\mathcal A}}
\newcommand{\Z}{${\mathcal{Z}}$}
\def\mZ{{\mathcal Z}}
\def\C{${\mathcal C}$}
\def\Cc{${\cal C}_{cl}$}
\def\mCc{{\cal C}_{cl}}
\def\mC{{\cal C}}
\def\B{${\cal B}$}
\def\mB{{\cal B}}
\def\F{${\cal F}$}
\def\mF{{\cal F}}
\def\Fr{$\mF_{\mrh}$}
\def\mFr{\mF_{\mrh}}
\def\mFP{\mF_{\mPH}}
\def\FP{$\mF_{\mPH}$}

\def\PM{$\mcl  P(\mfk M)$\ }
\def\mPM{\mcl P(\mfk M)}

\def\cl#1{${\cal #1}$}
\def\mcl#1{{\cal #1}}

\def\ben{$\beta_{\nu}$}
\def\mben{\beta_{\nu}}
\def\ber{$\beta_{\mrh}$}
\def\mber{\beta_{\mrh}}
\def\qr{$\rq_{\mrh}$}
\def\mqr{\rq_{\mrh}}
\def\pr{${\rm p}_{\mrh}$}
\def\mpr{{\rm p}_{\mrh}}
\def\qn{$\rq_{\nu}$}
\def\mqn{\rq_{\nu}}
\def\Tr{$T_{\mrh}$}
\def\mTr{T_{\mrh}}
\def\Trs{$T_{\mrh}^*$}
\def\mTrs{T_{\mrh}^*}
\def\Tn{$T_{\nu}$}
\def\mTn{T_{\nu}}
\def\Tns{$T_{\nu}^*$}
\def\mTns{T_{\nu}^*}

\def\Ty{$T_{\by}$}
\def\mTy{T_{\by}}

\def\P#1{$P_{#1}$}
\def\mP#1{P_{#1}}

\def\bcDp{${\boldsymbol{\mcl D}}_{r+}^1$}
\def\mbcDp{{\boldsymbol{\mcl D}}_{r+}^1}
\def\bcD{${\boldsymbol{\mcl D}}_r$}
\def\mbcD{{\boldsymbol{\mcl D}}_r}
\def\bcDF{${\boldsymbol{\mcl D}}(\mbF)$}
\def\mbcDF{{\boldsymbol{\mcl D}}(\mbF)}
\def\bcDFr{${\boldsymbol{\mcl D}_r}(\mbF)$}
\def\mbcDFr{{\boldsymbol{\mcl D}_r}(\mbF)}
\def\DdYa{$\mcl D_{ra}(\delta_Y)$}
\def\mDdYa{\mcl D_{ra}(\delta_Y)}
\def\DdX{$\mcl D_r(\delta_X)$}
\def\DdXa{$\mcl D_{ra}(\delta_X)$}
\def\DdXs{$\mcl D_{r*}(\delta_X)$}
\def\mDdX{\mcl D_r(\delta_X)}
\def\mDdXa{\mcl D_{ra}(\delta_X)}
\def\mDdXd{\mcl D_{rd}(\delta_X)}
\def\DdXd{$\mcl D_{rd}(\delta_X)$}
\def\mDdXs{\mcl D_{r*}(\delta_X)}
\def\DX{$\mcl D_r(X)$}
\def\DXa{$\mcl D_{ra}(X)$}
\def\DXd{$\mcl D_{rd}(X)$}
\def\DXs{$\mcl D_{r*}(X)$}
\def\mDX{\mcl D_r(X)}
\def\mDXa{\mcl D_{ra}(X)}
\def\mDXs{\mcl D_{r*}(X)}
\def\mAs{\mfkA^*}
\def\Ass{$\mfkA^{**}$}
\def\mAss{\mfkA^{**}}
\def\As{$\mfkA^*$}
\def\DAs{$D(A^*)$}
\def\DA{$D(A)$}
\def\HH{$\mH\oplus\mH$}
\def\mHH{\mH\oplus\mH}

\def\pEF{$\tilde{\mcl E_{\mbF}}$}
\def\mpEF{\tilde{\mcl E_{\mbF}}}
\def\EF{$\mcl E_{\mbF}$}
\def\mEF{\mcl E_{\mbF}}

\def\DGom{$\mcl D^{\mome}(G)$}
\def\mDGom{\mcl D^{\mome}(G)}
\def\DGomr{$\mcl D^{\mome}_r(G)$}
\def\mDGomr{\mcl D^{\mome}_r(G)}

\def\d#1,#2~{$d_{#2}{#1}$}
\def\md#1,#2~{d_{#2}{#1}}
\def\D#1,#2~{$D_{#2}{#1}$}
\def\mD#1,#2~{D_{#2}{#1}}
\def\Dfr{\D f,\mrh~}
\def\dfr{\d f,\mrh~}
\def\mDfr{\mD f,\mrh~}
\def\mdfr{\md f,\mrh~}
\def\Dfn{\D f,\nu~}
\def\dfn{\d f,\nu~}
\def\mDfn{\mD f,\nu~}
\def\mdfn{\md f,\nu~}
\def\Dhr{\D h,\mrh~}
\def\dhr{\d h,\mrh~}
\def\mDhr{\mD h,\mrh~}
\def\mdhr{\md h,\mrh~}
\def\mDhn{\mD h,\nu~}
\def\mdhn{\md h,\nu~}
\def\Dhn{\D h,\nu~}
\def\dhn{\d h,\nu~}

\def\mN#1,#2~{\|#1\|_{#2}}
\def\N#1,#2~{$\|#1\|_{#2}$}

\def\ad{{\rm ad}}
\def\madr{{\rm ad}_{\mrh}}
\def\madn{{\rm ad}_{\nu}}
\def\adr{${\rm ad}_{\mrh}$}
\def\adn{${\rm ad}_{\nu}$}
\def\madrs{{\rm ad}^*_{\mrh}}
\def\madns{{\rm ad}^*_{\nu}}
\def\adrs{${\rm ad}^*_{\mrh}$}
\def\adns{${\rm ad}^*_{\nu}$}

\def\Opsi{\Omega_0^\psi}
\def\omeAB#1{$\omega^{A\&B}_{#1}$}
\def\momeAB#1{\omega^{A\&B}_{#1}}

\def\OGF{$\mcl O_{F}(G)$}
\def\mOGF{\mcl O_{F}(G)}
\def\OWHr{$\mcl O_{\mrh}(\mGWH)$}
\def\mOWHr{\mcl O_{\mrh}(\mGWH)}
\def\OUr{$\mcl O_{\mrh}(\mfk U)$}
\def\OGr{$\mcl O_{\mrh}(G)$}
\def\OUn{$\mcl O_{\nu}(\mfk U)$}
\def\OGn{$\mcl O_{\nu}(G)$}
\def\OU{\cl O(\fk U)}
\def\OG{\cl O($G$)}
\def\mOUr{\mcl O_{\mrh}(\mfk U)}
\def\mOGr{\mcl O_{\mrh}(G)}
\def\mOUn{\mcl O_{\nu}(\mfk U)}
\def\mOGn{\mcl O_{\nu}(G)}
\def\mOU{\mcl O(\mfk U)}
\def\mOG{\mcl O(G)}
\def\Or{$\mcl O_{\mrh}$}
\def\mOr{\mcl O_{\mrh}}
\def\On{$\mcl O_{\nu}$}
\def\mOn{\mcl O_{\nu}}

\def\fk#1{$\frak #1$}
\def\mfk#1{\frak{#1}}

\def\mfkA{\mfk A}
\def\fkA{$\mfk A$}
\def\fkg{$\mfk g$}
\def\mfkg{\mfk g}
\def\fkgs{$\mfk g^*$}
\def\mfkgs{\mfk g^*}

\def\fTs{$\mfk T_s$}
\def\mfTs{\mfk T_s}
\def\fNn{$\mfk N_{\nu}$}
\def\mfNn{\mfk N_{\nu}}
\def\fNr{$\mfk N_{\mrh}$}
\def\mfNr{\mfk N_{\mrh}}
\def\fMr{$\mfk M_{\mrh}$}
\def\mfMr{\mfk M_{\mrh}}

\def\eps{$\epsilon$\ }
\def\meps{\epsilon}
\def\veps{$\varepsilon$\ }
\def\mveps{\varepsilon}
\def\ome{$\omega$}
\def\mome{\omega}
\def\mOme{\Omega}
\def\Ome{$\Omega$}
\def\gam{$\gamma$\ }
\def\mgam{\gamma}
\def\alp{$\alpha$}
\def\malp{\alpha}
\def\mphi{{\varphi}}
\def\rh{$\varrho$}
\def\mrh{\varrho}
\def\sg{$\sigma$}

\def\msg{\sigma}

\def\mlam{\lambda}
\def\lam{$\lambda$}
\def\t{$\tau$}
\def\mt{\tau}

\def\Xx{$X_{\xi}$}
\def\mXx{X_{\xi}}
\def\Xe{$X_{\eta}$}
\def\mXe{X_{\eta}}
\def\Xxe{$X_{[\xi,\eta]}$}
\def\mXxe{X_{[\xi,\eta]}}

\def\Ac{$\mcl A_{cl}$}
\def\mAc{\mcl A_{cl}}
\def\Cbs{$\mcl C_{bs}$}
\def\mCbs{\mcl C_{bs}}
\def\CG{${\mcl C}^G$}
\def\mCG{{\mcl C}^G}
\def\CGc{${\mcl C}^G_{cl}$}
\def\mCGc{{\mcl C}^G_{cl}}
\def\CGq{$\mcl C^G_{q}$}
\def\mCGq{{\mcl C}^G_{q}}
\def\GGc{${\mcl G}^G_{cl}$}
\def\mGGc{{\mcl G}^G_{cl}}
\def\GG{${\mcl G}^G$}
\def\mGG{{\mcl G}^G}
\def\nGcl{$\nu G$--classical\ }
\def\rGcl{$\mrh G$--classical\ }
\def\Gcl{$G$--classical\ }

\def\GWH{$G_{WH}$}
\def\mGWH{G_{WH}}

\def\hh#1,#2,#3~{$\hat {h_{\mfk#1}}(#2,#3)$}
\def\mhh#1,#2,#3~{\hat {h_{\mfk#1}}(#2,#3)}

\def\mh#1{{h}_{#1}}
\def\h#1{${h}_{#1}$}
\def\hSl{$\mh H^{Sl}$}
\def\mhSl{\mh H^{Sl}}

\def\ph#1,#2~{$\varphi_{#1}^{#2}$}
\def\mph#1,#2~{\varphi_{#1}^{#2}}
\def\cmrh#1~{{\varrho_{#1}}}
\def\crh#1~{$\varrho_{#1}$\ }

\def\pph#1,#2~{$\tilde{\varphi}_{#1}^{#2}$}
\def\mpph#1,#2~{\tilde{\varphi}_{#1}^{#2}}
\def\un#1,#2,#3~{${\rm u}_#1(#2,#3)$}
\def\mun#1,#2,#3~{{\rm u}_#1(#2,#3)}
\def\gQ#1,#2~{$g_\rQ(#1,#2)$}
\def\mgQ#1,#2~{g_\rQ(#1,#2)}
\def\taQ{$\tau^\rQ$}
\def\mtaQ{\tau^\rQ}
\def\mtQ#1,#2~{\tau^\rQ_{#1}#2}
\def\tQ#1,#2~{$\tau^\rQ_{#1}#2$}

\def\mv#1{{\bf v}_{#1}}
\def\w#1{${\bf w}_{#1}$}
\def\mw#1{{\bf w}_{#1}}
\def\vv#1,#2~{${\mathbf v}_{#1}(#2)$}
\def\mvv#1,#2~{{\mathbf v}_{#1}(#2)}
\def\vfn{${\mathbf v}_f(\nu)$}
\def\vvfn{${\mathbf{\check{v}}}_f(\nu)$}
\def\mvvfn{{\mathbf{\check{v}}}_f(\nu)}
\def\mvfn{{\mathbf v}_f(\nu)}
\def\vfr{${\mathbf v}_f(\mrh)$}
\def\vf{${\mathbf v}_f$}
\def\mvf{{\mathbf v}_f}
\def\mvfr{{\mathbf v}_f(\mrh)}
\def\vh{${\mathbf v}_h$}
\def\mvh{{\mathbf v}_h}
\def\Tpq{$\mcl T^p_q(M)$}
\def\Tpqx{$T^p_{qx}(M)$}
\def\mTpqx{T^p_{qx}(M)}
\def\mTpq{\mcl T^p_q(M)}
\def\mrTpq{T^p_q(M)}
\def\rTpq{$T^p_q(M)$}
\def\XN{$\mcl X(N)$}
\def\mXN{\mcl X(N)}
\def\XM{$\mcl X(M)$}
\def\mXM{\mcl X(M)}
\def\L#1~{$\pounds_{#1}$}
\def\mL#1~{\pounds_{#1}}
\def\dom{$\rd\mome$}
\def\mdom{\rd\mome}
\def\ip#1{$\boldsymbol{i_{#1}}$}
\def\mip#1{\boldsymbol{i_{#1}}}

\def\Lq{$L^2({\Bbb R},\rd q)$}
\def\mLq{L^2({\Bbb R},\rd q)}
\def\Wa{$W^*$\hbox{-}algebra}
\def\Wsa{$W^*$\hbox{-}subalgebra}
\def\Ca{$C^*$\hbox{-}algebra}
\def\Csa{$C^*$\hbox{-}subalgebra}
\def\Csas{$C^*$\hbox{-}subalgebras}
\def\rep{${}^*$--representation}
\def\Wrep{$W^*$--representation}
\def\autm{${}^*$-automorphism}
\def\autms{${}^*$-automorphisms\ }
\def\aut#1{${}^*$\=\,Aut\ #1}
\def\maut#1{{}^*$\=\,Aut$\ #1}
\def\Aut#1{$Aut(#1)$}
\def\mAut#1{Aut(#1)}

\def\Psib{$\Psi^\flat$}
\def\mPsib{\Psi^\flat}

\def\lb{\langle}
\def\rb{\rangle}

\def\plr{$p_{\leftrightarrow}$}
\def\mplr{p_{\leftrightarrow}}

\def\eequiv{\Leftrightarrow}
\def\imply{\Rightarrow}

\def\wrt{with respect to\ }
\def\om#1,#2~{$\omega_{#1}^{#2}$}
\def\mom#1,#2~{\omega_{#1}^{#2}}
\def\ommr{$\mome_{\mu,\hat\mrh}$}
\def\mommr{\mome_{\mu,\hat\mrh}}

\def\om{$\omega$}
\def\mom{\omega}

\def\prob{{\rm prob}}
\def\nbhd{neighbourhood\ }
\def\lub{{\rm l.u.b.}}

\def\eq{{equation}}

\def\noidt{\noindent}
\def\id {{\rm{id}}}

\def\emn#1~{{\em #1}\ind{#1}}
\def\emm#1~{{\bf #1}\ind{#1}}
\def\glss#1~{#1\glo{#1}}

\def\th{{\rm th}}

\newcommand{\bequ}{\begin{equation}}
\newcommand{\enqu}{\end{equation}}
\newcommand{\barr}{\begin{eqnarray}}
\newcommand{\earr}{\end{eqnarray}}

\newcommand{\glo}{\glossary}
\newcommand{\ind}{\index}
\newcommand{\rarw}{\rightarrow}
\renewcommand{\Im}{{\rm Im}}
\renewcommand{\Re}{{\rm Re}}

\swapnumbers \theoremstyle{plain}
\newtheorem{thm}{Theorem}[section] 
\newtheorem{defi}[thm]{Definition}    
\newtheorem{defs}[thm]{Definitions}   
\newtheorem{notat}[thm]{Notation}     
\newtheorem{prop}[thm]{Proposition}    
\newtheorem{lem}[thm]{Lemma}        
\newtheorem{lem*}[thm]{Lemma*}       
\newtheorem{pt}[thm]{}             
\newtheorem{corl}[thm]{Corollary}
\newtheorem{corls}[thm]{Corollaries}
\newtheorem{intpn}[thm]{\bf Interpretation}  
\newtheorem{rem}[thm]{{\bf Remark}}         
\newtheorem{exmp}[thm]{{\bf Examples}}      
\newtheorem{exm}[thm]{{\bf Example}}         
\newtheorem{ill}[thm]{{\bf Illustration}}    
\newtheorem{note}[thm]{{\bf Notes}}       
\newtheorem{noti}[thm]{{\bf Note}}         


\addcontentsline{toc}{section}{\protect\numberline{}{\noindent\it Preface}}

\centerline{\it\Large Preface}\vspace{1.2cm} The work contains a description and an analysis of
two different approaches determining the connections between quantal and classical theories.

The first approach associates with any quantum-mechanical system with finite number of degrees of
freedom a classical Hamiltonian system `living' in projective Hilbert space $P(\cal H)$, and it is
called here the `classical projection'.

The second approach deals with `large' quantal (= quantum mechanical) systems in the limit of
infinite number of degrees of freedom and with their corresponding `macroscopic limits' described
as classical Hamiltonian systems of the system's global (intensive) quantum observables.

The last part of this work contains a series of models describing interactions of the ``small''
physical (micro)systems with the ``macroscopic'' ones, in which these interactions lead to a
(macroscopic) change of some ``classical'' parameters of the large systems. These models connect,
in a specific way, the two classes of the systems considered earlier in this work by modeling
their mutual interactions leading to striking (i.e. theoretically impossible in the framework of
finite quantum systems) results.

The projective space $P(\cal H)$ of any complex Hilbert space $\cal H$ is endowed with a natural
symplectic structure, which allows us to rewrite the quantum mechanics of systems with finite
number of degrees of freedom in terms of a classical Hamiltonian dynamics. If a quantum-mechanical
system is associated with a continuous unitary representation $U(G)$ of a connected Lie group $G$
on $\cal H$, the orbits (possibly factorized in a natural way) of the projected action of $U(G)$
in $P(\cal H)$ are naturally mapped onto orbits of the coadjoint representation $Ad^*(G)$ of $G$.
These coadjoint orbits have a canonical symplectic structure which coincides with the one induced
from the structure of $P(\cal H)$. For important classes of physical systems these symplectic
spaces are either symplectomorphic to the `corresponding' classical phase spaces, or they are some
extensions of them (describing, e.g. particles with `classical spin'). Quantal dynamics is
projected onto these phase spaces in a natural way, leading to classical Hamiltonian dynamical
systems without any limit of Planck constant $\hbar\rightarrow 0$.

For a large (infinite) quantal system an automorphic group action of $G$ on the $C^*$-algebra
$\mfk A$ of its bounded observables enables us to define a macroscopic subsystem being a classical
Hamiltonian system of the same type as we obtained in the case of finite number of degrees of
freedom. There is a difference, however, between the interpretations of `classical projections'
and of these `macroscopic limits': The classical (mechanical) projection describes classical
mechanics of expectation values of quantal observables whereas the macroscopic limit describes a
quantal subsystem with classical properties - its observables are elements of a subalgebra ${\mfk
M}_G$ of the center $\mfk{Z}$ of the double dual ${\mfk A}^{\ast\ast}$ of $\mfk A$. Any state
$\omega$ on $\mfk A$ has a unique 'macroscopic limit' $p_M\omega$ which is represented by a
probability measure on the corresponding (generalized) classical phase space. This offers us a
possibility of deriving a classical (macroscopic) time evolution (which is, in general, in a
certain sense stochastic, cf. \cite[Sec.III.G]{bon10}) from the underlying reversible quantal
dynamics.

 A scheme of `macroscopic
quantization' is outlined, according to which a (nonunique) reconstruction of the infinite quantal
system $(\mfk A;\sigma_G)$ from its macroscopic limit is possible. By determining a classical
Hamiltonian function in the macroscopic limit of $(\mfk A;\sigma_G)$ we can define a `mean-field'
time evolution in the infinite system $(\mfk A;\sigma_G)$. Our definition of the 'mean-field'
evolutions extends the usual ones. The schemes and results developed in the work are applicable to
models in the statistical mechanics as well as in gauge-theories (in the `large N limit'). They
might be relevant also in general considerations of 'quantizations' and of foundations of quantum
theory.

The last Chapter of this work is devoted to the description of several models of interacting
`microsystems' with `macrosystems', mimicking a description of the `process of measurement in QM'.
In these models, certain `quantal properties' of the system, namely a (coherent) superposition of
specific vector states (eigenstates of a 'measured' observable), transform by the unitary
continuous time evolutions (for $t\rarw\infty$) into the corresponding `proper mixtures' of
macroscopically different states of the `macrosystems' occurring in the models.

 In this connection we shall shortly discuss the old `quantum measurement problem', which however,
in the light of certain experiments performed in the last decades and suggesting the possibilities
of quantum-mechanical interference of several macroscopically different states of a macroscopic
system, need not be at all a fundamental theoretical problem; this might mean that the often
discussed `measurement process' can be included into the presently widely accepted model of
quantum theory.\nl

{\bf Acknowledgemets:}\nl

 This work is a revised and completed version of the unpublished text: ``Classical Projections
 and Macroscopic Limits of
Quantum Mechanical Systems'' written roughly in the years 1985 - 1986. The author is indebted to
Klaus Hepp for his stimulations and the kind help with correcting many formulations of the
original text. During the work on the old text, encouragements by Elliott Lieb and by the late
Walter Thirring were also stimulating, and the discussions with the colleagues, the late Ivan
Korec, Milan Noga, Peter Pre\v snajder and in particular with Ji\v r\'i Tolar, were useful and the
author expresses his gratitude to all of them. Thankfulness for several times repeated
encouragements to publish the old text should be expressed to Nicolaas P. Landsman. Author's
thanks belong also, last but not least, to his colleague Vladim\'ir Balek for his kind help with
final formulation of important parts of this book.\nl

{\bf Technical notes:}\
\begin{itemize}
\item[(a)] This book  contains several technical concepts which are
not introduced here in details. The readers needing to get a brief
acquaintance with some additional elementary concepts and facts of
topology, differential geometry (also in infinite dimensions),
group theory, or theory of Hilbert space operators and theory of
operator algebras, could consult e.g. the appendices of the freely
accessible publication \cite{bon-EQM}, and the literature cited in
our Bibliography. Due to connections of many places in the text of
this book with the content of the work \cite{bon-EQM} it is
recommended to keep the cited \cite{bon-EQM} as a handbook. The
separate complementary text of \cite[Textbook]{bon-EQM} might be
also helpful. The frequent citations from \cite{bon-EQM} contain
usually references to specific places of the cited work.
\item[(b)] Two kinds of quotation marks are used: Either the ones
which stress some ``{\em standard expressions}'', or those which
indicate `{\em intuitive denotations}'. The difference between
these two is not, however, very sharp.
\item[(c)] Many symbols appearing in mathematical formulas are
introduced in various places of the text and repeatedly used in
the rest of the book. For easier revealing of their meanings, they
are included into Index and their first appearance in the text is
stressed, sometimes in a not quite usual manner, by {\bf boldface
form} of a part of the surrounding text.
\end{itemize}


\pagebreak
\tableofcontents

\hyperlink{index}{{\bf Index}}{\hfill{\bf \pageref{index}}}


\chapter{Introduction}\label{Ch1}

\section{Motivation and Summary}\label{sec;1.1}
\label{motiv}

\pt\label{1.1.1}{\rm Successful communication and manipulation with `objects' requires
construction of some adequate theoretical models ($\approx$ theories) of some classes of
`objects', resp. `phenomena'. Different phenomena might be described by different theoretical
schemes. These schemes should be, however, mutually consistent in the sense giving the same
results for phenomena lying in the common domain of applicability of different theories. If one of
the theories is considered to be `more general' then a second one, then the whole domain of
applicability of the second theory has to be contained in the domain of the first one. This is the
case of \emn quantum mechanics (QM)~, which is believed to be a `covering theory' of the more
special \emn classical mechanics (CM)~ - to the extent of measurement precision of apparatuses
determining of `classical systems'. Hence we can ask {\bf how to describe phenomena belonging to
the domain of applicability of CM in the framework of QM.} }\

\pt\label{1.1.2}{\rm Any single phenomenon, which is unambiguously and reproducibly determined by
a specification of an empirical situation is, however, expressible in terms of parameters (resp.
variables) occurring in \emn CM~: coordinates of positions and velocities of points distinguished
and measured by `macroscopic bodies' and various correlations between these variables. Hence also
any experimentally realizable situation described in QM (which need not be a consequence of laws
of CM, e.g. observation of spectra of atoms) is expressible in terms of CM (e.g. preparation of
sources of radiation and measurement of positions of spectral lines displayed on screens). Quantal
(:= quantum mechanical) phenomena are not only observed on a `background' and `from the point of
view' of quantities describing states of macroscopic bodies (resp. of such parameters, the
behaviour of which is adequately described by laws of CM), but also specific theoretical models
for description of such phenomena in the framework of QM are constructed under strong influence of
existing models in CM (e.g. the quantal models of atoms compared to classical planetary motions,
or, more generally, some systems of canonically conjugated observables in the sense of the
Hamiltonian CM correspond isomorphica1ly to a subset of quantal observables). Quantal models of
many systems, on the other hand, might be constructed from classical models of the same systems
(which are adequate in a certain range of conditions, e.g. classical gases in some intervals of
temperature and density) by a more or less standard procedure of `quantization', compare, e.g.
\cite{berez3,doeb&tol,fronsdal},\cite{nelson,nied&tolar,tolar}, and works quoted therein. The
(vaguely stated) question arising from these considerations is: {\bf What is a `physically
justified way' of correct determination of quantal models from their classical approximates~?}}\

\pt\label{1.1.3}{\rm One of the remarkable features of QM is the occurence of the {\rm universal}
(Planck) constant $\hbar$, which might be used to measure mutual `deviation' of quantal and
classical descriptions of a given physical system (we shall not discuss here the nontrivial
methodological question: how to determine a `physical system' and what is its dependence on
theoretical concepts used in the process of the determination). Consequently, an approximate
description of processes in the framework of QM that are characterized by some quantities S large
compared to the Planck constant (S being of the same physical dimension as $\hbar$) is often
reached in the limit of large values of $S\hbar^{-1}$ (`short wave asymptotics'). If, however, the
system described by QM has some features ('variables' etc.) which are adequately described by CM
too, then {\bf the description of this `classical subsystem' has to be contained in QM} with the
fixed value of Planck constant (i.e. the classical description should be exact consequence of QM
without any approximation procedure, which is often formally performed by the limit
$\hbar\rightarrow 0$\ \footnote{Consider here macroscopic quantal effects (e.g. superconductivity,
superfluidity) vanishing for $\hbar\rightarrow 0$.}). We shall introduce a standard procedure of
obtaining classical systems from quantal ones. Such a classical system is called here a `classical
projection' of the quantal system (contrary to the `classical limit' obtained in some way by
$\hbar\rightarrow 0$).}\

\pt\label{1.1.4}{\rm This work is considered as a conceptually and intuitively (however, not
always technically) simple way to give some insight into the indicated questions. Much more
complete and extensive overview of these and related technical topics is given in the recent book
\cite{lands3} by Landsman. Many relevant questions are discussed in the author's work
\cite{bon-EQM}, containing also a detailed discussion of possible extensions of the QM formalism
to its nonlinear versions; these nonlinear quantum motions are closely connected with the   theory
presented
 in our Chapter \ref{Ch6}, corresponding to the motions of a single ``microsystem'' moving in
the ``mean-field'' acting on it by interaction with infinite number of similar microsystems; the
dynamics of the whole infinite collection of ``microsystems'' is, however, linear. Such a
nonlinear quantum dynamics is also discussed by S.Weinberg in \cite{weinb}, whose work is also
discussed and reformulated in \cite[Sec.3.6]{bon-EQM}.

The mentioned work of S.Weinberg is not intrinsically consistent in the case of nonlinear motions
of nontrivial density matrices, resp. ``mixtures''. To obtain successful picture of nonlinear
quantum dynamics of ``mixed states'' together with their physically satisfactory quantal
interpretation, one has to introduce \emm two kinds of ``mixed states''~: The usual one used in
(linear) QM are described in the standard way by density matrices (called there ``elementary
mixtures''), and others are called ``genuine mixtures'' (or. also ``proper mixtures'') - these
correspond to the states which arose by a real `mixing' of different quantal states, as it appears
in classical statistical mechanics in ensembles of systems occurring in different states -
different points of the phase space of the described system; they are introduced in
\cite[Sec.2.1-e]{bon-EQM} and difference of these two kinds of  mixtures is illustrated e.g. in
\cite[Sec. 3.3-e]{bon-EQM}.

In the remaining sections of this introductory chapter it is specified briefly what we mean here
by QM, CM and by the `quantum theory of large systems'. The second chapter is devoted to a
detailed study of geometry of the \emn projective Hilbert space \PH~, where \emn \H\ is the
Hilbert space~ used in description of a quantal system. We emphasize there the natural \emm
symplectic structure~ on \PH, cf. e.g. \cite{arn1,mars&rati,bon-EQM,odzi&rati}. This structure is
used in Section~\ref{sec;2.3} to description of QM in terms of infinite dimensional CM, i.e. of
classical Hamiltonian field theory with, however, the standard quantum statistical
interpretation.}\

\pt\label{1.1.5}\rm The Chapter~\ref{Ch3} ``Classical Mechanical Projections'' is devoted to a
general construction of Hamiltonian CM from a given quantal system (provided that an
interpretation of its `basic quantities' is specified by a unitary representation $U(G)$ of a \emn
Lie group~ $G$; for Lie groups see e.g. \cite{bar&racz,bourb;Lie,mack3,pontrjag}). The scheme of
this construction is very simple: Take the \emn orbit $O_\mrh$~ $:= G.\mrh$ through a point
$\mrh\in\mPH$ of the action of $\mbs{U}(G)$ on \glss \PH~\ corresponding to the action of $U(G)$
on \H\ and restrict the natural symplectic form on \PH\ onto $O_\mrh$. For properly chosen \rh\
the orbit $O_\mrh$ is an immersed (and regularly embedded, cf. \cite[Proposition
2.1.5(iv)]{bon-EQM} completed by \cite{bon-orbit}) submanifold of $\mPH\subset\mTH$\ (cf.
\ref{1.2.3}), hence the restriction is well defined. The obtained two-form on the manifold
$O_\mrh$ might be degenerate, but after a natural factorization of the orbit we obtain a
symplectic manifold which is symplectomorphic to an orbit of the coadjoint representation
$Ad^*(G)$. Symplectic manifolds obtained in this way are interpreted as classical phase spaces. In
some cases, if the generator of time evolution (the Hamiltonian operator) belongs to the
generators of $U(G)$, we can obtain from the symplectic structure of \PH\ a \emn contact
structure~ on $O_\mrh$ which reproduces an `\emn extended phase space~' (odd dimensional) of
classical mechanics. If the Hamiltonian is not a generator of $U(G)$ (i.e. if $G$ is only a
`kinematical group' without representing any time evolution), the quantal dynamics might be in
some cases naturally projected onto the obtained classical phase space as a \emn globally
Hamiltonian~ \emn complete vector field~; this situation is analyzed in Section \ref{sec;3.3}.

Although such a construction of CM from QM is equally applicable to any quantal system (specified
by some $U(G)$), the interpretation of the obtained classical system depends on the specific
physical system, and also on the physical quantal state \rh\  from which the orbit $O_\mrh$ is
constructed. In any case, it is obtained a formal procedure for construction of `classical
projections' from arbitrary (finite) quantal systems.

Chapter \ref{Ch4} provides some simple examples of this formal procedure. In the subsection
\ref{4.1.6}, we obtain from a simple nonrelativistic quantal system with the potential energy $V$
the corresponding classical system (in the conventional sense) with a modified potential energy,
where the modification depends on the choice of the `initial state' $\rho\in\mPH$ (for the orbit
$O_\mrh$) and can be made arbitrarily small (in the sense of weak convergence of distributions to
the distribution $V$).

For a general time evolution, the orbits $O_\mrh$ are not invariant \wrt the quantal time
evolution, and also on various orbits of the same quantal system the projected classical
evolutions are mutually different. This brings in mind an idea of some stochastic time evolution
on a classical phase space reflecting the underlying quantal evolution.

 Such an idea is not, however, realizable for systems with finite number degrees of freedom
(briefly: finite systems) because their density matrices have not unique decomposition into convex
combinations of pure states $\mrh\in\mPH$. This is just a crude intuition which was not clearly
formulated and  realized in the following text.\footnote{Some more specific hints on this possible
classical stochastic evolutions from quantal time development could be found perhaps in
\cite{bon10}.}

\pt\label{1.1.6}\rm Quantal systems with infinite number of degrees of freedom (briefly: infinite
systems) are considered in the Chapter \ref{Ch5}. A physical motivation for such a consideration
connected with our investigation of the relations between QM and CM consists in the fact, that
'macroscopicality' and 'classicality' are almost synonyma: most of physical systems containing an
operationally well defined classical subsystem are compound of a large number (say: of the order
$10^{20}$ and more) of microscopic constituents (like atoms) and vice versa.\footnote{The
macroscopic quantal effects like superfluidity and superconductivity are additional effects
observed in these `\emn classical subsystems~' of the large quantal systems.}   Described
approximately as infinite quantal systems, these systems have some characteristic properties
distinguishing them from finite ones: the existence of nontrivial sets of `classical observables'
in given representations of observable algebra (this fact is a consequence of the existence of
various inequivalent unitary representations), the existence of quite a rich simplexes (in the
sense of Choquet) in the state space of the system allowing (in the presence of some additional
assumptions) unique decomposition of their elements into extremal elements etc. This enables us to
describe their '\emn classical subsystems~' directly in terms of the quantal description - hence
the name '\emn macroscopic limit~'. This means that, contrary to the case of finite
systems\footnote{where the quantal interpretation of classical quantities (i.e. expectation values
of generators of $U(G)$ in corresponding states) was different from the classical interpretation
(i.e. sharp values of corresponding classical generators)}, in the case of infinite systems
quantal and classical interpretations of the `macroscopic observables' coincide (at least on a
$G$-invariant subset of states): classical, resp. macroscopic quantities are represented by
operators belonging to the \emn center~ of the weak closure of the algebra of observables in some
representations.\footnote{The \emm center \bs{\mcl Z(\mfk A)}~ of a \Ca\ \fk A\ is the commutative
\Csa\ of \fk A\ consisting of all elements of \fk A,\ each commuting with all elements of \fk A:
$\mcl Z(\mfk A):=\{z\in\mfk A:z\dti x-x\dti z=0, \forall x\in\mfk A\}$.}

\pt\label{1.1.7}\rm The Chapter \ref{Ch5} is divided into two sections. In the first one we
consider the system consisting of denumerably infinite number of quantal subsystems, each of which
is described by a $G$-covariant representation of its algebra of bounded observables. To be more
specific, we consider a sequence of copies of the same finite system in the (infinite) complete
tensor product representation on a nonseparable Hilbert space $\mH_\Pi$. The representation $U(G)$
describing an elementary subsystem determines a unitary (discontinuous) representation $U_\Pi(G)$
on $\mH_\Pi$ which, in turn, determines an automorphism group $\sigma_G$ of the algebra ${\mfk
A}^\Pi$ of quasilocal observables of the infinite system. A natural definition of a classical
subsystem of the large quantal system (${\mfk A}^\Pi;\ \sigma_G$) appearing in this case can be
extended to the case of arbitrary systems ($\mfk A; \sigma_G$), as it is shown in Sec.5.2. The
arising classical (macroscopic) subsystem ($\mfk M; \sigma_G$) is naturally mapped into the
classical Poisson system ($G^*; Ad^*(G)$), or to its generalizations. \

\pt\label{1.1.8}\rm Chapter \ref{Ch6} is devoted to an application of Sec's 5.1 and 5.2:

It is shown that `mean-field' type time evolutions can be determined on a large quantal system
$(\mfkA; \msg_G)$ by specification of a Hamiltonian dynamics of a classical (macroscopic ) Poisson
system - the macroscopic limit of $(\mfkA; \msg_G)$.

This is a perhaps simplest example of (infinite-)long-range interactions in many body systems. The
correspondence between classical and quantum descriptions of systems appear there `\emm
selfconsistently~': The quantum theory of the entering `elementary subsystems' is built `on the
background' of the classical `environment' what is compound of the infinite collection of those
`elementary subsystems'.\footnote{Ideas of this kind could, perhaps, reconcile the basic idea of
Niels Bohr \cite{bohr3,bohr4} on fundamental role of a ``classical background'' in formulations of
QM with the postulate that QM is {\em the} basic theory.} The dynamics of a general class of such
systems is described in Section \ref{sec;6.3}, and the statistical thermodynamics of equilibrium
states is introduced in Section \ref{sec;6.4}.

 A slightly alternative approach to these quantum mean-field theories is described in the
papers \cite{bon1,bon2}.

\pt\label{1.1.9}\rm Finally, the Chapter \ref{Ch7} contains four exactly solved models of
interaction of a microscopic quantal system with a `macroscopic' one. Due to this interaction the
macroscopic quantal system changes its classical state to a different one. Such a change of a
macroscopic (classical) state can be interpreted as a change of a `pointer position,  hence these
models could be considered as models of `quantum measurements' in the sense of Klaus Hepp
\cite{hp-meas}. The change of the macroscopic state is reached in the limit $t\rarw\infty$ of
infinite time, and the convergence in the first three models is very slow.

In the last of the described models (in Section \ref{sec;7.6}) the `macroscopic' quantal system is
described as a {\bf finite collection} of `small' quantal systems. This leads to problems with an
unambiguous definition of `macroscopic states', since it is possible (formally, in this abstract
theory) to observe interference effects between such different `macroscopic states'. To make clear
the correspondence of quantum theory with observations, it would be necessary to introduce also
quantum  models of observation apparatuses used for detection of states of such a large but finite
`macroscopic system'. Some discussion on this problem (including reports of observations of
`macroscopic interference phenomena') appeared in literature in last decades, cf. e.g.
\cite{leggett1,leggett2,lands1,lands2,lands3,breuer,breuer1}. In the model of Sec. \ref{sec;7.6},
the (large but finite) `apparatus' radiates a Fermi particle escaping to infinity and, contrary to
the other above mentioned models of Chap. 7, {\em it converges very quickly} to the final `almost
macroscopic' state.

\pt\label{1.1.10}{\bf Bibliographical notes.}

\rm The canonical symplectic structure (in the case of finite dimensions) on complex projective
Hilbert spaces is described in \cite{arn1}; in context of QM it appeared, e.g. in
\cite{berez1,cant,rowe2,bon-EQM}. Orbits of $U(G)$ in the Hilbert space were introduced in the
special case of Heisenberg group $G$ in \cite{glaub}, and in general case in \cite{klaud1,perel1}
under the name `(generalized) coherent states'. John Klauder obtained CM on such orbits (or even
on more general submanifolds of Hilbert space) from the quantal Hamilton principle restricted to
corresponding orbits (resp. to `overcomplete sets of unit vectors'), see \cite{klaud1}. The orbits
$G.\mrh$ in \PH, and the functions $\nu\mapsto f_A(\nu):= Tr(\nu A)\ (\nu\in G.\mrh)$, named (in
the case of one-dimensional \rh) `covariant symbols' by Berezin \cite{berez2} or `lower symbols'
by Simon \cite{sim1}, were used for determination of bounds for quantum partition functions (see
\cite{lieb1,sim1}), in time dependent Hartree-Fock theory \cite{rowe2}, and also for description
of specific types of unitary representations of Galilean and Poincar\'e groups \cite{ali}. Some
essential properties of generalized coherent states are described in \cite{davies}. The natural
symplectic orbits of coadjoint representations was introduced in \cite{kiril}.

A further development of these (mathematical, as well as physical) ideas is also contained in the
work \cite{bon-EQM,bon-orbit}, which contains also a nonlinear extension of the formulation of QM.
This nonlinear extension is also compared in \cite[Sec.3.6]{bon-EQM} with the Weinberg attempt
\cite{weinb} to formulate a nonlinear version of QM.

 Some of the main ideas on connections between QM and CM leading to the present work are implicitly
contained already in the classical work \cite{weyl}. The idea and techniques used for transition
to infinite systems was gained mainly from works by Haag, Hepp, Lieb, Neumann, Ruelle and others
(see e.g. \cite{haag&kast,hp+lie1,neum2,ruelle1}, and for a review compare
\cite{bra&rob,bra&rob2,emch1}). A transition to macroscopic limit (`statistical quasiclassics') is
described in \cite{berez1} for a specific choice of the group $G$ and a mean-field type
interaction. A review of works on macroscopic limits ('large N limits') is given in \cite{yaffe}.
An attempt of the description of classical quantities of large quantal systems analogous to the
here presented one is described in works by Rieckers with collaborators \cite{rieck,duf&rieck},
and by Morchio with Strocchi \cite{morch&stroc,stroc}; see also the works
\cite{unner0,unner1,unner2,unner3} by Thomas Unnerstall. A preliminary outline of a part of this
work is contained in \cite{bon8}, and also in \cite{bon1,bon2}. The necessary mathematics can be
found in the cited monographs, cf. also Appendices in \cite{bon-EQM}.

An alternative way of description of thermodynamics and dynamics of quantum mean-field systems was
later proposed in the work of the group around R.F.Werner, see e.g. \cite{d+wer1}.

A new approach to the theoretical description of classical (macroscopic) systems in the framework
of quantum theory in a unique mathematical formalism is presented in a series of papers by
Jean-Bernard Bru and collaborators \cite{bru}.

\section{Quantum Mechanics}\label{sec;1.2}

\def\autor{{
\ref{I;int}\quad Introduction}}
\def\nazov{{
\ref{motiv}\quad Motivation and Summary}}

\pt\label{1.2.1}{\rm In formal schemes of all theories considered in this work, the basic concepts
are `states', `observables' and their transformations ascribed to a considered physical system. We
shall not discuss here details of the empirical meaning of these concepts. Roughly, \emm states~
are prepared by some standard empirical procedures and represent the situation, what has to be
measured, \emm observables~ describe (equivalence classes of) measuring apparatuses (i.e. the role
of their function in the theory) giving certain empirically obtained responses if applied to
states, and transformations include time evolution of the system in given conditions as well as
various changes of equivalent descriptions of the system (\emm symmetries~).

In this section, we shall outline a simple standard scheme of the formalism of nonrelativistic
(resp. Galilean-relativistic) quantum mechanics of finite systems (QM), i.e. the nonrelativistic
view on physical systems containing only finite number of their further indecomposable elementary
constituents (particles, spins,\dots).}\footnote{The concepts of ``system'', and ``physical
system'' are taken here to be as intuitively clear.}

\pt\label{1.2.2}\rm {\bf Observables:} A separable complex Hilbert space \H\ corresponds to any
physical system in QM. Let \LH\ denote the set of all bounded linear operators from \H\ to \H,
where the boundedness (equiv. continuity) is defined \wrt the norm of \H\ coming from the scalar
product $(x,y),\ (x,y\in\mH)$, which is linear in the second factor $y$. \emm Observables in QM~
(i.e. physical quantities empirically identifiable by some realizable(?) measuring devices) are
represented by selfadjoint operators on \H\ (in general unbounded). It is useful to consider along
with any selfadjoint operator $A$ (corresponding to an equally denoted observable $A$) its
spectral measure $E_A$ defined on Borel subsets of the real line \bR\ with values in projectors
$E_A(\bullet)$ in \LH, $E_A(\mbR) = I := id_\mH$ (:= the identity of the algebra \LH), cf.
\cite[Appendices B\,\&\! C]{bon-EQM}.\

It is important to stress here, that in the conventional QM of finite systems (atoms, molecules,
and finite collections of them) the set of observables contains {\bf the whole set \LH\ of
operators} representing these observables. Hence, the algebra \LH\ acts on \H\ by the \emm
irreducible~ manner (i.e. no nontrivial subspace of \H\ is by the actions of the whole \LH\ left
invariant). This also implies the impossibility, resp. inadequacy, of interpretation of the
``mixed states'' as representing some statistical mixture of systems occurring in the states
decomposing the corresponding ``mixture'' (cf. \ref{1.2.3}) in this QM of finite systems.

\pt\label{1.2.3}{\rm  \emm States~ in QM are conventionally
represented by \emn{density matrices}~, i.e. {\bf positive} \emn
trace class operators~ \rh\ on \H\ {\bf with unit trace} (=the
trace norm): $Tr(\mrh)=1$. Density matrices form a convex subset
in the \emn linear space \TH~\ of all trace class operators which
is closed in the \emm trace norm~ $\|A\|_1\ :=\ Tr\sqrt{A^*A}$.
Denote this \emm set of states \Ss~. The \emm extreme points of
\Ss ~\ are represented by the one-dimensional orthogonal
projectors $P_{{\it x}}\in\mLH$\ (projecting \H  onto
one-dimensional subspaces $\bx$\ containing {\it x},\ $ 0\neq{\it
x}\in\mH$). Any $\mrh\in\mSs$ can be expressed as a weak limit of
finite convex combinations of elements $P_j\in \mPH\  := \{P_{{\it
x}} : {\it x}\in\mH,{\it x}\neq 0\}$ of the \emm projective
Hilbert space \PH~. We can write

\bequ\label{1.2.3(1)} \mrh=\sum_j\lambda_jP_j,\quad \sum_j\lambda_j=1,\ \lambda_j\geq 0.
 \enqu
The states from \PH\ are called \emm pure states~. The decomposition\rref{1.2.3(1)}~ of an
arbitrary state \rh\ into pure states is highly nonunique if \rh\ does not belong to \PH, hence
the state-space \Ss\ is not a \emm simplex~, cf. \cite{choquet}, what is an important difference
\wrt  classical mechanics. This have important consequences for interpretation of the `mixed
states' described by density matrices $\mrh\not\in \mPH$: The nonunique decompositions\rref
1.2.3(1)~ show that {\bf these quantum states cannot be interpreted as representations of
statistical ensembles} each element of which (i.e. a copy of the considered physical system)
occurs in a definite pure state, because pure states appearing in certain mutually different
decompositions of the same density matrix are in general incompatible, i.e. they are eigenstates
of mutually noncommuting (hence simultaneously nonmeasurable) observables, cf.
\cite[1.5-b]{bon-EQM}.\footnote{\label{PRL}This point was important also in the discussion about
(im-)possibility of deducing the linearity of QM-time evolutions from mere quantal kinematics
together with the so called ``No-Signaling Condition'', cf. \cite{bon-PRL}.} We have $\mrh\in\mPH$
iff $\mrh^2 = \mrh$ and $\mrh\in\mSs$.}\

\pt\label{1.2.4}\rm Quantum theories are `intrinsically (or irreducibly) statistical', i.e.
experimentally verifiable assertions can be expressed in general in terms of probabilities only in
the frame of these theories. Results of repeated measurements of a given quantity (observable)
applied to the same state (which should be, however, repeatedly prepared for each single
measurement because of its unavoidable disturbance by the interaction with the measuring
apparatuses) have a nonzero dispersion for a general quantity. The \emm expectation value~ of
measured values of a given bounded observable (represented by the operator) $A=A^*\in\mLH$ in the
state (represented by the density matrix) $\mrh\in\mSs$ is in QM expressed by
 \bequ\label{1.2.4(2)}
 \mom_\mrh(A) := Tr(\mrh A).
\enqu

\noidt $\mom_\mrh$ can be considered here as a positive linear functional on \LH, which is \emm
normalized~ (i.e. $\mom_\mrh(I_\mH)=1$) and \emm normal~ (i.e. ultraweakly continuous), compare,
e.g. \cite{bra&rob,bra&rob2,sak1}; the set of all such functionals \om\ might be identified with
\Ss : to each \om\ corresponds a unique density matrix $\mrh =: \mrh_\mom$, for which $\mom =
\mom_\mrh$\ according to\rref 1.2.4(2)~. For an arbitrary selfadjoint (not necessarily bounded)
operator $A$, the probability of obtaining of its value in a Borel set $B\subset\mbR$, if measured
in the state $\mom\in\mSs$, is
\[ \centerline{ \emn$\mom(E_A(B))$~.}\]

Here $E_A: B\mapsto E_A(B),$ is the unique \emn projector valued measure~ of $A$, or its \emn
spectral measure~, characterizing any selfadjoint operator $A$, \cite[B\,\&\! C]{bon-EQM}. We
shall define also

\bequ\label{1.2.4(3)}\mom(A):=\int_\mbR\lambda\ \mom(E_A(\rd\lambda)) \enqu

\noidt if the integral converges absolutely. This is a generalization, resp. an alternative form
of\rref 1.2.4(2)~. If $\mom_x\in\mSs$ corresponds to $P_x\in\mPH$ and for a given $A=A^*$ the
quantity $\mom_x(A^2)$ is defined (i.e. is finite), then $x\in D(A)$\ (:= the domain of $A$), and
vice versa.\footnote{Let us remember here that no unbounded symmetric linear operator $A$ acting
on a Hilbert space \H\ can be defined on the whole space $\mH:\ D(A)\subsetneqq\mH$.}\

\pt\label{1.2.5}\rm Any observable $A$ determines a strongly continuous \emn one-parameter group~
$t\mapsto\exp(-itA)$ of unitary transformations of \H\ of which $A$ is its \emm generator~. This
induces a \emn weakly*-continuous ($\equiv$ w$^*$-continuous) group $\tau^A$~ of \emn
*-automorphisms~ of the von Neumann algebra \LH\ (cf. \cite[B.2.1(v)]{bon-EQM}),
$B\mapsto\tau^A_t(B):=e^{itA}Be^{-itA},\ B\in\mLH,\ t\in\mbR$, i.e. the functions

\bequ\label{1.2.5(4)} t\mapsto\mome(\tau^A_tB) := \mome(e^{itA}Be^{-itA}) \enqu

\noidt are continuous for all $B\in\mLH$ and all $\mom\in\mSs$. The observable $A$ represents in
this way a one-parameter group of {\bf symmetries} of the physical system. Conversely, any
w$^*$-continuous one-parameter group of *-automorphisms of \LH\ is given by an observable
(determined up to an arbitrary additive real constant) in the above described manner (see e.g.
\cite[Example 3.2.35]{bra&rob}). If $A$ is bounded, $t\mapsto\exp(-itA)$ is norm-continuous.

\pt\label{1.2.6}{\rm To obtain an empirical meaning of the formal scheme outlined above, it is
necessary to specify how to measure quantities corresponding to specific operators. As far as the
present author knows, this type of \emn interpretation~ for arbitrary selfadjoint operators was
not realized for any physical system (except, perhaps, of some systems consisting of spins only).
It might be, however, sufficient to ascribe a certain empirical meaning to `sufficiently many'
operators. We can use, for such an identifcation of operators and empirical manipulations, the
above mentioned connection between one-parameter groups of automorphisms $\tau^A$ and operators
$A$. We shall take into account, moreover, that also 'microscopic systems' described adequately in
the framework of quantum mechanics are only empirically specified by manipulations with
'macroscopic bodies', which are well described by CM. Let a physical system preserve its identity
if the surrounding macroscopic bodies undergo some group of motions. Then we obtain a group of
symmetry transformations of that system.\footnote{This is so called ``passive symmetry
transformation'', contrasted to the ``active'' one, when the `physical system' is moved in the
fixed environment; these two ways of understanding of transformations applied to a system are
mathematically equivalent.} To any one-parameter subgroup of such 'macroscopically determined'
transformations corresponds in our formalism a selfadjoint operator, which in turn corresponds in
some way (we shall not specify it here) to a measurable quantity connected with the macroscopic
motions. We shall assume (and this is really fulfilled for many finite systems) that the group $G$
obtained in this way is large enough to determine all the 'basic observables'; all the other
observables are supposed to be functions of these basic ones (see the following subsections).} \

\pt\label{1.2.7}{\rm We shall assume that a w*-continuous representation \sg\ of a connected Lie
group $G$ in the group of *-automorphisms of \LH\ is given and that the group $\{\msg_g \in
\maut\mLH: g\in G\}$ acts on \LH\ irreducibly: there is no nontrivial von Neumann subalgebra of
\LH\ which is left invariant by the all $\msg_g\ (g\in G)$. One-parameter subgroups of $G$ are in
bijective correspondence with elements $\xi$ of the Lie algebra \fk{g}\ of $G$ to which, in turn,
correspond selfadjoint generators $X_\xi$ of unitary groups determined by $\msg_{exp(t\xi)}$, cf.
\cite[A.4.8]{bon-EQM}.

Since the unitary operators $U(g)$ determined by automorphisms $\msg_g\ (g\in G)$\ via the
relation

\bequ\label{1.2.7(1)} U(g)^* B U(g) = \msg_g(B),\quad \forall B\in \mLH \enqu

\noidt are only defined up to a phase factor, in general case, the representation \sg\ leads only
to a \emm projective representation~ $g\mapsto U(g)$ of $G$\ in the unitary group of \H, i.e.

\bequ\label{1.2.7(2)} U(g_1g_2) = m(g_1,g_2) U(g_1) U(g_2),\enqu

\noidt where $m: G\times G\mapsto S^1$\ (:= the complex numbers of unit modulus) is a \emm
multiplier~ of the projective representation, cf. \cite[3.3.6]{bon-EQM}. Such a representation can
be always extended to a unitary representation of a group $G_m$, which is the \emm central
extension~  of $G$ \cite[§15.2, Thm. 1]{kiril} by the multiplicative group $S^1$ corresponding to
the multiplier $m$, \cite[1.5-c]{bon-EQM}. The group multiplication in $G_m$ (which can be
identified, as a set, with $G\times S^1)$\ is

\bequ\label{1.2.7(3)}(g_1;\lambda_1)(g_2;\lambda_2)= ( g_lg_2; m(g_1,g_2)\lambda_1\lambda_2)
,\quad \lambda_j\in S^1. \enqu

In the unitary extension of the projective representation $U(G)$ the elements of the center of
$G_m$ are represented by the numbers from $S^1$ ('phase factors') acting by multiplication of the
vectors $x \in\mH$. All the extensions $G_m$ of G (corresponding to various multipliers $m$) are
classified by the second cohomology group $H^2(G,S^1)$ of the group $G$ with values in $S^1$, for
details see \cite{kiril,varad}. We shall assume that the unitary representation $U(G_m)$
corresponding to the representation \sg\ of $G$ according to\rref{1.2.7(1)}~ can be (and really
is) chosen strongly continuous. In the following we shall usually write $G$ instead of $G_m$.

A natural consequence of irreducibility of \sg\ is the irreducibility of corresponding unitary
representation $U$. Hence, the weak-operator closure of the linear hull of the subset $\{U(g): g
\in G\}$ of \LH\ in the von Neumann algebra \LH\ is \LH\ itself.}\

\pt\label{1.2.8}{\rm The interpretation of $G$ as a group of (empirically defined) physical
symmetries of the system leads to a natural interpretation of generators $X_\xi\ (\xi\in G)$ of
the unitary representation $U$. Since any bounded operator is weakly approximated by linear
combinations $\sum\lambda_j U(g_j)$ we can hope to obtain some insight into possible
interpretations of other operators. The complete answer to this problem of interpretation needs,
probably, an analysis of possible interactions of the system under consideration with all other
systems, or, at least with systems which could be used in the role of measuring instruments. The
choice of $G$ together with (eventually) some other assumptions on the physical properties of the
system (e.g. the value of spin) might also determine the dimension of \H.

The proper choice of the representation of $G$\ depends on comparison of consequences of the
chosen 'interpretation $U$' with empirical data; this step contains, e.g. the choice of the
correct value of the Planck constant, if $G$ is the Heisenberg group (i.e. a \emn central
extension~ of the classical phase space $\mbR^{2n}$ considered as the commutative group of
translations).}\

\pt\label{1.2.9}\rm It will be further assumed that the time evolution of the system is either a
one-parameter subgroup of $G$, or it is separately defined as a one-parameter w*-continuous
subgroup \t\ of the group of *-automorphisms of \LH\: $t\mapsto \mt_t\in *\= aut(\mLH),\ \mt_{t+u}
= \mt_t\circ\mt_u\ (t,u\in\mbR),\ \mt_o:=$ identity. Note that for each automorphism $\malp\in *\=
aut(\mLH)$ there is some unitary $U_\malp\in\mLH$ such that for all $A\in\mLH:\ \malp(A)\equiv
U_\malp A\, U_\malp^*$, i.e. the automorphisms of \LH\ are \emn inner automorphisms~, cf. e.g.
\cite[Corollary 2.9.32]{sak1}.

\section{Classical Hamiltonian mechanics}\label{sec;1.3}

\pt\label{1.3.1}\rm  In this section, we shall outline the formal scheme of classical Hamiltonian
mechanics (CM) parallel to the exposition of QM in the preceding section. We shall restrict our
considerations to the case of systems with finite number of degrees of freedom. We shall use the
language of \emn differential geometry~ (for pedagogically well written course of differential
geometry we refer to \cite{fecko1}). A technically more complicated quantum theory of systems with
infinite number of degrees of freedom will be described later. For classical theory of infinite
systems, i.e. \emn classical field theory~, see the corresponding monographs, or also e.g.
\cite[II.5.5]{abr&mars}, \cite[Append.2]{arn1}, \cite{bon-EQM}.

\pt\label{1.3.2}\rm To any physical system there corresponds in CM a \emm symplectic manifold~
$(M;\mOme)$ (cf. \cite{arn1,abr&mars,kob&nom}). $M$ is here an (even dimensional) infinitely
differentiable Hausdorff second countable connected \emn manifold modeled by~ $\mbR^{2n}$ and
\Ome\ is a nondegenerate closed \emn two-form on $M$~, the \emm symplectic form~, cf. also
\cite[A.3]{bon-EQM}. Observables in CM are represented by real-valued functions $f$ on $M$; for
technical convenience, we shall assume usually $f$ to be infinitely differentiable, $f\in
C^\infty(M,\mbR).$ These observables constitute a real associative algebra $\mfk{F}(M)$ with
respect to the ordinary multiplication of functions: $f.g(x) := f(x)g(x)\ (f,g \in \mfk{F}(M),\
x\in M)$. This algebra has the natural complexification $\mfk{F}_\mbC(M)$.

\nl {\it Remark}: In the larger algebra $\mfk B(M)$ of all bounded Borel functions on $M$ we can
associate to any $f \in \mfk B(M)$ the projector-valued measure $E_f$ defined on Borel subsets of
\bR\ (for a real-valued $f$):

 $E_f:\ B \mapsto\chi_{f^{-1}(B)}$\ for any Borel $B\subset\mbR,$ where $\chi_N$ is the
 characteristic function of the Borel subset $N\subset M$.

 It is clear that $E_f(B) := \chi_{f^{-l}(B)}$  are
projectors in $\mfk B(M)$ and the association $B\mapsto E_f(B)$ is \sg-additive, with $E_f(\mbR) =
\chi_M$ = the unit element of $\mfk B(M)$. The real-valued Borel functions $f$ can be also
considered as selfadjoint operators on a Hilbert space $\mH:= L^2(\mbR,\mu)$ acting as the
multiplication operators, and $E_f$'s are their canonical spectral measures.

\pt\label{1.3.3}\rm \emm States~ in CM are probability  Borel measures $\mu$\ on $M,$ which form a
convex set ${\cal S}_{cl}$\ with extremal points consisting of all measures concentrated at
one-point sets in $M$, i.e. of all \emn Dirac measures~ on $M.$ Hence, \emn pure states~ are
identified with points $x \in M.$\ Any measure $\mu \in {\cal S}_{cl}$ has a {\em unique
decomposition} into (an integral of) Dirac measures, i.e. it is a \emm simplex~, contrary to the
state space of QM. This has serious consequences for different possibilities of statistical
interpretations of states in CM and  QM, cf. \ref{1.2.3}, also footnote \ref{PRL}.\

\pt\label{1.3.4}\rm According to CM, the disturbance of the state connected with the measurement
of arbitrary observables can be made negligibly small. Because of uniqueness of the decomposition
of an arbitrary state to its extremal components we can interpret any $\mu\in{\cal S}_{cl}$ as a
representative of a statistical ensemble of a large number of copies of the considered system,
each being in an (its own) pure state. Repeated measurements on the state $\mu$\ have to be
understood now as a repeated random choice (with probability corresponding to the probability
measure $\mu$\ on $M$) from the ensemble of a system appearing in a pure state $x\in M$ and
measuring precise values $f(x)$\ of observables $f\in\frak F(M)$ afterwards. For such a
measurement procedure the probability of finding the value of an observable $f$ in a Borel set
$B\subset\mbR$ is $\mu(E_f(B))$ (compare Remark in \ref{1.3.2}), where $\mu(f)$ for $f\in\mfk
B(M)$ means the integral of $f$ with the measure $\mu$ on $M$. The value $\mu(f)$ for $f\in\mfk
F(M)$ is then the expectation value of $f$ in the state $\mu\in{\cal S}_{cl}$. The mapping $\mu:
f\mapsto\mu(f)$ is a positive normalized linear functional on $\mfk F(M)$ (and also on $\mfk
B(M)$), which is continuous \wrt  the usual sup-norm on $\mfk B(M)$. Better continuity properties
have, e.g. functionals $\mu$\ which are absolutely continuous (as measures) \wrt  the natural
measure $\mOme^n$ on the \emn symplectic manifold~ $(M;\mOme)$.

\pt\label{1.3.5}{\rm A \emm symmetry~ of a system in CM is defined as a \emm symplectomorphism~
$F$ of $(M;\mOme)$, i.e. $F$ is such a diffeomorphism of $M$ onto itself which leaves the
symplectic form \Ome\ unchanged: $F^*\mOme = \mOme$, where $F^*$ is the pull-back on $M$, see e.g.
\cite{abr&mars}, or also \cite[A]{bon-EQM}. For $f\in\mfk{F}(M)$ let $F^*f:=f\circ F$; such an
action of $F$ onto the algebra $\mfk{F}(M)$ is an automorphism. It conserves, moreover, another
structure on $\mfk{F}(M)$ - the \emn Poisson algebra structure~ defined below.

Let $t\mapsto F_t$ be a one-parameter group of symmetries, which is differentiable \wrt
$t\in\mbR: F_{t+s}=F_t\circ F_s\ (t,s,\in\mbR)$ and the derivative

\bequ\label{1.3.5(1)}\left. \frac{\rd}{\rd t}\right|_{t=0} f(F_tx) =: \rd_xf(\msg_F) \enqu

\noidt exists for all $f\in\mfk{F}(M), \forall x\in M$, and the functions

\bequ\label{1.3.5(2)} \rd f(\msg_F) : x \mapsto \rd_xf(\msg_F)\ (\in\mbR) \enqu

\noidt are infinitely differentiable, $\rd f(\msg_F)\in\mfk F(M)$.
 Here $\msg_F$ is the vector
field on $M$ corresponding to the flow $x\mapsto F_tx,\ (x\in M)$. Let $\mfk X(M)$ be the set of
all infinitely differentiable vector fields on $M$. Let $i(\msg)\mOme$ be the one-form on $M$
defined by: $ i(\msg)\mOme(\varphi) := \mOme(\msg,\varphi)$ for any $\msg,\varphi\in\mfk X(M)$,
i.e. $i(\msg)\mOme$ is the \emm inner product~ \cite[A.3.10]{bon-EQM} of the vector field \sg\
with the two-form \Ome. For the vector field $\msg_F$ we have:

\bequ\label{1.3.5(3)}\rd i(\msg_F)\mOme = 0.\enqu

 Vector fields $\msg_F$ and the corresponding flows of
symplectomorphisms $F_t$ are called \emm locally Hamiltonian~. If there is $f_F\in\mfk F(M)$ such
that

\bequ\label{1.3.5(4)} i(\msg_F)\mOme = - \rd f_F\ \text{on}\ M,\enqu

\noidt then $\msg_F$ is (globally) Hamiltonian and $f_F$ is its \emm Hamiltonian
 function~. To any $f\in\mfk F(M)$, we can unambiguously define a
 \emm Hamiltonian vector field~ $\msg_f$\ with the Hamiltonian function $f$\ by the formula

 \bequ\label{1.3.5(5)} i(\msg_f)\mOme = - \rd f.\enqu

 Uniqueness of $\msg_f$ is a consequence of nondegeneracy of
\Ome. Two functions $f,g\in\mfk F(M)$ give the same vector field $\msg_f=\msg_g$ iff $f-g =
const.$ We can introduce now a \emn Lie algebra structure~ into $\mfk F(M),$ the structure of \emm
Poisson bracket~ multiplication: $(f; g)\mapsto\{f,g\}\in\mfk F(M)$ for all $f,g \in\mfk F(M).$ We
define

\bequ\label{1.3.5(6)} \{f,g\} := \mOme(\msg_f,\msg_g),\enqu

\noidt where $\msg_f$ (resp.$\msg_g$) is given in\rref 1.3.5(5)~. If we denote by $\mL\msg~,\
\msg\in\mfk X(M),$ the \emm Lie derivative~ \cite[A.3.7,A.3.8]{bon-EQM} in the direction of \sg\
of tensor fields on $M$ ($\mL\msg~$ acting on the differential forms has the expression
$\mL\msg~=i(\msg)\rd+\rd i(\msg)$), then, according to\rref 1.3.5(5)~:

\bequ\label{1.3.5(7)} \{f,g\}=\mL\msg_f~g=-\mL\msg_g~f.\enqu

The properties of \Ome\ are reflected in the following properties of the Poisson bracket:\nl

$\begin{array}{l} (i)\ \ \quad \{f,g+\lambda
h\}=\{f,g\}+\lambda\{f,h\}, \\
(ii)\ \quad \{f,g\}=-\{g,f\},
(bilinearity\ and\ antisymmetry\ of\ \mOme), \\
(iii)\quad \{f,\{g,h\}\}+\{g,\{h,f\}\}+\{h,\{f,g\}\}=0, (closedness\ \rd\mOme=0),\\
(iv)\ \quad \{f,g\cdot h\}=\{f,g\}\cdot h+g\cdot \{f,h\}, (derivation\ property\
\rref 1.3.5(7)~), \\
(v)\ \ \quad If\ \{f,g\}=0\quad \forall g\in\mfk F(M)\Rightarrow f\equiv const.\ (nondegeneracy\
of\ \mOme). \end{array}$

\nl It is not difficult to prove for the commutator of Hamiltonian vector fields:}

\bequ\label{1.3.5(8)} [\msg_f,\msg_g]=\msg_{\{f,g\}}.\enqu

\pt\label{1.3.6} {\rm In CM, all the observables are functions of points $x\in M,$ hence locally
can be expressed as functions of a finite number $2n$\ coordinate functions. In accordance with
the `philosophy' of \ref{1.2.6}, we shall look for an interpretation of a finite number of
observables which contain systems of coordinate functions for a neighbourhood of any point of $M$.
This can be naturally done, if $M$ is a homogeneous space of a connected \emn Lie group~ $G$ (cf.
\cite[A.4]{bon-EQM}) corresponding to a group of empirical manipulations with objects relevant to
the determination of the considered system. Since the symplectic structure \Ome\ on $M$ reflects
important physical properties of many physical systems, it is desirable for the group action on
$M$ to conserve this structure. In this case, one parameter subgroups of symmetries correspond to
Hamiltonian flows which can be physically interpreted.}

 \pt\label{1.3.7}{\rm From now on, we
shall assume that $(M;\mOme)$ is a \emn homogeneous space~ of a connected \emn Lie group~ $G$, on
which the group $G$ acts as an infinitely differentiable group of symplectomorphisms $F_g\ (g\in
G)$: $F^*_g\mOme=\mOme,\ F_{gh}=F_g\circ F_h\ (g,h,\in G),$\ and functions $g\mapsto f(F_gx)$ are
in $C^\infty(G,\mbR)$ for all $f\in\mfk F(M)$\ and all $x\in M.$ If $e\in G$ is the unit element
of $G$, then $F_e:= id_M.$ To any $\xi\in\mfk g$\ (:=  the \emm Lie algebra of $G$~)\ there is a
one-parameter group of symplectomorphisms $t\mapsto F_{\exp(t\xi)}$ of $M$ generated by the vector
field $\msg_\xi$\ (compare with\rref 1.3.5(1)~). If $[\xi,\eta]$ denotes the \emm commutator in
\fk g~, and $[\msg_\xi,\msg_\eta]\in\mfk X(M)$ the \emn commutator of vector fields~ on $M$, then
(see \cite[Proposition 4.1.26]{abr&mars})

\bequ\label{1.3.7(1)} [\msg_\xi,\msg_\eta]=-\msg_{[\xi,\eta]}. \enqu

Every homogeneous symplectic manifold has universal covering symplectic homogeneous manifold \wrt\
the universal covering group of $G$. On any simply connected homogeneous symplectic manifold of a
connected Lie group $G$, the functions \emn$f_\xi\ (\xi\in\mfk g)$~ determined up to additive
constants by the formula

\bequ\label{1.3.7(2)} i(\msg_\xi)\mOme = -\rd f_\xi \enqu

\noidt are defined globally on the manifold $M,\ f_\xi\in\mfk F(M),$ i.e. the vector fields
$\msg_\xi\ (\xi\in\mfk g)$ are globally Hamiltonian. We shall assume that this is the case for our
$(M;\mOme)$. Then arbitrary additive constants in the definitions of $f_\xi$'s can be chosen such
that the mapping $\xi\mapsto f_\xi$ from $\mfk g$ to \fk F(M) will be linear. Then

\bequ\label{1.3.7(3)} \{f_\xi,f_\eta\} = -f_{[\xi,\eta]}+ C(\xi,\eta), \enqu

\noidt where $C$ is a bilinear antisymmetric mapping from $\mfk g\times\mfk g$ to real constants
on $M$ called a \emm two-cocycle on \fk g~ with values in \bR. Any change of constants in
$f_\xi$'s (conserving the linearity of $\xi\mapsto f_\xi$) leads to an \emn equivalent cocycle~
$C'(\xi,\eta)= C(\xi,\eta)+ a([\xi,\eta]),$ where $a\in\mfk g^*$ (:= the dual of \fk g).
Equivalence classes of two-cocycles form the commutative (additive)
\[ \centerline{\emn 2-cohomology group $H^2(\mfk g,\mbR)$~} \]

\noidt of \fk g with values in \bR. This group is isomorphic to $H^2(G,S^1)$ if $G$ is simply
connected (compare \cite[Chap. 10.4.]{varad}). This isomorphism determines a canonical bijection
between classes of irreducible projective representations and symplectic transitive actions of a
simply connected Lie group. This bijection associates the class of all representations
corresponding to the given (similarity class of a) multiplier with the class of symplectic actions
with the corresponding (equivalence class of a) cocycle, compare also \cite{b-yc,haag&kast}. If
the \emn multiplier $m$ corresponds to the cocycle~ $C$ from\rref 1.3.7(3)~, then the \emn central
extension $G_m$~ of $G$ (cf. \cite[§15.2, Thm. 1]{kiril}) acts on $M$ in such a way, that

\bequ\label{1.3.7(4)} \{f_\xi,f_\eta\}= - f_{[\xi,\eta]}\quad\text{for\ all}\ \xi,\eta\in \mfk
g_m, \enqu

\noidt if the added vector fields act on $M$ trivially and constants in $f_\xi$'s are properly
chosen. If the action of $G$ on $M$ satisfies\rref 1.3.7(3)~ with $C\equiv 0,$ then it is called a
\emm Poisson action~ \cite{arn1}, and the symplectic manifold $M$ is called \emn exactly
homogeneous~ \cite{kiril}.}\

\pt\label{1.3.8}{\rm Any observable $f\in\mfk F(M)$ on the homogeneous symplectic manifold $M$
with globally defined Hamiltonian functions $f_\xi\ (\xi\in\mfk g)$ can be expressed as a function
of the 'basic observables $f_\xi$'. Hence measurement of any $f\in\mfk F(M)$ can be reduced to the
measurements of $f_\xi$'s. This does not make easier, however, of an ascribing a direct physical
(i.e. empirical) interpretation to an arbitrary $f\in\mfk F(M)$ and the situation is  similar to
that one of QM, see \ref{1.2.8}.}

\pt\label{1.3.9}{\rm A time evolution on $(M;\mOme)$ is defined in CM as a differentiable
one-parameter group of symplectomorphisms with a globally defined Hamiltonian function $h\in\mfk
F(M)$. This one-parameter group might be either a subgroup of $G$, or it is separately defined. In
each case the group $G$ might contain an invariance subgroup of $h$ - the symmetry group of the
dynamics (determining integrals of motion - conservation laws).}

\section{Quantum theory of large systems}\label{sec;1.4}

\pt\label{1.4.1}{\rm Models of systems with infinite number of degrees of freedom enter to quantum
theory when we want to describe either processes accompanied with changes of numbers of particles
(resp. quasiparticles) present in the physical system (what also occurs each time if we try to
describe quantal analogues of classical continuous media, resp. fields), or systems with actual
infinity of particles (the `thermodynamic limit' necessary e.g. for clear conceptual description
and abstract investigation of phase transitions). In standard models of infinite systems in
quantum theory the algebras of bounded observables (e.g, CCR or CAR algebras for infinite number
of degrees of freedom or algebras of spin systems on infinite lattices) have many mutually
unitarily inequivalent physically relevant representations as algebras of bounded operators in
some Hilbert spaces. These inequivalent representations might correspond e.g. to various states on
the algebra of observables representing situations with various values of some macroscopic--global
parameters of the large syatem. It often happens, moreover, that for description of some processes
(time evolution, symmetry transformations), we are not able to work in the framework of only one
(even faithful) representation. It is, consequently, useful to formulate theoretical scheme for
the quantum theory of large systems (QTLS) in a representation independent, algebraic language. As
basic sources of most of the here necessary mathematics and its application to description of
large quantal systems could be taken, e.g.
\cite{bra&rob,bra&rob2,davies,pedersen,najm,sak1,sewell,sewell3}; a very brief summary can be
found also in \cite[Sec.3.4]{bon-EQM}.}\

\pt\label{1.4.2}{\rm A \emm \Ca\ \fk A~\ \cite[B.2]{bon-EQM} (details on \Ca s can be found in
\cite{dix1,dix2,sak1,sak2,pedersen,tomita,takesI,bra&rob,bra&rob2,emch1}) corresponds to any
physical system in QTLS. \fk A\ is a Banach algebra over complex numbers with involution $x\mapsto
x^*, x\in\mfk A,$ and with special ($C^*$) property. This means that it is a norm-closed linear
space endowed with associative and distributive multiplication, and for any $x,y\in\mfk A,\
\lambda\in\mbC,$\ and with $\|x\|\geq 0$ - the norm of $x\in\mfk A$, it is: $
\|xy\|\leq\|x\|\cdot\|y\|,$\ the \emm involution~ $x\mapsto x^*$ is antilinear: $(x+\lambda
y)^*=x^*+\overline{\lambda}y^*$,\ where $\overline{\lambda}$\ is the complex conjugate of
$\lambda$,\ with $(xy)^*=y^*x^*$,\ $\|x^*\|=\|x\|\ (=0\ \text{iff}\ x=0),$ and the \emm
$C^*$-property~\ means:\ $\|x^*x\|=\|x\|^2,\ \forall x\in\mfk A$. \fk A\ is called \emm unital
\Ca~ if it contains unit element $e\in\mfk A: ex=xe=x,\ \forall x\in\mfk A$. Selfadjoint elements
$x = x^*\in\mfk A$\ represent bounded \emm observables~ of the system. The algebra \fk A\ is the
\emm algebra of observables~ of the system. For many interesting systems, \fk A\ is constructed as
a \emm $C^*$-inductive limit~ of a \emn net of local algebras~ of finite (sub)systems (see
\cite{bra&rob,emch1,sak1}, and specifically \cite[1.23]{sak1}); in this case, these finite systems
are interpreted e.g. as systems located in bounded space (-time) regions. Quasilocal algebras used
in QTLS have such a structure (see \cite[Definition 2.6.3]{bra&rob}). It will be useful in our
considerations to connect the \emn quasilocal structure of \fk A~\ with an action of a (usually
abelian) group \emn$\Pi$~ ({\bf $\Pi$ is an infinite set} - it might be a locally compact
noncompact group) on \fk A:\ For any $p\in\Pi$ let $\pi(p)\in$\aut{\fkA},
$\pi(p_1p_2)=\pi(p_1)\pi(p_2)\ (p_1,p_2\in\Pi)$. Let $\Pi$ act transitively on a noncompact
locally compact space $V$ and let to any bounded open subset $v\subset V$ (denote the set of all
such subsets by $\mcl B(V)$) corresponds a \Csa\ $\mfk A_v$ of \fk A, the \emm local subalgebra of
\fk A~ corresponding to $v\subset V$. If $v_1\subset v_2\subset V$, then $\mfk A_{v_1}\subset\mfk
A_{v_2}$. All the $\mfk A_v\ (v\in\mcl B(V))$\ have common unit $\equiv$ the unit $e:={\rm
id}_{\mfk A}$ of \fk A, and

\bequ\label{1.4.2(1)} \overline{\bigcup_{v\in\mcl B(V)}\mfk A_v} = \mfk A,\enqu

\noidt where the over-bar denotes the uniform closure. We assume further that $\pi(p)(\mfk
A_v)=\mfk A_{p\cdot v}$, where $p\cdot v:=\{\lambda'\in V: \lambda'=p\cdot\lambda, \lambda\in
v\}$, and $p\cdot\lambda$ denotes the action of $p\in\Pi$ on the point $\lambda\in V$. This action
is supposed continuous and bounded: $p\cdot\mcl B(V)\subset\mcl B(V)$ for all $p\in\Pi.$ We can
assume (for simplicity) that for mutually disjoint $v,u\in\mcl B(V),\ v\cap u=\emptyset,$ we have

\bequ\label{1.4.2(2)} [x,y]=0,\  \text{for all}\ x\in\mfk A_v,\  y\in\mfk A_u.\enqu

\noidt (The anticommutativity of Fermi systems can also be included, cf. \cite[Sec.
2.6]{bra&rob}).\footnote{Our formalism is built for the nonrelativistic situations. If the space
$V$ was the Minkowski space and our considerations were Einstein-Lorentz--relativistic, the
condition for the commutativity in\rref 1.4.2(2)~ would be the space--like separation instead of
the disjointness of the domains $u,v\subset V$.} We shall characterize this situation by saying
that the algebra \fk A is \emm quasilocal \wrt the action~ of the group $\Pi$. We shall use
another technical assumption, that all the local subalgebras $\mfk A_v$ are \Wa s: A \emm \Wa~ \fk
A is such a \Ca\ which is (isomorphic to a) Banach space topological dual of another B-space $\mfk
A_*$ called the \emm predual of \fk A~; such an \fk A is always unital and generated by its
projectors. \Wa s were introduced originally as weakly closed symmetric subalgebras of bounded
operators in a Hilbert space containing identity and named \emm von Neumann algebras~ after their
originator. }\

\pt\label{1.4.3}{\rm Mathematically defined \emm states~ on a \Ca\ \fk A are any positive
normalized linear functionals \ome\ on \fk A, i.e. such $\mome\in \mfk A^*$\, (:= the dual of \fk
A), that

\bequ\label{1.4.3(1)} \mome(x^*x)\geq 0,\quad \|\mome\|=1\ (=\mome({\rm id}_{\mfk A})).\enqu

 Not all mathematical states, however, can be used as adequate
 descriptions of physical situations. As physical states on a
quasilocal algebra \fk A\ are usually used \emm locally normal states~, i.e. such states \ome\ on
\fk A, the restrictions of which to all the local \Wsa s\  $\mfk A_v\ (v\in\mcl B(V))$ are \emn
$\msg(\mfk A_v,(\mfk A_v)_*)$-continuous~ (here $(\mfk A_v)_*$ is the predual Banach space of
$\mfk A_v$); the local normality of \ome\ means that the restriction of \ome\ to any $\mfk A_v$ is
expressible by a density matrix in a faithful $W^*-$representation of $\mfk A_v$. We shall denote
by \cS(\fk A)\ the set of all mathematical states on \fk A\ and by $\mS_{ph}:=\mS_{ph}(\mfk A)$\
the set of (properly defined) physical states of the system. The subset $\mS_{ph}(\mfk
A)\subset\mS(\mfk A)$ has to satisfy some natural requirements, e.g. invariance \wrt\
transformations of physical symmetries (cf. below), convexity, local normality and (eventually) to
form a stable face (see \cite[Sec.4.1]{bra&rob}).

The set \cS(\fk A) is convex and compact in the $w^*$-topology of $\mfk A^*$ (i.e. in \emn
$\msg(\mfk A^*,\mfk A)$-topology~). The set $\cal{ES}(\mfk A)$ of extreme points of \cS(\fk A)
consists of \emm pure states~ on \fk A: $\mome\in \cal{ES}(\mfk A)\Leftrightarrow$ $ \{
\mome=\frac{1}{2}\mome_1+\frac{1}{2}\mome_2\ (\mome_{1,2}\in\mS(\mfk A))\imply
\mome_1=\mome_2=\mome \}$. Although the decomposition of a general $\mome\in\mS(\mfk A)$ into its
extremal components ($\in\cal{ES}(\mfk A)$) is not unique if \fk A\ is noncommutative, there are
other physically relevant convex compact subsets of \cS(\fk A) (\emn Choquet simplexes~) allowing
unique extremal decompositions of their elements into extremal components of these simplexes, cf.
\cite{choquet,meyer} for basic mathematics, or also \cite[Ch.4]{bra&rob}, \cite[Ch.4]{pedersen},
\cite[Ch.3]{sak1} for broader contexts.}

\pt\label{1.4.4} \rm The expectation value of a bounded observable $x = x^*\in\mfk A$ in the state
$\mome\in\mS(\mfk A)$ (in accordance with comments in \ref{1.2.4}) is expressed by the value
$\mome(x)$ of the functional \ome\ on the element $x$. For calculations of probability
distributions of values of $x=x^*\in\mfk A$ in the states $\mome\in\mS(\mfk A)$\ it is used,
however, the spectral decomposition of $x$. If \fk A\ is a general \Ca, its selfadjoint elements
need not have their spectral resolutions in \fk A. The spectral resolutions in \fk A exist,
however, if \fk A is a \Wa, \cite{sak1}: $x=x^*\in\mfk A\imply x=\int_\mbR \mlam\,
E_x(\rd\,\mlam),\ E_x(B)^*=E_x(B)=E_x(B)^2\in\mfk A,\ B\subset\mbR\ \text{Borel}, \dots$, hence
$E_x$ is the projector valued spectral measure in the \Wa\ \fk A. Any \Ca\ is naturally embedded
into a \Wa\ - the bidual $\mfk A^{**}$ of \fk A, and any state $\mome\in \cal{S}(\mfk A)$ can be
uniquely extended to a state (equally denoted) $\mome\in\cal{S}_*(\mfk A^{**})$. For any state
$\mome\in\cal{S}(\mfk A)$, we can construct by the \emm GNS-algorithm~ corresponding \emn cyclic
representation~ $\pi_\mome$ of \fk A in a Hilbert space $\mH_\mome$ with a \emn cyclic vector~\,
$\mOme_\mome$ (i.e. the norm-closure $\overline{\pi_\mome(\mfk A)\mOme_\mome} = \mH_\mome$), cf.\
\cite{bra&rob,najm,sak1}, or also \cite[Textbook]{bon-EQM},  characterized (up to the unitary
equivalence) by

\bequ\label{1.4.4(1)} \mome(x)=(\mOme_\mome,\pi_\mome(x) \mOme_\mome),\ \ \forall x\in\mfk A.
\enqu

The representation $\pi_\mome$ is irreducible iff $\mome\in\cal{ES}(\mfk A)$. If we generalize the
concept of observables to all operators from the bicommutant $\pi_\mome(\mfk A)''$ in ${\cal
L}({\cal H}_\mome)$ (what is a \Wsa\ in $\mcl L(\mH_\mome)$), we can obtain spectral resolutions
of selfadjoint elements of \fk A\ in such (extended) representations and the corresponding
expressions for probability distributions, compare \ref{1.2.4}. In specific representations, we
can define also unbounded observables as such selfadjoint operators on $\mH_\mome$ the spectral
projectors of which belong to $\pi_\mome(\mfk A)''$, cf. \cite{sak1}.

We shall need later in this work to distinguish between states which are mutually \emn
macroscopically distinguishable~. Mathematically are such states mutually \emm disjoint~ together
with the mutual disjointness of their GNS representations. It might be useful, for a
characterization of this difference, to reproduce a theorem from \cite[Thm. 3.8.11]{pedersen}:

{\bf Theorem:} Let $\{\pi_1;\mH_1\},\ \{\pi_2;\mH_2\}$ be two nondegenerate representations of a
\Ca\ \fkA\ with their central supports (equiv. central covers) $s_1,\ s_2$, cf.
\cite[3.8.1]{pedersen}. The following conditions are equivalent:

 $\begin{array}{l} (i)\quad\  s_1 \bot s_2. \\
 (ii)\ \  ((\pi_1\oplus\pi_2)(\mfkA))''=\pi_1(\mfkA)''\oplus\pi_2(\mfkA)''.\\
 (iii)\ \ ((\pi_1\oplus\pi_2)(\mfkA))'= \pi_1(\mfkA)'\oplus\pi_2(\mfkA)'.\\
 (iv)\ \ \text{There are no unitarily equivalent subrepresentations of}\  \{\pi_1;\mH_1\}\
 \text{and}\
 \{\pi_2;\mH_2\}.\end{array}$\nl

  Here $\mfk C'$ for a subset $\mfk C\subset\mLH$\ denotes the commutant of \fk C\ in \LH:\
  $\mfk C':=\{B\in\mLH:\ [B,A]\equiv BA-AB = 0,\forall A\in\mfk C\}$,\ and
 $\mfk C'':=(\mfk C')'$. The representations $\pi_1,\ \pi_2$ satisfying the conditions of the
 Theorem are called mutually \emm disjoint representations~. If the GNS representations determined by the two states
 $\mome_1,\ \mome_2$:\
 $\{\pi_{\mome_1};\mH_{\mome_1}\},\ \{\pi_{\mome_2};\mH_{\mome_2}\}$,\ are mutually disjoint, then
 we call these two states also mutually \emm disjoint : $\mome_1\,\perp\,\mome_2$~.

\pt\label{1.4.5}{\rm An abstractly defined \emm symmetry~ of the system in QTLS is any \autm\ of
the algebra \fk A of bounded observables. Let $\tau$ be a representation of the group \bR\ as a
group of symmetries, i.e. a homomorphism $t(\in\mbR)\mapsto \tau_t\in \maut\mfk A,$ which is
'conveniently continuous', e.g. functions $t\mapsto \mome(\tau_tx)$ are continuous for all
$x\in\mfk A$ and all $\mome\in\mS_{ph}(\mfk A)$. It is often assumed, that the group $\tau$
corresponding to a one-parameter group of empirically defined transformations is $\msg(\mfk A,\mfk
A^*)$-continuous (i.e. $\mS_{ph}$ replaced by \cS(\fk A) in the last mentioned case), but this
assumption might be too stringent. Let \cS\ be a 'sufficiently large' subset of states containing
$\mS_{ph}$ and denote by \emn $\msg(\mfk A,\mS)$~ {\bf the topology on} \fk A~\ determined by
functions $x(\in\mfk A)\mapsto\mome(x)$ for all $\mome\in\mS.$ We shall assume, that $\tau$ is
$\msg(\mfk A,\mS)$-continuous in the sense:

\nl $\begin{array}{l} (i)\ \quad functions\ t\mapsto\mome(\tau_tx)\  are\
continuous\ for\ all\ \mome\in\mS,\ x\in\mfk A,\\
(ii)\quad  functions\ x\mapsto\tau_tx\  are\ \msg(\mfk A,\mS) - \msg(\mfk A,\mS) -  continuous\
for\ all\ t\in\mbR,\\
(iii)\ \ \mome\in\mS\Rightarrow\mome\circ\tau_t\in\mS\  for\ all\ t\in\mbR.
\end{array}$

\nl \noidt The last condition (iii) allows us to define a \sg(\cS,\fk A)-continuous group of
transformations of \cS\ by
\bequ\label{1.4.5(1)} \tau^*_t\mome:= \mome\circ\tau_t\quad (\text{for\
all}\ t \in\mbR),\ \mome\in\mS. \enqu

Any selfadjoint element $a\in\mfk A$ generates a \sg(\fk A,\cS)\ (i.e. \sg(\fk A,$\mfk
A^*$))-continuous\ group\ of\ inner\ \autms of \fk A, $\tau^a$, by

\bequ\label{1.4.5(2)} \tau^a_tx:=\exp(ita)x\exp(-ita),\quad\text{for all}\  x\in\mfk A. \enqu

A one-parameter group of \emm inner automorphisms~ of \fk A\ cannot represent some of physically
important global transformations of \emn quasilocal algebras~, e.g. \emn Euclidean or Poincar\'e
transformations~, cf. e.g. \cite[Ch.4,Thm.3]{emch1}. For a general (sufficiently continuous)
one-parameter group $\tau$\ of automorphisms of \fk A\ we can define a generator $\delta_\tau$ - a
densely defined \emm derivation on \fk A~,\ \cite{sak1,bra&rob}.\footnote{A densely defined linear
mapping $\delta:D(\delta)\subset\mfk A\rarw\mfk A$ is a \emm derivation on \fk A~ if it satisfies
the \emn Leibniz rule~: $\delta(xy)=\delta(x)y+x\delta(y)\ \forall\ x,y\in D(\delta)\subset\mfk
A$.} The connection of such generators with physically measurable quantities is in general in QTLS
less transparent then it is in QM or in CM. If the state $\mome\in\mS$ is $\tau-$invariant, i.e.
$\tau^*_t\mome\equiv\mome$, then there is unique weakly continuous unitary group $U^\mome$ acting
on $\mH_\mome$ (cf. \ref{1.4.4}) such that \cite{ruelle1}:
\bequ\label{1.4.5(3)}
\pi_\mome(\tau_tx)=U^\mome_{-t}\pi_\mome(x)U^\mome_t,\
U^\mome_t\mOme_\mome=\mOme_\mome\quad\text{for all}\ t\in\mbR.\enqu

\noidt Representations of \fk A\ in which $\tau$\ is unitarily implemented in the sense of the
first relation of\rref 1.4.5(3)~ are called $\tau-$\emn covariant representations~. In such
representations, the action of $\tau$ is given by an `observable' - the selfadjoint generator of
the corresponding unitary group acting on the Hilbert space of the representation. An
interpretation of such a generator might be dependent, however, on the choice of the covariant
representation.}

\newpage

\chapter{Geometry of the state space of quantum mechanics} \label{Ch2}

\section{Manifold structure of \PH}\label{sec;2.1}

\pt\label{2.1.1}{\rm Let \H\ be a complex separable Hilbert space with the \emn scalar product~
$(x,y)\in\mbC\ (x,y,\in\mH)$, which is linear in the second factor $y$. Let $\mPH\ := \mH/\mbC^*$
be the \emn factor-space of \H~\  by the multiplicative group $\mbC^*$\ of nonzero complex numbers
acting on \H\ by multiplications by scalars. Any element $\bx\in\mPH$ has the form

\bequ\label{2.1.1(1)}  \bx := \{{\it y}\in\mH:{\it y} =\lambda{\it x}, \lambda\in\mbC^*\},\
0\neq{\it x}\in\mH.\enqu

The natural topology on \PH\ is the \emn factor-topology~ coming from the norm-topology in \H.
This topological space \PH\ is the \emm projective Hilbert space~ of \H. The space \PH\ can be
considered as the set of all one-dimensional complex subspaces of \H, or the set of all
one-dimensional projectors $P_x\in\mLH\ (0\neq x\in\mH),\ P_x^*=P_x=P_x^2,\ P_xx=x,$\ with the
natural bijective correspondence $P_x\leftrightarrow\bx$. It is known that there is a natural \emm
K\"ahler structure~ on complex projective spaces. We shall describe it in some details in the case
of \PH.\footnote{Another, more intuitive and more detailed approach to the structure of quantum
state space can be found in \cite{beng&zycz}. For geometry and dynamics (also nonlinear) of
general - not only pure - states see also \cite[Sec.2.1]{bon-EQM}.}}

\pt\label{2.1.2}{\rm Let us define two natural (mutually equivalent) metrices (i.e. \emn distance
functions~) $\rd_1, \rd_2$ on \PH\ (as usual: $\|x\|^2:= (x,x),\ x\in\mH$):

\bequ\label{2.1.2(1)} \rd_1(\bx,\by):=
\sqrt{2}\inf\left\{\left\|\frac{x}{\|x\|}-e^{i\lambda}\frac{y}{\|y\|}\right\|:\ \lambda\in\mbR
\right\}, \enqu

\bequ\label{2.1.2(2)} \rd_2(\bx,\by):= \sqrt{2}\,\|P_x-P_y\|. \enqu

It is not difficult to see that

\bequ \rd_1(\bx,\by)=2\left(1-(Tr(P_xP_y))^{1/2}\right)^{1/2}. \enqu

In\rref 2.1.2(2)~, $\|A\|$ denotes the usual $C^*-$norm of the operator $A\in\mLH$; if
$|A|:=\sqrt{A^*A}\in\mLH$ is its \emn absolute value~, then one can prove

\bequ\label{2.1.2(3)} \begin{split}\sqrt{2}\,\rd_2(\bx,\by) &= Tr|P_x-P_y|=
2\left(1-Tr(P_xP_y)\right)^{1/2},\\
&=\left(1+(Tr(P_xP_y))^{1/2}\right)^{1/2}\rd_1(\bx,\by),
\end{split}
\enqu \noidt what proves the equivalence of $\rd_1$ and $\rd_2$.

 We shall examine now relations
between various natural topologies on \PH. We shall prove first }

\begin{lem}\label{2.1.3} The factor-topology on \PH\ coming from the Hilbert-space
norm-topology of \H\ is equivalent to the metric topology defined on \PH\ by the \emn distance
function~ $\rd_1$ (equiv.: by $\rd_2$).
\end{lem}
\begin{proof}
Let Pr:\,$x\mapsto\bx$ be the natural projection of \H\ onto \PH. The factor-topology on \PH\ is
generated by projections of open balls $B(x;\mveps):=\{y\in\mH:\|x-y\|<\mveps\}$ for $\mveps > 0$,
$x\neq 0$. But Pr\,$B(x;\mveps)=\{\by\in\mPH:\inf\{\|\mlam y-x\|:\mlam\in\mbC\}<\mveps\}$, and
$\inf\{\|\mlam y-x\|:\mlam\in\mbC\}=\|z_\by-x\|$ with $z_\by:= \frac{\overline{(x,y)}}{\|y\|^2}y$
if $y\neq 0$. Hence Pr\,$B(x;\mveps)=\{\by:\|z_\by-x\|<\mveps\}=\{\by:1-Tr(P_xP_y)<
\frac{\mveps^2}{\|x\|^2}\}=\{\by\in\mPH:\rd_2(\bx,\by)< \sqrt{2}\,\frac{\mveps}{\|x\|}\}$, which
is an open ball in the metric topology and the desired equivalence of topologies follows.
\end{proof}

\begin{prop}\label{2.1.4} All the following natural topologies on \PH\ are
mutually equivalent:
\item{(i)}\quad the factor-topology coming from the Hilbert space
norm-topology on \H;
\item{(ii)}\ \ the metric topology defined by the \emn distance functions~
on \PH\ from \ref{2.1.2};
\item{(iii)} the Hilbert-Schmidt topology of $\mfk H\subset\mLH$ of Hilbert--Schmidt operators;
\item{(iv)}\ the trace-norm topology of \TH;
\item{(v)}\quad \sg(\PH,\LH)-- topology;
\item{(vi)}\ \ \sg(\PH,\CH) -- topology.

[In (v), resp. (vi), the topologies are determined by the functions $\bx\mapsto Tr(P_xA)$ for all
$A\in\mLH$, resp. for all $A\in\mCH$:= the set of all compact operators on \H.]

\end{prop}
\begin{proof}
The equivalence of the first four topologies follows from the lemma \ref{2.1.3} and from the
formulas\rref 2.1.2(2)~,\rref 2.1.2(3)~, since the \emn Hilbert-Schmidt operator topology~ is
given by the norm

\bequ\label{2.1.4(1)} \|P_x-P_y\|^2_{HS}:= Tr(P_x-P_y)^2=2(1-Tr(P_xP_y))=[\rd_2(\bx,\by)]^2. \enqu

The equivalence of the trace-norm topology and the \emn $\msg(\mPH,\mCH)$-topology~ follows from
\cite[Proposition 2.16.15]{bra&rob}, and the `stronger' $\msg(\mPH,\mLH)$-topology coincides with
the\break $w^*$-topology from $\mLH^*$, which is `weaker' than the norm-topology of $\mLH^*$. The
last mentioned topology coincides on \PH\ with the trace-norm topology given by the metric
$\rd_2(\bx,\by)\propto Tr|P_x-P_y|$, what finishes the proof.
\end{proof}

\pt\label{2.1.5}{\rm We shall introduce now a manifold structure on \PH\ consistent with the
topology of \PH. Let for $0\neq x\in\mH$

\bequ\label{2.1.5(1)} N_\bx:=\{\by\in\mPH: Tr(P_xP_y)\neq 0\} \enqu

\noidt \ind{$N_\bx$} be an open \emn neighbourhood of $\bx\,\in\,\mPH$~, and let $[x]^\perp$\ be the complex
orthogonal complement of \bx\ in \H. We shall define the mapping $\Psi_x: N_\bx\mapsto [x]^\perp$\
by the formula

\bequ\label{2.1.5(2)} \Psi_x(\by):=\frac{\|x\|^2}{(x,y)}(I-P_x)y, \enqu

\noidt where $y\in\by$.}

\begin{prop}\label{2.1.6} The mapping $\Psi_x$ is a homeomorphism of $N_\bx$\
onto $[x]^\perp$ (with the norm-topology of \H). The set

\bequ\label{2.1.6(1)} \{(N_\bx;\Psi_x;[x]^\perp): 0\neq x \in\mH\} \enqu

\noidt is an \emm atlas on \PH~\ defining a complex-analytic \emn manifold structure~ consistent
with the topology of \PH\ (defined in \ref{2.1.1}).\end{prop}
\begin{proof}
Let $0\neq x\in\mH.$ For any $\by_j\in N_\bx$\ and any $y_j\in\by_j\ (j= 1,2)$\ it is
$\by_1\neq\by_2$\ iff $(x,y_2)y_1\neq (x,y_1)y_2$, hence $\Psi_x$ is injective. For any $z\in
[x]^\perp$\ and $y:= z+x$\ we have $\by\in N_\bx$ (since $x\neq 0$) and $\Psi_x(\by)=z,$ hence
$\Psi_x$\ is bijective. For $\|x\|=1$ and $z_j\in [x]^\perp,\ y_j:=z_j+x\ (j=1,2,)$\ the identity
\bequ 1-Tr(P_{y_1}P_{y_2})=\frac{1}{(\|z_1\|^2+1)(\|z_2\|^2+1)}
\left(\|z_1-z_2\|^2+\|z_2\|^2\,\|(1-P_{z_2})(z_1-z_2)\|^2\right)\enqu

\noidt implies the bicontinuity of $\Psi_x$. For $z\in\Psi_{x_1}(N_{x_1}\cap N_{x_2})$\ it is
$$ \Psi_{x_2}\circ\Psi_{x_1}^{-1}(z)=\|
x_2\|^2\frac{x_1+z}{(x_2,x_1+z)}-x_2 $$ and we see that the mapping

\bequ\label{2.1.6(2)} \Psi_{x_2}\circ\Psi_{x_1}^{-1}: \Psi_{x_1}(N_{\bx_1}\cap
N_{\bx_2})\rightarrow \Psi_{x_2}(N_{\bx_1}\cap N_{\bx_2}) \enqu

\noidt is a complex analytic function, compare e.g. \cite{bourb;manif,h-cartan}.
\end{proof}

\pt\label{2.1.7}{\rm Let \emn$T_\bx\mPH$~\ be the \emm tangent space of \PH\ at \bx~, elements of
which can be represented in the usual way (see e.g. \cite{abr&mars,3baby}) by (classes of mutually
tangent) differentiable curves at \bx. If $c$ is such a curve (i.e. $c:J\rightarrow \mPH$\ for an
open interval $J$ in \bR\ containing $0\in\mbR,\ c(0)=\bx$\ and $t\mapsto \Psi_\by(c(t))$ is
differentiable for $\bx\in N_\by$) denote by $\dot{c}_\bx:=\dot{c}(0)$ (or simply $\dot{c}$\ if
the point \bx\ is fixed) the corresponding equivalence class, $\dot{c}_\bx\in T_\bx\mPH$. With any
$x\in\bx$, we associate an identification of $T_\bx\mPH$ with $[x]^\perp$ by the mapping

\bequ\label{2.1.7(1)} T_\bx\Psi_x:\ T_\bx\mPH\rightarrow [x]^\perp,\ \dot{c}\mapsto
T_\bx\Psi_x(\dot{c}):=\left.\frac{\rd}{\rd t}\right|_{t=0}\Psi_x(c(t)). \enqu

In\rref 2.1.7(1)~, we identify, in the usual way, the tangent space $T_v[x]^\perp$\ of the linear
space $[x]^\perp$\ at any of its points $v\in [x]^\perp$\ with the base space $[x]^\perp$\ itself.
The mapping $T_\bx\Psi_x$ is a linear isomorphism for any $x\in\bx$, and also $T_\bx\Psi_{\lambda
x}=\lambda T_\bx\Psi_x\ (\lambda\in\mbC)$. The derivative in\rref 2.1.7(1)~ is taken \wrt\ the
Hilbert space norm in $[x]^\perp$.}

\pt\label{2.1.8}{\rm Let us mention two simple examples of the representation of elements
$\dot{c}\in T_\bx\mPH$\ by curves $c$\ and of the corresponding identification of $T_\bx\mPH$\
with\ $[x]^\perp$. Each vector $\dot{c}\in T_\bx\mPH$\ can be represented by a curve of any of the
following forms (the expressions written by bold typeface represent the projections to \PH\ of the
corresponding elements of \H, i.e. $z(\in\mH)\mapsto \bz\equiv P_z(\in\mPH)$):
\begin{eqnarray}
\label{2.1.8(1)} c_1(t) & := & \boldsymbol{\lambda x+ ty}\equiv P_{\lambda x+ ty}\quad
(\lambda\in\mbC,\ y\in\mH,\
x\in\bx),\quad\  t\in\mbR,\\
\label{2.1.8(2)} c_2(t) & := & \boldsymbol{\exp(itB)x}\equiv P_{\exp(itB)x}\quad (B=B^*\in\mLH,\
x\in\bx),\ t\in\mbR.
\end{eqnarray}

 If we denote corresponding tangent vectors by $\dot{c}_1$\ and
$\dot{c}_2$, then
\begin{eqnarray}
\label{2.1.8(3)} T_\bx\Psi_x(\dot{c}_1) &=& \lambda^{-1}(1-P_x)y,\\
\label{2.1.8(4)} T_\bx\Psi_x(\dot{c}_2) &=& i(1-P_x)Bx.
\end{eqnarray}

 Clearly $\dot{c}_1=\dot{c}_2$\ iff the right hand sides of\rref 2.1.8(3)~\
 and\rref 2.1.8(4)~\ coincide as vectors in $[x]^\perp$. This is
 the case if e.g. $y=i\mlam Bx$\ in\rref 2.1.8(3)~. The
 representants ($c_1$, or $c_2$, or \dots) of a given $\dot{c}$\ can be chosen
 in many various ways. We shall use notation:
 \[\centerline{\emn $v_x:= T_\bx\Psi_x(v)\in [x]^\perp$~}\]

 \noidt for $v\in T_\bx\mPH;\ v_{\mlam x}=\mlam v_x.$}

 \pt\label{2.1.9}{\rm We shall consider \PH\ as a {\bf real manifold} of the
 dimension $\dim\mPH=2\dim_\mbC\mH-2,$ (if \H\ is finite dimensional)
 where $\dim_\mbC$ means the complex dimension. On this manifold,
 we introduce a \emm metric $Q$~, i.e. a real-analytic
symmetric 2-covariant tensor field $\bx\mapsto Q_\bx$ defining an isomorphism between $T_\bx\mPH$
and its dual $T_\bx^*\mPH$ at any point $\bx\in\mPH$:

\bequ\label{2.1.9(1)} v\,(\in T_\bx\mPH)\mapsto Q_\bx(v,\cdot)\in T_\bx^*\mPH, \enqu

\noidt where the linear functional $Q_\bx(v,\cdot): w\, (\in T_\bx\mPH)\mapsto Q_\bx(v,w)\in\mbR$\
depends linearly on $v$, and for any $F\in T_\bx^*\mPH$\ there is a unique $v_F\in T_\bx\mPH$\
such, that $F=Q_\bx(v_F,\cdot).$ Let the metric be given by

\bequ\label{2.1.9(2)} Q_\bx(v,v) :=
\frac{2}{\|x\|^2}\,(v_x,v_x)=\frac{2}{\|x\|^2}\,\|v_x\|^2,\quad v_x:= T_\bx\Psi_x(v).\enqu

Since $v_{\mlam x}=\mlam v_x$, the definition does not depend on the choice of $0\neq x\in\bx$ in
the mapping $\Psi_x$. The nondegeneracy is a consequence of the Riesz theorem applied to the
Hilbert space $[x]^\bot$ and analyticity is also straightforward. From the bilinearity and
symmetry we have

\bequ\label{2.1.9(3)} Q_\bx(v,w)=\frac{2}{\|x\|^2}\, \Re(v_x,w_x),\quad\forall v,w\in
T_\bx\mPH.\enqu

It is possible to prove by straightforward calculations of lengths of differentiable curves in
\PH\ (compare also \cite{abr&mars,R&S}):}

\begin{prop}\label{2.1.10} The \emn metric $Q$~ from\rref 2.1.9(3)~ endows \PH\ with a
\emn distance function~ $\rd$ (calculated as the minimal length of differentiable curves joining
two points) different from $\rd_j,\ j=1,2$;\rref 2.1.2(1)~,\rref 2.1.2(2)~. Both the distance
functions $\rd_1,\ \rd_2$\ give (by differentiation) the metric $Q$ from\rref 2.1.9(3)~ on
\PH.\end{prop}

\section{Symplectic structure}\label{sec;2.2}

\pt\label{2.2.1}{\rm Let us define a \emn complex structure $J$ on \PH~\ induced by that of \H.
For each $\bx\in \mPH$ and $v\in T_\bx\mPH$, we define

\bequ\label{2.2.1(1)} Jv:=(T_\bx\Psi_x)^{-1}\circ i \circ(T_\bx\Psi_x)(v),\enqu

\noidt where $i$ is the multiplication by the imaginary unit $i\in\mbC$ in the complex subspace
$[x]^\bot\subset\mH.$ The definition\rref 2.2.1(1)~ of $J$ does not depend on the choice of
$x\in\bx$. Clearly: $(Jv)_x=i\,v_x.$  We define now a \emn two-form \Ome\ on \PH~:

\bequ\label{2.2.1(2)} \mOme_\bx(v,w):=Q_\bx(v,Jw),\ \forall \bx\in\mPH,\ v,w\in T_\bx\mPH.\enqu

We shall use charts $\Psi_x$ with $\|x\|=1$ in the following. In such a chart, the form \Ome\ is
written

\bequ\label{2.2.1(3)}  \mOme_\bx(v,w)= -2\, \Im (v_x,w_x). \enqu

 The just introduced structures lead to the standard \emn symplectic~, and also metric (known as the
``\emn Fubini-Study metric~'') structures on the space of pure quantum states \PH. If this both
structures are connected as in\rref 2.2.1(2)~ by a complex structure $J$ (coming, in this case,
from that in the underlying Hilbert space \H), we obtain a structure on the manifold \PH\ which is
called the \emm K\"ahler structure~.}

\begin{lem}\label{2.2.2}
The form \Ome\ is nondegenerate.
\end{lem}
\begin{proof}
If $\mOme_\bx(w,v)=0$ for all $w\in T_\bx\mPH,$ then also $\mOme_\bx(Jv,v)=2\, \|v_x\|^2=0,$ hence
$v=0$.
\end{proof}

\begin{lem}\label{2.2.3}
For any unitary transformation $U$\ of \H\ onto itself, the form \Ome\ is invariant \wrt the
projected mapping $\mbs U: \mPH\rightarrow\mPH,\, \bx\mapsto\mbs U(\bx):= \mbs U\bx,$ i.e.

\bequ\label{2.2.3(1)} (\mbs U^*\mOme)_\bx(v,w):=\mOme_{\mbs U\bx}(\mbs U_*v,
\mbs{U_*}w)=\mOme_\bx(v,w).\enqu

Here $\mbs U^*\mOme$ is the \emn pull-back~ of \Ome\ by $\mbs U$, and $\mbs{U_*}:
T_\bx\mPH\rightarrow T_{\mbs U\bx}\mPH$ maps the equivalence class $\dot{c}$ containing the curve
$c:t\mapsto c(t)$ at \bx\ (i.e. $\bx=c(0)$) into the class $\mbs U\mbs c$ containing the curve
$\mbs Uc: t\mapsto \mbs Uc(t)$ at $\mbs U(\bx)$.
\end{lem}
\begin{proof}
According to \ref{2.1.8}, the vector $v_x$ corresponds to the class containing the curve $c:
t\mapsto{\bf x+tv}_x$, hence the vector $(\mbs U_*v)_{Ux}$ corresponds to the class $\mbs U\mbs
c\in T_{\bf Ux}\mPH$ containing the curve $\mbs Uc:t\mapsto{\bf Ux+tUv}_x$, and since $U$
conserves orthogonality in \H\ we have

\bequ\label{2.2.3(2)} (\mbs U_*v)_{Ux}=Uv_x.\enqu

Substitution into the expression\rref 2.2.1(3)~ from\rref 2.2.3(2)~ gives the result.
\end{proof}

\begin{prop}\label{2.2.4} The two-form \Ome\ on \PH\ is closed: $\rd\mOme=0$;
it is a \emm symplectic form on \PH~, hence \emn strongly nondegenerate~ (cf.
\cite[A.3.14]{bon-EQM}).
\end{prop}
\begin{proof}
The skew symmetry and bilinearity is trivial and (strong) nondegeneracy is proved in Lemma
\ref{2.2.2}. The proof of closedness used in an appendix of the Arnold's book \cite[Appendix 3
B]{arn1} in the finite-dimensional case is literally applicable for any complex Hilbert space and
its projective space, because of the validity of Lemma \ref{2.2.3}.\footnote{For an alternative
proof valid also for unitary orbits of density matrices see \cite[Theorem 2.1.19]{bon-EQM}.} Hence
\Ome\ is symplectic.
\end{proof}

\pt\label{2.2.5}{\rm According to a \emn theorem by Wigner~, any bijective transformation \cbF\ of
\PH\ which conserves the \emn'transition probabilities'~, i.e.:

\bequ\label{2.2.5(1)}
 Tr(P_xP_y) =
Tr(\mcbF(P_x)\mcbF(P_y)),\quad\forall x, y \in\mH,\ x\neq 0\neq y,\  \enqu

\noidt can be extended to a transformation F of \H\ onto itself, which is either unitary or
antiunitary, compare \cite[3.2.1 and 3.2.14]{bra&rob}:

\bequ\label{2.2.5(2)} Tr(P_{Fx}P_{Fy}) = Tr(P_xP_y).\enqu

\noidt Such transformations conserve also distances and the metric Q, see \ref{2.1.2} and
\ref{2.1.9}. Bijections of \PH\ onto itself conserving the metric Q will be called the \emm Wigner
maps~.

On the other hand, \emn antiunitary transformations~ F of \H\ do not conserve the symplectic form
$\mOme:\mcbF_*\mOme=-\mOme.$ Transformations \cbF\ of \PH\ conserving \Ome\ are called \emm
symplectic transformations~.}

\begin{lem}\label{2.2.6}
Let \cbF\ be any symplectic transformation of \PH\ the restriction of which to $T_x\mPH$ for any
$\bx\in\mPH$\ (i.e. the mappings $\mcbF_*:T_\bx\mPH\rightarrow T_{\mcbF\bx}\mPH$) are complex
linear \wrt the complex structure J, cf. \ref{2.2.1}. Then \cbF\ can be extended to a unitary
transformation $F\in\mLH$.
\end{lem}
\begin{proof}
Symplecticity and complex linearity of \cbF\ give

\bequ\label{2.2.6(1)} Q_\bx(v,w)=-\mOme_\bx(v,Jw)=-\mOme_{\mcbF\bx}(\mcbF_*v,J\mcbF_*w)=
Q_{\mcbF\bx}(\mcbF_*v,\mcbF_*w), \enqu

i.e. $Q=\mcbF_*Q$, what implies the invariance of distances:
$$ \rd(\mcbF\bx,\mcbF\by)=\rd(\bx,\by),$$
which in turn implies the invariance of $Tr(P_xP_y)$. Hence \cbF\ can be extended either to a
unitary or to an antiunitary transformation. Since antiunitary transformations have nonsymplectic
projections in \PH, extension F of \cbF\ must be unitary.
\end{proof}

\begin{prop}\label{2.2.7}
Any symplectic isometry $\mcbF:\mPH\rightarrow\mPH$\ is an analytic diffeomorphism of \PH.
\end{prop}
\begin{proof}
\cbF\ is a symplectic Wigner map, hence extendable to a unitary $F\in\mLH.$ With the help of the
charts $\Psi_x$, analyticity follows for the projection \cbU\ of any unitary $U\in\mLH$. The same
considerations apply to the inverse map $\mcbF^{-1}$, and the assertion follows.
\end{proof}

\section{Quantum mechanics as a classical Hamiltonian field
theory}\label{sec;2.3}

\pt\label{2.3.1}{\rm After introducing the symplectic structure \Ome\ on the set \PH\ of all pure
states of conventional QM (compare Sec.\ref{sec;1.2}), we shall try to reformulate also other
concepts of QM into the form analogous to that of CM as it was outlined in Sec.\ref{sec;1.3}. It
will be shown that this is possible to a large extent. There are, however, certain important
differences. The main technical difference consists in infinite dimensionality of the 'phase
space' \PH\ what implies e.g. \emn nonexistence of a (Liouville) measure on \PH~, invariant \wrt\
all symplectic Wigner maps. The main physical difference consists, however, in the interpretation
of basic quantities in QM. This \emn difference between QM and CM~ does not vanish even for finite
dimensional Hilbert space \H.}

\pt\label{2.3.2}{\rm Let $A$ be a \emn selfadjoint operator~\footnote{\label{ftn2.3.2}A brief
review of the theory of unbounded operators is present in \cite[C]{bon-EQM}, or in
\cite[Textbook]{bon-EQM} in detail.} on the Hilbert space \H\ with \emn domain $D(A)\subset\mH$~.
Let \emn$PD(A)\subset\mPH$~ be the projection of $D(A)$ into \PH:

\bequ\label{2.3.2(1)} PD(A) := \{\bx\in\mPH:x\in D(A),\ x\in\bx\}.\enqu

Define a real-valued function $f_A$ on $PD(A)$:

\bequ\label{2.3.2(2)}  f_A(\bx) := Tr(P_xA)\equiv\frac{(x,Ax)}{\|x\|^2},\ 0\neq x\in\bx\in
PD(A).\enqu

The function $f_A$ determines the operator $A$ in an unambiguous way by the \emn polarization
identity~:

\bequ\label{2.3.2(3)} (x,Ay) = \frac{1}{4}\sum_{\mlam=\pm 1,\pm i}\mlam\|\mlam
x+y\|^2f_A({\mbs{\lambda x+y}}) .\enqu

For bounded $A\in\mLH$, the function $f_A: \mPH\rightarrow\mbR$ is real analytic. Since for
arbitrary selfadjoint $A,B\in\mLH$ there need not be any selfadjoint operator $C$ on \H\ such,
that $f_C := f_A\cdot f_B$ (:= pointwise multiplication of functions), the set of 'classical
observables' $f_A\ (A^*= A\in\mLH)$ does not form an associative algebra.

\nl {\it Remark}: Corresponding to the spectral decomposition of $A$,${}^{\ref{ftn2.3.2}}$ we have
the decomposition of $f_A$:

\bequ \label{2.3.2(4)}f_A(\cdot)=\int_\mbR \mlam E_A^f(\rd\mlam)(\cdot),\quad \text{where}\
E_A^f(B)(\bx):=Tr(P_xE_A(B))
 \enqu

\noidt for any Borel set $B\subset\mbR$, with  $E_A$ the \emn spectral measure of $A$~. Contrary
to the case of classical mechanics \ref{1.3.2}, the functions $\bx\mapsto E_A^f(B)(\bx)$ are not
characteristic (indicator) functions on \PH. The decomposition into characteristic functions
similar to that in \ref{1.3.2} does not correspond to any decomposition into quantal observables.}
\nl

\pt\label{2.3.3}\rm The function $f_\mrh(\bx):=Tr(P_x\mrh)$ might remind us of a probability
distribution on the ``phase space'' \PH\ representing a Gibbs ensemble in the sense of classical
statistical physics, cf. e.g. \cite{ruelle1,ruelle2}, or any textbook on statistical physics.

Although any density matrix \rh\ is uniquely reconstructed from the corresponding function
$f_\mrh(\bx)$ on the phase space \PH\ with the help of\rref 2.3.2(3)~, the function $f_\mrh$
cannot be interpreted as a \emn probability distribution~ of systems  occurring in the pure states
$P_x\equiv\bx$ in a statistical ensemble described by \rh. The function $f_\mrh$ is interpreted to
give the probability $f_\mrh(\bx)$ of positive result (i.e. of the number = 1) by measuring of the
observable $P_x$ (with just two possible outcomes $\in\{0,1\}$ of any of its measurements) in the
state \rh. Because of the existing nonuniqueness of decompositions of \rh\ into pure states,
mentioned in \ref{1.2.3}, a \emn classical interpretation~ of any probability measure on \PH\
representing \rh\ would be inadequate in general. In the following, we shall restrict our
attention (mainly) to pure states.\footnote{A certain, more detailed, account of the geometry and
interpretation questions of the set of density matrices is given in \cite[2.1-e]{bon-EQM}.}

For a quantal observable $A$, the numbers $f_A(\bx)$ are interpreted as \emn expectation values~
for (real valued) results of measurements of the observable $A$ in the state $\bx\in\mPH$. Also
the functions $f_A$ will be called here `the (quantal) observables'.

\pt\label{2.3.4}{\rm In the setting of this section, it is natural to define a symmetry of the
system as a \emn symplectic isometry of \PH~. According to the Sec.\ref{sec;2.2}, any such
symmetry can be extended to a unitary transformation of \H. Let $t\mapsto \mcbF_t$ be a
one-parameter group of symplectic isometries of \PH\ which is weakly continuous, i.e. the
functions

\bequ\label{2.3.4(1)} t\mapsto \mbs{F_t}\bx,\quad \forall \bx\in\mPH \enqu

\noidt are continuous. Such a group can be extended to a \emn weakly continuous unitary group~ on
\H\ (compare \cite[3.2.35]{bra&rob}), which corresponds to uniquely defined selfadjoint operator
$A$ on \H\ (by Stone's theorem, \cite[C3 \& Textbook]{bon-EQM}). In this way, for the group
$\mcbF_t$, we obtain the expression:

\bequ\label{2.3.4(2)} \mcbF_t\bx={\boldsymbol{\exp(-itA)x}},\ i.e. \
\mcbF_t\bx=\exp(-itA)P_x\exp(itA)\in\mPH.\enqu

The operator $A$ in\rref 2.3.4(2)~ is defined by $\mcbF_t$\ up to an additive real constant
multiple of identity $I$ \  of \LH, i.e. any other $A'$ satisfying\rref 2.3.4(2)~ has the form
$A'=A+\mlam I,\ (\mlam\in\mbR)$. Conversely, any selfadjoint operator $A$ on \H\ determines,
according to\rref 2.3.4(2)~, a weakly continuous one-parameter group of \emn symplectic
isometries~ of \PH. The flow $\mcbF_t$ and its unitary extension $F_t := \exp(-itA)$\ are related
by

\bequ \label{2.3.4(3)}\mcbF_t(P_x) = P_{F_tx} = F_tP_xF_{-t},\quad P_x\in \mPH.\enqu

The functions\rref 2.3.4(1)~ for specific $\bx$'s are differentiable if the corresponding
generator $A$ has domain $D(A)$ containing $x\in\bx : x\in D(A).$ If $A\in\mLH,$ then
functions\rref 2.3.4(1)~ are analytic in $t\in\mbC,\ \forall\ \bx\in\mPH.$ It is clear from the
group property of $t\mapsto\mcbF_t$, that differentiability of\rref 2.3.4(1)~ in any point \bx\
for $t = 0$ implies differentiability on the whole curve\rref 2.3.4(1)~, i.e. for all $t\in\mbR$.}

\pt\label{2.3.5}{\rm We have obtained a set of differentiable curves lying densely in \PH\ for any
one-parameter weakly continuous group $\mcbF_t$\ of symmetries of (\PH,\,\Ome) (since $PD (A)$ is
dense in \PH\ for any selfadjoint $A$). For $\bx\in PD(A)$ ($A$ is a generator of $F_t$), the
curve\rref 2.3.4(1)~ determines a vector $\msg_A(\bx) \in T_{\bx}\mPH$. The set of vectors
$\msg_A(\bx)$ is defined for $\bx\in PD(A)$ only, and for unbounded $A$ it is not a differentiable
vector field on \PH\ (it is differentiable only in directions of some curves lying densely in
$PD(A)$, and in \PH). We shall call it, nevertheless, `\emn the vector field $\msg_A$~'. Its value
in \bx\ is expressed in $[x]^\bot$ according to\rref 2.1.8(4)~:

\bequ\label{2.3.5(1)} T_\bx\Psi_x (\msg_A(\bx)) = -i(I - P_x )Ax\quad \text{for}\quad x\in
D(A).\enqu

 For $A\in\mLH$, $\msg_A$ is an analytic vector field on \PH. But also
 for an unbounded $A$, the vector field
$\msg_A$  determines its flow $\mcbF_t=:\mcbF^A_t$ uniquely: it can be integrated along a densely
in \PH\  lying set of differentiable curves (this is just the solution of Schr\"odinger equation
with the Hamiltonian $A$), and afterwards the obtained (densely defined) flow extended to the
whole \PH\ by continuity.}

\pt\label{2.3.6}{\rm  Let $x\in D(A)\cap D(B)$ for two selfadjoint operators $A$\ and $B$\ on \H\
and $\|x\|=1$. Then the value of the symplectic form \Ome\ on vectors $\msg_A(\bx)$\ and
$\msg_B(\bx)$ is, according to\rref 2.2.1(3)~ and\rref 2.3.5(1)~,

\bequ\label{2.3.6(1)} \mOme_\bx(\msg_A,\msg_B)=-2\, \Im (Ax,(I-P_x)Bx). \enqu

If, moreover, $Bx\in D(A)$ and $Ax\in D(B)$\ (e.g. if $A$ and $B$ have a common invariant set
$D\subset D(A)\cap D(B)$ and $x\in D$), then we can write

\bequ\label{2.3.6(2)} \mOme_\bx(\msg_A,\msg_B) = i\,Tr(P_x[A,B])\enqu

 \noidt where $[A,B]:=AB-BA.$

Let $f$ be a real-valued function defined on a dense subset \bD\ of \PH. Let $c$\ be a
differentiable curve in \PH\ at $\bx\in\bD$\ such, that $c(t)\in\bD$\ for some open interval of
reals $t$ containing $t=0,\ c(0)=\bx.$ Let $\dot{c}\in T_\bx\mPH$\ be the corresponding tangent
vector. Denote

\bequ\label{2.3.6(3)} \rd_\bx f(\dot{c}) :=\left. \frac{\rd}{\rd t}\right|_{t=0}f(c(t)) \enqu

\noidt if the derivative on the \rhs\ exists. Assume, that\rref 2.3.6(3)~ is well defined for a
dense set of vectors $\dot{c}\in T_\bx\mPH.$ The function

\bequ\label{2.3.6(4)} \rd_\bx f:\dot{c}\mapsto \rd_\bx f(\dot{c}) \enqu

\noidt is linear. If it is bounded, it can be extended by continuity to the whole $T_\bx\mPH,$
hence it defines an element $\rd_\bx f\in T^*_\bx\mPH$ which will be called the \emm exterior
differential~ of $f$ in \bx.}

\begin{prop}\label{2.3.7}
Let $A$ be a selfadjoint operator on \H, $f_A$\ is given by\rref 2.3.2(2)~, and the corresponding
vector field $\msg_A$\ is defined in \ref{2.3.5}. Then, for any $\bx\in PD(A),$ the exterior
differential $\rd_\bx f_A\in T_\bx^*\mPH$ exists, and for all $v\in T_\bx\mPH$\ we have

\bequ\label{2.3.7(1)} \mOme_\bx(\msg_A(x),v)=-\, \rd_\bx f_A(v),\qquad \forall \bx\in PD(A).\enqu
\end{prop}
\begin{proof} Let $\{v_x\}^\bot$\ be defined according to\rref
2.1.9(2)~ for $v\in T_\bx\mPH.$ Define the selfadjoint $B(v)\in\mLH$:

\bequ\label{2.3.7(2)} B(v)y := i(v_x,y)x - i(x,y)v_x, \quad \forall y\in\mH. \enqu

Assume $\|x\|=1.$ Then, according to\rref 2.1.8(2)~\ and\rref 2.1.8(4)~, the curve

\bequ\label{2.3.7(3)} t\mapsto c_v(t):= {\mbs\exp(\bf{itB(v)})\bx} \enqu

\noidt corresponds to $v=\dot{c}_v.$\ Let $v$ be such that $v_x\in D(A).$ Then it is seen that
$\rd_\bx f_A(v)$\ defined in\rref 2.3.6(3)~\ exists and has the form

\bequ\label{2.3.7(4)} \rd_\bx f_A(v)=-i\, Tr(P_x[B(v),A])=-\mOme_\bx(\msg_A(\bx),v), \enqu

\noidt where, in the second equality, we used\rref 2.3.6(2)~\ and $\msg_{B(v)}(\bx)=-v.$ Because
$(I-P_x)D(A)\subset D(A)$ is dense in $\{x\}^\bot$, we have proved\rref 2.3.7(1)~\ for a dense
linear subset $D\subset T_\bx\mPH,\ v\in D.$ The boundednes  is clear either from our
construction, or from the boundednes of the \lhs\ of\rref 2.3.7(1)~\ for a well defined
$\msg_A(\bx).$
\end{proof}

\pt\label{2.3.8}{\rm We can see from the proposition \ref{2.3.7}, how to reconstruct the vector
field $\msg_A$\ from $f_A$\ with the help of the symplectic form \Ome. Hence, $\msg_A$\ is
globally Hamiltonian vector field on (the dense subset of) \PH\ corresponding to the Hamiltonian
function $f_A$\ (compare with \ref{1.3.5} - up to domain differences).

Let two selfadjoint $A,B$\ have a common dense domain $D\subset D(A)\cap D(B).$ Then the function
(the Poisson bracket)

\bequ\label{2.3.8(1)} \bx\mapsto \{f_A,f_B\}(\bx):=\mOme_\bx(\msg_A,\msg_B),\ \bx\in PD, \enqu

\noidt is densely defined. If, moreover, the operator $i[A,B]$ is selfadjoint and $D$ is its
core\footnote{A \emm core $D\subset\mH$ of a closable operator~ $C$ is such a subset $D\subset
D(C)\subset\mH$, that the closure of the restriction $\overline{C\upharpoonright D}=\overline{C}$,
cf. also \cite[C1]{bon-EQM}.} then, according to\rref 2.3.6(2)~, we have

\bequ\label{2.3.8(2)} \{f_A,f_B\} = f_{i[A,B]}.\enqu

Remember that this is a quantummechanical formula corresponding to\rref 1.3.5(8)~.}

\pt\label{2.3.9}{\rm Assume that a weakly continuous unitary representation $U$\ of a connected
Lie group $G$\ in the Hilbert space \H\ is given:

\bequ\label{2.3.9(1)} U(g_1g_2)=U(g_1)U(g_2),\ g_1,g_2\in G.\enqu

Then $U$ is projected onto a weakly continuous realization of $G$\ by a group of symplectic
isometries $\mbs U(g)\ (g\in G)$ of $(\mPH,\mOme).$ To any element $\xi$ of the Lie algebra
\fk{g}\ of $G$\ corresponds the selfadjoint  generator $X_\xi$ of the one-parameter subgroup
$U(\exp(t\xi)):$

\bequ\label{2.3.9(2)} X_\xi x:= i\, \left.\frac{\rd}{\rd t}\right|_{t=0}U(\exp(t\xi))x,\ x\in
D(X_\xi), \enqu

\noidt and $U(\exp(t\xi))=\exp(-itX_\xi).$ By a use of the \emm adjoint representation~ $Ad:
G\rarw \cal{L}(\mfk{g}),$

$$ Ad(g)\xi := \left.\frac{\rd}{\rd t}\right|_{t=0}[g
\exp(t\xi)g^{-1}]$$

\noidt we obtain:

\bequ\label{2.3.9(3)} [X_\xi,X_\eta]:= X_\xi X_\eta - X_\eta X_\xi = i\,X_{[\xi,\eta]}.\enqu

The mapping $\xi\mapsto X_\xi$\ is linear. It is known (compare \cite{bar&racz}), that the \emn
G{\aa}rding domain $\mcl D_G$~, as well as the \emn analytic domain $\mcl A_G$~\ of the
representation $U(G)$\ are common dense invariant sets of all the generators $X_\xi\ (\xi\in \mfk
g)$ and they are also common cores of all these selfadjoint operators (cf. also \ref{3.1.1}). Let
us {\bf define the vector fields} \emn$\msg_\xi\ (\xi\in\mfk g)$~ {\bf on} \emn$P\mcl
D_G\subset\mPH$~\ corresponding to the flows $\mbs U(\exp(t\xi))$ on \PH\ according to the
definition of $\msg_A$ in \ref{2.3.5}. Let $f_\xi(\bx) := Tr(P_xX_\xi)$\ for $x\in\mcl D_G.$ Then
\ref{2.3.8} is applicable to these quantities. All the formulas of \ref{1.3.7} are valid on $P\mcl
D_G.$ Difference w.r.t. the classical case is that neither \PH\ nor $P\mcl D_G$ are homogeneous
spaces even for irreducible $\mbs U(G).$}

\pt\label{2.3.10}{\rm Up to now, we used \emn charts $(N_\bx;\Psi_\bx;[x]^\bot)$~ for
identification of $T_\bx\mPH$\ with $[x]^\bot$, and for each point $\bx\in\mPH$\ it was used its
own chart. Let us rewrite now the evolution equation corresponding to the one-parameter flow
$\mcbF_t^A$\ on \PH\ generated by the Hamiltonian  $A$,\ i.e. the \emn Schr\"odinger equation~

\bequ\label{2.3.10(1)} i\frac{\rd}{\rd t}x(t)= Ax(t),\ x(0):=x\in\mH,\enqu

\noidt projected onto \PH, with the help of the chart $(N_\by;\Psi_y;[y]^\bot),\ \bx\in N_\by.$
Let us denote by $c: t\mapsto c(t)$ a differentiable curve in \PH, and by $\dot{c}(t)$ its tangent
vector : $\dot{c}(t)\in T_{c(t)}\mPH.$\ The curve $c$ will be a solution of our problem, if for
some $\bx\in\mPH:\ c(t)=\mcbF_t^A\bx$\ for all $t\in\mbR.$ For $\bx\in PD(A),$ we then obtain by
differentiation

\bequ\label{2.3.10(2)} \dot{c}(t)=\msg_A(c(t)),\enqu

\noidt which is an abstract form of Hamilton equations on \PH\ corresponding to the Hamiltonian
function $f_A$, cf.\rref 2.3.7(1)~. The correspondence with\rref 2.3.10(1)~ consists in that, that
$c(t)=\bx(t)$\ if $c(0)=\bx,$ where $x(t)\, (\in\bx(t))$ is the solution of\rref 2.3.10(1)~ with
the initial value $x\in\bx$. Let us fix $y\in\mH,\ \|y\|=1,$ and choose the chart
$(N_\by;\Psi_y;[y]^\bot)$ defined in\rref 2.1.5(2)~. Denote

\bequ\label{2.3.10(3)} \Psi(t) := \Psi_y(c(t))\ \text{for a curve c in}\ N_\by.\enqu

\noidt The curve $\Psi$ in $[y]^\bot$\ will correspond to a solution $c$ of\rref 2.3.10(2)~ iff it
satisfies the equation

\bequ\label{2.3.10(4)} i\,\frac{\rd}{\rd t}\Psi(t) = [A-(y,A(y+\Psi(t)))](y+\Psi(t)),\ \Psi(0)\in
[y]^\bot.\enqu

The equation\rref 2.3.10(4)~ describes the wanted projection of\rref 2.3.10(1)~ onto \PH\ in the
chart $\Psi_y$. It is a \emn nonlinear (field-) equation~ in the Hilbert space $[y]^\bot$, in
which different vectors correspond to different physical states.

If we denote by $v_y$\ the representative of a vector $v\in T_\bx\mPH$\ for $\bx\in N_\by$\ in the
chart $\Psi_y\ (\|y\|=1),$ then the symplectic form \Ome\ in this chart has the expression:

\bequ\label{2.3.10(5)} \mOme_\bx(v,w)=-2\,Tr(P_xP_y)\,\Im(v_y,(I-P_x)w_y).\enqu

\noidt Remember, that $v_y,w_y\in [y]^\bot := (I-P_y)\mH.$

Let us write $f^t:= f\circ\mcbF^A_t$\ for any differentiable function $f$\ on \PH. Then, for
$\bx\in PD(A)$, we obtain the wanted form of the Schr\"odinger equation:

\bequ\label{2.3.10(6)} \frac{\rd}{\rd t}f^t(\bx)=\{f_A,f^t\}(\bx) :=
\mOme_\bx(\msg_A,\msg_f),\enqu

\noidt where $\msg_f$\ is a vector field defined on the whole \PH\ by

\bequ\label{2.3.10(7)} \mOme_\bx(\msg_f(\bx),v):=-\rd_\bx f(v),\ \forall v\in T_\bx\mPH.\enqu

The equation\rref 2.3.10(6)~ has the form  of evolution equation of classical mechanics in terms
of Poisson brackets.}

\pt\label{2.3.11}\rm Let us add a note concerning possible generalizations of the here presented
dynamics. Since \PH\ is a symplectic manifold, more general Hamiltonian evolutions can be defined
on it than the evolutions corresponding to linear Schr\"odinger equations \rref 2.3.10(1)~. We can
choose instead of the function $f_A: \mPH\rarw\mbR$ as a (`classical') Hamiltonian an arbitrary
`sufficiently differentiable' function $h: \mPH\rarw\mbR$. Then we obtain from the corresponding
Hamiltonian dynamics on the infinite dimensional symplectic manifold \PH\ evolution of QM-vector
states in \H, which cannot be described (in general) by a linear Schr\"odinger equation. This
situation is described in many details in \cite{bon-EQM}.



\chapter{Classical mechanical projections of QM}\label{Ch3}

\section{Orbits of Lie group actions on \PH}\label{sec;3.1}

\pt\label{3.1.1}{\rm Let $U$ be a weakly continuous unitary representation of a connected Lie
group $G$ in the Hilbert space \H, and $X_\xi$ be the selfadjoint generator of the one-parameter
subgroup of $U(G)$ corresponding to an arbitrary element $\xi$ of the Lie algebra \fkg\ of $G$, as
it was defined in\rref 2.3.9(2)~. {\bf Let} \emn$\mcl D_G\subset\mH$~ be the \emn G{\aa}rding
domain of $U(G)$~ \cite[11.1.8.]{bar&racz}, i.e. a dense $U(G)$-invariant set of vectors
$x\in\mH$, for which the functions $g\mapsto U(g)x\ (g\in G)$ are infinitely differentiable. {\bf
We shall  denote by} \emn$\mcl A_G\ (\subset\mH)$~ the dense set of \emn analytic vectors of
$U(G)$~ invariant \wrt the action of $U(G).$ For $x\in\mcl A_G$, not only the functions $g\mapsto
U(g)x$ are real analytic (resp. the functions $t\mapsto U(\exp(t\xi))x$ are complex analytic in a
\nbhd\footnote{In the following, if not explicitly mentioned different, the word `neighbourhood'\
in a topological space will mean `an open neighbourhood'.} of real axis for any $\xi\in\mfk g$) in
the norm of \H, but also $\mcl A_G$ is invariant and analytic \wrt the \emm Lie algebra $U(\mfk
g)$ of generators~ $X_\xi\ (\xi\in\mfk g)$; for $x\in\mcl A_G$, also $X_\xi x\in\mcl A_G\ (\forall
\xi\in U(\mfk g))$ and for any basis $\{X_j\in U(\mfk g): j=1,2,\dots \rd:=\dim G\}\subset U(\mfk
g)$ and $x\in\mcl A_G$ there is some $t\neq 0$ such that

\bequ\label{3.1.1(1)}\sum_{n=0}^\infty\frac{|t|^n}{n!}\sum_{j_1,\dots j_n=1}^{\rd}\|X_{j_1}\dots
X_{j_n}x\|<\infty,\enqu

\noidt compare \cite[Chap.11.\S 3]{bar&racz}.

{\bf Let} \emn$\mbs U(G)$~ be the projection of $U(G)$ onto \PH, i.e. $\mbs U(G)$ is a realization
of $G$ in a continuous group of symplectic isometries of $(\mPH,\mOme)$.\footnote{Remember that if
$\bx\equiv P_x\in\mPH$, then $\mbs U(g)\bx\equiv P_{U(g)x}\equiv U(g)P_xU(g^{-1})$.}  For any
$\bx\in\mPH$, {\bf define the orbit} \emn$O_\bx := G\cdot\bx$~ (we shall use also the notation
\emn$g\cdot\bx := \mbs U(g)\bx$~):

\bequ\label{3.1.1(2)} O_\bx=O_{g\cdot\bx} := \{\bz\in\mPH:\bz=g\cdot\bx,\ g\in G\}.\enqu

{\bf Let} \emn$K^\circ_\bz:=K^\circ:=\{h\in G:\mbs U(h)\bz=\bz\}$~ {\bf be the stability (or
`isotropy') group} of the point \emn$\bz\in O_\bx=O_\bz$~. Because \PH\ is a Hausdorff space and
$\mbs U$ is continuous, the group $K^\circ$ is closed, hence it is a Lie subgroup of $G$. The
space $G/K^\circ$ of left cosets $gK^\circ\subset G$ is an analytic manifold (with the analytic
structure coming from $G$ via the natural projection, \cite[Ch.II,Theorem 4.2]{helgas}) and it is
bijectively and continuously mapped onto $O_\bz$ by the mapping \emn$\mbs u:\ m := gK^\circ\mapsto
\mbs U(g)\bz$~. The orbit is not, in general, closed in \PH\ and it need not be a submanifold of
\PH, cf. also \cite[Proposition 2.1.5]{bon-EQM}\&\cite{bon-orbit}. The mapping $\mbs u$ induces,
however, a manifold structure on $O_\bz$ from the analytic manifold $G/K^\circ.$ This manifold
structure is not in general consistent with the relative topology of $O_\bz$ in \PH. If the map is
differentiable, then we have:}

\begin{prop}\label{3.1.2}
Let $\mbs u$ (defined as above) be continuously differentiable in a neighbourhood of a point $m\in
G/K^\circ,$ where $K^\circ:=\{h\in G:\mbs U(h)\bz=\bz\}$. Then there is a \nbhd\ $N_m$ of $m$
such, that the restriction of $\mbs u$ on $N_m$ is a diffeomorphism of $N_m$ onto the submanifold
$\mbs u(N_m)$ of \PH. If $\bz\in P\mcl D_G$ (resp. $\bz\in P\mcl A_G$), then each point $m\in
G/K^\circ$ has a \nbhd\ $N_m$, which is $C^\infty-$diffeomorphic (resp. analytically
diffeomorphic) to $\mbs u(N_m)$, with the submanifold structure from \PH; in this case, the orbit
$O_\bz$ is an immersed submanifold of \PH.
\end{prop}
\begin{proof}
Bijectivity of $\mbs u:G/K^\circ\rarw O_\bz$ and differentiability in a \nbhd of $m$ imply, that
the tangent mapping $T_m\mbs u: T_m(G/K^\circ)\rarw T_{\mbs u(m)}\mPH$ is an isomorphism onto a
finite dimensional subspace of the tangent space of \PH\ at $\mbs u(m)$.  Since each finite
dimensional real subspace of a Banach space is \emn complementable~, the restriction of $\mbs u$
to a \nbhd\ is an immersion. Hence, there is a \nbhd\ $N_m$ of $m$ satisfying the first statement,
compare \cite[p. 549]{3baby}. The rest is a consequence of the invariance of $P\mcl D_G$ and
$P\mcl A_G$ as well as of the inverse mapping theorem, see also \cite{bourb;manif}.
\end{proof}

\pt\label{3.1.3}

\rm We shall assume in the following that $\bz\in P\mcl A_G$, for the orbit $O_\bz$ which we shall
consider. Many of the following considerations are valid, however, also for orbits passing through
$\bz\in P\mcl D_G.$ Let $\msg_\xi\ (\xi\in\mfk g)$ be the (densely defined) vector field on \PH\
corresponding to the generator $X_\xi$ according to \ref{2.3.9} and \ref{2.3.5}. According to the
definition of $O_\bz$, for any $\bx\in O_\bz,$ the vectors $\msg_\xi(\bx)\ (\xi\in\mfk g)$ are
well defined, they span $T_\bx O_\bz$ and depend analytically on $\bx\in O_\bz$. (Note: Here and
in the following, we use without comments the topology on $O_\bz$ inherited from $G/K^\circ$ via
the mapping $\mbs u$ introduced in \ref{3.1.1}) Let $K_\bx^\circ$ be the stability subgroup of $G$
at the point $\bx\in O_\bz$, and let {\bf its Lie algebra} be \emn $\mfk k^\circ_\bx$~. Then the
Lie algebra $\mfk g$ of $G$ is the direct sum
 \bequ\label{3.1.3(1)} \mfk g=\mfk
m_\bx^\circ \oplus \mfk k_\bx^\circ \enqu

\noidt{\rm of two vector spaces (the choice of $\mfk m_\bx^\circ\subset\mfk g$ is nonunique). If
$\{\xi_j\in\mfk g: j=1,2,\dots n:=\dim O_\bz\}$ is a basis of $\mfk m_\bx^\circ,$ then
$\msg_{\xi_j}$ span tangent spaces to $O_\bz$ in any point \by\ lying in some \nbhd of \bx\ in
$O_\bz.$ Then integral curves of $\msg_{\xi_j}\ (j=1,2,\dots n)$ can be used to introduce a
natural coordinate system on $O_\bz$\ in a \nbhd of \bx\ (see \cite[Ch.II. Lemma 4.1]{helgas}). In
these coordinates, the point}

\bequ\label{3.1.3(2)} \by(\verb"t"):=\mbs U(\exp(t_1\xi_1+t_2\xi_2+\dots+t_n\xi_n))\bx\in O_\bz
\enqu

\noidt{\rm corresponds to the point $\verb"t"\in\mbR^n.$ We would like to interpret physically the
coordinates as possible values of `quantities' $\xi_j$ (where the choice of lengths of vectors
$\xi_j$ corresponds to a choice of units). If we, however, take such a point of view that only the
expectation values}

\bequ\label{3.1.3(3)} F_\bx(\xi):= Tr(P_xX_\xi),\ \xi\in\mfk g, \enqu

\noidt {\rm of quantal observables $X_\xi$ in states $\bx\in\mPH$ are measurable, then, for a
general orbit $O_\bz$ and a group $G$, not all values $\verb"t"\in\mbR^n$ (neither all $\verb"t"$
in any open \nbhd of $0\in\mbR^n$) are physically distinguishable. From this point of view the
most natural coordinates of $\bx\in P\mcl A_G$ are just the values $F_\bx(\xi)$ for a conveniently
chosen subset of $\xi\in\mfk g.$ These values need not distinguish points of a \nbhd of $\bx\in
O_\bz:$}

\bequ\label{3.1.3(4)} \left.\frac{\rd}{\rd
t}\right|_{t=0}F_{\exp(t\eta)\cdot\bx}(\xi)=F_\bx([\xi,\eta]),\quad \xi,\eta\in\mfk g, \enqu

\noidt{\rm (compare\rref 2.3.9(3)~), and the derivative might be zero for some nonvanishing
$\eta\in\mfk m_\bx^\circ$ and for all $\xi\in\mfk g.$ If it is the case, then the derivative in
\rref 3.1.3(4)~ vanish on the whole curve $t\mapsto \exp(t\eta)\cdot\bx\ (t\in\mbR).$ This is
easily seen with a help of the next Lemma, cf. Proposition \ref{3.1.6}:}

\begin{lem}\label{3.1.4}
For all $g\in G$ and $\xi\in\mfk g$, we have:

\bequ\label{3.1.4(1)} U(g)X_\xi U(g^{-1}) = X_{Ad(g)\xi}, \enqu

\noidt where the adjoint representation $Ad$ of $G$ is defined in \ref{2.3.9}. \end{lem}
\begin{proof}
According to the definition of $Ad$, the curve $t\mapsto g\,\exp(t\xi) g^{-1}$ at the identity $e$
of $G$ determines the tangent vector $Ad(g)\xi\in T_e G$, and this one, in turn, according to the
definition of the Lie algebra $\mfk g$, determines a unique curve $t\mapsto\exp(t Ad(g)\xi)$\ in
$G$ at $e$. Hence,
\bequ\label{3.1.4(2)} g\,\exp(t\xi)\, g^{-1} = \exp(t Ad(g)\xi)\quad\forall
t\in\mbR, g\in G, \xi\in \mfk g.\enqu

\noidt From the definition\rref 2.3.9(2)~ of the generators $X_\xi$ of the representation $U(G)$,
we then obtain
\bequ\label{3.1.4(3)} U(g)\exp(-itX_\xi)U(g^{-1})=U(g\,\exp(t\xi) g^{-1})=
U(\exp(t\,Ad(g)\xi)), \enqu

\noidt and after differentiation at $t=0$ we obtain\rref 3.1.4(1)~. \end{proof}

\pt\label{3.1.5} {\rm Suppose now that $F_\bx([\xi,\eta])=0$ for all $\xi\in\mfk g$ at some
$\bx\in O_\bz$. Substitution of $\exp(t\eta)\cdot\bx$ to the place of $\bx$\ gives according to
the preceding lemma:}
\barr \label{3.1.5(1)} F_{\exp(t\eta)\cdot\bx}([\xi,\eta])&=&
Tr(U(\exp(t\eta))P_xU(\exp(-t\eta))X_{[\xi,\eta]}) \\
\label{3.1.5(2)}
&=& -i\,Tr\left(P_xU(\exp(-t\eta))[X_\xi,X_\eta]U(\exp(t\eta))\right) \\
\label{3.1.5(3)}&=& -i\,Tr\left(P_x[U(\exp(-t\eta))X_\xi U(\exp(t\eta)),X_\eta]\right) \\
\label{3.1.5(4)}&=& -i\,Tr\left(P_x[X_{Ad(\exp(-t\eta))\xi},X_\eta]\right) \\
\label{3.1.5(5)}&=& F_\bx([Ad(\exp(-t\eta))\xi,\eta]). \earr

{\rm  We used\rref 2.3.4(3)~ in\rref 3.1.5(1)~, it was used the formula\rref 2.3.9(3)~ in\rref
3.1.5(2)~, we considered commutativity of $U(\exp(t\eta))$ with $X_\eta$ in\rref 3.1.5(3)~, and in
the last step the Lemma \ref{3.1.4} was used. According to the assumption, the expression\rref
3.1.5(5)~ vanishes for all $\xi\in\mfk g,$ since $Ad(g): \mfk g\rarw\mfk g.$ Hence we have
obtained:}

\begin{prop}\label{3.1.6}
For all $\bx\in P\mcl A_G,\ \xi,\eta\in\mfk g$, and all $t\in\mbR$, it is

\bequ\label{3.1.6(1)} \frac{\rd}{\rd
t}F_{\exp(t\eta)\cdot\bx}(\xi)=F_\bx([Ad(\exp(-t\eta))\xi,\eta]).
 \enqu

\noidt If, in particular, the derivative vanishes for all $\xi\in\mfk g$
 at one value of $t\in\mbR$, then it vanishes for all $\xi\in\mfk
 g$ at all $t\in\mbR$, for the given $\eta$.
\end{prop}

\pt\label{3.1.7}{\rm From the preceding considerations, we see that the numbers $F_\bx(\xi)$
cannot distinguish points \bx\ on the integral curves of the vector fields $\msg_\eta$ passing
through \bx\ iff $F_\bx([\xi,\eta])=0$ for all  $\xi\in\mfk g$. Physical states lying on such
curves should be identified mutually, if we could measure only expectations of observables
$X_\xi,\ (\xi\in\mfk g)$. Such an identification of points of orbits $O_\bz\ (\bz\in P\mcl A_G)$
will be performed in the next section. After the identification, we obtain from each orbit an
even-dimensional manifold endowed with canonical symplectic structure obtained from the symplectic
structure \Ome\ on \PH.}

\pt\label{3.1.8}{\rm Note that, for an irreducible representation $U(G)$, there can occur in
$P\mcl A_G$ mutually nonhomeomorphic orbits. But any such an orbit $O_\bz$, if it is considered in
the Hilbert space \H\ as the union of equivalence classes $\bx=\{z\in\mH: z=\mlam
x,\mlam\in\mbC\}\subset\mH$ for all $\bx\in O_\bz$, contains total sets of vectors in \H. Such
`\emn overcomplete families of vectors~' in \H\ were discussed e.g. in
\cite{berez2,davies,klaud1,perel1} and they are interesting from the point of view of
representation theory, as it is explained e.g. in \cite{dix2}, and used in \cite{ali}.}

\section{Classical phase spaces from the quantal state
space}\label{sec;3.2}

\pt\label{3.2.1}{\rm We have constructed orbits $O_\bz$ of the action of $G$, $\mbs U(G)$, on \PH\
from pure states of conventional QM. We shall construct now \emn symplectic homogeneous spaces~ of
$G$ from these orbits, of which the symplectic structure is a canonical restriction of the form
\Ome\ defined on \PH\ in Sec. \ref{sec;2.2}. The obtained symplectic manifolds are all
symplectomorphic to the orbits of $G$ in the \emn coadjoint representation $Ad^*(G)$~ on the space
$\mfk g^*$ dual to the Lie algebra $\mfk g$ endowed with the natural \emn Kirillov-Kostant
symplectic form~. }

\pt\label{3.2.2}{\rm Let \emn $\mOme^\circ$~ denotes the restriction of the form \Ome\ onto the
immersed submanifold $O_\bz\ (\bz\in P\mcl A_G)$ of \PH. Since the vector fields $\msg_\xi\
(\xi\in\mfk g)$ span $T_\bx O_\bz$ at each point $\bx\in O_\bz$, the form $\mOme^\circ$ is
uniquely defined by its values on vectors $\msg_\xi(\bx)\ (\xi\in\mfk g, \bx\in O_\bz)$:

\bequ\label{3.2.2(1)} \mOme^\circ_\bx(\msg_\xi,\msg_\eta):=\mOme_\bx(\msg_\xi,\msg_\eta)=
i\,Tr(P_x[X_\xi,X_\eta]), \enqu

\noidt where we used formula\rref 2.3.6(2)~ and the restrictions of the fields $\msg_\xi$ onto
$O_\bz$ are equally denoted as the unrestricted fields. According to the definition\rref 3.1.3(3)~
and with the use\rref 2.3.9(3)~, we can write

\bequ\label{3.2.2(2)} \mOme^\circ_\bx(\msg_\xi,\msg_\eta)= -F_\bx([\xi,\eta]).\enqu

\noidt If we denote by
 \[ \mbs u_\circ: O_\bz=\mbs u(G/K^\circ)\rarw \mPH \] 

\noidt the inclusion of the orbit into \PH, then the form $\mOme^\circ$ is simply the pull-back of
\Ome\ by $\mbs u_\circ$:

\bequ\label{3.2.2(3)} \mOme^\circ = \mbs u_\circ^*\,\mOme. \enqu

Since exterior derivative commutes with any pull-back, e.g. \cite[p.204]{3baby}, we see that the
\emn two-form $\mOme^\circ$ on $O_\bz$ is closed~.  It is clear from\rref 3.2.2(2)~, that
$\mOme^\circ$ is degenerate iff for some  $\eta\neq 0$ and for all $\xi\in\mfk g$ the term
$F_\bx([\xi,\eta]) = 0$ for some\ \bx\ in the orbit. This is, however, the situation discussed in
\ref{3.1.7}.}

\pt\label{3.2.3}{\rm The mapping $F_\bx: \mfk g\rarw \mbR, \xi\mapsto F_\bx(\xi)$ is linear
because of linearity of $\xi\mapsto X_\xi$, hence $F_\bx\in\mfk g^*$ for any $\bx\in O_\bz$.
Define the action of $G$ on the functionals $F_\bx\ (\bx\in O_\bz)$ by

\bequ\label{3.2.3(1)} g\cdot F_\bx := F_{g\cdot\bx},\ \text{for all}\  g\in G.\enqu

Then analogous computations to those in \ref{3.1.5} lead to:

\bequ\label{3.2.3(2)} F_{g\cdot \bx}(\xi)=F_\bx(Ad(g^{-1})\xi), \ \text{what\ means:}\, g\cdot
F_\bx= Ad^*(g)F_\bx.\enqu

Let now $K_\bx$ be, as above, the stability subgroup of $G$ of the coadjoint action at the point
$F_\bx\in\mfk g^*.$ Since $Ad^*$ is continuous, $K_\bx$ is closed. Let $\mfk k_\bx$ be the Lie
algebra of $K_\bx.$ Then it is clear, that:}

\begin{lem}\label{3.2.4} Let $\bx\in P\mcl A_G,\ \by:= g\cdot\bx.$
Then $K_\by= g K_\bx g^{-1},\ \mfk k_\by= Ad(g)\mfk k_\bx, $  and $K_\bx^\circ\subset K_\bx$ for
all \bx\ and all $g\in G.$ It is $\xi\in\mfk k_\bx$ iff

\bequ\label{3.2.4(1)} F_\bx([\xi,\eta]) = 0,\quad \forall \eta\in\mfk g.\enqu
\end{lem}

{\rm A trivial consequence of this is, according to\rref 3.2.2(2)~, the}

\begin{prop}\label{3.2.5}
$\mOme_\bx^\circ(\msg_\xi,\msg_\eta) = 0$\ for all\ $\eta\in\mfk g$\ iff\ $\xi\in\mfk k_\bx.$
\end{prop}

\pt\label{3.2.6}{\rm We can decompose $O_\bz$ into {\bf equivalence classes} \bequ\label{3.2.6(1)}
[\bx]:=\{g\cdot\bx: g\in K_\bx\},\ x\in O_\bz\quad (\bz\in P\mcl A_G).\enqu

The action of $G$ on $O_\bz$ is analytic, and \emm$[\bx]$ are analytic submanifolds of $O_\bz$~
(if $O_\bz$ is endowed with the topology of $G/K_\bx^\circ$) which are mutually diffeomorphic for
all $\bx\in O_\bz.$ Hence $O_\bz$ can-be considered as a fibred manifold with a typical fibre
diffeomorphic  to \emn$K_\bz\cdot\bz=[\bz]$~, which is in turn diffeomorphic to
$K_\bx/K_\bx^\circ\ (\bx\in O_\bz).$  Let us denote the {\bf base space} by \emn$M=M_\bz$~:

\bequ\label{3.2.6(2)} M:= M_\bz := \{[\bx] : \bx\in O_\bz\}, \enqu

\noidt which is endowed with the natural factor topology given by the continuity and openness
condition on the projection
\bequ\label{3.2.6(3)} p_M: O_\bz \rarw M_\bz,\ \bx\mapsto p_M(\bx) :=
[\bx].\enqu

From the definitions\rref 3.1.3(3)~ of $F_\bx$ and of the action of $G$ on $F_\bx$ in\rref
3.2.3(1)~, we see that $[\bx]$ are exactly those subsets of $O_\bz,$ on which expectations of all
the observables $X_\xi\ (\xi\in\mfk g)$ remain constant. }

\begin{lem}\label{3.2.7}
$\mOme^\circ_{h\cdot\bx}(\msg_\xi,\msg_\eta) = \mOme^\circ_\bx(\msg_\xi,\msg_\eta)$\ for all $h\in
K_\bx$\ and all $\eta, \xi\in\mfk g.$
\end{lem}
\begin{proof}
Immediate from\rref 3.2.2(2)~ and the definition of $K_\bx.$
\end{proof}

\pt\label{3.2.8} {\rm Let $p_{M*} := Tp_M:\ TO_\bz\rarw TM_\bz$ be the tangent mapping
corresponding to the natural projection\rref 3.2.6(3)~. For a general \emn vector field \sg\ on
$O_\bz$~,\ the vectors $Tp_M\msg(\bx)$ are mutually different for various choices of $\bx\in
[\bz].$ Let, however, $t\mapsto g(t)$ be any differentiable curve in $G$. Then curves $t\mapsto
g(t)\cdot\bx$ and $c_h: t\mapsto g(t)h\cdot\bx$ for any $h\in K_\bx$ are projected by $p_M$ onto
the same curve $t\mapsto [g(t)\cdot\bx]$ in $M_\bz.$ This is true due to the validity of

\bequ\label{3.2.8(1)} [g\cdot\bx] = K_{g\cdot\bx} g\cdot\bx=gK_\bx g^{-1}g\cdot\bx=
gK_\bx\cdot\bx,\enqu

\noidt for all $g\in G$,

\bequ\label{3.2.8(2)} [gh\cdot\bx]=ghK_\bx\cdot\bx=gK_\bx\cdot\bx=[g\cdot\bx],\ \forall h\in
K_\bx,\ g\in G.\enqu

Hence tangent vectors $\dot{c}_h\in T_{h\cdot\bx}O_\bz$ corresponding to the curves $c_h$ with
$g(t=0):= e$ have identical projections $Tp_M(\dot{c}_h)=Tp_M(\dot{c}_e)\in T_{[\bx]}M_\bz$ for
all $h\in K_\bx.$ If we set $g(t):=\exp(t\xi),$ i.e. $\dot{c}_h =\msg_\xi(h\cdot\bx),$ then we
have obtained:

\begin{lem}\label{3.2.9} All the vector fields $\msg_\xi\
(\xi\in\mfk g)$ on $O_\bz$ are projected onto unambiguously defined (analytic, if $\bz\in P\mcl
A_G$) vector fields $\msg^M_\xi$ on $M_\bz$:

\bequ\label{3.2.9(1)} \msg^M_\xi([\bx]) := Tp_M\msg_\xi(h\cdot\bx) \enqu

\noidt for all $h\in K_\bx$.\end{lem} }

\begin{prop}\label{3.2.10} There is a unique symplectic form
$\mOme^M$ on $M_\bz$ satisfying

\bequ\label{3.2.10(1)} \mOme^M_{[\bx]}(\msg_\xi^M,\msg_\eta^M) =
\mOme^\circ_\bx(\msg_\xi,\msg_\eta)=(p^*_M\mOme^M)_\bx(\msg_\xi,\msg_\eta) \enqu

\noidt for all $\xi,\eta\in\mfk g$ and all $\bx\in O_\bz.$ $p^*_M$ in\rref 3.2.10(1)~ is the \emn
pull-back~ corresponding to the projector $p_M$ (compare \cite{abr&mars}, resp. also
\cite[A.3.11]{bon-EQM} for the definition).\end{prop}
\begin{proof}
The first equality can be considered as a definition of a two-form $\mOme^M$, which is correct due
to two preceding lemmas and the fact, that vectors $\msg_\xi^M(p_M\bx)\ (\xi\in\mfk g)$ contain a
basis of $T_{[\bx]}M_\bz:\ \msg^M_\eta(p_M\bx)=0$ implies $\eta\in\mfk k_\bx$ and $M_\bz$ is
diffeomorphic to $G/K_\bx.$ This ensures also the uniqueness of $\mOme^M$. The second equality is
a consequence of the definition\rref 3.2.9(1)~ of $\msg^M_\xi$ and it shows, how $\mOme^\circ$ can
be reconstructed from $\mOme^M$.

The bilinearity of $\mOme^M$ follows from linearity of the mapping $T_\bx\,p_M$ and the
bilinearity of $\mOme^\circ$, antisymmetry is trivial and closedness holds due to commutativity of
the exterior derivative with the pull-bacs: ${\rm d}p^*_M=p^*_M{\rm d}$, and due to closedness of
$\mOme^\circ.$ Nondegeneracy follows from\rref 3.2.2(1)~ and \ref{3.2.4}, which completes the
proof.
\end{proof}

\pt\label{3.2.11}{\rm As it was pointed out, the manifold $M:= M_\bz\ (\bz\in P\mcl A_G)$ is
diffeomorphic to $G/K_\bz$, where $K_\bz$ is the stability group of the point $F_\bz\in\mfk g^*$
\wrt\ the coadjoint representation of $G$. On the other hand, the form $\mOme^M$ on $M$ has the
expression

\bequ\label{3.2.11(1)} \mOme^M_{[\bz]}(\msg_\xi^M,\msg^M_\eta)=- F_\bz([\xi,\eta]),\enqu

\noidt which follows from\rref 3.2.2(2)~. This is, up to the sign, the canonical symplectic form
on the orbit of $Ad^*(G)$ passing through $F_\bz$ and diffeomorphic to $G/K_\bz.$ Hence the
symplectic manifold $(M;\mOme^M)$ is symplectomorphic to a \emn Kirillov-Kostant  symplectic
orbit~, compare \cite{kiril}. This manifold is here interpreted as a \emn classical phase space~
obtained by the above described canonical procedure from a given quantal system, in which \emn
interpretation of observables~ is (at least partly) determined by a Lie group action $U(G)$. This
action is projected on the coadjoint action $Ad^*(G)$ on $M,$ see\rref 3.2.3(2)~. Almost obvious
is the following}

\begin{prop}\label{3.2.12} The vector fields $\msg^M_\xi\ (\xi\in \mfk g)$
are \emn globally Hamiltonian vector fields~ on the symplectic manifold $(M;\mOme^M)$
corresponding to Hamiltonian functions

\bequ\label{3.2.12(1)} f_\xi:\ [\bx]\mapsto f_\xi([\bx]):= F_\bx(\xi).\enqu

\noidt They generate Hamiltonian flows $F_t^\xi$ on $M$:

\bequ\label{3.2.12(2)} F^\xi_t: [\bx]=p_M\bx\mapsto F_t^\xi(p_M\bx):= p_M\left(\mbs
U(\exp(t\xi))\bx\right).\enqu
\end{prop}
\begin{proof}
From the definition of $\msg_\xi^M$ in \ref{3.2.8} and \ref{3.2.9} the relation\rref 3.2.12(2)~
follows. Differentiation of $f_\xi$ according to\rref 3.1.6(1)~ and\rref 3.2.2(2)~ gives

\bequ\label{3.2.12(3)} \rd f_\xi=-i(\msg_\xi^M)\mOme^M,\enqu

\noidt compare\rref 1.3.5(4)~. This proves the first statement.
\end{proof}

{\rm With the usual definition of Poisson brackets on $(M;\mOme^M)$, we obtain the obvious
(compare also\rref 1.3.7(3)~ +\rref 1.3.7(4)~)}

\begin{lem}\label{3.2.13} $\{f_\xi,f_\eta\}= - f_{[\xi,\eta]}$ for
all $\xi,\eta\in \mfk g.$
\end{lem}

\pt\label{3.2.14}{\rm This shows, that the action of $Ad^*(G)$ is \emn strictly Hamiltonian~.
Since for the generators of $U(G)$ in \H\ we have $X_{[\xi,\eta]}=-i[X_\xi,X_\eta],$\rref
2.3.9(3)~, the Lemma \ref{3.2.13} establishes the usual \emn correspondence between classical and
quantal observables~ associated with generators of the group action. }

\section{Classical mechanical projections of quantal dynamics}\label{sec;3.3}

\pt\label{3.3.1}{\rm Let the time evolution of a given system in QM be described by a one
parameter subgroup of $U(G)$ corresponding to an element $\chi\in\mfk g$. Then, for a given
$\bz\in\mPH$, the flow $\mbs U(\exp(t\chi))$ leaves the orbit $O_\bz$ invariant. If $\bz\in P\mcl
A_G,$ then this flow is projected onto the Hamiltonian flow on $M_\bz$ generated by the
Hamiltonian function $f_\chi$ with the corresponding Hamiltonian vector field $\msg^M_\chi$, as it
was described above. Models one frequently encounters are, however, in which the time evolution is
given by a one parameter group of unitaries $U_A(\mbR):$

\bequ\label{3.3.1(1)} U_A: \ t\mapsto U_A(t) := \exp(-itA),\ A=A^*,\enqu

\noidt where the generator $A$ has not the form $X_\chi$ for any $\chi\in\mfk g.$ The orbits
$O_\bz$ are then in general not invariant \wrt\ the action of $U_A(\mbR).$ We shall be interested
here in the question whether and how such an action $\mbs U_A(\mbR)$ can be projected onto a
Hamiltonian flow on $M_\bz$.}

\pt\label{3.3.2}{\rm Let $A$ be any selfadjoint operator on \H\ and $E_A$ the corresponding
projector-valued measure on \bR. Assume that $\bz\in P\mcl A_G$\ (defined in \ref{3.1.1}) and that
$O_\bz:= \mbs U(G)\bz$ is contained in the form domain of $A$, i.e. the integral in

\bequ\label{3.3.2(1)} f_A(\bx) := Tr(P_xA) := \int_\mbR\mlam\, Tr(P_xE_A(\rd\mlam)) \enqu

\noidt converges absolutely for all $\bx\in O_\bz.$ In an analogy with the constructions of the
preceding sections, the function $f_A$ will be considered as a candidate for a \emn classical
observable corresponding to the quantal observable~ $A.$ We shall require that

\bequ\label{3.3.2(2)} f_A \in C^\infty(O_\bz).\enqu

This requirement is fulfilled in the following situation:}

\begin{lem}\label{3.3.3} {\bf Let} \emn $\mcl E(\mfk g)$~ be the linear space
of all \emn polynomials in selfadjoint generators~ $X_\xi\ (\xi\in\mfk g)$ of $U(G)$ with complex
coefficients. Assume that for a fixed $\bz\in P\mcl A_G$ and for any $\bx\in O_\bz$ and any
$E\in\mcl E(\mfk g)$ there is an open \nbhd $N(\bx,E)$ of the identity $e\in G$ such, that the
function

\bequ\label{3.3.3(1)} g\mapsto \| AEU(g)x \|\quad (x\in\bx)\enqu

\noidt is uniformly bounded on $N(\bx,E).$ Here $A$ is a given symmetric operator on \H\
containing $\mcl E(\mfk g)U(G)z := \{EU(g)z: E\in\mcl E(\mfk g),\ g\in G\}$ in its domain $D(A),\
z\in\bz.$ Set $f_A(\bx):= (x,Ax)$ for $\|x\|=1,\ x\in\bx\in O_\bz.$ Then $f_A$ is infinitely
differentiable on $O_\bz.$
\end{lem}
\begin{proof}
It suffices to prove infinite differentiability of the function $g\mapsto f_A(g\cdot\bx)$ defined
on $G.$ For any $E_1, E_2 \in\mcl E(\mfk g)$ the functions $g\mapsto E_jU(g)x\ (j=1,2)$ are
norm-analytic according to\rref 3.1.1(1)~, see also \cite{bar&racz}. Consequently, the function

\bequ\label{3.3.3(2)} (g_1;g_2)\mapsto (E_1U(g_1)x, AE_2U(g_2)x)\ \text{from}\ G\times G\
\text{to}\ \mbC\enqu

\noidt is infinitely differentiable in each variable $g_1,\, g_2$ separately and any partial
derivative (in the direction of some one parameter subgroup of $G$) has the form\rref 3.3.3(2)~
(with some other $E_j$'s). To prove differentiability of

\bequ\label{3.3.3(3)} g\mapsto (E_1U(g)x, AE_2U(g)x),\enqu

\noidt it suffices to prove simultaneous continuity of all functions of the form\rref 3.3.3(2)~ in
both variables $g_1, g_2$. It follows, however, from the assumption of uniform boundedness  on
$N(\bx,E_2)$, analyticity of $U(g)x$ \wrt $U(G)$ and continuity of $U(g)$:

\barr \label{3.3.3(4)} & \left|(E_1U(g_1)x, AE_2U(g_2)x) - (E_1x,
AE_2x)\right| \leq \nonumber \\
& \|E_1(U(g_1)-I)x\|\cdot\|AE_2U(g_2)x\| + \|E_2(U(g_2) - I)x\|\cdot\|AE_1x\|.\earr

This concludes the proof.
\end{proof}

\pt\label{3.3.4}{\rm If the assumptions of the preceding lemma are valid for $A,$ the explicit
expressions for the partial derivatives $\partial_\xi f_A$ along the curves $t\mapsto
exp(t\xi)\cdot\bx$ have the form $(\|x\|=1,\ \xi,\eta\in \mfk g)$:

\barr\label{3.3.4(1)}
\partial_\xi f_A(\bx) &=& 2\,\Im(x, AX_\xi x),\\
\label{3.3.4(2)}\partial_\eta\partial_\xi f_A(\bx) &=& 2\,\Re[(X_\xi x, AX_\eta x) - ( x, AX_\xi
X_\eta x)],
 \earr

 \noidt and similarly for higher derivatives. For these expressions,
 we shall use also forms which are literally valid only if the
 set $\mcl E(\mfk g)U(G)x$ is mapped by $A$ into $\mcl D_G$:

 \barr\label{3.3.4(3)} \partial_\xi f_A(\bx) &=:& i\, Tr(P_x[X_\xi,
 A]), \\
 \label{3.3.4(4)}\partial_\eta\partial_\xi f_A(\bx) &=:& i^2\,
 Tr(P_x[X_\eta,[X_\xi,A]]),
 \earr
 \noidt etc. Also in more general cases, we shall write
 symbolically

 \bequ\label{3.3.4(5)} i\, Tr(P_x[X_\xi, A]) :=
 f_{i[X_\xi,A]}(\bx) := \partial_\xi f_A(\bx).\enqu

 The Definition \ref{3.3.6} (ii) deals with such symbols.}

 \begin{exmp}\label{3.3.5}Assumptions of the Lemma \ref{3.3.3} are satisfied,
 e.g. for\nl
 (i)\ \ all bounded operators $A=A^*\in\mLH,$\nl
 (ii)\ all symmetric operators $A\in\mcl E(\mfk g).$
 \end{exmp}
\begin{defs}\label{3.3.6}
(i)\label{3.3.6(i)} Let $A$ be a symmetric operator on \H\ with $O_\bz\subset D(A)$ for some
$\bz\in P\mcl A_G$ and let $f_A: \bx\mapsto f_A(\bx) := Tr(P_xA)$ be infinitely differentiable on
$O_\bz$. Let $K_\bx$ be the stability group of $F_\bx\in \mfk g^*,\ F_\bx(\xi):= Tr(P_xX_\xi),$
\wrt  the coadjoint representation of $G$ and $[\bx] := K_\bx\cdot\bx\ (\bx\in O_\bz).$ If

\bequ\label{3.3.6(1)} f_A([\bx]) := f_A(\bx) = f_A(h\cdot\bx),\quad\forall h\in K_\bx, \forall
\bx\in O_\bz,\enqu

\noidt the operator A will be called a \emm U(G)-classical operator on $O_\bz$~ or simply a \emm
\bz-classical operator~.\nl

\noidt (ii)\label{3.3.6(ii)} Let $\mbs A:= A_1A_2\dots A_n$ be formal product of some selfadjoint
operators $A_j^*=A_j,\ j=1,2,\dots n$. Let $A_0:= I.$ Suppose, that for some $j\in\{0,1,2,\dots
n\}$ the products $A_{j+1}\dots A_n$ and $A_jA_{j-1}\dots A_1A_0$ are well defined operators with
$U(G)z\ (0\neq z\in\bz)$ lying in the intersection of their domains. Denote then (with $x\in\bx,
\|x\|=1, \bx\in O_\bz$)

\bequ\label{3.3.6(2)} f_{\mbs A}(\bx):=f_{A_1A_2\dots A_n}(\bx) := (A_jA_{j-1}\dots A_1 x,\,
A_{j+1}A_{j+2}\dots A_n x).\enqu

For any other $j\in\{1,\dots n\}$ satisfying these conditions the values in\rref 3.3.6(2)~ will be
the same. If $f_{\mbs A}\in C^\infty(O_\bz)$ and if\rref 3.3.6(1)~ is valid (with $\mbs
A\hookrightarrow A$) for $f_{\mbs A}$, then $\mbs A$ will be called a \emm generalized
\bz-classical operator~. The same name will be given to any formal complex finite linear
combination $\mbs B$ of generalized \bz-classical operators $\mbs A^\tau := A_1^\tau A_2^\tau\dots
A_{n_\tau}^\tau$:

\bequ\label{3.3.6(3)} \mbs B:= \sum_\tau\mlam_\tau\mbs A^\tau,\enqu

\noidt and we shall set

\bequ\label{3.3.6(4)} f_{\mbs B}([\bx]) := f_{\mbs B}(\bx):= \sum_\tau\mlam_\tau\,f_{\mbs
A^\tau}(\bx).\enqu \nl

The adjective `generalized' will be sometimes omitted.
\end{defs}

\begin{exmp}\label{3.3.7}
\begin{itemize}
\item[(i)] All the generators $X_\xi\ (\xi\in\mfk g)$ are
\emn \bz-classical~ for all $\bz\in P\mcl A_G.$
\item[(ii)] If, for some $\bz\in\mPH: K_\bz=K^\circ_\bz$, (cf.
\ref{3.1.1}) and $f_A\in C^\infty(O_\bz),$ then $A$ is \bz-classical.
\item[(iii)] If $A$ is \bz-classical and $X_\xi,\dots X_\chi \in
U(\mfk g),$ then all the symbols $[X_\xi,[X_\eta,\dots [X_\chi, A]\dots]]$ represent generalized
\bz-classical operators.  We can see this from \ref{3.3.4} and\rref 3.2.8(2)~:
$$ f_A(g\cdot \bz)=f_A([g\cdot\bz])=f_A([gh\cdot \bz])$$
and differentiations and induction give the result.
\item[(iv)] Let $f_A\in C^\infty(O_\bz)$ and all the $K_\bx$ be
symmetry groups of the observable $A: U(h^{-1})AU(h)=A$ for all $h\in K_\bx$ and all $\bx\in
O_\bz$ (e.g. if $K_\bz$ is a normal subgroup of $G$ and $K_\bz$ is a symmetry group of $A$). Then
$A$ is \bz-classical.
\end{itemize}
\end{exmp}

\pt\label{3.3.8} {\rm If $A$ is \emn\bz-classical~, then the function $f_A$ can be considered as a
function on $M_\bz$ according to\rref 3.3.6(1)~ and then $f_A\in C^\infty(M).$ Denote by
$\msg_A^M$ the \emn Hamiltonian vector field on $M$~ corresponding to the Hamiltonian function
$f_A: m\mapsto f_A(m), m \in M.$ Choose a system $\msg_j\ (j=1,\dots dim M)$ of vector fields on
$M$ forming a basis of $T_mM$ for all $m$ in a \nbhd of $m_0\in M.$ Since the symplectic form
$\mOme^M$\  is nondegenerate, the inverse matrix to $\mOme^M_m(\msg_j,\msg_k)$ with elements
$\mOme_M^{jk}(m)\ (j,k = 1,2,\dots \dim M)$ exists:

\bequ\label{3.3.8(1)} \sum_i\mOme^{ji}_M(m)\mOme^M_m(\msg_i,\msg_k)=
\sum_i\mOme^M_m(\msg_k,\msg_i)\mOme_M^{ij}(m)=\delta_{jk}. \enqu

From the connection between Hamiltonian vector fields and corresponding Hamiltonian functions, we
obtain:

\bequ\label{3.3.8(2)} \msg_A^M(m)=\sum_{j,k}\mOme_M^{jk}(m)\rd_mf_A(\msg_k)\msg_j(m).\enqu

For Poisson brackets of functions $f_A$ and $f_B$ on $M$ corresponding to \bz-classical operators
$A$ and $B,$ we obtain with a help of\rref 3.3.8(1)~:}

\bequ\label{3.3.8(3)} \{f_A,f_B\}(m):= \mOme^M_m(\msg_A^M,\msg^M_B) =
-\,\sum_{j,k}\rd_mf_A(\msg_j)\mOme_M^{jk}(m)\rd_mf_B(\msg_k).\enqu

{\rm If $\msg_j$ are Hamiltonian vector fields corresponding to generators $X_j\in U(\mfk g),$
then we obtain according to\rref 3.3.4(5)~

\bequ\label{3.3.8(4)} \rd_mf_A(\msg_j)=f_{i[X_j,A]}(m) \enqu

\noidt and the Poisson bracket\rref 3.3.8(3)~ has the form

\bequ\label{3.3.8(5)} \{f_A,f_B\}(m)=-\,\sum_{j,k}f_{i[X_j,A]}(m) \mOme^{jk}_M(m) f_{i[X_k,B]}(m).
\enqu

If the operator $B$ is one of the generators of $U(G),\ B:= X\in U(\mfk g),$ then the Poisson
bracket\rref 3.3.8(5)~ has the expression:

\bequ\label{3.3.8(6)} \{f_A,f_X\}(m) = i\,Tr(P_x[A,X])= f_{i[A,X]}(m),\enqu

\noidt where $x\in\bx\in[\bx] := m\in M.$ The results\rref 3.3.8(5)~ and\rref 3.3.8(6)~ have to be
compared with \ref{3.2.14}. If the orbit $O_\bz$ coincides with the manifold $M:= M_\bz,$ then the
vector field  $\msg_A^M$ in\rref 3.3.8(2)~ is the skew-orthogonal projection of $\msg_A$ (from
\rref 2.3.5(1)~) onto $M$, the skew-orthogonality being defined by the form \Ome\ on \PH, see Sec.
\ref{sec;2.2}.}

\pt\label{3.3.9}{\rm The unitary group $U_A: t\mapsto U_A(t):= \exp(-itA)$ does not leave the
orbit $O_\bz$ invariant for a general selfadjoint \bz-classical operator $A.$ Then we would like
to compare the classical Hamiltonian evolution on $M_\bz$ generated by $f_A$ (with the flow
$F_t^A$) and the quantal evolution on \PH\ described by the flow $\mbs U_A(t).$ From the point of
view of this work, the 'quantities of interest' are generators of the representation $U(G)$. The
evolutions of the corresponding functions $f_X\ (X = X^*\in U(\mfk g))$ are described by

\bequ\label{3.3.9(1)} \frac {\rd}{\rd t}f_X^t=\{f_A,f_X^t\} = f^t_{i[A,X]} \enqu

\noidt in both cases of the classical flow $F^A_t$ as well as of the quantal evolution $\mbs
U_A(t),$ compare \ref{3.2.12},\rref 2.3.10(6)~ and\rref 3.3.8(6)~. The difference is between the
two cases in the meaning of $f^t$:

\begin{itemize}
\item[(i)] In the case of the flow $F^A_t$ on $M$ for any $f\in
C^\infty(M),$ we define
\bequ\label{3.3.9(2)} f^t(m) := f(F_t^Am),\quad m\in M, \enqu

\noidt and the flow $F_t^A$ has to be determined from\rref 3.3.9(1)~\ ($\forall X^*=X\in U(\mfk
g)$).

\item[(ii)] In the quantal case, we have given the flow $\mbs U_A$
on \PH\ and for functions $f$ on (the dense $\mbs U_A$-invariant subset of) \PH\ we set
\bequ\label{3.3.9(3)} f^t(\bx):= f(\mbs U_A(t)\bx).\enqu

The functions $f_B$ for any \bz-classical $B$ are defined in the both cases by the formula
\bequ\label{3.3.9(4)} f_B(\bx):= Tr(P_xB). \enqu

The `classical $f_B$' is the restriction of the `quantal $f_B$' to the manifold $M:= M_\bz.$ The
classical flow is the specific kind of restriction of the flow $\mbs U_A$ onto $M$\ (compare\rref
3.3.8(2)~ and the note in the last sentence of \ref{3.3.8}).
\end{itemize}

Although the rules for computation of the functions

\bequ\label{3.3.9(5)} t \mapsto f_X(F_t^A[\bx]),\ [\bx] \in M_\bz,\ X=X^*\in U(\mfk g),\enqu

and

\bequ\label{3.3.9(6)} t\mapsto f_X(\mbs U_A(t)\bx),\ \bx\in [\bx]\in M_\bz,\ X=X^*\in U(\mfk g),
\enqu

\noidt seem to be very similar, the mutually corresponding functions from\rref 3.3.9(5)~ and\rref
3.3.9(6)~ might be radically different for an abstractly defined selfadjoint (\bz-classical)
operator $A$. We shall give in the next chapter an example, in which both the functions from\rref
3.3.9(5)~ and\rref 3.3.9(6)~ (given by the same $X\in U(\mfk g)$ and with the same initial
condition $\bx\in O_\bz$) are periodic with different periods (and, moreover, with mutually
different dependence of these periods on the initial condition \bx); the corresponding orbits in
$\mfk g^*$:

\bequ\label{3.3.9(7)} \{ F^{cl}_{[\bx]}(t): t\in\mbR\}\subset \mfk g^*,\ \text{with}\
F^{cl}_{[\bx]}(t): \xi\mapsto f_{X_\xi}(F_t^A[\bx]),\ \xi\in \mfk g,\enqu

\noidt and the orbit

\bequ\label{3.3.9(8)} \{F^q_\bx(t): t\in\mbR\}\subset \mfk g^*,\ \text{where}\ F^q_\bx(t): \mfk
g\rarw \mbR,\ \xi\mapsto F^q_\bx(t)(\xi):= f_{X_\xi}(\mbs U_A(t)\bx),\enqu

\noidt are mutually different closed curves in $\mfk g^*,$ see \ref{4.1.10}. }

\pt\label{3.3.10}{\rm  We expect, contrary to the above mentioned example, that in certain
situations the para\-me\-tri\-zed curves in $\mfk g^*$ defined in\rref 3.3.9(5)~ and\rref
3.3.9(6)~ will be in some sense close one to another, at least for not too large times $t\in\mbR.$
We mean namely such situations, in which $A$ is the Hamiltonian operator of a `realistic' quantal
model and the initial condition \bx\ leads to subsequent evolution $\mbs U_A(t)\bx,$ which is
sufficiently well approximated by laws of CM. For some estimates in these directions, they might
be useful Taylor expansions of the functions in\rref 3.3.9(5)~ and\rref 3.3.9(6)~ in the initial
point $t=0$. Set, as usual,

\bequ\label {3.3.10(1)} \{f_A,f_X\}^{(n)}:= \{f_A,\{f_A,f_X\}^{(n-1)}\},\quad
\{f_A,f_X\}^{(0)}:=f_X,\ \text{for}\ n\in\mbZ_+,\enqu

\noidt and also the corresponding notation for multiple commutators for operators. Then we have
expressions for derivatives

\bequ\label{3.3.10(2n)} \left.\frac{\rd^n}{\rd t^n}\right|_{t=0}f_X(F^A_t[\bx]) =
\{f_A,f_X\}^{(n)}(\bx),\enqu

\noidt and

\bequ\label{3.3.10(3n)}\left.\frac{\rd^n}{\rd t^n}\right|_{t=0}f_X(\mbs U_A(t)\bx)=i^n\,
Tr(P_x[A,X]^{(n)}) =: f_{i^n[A,X]^{(n)}}(\bx).\enqu

The \rhs\ of\rref 3.3.10(2n)~ can also be expressed as a polynomial in expectation values of
quantal observables in the initial state \bx\ by multiple application of\rref 3.3.8(5)~. To make
these formulae clearly applicable it is  necessary  to have some assumptions on the domain of $A$,
e.g. let $A$ be \bz-classical with $O_\bz$ in its invariant analytic domain, $\bx\in O_\bz$ and
$A^nx\in\mcl A_G$\ (:= the analytic domain of $U(\mfk g))$ for all $n\in\mbZ_+.$ If these
assumptions are fulfilled, then the identity of functions\rref 3.3.9(5)~ and\rref 3.3.9(6)~ (for
given $X\in U(\mfk g)$ and $\bx\in O_\bz$) is equivalent to the equality of the right hand sides
in\rref 3.3.10(2n)~ and\rref 3.3.10(3n)~ for all $n\in\mbZ_+$. This equality holds for any such
$A$ for $n=0,1.$ The equality in higher orders is essentially dependent on the choice of $A$.

Content of this subsection is closely related to the investigation of $\hbar\rarw 0$ limit of
quantal correlation functions in the work by Hepp \cite{hp3}, cf. also \ref{4.1.8} -
\ref{4.1.10}.}

\pt\label{3.3.11}{\bf Extended phase spaces:} {\rm If the one-parameter group of time evolution is
included into $G$ as a subgroup, the reduction of the orbits $O_\bz$ to the symplectic manifolds
$M_\bz$ can be sometimes replaced by a natural procedure of a reduction of $O_\bz$ to odd
dimensional manifolds of the dimension $2n + 1,$ if the dimension of the corresponding classical
phase space is equal to $2n$. In this case, the restriction of the form $\mOme^\circ$ to such a
manifold is degenerate, of the rank $2n$. Such odd dimensional manifolds with a given closed
two-form of the maximal rank are called {\bf contact manifolds}. Usage of contact manifolds in  CM
is convenient for a natural possibility of passing to moving reference frames. Another situation,
in which they are useful is that of time dependent Hamiltonians, cf. \cite[Ch. 5]{abr&mars}, and
also \cite[Sec. 18.5]{fecko1}.

Sometimes it is useful to describe mechanical systems in CM by symplectic manifolds which are of
the dimension higher by $2$ than the usual ones. Any symplectic manifold can be extended to a \emm
contact manifold~ and any contact manifold can be extended to a symplectic manifold, each time
increasing the dimension by one.

We shall not try to give here the theory of these situations. For generalities on such structures
cf. e.g. \cite{abr&mars,arn1}. Some cases will be mentioned in the following chapter. }

\newpage

\chapter{Examples of classical mechanical projections}\label{Ch4}

\section{The Heisenberg group (CCR)}\label{sec;4.1}

\pt\label{4.1.1}{\rm A physical system consisting of the finite number $N$ of nonrelativistic
(apriori mutually distinguishable) point particles is described in the conventional QM by an
infinite dimensional unitary irreducible representation of the $2n+1$ - dimensional \emm Heisenberg
group~ $G$ ($n := N\nu,\ \nu$ is the dimension of the one-particle configuration space); cf. also
\cite[Sec. 3.3-b]{bon-EQM}. The Heisenberg group $G$ is a \emn central extension~ by \bR\ of the
commutative group $\mbR^{2n}$ (which can be identified with the classical flat phase space
$\mbR^{2n}= T^*\mbR^n$), compare \cite{varad} and \cite{zelob&stern}. The (scalar multiples of
the) selfadjoint generators $X_j,\ j=1,2,\dots 2n,$ of the representation correspond to basic
'kinematical' observables of the system. The choice of $X_j's$ is conveniently made in such a way,
that on corresponding domains (e.g. on $\mcl D_G$) the commutation relations (CCR) are fulfilled:

\begin{subequations}\bequ\label{4.1.1(1)} [X_j,X_k] = i\, S_{jk}X_0\quad
\text{for}\quad j,k=1,2,\dots 2n;\enqu
 \bequ\label{4.1.1(1a)}[X_j,X_0] = 0,\quad j=1,2,\dots
2n.\enqu
\end{subequations}

\noidt Here the elements $S_{jk}$ of the $2n\times 2n$ real matrix $S$ are defined:

\bequ\label{4.1.1(2)} S_{j\ j+n}= -S_{j+n\ j}=1,\ j=1,2,\dots n,\ S_{jk}=0\quad
\text{otherwise}.\enqu

\noidt Hence $S^{-1}=S^T=-S$ where $S^T$ is the transposed matrix to $S$. From\rref 4.1.1(1a)~ we
see, that

\bequ\label{4.1.1(3)} X_0=\hbar I,\quad (I\ \text{is\ the\ identity\ of}\ \mLH).\enqu

\noidt The parameter $\hbar\in\mbR,\ \hbar\neq 0,$ ($\hbar$ := the `\emn Planck constant~', if its
value is chosen properly) classifies all infinite-dimensional unitary irreducible representations
of $G$; representations corresponding to various values of $\hbar$\ are mutually inequivalent,
\cite{zelob&stern}. Setting

\bequ\label{4.1.1(4)} Q_j:= X_j,\ P_j:=X_{j+n}\quad \text{for}\ j=1,2,\dots n\enqu

\noidt we obtain from\rref 4.1.1(1)~ the usual form of the \emm canonical commutation relations
(CCR)~. There is only one physically admissible choice of the constant $\hbar$: it is the Planck
constant divided by $2\pi$ (its numerical value depends on a choice of physical units for
determination of which it is necessary to consider also dynamics). Operators $Q_j$ (resp. $P_j$)
are interpreted to correspond to observables called 'coordinates of the configuration' (resp.
'coordinates of the linear momentum'), in a cartesian basis. Note, that this representation of $G$
can be considered as a projective representation of $\mbR^{2n}$, as it was described in
\ref{1.2.7}.}

\pt\label{4.1.2}{\rm The Schr\"odinger form of the above mentioned representation of $G$ consists
of the realization of the Hilbert space \H\ of the representation as $L^2(\mbR^n,\rd^nq)\ (\rd^nq$
is the Lebesgue measure) and the action of $X_j$'s can be defined on such $\mphi \in
L^2(\mbR^n,\rd^nq)$, which belong to Schwartz test functions:

\begin{subequations}\label{4.1.2(1)}
\bequ\label{4.1.2(1a)} (X_j\mphi)(q_1,q_2,\dots q_n):= q_j\,\mphi(q_1,q_2,\dots q_n) \enqu

\noidt and

\bequ\label{4.1.2(1b)} (X_{j+n}\mphi)(q_1,q_2,\dots q_n):= -i\hbar\frac{\partial}{\partial
q_j}\mphi(q_1,q_2,\dots q_n)\enqu
\end{subequations}

\noidt for $j=1,2,\dots n.$ An equivalent realization of CCR is obtained by an arbitrary unitary
transformation $U$ of \H\ onto itself, e.g. by the scaling $U := U_\mlam\ (\mlam\in \mbR_+\setminus
\{0\})$:

\bequ\label{4.1.2(2)} (U_\mlam \mphi)(\mbbs q):= \mlam^{n/2} \mphi(\mlam\mbbs q).
 \enqu

\noidt It is $U_\mlam^{-1}=U_{1/\mlam}$ and we have:

\bequ\label{4.1.2(3)} X'_j:= U_\mlam X_j U^{-1}_\mlam = \mlam X_j,\ X'_{j+n}:= U_\mlam X_{j+n}
U^{-1}_\mlam =\frac{1}{\mlam}X_{j+n},\quad j=1,2,\dots n.\enqu

These transformations are useful for taking limits $\hbar\rarw 0,$ compare \cite{hp3,bon-m} and
also our \ref{4.1.8}. }

\pt\label{4.1.3}{\rm Let $X\cdot S\cdot x := X_jS_{jk}x_k$ with summation over $j,k=1,2,\dots 2n,$
where $x_k\in\mbR$ for all $k$. Let $W_x\ (x\in\mbR^{2n})$ be unitary operators of the above
mentioned projective representation (cf. \ref{1.2.7}):

\bequ\label{4.1.3(1)} W_x:= \exp\left(\frac{i}{\hbar}X\cdot S\cdot x\right).\enqu

\noidt From\rref 4.1.1(1)~ we obtain

\bequ\label{4.1.3(2)} W_x^{-1}X_j W_x = X_j + x_jI, \enqu

\bequ\label{4.1.3(3)} W_{x+x'}=\exp\left(\frac{i}{\hbar}x\cdot S\cdot x'\right) W_x W_{x'}.\enqu

Let us mention here, that the multiplier in\rref 4.1.3(3)~ is determined by the standard
symplectic form $\mOme^{cl}$ on the classical flat phase space $\mbR^{2n}$; setting $q_j:= x_j,\
p_j:= x_{j+n}$ for $j=1,2,\dots n,$ it is

\bequ\label{4.1.3(4)} \mOme^{cl}:= \sum_{j=1}^n \rd p_j\wedge \rd q_j,\enqu

\bequ\label{4.1.3(5)} x'\cdot S\cdot x = \mOme^{cl}(x,x').\enqu }

\pt\label{4.1.4}{\rm {\bf Let} $\mphi\in\mcl A_G:=$ the analytic domain of $U(G),\ \|\mphi\|=1$,\,
\emn$\mphi_x:=W_x\mphi\ (x\in\mbR^{2n})$~. {\bf Let} \emn$P^\mphi_x\in\mPH$~ be the corresponding
projectors, $Tr(P^\mphi_xA):= (\mphi_x, A\mphi_x)\ (A\in\mLH)$\ and $P^\mphi_0 = P_\mphi.$
From\rref 4.1.3(2)~ one has

\bequ\label{4.1.4(1)} Tr(P^\mphi_xX_j) = Tr(P_\mphi X_j) + x_j.\enqu

Hence the {\bf mapping} \emn$P^\mphi: x\mapsto P^\mphi_x$~ is a bijection of $\mbR^{2n}$ {\bf onto
the orbit} \emn$O_\mphi :=\{ P^\mphi_x: x\in\mbR^{2n}\}$~ and it is continuous if $O_\mphi$ is
taken in the relative topology from \PH. Due to absolute continuity of spectra of all $X_j\
(j=1,2,\dots 2n)$ \wrt\ the Lebesgue measure on \bR\ the function $x\mapsto (\mphi, W_x\mphi)$
converges to zero with $|x|\rarw \infty$\ and $|(\mphi,W_x\mphi)|=1$ iff $x=0.$ Consequently, the
mapping $P^\mphi$ is also open (i.e. any open set is mapped to an open set), hence it is a regular
$C^\infty$-embedding of $\mbR^{2n}$ into \PH; with our choice of $\mphi\in\mcl A_G,\ P^\mphi$ is
even an analytic embedding into \PH.}

\pt\label{4.1.5}{\rm Let $\msg_j$ denote the vector field on $O_\mphi$ corresponding to the
generator $\frac{1}{\hbar}X_j\ (j=1,2,\dots 2n).$ We shall denote by $\mOme^\mphi$ the restriction
of the symplectic form \Ome\ on \PH, \ref{2.2.1}, onto $O_\mphi$. The form $\mOme^\mphi$ is
nondegenerate, since for the values $\mOme^\mphi_x$ of $\mOme^\mphi$ in any point $\mbbs\mphi_x
\in O_\mphi$ we have:

\bequ\label{4.1.5(1)} \mOme^\mphi_x(\msg_j,\msg_k)=\frac{i}{\hbar^2}
Tr(P^\mphi_x[X_j,X_k])=-\frac{1}{\hbar} S_{jk} \enqu

\noidt and $\det S=1.$ Hence $M_\mphi =O_\mphi$ in this case. Let
$f_{X_j}(x):=f_{X_j}(\mbbs\mphi_x):=Tr(P_x^\mphi X_j)\ (x\in \mbR^{2n})$ be the classical
observable corresponding to $X_j.$ From\rref 4.1.4(1)~ we see, that a unique $\mbbs\mphi_0\in
O_\mphi$ can be chosen such, that

\bequ\label{4.1.5(2)} Tr(P_{\mphi_0}X_j)=0\ \text{for all}\ j=1,2,\dots 2n. \enqu

\noidt In the following, we shall take $\mphi := \mphi_0$ according to\rref 4.1.5(2)~. Then

\bequ\label{4.1.5(3)} f_{X_j}(x) = x_j,\ j=1.2.\dots 2n. \enqu

From\rref 4.1.5(1)~ and\rref 4.1.3(5)~ we see, that in the coordinates\rref 4.1.5(3)~ the form
$\mOme^{cl} := \hbar \mOme^\mphi$ is identical with $\mOme^{cl}$ defined earlier. Hence the
brackets

\bequ\label{4.1.5(4)} \{f_{X_j},f_{X_k}\}(x) := \hbar\, \mOme^\mphi_x(\msg_j,\msg_k) = - S_{jk}
\enqu

\noidt are exactly the \emm classical Poisson brackets on $\mbR^{2n}$~, cf. also \cite{bon4}. The
Hamiltonian vector fields on $O_\mphi$ corresponding to the Hamiltonian functions $f_{X_j}$ are
$\msg_j$ with flows $\exp(-\frac{i}{\hbar}tX_j)$. This recovers on $O_\mphi$ the standard
classical kinematics from the geometry of \PH\ and the CCR. }

\pt\label{4.1.6}{\rm Let us look now on the dynamics on $O_\mphi$ generated by the Hamiltonian
operator

\bequ\label{4.1.6(1)} A:= A_V :=\frac{1}{2}\sum_{jk=1}^n a_{jk}P_jP_k + V(Q) \enqu

\noidt from the point of view of the Sec.\ref{sec;3.1}\ (see\rref 4.1.1(4)~ for the notation).
Here \bbs a\ $\equiv\{a_{jk}\}$ is a real symmetric positive matrix and $V$ is a real distribution
on $\mbR^n$\ chosen such, that the operator $A$ is $\mphi$-classical, Def.\,\ref{3.3.6(i)}. The
quantal dynamical group is $\exp(-\frac{i}{\hbar}tA)$ and the corresponding classical projection (= classical mechanical projection)
$F^A_t$ on $O_\mphi$ is given by the Hamiltonian function

\bequ\label{4.1.6(2)} f_A(x):= Tr(P^\mphi_x A).\enqu

\noidt From\rref 4.1.4(1)~ we obtain (with $(q;p) := x)$:

\bequ\label{4.1.6(3)} f_A(q,p)=\frac{1}{2}\sum_{jk=1}^n a_{jk}p_jp_k + Tr(P_\mphi V(Q+q)) +
\frac{1}{2}\sum_{jk=1}^n a_{jk}Tr(P_\mphi P_jP_k),\enqu

\noidt where we write $V(Q+q):=W_x^{-1}V(Q)W_x$. The potential term in the realization\rref
4.1.2(1)~ is rewritten as

\begin{subequations}
\bequ\label{4.1.6(4)} V_\mphi(q):= Tr(P_\mphi V(Q+q))=\int_{\mbR^n} |\mphi(q')|^2\, V(q+q')\,
\rd^nq',\enqu

\noidt or as a \emn convolution~ $(\tilde{\mphi}(q):=\mphi(-q))$:

\bequ\label{4.1.6(4a)} V_\mphi(q)=|\tilde{\mphi}|^2\ast V(q)=: \mrh_\mphi\ast V(q).\enqu
\end{subequations}

\noidt This `smearing' of the potential energy by a density $\mrh_\mphi$ is the only difference
between the classical projections in the case of $G$ :=(the Heisenberg group) and the usual
classical limit with the 'unsmeared' potential energy $V(q)$ (up to the unessential additive
constant term in\rref 4.1.6(3)~).}

\begin{note}\label{4.1.7}

(i)\ The quantal correlation functions are constant on the orbits $O_\mphi$; e.g.
\bequ\label{4.1.7(1)} Tr(P^\mphi_x(X_j-x_j)(X_k-x_k)) = Tr(P_\mphi X_jX_k),\ \text{for all}\ j,k,\
\text{and for all}\ x\in\mbR^{2n},\enqu

(ii) If the Hamiltonian operator $A$ is quadratic in all the generators $X_j$:

\bequ\label{4.1.7(2)} A:=\frac{1}{2}\, h^{jk}X_jX_k\imply f_A(x)=\frac{1}{2}\, h^{jk} x_jx_k +
const,\enqu

\noidt i.e. in this case the usual classical limit coincides with the classical projections. This
situation is analyzed in Sec. \ref{sec;4.2}.
\end{note}

\pt{\rm {\bf On the \emn limit $\hbar\rarw 0$~.}\label{4.1.8}

All the previous results and considerations are equally valid for any nonvanishing value of the
parameter $\hbar.$ Any change of the value of the parameter $\hbar$ might be interpreted from the
point of view of mathematics, either as a change of the representation $U_\hbar(G)$ of the
Heisenberg group $G$ to an inequivalent one leaving the correspondence of the generators
$\frac{1}{\hbar}\,X_j\in U_\hbar(\mfk g)$ to fixed elements $\xi_j\in\mfk g$ of the Lie algebra
unchanged, or as a change of the basis $\{\xi_j\}$ in \fk g\ into $\{\mlam\xi_j\}$ (corresponding
to a `reinterpretation' (i.e. change of units) of parameters $x$ occurring in\rref 4.1.3(1)~),
leaving the choice of the representation fixed.

Let a physical interpretation of the generators $X_j$ be fixed (compare Sec. \ref{sec;1.2}),
leaving the value of $\hbar$ unspecified. If some empirical system is adequately described by QM
with the given interpretation of $X_j$'s, for some value of $\hbar,$ then this value $\hbar$ is
for the system unique (independently on any choices of generators of the evolution in time -
consider, e.g. the occurrence of $\hbar$  in uncertainty relations). If two such systems could
form one composite system the mutually noninteracting parts of which they are, then the value of
$\hbar$ for both systems is the same (interpretation of $X_j$'s fixed!), since each of the
subsystems taken separately determines $\hbar$ for the whole system (we have now a $2(n_1+n_2)+1$
-dimensional Heisenberg group, if the subsystems have $n_1$, resp. $n_2$ degrees of freedom).

 These considerations show, that any \emn change of the value of
 $\hbar$~
 - if physically interpreted - has to be connected with a \emn change of
interpretation of the generators~ $X_j\in U_\hbar(\mfk g)$. We obtain an example of such a
reinterpretation, if we describe a system consisting of a large number of particles: in a
description of the center of mass motion we can deal instead of with center of mass coordinates
and total linear momenta (which satisfy CCR with the experimental value of Planck constant) rather
with center of mass coordinates and averaged momenta per a particle.

If we keep the interpretation of $X_j$'s fixed, then for various values of $\hbar$ we obtain
different theories. We shall describe a transition of $\hbar\rarw 0$  in the context of the
classical projections of QM. Let us write $\mlam^2\hbar\ (\mlam\in (0,1])$ instead of $\hbar$ in
all formulas of the subsections \ref{4.1.1} - \ref{4.1.7}. Let $X_j(\mlam)$ be the Schr\"odinger
realizations of the CCR-generators in $L^2(\mbR^n)=:\mH$ and let us apply to them the
transformation $U_\mlam$ from\rref 4.1.2(3)~ for each value of \lam. {\bf Let us denote}
\emn$X^\mlam_j:= U_\mlam X_j(\mlam)U_\mlam^{-1}$~. We obtain:

\bequ\label{4.1.8(1)} Q_j^\mlam\mphi(q)=\mlam q_j\mphi(q),\
P_j^\mlam\mphi(q)=-i\,\mlam\hbar\frac{\partial}{\partial q_j}\mphi(q),\enqu

\noidt where \[ \centerline{\emn$Q_j^\mlam:= X_j^\mlam$~,\ \emn$P_j^\mlam:=X_{j+n}^\mlam$~\
$(j=1,2,\dots n).$}\] Let us fix $\mphi=\mphi_0\in\mH$ according to\rref 4.1.5(2)~, which will be
held unchanged for all the values of \lam. Let \emn$W^\mlam$~ be the unitary representation from
\ref{4.1.3}:

\bequ\label{4.1.8(2)} W^\mlam_x :=\exp\left(\frac{i}{\mlam^2\hbar}\,X^\mlam\cdot S\cdot
x\right).\enqu

\noidt Let $\mphi^\mlam_x:=W^\mlam_x\mphi,$ i.e. for $x:= (q;p)\in\mbR^{2n}$ we have

\bequ\label{4.1.8(3)} \mphi^\mlam_x(q')=\exp\left(-\frac{i}{2\mlam^2\hbar}q\cdot
p\right)\exp\left(\frac{i}{\mlam\hbar}q'\cdot p\right)\mphi(q'-\frac{q}{\mlam});\ q,p,q'\in
\mbR^n.\enqu

Let \[\centerline{\emn$\mbs P^{(\mlam)}_x$~ be the projector onto $\mphi^\mlam_x$,\ $\mbs
P_0^{(\mlam)}=\mbs P_\mphi\equiv P_\mphi$\ for all \lam.}\] The correlations of all orders are for
any \lam\ independent of $x$:

\bequ\label{4.1.8(4)} Tr\left(\mbs P^{(\mlam)}_x(X_j^\mlam-x_j)(X_k^\mlam-x_k)\dots
(X_r^\mlam-x_r)\right) = Tr(\mbs P_\mphi X_j^\mlam X_k^\mlam \dots X_r^\mlam).\enqu

The \rhs\ of\rref 4.1.8(4)~ is proportional to $\mlam^s$, where $s$ is the number of $X^\mlam$
in the \rhs\ of\rref 4.1.8(4)~. From this we see that the \[\centerline{\emm algebra $\mcl E(\mfk
g)^\mlam$~ of quantal observables consisting of polynomials in $X^\mlam$}\] is mapped onto a
set of functions on $O_\mphi^\mlam := \mbs W^\mlam(G)\mbs\mphi:$

\bequ\label{4.1.8(5)} f^\mlam: E\mapsto f^\mlam_E(x):= Tr(\mbs P_x^{(\mlam)}E),\ E\in\mcl E(\mfk
g)^\mlam;\ F^\mlam_{X_j}(x) = x_j,\enqu

\noidt and this mapping $f^\mlam$ becomes in the limit $\mlam\rarw 0$ a homomorphism of
associative algebras.

For the generator $f^\mlam_A$ of the `projected' evolution in time, corresponding to the quantal
generator\rref 4.1.6(1)~, i.e., for each \lam, to the operator

\bequ\label{4.1.8(6)} A^\mlam:=\frac{1}{2}\,\sum_{jk=1}^n a_{jk}P_j^\mlam P_k^\mlam +
V(Q^\mlam),\enqu

\noidt we obtain:

\bequ\label{4.1.8(7)} f_A^\mlam(q,p) := Tr(\mbs P^{(\mlam)}_xA^\mlam) =\frac{1}{2}\,\sum_{jk=1}^n
a_{jk} p_jp_k + \mrh^\mlam\ast V(q)+b_\mlam.\enqu

\noidt Here $b_\mlam$ is a constant depending on \lam\ as $O(\mlam^2)$ and

\bequ\label{4.1.8(8)} \mrh^\mlam(q):= \mlam^{-n}|\mphi(-\frac{q}{\mlam})|^2\enqu

\noidt is a normalized density on $\mbR^n$, which weakly converges to the Dirac $\delta$-function
with $\mlam\rarw 0.$ A comparison of flows on $O^\mlam_\mphi\ (= \mbR^{2n})$ generated by
$f^\mlam_A$ for various \lam\ is not easy for given $V$ and $\mphi$ in general.}

\pt\label{4.1.9}{\rm Let $U^\mlam_A(t) := \exp\left(-\frac{it}{\mlam^2}A^\mlam\right)$ be the time
evolution group corresponding to the  generator\rref 4.1.8(6)~\ (we set $\hbar=1$). Let

\bequ\label{4.1.9(1)} x^\mlam_j(t,x):= Tr(U^\mlam_A(t)\mbs
P_x^{(\mlam)}U^\mlam_A(-t)X_j^\mlam)=f^\mlam_{X_j}(\mbs U_A^\mlam(t)\mbs \mphi^\mlam_x)\enqu

\noidt be time-evolved quantal expectations of the `canonical' observables $X^\mlam_j$ with
initial values $\mbs\mphi^\mlam_x\in O_\mphi^\mlam$\ (the mapping $f^\mlam$ from\rref 4.1.8(5)~ is
here extended to a mapping into functions on \PH). The well-known Ehrenfest's equations are
certain equalities including the functions\rref 4.1.9(1)~ and their time-derivatives, which have
an analogous form to that of equations of motion of CM, being in the same time exact consequences
of QM. We can write them in the form (with $x^\mlam:= (q_1^\mlam,\dots q_n^\mlam, p_1^\mlam,\dots
p_n^\mlam)$ and summation is over $1,2,\dots n$):

\bequ\label{4.1.9(2)} \frac{\rd}{\rd t} q_j^\mlam(t,x)=
a_{jk}p_k^\mlam(t,x)=\frac{\partial}{\partial p_j}f_A^\mlam(x^\mlam(t,x)),\enqu

\bequ\label{4.1.9(3)} \frac{\rd}{\rd t}p_j^\mlam(t,x)=- Tr\left(U_A^\mlam(t)\mbs
P_x^{(\mlam)}U^\mlam_A(-t)\frac{\partial V}{\partial q_j}(Q^\mlam)\right).\enqu

\noidt Here $f^\mlam_A$ is the classical Hamiltonian function corresponding to the quantal
generator $A^\mlam$. The corresponding equations for the classical projection on $O^\mlam_\mphi$
are of the form:

\bequ\label{4.1.9(4)}\frac{\rd}{\rd t} q_j^\mlam(t,x)_{cl}=
a_{jk}p_k^\mlam(t,x)_{cl}=\frac{\partial}{\partial p_j}f_A^\mlam(x^\mlam(t,x)_{cl}), \enqu

\bequ\label{4.1.9(5)}\frac{\rd}{\rd t}p_j^\mlam(t,x)_{cl}= -\frac{\partial}{\partial
q_j}\left(\mrh^\mlam\ast V(q^\mlam(t,x)_{cl})\right)=-\frac{\partial}{\partial q_j}
f^\mlam_A(x^\mlam(t,x)_{cl}),\enqu

\noidt with $f^\mlam_A$ from\rref 4.1.8(7)~. We shall rewrite\rref 4.1.9(3)~ into a form similar
to\rref 4.1.9(5)~.\footnote{It is left to the reader's assessment, whether the forthcoming
reformulation could be helpful for better understanding of the "classical limit $\hbar\rarw 0$" of
the dynamics.} Let $y\in\mbR^{2n}$ and $W^\mlam_y$ as in\rref 4.1.8(2)~. Inserting $W^\mlam_{\pm
y}$ into the trace in\rref 4.1.9(3)~ we obtain:

\bequ\label{4.1.9(6)} \frac{\rd}{\rd t}p^\mlam_j(t,x)=-\mrh^\mlam(y,t,x)\ast\frac{\partial
V}{\partial q_j}(q^y)=-\left(\frac{\partial}{\partial q_j}\mrh^\mlam(y,t,x)\right)\ast
V(q^y),\enqu

\noidt where $q^y:=(y_1,\dots y_n)$ and

\bequ\label{4.1.9(7)} \mrh^\mlam(y,t,x)(q):=
\frac{1}{\mlam^n}\left|(W^\mlam_{-y}U^\mlam_A(t)W^\mlam_x\mphi)(-\frac{q}{\mlam})\right|^2. \enqu

\noidt Since the \rhs\ of\rref 4.1.9(6)~ is independent of $y\in\mbR^{2n}$, we can insert there
$y:= x^\mlam(t,x)$ and obtain a formal analogy with\rref 4.1.9(5)~. We expect, that the difference
$\mrh^\mlam(y,t,x)-\mrh_\mphi^\mlam$ (compare\rref 4.1.6(4a)~) will converge to zero with
$\mlam\rarw 0$ in the sense of distributions uniformly on compacts in $t$, if $y:= x^\mlam(t,x)$,
and also for $y := x^\mlam(t,x,)_{cl},$ with some reasonable choice of $V$. This conjecture is
based on the results of \cite{hp3}. (The convergence holds for each fixed $t\in\mbR$ for
$y=x^\mlam(t,x)).$\footnote{This fact was kindly announced to the author by Prof. Klaus Hepp (in
1985).}) }

\begin{exm}\label{4.1.10}{\rm We shall give here an elementary example showing possible differences
between a quantal time-evolution and its classical projection. We shall notice also the behaviour
of these evolutions in the limit of vanishing \lam. In the formalism introduced above, let
$\mphi\in L^2(\mbR,\rd q)$ represents a '\emn minimal wave packet~':

\bequ\label{4.1.10(1)} \mphi(q):= \pi^{-\frac{1}{4}}\exp\left(-\frac{1}{2}q^2\right), \enqu

\noidt and choose $\mphi^\mlam_z:= W^\mlam_z\mphi$ with $z:= q-ip,\ W^\mlam_z:=
\exp[i\mlam^{-2}(Q^\mlam p-P^\mlam q)].$ Let the generator $A^\mlam$ of quantal time-evolution be

\bequ\label{4.1.10(2)} A^\mlam := a(\mlam)P_\mphi,\enqu

\noidt where $a(\mlam)$ is some real function. Then the classical Hamiltonian function on the
orbit $O^\mlam_\mphi$ of the Heisenberg group in $L^2(\mbR)$ is

\bequ\label{4.1.10(3)} f_A^\mlam(z):= Tr(\mbs
P_z^{(\mlam)}A^\mlam)=a(\mlam)\exp\left(-\frac{\overline{z}z}{2\mlam^2}\right) \enqu

\noidt with $\overline{z}$ being the complex conjugate of $z\in\mbC$. We are interested in the
comparison of solutions of classical Hamiltonian equations on  $O_\mphi,\ z^\mlam(t,z)_{cl},$ and
the corresponding quantal expectations:

\bequ\label{4.1.10(4)} z^\mlam(t,z):= Tr(U_A^\mlam(t)\mbs P_z^{(\mlam)}U_A^\mlam(-t)Z^\mlam),\quad
Z^\mlam:=Q^\mlam-iP^\mlam,\enqu

\noidt with the same initial values $z=q-ip.$ Elementary calculations give:

\bequ\label{4.1.10(5)} z^\mlam(t,z)=\left[1-\frac{1}{a(\mlam)}f^\mlam_A(z)\right]z +
\frac{1}{a(\mlam)}f^\mlam_A(z)\exp\left(-\frac{it}{\mlam^2}a(\mlam)\right)z,\enqu

\bequ\label{4.1.10(6)} z^\mlam(t,z)_{cl}=\exp\left(-\frac{it}{\mlam^2}f^\mlam_A(z)\right)z.\enqu

We see that\rref 4.1.10(5)~ and\rref 4.1.10(6)~ describe motions on mutually tangent circles in
\bC\ with different radii and different dependence of frequencies on initial conditions as well as
on the parameter \lam. For $\mlam\rarw 0$ the quantum evolution vanishes independently on the
`renormalization' $a(\mlam)$. For slowly varying $a(\mlam)$, the classical evolution vanishes too,
but the way of this vanishing looks qualitatively differently. If, e.g. $a(\mlam) =
\mlam^2\exp\left({b}/{2\mlam^{2}}\right),\ b>0,$ then $\overline{z}z = b$ is a critical value for
$\mlam\rarw 0.$}\end{exm}

\section{Extension of CCR by a quadratic generator}\label{sec;4.2}

\pt\label{4.2.1}{\rm All the orbits $O_\mphi$ occurring in Sec.\ref{sec;4.1} were mutually
homeomorphic (and homeomorphic to $\mbR^{2n}$). In this section, we shall give examples of
irreducible representations $U(G)$ of some Lie groups $G$ in a Hilbert space \H\ containing
various mutually nonhomeomorphic orbits $O_j := \mbs U(G)\mbs\mphi_j$ in \PH, $(j = 1,2,\dots)$.
Let $G$ be a connected Lie group containing the $2n + 1$ - dimensional Heisenberg group $G_n$ as
an invariant (i.e. normal) subgroup ($G$ will be specified later). Let $U$ be such a unitary
continuous representation of $G,$ the restriction of which to $G_n$ coincides with the irreducible
representation described in Sec.\ref{sec;4.1} with $\hbar = 1 = \mlam.$ With the notation of the
previous section, for $m = 1,2,\dots K,\ A_m\in U(\mfk g),$ set

\bequ\label{4.2.1(1)} A_m := \frac{1}{2}\, h_m^{jk}X_jX_k,\quad (\text{summation\ over}\
j,k=1,2,\dots 2n),\enqu

\noidt with any $h_m$ a real symmetric $2n\times 2n$ - matrix; the formally defined operator $A_m$
is symmetric on the G{\aa}rding domain of $U(G_n)$. From\rref 4.1.1(1)~ we have commutation
relations (cf. also \ref{4.1.3}):

\bequ\label{4.2.1(2)} [X_j,X_k]= i\,S_{jk}I,\ [X_j,A_m]=i\,S_{jk}h_m^{kl}X_l=: i\,(S\cdot h_m\cdot
X)_j,\enqu

\bequ\label{4.2.1(3)} [A_m,A_k]=\frac{i}{2}\, X\cdot(h_m\cdot S\cdot h_k - h_k\cdot S\cdot
h_m)\cdot X,\quad m,k=1,2,\dots K.\enqu

Assume that for any $m,k$ there are reals $c_{mk}^j$ such, that

\bequ\label{4.2.1(4)} h_m\cdot S\cdot h_k - h_k\cdot S\cdot h_m\ =\ \sum^K_{j=1} c_{mk}^j
h_j,\quad m,k=1,2,\dots K.\enqu

\noidt then the linear hull of the operators $X_j\ (j=1,2,\dots 2n),\ A_m\ (m=1,2,\dots K)$ and
$I:= I_\mH := id_\mH$\ forms the Lie algebra $U(\mfk g).$ We have (cf. also \cite[Proposition
3.3.12]{bon-EQM}):}

\begin{prop}\label{4.2.2} Let $U(\mfk g)$ be the above defined
representation of a Lie algebra \fk g\ in \H\ and let $G$ be the corresponding simply connected
Lie group with the Lie algebra \fk g. Then the representation $U(G_n)$ of the Heisenberg group
$G_n$ has a unique extension to the representation $U(G)$ of $G$ in \H\ such, that the closures of
the operators $X_j\ (j=1,2,\dots 2n),\ A_m\ (m=1,2,\dots K)$ and $I_\mH$ are selfadjoint
generators of $U(G)$ corresponding to basis vectors in \fk g\ according to\rref 2.3.9(2)~. In
particular the operators $A_m\ (m = 1,2,\dots K)$ are essentially selfadjoint on the G{\aa}rding
domain of $U(G_n).$
\end{prop}
\begin{proof}
The \emm G{\aa}rding domain of $U(G_n)$~ is a common dense invariant domain of all the operators
in $U(\mfk g)$. According to a Nelson's theorem (see \cite[Theorem 11.5.2.]{bar&racz}) it suffices
to prove essential selfadjointness of the operator $\Delta$,

\bequ\label{4.2.2(1)} \Delta:=\sum_{j=1}^{2n}X_j^2 + \sum_{m=1}^K A^2_m, \enqu

\noidt on the invariant domain. First we shall choose $m:=(j;k)$ with $j,k=1,2,\dots 2n$ and

\bequ\label{4.2.2(2)} A_m:= A_{(j;k)} := \frac{1}{2}(X_jX_k+X_kX_j).\enqu

\noidt In this case the operator $\Delta$ in\rref 4.2.2(1)~ can be expressed in the form

\bequ\label{4.2.2(3)} \Delta = \frac{3}{2} nI_\mH + \sum_{j=1}^n (P^2_j+Q_j^2)(I_\mH
+\sum_{k=1}^n(P^2_k+Q_k^2)),\enqu

\noidt where we used CCR and the notation\rref 4.1.1(4)~. From the known properties of the
Hamiltonians $P_j^2+ Q_j^2$ of independent linear oscillators, we conclude (with a help, e.g., of
\cite[Theorem VIII.33]{R&S}) that $\Delta$ is essentially se1fadjoint.

{\bf Denote the Lie algebra} generated by $X_j$'s and $A_{(j;k)}\ (j,k = 1,2,\dots 2n)$  by \emn
$\mfk g_{max}$~ and the corresponding {\bf simply connected group} by \emn $G_{max}$~. Any $A_m$
of the form\rref 4.2.1(1)~ is a linear combination of $A_{(j;k)}$'s. Consequently, any Lie algebra
$U(\mfk g)$ from \ref{4.2.1} is a subalgebra of $U(\mfk g_{max})$ and the corresponding group $G$
is a subgroup of $G_{max}$. From this just proved integrability of $U(\mfk g_{max})$ to a unitary
representation $U(G_{max})$, it follows integrability of $U(\mfk g)$ for any \fk g\ introduced in
\ref{4.2.1}. This implies the selfadjointness of\rref 4.2.2(1)~ with arbitrary $A_m$ of the
form\rref 4.2.1(1)~ and this, in turn, implies uniqueness of $U(G).$
\end{proof}

\pt\label{4.2.3}{\rm In this section we shall restrict our attention to the cases of
representations $U(G)$ obtained from $U(G_n)$ by addition of only one generator $A := A_m$ of the
form\rref 4.2.1(1)~ in the manner described above. Let $h$ be any nonzero real symmetric matrix
with elements $h_{jk}\ (j,k =1,2,\dots 2n)$ and let

\bequ\label{4.2.3(1)} A := \frac{1}{2}\,h^{jk}X_jX_k \enqu

\noidt denote here the selfadjoint operator corresponding to the \rhs of\rref 4.2.3(1)~. According
to\rref 4.2.2(2)~ the operators $X_j\ (j=1,2,\dots 2n),\ A$\ and $I_\mH$\ are selfadjoint
generators of an irreducible unitary representation $U(G)$ of a $2n + 2$ - dimensional connected
Lie group $G$ containing $G_n$ as a normal subgroup. The restriction $U(G_n)$ of $U(G)$ is
irreducible, too. Let \emn$\mbs U(G)$~ be the realization of $G$ in \PH\ obtained by the natural
projection of $U(G),$ \ref{2.3.9}. We shall investigate infinitely differentiable orbits of $\mbs
U(G)$ in \PH.}

\begin{lem}\label{4.2.4} Let U(G) be as in \ref{4.2.3} and $\mcl D_G$ be a dense invariant subset in
\H\ consisting of infinitely differentiable vectors of $U(G)$, e.g. $\mcl D_G$ is the G{\aa}rding
subspace for U(G)(\cite[11.1.8]{bar&racz}). Let $\mphi\in\mcl D_G,\ \|\mphi\|= 1$ and $O_\mphi :=
\mbs U(G)\mbs\mphi$ be the immersed submanifold of \PH\ according to \ref{3.1.2}. The orbit
$O_\mphi$ is 2n-dimensional iff there is an element $C_\mphi\in U(\mfk g)$,

\bequ\label{4.2.4(1)} C_\mphi := c^j_\mphi X_j- A,\enqu

\noidt such, that $\mphi$ is its eigenvector:

\bequ\label{4.2.4(2)} C_\mphi\mphi = \mlam\mphi\quad for\ some\ \mlam\in\mbR.\enqu

\noidt If\rref 4.2.4(1)~ with\rref  4.2.4(2)~ is the case, then $O_\mphi=\mbs U(G_n)\mbs\mphi.$
\end{lem}
\begin{proof}
The tangent space to $O_\mphi$ at $\mbs\mphi$ is the linear hull of vectors $\msg_j(\mbs\mphi)\
(j=1,2,\dots 2n)$\ (see \ref{4.1.5}) and $\msg_A(\mbs\mphi)$\ (see e.g. \ref{2.3.5} and
Sec.\ref{sec;3.2}). According to \ref{4.1.5}, all the $\msg_j$'s are linearly independent. Hence
$O_\mphi$ is 2n-dimensional iff

\bequ\label{4.2.4(3)} \msg_A(\mbs\mphi)=c_\mphi^j\msg_j(\mbs\mphi)\enqu

\noidt for some reals $c^j_\mphi$. According to\rref 2.3.5(1)~, the equation\rref 4.2.4(2)~
implies\rref 4.2.4(3)~. Assuming\rref 4.2.4(3)~, we have in the standard identification of
$T_{\mbs\mphi}\mPH$ with $[\mphi]^\bot$ by the help of $\Psi_\mphi$\ (see \ref{2.1.7} and\rref
2.1.8(4)~):

\bequ\label{4.2.4(4)} (I-P_\mphi)(c^j_\mphi X_j - A)\mphi=0\imply (c_\mphi^jX_j-A)\mphi=\mlam\mphi
\enqu

\noidt with $\mlam := \mlam(\mphi) := Tr(P_\mphi(c^j_\mphi X_j - A))$.
\end{proof}

\pt\label{4.2.5}{\rm Let $C := C_\mphi$ have the form\rref 4.2.4(1)~ and let\rref 4.2.4(2)~ be
fulfilled. Then $C$ satisfies the system of linear equations:

\bequ\label{4.2.5(1)} Tr(P_\mphi[C,X_j])=0,\ j= 1,2,\dots 2n,\enqu

\bequ\label{4.2.5(2)} Tr(P_\mphi[C,A]) = 0,\enqu

\noidt where\rref 4.2.5(2)~ follows from\rref 4.2.5(1)~. The equations\rref 4.2.5(1)~ have unique
solution $C$ of the form\rref 4.2.4(1)~ for any $\mphi\in\mcl D_G$, even if the relation\rref
4.2.4(2)~ is not fulfilled:

\bequ\label{4.2.5(3)} c_\mphi^j = h^{jk}\, Tr(P_\mphi X_k).\enqu

\noidt The corresponding operator $C_\mphi$ represents the generator of the isotropy subgroup
$K_{\mbs\mphi}\subset G$ at the point $F_{\mbs\mphi}\in\mfk g^*$ in the $Ad^*(G)$-representation;
here  $F_{\mbs\mphi}(\xi):= Tr(P_\mphi X_\xi)$\ for $\xi\in\mfk g,$ compare \ref{3.2.3} and
\ref{3.2.6}. From\rref 4.2.5(3)~ and\rref 4.2.4(1)~, we have immediately:}

\begin{lem}\label{4.2.6} If $\mbs\mphi\in O_\mphi$ is chosen such
that $Tr(P_\mphi X_j)=0$ for all $j=1,2,\dots 2n,$ then $C_\mphi =-A.$
\end{lem}

\begin{prop}\label{4.2.7}
The orbit $O_\mphi$ of $\mbs U(G)$ is 2n-dimensional iff it contains an eigenprojector $P_\mphi$
of $A$, i.e. iff for some $\mbs\mphi\in O_\mphi$ it is

\bequ\label{4.2.7(1)} Tr(P_\mphi A^2) = \left(Tr(P_\mphi A)\right)^2.\enqu
\end{prop}
\begin{proof}
In any orbit lying in $\mcl D_G$ there is a point $\mbs\mphi$ satisfying the conditions of the
Lemma \ref{4.2.6}, compare \ref{4.1.5}. The assertion is an immediate consequence of the Lemmas
\ref{4.2.6} and \ref{4.2.4}.
\end{proof}
\begin{corls}\label{4.2.8}
(i)\quad Let $\mphi_0\in U(G)\mphi$ satisfy\rref 4.1.5(2)~. If $\mphi$ is an eigenvector of $A$,
then also $\mphi_0$ is an eigenvector of $A$, if $\mphi\in\mcl D_G$.

(ii)\ If $\mphi_0\in\mcl D_{G_{max}}$ satisfies\rref 4.1.5(2)~, then those relations are satisfied
by all the vectors

\bequ\label{4.2.8(1)} \mphi_t^A:= \exp(-itA)\mphi_0 \enqu

\noidt for all $t\in\mbR$ and all the choices of $A$;\rref 4.2.3(1)~.
\end{corls}
\begin{prop}\label{4.2.9}
For any choice of $A$ in \ref{4.2.3} there is in \PH\ a 2n+1-dimensional orbit of the
corresponding representation $\mbs U(G)$\ (defined in \ref{4.2.3}), which is an infinitely
differentiable immersed submanifold of \PH.
\end{prop}
\begin{proof}Remember that any $A$ is an unbounded selfadjoint
operator. Let $U_\pi :=P_+-P_-$ be the `parity operator' defined by\footnote{In $\mH\equiv
L^2(\mbR^n,\rd^n x)$, it is defined as $[U_\pi \psi](x):=\psi(-x),\ \forall \psi\in \mH,\
x\in\mbR^n$.}

\bequ\label{4.2.9(1)} U_\pi^*=U_\pi^{-1}=U_\pi,\ U_\pi X_jU_\pi = - X_j\quad (j=1,2,\dots
2n),\enqu

\noidt and $P_+$ (resp.$P_-$) are corresponding orthogonal projectors,

\bequ\label{4.2.9(2)} P_+ +P_-=I_\mH.\enqu

\noidt Choose a dense invariant linear subset $\mcl D_G$ of \H\ consisting of infinitely
differentiable vectors of $U(G)$ such, that (as usually)

\bequ\label{4.2.9(3)} U_\pi\mcl D_G\subset\mcl D_G,\ \text{hence}\ P_\pm\mcl D_G\subset\mcl
D_G.\enqu

\noidt This condition implies, that $P_+\mcl D_G$\ (resp. $P_-\mcl D_G$) is dense in $\mH_+:=
P_+\mH$\ (resp. in $\mH_-:=P_-\mH$). For any $\mphi\in \mcl D^+_G\cup \mcl D^-_G$\ (with $\mcl
D^\pm_G:=P_\pm\mcl D_G$), the assumption of \ref{4.2.6} is fulfilled due to\rref 4.2.9(1)~. If
$\mphi$ is not an eigenvector of $A$, then the orbit $O_\mphi$ is 2n+1-dimensional. Assume, that
$A\mphi=\mlam\mphi$. Let $\mphi\in\mcl D^+_G,$ for definiteness. Since $A$ is $U_\pi$-invariant:
$[A,U_\pi]=0,$ its spectral measure $E_A$ commutes with projectors $P_\pm$. Denote for any Borel
set $B\subset \mbR$

\bequ\label{4.2.9(4)} E^\pm_A(B):= P_\pm E_A(B),\ \text{hence}\ E_A=E_A^++E_A^-,\enqu

\noidt and $E_A^+$ is the spectral measure of the restriction of $A$ to the $U(G)$-invariant
(infinite dimensional) subspace $\mH_+$ of \H. Due to unboundedness of $A$, we can assume that the
subspace $(P_+-E_A^+(\{\mlam\}))\mH$ of $\mH_+$ is nonempty; here $E_A^+(\{\mlam\})$ is the
eigenprojector of $P_+A$ corresponding to the eigenvalue \lam. Choose a nonzero vector

\bequ\label{4.2.9(5)} \mphi'\in (P_+-E_A^+(\{\mlam\}))\mH \enqu

\noidt and assume the normalization $\|\mphi\|=\|\mphi'\|=1.$ Let $\chi
:=\frac{1}{\sqrt{2}}(\mphi'+\mphi)$. Since $\mcl D_G^+$ is dense in $\mH_+$, we can find for
arbitrarily small $\delta >0$ a vector $\mphi_0$:

\bequ\label{4.2.9(6)} \mphi_0\in\mcl D_G^+:\ \|\mphi_0-\chi\|^2<\delta,\ \|\mphi_0\|=1.\enqu

\noidt With $\delta<2-\sqrt{2},$ the vector $\mphi_0$ cannot be an eigenvector of $A$ and,
moreover, it satisfies\rref 4.1.5(2)~. Hence the corresponding orbit $O_{\mphi_0}$ is
$2n+1$-dimensional. The manifold structure was proved in \ref{3.1.2}.
\end{proof}

\pt\label{4.2.10}{\rm Let $O_\mphi$\ (with $\mphi\in\mcl D_G$) be a 2n+1-dimensional orbit of
$\mbs U(G)$ and let $\mOme^\circ$ be the restriction of the standard symplectic form \Ome\ on \PH\
onto $O_\mphi$, compare \ref{3.2.2}. According to the previous results (Sec.'s \ref{sec;3.2} and
\ref{sec;4.1}), $\mOme^\circ$ is a closed two-form of the maximal rank 2n, hence it is a contact
two-form on $O_\mphi$\ (see, e.g. \cite[Chap. 5.1.]{abr&mars}). The equations\rref 4.2.5(1)~
determine the characteristic line-bundle of $\mOme^\circ$ in terms of operators $C=C_\mphi$
corresponding to generators of stability groups of $F_\mphi\in\mfk g^*$ (see \ref{4.2.5}) \wrt\
$Ad^*(G)$. The characteristic line bundle of $\mOme^\circ$ is integrable, determining a regular
foliation of $O_\mphi$. The factorization of $O_\mphi$ \wrt\ this foliation is the symplectic
manifold $M_\mphi$\ (symplectomorphic to the classical phase space $T^*\mbR^n$) as it was
constructed in Sec. \ref{sec;3.2} (for definition of the \emm cotangent bundle $T^*(M)$~ of a
general manifold $M$ see e.g. \cite[A.3.6 Definitions (v)]{bon-EQM}). }

\pt\label{4.2.11}{\rm The quantal and classical evolutions corresponding to the generator $A$, cf.
\ref{4.2.3}, (resp. to the Hamiltonian function $f_A$) coincide in our examples in the sense of
\ref{3.3.9}+\ref{3.3.10}, independently of the dimension (= 2n or 2n+l) of the orbit $O_\mphi$.
The time-evolved quantal states remain all the time on the orbit $O_\mphi$. We might be interested
also in time evolution of other quantities than (the expectations of) $X_j$ in the quantal
interpretation. According to \ref{4.1.7}(i), in the case of $\dim O_\mphi = 2n = \dim M_\mphi$,
any '\emn spreading of the wave packet~' does not occur. The situation is different, however, on
2n+l - dimensional orbits. For various $\mphi_j\ (j= 1,2)$ corresponding to distinct quantal
states in the same leaf $[\mbs\mphi]\in M_\mphi$ we have in general (cf. \ref{4.1.4} for notation)

\bequ\label{4.2.11(1)} Tr(P^{\mphi_1}_x(X_j-x_j)(X_k-x_k)) \neq
Tr(P^{\mphi_2}_x(X_j-x_j)(X_k-x_k)).\enqu

\noidt This is the case of e.g., free particle motions. This fact makes a certain difference
between classical and quantal interpretations of the `extended phase spaces' $O_\mphi\ (\dim
O_\mphi= 2n+1)$. This will be briefly discussed later on, in \ref{4.3.5}.}

\section{Notes on other examples}\label{sec;4.3}

\pt\label{4.3.1}{\rm By the method developed in our Chap. \ref{Ch3}, we can construct from an
arbitrary continuous unitary representation $U(G)$ of a Lie group $G$ `classical phase spaces',
which are diffeomorphic (and even symplectomorphic) to orbits of $Ad^*(G)$. It can be shown,
\cite[15.2]{kiril}, that any symplectic homogenous space of any connected Lie group G is a \emn
covering symplectic space~ of either an orbit of Ad*(G), or an orbit of $Ad^*(G_1)$, where $G_1$
is a \emn central extension of $G$ by \bR~\ - see also \ref{1.3.7}. On the other hand, unitary
continuous representations of $G$ can be constructed from orbits of $Ad^*(G)$, \cite{kiril}.
Considerations in Sec.'s \ref{sec;1.2} and \ref{sec;1.3} show reasons for modeling state spaces of
CM-systems as homogeneous symplectic spaces of some Lie groups, at least for 'basic' or
'elementary' physical systems. In this section we shall outline further examples of obtaining
CM-systems from unitary group representations which suggest, that all generally accepted models of
'elementary' finite dimensional CM - systems could be obtained in this way.}

\pt\label{4.3.2}{\rm {\emm Classical spin from SO(3)~:}

Let $U$ be a (projective) irreducible representation of the compact Lie group SO(3) - the
connected component of the 3-dimensional orthogonal group O(3) of orthogonal transformations of a
3-dimensional Euclidean space $E_3$. The representation space $\mH= \mbC^{2J+1}\ (J=\frac{n}{2},\
n\in\mbZ_+)$\ is finite dimensional. Generators $Y_k\ (k=1,2,3)$ of $U$ corresponding to rotations
around orthogonal axes in $E_3$ satisfy the commutation relations (with
$\meps_{jkm}=-\meps_{kjm}=-\meps_{jmk},\ \meps_{123}=1)$:

\bequ\label{4.3.2(1)} [Y_k,Y_m]=i\,\meps_{kmj}Y_j.\enqu

Choose any nonzero $\mphi\in\mH$ and form the orbit $O_\mphi := \{\mbs U(g)\mbs\mphi : g\in
SO(3)\}$. Let us denote by $Y_\xi$ the generator of $t\mapsto U(\exp(t\xi))$ corresponding to an
element $\xi$ of the Lie algebra $\mfk g:= so(3)$. We are interested in the $Ad^*(SO(3))$-action
onto $F_\mphi\in so(3)^*,$ where

\bequ\label{4.3.2(2)} F_\mphi: \xi\mapsto F_\mphi(\xi):= Tr(P_\mphi Y_\xi),\ \xi\in so(3).\enqu

\noidt Generators $C_\mphi := c^j_\mphi Y_j$ of one-parameter subgroups of the isotropy group of
$F_\mphi$ are just all nonzero solutions of equations

\bequ\label{4.3.2(3)} Tr(P_\mphi[Y_k,C_\mphi]) = 0,\quad k=1,2,3.\enqu

\noidt With $y^k := y^k(\mbs\mphi):= Tr(P_\mphi Y_k)$, the only linearly independent solution
$C_\mphi$ of\rref 4.3.2(3)~ can be written:

\bequ\label{4.3.2(4)} C_\mphi = y^k(\mbs\mphi)Y_k.\enqu

One could easily check that $C_\mphi=0$ in\rref 4.3.2(4)~ for some $\mphi$, iff $C_\mphi =0$ for
all $\mphi\in\mH,$ iff $J=0$\ (i.e. $\dim_\mbC\mH=1$), iff the matrix of the homogeneous
equations\rref 4.3.2(3)~ is identically zero. In all other cases the rank of the matrix of the
system\rref 4.3.2(3)~ equals to 2. For $J=0$ the corresponding classical phase space degenerates
to a point: this corresponds to the traditional point of view according to of which spin does not
occur in classical mechanics.

For orbits $O_\mphi$ in representations with $J\neq 0$ we have two possibilities: \nl

(i)\ The vector $\mphi$ is an eigenvector of $C_\mphi$ and the orbit $O_\mphi$ is two-dimensional
(any generator $Y\in U(so(3))$ which is linearly independent of $C_\mphi$ cannot be a solution of
\rref 4.3.2(3)~: $Tr(P_\mphi [Y_k,Y]) =0$ for $k =1,2,3$ implies $Y = \mlam C_\mphi$; hence $Y$
linearly independent of $C_\mphi$ generate two-dimensional tangent space to $O_\mphi$ at
$\mbs\mphi$).\nl

(ii) If $\mphi$ is not an eigenvector of $C_\mphi$, then the generator $C _\mphi$ generates a
one-dimensional submanifold of $O_\mphi$ diffeomorphic to a circle $S^1\ (C_\mphi$ generates the
\emn isotropy subgroup~ of $SO(3)$ at $F_\mphi,$ which is closed, hence compact). In this case
$O_\mphi$ is 3-dimensional.

Note that for $J=\frac{1}{2}$ only the possibility (i) occurs, since $\mH=\mbC^2$ and
$\dim_\mbR\mPH=2.$

 It can be easily shown that in the both cases the corresponding
classical phase space $M_\mphi$ (in the case (i) identical with $O_\mphi $) is diffeomorphic to
the sphere $S^2$ in $so(3)^*$ with coordinates

\bequ\label{4.3.2(5)} y^k: F_\mphi\mapsto y^k(\mbs\mphi) := Tr(P_\mphi Y_k),\ k=1,2,3;\quad
\mbs{\mphi}\in O_\mphi.\enqu

Let $t\in\mbR$ and let $\tau\in\mbR^3$ be any unit vector:\ $\sum_k(\tau^k)^2=1.$ Let
$y(\mbs\mphi)\in\mbR^3$ be given by coordinates $y^k$ in\rref 4.3.2(5)~ and $\tau\cdot
y:=\sum_k\tau^k y^k.$ Using\rref 4.3.2(1)~ we obtain:

\bequ\label{4.3.2(6)} y^k(\exp(-it\tau^jY_j)\mbs\mphi)=y^k(\mbs\mphi)\cos t+
\meps_{kjm}\tau^jy^m(\mbs\mphi)\sin t+2\tau^k\tau\cdot y(\mbs\mphi)\sin^2\frac{t}{2},\enqu

\noidt what gives an explicit expression for the sphere $S^2\subset so(3)^*.$ The $r_\mphi:=\
${\em radius of the sphere}\ is equal to the length of $y(\mphi)$,

\bequ\label{4.3.2(7)} |y(\mbs\mphi)|^2=y(\mbs\mphi)\cdot y(\mbs\mphi)=r_\mphi^2.\enqu

In the case (i) the values of\rref 4.3.2(7)~ might be only the numbers $J^2, (J-1)^2,
(J-2)^2,\dots$, i.e. the orbits $O_\mphi \subset\mPH$ are mapped by the association
$\mbs\mphi\mapsto F_\mphi$ (cf.\rref 4.3.2(2)~) onto a finite-number of $[J+1]$ distinct spheres
in the three-dimensional linear space $so(3)^*$ (here $[k]$ is the integer part of $k\in\mbR_+$;
if $J\in\mbZ_+$\ one of the spheres degenerates into a point). But \PH\ is a connected manifold
and the mapping $\mbs\mphi\mapsto F_\mphi$ is continuous, hence for $J\geq 1$ also the cases (ii)
occur and the numbers\rref 4.3.2(7)~ acquire values from a whole interval of $\mbR_+$, if
$\mbs\mphi$ runs over \PH.

Let us write explicitly the symplectic form $\mOme^M$ on the phase space $M_\mphi = S^2.$ In terms
of coordinate functions $y_k$ from\rref 4.3.2(5)~, we obtain in the region where
$y_3(\mbs\mphi)\neq 0$ (indices are written down for convenience):

\bequ\label{4.3.2(8)} \mOme^M=-\,\frac{1}{y_3}\rd y_1\wedge \rd y_2,\quad
y_3^2:=r_\mphi^2-y_1^2-y^2_2.\enqu

The Poisson bracket of these coordinate functions is

\bequ\label{4.3.2(9)} \{y_k,y_m\}=-\,\meps_{kmj}y_j.\enqu

The sphere $S^2$ with this symplectic structure is interpreted as the phase space of an (isolated)
\emn classical spin~. It is an example of a compact symplectic manifold.}

\pt\label{4.3.3}{\rm We can construct now certain combinations of the previous example with those
of Sec.'s \ref{sec;4.1} and \ref{sec;4.2}. Let us distinguish generators $X_j$ of the
representation of $6N+1$ - dimensional Heisenberg group corresponding to coordinates of positions
and momenta of $N$ individual particles. Denote them $Q_j^a, P_j^a\ (a = 1,2,\dots N; j = 1,2,3)$
with CCR in the form

\bequ\label{4.3.3(1)}  [Q_j^a, P_k^b]=i\,I\delta_{ab}\delta_{jk},\ [Q_j^a,Q_k^b] = [P_j^a,P_k^b] =
0,\enqu

\noidt for all $a,b=1,2,\dots N;\ j,k =1,2,3.$ Now we define operators of orbital momenta (no
summation over indices $a, b$):

\bequ\label{4.3.3(2)} Y_j^a:=\meps_{jkm}Q_k^aP_m^a,\ Y_j:=Y_j^{tot}:=\sum_aY_j^a \enqu

\noidt satisfying\rref 4.3.2(1)~ (up to domain specifications) for any upper index ($a$, or
$tot$). Relations\rref 4.2.1(2)~ have now the form:

\bequ\label{4.3.3(3)} [Y_j^a,Q_k^b]=i\,\delta_{ab}\meps_{jkm}Q_m^a,\
[Y_j^a,P_k^b]=i\,\delta_{ab}\meps_{jkm}P_m^a,\enqu

\bequ\label{4.3.3(4)} [Y_j,Q_k^a]=i\,\meps_{jkm}Q_m^a,\ [Y_j,P_k^a]=i\,\meps_{jkm}P_m^a,\
[Y_j,Y_k^a]=i\,\meps_{jkm}Y_m^a.\enqu

Let us first consider the Lie algebra $\mfk g_0$ represented by generators $Q_j^a, P_j^a$ and
$Y_j\ (j=1,2,3;\ a=1,2,\dots N)$ of the representation $U(G_0)$ of the corresponding group $G_0$,
compare Proposition \ref{4.2.2}. We see that $G_0$ is a semidirect product of of $SU(2)$ with the
Heisenberg group $G_{3N}$ (with the notation from \ref{4.2.1}), $G_0 = SU(2)\ltimes G_{3N}$, where
the Heisenberg group is a normal subgroup. Let us investigate the orbits $O_\mphi := \mbs
U(G_0)\mbs\mphi\subset \mbs{\mcl D}_{G_0}$ in \PH\ and the corresponding classical phase spaces
$M_\mphi$. Since any $O_\mphi$ is a homogeneous space of $G_0$, it can be generated from a point
$\mbs\mphi$ satisfying (see \ref{4.1.5})

\bequ\label{4.3.3(5)} Tr(P_\mphi X_j) = 0,\ \text{for\ all}\ j = 1,2,\dots 6N. \enqu

The local structure of $O_\mphi$ is most easily seen in a \nbhd\ of such $\mbs\mphi.$ The isotropy
group of $F_\mphi\in\mfk g^*_0$\ (see\rref 3.1.3(3)~) \wrt\ $Ad^*(G_0)$ has the Lie algebra
generated by such $C\in U(\mfk g_0),$ which are solutions of the system

\bequ\label{4.3.3(6)} Tr(P_\mphi[C,X_j])=0\ (j=1,2,\dots 6N),\quad Tr(P_\mphi[C,Y_k])=0\
(k=1,2,3).\enqu

\noidt The corank of the matrix of this homogeneous system is:\nl

(i) equal to 3 iff $Tr(P_\mphi Y_k)=0$ for all $k=1,2,3$; in this situation there might occur
cases with $\dim O_\mphi = 6N,\ 6N+2,\ 6N+3$ corresponding to such $\mphi$, for which $Y\mphi = 0$
for all $Y\in U(so(3)),$ (resp. $Y\mphi=0$ for just one linearly independent $Y\in U(so(3))$,\
resp. $Y\mphi\neq 0$ for all $Y\neq 0$); as an example of the case of $\dim O_\mphi = 6N+3$ we can
take $\mphi$ for $N = 1$ in Schr\"odinger realization of CCR:

\bequ\label{4.3.3(7)} \mphi(q):=\mphi(q_1,q_2,q_3):= c\,q_1q_2q_3\,\exp(-q_1^2-q_2^2-q_3^2),\
c:=\left(\frac{2^5}{\pi}\right)^{\frac{3}{4}},\enqu

\noidt corresponding to the value $J=3$ of the total momentum. In all these cases of various
values of $\dim O_\mphi$ the corresponding symplectic spaces $M_\mphi$ are homeomorphic to
$T^*\mbR^{3N}=\mbR^{6N}.$

(ii) equal to $1$ in all other cases; now all the solutions $C$ of\rref 4.3.3(6)~ are proportional
to $C_\mphi$ of the form\rref 4.3.2(4)~. If $\mphi$ is an eigenvector of $C_\mphi$, then $\dim
O_\mphi=\dim M_\mphi=6N+2.$ In the remaining case it is $\dim O_\mphi=6N+3$ and $\dim M_\mphi =
6N+2$. If $\mphi$ is proportional to $C_\mphi\mphi$, the orbit $O_\mphi$ is the fiber-bundle with
base $\mbR^{6N}$ and typical fiber $S^2;$ if $\mphi$ is not an eigenvector of $C_\mphi,$ then the
fiber on $R^{6N}$ is the whole group $SO(3)$. In the both cases the phase space is $T^*\mbR^{3N}$
fibered by two dimensional spheres $S^2$ with the canonical symplectic form from \PH\  being the
sum of the canonical form on $T^*\mbR^{3N}$ and that on $S^2$ described in\rref 4.3.2(8)~.

Let us take now all the operators $Q_j^a,\ P_j^a,\ Y_j^a\ (a=1,2,\dots N; j=1,2,3)$ as generators
of the considered representation $U(G)$\ (now $G$ is semi direct product of the Heisenberg group
$G_{3N}$ and of the direct product of $N$ copies of the group $SU(2)$). The orbits and
corresponding phase spaces arising from the action of this group $G$ on \PH\ with $\mH =
L^2(\mbR^{3N}) = L^2(\mbR^3)\otimes L^2(\mbR^3)\otimes\dots L^2(\mbR^3)$\ (N-tuple tensor product)
can be constructed as N-tuple direct product manifolds; each of the multipled manifolds can be
obtained by the above described procedure with N = 1.

Examples of classical systems obtained in this subsection include systems of several
nonrelativistic spinning particles. Here the '\emn classical spin~' was obtained from quantal {\bf
orbital} momentum.}

\pt\label{4.3.4}\rm The groups which are, perhaps, physically most important ones, are Galilean
and Poincar\'e groups. Because of relative complexity of any complete exposition of these
important examples, we shall restrict our present exposition to several notes and remarks. For
more detailed nice exposition see e.g. in \cite{varad}.

(i) {\bf The \emn Galilean group~.}

This group realizes the nonrelativistic (better: Galilean relativistic) conception of relative
positions and motions of mechanical systems (particles, bodies etc.). It is a ten parameter Lie
group, the parameters of which can be chosen to describe time and space translations (4
parameters), space rotations (3 parameters) and transition to uniformly moving systems (3
coordinates of a velocity). Any unitary (vector) representation of this group cannot be, however,
interpreted in terms of really observed physical systems, see e.g. \cite[Sec.XII.8]{varad}.
Physically interpreted projective representations correspond to multipliers $m_\tau$ of the
Galilean group characterized by a real parameter $\tau$ - the mass of the system. Let us denote by
$G$ the \emn central extension~ (cf. \cite{varad,kiril}, resp. also \cite[Note 3.3.6]{bon-EQM}) of
(the covering group of) the Galilean group by \bR\ corresponding to a multiplier $m_\tau$ with
$\tau\neq 0$\ (all such groups are mutually isomorphic). Orbits of $Ad^*(G)$ (described e.g. in
\cite{alonzo}) are just one particle phase spaces obtained in our subsection \ref{4.3.3}. Unitary
representations of $G$, in which the central subgroup \bR\ acts by a multiplication by constants,
correspond to physically interesting projective representations of the Galilean group. Irreducible
representations of $G$ describe one-particle systems. The projected orbits $O_\mphi$ of these
representations are either seven or nine or ten dimensional (this is a consequence of \ref{4.3.3},
\ref{4.2.7} and absolute continuity of the spectrum of the time-evolution generator
$P_1^2+P_2^2+P_3^2$ of $U(G)$). In the cases $\dim O_\mphi = 7\ {\rm or}\ 9$ the manifolds
$O_\mphi$ with the two-form $\mOme^\circ$ (cf. \ref{3.2.2}) are just contact manifolds of the
extended phase spaces, $\dim O_\mphi = \dim M_\mphi + 1.$

(ii) {\bf The \emn Poincar\'e group~.}

Let now $G$ be the ten-parameter covering group of the Poincar\'e group. Physical interpretation
of the parameters is the same as that of the corresponding parameters of the Galilean group. In
the present case of $G$, however, the conception of Galilean relativity is replaced by the
conception of Einstein relativity of mechanical motions. Since the second cohomology group of $G$
is now trivial, we have to deal with unitary (vector) representations of $G$ only. The orbits of
the coadjoint action of $G$ corresponding to phase spaces of particles with nonvanishing masses
have the same topological and symplectic structure as in the case (i). The action $Ad^*(G)$ is,
however, different from that of the Galilean case; with this are connected also different
interpretations of coordinates determined by the mutually corresponding generators in cases (i)
and (ii). The dimensionality of orbits $O_\mphi$ of unitary irreducible representations $\mbs
U(G)$ corresponding to nonzero masses is the same as in (i). Also here, we obtain 7- and
9-dimensional contact manifolds the contact two-form $\mOme^\circ$ on them coincides with the
standard two-form of classical relativistic mechanics (which, in the case of $\dim O_\mphi = 7,$
comes from the restriction of $\rd p_\mu\wedge\rd q^\mu$ defined on $T^*\mbR^4$ onto the
submanifold $p_0^2-\sum_j p_j^2 = (mass)^2)$.

\begin{rem}\label{4.3.5}\rm Any symplectic manifold can be trivially extended to
a contact manifold by taking the direct product with \bR. If M is a symplectic phase space of some
physical system, then the added dimension in $\mbR\times M$ can be interpreted as the 'time
variable' $t.$ Let \Ome\ be the symplectic form on $M,\ \pi: \mbR\times M\rarw M$ be the canonical
projection and $\msg_A$ the Hamiltonian vector field on $M$ with Hamiltonian function $f_A$, i.e.
$i(\msg_A)\mOme = -\rd f_A.$ The contact two-forms $\mOme^\circ :=\pi^*\mOme,$ resp. $\mOme^A:=
\mOme^\circ - \rd f_A\wedge\rd t$ on the manifold $\mbR\times M$ have characteristic vector fields
$\delta_t$ (defined by $\rd t(\delta_t) = 1$ and $\rd f(\delta_t) =0$ for any function $f$ of the
form $f := \pi^* f',$ where $f'\in C^\infty(M)$), resp. $\msg_A^\circ := \pi^*\msg_A+\delta_t$
(with the identification $T(\mbR\times M) = T\mbR\times TM$ in the sense of vector bundle
isomorphisms). Clearly $\pi_*\msg_A^\circ=\msg_A.$ For a time-independent vector field $\msg_A$
this procedure is trivial, if we have no possibility to distinguish various points of the fibres
$\mbR = \pi^{-l}(x)\ (x \in M)$ by some measurements, i.e. if time is homogeneous \wrt\ the
considered physical system. This is the case of classical mechanics determined by $(M;\mOme)$ and
$f_A\in C^\infty(M).$

 The situation is different for contact orbits $O_\mphi\subset \mPH.$ Each point
of $O_\mphi$ corresponds to a quantum mechanically clearly distinguishable physical state: by
measuring of also quantities other than expectations of generators of $U(G),$ we can empirically
distinguish various points of the same fibre, on which
 all the expectations of the generators in $U(\mfk g)$ are constant. This
fact breaks, in a certain sense, the homogeneity of time on contact orbits of the representations,
which contain also time evolution of the system as a one parameter subgroup.
\end{rem}

\pt\label{4.3.6} \emm Identical particles~.

 \rm If the physical system consists of $N$
mutually distinguishable, but otherwise equal subsystems, it is described in QM by the $N$-fold
tensor product Hilbert space $\mH_N := \mH\otimes \mH \otimes\dots \otimes\mH$ with the Hilbert
space \H\ describing a single subsystem. If the 'basic observables' of a single subsystem are
determined by a representation $U(G)$ in \H, observables of the whole compound system might be
determined by the representation $U_N$ of the $N$-fold direct product group\ $G_N := G\times
G\times\dots\times G,$\ i.e. for $\mphi := \mphi_1\otimes\mphi_2\otimes\dots \otimes\mphi_N\in
\mH_N, \ \mphi_j\in\mH,$ we set $U_N(g_1\times g_2\times\dots\times g_N)\mphi :=
U(g_1)\mphi_1\otimes U(g_2)\mphi_2\otimes\dots U(g_N)\mphi_N$ for all $g_j\in G,$ and extend $U_N$
onto $\mH_N$ by linearity and continuity. This is the case, e.g. of the example in
Sec.\ref{sec;4.1}. Then we can construct in the usual way orbits $O_\mphi := \mbs
U_N(G_N)\mbs\mphi$ in $P(\mcl H_N)$ and corresponding symplectic manifolds $M_\mphi$. We shall
write also $U(G_N):=U_N(G_N).$

In physics, however, `equal (micro-)subsystems' are \emn indistinguishable~. If the $N$ subsystems
are indistinguishable (identical), then for any permutation $\pi\in\Pi_N$ (:= the permutation
group of $N$ elements) the product-vectors $\mphi :=
\mphi_1\otimes\mphi_2\otimes\dots\otimes\mphi_N$ and $\pi\cdot\mphi :=
\mphi_{\pi(1)}\otimes\mphi_{\pi(2)}\otimes\dots\otimes\mphi_{\pi(N)}$, as well as their linear
combinations (the permutations $\pi\in\Pi_N$ act here also as linear operators on $\mH_N$)
 are physically indistinguishable. There were discovered in the particle and statistical physics
 two kinds of particles: \emm Bose particles~ - \emn bosons~ (e.g. photons, mesons) specified
 by their integer particle spin, and \emm Fermi
 particles~ - \emn fermions~ (e.g. electrons, protons, neutrinos) having  half-integer spins.
 Collections of $N$ \emn identical particles~ of each
 of these kinds behave according of their own specific `statistics': Bose, resp. Fermi statistics.
The two `statistics' are formalized by two different symmetry properties of multiparticle wave
functions of corresponding collections of particles. In the case of Bose (resp. Fermi) statistics
the only physically realizable states correspond to \emn totally symmetric~ (resp. \emn totally
antisymmetric~) vectors $\mphi\in \mH_N$:

\begin{subequations}\label{4.3.6(1)}
\bequ\label{4.3.6(1+)} \pi\cdot\mphi = \meps_+(\pi)\mphi,\quad \meps_+(\pi):= 1,\quad \text{for
all}\  \pi\in\Pi_N,\enqu

\noidt in the case of \emm Bose statistics~, resp.

\bequ\label{4.3.6(1-)} \pi\cdot\mphi=\meps_-(\pi)\mphi,\quad \meps_-(\pi):=\pm 1:=\text{parity\
of}\ \pi\in\Pi_N.\enqu

\end{subequations}
\noidt in the case of \emm Fermi statistics~.\footnote{This relation between spin and statistics
can be obtained as a consequence of mathematical axiomatics of relativistic quantum field theory,
cf. e.g. \cite{str&wight}.}

Let $P_+$\ (resp.\ $P_-$) be the orthogonal projector in $\mH_N$\ onto the subspace $\mH^+_N$\
(resp. $\mH_N^-$) of the totally symmetric\rref 4.3.6(1+)~ (resp. totally antisymmetric\rref
4.3.6(1-)~) vectors. Now we intend to project the above mentioned orbits $O_\mphi\subset P(\mH_N)$
into $P(\mH_N^+),$ resp. into $P(\mH_N^-).$ To make the procedure more transparent we shall divide
it to more steps then it is, perhaps, necessary. For a $U(G_N)$-analytic vector $\mphi\in \mH_N\
(\mphi\neq 0)$ let $\widetilde{O}_\mphi\ := U(G_N)\mphi$, so that $O_\mphi :=
P\widetilde{O}_\mphi.$  We shall denote by $P: \mH'\rarw P({\cal H'}),\ \mphi\mapsto P_\mphi$\,,\
the natural projection in all the cases of $\mH' := \mH_N,\ \mH_N^+,\ \mH_N^-.$ Let

\bequ\label{4.3.6(2)} \widetilde{O}^+_\mphi := P_+\widetilde{O}_\mphi\, ,\ \widetilde{O}^-_\mphi
:= P_-\widetilde{O}_\mphi\ \text{be\ subsets\ of}\ \mH_N^+\ (\text{resp.}\ \mH_N^-).\enqu

Assume, for definiteness, that $P_+\mphi\neq 0$, and {\sl concentrate ourselves to the Bosonic
case} (the formal procedures are similar with the fermions).  Let $K^\mphi$ be the stability group
of $\mphi$ \wrt\ $U(G_N).$ Considerations similar to those of Sec.\ref{sec;3.1} show that
$\widetilde{O}_\mphi$, as an immersed submanifold of $\mH_N,$ is diffeomorphic to $G_N/K^\mphi.$
We shall consider $\widetilde{O}_\mphi$ with the differentiable manifold structure of
$G_N/K^\mphi.$ The restricted mapping of $P_+$:

\bequ\label{4.3.6(3)} P^\mphi_+:\ \widetilde{O}_\mphi\rarw\mH^+_N,\ \psi\mapsto P_+\psi, \
\psi\in\widetilde{O}_\mphi, \enqu

\noidt is (infinitely) differentiable. Hence the set

\bequ\label{4.3.6(4)} \widetilde{O}_{\mphi +}^\circ := (P^\mphi_+)^{-1}(0)\subset
\widetilde{O}_\mphi \enqu

\noidt is closed in $\widetilde{O}_\mphi$\,, and
\[\centerline{\emn$\widetilde{O}_{\mphi +} :=
\widetilde{O}_\mphi\setminus\widetilde{O}_{\mphi +}^\circ$~ is a submanifold of
$\widetilde{O}_\mphi$.}\] Each point of \emn$P^\mphi_+\widetilde{O}_{\mphi +}$~ has a well defined
projection into $P(\mH_N^+)$ and the mapping \emn$PP^\mphi_+$~,

\bequ\label{4.3.6(5)}  PP^\mphi_+: \widetilde{O}_{\mphi +}\rarw P(\mH_N^+),\ \mphi'\mapsto
PP^\mphi_+\mphi' := \ \{\mlam P^\mphi_+\mphi':\ \mlam\in\mbC\}\in P(\mH_N^+),\enqu

\noidt is real analytic. The number \emn$rg(\mphi')\in\mbZ_+\ (\mphi'\in\widetilde{O}_{\mphi
+})$~:

\bequ\label{4.3.6(6)} rg(\mphi') := rank\ T_{\mphi'}(PP^\mphi_+), \enqu

\noidt where $T_{\mphi'}$ is the tangent mapping in an arbitrarily chosen point $\mphi'\in
\widetilde{O}_{\mphi +}$, is given in some charts on $\widetilde{O}_{\mphi +}$ around $\mphi'$ and
on $P(\mH_N^+)$ around $PP^\mphi_+\mphi'$ as the dimension of the vector space\footnote{This
vector space is, as could be seen from the formula, the image of the tangent space
$T_{\mphi'}\widetilde{O}_{\mphi +}$ by the tangent map of the mapping $PP_+^\mphi$.}
$T_{\mphi'}(PP_+^\mphi)[T_{\mphi'}\widetilde{O}_{\mphi +}]$\ (which is, roughly speaking, the
maximal rank of submatrices of the mapping $T_{\mphi'}(PP^\mphi_+)$\ in these charts with
nonvanishing determinants). The function $\mphi'\mapsto rg(\mphi')$ is lower semicontinuous, and
possesses only finite number of values. Hence for $m_\mphi := \max\{rg(\mphi'):\
\mphi'\in\widetilde{O}_{\mphi +}\}$ the subset \emn$\widetilde{O}_{\mphi +}^m$~ of
$\widetilde{O}_{\mphi}$ defined by:

\bequ\label{4.3.6(7)} \widetilde{O}_{\mphi +}^m := rg^{-1}(m_\mphi) := \{\mphi'\in
\widetilde{O}_{\mphi +}: rg(\mphi')=m_\mphi\},\enqu

\noidt is open, hence it is a submanifold of $\widetilde{O}_{\mphi}$. We can assume that $\mphi$
was chosen such, that $\mphi\in\widetilde{O}_{\mphi +}^m.$ Let, for any
$\psi\in\widetilde{O}_{\mphi +}^m,$ the \emn$\mfk k_0^\psi$~ $\subset$ \emn$\mfk g_N$~\ (:= the
\emm Lie algebra of $G_N$~) be the linear space consisting of those generators $\xi\in\mfk g_N,$
for which

\bequ\label{4.3.6(8)} T_\psi(PP_+^\mphi)X_\xi\psi := i\,\left.\frac{\rd}{\rd t}\right|_{t=0}
PP^\mphi_+\exp(-itX_\xi)\psi = 0.\enqu

Clearly, $\dim \mfk k_0^{\psi}=\dim G_N-m_\mphi$ is constant for all $\psi\in\widetilde{O}_{\mphi
+}^m.$ The equation \rref 4.3.6(8)~ is equivalent to the equation

\bequ\label{4.3.6(9)} (I_\mH-P_{\psi^{(+)}})P_+X_\xi\psi = 0,\ \text{with}\ \psi^{(+)}:=
P_+^\mphi\psi.\enqu

By the relation \emn$\psi^{(\pm)}\in\mH^\pm_N$~ is {\bf defined} {\em the completely symmetric
(resp. antisymmetric) part of the vector} $\psi\in\mH_N$. Let
\[\centerline{$\mfk m^\mphi_0$ be a
complementary subspace in $\mfk g_N$ to $\mfk k_0^{\mphi}$.}\]

Since the mapping $PP^\mphi_+$ restricted to $\widetilde{O}_{\mphi +}^m$ is smooth and of constant
rank $m_\mphi$, it is a \emn subimmersion~ (compare \cite[5.10.6.]{bourb;manif}), hence {\bf there
is a manifold} \emn$Z^m_{\mphi +}$~ of dimension $m_\mphi$ and a \emn submersion~ $s_+^\mphi:
\widetilde{O}_{\mphi +}^m\rarw Z^m_{\mphi +}$ as well as an \emn immersion~ $i_+^\mphi:Z^m_{\mphi
+}\rarw P(\mH_N^+)$ such, that

\begin{subequations}\label{4.3.6(10)}
\bequ\label{4.3.6(10a)} PP_+^\mphi = i_+^\mphi\circ s^\mphi_+\ \text{on}\ \widetilde{O}^m_{\mphi
+}.\enqu

This means, that the image $PP^\mphi_+(\widetilde{O}^m_{\mphi +})\subset P(\mH_N^+)$ can be
considered as an \emn immersed submanifold~ (with possible \emn selfintersections~) of
$P(\mH_N^+):$

\bequ\label{4.3.6(10b)} PP_+^\mphi(\widetilde{O}^m_{\mphi +}) =i^\mphi_+(Z^m_{\mphi +}).\enqu
\end{subequations}

A basis of the tangent space to $Z^m_{\mphi +}$ is generated in the point $\nu:=s_+^\mphi(\mphi)$
by curves $t\mapsto s^\mphi_+(\exp(-itX_\xi)\mphi$ with $\xi\in \mfk m^\mphi_0$. The image by
$T_\nu i_+^\mphi$ of this tangent space in $T_{\mbs{\mphi^{(+)}}}P(\mH_N^+)$ is generated by
vectors which, in the chart $\Psi_{\mphi^{(+)}}$ (see \ref{2.1.5},\ \ref{2.1.8}), have the form

\bequ\label{4.3.6(11)} T_{\mbs{\mphi^{(+)}}}\Psi_{\mphi^{(+)}}(v_\xi) :=
-i\,(I-P_{\mphi^{(+)}})P_+X_\xi\mphi,\quad \xi\in \mfk m^\mphi_0.\enqu

The values of the symplectic form \Ome\ on $P(\mH_N)$ on these vectors are:

\bequ\label{4.3.6(12)} \mOme_{\mbs{\mphi^{(+)}}}(v_\eta,v_\xi)=-2\|\mphi^{(+)}\|^{-2}
\Im(P_+X_\eta\mphi,(I-P_{\mphi^{(+)}})P_+X_\xi\mphi).\enqu

The pull-back of \Ome\ by $i^\mphi_+$ makes $Z^m_{\mphi +}$ a manifold endowed with a canonical
two-form. It is known, that the factorization of the subimmersion $PP^\mphi_+$ (together with the
choice of the manifold $Z^m_{\mphi+}$) can be chosen in a canonical way, see
\cite[5.10.7]{bourb;manif}. We assume here, that the mapping $s^\mphi_+$ is onto (i.e.
surjective), what is possible, because any submersion is an open mapping. The form
$i^{\mphi*}_+\mOme$ on $Z^m_{\mphi+}$ is closed. The subset of $Z^m_{\mphi+}$ on which the form
$i_+^{\mphi*}\mOme$ has its maximal rank is an open set, hence a submanifold $Z_{\mphi+}$  of
$Z^m_{\mphi+}$. Denote by $\mOme^\circ_+$  the restriction of $i_+^{\mphi*}\mOme$ onto
$Z_{\mphi+}$. Since $\rd \mOme_+^\circ = 0,$ the \emn characteristic bundle~ of $\mOme^\circ_+$\
(consisting of vector fields on $Z_{\mphi+}$ annihilating the form $\mOme^\circ_+$) is an \emn
integrable subbundle~ of $TZ_{\mphi+}$, see e.g. \cite[5.1.2]{abr&mars}, determining a \emn
natural foliation~ of $Z_{\mphi+}$; any \emn leaf of this foliation~ is an immersed connected
submanifold of $Z_{\mphi+}$. Let $M_\mphi^+$ be the factor space obtained from $Z_{\mphi+}$ by its
decomposition into the leaves of this foliation and let $p^+_M: Z_{\mphi+}\rarw M^+_\mphi$ be the
natural projection. If the equivalence relation on $Z_{\mphi+}$ given by classes identical with
leaves $[p_M^{+}]^{-1}(x)\ (x\in M^+_\mphi)$ is regular (see \cite[5.9.5]{bourb;manif}), then
there is unique manifold structure on $M^+_\mphi$ such that $p^+_M$ is a submersion. In this case
there is, on the malnifold $M^+_\mphi$, a unique symplectic form $\mOme^M_+$ satisfying

\bequ\label{4.3.6(13)} p^{+*}_M\mOme_+^M = \mOme^\circ_+.\enqu

 The Proposition \ref{3.2.10} is a special case of this assertion. \nl

 \noidt{\bf Note:} In the above presented construction of the symplectic manifold
$(M_\mphi^+,\mOme_+^M)$, we did not use any specific properties of the projector $P_+$ and of the
group action $U(G_N)$. These properties enter in constructions of specific orbits.

\pt\label{4.3.7} \rm We shall specify here the previous construction to the case of $G_N$ :=
$N$-fold direct product of 2n+l-dimensional Heisenberg group $G$ with infinite-dimensional unitary
irreducible representations $U$ in \H. The linear space $U_N(\mfk g_N)$ is spanned by
elements\footnote{For $\mfk g_N=\bigoplus_{j=1}^N \mfk g^{(j)},\ \mfk g^{(j)}$ are copies of \fk
g, one has $\xi:=\sum_{j=1}^N \xi_j$\ with $\xi_j\in\mfk g^{(j)},\  X^j_\xi:=X_{\xi_j}\in U(\mfk
g)$.}

\bequ\label{4.3.7(1)} X_\xi:=\sum_{j=1}^N X^j_\xi\quad  \text{with\ any}\ X_\xi^j\in U(\mfk g),\
\xi\in\mfk g_N,\enqu

where the index $j$ has the following meaning: If $\mphi\in\mH_N$ has the form

\bequ\label{4.3.7(2)} \mphi := \mphi_1\otimes\mphi_2\otimes\dots\otimes\mphi_N,\enqu

then the linear operator $X^j$ on $\mH_N$ corresponds to an (equally denoted) operator on \H\ by:

\bequ\label{4.3.7(3)} X^j\mphi := \mphi_1\otimes\mphi_2\otimes\dots\otimes
X^j\mphi_j\otimes\mphi_{j+1}\otimes\dots\otimes\mphi_N.\enqu

(No summation! In this subsection all sums are explicitly indicated.)

Let us work in the Schr\"odinger realization of CCR, i.e. $\mH = L^2(\mbR^n),\ \mH_N =
L^2(\mbR^{Nn})$ and operators $X^k_j\ (k = 1,2,\dots N;\ j= 1,2,\dots 2n)$ acting on the k-th copy
of $L^2(\mbR^n)$ are chosen as in\rref 4.1.2(1)~. Let $\mphi\in\mH_N$ be given by\rref 4.3.7(2)~
with $\mphi_j\in L^2(\mbR^n),\ \supp \mphi_j\cap \supp \mphi_k = \emptyset\ (j\neq k)$ and such,
that there is a \nbhd\ of unity $e\in G$ so that for any $g_j\ (j=1, 2,\dots N)$ in this \nbhd\
also $U(g_j)\mphi_j$ and $U(g_k)\mphi_k\ (j\neq k)$ have disjoint supports. We assume, moreover,
that $\mphi$ is a smooth function on $\mbR^{Nn}.$ With these assumptions, we obtain from\rref
4.3.6(12)~ in a \nbhd\ of the point $s^\mphi_\pm(\mphi)$ on $Z^m_{\mphi\pm}$\ (the following
result shows that the mappings $PP^\mphi_\pm$ have at $\mphi$ the maximal rank):

\bequ\label{4.3.7(4)} \mOme_{\mbs{\mphi^{(\pm)}}}(v_\eta,v_\xi)=i\,\sum_{j=1}^N
(\mphi_j,[X^j_\eta,X^j_\xi]\mphi_j),\enqu

\noidt where we assumed for all the $j:\ \|\mphi_j\|=1,$ and $X_\eta,\ X_\xi$ in\rref 4.3.6(12)~
are of the form\rref 4.3.7(1)~. The expression\rref 4.3.7(4)~ shows, that $Z_{\mphi\pm} =
M^\pm_\mphi$ is a $2Nn$-dimensional symplectic manifold. This means, that $Z_{\mphi\pm}$ for both
signs are locally diffeomorphic (and symplectomorphic) to $M_\mphi = O_\mphi = \mbR^{2Nn}\subset
P(\mH_N)$\ (Section \ref{sec;4.1}). In a \nbhd\ of $\mbs{\mphi'}\in O_\mphi$, the functions
$\mbs{\mphi'}\mapsto Tr(P_{\mphi'} X^j_k)\ (j=1,2,\dots N;\ k=1,2,\dots 2n)$ are symplectic
coordinates. Similarly, in a \nbhd\ of $s_\pm^\mphi(\mphi)$ the functions

\bequ\label{4.3.7(5)} f_k^j:\ s_\pm^\mphi(\mphi')\mapsto Tr(P_{\mphi'}X_k^j)=(\mphi'_j,
X_k^j\mphi'_j),\ j=1,2,\dots N;\ k=1,2,\dots 2n,\enqu

\noidt are symplectic coordinate functions on $Z_{\mphi\pm}$.

Let us assume now, that $\mphi_j$'s in\rref 4.3.7(2)~ have the form

\bequ\label{4.3.7(6)} \mphi_j := W_{x^{(j)}}\mphi_0\ \text{for\ some}\ \mphi_0\in L^2(\mbR^n),\
x^{(j)}\in\mbR^n,\enqu

\noidt assuming $\mphi_0$ to be smooth with compact support, and $x^{(j)}\neq x^{(k)}\,(j\neq k)$
such that $\mphi_j, \mphi_k$ have mutually disjoint supports, see \ref{4.1.3} for the notation. On
the orbit $\widetilde{O}_\mphi$ in $\mH_N$, there is also the point

\bequ\label{4.3.7(7)} (\otimes \mphi_0)^N:=\mphi_0\otimes\mphi_0\otimes\dots\otimes\mphi_0.\enqu

Choose now $\mphi$ equal to\rref 4.3.7(7)~ and calculate the values of\rref 4.3.6(12)~ in the
points $\mbs\mphi_\pm\in P(\mH_N^\pm).$ In the antisymmetric case we obtain zero, since $P_-\mphi
=: \mphi_- =0$\ (hence  $\mphi\in\widetilde{O}^\circ_{\mphi-}$,\rref 4.3.6(4)~, and $PP_-\mphi$
{\it is not defined}).

In the case of Bose statistics we have:

\bequ\label{4.3.7(8)} \mOme_{\mbs\mphi^{(+)}}(v_\eta,v_\xi) = \frac{i}{N}\sum_{j=1}^N\sum_{k=1}^N
(\mphi_0,[X_\eta^k,X_\xi^j]\mphi_0),\enqu

\noidt where $X_\eta^j\ (j=1,2,\dots N)$ should be considered as operators in $L^2(\mbR^n),$
ignoring the definition\rref 4.3.7(3)~: they act on $L^2(\mbR^n)$ regardless of its order in the
tensor product forming the whole Hilbert space $\mH_N$. The rank of the form\rref 4.3.7(8)~ equals
to $2n$ and the point $PP_+^\mphi(\mphi)$ does not belong to $i_+^\mphi(Z_{\mphi+})$ for $N\geq
2$, i.e.\ $\mphi$ is not mapped by $s^\mphi_+$ into the symplectic manifold $Z_{\mphi+}$. We see
that, although locally symplectomorhic to $R^{2Nn}$, the both classical phase spaces $Z_{\mphi-}$
and $Z_{\mphi+}$ of identical particles are globally different from the standard cotangent bundle
$T^*\mbR^{Nn}:$ in classical projections the Pauli exclusion principle holds for identical
particles, regardless to the kind of their statistics.

\pt\label{4.3.8} \rm With the notation from \ref{4.3.6}, let \emn$V_N(G)$~ be the unitary
representation of $G$ in $\mH_N$\ (reducible for $N\geq 2$) defined as the diagonal part of $U_N$:

\bequ\label{4.3.8(1)} V_N(g) := U_N(g\times g\times\dots \times g),\ \text{for all}\ g\in G.\enqu

The {\bf Lie algebra} \emn$V_N(\mfk g)$~ is generated by the basis of the form\rref 4.3.7(1)~ with
$X_\xi^j=X_\xi^k$ (considered as operators in \H) for all $j,k=1,2,\dots N,\ \xi\in\mfk g.$ Such
operators $X_\xi\in V_N(\mfk g)$ commute with projectors $P_\pm$. Hence $V_N$ leaves the subspaces
$\mH_N^+$ and $\mH_N^-$ invariant, and we can obtain the classical projections of this
'macroscopic' (for large $N$) subsystem in the standard way, (Sec.\ref{sec;3.2}); the obtained
classical phase spaces are orbits of $Ad^*(G)$ with their canonical symplectic structure - there
is no difference in the kinds of statistics, from the point of view of kinematics.

In trying to extend our constructions to systems consisting of infinite number $N\rarw\infty$ of
equal (or identical) subsystems, we meet the problems of divergence of 'global (or collective)
observables' $X^N_\xi := X_\xi$ and of discontinuity of the resulting representation $V_\infty$ of
G. We give a formalization of this 'large N limit' in the next Section \ref{sec;5.1}, and in the
Sec. \ref{sec;5.2} we outline a possible generalization of obtaining classical subsystems of
collective observables from infinite quantal systems. We shall not take any care of statistics of
subsystems, what could be motivated by results of the last two subsections: the statistics seems
to have no essential influence upon the classical phase spaces of systems of identical particles.

\newpage

\chapter{Macroscopic limits}\label{Ch5}

\section{Multiple systems}\label{sec;5.1}

\pt\label{5.1.1} {\rm We shall construct in this section classical subsystems of a large quantal
system. We shall assume here that the large system consists of infinite number of copies of a
finite subsystem of the type dealt with in preceding sections. The infinite "macroscopic" system is obtained as an inductive limit of a net of systems consisting of an increasing number of copies of the mentioned finite systems. The symmetry group $G$ of a single finite
subsystem is then also a symmetry group of the large system. An essential formal difference \wrt
the systems discussed in preceding sections is that the action of $G$ on the large system is not
described by a continuous unitary representation, hence we cannot introduce generators
corresponding to one-parameter subgroups of $G$ as operators in some Hilbert space.}

\pt\label{5.1.2}\rm To make the following considerations more intuitive, let us come back for a
while to finite systems consisting of $N$ equal subsystems. Let the unitary representation
$V_N(G)$ and its generators \emn$X^N_\xi := X_\xi\ (\xi\in\mfk g)$~ be defined as in \ref{4.3.7}
and \ref{4.3.8}, esp. in\rref 4.3.7(1)~. Then

\bequ\label{5.1.2(1)} [X_\xi^N,X_\eta^N] = i\,X^N_{[\xi,\eta]}\quad (\xi,\eta\in\mfk g)\enqu

\noidt and the restriction to the orbit \emn$O^N_\mphi := \mbs V_N(G)\mbs\mphi$~\
$(\mphi\in\mH_N)$ of the canonical symplectic form \emn$\mOme^N$~ on $P(\mH_N)$ is determined by

\bequ\label{5.1.2(2)} \mOme^N_{\mbs\mphi}(\msg_\xi,\msg_\eta) = i\,
Tr(P_\mphi[X^N_\xi,X^N_\eta]),\ (\xi,\eta\in\mfk g).\enqu

\noidt Here $\msg_\xi$ is the vector field on $P(\mH_N)$ corresponding to the unitary flow

\bequ\label{5.1.2(3)} (t;\mphi)\mapsto \exp(-itX^N_\xi)\mphi,\quad \mphi\in\mH_N,\ t\in\mbR.\enqu

For $N\rarw\infty$, the operators $X^N_\xi$ diverge and $V_N(G)$ does not converge to any
continuous unitary representation - compare the next subsection. Let

\bequ\label{5.1.2(4)} X_{\xi N} := \frac{1}{N} X^N_\xi,\quad \xi\in\mfk g,\ N=1,2,\dots.\enqu

In terms of \cite{hp+lie1} $X^N_\xi$\ (resp. $X_{\xi N}$) are `extensive (resp. intensive)
observables' but, contrary to \cite{hp+lie1}, they can be unbounded in our case. The limits for
large $N$ of $X_{\xi N}$'s could exist in some convenient sense, but they are not generators of
any unitary representation of the group $G.$ Due to the commutation relations

\bequ\label{5.1.2(5)} [X_{\xi N},X_{\eta N}]=\frac{i}{N} X_{[\xi,\eta]N}, \enqu

\noidt the limits of $X_{\xi N}\ (\xi\in\mfk g)$ will be mutually commuting operators. To obtain
correct classical commutation relations (i.e. the Poisson brackets, see \ref{1.3.5}) for functions
$f_{\xi N}$ on the orbits $O^N_{\mbs x}\ (x\in\mH_N),$

\bequ\label{5.1.2(6)} f_{\xi N}:\ \mbs x\mapsto f_{\xi N}(\mbs x):= Tr(P_xX_{\xi N}), \enqu

\noidt in the limit $N\rarw \infty$, the two-form $\mOme^N$ from\rref 5.1.2(2)~ should be
'renormalized'. We define

\bequ\label{5.1.2(7)} \mOme_N := \frac{1}{N}\mOme^N.\enqu

The form $\mOme_N$\ (if restricted onto the symplectic manifold $M^N_{\mbs x}$ obtained from
$O^N_{\mbs x}$ as in Sec.\ref{sec;3.2}) associates with the Hamiltonian function $f_{\xi N}$ the
vector field $\msg_\xi$ (restricted to $M^N_{\mbs x}$) given by the flow\rref 5.1.2(3)~. It is

\bequ\label{5.1.2(8)} \mOme_{N\bullet}(\msg_\xi,\msg_\eta)=i\,Tr(P_\bullet[X_\xi^N,X_{\eta N}]) =
- Tr(P_\bullet X_{[\xi,\eta]N}). \enqu

 We intend to develop a corresponding formalism for infinite
systems, i.e. a suitable one for the work in the limit $N =$ 'actual infinity'.

\pt\label{5.1.3}\rm  Let $U(G)$ be a continuous unitary representation of a connected Lie group
$G$ on a separable Hilbert space \H. We shall use notation of Chap. \ref{Ch4} for concepts related
to $U(G)$. Let $\Pi$ be an index set (of arbitrary cardinality) and $\mH_j\ (j\in\Pi)$ be copies
of \H. Let us fix unitary maps

\bequ\label{5.1.3(1)} u_j: \mH \rarw \mH_j,\quad j\in \Pi,\enqu

\noidt of \H\ onto $\mH_j$'s. Let \ind{$\mH_\Pi$}

\bequ\label{5.1.3(2)} \mH_\Pi := \bigotimes_{j\in\Pi} \mH_j \enqu

\noidt be the tensor product defined according to von Neumann \cite{neum2} and  known as \emn
CTPS~ (:= \emm complete tensor product space~ - see also notes in the text in \ref{5.1.4} below
and \cite{emch1,sak1,bon-disert}). For $\mphi_j\in\mH_j$ let

\bequ\label{5.1.3(3)} \Phi := \bigotimes_{j\in\Pi} \mphi_j \enqu

\noidt be a \emn product-vector in $\mH_\Pi$~. For any linear densely defined operator $A$ on \H\
(with domain $D(A)\subset\mH$) and for $\mphi_j\in\mH_j $ such that $u_j^{-1}\mphi_j\in D(A)$ let
\emn $\pi_j(A)$~ be the operator on \emn$\mH_\Pi$~ determined by

\bequ\label{5.1.3(4)} \pi_j(A)\Phi := \left(\bigotimes_{k\in\Pi\setminus\{j\}}\mphi_k\right)
\otimes(u_jAu_j^{-1}\mphi_j).\enqu

\noidt Symbolically: $\pi_j(A) := I_1\otimes I_2\otimes\dots \otimes I_{j-1}\otimes A\otimes
I_{j+1}\otimes\dots, $
 if $\Pi = \mbZ_+\setminus\{0\}.$

 Unitary group action $U_\Pi$ of $G$ on $\mH_\Pi$ is determined by

 \bequ\label{5.1.3(5)} U_\Pi(g)\Phi :=
 \bigotimes_{j\in\Pi}(u_jU(g)u_j^{-1}\mphi_j).\enqu

 \noidt For $|\Pi|$ (:= the cardinality of $\Pi$) finite, the representation $U_\Pi$ is strongly
 continuous with generators

\bequ\label{5.1.3(6)} X_\xi^\Pi := \sum_{j\in\Pi} \pi_j(X_\xi),\ \xi\in\mfk g.\enqu

$U_\Pi$ is not weakly continuous in the case of infinite $\Pi:$ If $\mphi\in\mH$ is not an
eigenvector of $X_\xi,\ \mphi_j := u_j\mphi$ for all $j\in\Pi,\ \|\mphi\|=1$ and $\Phi$ is the
corresponding product-vector\rref 5.1.3(3)~ in $\mH_\Pi$, then $\|\Phi\|=1$ and

\bequ\label{5.1.3(7)} (\Phi,\ U_\Pi(\exp(t\xi))\Phi) = 0 \enqu

\noidt for all sufficiently small $|t|\neq 0, t\in\mbR$, since
\bequ\label{5.1.3(8)} |(\mphi_j,
u_j\exp(-itX_\xi)u_j^{-1}\mphi_j)| = |(\mphi,\exp(-itX_\xi)\mphi)|< 1\quad \text{if}\quad
e^{-itX_\xi}\mphi\neq \mlam\mphi,\enqu

\noidt for any $\mlam\in\mbC$, i.e. the function in\rref 5.1.3(7)~ is discontinuous at $t=0$.

\pt\rm {\bf Notes on the structure of CTPS.}\label{5.1.4} \nl

 We shall not give here a thorough definition of \emm CTPS~. We shall assume
that the definitions of (convergence and quasiconvergence of)\ infinite products and sums of
complex numbers as well as of the scalar product in $\mH_\Pi$  according to \cite{neum2} are known
to the reader. Let $z\in\mbC^\Pi$, i.e. $z$ is a function \bequ\label{5.1.4(1)} z:
\Pi\rarw\mbC,\quad j\mapsto z_j.\enqu

Assume that $|z_j|= 1$ for all $j\in\Pi$ and define a unitary operator $U_z$ on $\mH_\Pi$ by its
linear action on product vectors\rref 5.1.3(3)~\ (the set of which is total in $\mH_\Pi$) given by

\bequ\label{5.1.4(2)} U_z\Phi:=\bigotimes_{j\in\Pi} (z_j\mphi_j).\enqu

 Let $\{\mphi^n: n\in\mbZ_+\}$ be an orthonormal basis in \H. Let
 $a,b\in\mbZ_+^\Pi$
with components $a_j,b_j\in\mbZ_+\ (j\in\Pi)$, and set

\bequ\label{5.1.4(3)} \mphi_j^a:= u_j(\mphi^{a_j})\in\mH_j,\
\Phi^a:=\bigotimes_{j\in\Pi}\mphi^a_j.\enqu

For $a\neq b$, the vectors $\Phi^a$ and $\Phi^b$ are mutually orthogonal: $(\Phi^a,\Phi^b)=0.$ Let
$\Phi :=\Phi^a$ for some $a$ (this can be done so for any normalized product-vector
$\Phi\in\mH_\Pi$ by a choice of the identifications $u_j,\ j\in\Pi,$ of $\mH_j$ with \H). The
vectors $\Phi^b,$ for which $b_j=a_j$ for all $j\in\Pi\setminus J_b$, $b_j\in\mbZ_+$ for all $j\in
J_b$, where $J_b$ runs over all {\em finite} subsets of $\Pi,$\footnote{i.e. all the vectors
$\Phi^b$ for which $b_j\neq a_j$ for finite number of indices $j\in\Pi$ only}  form an orthonormal
basis in a closed subspace of $\mH_\Pi$ {\bf denoted by} \emn$\mH^\Phi_\Pi$~ and called \emm ITPS~
(\emm incomplete tensor product space~). {\bf Let} \emn$P_\Phi$~ be the orthogonal projector in
$\mH_\Pi$ onto $\mH_\Pi^\Phi.$ For two arbitrary product vectors $\Phi,\Psi\in\mH_\Pi$ the
projectors $P_\Phi$ and $P_\Psi$ are either orthogonal or equal. For any $U_z$ from\rref 5.1.4(2)~
we have

\bequ\label{5.1.4(4)} U_zP_\Psi U^*_z = P_{U_z\Psi}, \enqu

\noidt and the product vectors $\Psi$ and $U_z\Psi$ are \emm weakly equivalent~. If $P_\Phi
\Psi=\Psi$ (hence $P_\Psi \Phi=\Phi)$, then $\Phi$ and $\Psi$ are \emm(strongly) equivalent~. The
set of all product vectors $\Phi$ weakly equivalent to a product vector $\Psi$ form a total set in
a closed subspace of $\mH_\Pi$ with the {\bf orthogonal projector} \emn$P_\Psi^w$~. Clearly,
$P_\Psi^w$ is the sum of all such $P_\Phi$, which correspond to mutually \emn strongly
inequivalent~ product vectors $\Phi$, all of them being weakly equivalent to $\Psi$. The sum of
all mutually strongly inequivalent $P_\Psi$ (we use an obvious licence in language) is the unit
operator in $\mH_\Pi$. \footnote{We shall use sometimes projectors instead of the corresponding
subspaces.}

Let \emn$\mfk A^\Pi$~ denotes the $C^*$-subalgebra of the algebra of all bounded operators on
$\mH_\Pi$ (denoted by \emn $\mcl L(\mH_\Pi)$~), generated by the elements

\bequ\label{5.1.4(5)} \{\pi_j(A)\in \mcl L(\mH_\Pi):A\in\mLH,\ j\in\Pi\},\enqu

\noidt where \emn \LH~ is the algebra of all bounded operators on the Hilbert space \H.

\noidt For any $x\in\mfk A^\Pi$, the following relations are valid, \cite{neum2}:

\bequ\label{5.1.4(6)} [x, P_\Psi] = [x, U_z] = 0\ \text{for all}\ U_z, \text{and for all}\
P_\Psi,\enqu

\noidt with $U_z$ from\rref 5.1.4(2)~. If $p$ is another orthogonal projector in $\mcl
L(\mH_\Pi)$, and for some product-vector $\Psi$ it is $pP_\Psi= p$, then

\bequ\label{5.1.4(7)}\quad \text{if}\ [x,p] = 0\ \text{for all}\ x\in\mfk A^\Pi\Rightarrow  p =
P_\Psi\ \text{or}\ p = 0,\enqu

\noidt i.e. irreducibility of the action of $\mfk A^\Pi$ in each $\mH^\Psi_\Pi$. The weak closure
of $\mfk A^\Pi$ in $\mcl L(\mH_\Pi)$ consists of all elements $x\in\mcl L(\mH_\Pi)$ satisfying
\rref 5.1.4(6)~. The action of $\mfk A^\Pi$ in $\mH^\Psi_\Pi$ is a representation of this \Ca.
Such representations (all irreducible and faithful) for two product vectors are unitarily
equivalent iff these vectors are weakly equivalent. The \emn center~ of the weak closure of $\mfk
A^\Pi$ in $\mcl L(\mH_\Pi)$ is generated by the projectors $P_\Psi^w$. {\bf Denote} this weak
closure by \emn$\mfk B^\#$~ and by \emn$\mfk Z^\#$~ its center: $x\in\mfk Z^\# \subset \mfk B^\#$
iff $[x,y]=0$ for all $y\in\mfk B^\#$.

\begin{prop}\label{5.1.5} The mapping

\bequ\label{5.1.5(1)} \msg:\ G\rarw{}^*\= Aut\,\mfkA^\Pi,\ g\mapsto \msg_g, \enqu

\noidt defined by (see\rref 5.1.3(5)~)

\bequ\label{5.1.5(2)} \msg_g(x) := U_\Pi(g) x U_\Pi(g^{-1}),\ \forall x\in\mfk A^\Pi,\ g\in G,
\enqu

\noidt is a group homomorphism of $G$ into the group $\maut\mfk A^\Pi$ of \autm s of the \Ca\
$\mfk A^\Pi$. For any normalized vector $\Psi\in\mH_\Pi$ define the vector state \emn$\mome^\Psi$~
on $\mfk A^\Pi$ by

\bequ\label{5.1.5(3)} \mome^\Psi:\ x\mapsto \mome^\Psi(x)\ := (\Psi, x\Psi).\enqu

The functions

\bequ\label{5.1.5(4)} g\mapsto \mome^\Psi(\msg_g(x)) \enqu

\noidt for any $x\in\mfk A^\Pi$ and any $\Psi\in\mH_\Pi$ are continuous functions from $G$ to \bC.
\end{prop}
\begin{proof}
The mapping $A\mapsto U(g)AU(g^{-1})$ is a \autm\ of  \LH,\ $A\in\mLH$. Since $\mfk A^\Pi$ is
generated by elements $x := \pi_j(A)\ (j\in\Pi,\ A\in\mLH)$ defined in\rref 5.1.3(4)~\ (i.e. $\mfk
A^\Pi$\ is the norm-closure of finite linear combinations of finite products of such elements),
the first statement follows from the definition\rref 5.1.3(5)~ of $U_\Pi$. The functions\rref
5.1.5(4)~\ are continuous for all $x = \pi_j(A)$ and for all product states $\mome^\Psi$ (i.e.
states corresponding via\rref 5.1.5(3)~\  to product vectors $\Psi$ of the form\rref 5.1.3(3)~).
The set of product vectors is total in $\mH_\Pi$ and any \autm\ of a \Ca\ is norm-continuous.
These facts imply by standard considerations validity of the last statement.
\end{proof}

\begin{noti}\label{5.1.6} Due to weak discontinuity of $U_\Pi$, the
second statement of \ref{5.1.5}\ is not valid if $\mfk A^\Pi$ would be replaced by its weak
closure $\mfk B^\#$ in $\mcl L(\mH_\Pi)$. This can be seen by setting $\Psi:=\Phi^a$ from\rref
5.1.4(3)~\ with $a_j := 0$ (for all $j\in \Pi$) and with a choice $\mphi^0\in\mH$ such that it is
not an eigenvector of the generator $X_\xi$ of $U(G)$ for some $\xi\in\mfk g$. Then, setting $x :=
P_\Psi^w\in\mfk B^\#$ in\rref 5.1.5(3)~, the function

\bequ\label{5.1.6(1)} t\mapsto \mome^\Psi(\msg_{\exp(t\xi)}(P^w_\Psi)) \enqu

\noidt is discontinuous at $t = 0$ : For $t = 0$ its value equals to $1$, but for arbitrarily
small nonzero values of $t\in\mbR$ the values of\rref 5.1.6(1)~ are found to be zero.\end{noti}

\pt\label{5.1.7}\rm To simplify notations, we shall set $\Pi := \mbZ_+\setminus\{0\}$ for the rest
of the present section. For a densely defined linear operator $A$ on \H\ with domain $D(A)$, let

\bequ\label{5.1.7(1)} D^\Pi(A):=\bigotimes_{j\in\Pi} u_jD(A) \enqu

\noidt be the linear subset of \emn$\mH_\Pi$~ consisting of finite linear combinations of product
vectors $\Phi$,\rref 5.1.3(3)~, with $\mphi_j\in u_j D(A)\ (j\in\Pi)$. \emn$D^\Pi(A)$~ is not, in
general, dense in $\mH_\Pi$. Let

\bequ\label{5.1.7(2)}  A_N :=\frac{1}{N}\sum_{j=1}^N \pi_j(A),\quad (N\in\Pi), \enqu

\noidt be (densely defined) operators on $\mH_\Pi$, a common domain of which contains $D^\Pi(A)$.
{\bf Let} \emn$D_\Pi(A)$~ be the set of vectors $\Psi\in\mH_\Pi$ such, that

\bequ\label{5.1.7(3)} A_\Pi\Psi :=\ norm-\lim_{N\rarw\infty} A_N\Psi \enqu

\noidt exists in $\mH_\Pi$. The set $D_\Pi(A)$ is a nonzero linear subset of $\mH_\Pi$: for
$\mphi\in D(A)$ and $\mphi_j:= u_j\mphi\
 (j\in \Pi)$, the product vector $\Phi$ from\rref 5.1.3(3)~\ belongs
 to $D_\Pi(A)$.
Let $\{\mphi^n: n\in\mbZ_+\}\subset D(A)$ be an orthonormal basis in \H\ and, for some
$a\in\mbZ_+^\Pi,$ let $\Phi^a$ defined according to\rref 5.1.4(3)~\ belongs to $D_\Pi(A)$. Then,
for $b\in \mbZ_+^\Pi$ differing from $a$ in at most finite number of components, it is $\Phi^b\in
D_\Pi(A)$. With $\Psi:=\Phi^a,$ such vectors $\Phi^b$ form an orthonormal basis in $\mH_\Pi^\Psi,$
hence \emn$P_\Psi D_\Pi(A)$~ is dense in \emn$\mH_\Pi^\Psi$~, and\rref 5.1.7(3)~ give a densely
defined operator on $\mH_\Pi^\Psi$. For any product vector $\Psi\in D_\Pi(A)$, let us define a
densely defined operator on $\mH_\Pi^\Psi$:

\bequ\label{5.1.7(4)}  A^\Psi := P_\Psi A_\Pi P_\Psi = P_\Psi A_\Pi. \enqu

The second equality is a consequence of the obvious commutativity of $A_\Pi$ with $P_\Psi$ for any
product vector $\Psi\in D_\Pi(A).$ The restriction of $A_\Pi$ to the subspace $\mH^\Psi_\Pi$
(which clearly is a linear, not densely defined operator on $\mH_\Pi$) will be denoted by
\emn$A^\Psi_\Pi$~, or simply \emn$A^\Psi$~ ($\Psi\in D_\Pi(A)$). Now it is easy to prove

\begin{lem}\label{5.1.8} For a densely defined operator $A$ on \H, let $\Psi\in
D_\Pi(A)$ be a product vector in $\mH_\Pi.$ Then $A^\Psi = \mlam P_\Psi$ for some $\mlam\in\mbC$,
on $D_\Pi(A).$
\end{lem}
\begin{proof} Since $\Psi\in D_\Pi(A)$ is a product vector, it is also
$\Psi\in D^\Pi(A)$. We shall assume that $\Psi$ is normalized. Then it can be written in the form

\bequ\label{5.1.8(1)} \Psi = \bigotimes_{j=1}^\infty \mphi_j,\ \text{with}\ u_j^{-1}\mphi_j\in
D(A)\ \text{for}\ j=1,2,\dots,\enqu

\noidt where each $\mphi_j\ (j\in\Pi)$ is normalized in $\mH_j:\ \|\mphi_j\|^2 = (\mphi_j,\mphi_j)
= 1.$ Let $\Psi_k\in D^\Pi(A)\ (k=1,2)$ be such product vectors in $\mH_\Pi^\Psi$ which differ
from\rref 5.1.8(1)~ at most in the first $n$ factors $\mphi_j$. Such vectors $\Psi_k,$ with
$n\in\Pi,$  form a total set in $\mH_\Pi^\Psi$. We have

\begin{eqnarray}\label{5.1.8(2)}
(\Psi_1, A^\Psi\Psi_2) &=& \lim_{N\rarw\infty} \frac{1}{N}\left(\sum_{j=1}^n (\Psi_1,
\pi_j(A)\Psi_2)+\sum_{j=n+1}^N (\Psi_1,\pi_j(A)\Psi_2)\right)=\nonumber
\\ &=& \lim_{N\rarw\infty}\frac{1}{N}\sum_{j=n+1}^N
(\mphi_j,u_jAu_j^{-1}\mphi_j)(\Psi_1,\Psi_2)=\nonumber \\ &=&
\lim_{N\rarw\infty}\frac{1}{N}\sum_{j=1}^N
(\Psi,\pi_j(A)\Psi)(\Psi_1,\Psi_2)=(\Psi,A^\Psi\Psi)(\Psi_1,\Psi_2).
\end{eqnarray}

\noidt By linearity, the obtained relation extends to all $\Psi_k\in P_\Psi D_\Pi(A).$ On that
domain, we obtain

\bequ\label{5.1.8(3)} A^\Psi = Tr(P_\Psi^\circ A^\Psi)P_\Psi = Tr(P_\Psi^\circ A_\Pi) P_\Psi,
\enqu

\noidt where $P_\Psi^\circ$ is the projector onto the one-dimensional subspace of $\mH_\Pi^\Psi$
spanned by the vector $\Psi$.
 \end{proof}

\noidt {\bf Note:} Since $A^\Psi$ is bounded on $\mH_\Pi$\  (if $\Psi\in D_\Pi(A)$\ is a
product-vector), we shall extend this operator to the whole $\mH_\Pi$ by continuity and we shall
denote this extension by the same symbol, hence: $A^\Psi \in \mcl L(\mH_\Pi)$.

\begin{prop}\label{5.1.9} Let $\Psi\in$ \emn $D_\Pi(\mfk g)$~ be an
arbitrary vector from

\bequ\label{5.1.9(1)} D_\Pi(\mfk g) := \bigcap_{\xi\in\mfk g} D_\Pi(X_\xi), \enqu

\noidt in the notation of \ref{5.1.3} and \ref{5.1.7}. Then $U_\Pi(g)\Psi\in D_\Pi(\mfk g),$ for
all $g\in G$. In particular, with $g\cdot \Psi := U_\Pi(g)\Psi$, we have for product-vectors
$\Psi\in D_\Pi(\mfk g)$:

\bequ\label{5.1.9(2)} X_\xi^{g\cdot \Psi} = Tr(P^\circ_{g\cdot\Psi}X_{\xi\Pi}) P_{g\cdot\Psi} =
Tr(P^\circ_\Psi X_{Ad(g^{-1})\xi\Pi}) P_{g\cdot\Psi}.\enqu
\end{prop}

\begin{proof}
According to Lemma \ref{3.1.4}, $U(g)X_\xi U(g^{-1})=X_{Ad(g)\xi}$ for any $\xi\in\mfk g$. Then,
according to \ref{5.1.3}, we have also

\bequ\label{5.1.9(3)} U_\Pi(g^{-1})\pi_j(X_\xi)U_\Pi(g)=\pi_j(X_{Ad(g^{-1})\xi}). \enqu

For $\Psi\in D_\Pi(\mfk g)$ there exist $X^\Psi_\xi$ for all $\xi\in\mfk g.$ Because of continuity
of unitary operators $U_\Pi(g)$ for any fixed $g\in G$, there exist also the limits

\bequ\label{5.1.9(4)} \lim_{N\rarw \infty} U_\Pi(g)X_{Ad(g^{-1})\xi N}\Psi =
U_\Pi(g)X^\Psi_{Ad(g^{-1})\xi}\Psi  \enqu

\noidt for all $\xi\in\mfk g.$ Rewriting the expression on the \lhs\ of\rref 5.1.9(4)~ we get

\bequ\label{5.1.9(5)} U_\Pi(g)X_{Ad(g^{-1})\xi N}\Psi=\frac{1}{N}\sum_{j=1}^N
U_\Pi(g)\pi_j(X_{Ad(g^{-1})\xi})\Psi= X_{\xi N}U_\Pi(g)\Psi.\enqu

\noidt This shows that the limit of the \rhs of\rref 5.1.9(5)~ for large $N$ exists for any
$\xi\in\mfk g,$ what proves the first assertion. The proof of the second assertion is a corollary
of the proof of the first one for the case of a product vector $\Psi\in D_\Pi(\mfk g)$, obtained
from\rref 5.1.8(3)~.
\end{proof}

\pt\label{5.1.10}\rm For a product vector $\Psi\in D_\Pi(\mfk g),$ let \emn$\mome^\Psi$~ be the
corresponding state on $\mfk A^\Pi$ defined in\rref 5.1.5(3)~. We shall denote the obvious
extension of this state to the unbounded observables $X_{\xi N}\ (N\in\Pi)$ by the same symbol.
Then we have

\bequ\label{5.1.10(1)} \lim_{N\rarw\infty} \mome^\Psi(X_{\xi N}) = Tr(P^\circ_\Psi X_{\xi\Pi}) =:
\mome^\Psi (X_{\xi\Pi}). \enqu

\noidt We see that the value of expressions in\rref 5.1.10(1)~ can be interpreted as the value of
the intensive (unbounded) observable $X_{\xi\Pi}$ in the state $\mome^\Psi$. Define the linear
functional \emn$F_\Psi\in\mfk g^*$~ by

\bequ\label{5.1.10(2)} F_\Psi: \xi\mapsto F_\Psi(\xi) := Tr(P^\circ_\Psi X_{\xi \Pi}),\ \text{for\
product\ vectors}\ \Psi\in D_\Pi(\mfk g).\enqu

According to\rref 5.1.9(2)~, the action $g\cdot F_\Psi := F_{g\cdot\Psi}$\ of $G$ coincides with
the $Ad^*(G)$-action:

\bequ\label{5.1.10(3)} (g\cdot F_\Psi)(\xi) = F_{g\cdot\Psi}(\xi) = F_\Psi(Ad(g^{-1})\xi) =
(Ad^*(g)F_\Psi)(\xi).\enqu

According to \ref{5.1.9}, the set of product vectors in $D_\Pi(\mfk g)$ is $U_\Pi(G)$-invariant,
hence any point of the orbit $G\cdot F_\Psi$ has the form\rref 5.1.10(2)~.

Define the group homomorphism $\msg^*$ of $G$ into the group of affine transformations of the
state-space $\mcl S(\mfk A^\Pi):$

\bequ\label{5.1.10(4)} \msg^*:\ G\rarw \msg^*_G\,,\ g\mapsto \msg^*_g\,,\ \text{where}\
(\msg^*_g\mome)(x) := \mome(\msg_{g^{-1}}(x)) \enqu

\noidt for all $g\in G,\ \mome\in\mcl S(\mfk A^\Pi)\ \text{and} \ x\in\mfk A^\Pi\ \text{with}\
\msg_g$ defined in\rref 5.1.5(2)~. Let $\Psi\in D_\Pi(\mfk g)$ be a product vector and

\bequ\label{5.1.10(5)}  O_\Psi :=\{\msg^*_g\mome^\Psi: g\in G\}\subset \mcl S(\mfk A^\Pi) \enqu

\noidt be the orbit through $\mome^\Psi$ of the action $\msg^*_G$. For $\mome\in$ \emn$ O_\Psi$~
let

\bequ\label{5.1.10(6)} F_\mome\in\mfk g^*: \ F_\mome(\xi):= \mome(X_{\xi\Pi}).\enqu

Let us write also $g\cdot\mome := \msg^*_g\mome.$ Clearly $g\cdot\mome^\Psi := \mome^{g\cdot
\Psi}$. According to\rref 5.1.10(3)~, the mapping \emn $\mbs F$~ from the state space into the
dual $\mfk g^*$ of the Lie algebra:

\bequ\label{5.1.10(7)} \mbs F:\ O_\Psi\rarw \mfk g^*, \ \mome\mapsto \mbs F(\mome):= F_\mome,
\enqu

\noidt maps the orbit $O_\Psi$ onto an orbit of $Ad^*(G).$ Let
\[\centerline{\emn$[\mome] := \mbs
F^{-1}(F_\mome)$~, for $\mome\in O_\Psi$, be equivalence classes in $O_\Psi$.}\]

\noidt The corresponding factor space \emn$M_\Psi$~ is mapped by $\mbs F$ (which is constant on
classes $[\mome]$) bijectively onto the orbit $G\cdot F_\Psi$. The last orbit is endowed by the
\emn Kirillov-Kostant symplectic structure~. The functions $f_\xi$ on $M_\Psi$:

\bequ\label{5.1.10(8)} [\mome] \mapsto f_\xi(\mome) := \mome(X_{\xi \Pi}),\ \mome\in O_\Psi,\
\xi\in\mfk g, \enqu

\noidt are the Hamiltonian functions generating the flows

\bequ\label{5.1.10(9)} (t;[\mome])\mapsto [\exp(t\xi)\cdot\mome]. \enqu

\noidt Corresponding Poisson brackets are:

\bequ\label{5.1.10(10)} \{f_\xi,f_\eta\}([\mome]) = -F_\mome ([\xi,\eta]),\ \xi,\eta\in\mfk g,
\enqu

\noidt compare e.g.\rref 3.2.2(2)~. Here it is assumed that $M_\Psi$ is endowed by the manifold
structure of the $Ad^*(G)$-orbit $\mbs F(M_\Psi)$. We have obtained here classical phase spaces
from equivalence classes of states in $\mcl S(\mfk A^\Pi)$ determined by the group action
$\msg_G^*.$ Although the construction is formally parallel to that in the case of finite systems,
there are certain physically significant differences in the interpretation, as mentioned in
\ref{1.1.6}.

\pt\label{5.1.11}\rm {\bf Let} \emn$P_G$~ be the orthogonal projector in \LHp\ onto the subspace
of \Hp\ spanned by all product vectors $\Psi\in D_\Pi(\mfk g).$ The operator $P_G$ is equal to the
sum of all mutually orthogonal projectors $P_\Psi^w$ corresponding to the product vectors $\Psi\in
D_\Pi(\mfk g)$, as is seen from\rref 5.1.8(3)~ and obvious commutativity of any $A_\Pi$ with all
the $U_z,$\rref 5.1.4(2)~.\ Hence
\[\centerline{\emn$P_G\in \mfk Z^\#$~ := the center of \emn$\mfk
B^\# := (\mfk A^\Pi)''$~ := the weak operator closure of $\mfk A^\Pi$ in \LHp }\]

\noidt (commas denote here the double commutant). The mapping

\bequ\label{5.1.11(1)} \rho:\ \mfk A^\Pi\rarw P_G\mfk B^\#,\quad x\mapsto P_Gx, \enqu

\noidt is a *-representation of the \Ca\ $\mfk A^\Pi$ in the Hilbert space $P_G\mHp.$

The \emn representation $\rho$~\ can be uniquely extended to a \emm$W^*$-representation~ of the
$W^*$-algebra (i.e. abstract von Neumann algebra) $(\mfk A^\Pi)^{**}:=$ the double dual of $\mfk
A^\Pi$, see \cite[1.21.13]{sak1}. (The unique extensions of mappings from a \Ca\ to mappings from
its double dual will be usually denoted by the same symbols used for the original mappings.) The
image of $(\mfk A^\Pi)^{**}$ under $\rho$\ is $P_G\mfk B^\#$. Let
\[\centerline{ \emm $s_G\in\mfk
Z$~ := the center of $(\mfk A^\Pi)^{**}$,}\]

\noidt be the support of $\rho$, i.e. $(I - s_G)(\mfk A^\Pi)^{**}$ is the kernel of \emn$\rho$~
($I$ is here the identity of $(\mfk A^\Pi)^{**}$). The restriction \emn$\rho_G$~ to $s_G(\mfk
A^\Pi)^{**}$ of $\rho$\ is an isomorphism of $W^*$-algebras (which is \sg\ - \sg\ continuous, see
\cite[1.21.13+4.1.23]{sak1}). {\bf Let $\mcl S_{\mfk g}\subset \mcl S(\mfk A^\Pi)$ consists of
such states $\omega$, the central supports $s_\mome\in\mfk Z$ of which are contained in $s_G$},
i.e. $s_\mome s_G = s_\mome$ (the \emm central support~ of a state is defined as the central
support, equiv. \emm central cover~ - cf.
\cite[3.8.1]{pedersen},\cite{takesI},\cite[1.14.2]{sak1}, of the extension to $(\mfk A^\Pi)^{**}$
of the corresponding cyclic representation of $\mfk A^\Pi$). The set \emn$\mcl S_{\mfk g}$~ will
play an important role in the following.

The automorphisms $\msg_g\ (g\in G)$,\rref 5.1.5(2)~, have unique extensions to automorphisms of
the \Wa\ $(\mfk A^\Pi)^{**}$, which are \sg\ - \sg\ and also norm - norm continuous,
\cite[1.21.13]{sak1}. The $\msg_g$ can be understood also as an (uniquely defined) automorphism of
the von Neumann algebra $\mfk B^\#$.\ Due to Proposition \ref{5.1.9},\ it is

\bequ\label{5.1.11(2)} \msg_g(P_G) = P_G\ \text{for all}\ g\in G,\enqu

\noidt hence also

\bequ\label{5.1.11(3)} \msg_g(s_G)= s_G,\quad g\in G.\enqu

Let us keep the {\bf notation} \emn$X_{\xi\Pi}\ (\xi\in\mfk g)$~ for the closures of the
restrictions to $P_G\mHp$ of operators denoted previously by the same symbols. According to\rref
5.1.9(2)~, all the $X_{\xi\Pi}$'s have in $P_G\mHp$ a common complete orthonormal set (a basis) of
eigenvectors consisting of product vectors $\Psi\in D_\Pi(G),$ with real eigenvalues. Hence, they
form a set of mutually commuting selfadjoint operators on $P_G\mHp$. Let $E^\#_{\xi\Pi}(B)\ (B :=$
any Borel subset of \bR) be projectors forming their spectral measures $E^\#_{\xi\Pi}.$ All these
projectors belong to $P_G\mfk Z^\#$, since any $X_{\xi\Pi}\ (\xi\in\mfk g)$ is a constant on each
$P_\Psi^w < P_G.$ Define

\bequ\label{5.1.11(4)} E_{\xi\Pi}(B) := \rho_G^{-1}[E^\#_{\xi\Pi}(B)] \in s_G\mfk Z\quad \text{for
all}\ \xi\in\mfk g\ \text{and Borel}\ B\subset \mbR.\enqu

Any \emn$E_{\xi\Pi}\ (\xi\in\mfk g)$~ is a resolution of identity in the \Wa\ $s_G\mfk Z.$ Let us
define also

\begin{eqnarray}\label{5.1.11(5)} E_{\xi\Pi}'(B) &:=& E_{\xi\Pi}(B),\ \text{if}\ B\
\text{does\ not\ contain the zero}\ 0\in\mbR,\\ \nonumber &:=& E_{\xi\Pi}(B)+ I- s_G,\ \text{if}\
0\in B.
\end{eqnarray}

\noidt Here $I$ is the identity of $\mfk Z$. Then $E_{\xi\Pi}'\ (\xi\in\mfk g)$ is a resolution of
identity in $\mfk Z$.

\begin{defi}\label{5.1.12} Let \emn$\mfk M_G$~ be the \Wsa\ of $\mfk Z$ generated
by projectors $E_{\xi\Pi}(B)\ (\xi\in\mfkg,\ B$\ -\ Borel\ in\ \bR) and by I. $\mfk M_G$ is called
the \emm algebra of $G$-macroscopic observables~ of the system $(\mfk A^\Pi,\msg_G)$, or simply
the \emm($G$-)macroscopic algebra~. Let \emn$\mfk N_G$~ $:= s_G\mfk M_G$ be the \Wsa\ of $\mfk
M_G$ generated by projectors $E_{\xi\Pi}(B)$ and called the \emm algebra of G-definiteness~ of
$(\mfk A^\Pi,\msg_G),$ or sometimes also the (G-)macroscopic algebra, if there will be no
confusion possible.
\end{defi}

\begin{lem}\label{5.1.13} Let $\xi_j\ (j = l,2,\dots n := \dim G)$ form a basis
 in \fk g. For $\mlam\in\mbR^n$ let $F:= \sum_j \mlam_j F_j\in\mfk
 g^*$ expressed in the corresponding dual basis $\{F_j\}\subset \mfk g^*.$ Let

 \bequ\label{5.1.13(1)} E_{\mfk g}(F):=
 E_{\xi_1\Pi}(\mlam_1)E_{\xi_2\Pi}(\mlam_2)\dots
 E_{\xi_n\Pi}(\mlam_n)\in \mfk N_G. \enqu

 The projectors \emn$E_{\mfk g}(F)\ (F\in\mfk g^*)$~ do not depend on
 a  specific choice of the basis in \fk g\ and they are all
 minimal projectors in $\mfk N_G$. Here $E_{\xi\Pi}(\mlam):= E_{\xi\Pi}(\{\mlam\})$, and
 $E_{\mfk g}(F):= E_{\mfk g}(\{F\})$.
\end{lem}

\begin{proof} The restriction of the mapping $\rho_G$ to $\mfk
N_G$ is a $W^*-$isomorphism of $\mfk N_G$ into $P_G\mfk Z^\#\subset \mfk B^\#.$ Let
$\Psi\in\rho_G(E_{\mfk g}(F))\mH_\Pi.$ From linearity of the mapping $\xi\mapsto X_{\xi\Pi}\ for\
\xi=\sum\tau_j\xi_j,$ we have

\bequ\label{5.1.13(2)} X_{\xi\Pi}\Psi=\sum_j\tau_jX_{\xi_j\Pi}\Psi
=\sum_j\tau_j\mlam_j\Psi=F(\xi)\Psi.\enqu

\noidt The second equality is due to the definition of $E_{\xi\Pi}(\mlam_j)$ as the projector
corresponding to the eigenvalue $\mlam_j\in\mbR$ of $X_{\xi\Pi}$ (we write $\mlam_j$ in the place
of the one-point set $\{\mlam_j\}$ for simplicity). The last equality in\rref 5.1.13(2)~ is due to
definition of the dual basis and shows the stated independence of $E_{\mfk g}(F)$ on the choice of
a basis.

Let
\[\centerline{\emn$E_{\mfk g}^\#(F) := \rho_G(E_{\mfk g}(F))$~.}\]

\noidt {\bf Any projector} \emn$E^\#_{\mfk g}(B)$~ is a sum (uncountable - in general, see also
\cite[1.13.4]{sak1}) of projectors $E_{\mfk g}^\#(F)\ (F(\xi)\in B)$. The algebra $\rho_G(\mfk
N_G)$ is the double commutant of the set

\bequ\label{5.1.13(3)} \{E^\#_{\mfk g}(F)\ :\ F\in\mfk g^*\},\enqu

\noidt according to the bicommutant theorem by von Neumann taken in the algebra $\mcl
L(P_G\mH_\Pi)$ of bounded operators on $P_G\mH_\Pi$. All the projectors $E^\#_{\mfk g}(F)$ in
\rref 5.1.13(3)~ are mutually orthogonal. The commutant of\rref 5.1.13(3)~ contains all the
orthogonal projectors $p\leq E^\#_{\mfk g}(F)$. But any nonzero orthogonal projector $q <
E^\#_{\mfk g}(F)$ (strict inequality!) cannot commute with all such $p$'s. Hence $E^\#_{\mfk
g}(F)$ is minimal in $\rho_G(\mfk N_G)$ and $E_G(F)$ is minimal in $\mfk N_G$ for any $F\in\mfk
g^*$. Since $E_{\xi\Pi}(\mbR) = s_G$ (= the identity of $\mfk N_G)$ is a sum of $E_{\mfk g}(F)$'s
and $\mfk N_G$ is commutative, the set of all the $E_{\mfk g}(F)$'s exhausts the set of all the
minimal projectors in $\mfk N_G$.\end{proof}

\pt\label{5.1.14}\rm Any state $\mome\in\mcl S(\mfk A^\Pi)$ on the algebra of bounded observables
of our system has unique extension to a normal state on the algebra $(\mfk A^\Pi)^{**}$ and its
restriction to $\mfk M_G$ is a normal state $\mome\in\mcl S(\mfk M_G)$.  Any normal state on $\mfk
M_G$ can be obtained in this way, \cite[1.24.5]{sak1}. {\bf Let} \emn$\mcl M$~ be the \emm
spectrum space of $\mfk M_G$~, i.e. the compact set of all pure states on $\mfk M_G$ endowed with
the induced topology from the $w^*$-topology of its dual $\mfk M^*_G$. Then $\mfk M_G$ is
isomorphic (denoted by $\sim$) to the \Ca\ $C(\mcl M)$ of all complex valued continuous functions
on $\mcl M$ (by a Gel'fand-Najmark theorem, cf. \cite[§16.2
Thm.1]{najm},\cite[Thm.2.1.11A]{bra&rob}): $x\ (\in\mfk M_G) \leftrightarrow \hat{x}\ (\in C(\mcl
M))$. An element $x\in\mfk M_G$ is an orthogonal projector iff the corresponding element
\emn$\hat{x}\in C(\mcl M)$~ is characteristic function of some Borel subset $B$ of $\mcl M$, i.e.
\[\centerline{ $\hat{x}(m) = \chi_B(m)$ for all $m\in \mcl M$.}\]

\noidt A pure state $m\in \mcl M$ is normal, iff the characteristic function $\chi_{\{m\}}$ of the
one-point set $\{m\}$ is continuous, $\chi_{\{m\}}\in C(\mcl M)$. This means, that \emn normal
pure states~ on $\mfk M_G$ are just the isolated points of $\mcl M$. The corresponding projectors
$\chi_{\{m\}}$ are \emn minimal projectors~ in $\mfk M_G \sim C(\mcl M)$. The spectrum space $\mcl
M$ is Hausdorff and the family of {\bf clopen} (i.e. closed and open) {\bf sets} forms a basis of
the topology of $\mcl M$, cf. \cite{sak1}. Hence, any minimal projector in $C(\mcl M)$ is of the
form $\chi_{\{m\}}$. \ind{clopen set}

Any state $\mome\in \mcl S(\mfk M_G)$ is represented by a probability Baire (i.e. regular Borel)
measure on $\mcl M$ and any such measure $\mu_\mome$ represents a state on $\mfk M_G:
\mome(x)=\mu_\mome(\hat{x})$, where $x$ in the \lhs\ is an element of the abstract algebra $\mfk
M_G$ and $\hat{x}$ in the \rhs\ denotes the corresponding function $\hat{x}\in C(\mcl M)$. Any
pure state $m\in \mcl M$ corresponds to the Dirac measure $\delta_m$.

\pt\label{5.1.15}\rm The algebra $\mfk M_G$ (and also $\mfk N_G$) is $\msg_G-$invariant:

\bequ\label{5.1.15(1)} \msg_g x\in \mfk N_G\ \text{for\ all}\ g\in G\ \text{and\ any}\ x \in \mfk
N_G. \enqu

\noidt This is a consequence of the relation (compare the proof of \ref{5.1.9})

\bequ\label{5.1.15(2)}U_\Pi(g) X_{\xi\Pi}U_\Pi(g^{-1})=X_{Ad(g)\xi\Pi} ,\quad (g\in G,\ \xi\in\mfk
g),\enqu

\noidt what implies

\bequ\label{5.1.15(3)}\msg_g[E_{\mfk g}(B)] = E_{Ad(g)\xi\Pi}(B) \quad(g\in G\ \text{and\ Borel}\
B\subset \mbR),\enqu

\noidt due to uniqueness of spectral measures of selfadjoint operators and also due to continuity
properties of the used mappings. From\rref 5.1.15(3)~, we obtain immediately (by calculation of
the eigenvalues of $X_{\xi\Pi}$):

\bequ\label{5.1.15(4)} \msg_g[E_{\mfk g}(F)] = E_{\mfk g}(Ad^*(g)F),\quad (g\in G,\ F\in\mfk
g^*).\enqu

\noidt This specifies, according to \ref{5.1.13} and \ref{5.1.14}, the action of $G$ on the set of
all normal pure states on $\mfk N_G$. The remaining normal pure state on $\mfk M_G$ corresponds to
the $\msg_G$-invariant minimal projector $I-s_G$. Hence, $\msg_G$ acts on $\mfk M_G$ as a group of
$W^*-$automorphisms and \emn$\msg_G^*$~ acts on $\mcl M$ (resp. on $\mcl S(\mfk M_G))$ as a group
of homeomorphisms (resp. a group of continuous affine transformations). As a consequence, the
orbits

\bequ\label{5.1.15(5)} O_\mome :=\{\msg^*_g\mome:\ g\in G\}\subset \mcl S(\mfk A^\Pi) \enqu

\noidt are canonically mapped onto orbits of $\msg^*_G$ in $\mcl S(\mfk M_G)$ consisting of normal
states on $\mfk M_G$. By this mapping orbits consisting of vector states $\mome^\Psi$\ are
\footnote{Where $\Psi\in\mH_\Pi$ such that there is an $F\in\mfk g^*$ satisfying: $E_{\mfk
g}^\#(F)\Psi = \Psi$.\label{5.1.15(6)}} mapped onto orbits in $\mcl M$. The functions

\bequ\label{5.1.15(7)} \msg^*_m:\ G\rarw \mcl M,\ g\mapsto \msg_m^*(g):=\msg_g^*m,\ (m\in \mcl
M)\enqu

\noidt are not continuous in the given topology on $\mcl M$,\ \ref{5.1.14}. The orbits of
$\msg^*_G$ consisting of normal pure states on $\mfk M_G$ are, due to\rref 5.1.15(4)~, bijective
images of (some) orbits of $Ad^*(G)$ in $G^*$. It is also clear that the {\bf normal pure states
on $\mfk M_G$ form a  $G$-invariant subset} \emn $\mcl M_*$~ of all states $\mcl S(\mfk M_G)$ on
$\mfk M_G$:

\bequ\label{5.1.15(8)} \msg^*_G\mcl M_*=\mcl M_*,\ i.e.\ m\in \mcl M_*\imply \msg^*_g m\in \mcl
M_*\ \text{for all}\ g\in G\ (\msg^*_e m\equiv m).\enqu

\begin{prop}\label{5.1.16} Let $p=p^*=p^2\in\mfk M_G$ be any projector and \ind{$p\mfk g^*$}

\bequ\label{5.1.16(1)} p\mfk g^* := \{F\in\mfk g^*:\ 0\neq E_{\mfk g}(F)\leq p\}. \enqu

\noidt Let $J\subset \mfk g^*$ be a finite set and let by \emn$p_J$~ be denoted

\bequ\label{5.1.16(2)} p_J := \sum_{F\in J} E_{\mfk g}(F),\ \text{for\ any\ finite}\ J\subset\mfk
g^*.\enqu

\noidt Denote further for any subset $\mbs K\subset\mfk g^*$: \ind{$\mbs c(\mbs K)$}

\bequ\label{5.1.16(3)} \mbs c(\mbs K) := \lub\{p_J:\ J\subset \mbs K,\ J\ \text{finite}\}.\enqu
 \noidt Assume $ps_G=p.$\nl

\noidt Then the following assertions are fulfilled:\nl

(i)\ $p=\mbs c(p\mfk g^*)$,\quad and\quad  (ii)\ $\mcl M=\overline{\mcl M}_*:= \text{the\ closure\
of}\ \mcl M_*.$
\end{prop}
\begin{proof} The projector $s_G$ is constructed in such a way that
$\rho_G(s_G)=P_G$ and $P_G= E^\#_{\xi\Pi}(\mbR)$ for any $\xi\in \mfk g.$ Since $\rho_G$ is an
isomorphism of $\mfk N_G=s_G\mfk M_G$ into $\mfk Z^\#$, \ref{5.1.11}, it is $s_G= \mbs c(s_G\mfk
g^*)$. Hence, for any projector $q=qs_G$ in $\mfk M_G$, there is a nonzero minimal projector
$E_{\mfk g}(F_\circ)=E_{\mfk g}(F_\circ)q$, if $q$ is nonzero. Let $q:=p-\mbs c(p\mfk g^*)\ (\geq
0,$ according to the definition\rref 5.1.16(3)~) and assume that $q\neq 0$. Let $0\neq E_{\mfk
g}(F_\circ)=qE_{\mfk g}(F_\circ)$.  But $E_{\mfk g}(F_\circ)\leq p$, hence $E_{\mfk
g}(F_\circ))\mbs c(p\mfk g^*)= E_{\mfk g}(F_\circ).$ This ia a contradiction, since $q$ is
orthogonal to \emn$\mbs c(p\mfk g^*)$~. Hence $q = 0,$ what proves (i).

 Any projector in $\mfk M_G$ is represented in $C(\mcl M)$
by the characteristic function of a \emm clopen set~, and conversely, the characteristic function
of a clopen set in $\mcl M$ represents by Gel'fand isomorphism a projector in $\mfk M_G$,
\ref{5.1.14}. The minimal projector $E_{\mfk g}(F)$ corresponds to the one-point clopen set
$\{m_F\}$ containing $m_F\in \mcl M_*.$  The union of all $\{m_F\}\ (F\in\mfk g^*)$ is an open
subset the closure of which is clopen, since $\mcl M$ is a Stonean space, see \ref{5.1.14}, and
\cite{sak1}. According to (i), it is the support of characteristic function corresponding to $s_G
= \mbs c(s_G\mfk g^*)= \mbs c(\mfk g^*).$ The projector $s_G$ is the unit element in $\mfk N_G$
and the projector $I-s_G$ is minimal. This shows that the sum of the characteristic functions
corresponding to $s_G$ and $I-s_G$ is the characteristic function of the whole $\mcl M$, i.e.
$\mcl M$ is the union of a one-point set \emn $\{m_\circ\}$~ corresponding to $I-s_G$ and of the
closure of

\bequ\label{5.1.16(4)}  \mcl N_*:= \mcl M_*\setminus \{m_\circ\}=\{m_F:F\in\mfk g^*\},\enqu

\noidt where we set $\{m_F\} := \emptyset :=$ the empty set, if $E_{\mfk g}(F)=0.$\ This is (ii).
\end{proof}

\noidt {\bf Notation:} Let us introduce, for following usage, some further concepts. Let

\bequ\label{5.1.16(5)} \mu^\mome_\xi: B\rarw \mu^\mome_\xi(B):=\mome(E_{\xi\Pi}(B)),\ \text{for\
any}\ \mome\in\mcl S(\mfk M_G)\ \text{and\ Borel}\ B\subset\mbR,\enqu

\noidt be a \emn finitely additive Borel measure~ on \bR. For mutually dual bases $\{\xi_j:\
j=1,2,\dots n\}$\ in $\mfk g$\ and $\{F_j:\ j=1,2,\dots n\}$ in $\mfk g^*$\ define\ \emn$\mu_{\mfk
g}^\mome$~\ on $\mfk g^*$\ by:
 \bequ\label{5.1.16(6)} \mu_{\mfk
g}^\mome(\mbs B):=\mome(E_{\xi_1\Pi}(B_1)E_{\xi_2\Pi}(B_2)\dots E_{\xi_n\Pi}(B_n))\ \text{for}\
\mbs B:=\{F\in\mfk g^*:\ F(\xi_j)\in B_j\}.\enqu

\noidt If\ $\xi\in L^1(\mu^\mome_{\mfk g}, \mfk g^*)\ \text{with} \ \xi\in(\mfk g^*)^*=\mfk g,$\
then

\bequ\label{5.1.16(7)} \mome(X_{\xi\Pi}):=\mu_{\mfk g}^\mome(\xi)=\int
\mlam\mu^\mome_\xi(\rd\mlam).\enqu

\begin{lem}\label{5.1.17} The image by the natural map defined in \ref{5.1.14} of any
factor state $\mome\in\mcl S(\mfk A^\Pi)$ into $\mS(\mfk M_G)$ is an equally denoted pure state
$\mome\in \mcl M_*$ (:= the set of all normal pure states on $\mfk M_G$).
\end{lem}
\begin{proof} The \emn canonical cyclic representation
$\{\pi_\mome,\mH_\mome,\mphi_\mome\}$~ of $\mfk A^\Pi$\ (here $\mphi_\mome$ is the cyclic vector
in the Hilbert space $\mH_\mome$ for the representation $\pi_\mome$ such, that

\bequ\label{5.1.17(1)} \mome(x)=(\mphi_\mome,\pi_\mome(x)\mphi_\mome)\enqu

\noidt for all $x\in\mfk A_\Pi$) corresponding to a \emm factor state~ $\mome\in\mcl S(\mfk
A^\Pi)$ has trivial center. Hence, any projector in the center of the commutant $\pi_\mome(\mfk
A^\Pi)'$ is trivial. The canonical extension to $(\mfk A^\Pi)^{**}$ (i.e. unique $W^*-$continuous)
of $\pi_\mome$ maps the bidual $(\mfk A^\Pi)^{**}$ onto the double commutant $\pi_\mome(\mfk
A^\Pi)''$ by which $\mfk M_G\subset \mfk Z$ is mapped into the center $\pi_\mome(\mfk Z)$ of this
bicommutant. Since $\pi_\mome(\mfk Z)\subset \pi_\mome(\mfk A^\Pi)'$, any projector in
$\pi_\mome(\mfk M_G)$ is trivial. The corresponding $\mome\in\mcl S(\mfk M_G)$ is expressed by
\rref
 5.1.17(1)~ for $x\in\mfk M_G.$ This \ome\ is normal:\
 $\mome\in\mcl S_*(\mfk M_G)$, hence there exists a unique
 projector $s_\mome$ in (the center of) $\mfk M_G$ such, that

 \bequ\label{5.1.17(2)} \mome(x)=\mome(xs_\mome),\ \text{for all}\
 \{ x\in\mfk M_G: \mome(x^*x)=0\}\ \imply x=x(I-s_\mome).\enqu

 \noidt Hence for any nonzero projector $s\leq s_\mome$ one has
 $\mome(s)\neq 0$\ and\ $\pi_\mome(s)=I_\mome :=$\ the identity of $\mcl L(\mH_\mome)$. From this
 follows $\mome(s_\mome-s)=0$ and  $s_\mome-s=(s_\mome-s)(I-s_\mome)=0,$ so
 that $s_\mome$ is a minimal projector in $\mfk M_G$. This proves
 that $\mome\in \mcl M_*$.
\end{proof}

\pt\label{5.1.18}\rm For any state $\mome\in\mcl S_{\mfk g}$, the measures $\mu^\mome_\xi\
(\xi\in\mfk g)$ are probability (\sg-additive) regular Borel measures on \bR, due to normality of
$\mome\in\mcl S(\mfk M_G)$,\rref 5.1.16(5)~. Define the subset $\mcl S_{\mfk g}^d\subset \mcl
S(\mfk A^\Pi)$:

\bequ\label{5.1.18(1)} \mcl S_{\mfk g}^d:=\{\mome\in\mcl S_{\mfk g}:\ \mome(X_{\xi\Pi})\ \text{is\
finite\ for\ all}\ \xi\in\mfk g\},\enqu

\noidt where $\mome(X_{\xi\Pi})$ is defined in\rref 5.1.16(7)~. Due to\rref 5.1.15(3)~, the set
$\mcl S_{\mfk g}^d$ is $\msg_G^*-$invariant. For any $f\in L^1(\mbR,\mu_\xi^\mome)$ define

\bequ\label{5.1.18(2)} \mome(f(X_{\xi\Pi})):=\int_\mbR f(\mlam)\,\mome(E_{\xi\Pi}(\rd\mlam)).\enqu

Any state $\mome\in\mcl S_{\mfk g}$ which is mapped into $\mcl M_*$, e.g. any pure state
$\mome\in\mcl S_{\mfk g}$, belongs to $\mcl S_{\mfk g}^d$ and, moreover,

\bequ\label{5.1.18(3)} \mome(X^2_{\xi\Pi})=[\mome(X_{\xi\Pi})]^2\ \text{for\ all}\ \xi\in\mfk
g.\enqu

Denote $F_\mome(\xi):=\mome(X_{\xi\Pi})$ for $\mome\in\mcl S^d_{\mfk g}$. The mapping

\bequ\label{5.1.18(4)} \mbs F:\ \mS^d_{\mfk g}\rarw \mfk g^*,\ \mome\mapsto\mbs F(\mome):=
F_\mome;\ F_\mome(\xi):=\mome(X_{\xi\Pi}),\quad \xi\in\mfk g,\enqu

\noidt maps orbits of $\msg^*_G$ in $\mcl S^d_{\mfk g}$ onto orbits of $Ad^*(G)$ in $\mfk g^*$.
Let $\mome\in\mcl S^d_{\mfk g}$ {\bf and \emn$O_\mome := \msg^*_G\mome$~ be the corresponding orbit}. If
\rref 5.1.18(3)~ is valid for $\omega$ then it is valid for all the states in $O_\mome$, as it is
seen from\rref 5.1.15(3)~. We shall call orbits $O_\mome\subset \mcl S^d_{\mfk g}$ satisfying\rref
5.1.18(3)~ the \emm G-macroscopically pure orbits~, and similarly for single states; simply, we
shall use also \emm (G-)pure orbits~ (resp. \emm G-pure states~). The
\[\centerline{ {\bf set of
all G-pure states will be denoted by} \emn$\mcl E_{\mfk g}\ (\subset \mcl S^d_{\mfk g})$~.}\]

\noidt The state $\mome\in\mcl E_{\mfk g}$ need not be a (pure) state in $\mcl{ES}(\mfk A^\Pi)$ or
in $\mcl{ES}_{\mfk g}$. But the following assertion is valid:

\begin{prop}\label{5.1.19} For $\mome\in\mcl S_{\mfk g}^d$ and its
canonical image $\mome\in\mcl S(\mfk M_G)$ the following statements are equivalent:\nl

\quad\quad (i)\ $\mome\in \mcl E_{\mfk g};$\quad (ii)\ $\mome\in \mcl M_*.$
\end{prop}
\begin{proof} The implication (ii) $\imply$ (i) is clear. Let $\mome\in \mcl E_{\mfk g}$
and let $\mu_\mome$ be the Baire measure on $\mcl M$ corresponding to $\mome\in\mcl S_*(\mfk
M_G)$. We shall prove that $\mu_\mome$ is concentrated on a one point set $\{\mome\}\subset\mcl
M_*.$ Let $B_n\subset\mbR,\ n\in\mbZ_+,$ be an increasing absorbing sequence of Borel sets, i.e.
for any bounded Borel $B\subset\mbR$ there is some $n_B\in\mbZ_+$ that for all $n\geq n_B$ it is
$B\subset B_n$, and $B_n\subset B_{n+1}$ for $n\in\mbZ_+$. If $f: \mbR\rarw\mbC$ is any Borel
function which is uniformly bounded on each bounded Borel subset $B$
 of \bR, then

 \bequ
 E_{\xi\Pi}(B)f(X_{\xi\Pi}):=\int_Bf(\mlam)E_{\xi\Pi}(\rd\mlam)=f(E_{\xi\Pi}(B)X_{\xi\Pi})
 \enqu

 \noidt is a well defined element of $\mfk N_G,$
 \cite[1.11.3]{sak1}. Since \ome is normal,
 we can write for such `locally finite' functions
 $f\in L^1(\mbR,\mu_\xi^\mome):$

 \barr
 \mome(f(X_{\xi\Pi})) &=& \lim_{n\rarw\infty}
 \mome(E_{\xi\Pi}(B_n)f(X_{\xi\Pi})),\label{5.1.19(2)}\\
 \mome(E_{\xi\Pi}(B)f(X_{\xi\Pi})) &=& \int_{\mcl M}
 m(E_{\xi\Pi}(B)f(X_{\xi\Pi}))\mu_\mome(\rd m),
 \label{5.1.19(3)}\\
 m(E_{\xi\Pi}(B)f(X_{\xi\Pi})) &=& m(E_{\xi\Pi}(B))
 m(E_{\xi\Pi}(B)f(X_{\xi\Pi})); \label{5.1.19(4)}
 \earr

\noidt in\rref 5.1.19(4)~ we have used the character-property of $m\in \mcl M:= \mcl{ES}(\mfk
M_G)$.

For $n\in\mbZ_+,$ the function $\chi_{\xi n}:\ m\mapsto m(E_{\xi\Pi}(B_n))$ is continuous
characteristic function of a clopen set $\mcl M_{\xi n}\subset \mcl M$. From the monotonicity
property of spectral measures, we have $\mcl M_{\xi(n+1)}\supset \mcl M_{\xi n}.$ The union

\bequ\label{5.1.19(5)} \bigcup_{n\in\mbZ_+} \mcl M_{\xi n} =: \mcl M_\xi \enqu

\noidt is open, hence measurable together with all the $\mcl M_{\xi n}.$ We see from\rref
5.1.19(2)~,\rref 5.1.19(3)~ and\rref 5.1.19(4)~ that $\mu_\mome$ is concentrated on $\mcl M_\xi$:

\bequ\label{5.1.19(6)} \mu_\mome(\mcl M_\xi)=\mu_\mome(\mcl M)=1,\quad \forall \xi\in\mfk g;\enqu

\noidt it suffices to set for $f$ a (nonzero) constant function. But

\bequ\label{5.1.19(7)} \mcl M_{\mfk g} := \bigcap_{\xi\in\mfk g} \mcl M_\xi=\bigcap_{j=1}^n \mcl
M_{\xi_j} \supset \mcl M_*\setminus\{m_\circ\} = \mcl N_*,\enqu

\noidt where $\{\xi_j:\ j=1,2,\dots n\}$ is a basis of $\mfk g$, \ref{5.1.13}, and $\mu_\mome$ is
concentrated on states in $\mcl M_{\mfk g}$,

\bequ\label{5.1.19(8)} \mu_\mome(\mcl M_{\mfk g}) = \mu_\mome(\mcl M)=1,\ \text{for\ any}\
\mome\in\mcl S_*(\mfk M_G).\enqu

Let

\bequ\label{5.1.19(9)} F_\xi:\ \mcl M_\xi\rarw \mbR,\ m\mapsto  F_\xi(m):=F_m(\xi):=\lim_n
m(E_{\xi\Pi}(B_n)X_{\xi\Pi}),\enqu

\noidt what is a bounded continuous function on each $\mcl M_{\xi n}$, and due to monotonicity it
is continuous on the whole $\mcl M_\xi$. For $f$ in\rref 5.1.19(2)~ we have:

\bequ\label{5.1.19(10)} m(E_{\xi\Pi}(B_n)f(X_{\xi\Pi})) = f(F_m(\xi))\ \text{for}\ m\in \mcl
M_{\xi n},\enqu

\noidt if for $\mlam=F_m(\xi)$ the value $f(\mlam)$ is defined. From\rref 5.1.19(3)~, one sees
that the functions $m\mapsto \chi_{\xi n}(m)f(F_m(\xi))$ are in $L^1(\mcl M,\mu_\mome).$  By an
application of the Beppo-Levi theorem to their absolute values, we obtain:\nl

\noidt The functions $F^*_\xi f\in L^1(\mcl M,\mu_\mome);$ here it is

\bequ\label{5.1.19(11)} F^*_\xi f := f\circ F_\xi:\ \mcl M_\xi\rarw \mbR,\ m\mapsto
f(F_\xi(m))=f(F_m(\xi)).\enqu

\noidt We have used here\rref 5.1.19(2)~ and\rref 5.1.19(3)~. After a subsequent application of
the Lebesgue dominated convergence theorem we arrive at:

\bequ\label{5.1.19(12)} \mome(f(X_{\xi\Pi})) = \mu_\mome(F^*_\xi f) := \int_{\mcl M}
f(F_m(\xi))\,\mu_\mome(\rd m).\enqu

The relation\rref 5.1.18(3)~ is valid due to (i). This means that the functions $f_1(\mlam) :=
\mlam, f_2(\mlam) := \mlam^2,\ (\mlam\in\mbR),$ are both in $L^1(\mbR,\mu_\xi^\mome)$ for all
$\xi\in\mfk g$ and for $f := f_j\ (j= 1,2)$\rref 5.1.19(12)~ is valid. Hence $F_\xi\in L^2(\mcl
M,\mu_\mome)$ for all $\xi\in\mfk g$ and, due to\rref 5.1.18(3)~, we have

\bequ\label{5.1.19(13)} (F_\xi,F_\xi)= (F_\xi,\mbs 1)(\mbs 1,F_\xi),\ \text{for all}\ \xi\in\mfk
g.\enqu

\noidt The brackets denote here here the scalar product in $L^2(\mcl M,\mu_\mome)$ and $\mbs 1\in
L^2(\mcl M,\mu_\mome)$ is the function identically equal to one: $\mbs 1(m):= 1$\ for\ all\ $m\in
\mcl M.$ Applying the Schwarz inequality to\rref 5.1.19(13)~, we obtain:

\bequ\label{5.1.19(14)} F_\xi= const.\ =\ (\mbs 1,F_\xi)\mbs 1=F_\mome(\xi)\mbs 1,\
\mu_\mome\=a.e.\ \text{for all}\  \xi\in\mfk g.\enqu

\noidt This means that {\bf the function}

\bequ\label{5.1.19(15)} F_{\mfk g}:\ \mcl M_{\mfk g}\rarw \mfk g^*,\ m\mapsto F_m,\enqu

\noidt {\bf is constant $\mu_\mome$-almost everywhere}, too.  The restriction of \emn$F_\mfkg$~ to
the set of normal states $\mcl N_*$ separates points in $\mcl N_*$ according to \ref{5.1.13} and
\rref 5.1.19(9)~. Hence the set $F_{\mfk g}^{-1}(F_\mome)\subset \mcl M_{\mfk g}$ contains at most
one $m\in\mcl N_*$. Due to continuity of $F_{\mfk g},$ the set $F_{\mfk g}^{-1}(F_\mome)$ is
closed in $\mcl M_{\mfk g} = \mcl M_{\mfk g}^\circ$ (:= the interior of $\mcl M_{\mfk g}$), what
implies measurability of $F^{-1}_{\mfk g}(F_\mome)$. Due to\rref 5.1.19(14)~:

\bequ\label{5.1.19(16)} \mu_\mome(F^{-1}_{\mfk g}(F_\mome)) = \mu_\mome(\mcl M)= 1.\enqu

 It is known, see e.g. \cite{sak1}, that for any
$\mome\in\mcl S_*(\mfk M_G)$ there is a unique projector $s_\mome\in\mfk M_G$ such that $\mome(x)
= \mome(xs_\mome)$\ for\ all\ $x\in \mfk M_G$ and $\mome(x^*x) = 0$ implies $xs_\mome = 0.$ The
characteristic function in $C(\mcl M)$ corresponding to $s_\mome$ is supported by the clopen set
${\rm supp}\, \mu_\mome\subset \mcl M.$ Since it is nonempty, it contains some $m\in \mcl
M_*\setminus\{m_\circ\}=\mcl N_*,$ and all these $m$'s are contained in $F_{\mfk g}^{-1}(F_\mome)$
due to\rref 5.1.19(7)~. Hence the clopen set ${\rm supp}\,\mu_\mome$ contains exactly one point of
$\mcl M_*$ which means, according to Proposition \ref{5.1.16}, that ${\rm supp}\,\mu_\mome$ is a
one point subset of $\mcl M_*$ and $s_\mome = E_{\mfk g}(F_\mome).$ This proves the implication
(i) $\imply$ (ii).
\end{proof}

\begin{corl}\label{5.1.20} $\msg^*_G\mcl E_{\mfk g} = \mcl E_{\mfk
g}$, i.e. $\mcl E_{\mfk g}$ is $\msg^*_G$-invariant\ (:= '$G$-invariant').
\end{corl}
\begin{proof} According to\rref 5.1.15(4)~ and Lemma \ref{5.1.13},
the set $\mcl N_*$ is $G$-invariant. The action of $G$ (via $\msg^*_G$) commutes with the mapping
$\mome\ (\in\mcl S(\mfk A^\Pi))\mapsto\mome\ (\in\mcl S_*(\mfk M_G)).$ Then the result is
immediate after an application of \ref{5.1.19}.
\end{proof}

\begin{prop}\label{5.1.21} For any $\mome\in\mcl S_*(\mfk M_G),$ the corresponding
probability Radon measure $\mu_\mome$ on $\mcl M$ is supported by $\mcl M_*$:
\bequ\label{5.1.21(1)} \mu_\mome(\mcl M_*)=\mu_\mome(\mcl M)=1.\enqu
\end{prop}
\begin{proof} We can assume that $s_Gs_\mome = s_\mome$ for the support
projector $s_\mome$ of \ome. We have, according to \ref{5.1.16} (i), $s_\mome = \mbs c(s_\mome\mfk
g^*).$ Due to normality of \ome, it is

\bequ\label{5.1.21(2)} 1=\mome(s_\mome)= \lub\{\mome(p_J):\ p_J:=\sum_{F\in J} E_{\mfk g}(F),\
\text{finite}\ J\subset s_\mome\mfk g^*\}.\enqu

Let $m_F\in \mcl M_*\ (F\in\mfk g^*,\ E_{\mfk g}(F)\neq 0)$\ be defined by $m_F(E_{\mfk g}(F))=1.$
For any subset $\mbs K\subset\mfk g^*,$ the open set (which is clopen for finite $\mbs K$)

\bequ\label{5.1.21(3)} \mcl M(\mbs K):=\{m_F\in \mcl M_*:\ F\in\mbs K\},\ m_F\ \text{is void if}\
E_{\mfk g}(F)=0,\enqu

\noidt is $\mu_\mome-$measurable. But $\mome(p_J)=\mu_\mome(\mcl M(J)),$ and $\mu_\mome$\ is
regular. Hence,
\bequ\label{5.1.21(4)} 1=\lub\{\mu_\mome(\mcl M(J)):\ J\subset s_\mome\mfk g^*\
\text{finite}\}\leq\mu_\mome(\mcl M(s_\mome\mfk g^*))\leq\mu_\mome(\mcl M_*)\leq 1,\enqu

\noidt what proves\rref 5.1.21(1)~.
\end{proof}

\begin{lem}\label{5.1.22} Any uniformly bounded function on $\mM_*$
with values in \bC\ can be uniquely extended to a continuous function on $\mcl M$, i.e. the
spectrum space $\mcl M$ of $\mfk M_G$ is the \emn Stone-\v Cech compactification~ of the discrete
space $\mM_*$ of \emn normal pure states on $\mfk M_G$~.
\end{lem}

\begin{proof} Since \Ms\ is discrete, $C(\mMs)$ consists of all
bounded complex valued functions on \Ms. The Stone-\v Cech compactification of a normal
topological space \cS\  is a compact Hausdorff space $\mS'$ and a
homeomorphism $\tau$ of \cS\ into $\mS'$ such, that $\tau(\mS)$ is dense in $\mS'$ and any $f\in
C(\mS)$ can be continued to some $\tilde{f}\in C(\mS')$. It is clear, that the continuation
$\tilde{f}$ is uniquely determined by $f$.

Let $f\in C(\mMs),\ f\geq 0.$ For any $\iota\in [0,\|f\|]$\ (:= closed interval in \bR) define
(cf.\rref 5.1.16(3)~) $p_\iota:= 0$ for $\iota\not\in sp(f)$\ and (let $f(m_\circ)=0$):

\bequ\label{5.1.22(1)} p_\iota:= \mbs c(\{F_m\in\mfk g^*:\ f(m)=\iota,\ m\in\mMs\}),\quad \iota\in
sp(f),\enqu

\noidt where $sp(f)$ denotes the spectrum of $f$. For any finite subset $J\subset sp(f)$\ define

\bequ\label{5.1.22(2)} x_J :=\sum_{\iota\in J} \iota\, p_\iota\ \in\ \mfk M_G.\enqu

\noidt The finite subsets $J$ of $sp(f)$\ are directed by inclusion and the net $\{x_J:\
\text{finite}\ J\subset sp(f)\}$ is increasing. Any increasing net of selfadjoint elements of a
\Wa\ \fk M\ converges to its least upper bound in \fk M, \cite[1.7.4]{sak1}. Let $x_f\in\mfk M_G$
be the limit of $\{x_J\}$. We claim that the function $\tilde{f}\in C(\mcl M),\ \tilde{f}(m) :=
m(x_f)$ coincides with $f$ on \Ms.

{\bf Let} \emn$m\in\mMs$~. Then, due to normality of $m$, \bequ\label{5.1.22(3)}
\tilde{f}(m)={\lub}\{m(x_J):\ \text{finite}\ J\subset sp(f)\} = \lub\,\{\sum_{\iota\in J} \iota\,
m(p_\iota):\ \text{finite}\ J\subset sp(f)\}.\enqu

But $m\in\mMs$ lies in support of the characteristic function $m\mapsto m(p_\iota)$ iff
$f(m)=\iota,$ compare \ref{5.1.16}. Hence $\tilde{f}(m)=f(m)$, what we intended to prove.
\end{proof}

\begin{lem}\label{5.1.23} For a finitely additive probability
measure $\mu$ on $\mfk g^*$ (without any specification of a $\Sigma$-algebra of measurable subsets
in $\mfk g^*$) supported by $s_G\mfk g^*$,\rref 5.1.16(1)~, the following assertions are
equivalent:

(i) $F^*_{\mfk g}\mu=\mu_\mome$ on $\mcl N_*$\ for some $\mome\in\mcl S_*(\mfk N_G),\ i.e. \
\mu=\mu_\mome\circ F_{\mfk g}^{-1}$.

(ii) $\mu$ is supported by a countable subset of $\mfk g^*.$

If these conditions are fulfilled, $\mu$ is \sg-additive. Any $\mu_\mome$ ($\mome\in\mcl S_*(\mfk
N_G)$)
 is of the form  $F^*_{\mfk g}\mu$ for
some \sg-additive probability Borel measure $\mu$ on $\mfk g^*$ with at most countable supporting
set in $s_G\mfk g^*.$
\end{lem}
\begin{proof} \Ns\ is mapped bijectively by $F_{\mfk g}$ onto
$s_G\mfk g^*$\ and $\mu_\mome$ is supported by \Ns\ for all $\mome\in\mSs(\mfk N_G)$. Hence (i) is
fulfilled for $\mu:=\mu_\mome\circ F_{\mfk g}^{-1}$. Complete additivity of $\mu_\mome$ (what is a
consequence of normality of \ome) leads then to the expression

\bequ\label{5.1.23(1)} \mu_\mome = \sum_{m\in\mNs}\mome(E_{\mfk g}(F_m))\,\delta_m,\ (\delta_m:=\
\text{Dirac\ measure\ at}\ m).\enqu

\noidt Hence at most countable number of coefficients $\mome(E_{\mfk g}(F_m))\neq 0.$ This proves
(i) $\imply$ (ii)\ as well as the last assertion of the Lemma. Let

\bequ\label{5.1.23(2)} \mu= \sum_{j\in\mbZ_+}\mlam_j\delta_{F_j},\ \text{with}\ \mlam_j\geq
0,\sum_j\mlam_j=1,\ F_j\in s_G\mfk g^*.\enqu

Then $F^*_{\mfk g}\mu:=\mu\circ F_{\mfk g}$\ is a Baire measure on \Ns, hence represents a
(normal) state \ome\ on $\mfk N_G$. The \sg-additivity is clear.
\end{proof}
\begin{lem}\label{5.1.24} Let, for $\mome\in\mcl  S(\mfk N_G),\
\mu_{\mfk g}^\mome$
 be the additive function of
Borel subsets of $\mfk g^*$ defined in\rref 5.1.16(6)~. Then $\mu^\mome_{\mfk g}$ has a unique
extension to a finitely additive probability measure on the set of all subsets of $s_G\mfk g^*$.
Conversely, any finitely additive probability measure on $s_G\mfk g^*$is of the form
$\mu^\mome_{\mfk g}$ for some $\mome\in\mcl S(\mfk N_G).$
\end{lem}
\begin{proof} For any subset $\mbs K\subset\mfk g^*$ define,\rref
5.1.16(3)~,

\begin{equation}\label{5.1.24(1)}
E_{\mfk g}(\mbs K) := \mbs c(\mbs K) := \sum_{F\in \mbs K}E_{\mfk g}(F).
\end{equation}

Then $E_{\mfk g}(\mbs K)$ is a projector in $\mfk N_G$ and we can define

\bequ\label{5.1.24(2)} \mu^\mome_{\mfk g}(\mbs K):=\mome(E_{\mfk g}(\mbs K))\ \text{for\ any}\
\mbs K\subset\mfk g^*\ \text{and\ any}\ \mome\in\mcl S(\mfk N_G).\enqu

It is easily to see that $\mu^\mome_{\mfk g}$ in\rref 5.1.24(2)~ is the desired unique extension.
For the proof of the second assertion, choose any finitely additive probability measure $\mu$ on
$s_G\mfk g^*,$ $\mu$ defined on all subsets $\mbs K$ of $s_G\mfk g^*$. Define a positive linear
functional on $\mfk N_G$,\ $\mome_\mu$, by its values on all projectors:

\bequ\label{5.1.24(3)} \mome_\mu(E_{\mfk g}(\mbs K)) := \mu(\mbs K),\enqu

\noidt compare \ref{5.1.16}. The von Neumann algebra $\mfk N_G$ is generated by the set of all its
projectors and\rref 5.1.24(3)~ defines uniquely a state on $\mfk N_G$.
\end{proof}

\pt\label{5.1.25}\rm Let us look what measures $\mu^\mome_\fg$ correspond to pure states $\mome\in
\mcl M$, which are not normal. From the character property of pure states we have
$\mome(E_\fg(\mbs K_1\cap\mbs K_2))=\mome(E_\fg(\mbs K_1)E_\fg(\mbs K_2))=\mome(E_\fg(\mbs
K_1))\mome(E_\fg(\mbs K_2))$ what together with finite additivity gives:

\bequ\label{5.1.25(1)} \mbs K\subset\fgs\imply \mu_\fg^\mome(\mbs K)\in\{0,1\}.\enqu

\noidt Remember that $\supp\ \mu^\mome_\fg\subset s_G\fgs.$ Any finitely additive measure $\mu$ on
$s_G\fgs$ satisfying\rref 5.1.25(1)~ corresponds to a pure state $\mome_\mu\in \mcl M$. It
determines also an \emn ultrafilter~ on $s_G\fgs$ consisting of all subsets $\mbs K$ for which it
is $\mu(\mbs K)= 1.$ This is clearly a bijection between the set of all ultrafilters on $s_G\fgs$
and the {\bf set of pure states}
\[\centerline{\emn$\mcl{ES}(\mfk N_G)$~=:\ \emn$\mcl N$~.}\]

 Remember that to any $m\in \mcl M$
corresponds the Dirac measure $\delta_m$ on $\mcl M$ which is concentrated at a point $m$. For a
nonnormal $m$ the measure $\mu^m_\fg$ is not concentrated at any point in $\fgs$.

\pt\label{5.1.26}\rm Let us keep in mind that we have associated with any state $\mome\in\mS(\mfk
A^\Pi)$  a state (equally denoted) $\mome\in\mS(\mfk M_G)$ which is the restriction to $\mfk M_G$
of the unique $w^*$-continuous extension to $(\mfk A^\Pi)^{**}$ of $\mome\in\mS(\mfk A_\Pi)$. Such
an $\mome\in\mS(\mfk M_G)$ is necessarily normal: $\mome\in\mSs(\mfk M_G)$, and the corresponding
measure $\mu^\mome_\fg:=\mu_\mome\circ F^{-1}_\fg$ is purely atomic, \ref{5.1.23}. This reflects
that fact that the described procedure maps into $\mS(\mfk N_G)$ only such states on $(\mfk
A^\Pi)^{**}$ which are describable by density matrices in ${\cal L}(P_G\mH_\Pi)$.

 Any state on $\mfk M_G$ can be, on the other
hand, extended to some states on $(\mfk A^\Pi)^{**}$ (not normal - in general) and these determine
their restrictions to $\mfk A^\Pi$ considered as a subalgebra of its bidual. In this way, we can
obtain also those states on $P_G\mfk A^\Pi$ which are not expressible by density matrices. Hence
to general finitely additive probability measures on \Ns\ 'correspond', in some many-to-many way,
arbitrary states on $P_G\mfk A^\Pi$. We intend now to change our ascription of states on $\mfk
M_G$ to arbitrary states on $\mfk A^\Pi$ in such a way, that any state on $P_G\mfk A^\Pi$ will be
mapped into $\mS(\mfk N_G)$\ (and not onto $m_\circ\in \mM_*$ as before).

\pt\label{5.1.27}\rm {\bf Quasilocal structure of $\mfk A^\Pi$:} The algebra $\mfk A^\Pi$ has a
natural \emn quasilocal structure~ in the sense of \ref{1.4.2}. It is generated by \emn local
algebras~ $\mfk A_v := \mfk A^N\ (N\in\Pi),$\rref 1.4.2(1)~, where, in the notations of
\ref{5.1.3}, $\mfk A^N$ is generated by $\pi_j(y)\ (y\in\mLH,\ j= 1,2,\dots N)$ and is isomorphic
to $\mcl L(\mH_N)$. $\mcl H_N$ is here the $N-$fold tensor product of the Hilbert space \H,\rref
5.1.3(2)~. Denote by $\mfk A_L$ the set of all finite linear combinations of finite products of
arbitrary elements $y\in\mfk A^N$ for any finite $N$. The algebra $\mfk A_L:=\cup_{\text{finite}\
N}\mfk A^N$ is called the 'local algebra' and its elements are '\emn local observables~'. The norm
closure of $\mfk A_L$ is $\mfk A^\Pi$ = the \emn algebra of quasilocal observables~ of our system.

A locally normal state $\mome\in\mS(\mfk A^\Pi)$, i.e. a state the restriction of which to any
local subalgebra $\mfk A^N$ is normal (cf. \ref{1.4.3}), can be calculated (with a use of natural
isomorphisms) on all the elements $x\in \mfk A^N\subset \mfk A^\Pi$ with the help of density
matrices $\rho_\mome^N$ on $\mcl H_N\ (N=1, 2, \dots)$ via the usual formula 
\bequ\label{5.1.27(1)}
\mome(x)= Tr(\rho_\mome^Nx),\quad x\in\mcl L(\mH_N),\enqu

\noidt where we have identified $\mfk A^N$\ with $\mcl L(\mH_N)$. Let $\mS_L(\mfk A^\Pi) =: \mS_L$
denotes the set of all locally normal states on $\mfk A^\Pi$. The states expressible (globally) by
a density matrix in the defining representation of $\mfk A^\Pi$ in $\mH_\Pi$ are locally normal.
$\mfk A^\Pi$ is simple, \cite[2.6.20]{bra&rob}.

\begin{exm}\label{5.1.28}\rm We shall illustrate here the fact that a strongly
continuous one parameter group of unitaries $\exp(itP)$ acting on a Hilbert space \H\ need not be
continuous in certain other representations of \LH.

Let \fkA := \LH\ be the considered \Wa, \H := $L^2(\mbR)$, and $Q$ (resp. $P$) be the selfadjoint
operator on \H\ defined on $\mphi\in C^1_0(\mbR)$ by $(Q\mphi)(\mlam) := \mlam\mphi(\mlam)$ (resp.
$(P\mphi)(\mlam):= -i\,\frac{\rd}{\rd\mlam}\mphi(\mlam)), \mlam\in\mbR.$ Let $\mfk M$ be the
maximal commutative \Wa\ in \LH\ generated by $\exp(itQ),\ t\in\mbR$. Let $\chi_\mlam$ be the pure
state on $\mfk M$ determined by
\bequ\label{5.1.28(1)} \chi_\mlam(\exp(itQ)) := \exp(it\mlam),\
t\in\mbR.\enqu

\noidt Let $\mome_\mlam$ be an extension of $\chi_\mlam$ onto the whole \Wa\ \fkA. We claim that
the function

\bequ\label{5.1.28(2)} t\mapsto \mome_\mlam(\exp(itP)),\ t\in\mbR, \enqu

\noidt is discontinuous, hence the group $\pi_\mlam(\exp(itP))$ of unitaries in the cyclic
representation $\pi_\mlam$\ of \fkA\ corresponding to the state $\mome_\mlam\in\mS(\mfkA)$ is not
strongly continuous. Since $\chi_\mlam$ is pure, it is a character on $\mfk M$. Consequently for
any projector $q\in\mfk M$ it is

\bequ\label{5.1.28(3)} \chi_\mlam(q)=[\chi_\mlam(q)]^2,\ i.e.\ \chi_\mlam(q)\in\{0,1\},\
q^*=q^2=q\in\mfk M.\enqu

\noidt We obtain from the \emn Schwarz inequality~, for any $x\in\mfkA$,

\bequ\label{5.1.28(4)} \mome_\mlam(x)=\mome_\mlam(qx)=\mome_\mlam(xq),\ \text{for all}\
\{q=q^*=q^2:q\in\mfk M,\ \chi_\mlam(q)=1\}.\enqu

Any element $z\in\mfk M$ can be expressed as a norm limit of finite linear combinations of
projectors $q\in\mfk M$. Since the product $z\mapsto xz$ is norm-continuous  and the state
$\mome_\mlam$ is also continuous in the norm of \fkA, we obtain from\rref 5.1.28(4)~:

\bequ\label{5.1.28(5)} \mome_\mlam(xz-zx)\equiv \mome_\mlam([x,z])=0,\quad\forall x\in\mfkA,\
\text{and}\ \forall z\in\mfk M.\enqu

Due to CCR we have \bequ\label{5.1.28(6)} [\exp(itP),\exp(i\tau Q)]=(e^{it\tau}-1)\,\exp(i\tau
Q)\exp(itP),\enqu

\noidt and after the substitution to\rref 5.1.28(5)~: \bequ\label{5.1.28(7)}
0=(e^{it\tau}-1)\,\mome_\mlam(\exp(i\tau Q)\exp(itP)).\enqu

\noidt The relation\rref 5.1.28(7)~ is valid for all real $t$ and $\tau$. From an application
of\rref 5.1.28(5)~ to
$$\mome_\mlam\left(\exp(-i\tau
Q)[\exp(itP),\exp(i\tau Q)]\right)=\mome_\mlam\left(\exp(-i\tau Q)\exp(itP)\exp(i\tau
Q)\right)-\mome_\mlam(\exp(itP)) $$

\noidt we obtain the invariance of $\mome_\mlam$\ \wrt\ the group $\msg^*$ of affine isometries of
$\mS(\mfkA)$,
\bequ\label{5.1.28(8)} \msg^*_\tau\mome(x):=\mome(\exp(-i\tau Q) x \exp(i\tau Q)),\
\tau\in\mbR,\ x\in\mfkA.\enqu

\noidt This leads, together with the formula\rref 4.1.3(2)~, to \bequ\label{5.1.28(9)}
\mome_\mlam(\exp(itP)) = e^{it\tau}\,\mome_\mlam(\exp(itP))\quad\text{for all}\
t,\tau\in\mbR,\enqu

\noidt what implies the discontinuity of\rref 5.1.28(2)~.

The obtained formulas show also uniqueness of the extension $\mome_\mlam$ of $\chi_\mlam$ to the
CCR-subalgebra of \fkA\ defined as the norm closed algebra generated by $\exp(i\tau Q)$ and
$\exp(itP)\ (t,\tau\in\mbR)$.
\end{exm}

\pt\label{5.1.29}\rm We shall now change the definition \ref{5.1.12} of the macroscopic algebra of
the system $(\mfkA^\Pi,\msg_G)$ in such a way that a larger subset of states from $\mS(\mfkA^\Pi)$
will be mapped onto probability measures on \fkgs\ than it was before, according to the ascription
from \ref{5.1.14}. In the notations from \ref{5.1.3}\ and \ref{5.1.4}, let

\bequ\label{5.1.29(1)} X_{\xi N} := \frac{1}{N}\sum_{j=1}^N \pi_j(X_\xi),\ N=1,2,\dots,\
\xi\in\mfkg.\enqu

\noidt Then the elements $\exp(itX_{\xi N})\in\mfkA^N\ (t\in\mbR)$ are represented in the defining
representation of $\mfkA^\Pi$ in $\mH_\Pi$ by strongly continuous groups converging with
$N\rarw\infty$ in the strong operator topology on the $G-$invariant subspace $P_G\mH_\Pi$ of
$\mH_\Pi$ to strongly continuous central subgroups of $P_G\mfk B^\#$,

\bequ\label{5.1.29(2)} s\=\lim_N\exp(itX_{\xi N})P_G = \exp(itX_{\xi\Pi})P_G,\enqu

\noidt see \ref{5.1.7},\ \ref{5.1.8}\ and \ref{5.1.11}. The algebra $\mfk M_G$ of macroscopic
observables was built from spectral projectors $E^\#_{\xi \Pi}$ of $X_{\xi\Pi}$'s mapped into the
center $\mfk Z$ of the bidual $(\mfkA^\Pi)^{**}$. We want to generalize this construction. We
shall identify the bidual $(\mfkA^\Pi)^{**}$ with the weak closure of the \emn universal
representation~ of $\mfkA^\Pi$ (cf. \cite[Def.1.16.5]{sak1},\cite[3.7.6]{pedersen}). {\bf Let
\emm$p_G$~ be the $\lub$ of all such projectors $p\in\mfk Z$,} for which the limits in
$\msg((\mfkA^\Pi)^{**},(\mfkA^\Pi)^*)$-topology:

\bequ\label{5.1.29(3)}\exp(itX_{\xi\Pi})p_G := \msg\=\lim_N\exp(itX_{\xi N})p_G,\quad\forall
\xi\in\mfkg,\enqu

\noidt exist (with $p_G\hookrightarrow p$).
The symbol $X_{\xi\Pi}$ denotes here a selfadjoint operator acting on the  subspace $p_G\mH_u$ of
the space $\mH_u$ -- the space of universal representation of $\mfkA^\Pi$. Here it is assumed, of
course, that the groups $t\mapsto \exp(itX_{\xi N})p_G$ are strongly continuous for all $N\in\Pi.$
It is clear from the definitions of $X_{\xi N}$ and $\msg_G$, \ref{5.1.5}, that

\bequ\label{5.1.29(4)} \msg_G(p_G) = p_G.\enqu

The convergence in\rref 5.1.29(3)~ means the convergence $X_{\xi N}\rarw X_{\xi \Pi}$ of
selfadjoint operators on $p_G\mH_u$ in the strong-resolvent sense, \cite{R&S}. From

\bequ\label{5.1.29(5)} [\exp(itX_{\xi N}),\
y]=\left(\exp\left(\frac{it}{N}\sum_{j=1}^K\pi_j(X_\xi)\right)\, y\,
\exp\left(-\frac{it}{N}\sum_{k=1}^K\pi_k(X_\xi)\right)-y\right)e^{itX_{\xi N}},\enqu

\noidt which is valid for all $y\in\mfk A^K\ (K\in\Pi)$ and $\xi\in\mfkg,\ t\in\mbR,$ as well as
from the assumed continuity of $p_G\exp(itX_{\xi N})$ we conclude that the limit
$p_G\exp(itX_{\xi\Pi})\in(\mfkA^\Pi)^{**}$ belongs to the center $\mfk Z$ of $(\mfkA^\Pi)^{**}$.
Let now the \emm $\Pi$-macroscopic algebra of $G$-definiteness~ of $(\mfkA^\Pi,\msg_G)$ be {\bf
defined as the von Neumann subalgebra \emn$\mfk N_G^\Pi$~ of the center $\mfk Z$} generated by all
the spectral projectors $E_{\xi\Pi}(B)$\ (Borel $B\subset\mbR$ and $\xi\in\mfkg$) of operators
$X_{\xi\Pi}$ in $p_G\mH_u$\ (we hope that no confusion arises from the keeping an old notation for
new objects!). The {\bf algebra \emn$\mfk M_G^\Pi$~ is obtained from $\mfk N_G^\Pi$ by adjoining
to it the identity $I$ of $\mfk Z$}; it will be called the \emm $\Pi G$-macroscopic algebra~ of
$(\mfkA^\Pi,\msg_G)$. The relation between $\mfk M_G^\Pi$ and the previously introduced $\mfk M_G$
, \ref{5.1.12}, is clear without any proof:

\begin{lem}\label{5.1.30} $\mfk N_G=s_G\mfk N_G^\Pi = s_G\mfk
M_G^\Pi = s_G\mfk M_G,$\ where the projector $s_G\in\mfk Z$ was introduced\nl\hspace*{3cm} in
\ref{5.1.11}.
\end{lem}

\pt\label{5.1.31}\rm We shall use concepts and notations connected with the usage of $\mfk
M_G^\Pi$ in analogy to those connected with $\mfk M_G$, as they were introduced above. Let e.g.,
$B\subset \mfkg^*$ be a Borel set (\wrt\ the usual topology of a finite dimensional vector space),
and $\xi_j\ (j=1,2,\dots n)$ form a basis of \fkg. Let

\bequ\label{5.1.31(1)} \xi_jB:=\{\mlam\in\mbR:\ \mlam=F(\xi_j),\ F\in B\} \enqu

\noidt be the projection of $B$ onto the j-th coordinate axis of the dual frame. If $B$ has the
form

\bequ\label{5.1.31(2)} B= \{F \in\mfkg^*: F(\xi_j)\in\xi_j B,\ \forall j\in\{1,2,\dots n:= \dim
G\}\},\enqu

\noidt then we set

\bequ\label{5.1.31(3)}E^\Pi_\mfkg(B):= E_{\xi_1\Pi}(\xi_1B)E_{\xi_2\Pi}(\xi_2B)\dots
E_{\xi_n\Pi}(\xi_nB).\enqu

\noidt The \Wa\ $\mfk M_G^\Pi$\ is {\bf generated by the projectors} \emn$E^\Pi_\mfkg(B)$~
from\rref 5.1.31(3)~ \&\rref 5.1.31(2)~, and by the unit $I\in\mfk Z.$

The {\bf algebra \emn$\mfk M_G^\Pi$~, contrary to $\mfk M_G$, cannot be built from
projectors}\rref 5.1.31(3)~ corresponding to one point sets $B:= \{F\}\ (F\in\mfkgs)$ only. This
can be seen as follows: Choose a probability Borel measure $\mu$ on $s_G\mfkgs$, in the old
notation from \ref{5.1.16}, such that any point (Dirac) measure is singular \wrt\ it:\
$\mu(\{F\})=0$ for all $F\in\mfkgs$. Choose a product vector $\Psi(F)\in E^\#_\mfkg(F)P_G\mH_\Pi$,
one and only one for each such $F\in\mfkgs$, for which $E^\#_\mfkg(F)\neq 0$. Denote \break by
$\mome^F:=\mome^{\Psi(F)}$ the corresponding state on $\mfkA^\Pi$. Assume, that all the functions

\bequ\label{5.1.31(4)} F\mapsto \mome^F(x),\ x\in\mfkA^\Pi,\enqu

\noidt are $\mu-$measurable. This last assumption is trivially fulfilled, if $\mu$ is concentrated
on an $Ad^*(G)$ orbit $G\cdot F\subset\mfkgs$ and $\mome^{g\cdot F}:= \msg^*_g\mome^F$. Define
then the state $\mome_\mu\in\mcl S(\mfkA^\Pi)$ by

\bequ\label{5.1.31(5)} \mome_\mu(x):=\int_{\mfkgs}\mome^F(x)\,\mu(\rd F).\enqu

In this way, we can construct {\bf states \emn$\mome_\mu$~ the central supports \emn$s_\mu\in\mfk
Z$~ of which are contained in} $p_G,\ s_\mu p_G=s_\mu$, but $s_\mu s_G=0$, as well as $s_\mu
E^\Pi_\mfkg(F)=0$ for all $F\in\mfkgs,$ in $\mfk M_G^\Pi$.

The last considerations show to us that $\mfk M_G$ and $\mfk M_G^\Pi$ are different from one
another. The \Wsa\ of $\mfk M_G^\Pi$ generated by all $E^\Pi_\mfkg (F):= E_\mfkg^\Pi(\{F\})\
(F\in\mfkgs)$ is naturally isomorphic to $\mfk M_G$. Hence $\mfk M_G^\Pi$ is larger than $\mfk
M_G$ which can be  injected into $\mfk M_G^\Pi$ via the last mentioned isomorphism.

\pt\label{5.1.32}\rm {\bf Let us now introduce the mapping} \emn$p_M$~:

\bequ\label{5.1.32(1)} p_M:\ \mS(\mfkA^\Pi)\rarw\mS_*(\mfk M_G^\Pi),\ \mome\mapsto p_M\mome,\enqu

\noidt {\bf where \emn$p_M\mome$~ is the restriction} to $\mfk M_G^\Pi$ of the canonical extension
of the state $\mome\in\mS(\mfkA^\Pi)$ to the normal state on $(\mfkA^\Pi)^{**}$. Any state
$\mome\in\mS(\mfkA^\Pi)$\ (resp. $\mome\in\mS(\mfk M_G^\Pi)$) can be uniquely decomposed as

\bequ\label{5.1.32(2)} \mome = \mome(p_G)\, p_G\mome + \mome(I-p_G)\, \mome_\circ,\enqu

\noidt {\bf where the symbols \emn$p_G\mome(x)$~ and \emn$\mome_\circ(x)$~ are given by} \bequ
p_G\mome(x):=\frac{1}{\mome(p_G)}\mome(x\,p_G),\quad
\mome_\circ(x):=\frac{1}{\mome(I-p_G)}\mome(x(I-p_G)).\nonumber\enqu

\noidt Hence for $\mome(I-p_G)\neq 0$ it is \[\centerline{$p_M\mome_\circ=$\emn$m_\circ$~\,:= the
{\bf pure state in $\mS(\mfk M^\Pi_G)$ supported} by the minimal projector $I-p_G\in\mfk
M_G^\Pi$.}\]

 Let
\bequ\label{5.1.32(3)} \mS^\Pi_\mfkg :=\{\mome\in\mS(\mfkA^\Pi):\ \mome(p_G)=1\}.\enqu

\noidt {\bf In other words:} \emn$\mS_\mfkg^\Pi = p_G\mS(\mfkA^\Pi)$~. For $\mome\in\mS_\mfkg^\Pi$
one has $p_M\mome\in\mSs(\mfk N_G^\Pi)$. Conversely, each state  in $\mSs(\mfk N_G^\Pi)$ is of the
form $p_M\mome$ for some states $\mome\in \mS_\mfkg^\Pi.$

\begin{lem}\label{5.1.33} The projector-valued additive function
of intervals in \fkgs\ introduced in\rref 5.1.31(3)~ can be extended to a unique {\bf
projector-valued measure} \emn$E^\Pi_\mfkg$~$:\ B\mapsto E^\Pi_\mfkg(B)$\ defined on all Borel
sets $B$ in \fkgs.
\end{lem}
\begin{proof} The mapping

\bequ\label{5.1.33(1)} w:\ \mfkg\rarw\mcl{L}(p_G\mH_u),\ \xi\mapsto
w(\xi):=\exp(iX_{\xi\Pi})p_G,\enqu

\noidt see\rref 5.1.29(3)~, is strongly continuous unitary representation of the abelian group
\fkg\ (group multiplication is here the vector addition) in the subspace $p_G\mH_u$ of the Hilbert
space $\mH_u$ of the universal representation of $\mfkA^\Pi$. This can be seen with a help of
linearity of the mapping

\bequ\label{5.1.33(2)} \xi\mapsto X_{\xi\Pi},\quad \xi\in\mfkg,\enqu

According to the \emn SNAG-theorem~ (\cite[Chap.
X]{riesz&nagy},\cite[Thm.VIII.12]{R&S},\cite[Chap. IV]{GRS}), there is unique projection measure
$E_\mfkg^\Pi$\ on the dual group $\hat{\mfkg}$ of \fkg\ representing this unitary representation
in the standard fashion. The linear space $\mfkgs$ can be identified with the group $\hat{\mfkg}$
of characters by the bijection associating with any $F\in\mfkgs$\ the character $\xi\mapsto
\exp(iF(\xi))$ on \fkg. It is clear that the restriction of $E_\mfkg^\Pi$ on intervals in \fkgs\
coincides with\rref 5.1.31(3)~.
\end{proof}
\begin{lem}\label{5.1.34} All the nonzero projectors of the form
$E_\mfkg^\Pi(F):= E_\mfkg(\{F\}),\ F\in\mfkgs$, are minimal projectors in $\mfk N_G^\Pi$ and all
minimal projectors in $\mfk N_G^\Pi$ are of this form.
\end{lem}
\begin{proof} Let $q\in\mfk N_G^\Pi$ be a minimal projector.
Since $q\in\mfk Z$, there is a state $\mome\in\mS(\mfkA^\Pi)$, the central projector of which is
$s_\mome\leq q$. Choose such an \ome. Then $\mome(x)=\mome(xs_\mome)=\mome(xq)$ for all
$x\in\mfkA_\Pi$, and due to continuity properties of products in $(\mfkA^\Pi)^{**}$ as well as of
the normal extension $\mome\in\mS_*((\mfkA^\Pi)^{**}),$ the same is true for all
$x\in(\mfkA^\Pi)^{**}.$ The minimality of $q$ in $\mfk N_G^\Pi$ implies that one of the following
possibilities (i) or (ii) is valid

\bequ\label{5.1.34(1)} {\rm (i)\ } qE_\mfkg^\Pi(B)=q,\quad\quad {\rm (ii)\ } qE_\mfkg^\Pi(B)=0
\enqu

\noidt for any Borel $B\subset\mfkgs.$ Let us define a probability Borel measure $\mu^\mome_\mfkg$
on \fkgs\ corresponding to the $\mome\in\mS_\mfkg^\Pi$:

\bequ\label{5.1.34(2)} \mu^\mome_\mfkg(B):= p_M\mome(E_\mfkg^\Pi(B)),\ \text{for all Borel}\
B\subset\mfkgs.\enqu

We see from\rref 5.1.34(1)~ that for the chosen \ome\ the values of $\mu^\mome_\mfkg$ lie in the
two point set $\{0,1\}\subset\mbZ_+$. Each of the projection measures $E_{\xi\Pi}\ (\xi\in\mfkg)$
and $E^\Pi_\mfkg$ are \sg-additive, hence $\mfk N_G^\Pi$ is generated by those $E_\mfkg^\Pi(B)$
which correspond to {\em bounded} Borel subsets $B$ of \fkgs. Hence $\mu^\mome_\mfkg$ is
concentrated on a compact subset of \fkgs:\ $\mu_\mfkg^\mome(B_\circ)=1$ for some compact
$B_\circ$. The \sg-additivity of $\mu^\mome_\mfkg$ implies then that $\mu^\mome_\mfkg$ is
concentrated on a one-point set $F_\mome\in B_\circ$:

\bequ\label{5.1.34(3)} \mu^\mome_\mfkg(\{F_\mome\})= p_M\,\mome(E_\mfkg^\Pi(F_\mome))=1.\enqu

\noidt This implies $s_\mome\leq E_\mfkg^\Pi(F_\mome),$ and, due to\rref 5.1.34(1)~ and due to our
choice of $s_\mome$:

\bequ\label{5.1.34(4)} q\leq E_\mfkg^\Pi(F_\mome).\enqu

According to the definition of $\mfk N_G^\Pi$ in \ref{5.1.29}, $q$ can be approximated, in
$\msg(\mfk N_G^\Pi,\mfk N_{G*}^\Pi)$ topology, by a net $j\mapsto E_\mfkg^\Pi(B_j)$, where
$F_\mome\in B_j$ for all $j$, due to\rref 5.1.34(4)~. Coming to the Gel'fand representation
$C(\mcl N^\Pi)$ of $\mfk N_G^\Pi$ and considering that {\bf clopen sets in the spectrum space}
\emn$\mcl N^\Pi$~ {\bf form a basis of topology}, e.g. \ref{5.1.14} and \cite{sak1}, we see that
the sets in $\mcl N^\Pi$ corresponding to the projectors $E^\Pi_\mfkg(B_j)$ have in their
intersection exactly one point $m_q \in \mcl N^\Pi$ corresponding to the minimal projector $q$.
All of $E_\mfkg^\Pi(B_j)$ contain, however, also $E^\Pi_\mfkg(F_\mome).$ This proves that
$q=E_\mfkg^\Pi(F_\mome)$.
\end{proof}

\begin{lem}\label{5.1.35} If $\mome\in\mS(\mfkA^\Pi)$ is pure or
factor state, then also $p_M\mome\in\mS_*(\mfk M_G^\Pi)$ is pure.
\end{lem}
\begin{proof}
Verbally the same proof as that of \ref{5.1.17}, with $\mfk M_G\hookrightarrow\mfk M_G^\Pi.$
\end{proof}

\begin{defs}\label{5.1.36} The \emm generalized G-macroscopic phase space~ is
the {\bf topological space} \emn$\mcl M^G:=\mfkgs\cup\{m_\circ\}$~ consisting of the finite
(n-)dimensional topological vector space \fkgs\ with the canonical symplectic forms defined on
each orbit of the $Ad^*(G)$-action and of an isolated point $m_\circ$. A \emm state on $\mcl M^G$~
is any probability \sg-additive Borel measure $\mu$ on $\mcl M^G$.  We shall associate with any
$\mome\in\mS(\mfk A^\Pi)$ the \emm G-macroscopic state~ on $\mcl M^G$ determined by the measure
 \bequ\label{5.1.36(1)}
\mu_\mfkg^\mome(B):=\mome(E_\mfkg^\Pi(B\setminus\{m_\circ\})) + \mome(I-p_G)\delta_{m_\circ}(B),\
\text{for\ Borel}\ B\subset \mcl M^G,\enqu

\noidt where on the \rhs\ \ome\  means the normal extension of the state on $\mfk A^\Pi$ to a
normal state on $(\mfk A^\Pi)^{**}$ and $\delta_m\ (m \in \mcl M^G)$ means the Dirac measure
concentrated on $\{m\}$. It is clear that every normal state $\mome\in\mS(\mfk M_G^\Pi)$ can be
transformed also into a state on $\mcl M^G$ by the formula\rref 5.1.36(1)~ and that its image
$\mu^\mome_\mfkg$ uniquely determines $\mome\in\mS_*(\mfk M_G^\Pi).$ It is also clear that the
state on $\mcl M^G$ corresponding to $p_M\mome$, \ref{5.1.32}, in this way, coincides with
$\mu_\mfkg^\mome$. The association $\mome\mapsto \mu^\mome_\mfkg$ is \emm G-equivariant~, i.e.

\bequ\label{5.1.36(2)} \mu_\mfkg^{g\cdot\mome}=\msg_g^*\mu^\mome_\mfkg,\ with\
g\cdot\mome:=\msg_g^*\mome,\ g\in G,\enqu

\noidt and with

\bequ\label{5.1.36(3)} \msg_g^*\mu(B):=\mu(Ad^*(g^{-1})(B\setminus\{m_\circ\}))
+\mu(\{m_\circ\}\cap B),\enqu

\noidt for all $g\in G$ and all Borel $B\subset \mcl M^G$. We shall use also $g\cdot \mu :=
\msg^*_g\mu.$  This follows from the transformation properties of $X_{\xi\Pi}$'s and from

\bequ\label{5.1.36(4)} \msg_g E^\Pi_\mfkg (B)= E_\mfkg^\Pi(Ad^*(g)B),\enqu

\noidt compare \ref{5.1.15}.

Let us redefine some symbols introduced in \ref{5.1.18}. Let,\rref 5.1.32(3)~,

\bequ\label{5.1.36(5)} \mS^d_\mfkg :=\{\mome\in\mS_\mfkg^\Pi:\ \xi\in L^1(\mcl
M^G,\mu^\mome_\mfkg),\ \forall \xi\in\mfkg\},\enqu

\noidt where $\xi$ is considered as the linear function $F\mapsto \xi(F):= F(\xi)$ on $\mfkgs\
(\ni F)$. Similarly, we shall define now \emm G-macroscopically pure states~ to be elements
$\mome\in\mcl E_\mfkg\subset\mS(\mfk A^\Pi),$ where

\bequ\label{5.1.36(6)} \mcl E_\mfkg := \{\mome\in\mcl S_\mfkg^d:\
\mu^\mome_\mfkg(\xi^2)=[\mu^\mome_\mfkg(\xi)]^2,\ \forall \xi\in\mfkg\}.\enqu

Using\rref 5.1.19(13)~ and\rref 5.1.19(14)~, we can see that the Proposition \ref{5.1.19} can be
replaced by: $\mome\in\mcl E_\mfkg\Leftrightarrow\mu_\mfkg^\mome =\delta_F$ for some $F\in\mfkgs.$
\end{defs}

\begin{defs}\label{5.1.37} A \emm Poisson manifold~ $\mcl M$ is a differentiable
$C^\infty$-manifold endowed with a bilinear mapping $(f;g)\mapsto \{f,g\}$ of couples of
infinitely differentiable real functions $f,g \in C^\infty(\mcl M,\mbR)$ into $C^\infty(\mcl
M,\mbR)$, the \emm Poisson bracket~, satisfying properties \ref{1.3.5}\ (i)+(ii)+(iii)+(iv), i.e.
the nondegeneracy \ref{1.3.5}(v) is {\em not} required. Due to \ref{1.3.5}(iv), the Poisson
bracket $\{f,g\}$ depends on $\rd f\ and\ \rd g$ only, and can be uniquely expressed as the value
of a two-contravariant \emm tensor field $\mbs\mlam$~ on these one forms:

\bequ\label{5.1.37(1)} \mbs\mlam(\rd f,\rd g):=\{f,g\}. \enqu

\noidt To any $f\in C^\infty(\mcl M,\mbR)$ corresponds then a unique vector field $\msg_f$ on
$\mcl M$ satisfying:

\bequ\label{5.1.37(2)} \rd g(\msg_f) := \mbs\mlam(\rd f,\rd g),\ \text{for all}\  g\in
C^\infty(\mcl M,\mbR). \enqu

\noidt $\msg_f$ is the \emm Hamiltonian vector field~ on $\mcl M$ with the \emn Hamiltonian
function~ $f$.

With $\mcl M:= \mfkgs$, the cotangent space $T^*_F\mfkgs$ can be naturally identified, for any
$F\in\mfkgs$, with the Lie algebra \fkg\ of $G$. Then, with this identification, $\rd_Ff\in\mfkg$
for any $f$ and $F$. Then the Poisson bracket

\bequ\label{5.1.37(3)} \{f,g\}(F) :=\ -F([\rd_Ff,\rd_Fg]), \enqu

\noidt where on the \rhs is the value of $F\in\mfkg$ on the Lie algebra commutator in \fkg,
defines a natural \emn Poisson structure~ on \fkgs. In this way, also $\mcl M^G$ is, naturally, a
Poisson manifold. Hamiltonian vector fields $\msg_f$ are tangent to orbits of $Ad^*(G)$-action of
G on \fkgs\ at any point $F\in\mfkgs$, compare \cite{marle}. The restriction of the Poisson
structure\rref 5.1.37(3)~ to any $Ad^*(G)$-orbit is the canonical symplectic structure on it.
\end{defs}

\begin{thm}\label{5.1.38} Let the system $(\mfkA^\Pi,\msg_G)$ be defined
by\rref 5.1.4(5)~ and\rref 5.1.5(2)~. Let $\mfk M^\Pi_G$  be the commutative $\msg_G$-invariant
\Wsa\ of \emn$\mfk Z$~ ({\bf := the center of} $(\mfkA^\Pi)^{**}$) defined in \ref{5.1.29}. Let
$p_M:\mS(\mfkA^\Pi)\rarw \mSs(\mfk M_G^\Pi)$ be the mapping\rref 5.1.32(1)~. We shall write also
\[ \centerline{ \emn$p_M\mome :=\mu^\mome_\mfkg$~,\rref 5.1.36(1)~,}\]

\noidt due to the existence of canonical embedding of $\mSs(\mfk M_G^\Pi)$ into the space of
probability Radon \emn measures on $\mcl M^G$~. Then:
\begin{enumerate}
\item[(i)]\ $p_M$ is affine,
$\msg((\mfkA^\Pi)^*,(\mfkA^\Pi)^{**})-\msg((\mfk M_G^\Pi)^*,\mfk M_G^\Pi)$-continuous surjection
onto $\mSs(\mfk M_G^\Pi)$ := the set of all normal states on $\mfk M_G^\Pi$;

\item[(ii)]\ $p_M$ is $G$-equivariant,\rref 5.1.36(2)~;

\item[(iii)]\ Let $\mS_F:=
\{\mome\in\mS(\mfkA^\Pi):\mu_\mfkg^\mome=\delta_F\}$,\ (here $F\in\mfkgs,\ \delta_F$ is the Dirac
measure concentrated at $F$). Then $\mS_F\subset\mcl E_\mfkg$,\rref 5.1.36(6)~, and $\mS_F$ is a
weakly closed convex \emm face~\footnote{A {\em face} S of a compact convex set $K$ is defined to
be a subset of $K$ with the property that if $\mome=\sum_{i=1}^n\mlam_i\mome_i$ is a convex
combination of elements $\mome_i\in K$ such that $\mome\in S$ then $\mome_i\in S,\ \forall\ i=
1,2,\dots n$.}\ in $\mS(\mfk A^\Pi)$;

\item[(iv)]\ $\mome\in\mcl E_\mfkg$ implies $\mu^\mome_\mfkg =\delta_F$
for $F=F_\mome\in\mfkgs$, and for any factor-state $\mome\in\mS(\mfkA^\Pi)$ it is
$\mu^\mome_\mfkg=\delta_m$ for some $m\in \mcl M^G$;

\item[(v)]\ Let $\mu_\mome$ be the canonical measure on the spectrum space
$\mcl M^\Pi$ of $\mfk M_G^\Pi=C(\mcl M^\Pi)$ corresponding to the state $p_M\mome\in\mSs(\mfk
M^\Pi_G),\ \mome\in\mS(\mfkA^\Pi).$  Then there is a canonically defined $\mu_\mome-$measurable
function $\hat{\mome}(m)=: \mome_m$ (spaces are taken with their $w^*$-topologies) such that the
{\bf restriction} \emn$r_M\mome_m$~ $=m\in \mcl{ES}(\mfk M_G^\Pi)$:\
\[\centerline{\emn$r_M$~$:\
(\mfkA^\Pi)^{***}\rarw (\mfk M_G^\Pi)^* \; \; \text{is the natural restriction}$}\]

\noidt and \bequ\label{5.1.38(1)} \mome(x)=\int_{\mcl M^\Pi}\mome_m(x)\mu_\mome(\rd m)\ \text{for\
any}\ x\in(\mfkA^\Pi)^{**}\ [\supset\mfkA^\Pi].\enqu
\end{enumerate}
\end{thm}
\begin{proof}
(i) is clear from the definition of $p_M$, compare also \cite[4.1.36]{bra&rob}. (ii) is a
rephrasing of\rref 5.1.36(2)~. Since $\delta_F$ corresponds to a pure state on $\mcl M^G$ and
$p_M$ is affine, $\mS_F$ is a face. Closedness of $\mS_F$ follows from the continuity of $p_M$,
and convexity is clear. The  rest of (iii) is contained in the concluding remark of \ref{5.1.36}
which implies also the first statement of (iv). A proof of the second statement of (iv) is an easy
adaptation of that of \ref{5.1.17} for the case of factor states. It remains to prove (v):

{\bf Let} \emn$\tilde{\mome}\in\mSs((\mfkA^\Pi)^{**})$~ {\bf be the unique normal extension of}
$\mome\in\mS(\mfkA^\Pi)$ and $(\pi_\mome,\mH_\mome,\Omega_\mome)$ be the corresponding cyclic
representation of $(\mfkA^\Pi)^{**}$. {\bf Denote by} \emm$\hat{\mu}_\mome$~ the \emm orthogonal
measure~ (cf. \cite[4.1.20]{bra&rob}) on $\mS((\mfkA^\Pi)^{**})$ corresponding to the canonical
decomposition of $\tilde{\mome}$ \wrt\ the subalgebra $\pi_\mome(\mfk M_G^\Pi)$ of the center of
$\pi_\mome((\mfkA^\Pi)^{**})$, compare \cite[4.1.25]{bra&rob}:

\bequ\label{5.1.38(2)}\tilde{\mome}(x)=\int \mphi(x)\,\hat{\mu}_\mome(\rd\mphi),\ \text{for all}\
x\in(\mfkA^\Pi)^{**}. \enqu

The {\bf mapping}
\[\centerline{$y\ (\in\mfk M_G^\Pi)\mapsto \hat{y}\ (\in C(\mS((\mfkA^\Pi)^{**})))$,\
\emn$ \hat{y}(\mphi):=\mphi(y)$~,}\]

\noidt restricted to the subalgebra $p_\mome\mfk M_G^\Pi$\ (which is isomorphic to $\pi_\mome(\mfk
M_G^\Pi)$ for the {\bf uniquely determined projector} \emn$p_\mome\in\mfk M^\Pi_G$~) provides an
isomorphism of the \Wa s $p_\mome\mfk M^\Pi_G$ and $L^\infty(\hat{\mu}_\mome)$, \cite[Chap.
I.9]{gamelin} and \cite[4.1.22]{bra&rob}. Hence, for $y_j\in p_\mome\mfk M_G^\Pi\ (j=1,2)$ we have

\bequ\label{5.1.38(3)} \widehat{(y_1y_2)}(\mphi)=\hat{y_1}(\mphi)\hat{y_2}(\mphi),\ \text{for}\
\mphi\in\supp\hat{\mu}_\mome. \enqu

Clearly, $\mphi(y)=0$\ for $y\in (I-p_\mphi)\mfk M^\Pi_G$ and $\mphi\in\supp\hat{\mu}_\mome$. This
fact together with\rref 5.1.38(3)~ implies that the restriction $r_M\mphi$ is a pure state on
$\mfk M^\Pi_G$ for $\mphi\in\supp \hat{\mu}_\mome,\ r_M\mphi =: m_\mphi\in \mcl M^\Pi$.
 The $w^*$-topology of the state space is Hausdorff and
the clopen  sets form a basis of the topology of $\mcl M^\Pi$. This and the isomorphism of
$L^\infty(\hat{\mu}_\mome)$ with $p_\mome\mfk M_G^\Pi$ imply that the restriction of the mapping
$r_M$ onto $\supp\hat{\mu}_\mome$ is a bijection onto \bequ\label{5.1.38(4)} \supp p_\mome:=\{m\in
\mcl M^\Pi:\ m(p_\mome)=1\}. \enqu

\noidt Denote $\mome_m:=\mphi$\ iff\ $r_M\mphi = m\in \mcl M^\Pi$. Let $\mu_\mome$ be the image of
$\hat{\mu}_\mome$\ under $r_M$: \bequ\label{5.1.38(5)} \mu_\mome := \hat{\mu}_\mome\circ
r_M^{-1}\quad \text{is \ a\ regular\ Borel\ measure\ on}\ \mcl M^\Pi.\enqu

Since $p_M\mome=r_M\tilde{\mome},\ \mu_\mome$ is the measure specified in (v). The measurability
of the function $\hat{\mome}:\ m\mapsto\mome_m$ defined on $\supp p_\mome=\supp \mu_\mome$ is
clear, compare \cite[4.1.36]{bra&rob}. The integral in\rref 5.1.38(1)~ is then another form of
\rref 5.1.38(2)~. This concludes the proof.
\end{proof}
\begin{noti}\label{5.1.39} {\bf Let} \emn$r_\mfkA$~:\ $\mS((\mfkA^\Pi)^{**})\rarw
\mS(\mfkA^\Pi)$ be {\bf the restriction mapping.} {\bf Let} $e_*:\ \mS(\mfkA^\Pi)\rarw
\mS_*((\mfkA^\Pi)^{**})$ {\bf be the normal extension}, \emn$e_*\mome=\tilde{\mome}$~. Then
$p_M=r_M\circ e_*.$ For a general $\mphi\in\mS((\mfkA^\Pi)^{**}),$ it is \bequ\label{5.1.39(1)}
r_M\mphi\neq (p_M\circ r_\mfkA)\mphi.\enqu

Since $\mome_m\in\supp\hat{\mu}_\mome$ need not be normal, the inequality\rref 5.1.39(1)~ holds
also for $\mphi=\mome_m$ in general. The open question is, however, whether (under some
conditions) $(p_M\circ r_\mfkA)\mome_m\in \mcl M^\Pi=\mcl{ES}(\mfk M_G^\Pi)$ or, at least, when
the canonical measure corresponding to $(p_M\circ r_\mfkA)\mome_m\in \mS(C(\mcl M^\Pi))$ is
concentrated on a set $F_\mfkg^{-1}(F)$ for some $F\in \mcl M^G$, {\bf where}
\[\centerline{\emn$F_\mfkg :\mcl
M^\Pi\rarw \dot{\mcl M}^G$~ is the natural mapping defined according to $\rref
5.1.19(9)~\wedge\!\!\rref 5.1.19(15)~$,}\]

\noidt and $\dot{\mcl M}^G$ {\bf is the one-point compactification of} $\mcl M^G$. Let us write
down the {\bf definition of \emn$F_\mfkg$~ explicitly} (see also proof of \ref{5.1.19}):\nl

\noidt (*)\quad Let $E_\mfkg^\Pi$ be the projection-valued measure defined on Borel subsets of
\fkgs\ with values in $\mfk Z$, as determined in \ref{5.1.29} and in \ref{5.1.33}. {\bf Let}
\emn$\dot{\mfkg}^*:= \mfkgs\cup\{\infty\}$~ be the {\bf one-point compactification} of \fkgs\ and
\emn$\dot{\mcl M}^G:=\dot{\mfkg}^*\cup\{m_\circ\}$~, where \emm$m_\circ$ is an isolated point~.
{\bf Let} \emn$\mcl M^\Pi:=\mcl{ES}(\mfk M_G^\Pi)$~ be the {\bf spectrum space of the algebra}
$\mfk M_G^\Pi=C(\mcl M^\Pi)$\ generated by projectors $E^\Pi_\mfkg(B)$\ (Borel $B\subset \mfkgs$),
i.e. by continuous functions $m\mapsto m(E^\Pi_\mfkg(B))$, \emn$m\in\mcl Z:=\mcl{ES}(\mfk Z)$~.
{\bf Define the (continuous) mapping $F_\mfkg:\ \mcl M^\Pi\rarw \dot{\mcl M}^G$ by}
\begin{enumerate}
\item[(i)] $F_\mfkg(m)\in\mfkgs$ iff there is a {\em bounded}
Borel $B\subset \mfkgs$ such that

\bequ\label{5.1.39(2)}\mome_m(E_\mfkg^\Pi(B))\equiv m(E_\mfkg^\Pi(B)) = 1, \enqu

\noidt and, in this case, $F_\mfkg(m)(\xi):=m(X_{\xi \Pi}E_\mfkg^\Pi(B))$ for all $\xi \in\mfkg$.
Here $X_{\xi\Pi}$ are defined in \ref{5.1.29}. The character property of $m$ ensures independency
of $F_\mfkg(m)$ on $B$ satisfying\rref 5.1.39(2)~.

\item[(ii)] $F_\mfkg(m):= m_\circ$ iff $m(I-p_G)=1,$ i.e. iff
$m=m_\circ\in \mcl M^\Pi$, \ref{5.1.32}.

\item[(iii)] $F_\mfkg(m):=(\infty)$ iff $m(I-p_G)=0$ and\rref
5.1.39(2)~ is false for all {\em bounded} Borel subsets $B\subset \mfkgs: m(E_\mfkg^\Pi(B))=0.$
\end{enumerate}

The same definition applied to all $m\in \mcl Z$ leads to the mapping

\bequ\label{5.1.39(3)} F_\mfkg\circ r_M:\ \mcl Z\rarw \dot{\mcl M}^G,\quad \mfk Z:= C(\mcl
Z),\enqu

\noidt which is continuous on the whole $\mcl Z$.

The mapping $F_\mfkg\circ r_M$ determines the projectors $E_\mfkg^\Pi(B)$. For bounded $B$ we have
($\overline{B}:=\ closure,\ B^\circ:=\ interior$)\footnote{The relation\rref 5.1.39(4)~ has been
proved in the assumption that any projector $p\in\mfk M_G^\Pi$ is of the form $p= E_\mfkg^\Pi(B)$
for some $B\subset \mfkgs,$\ if $p(I-p_G)=0$.}

\bequ\label{5.1.39(4)} m(E_\mfkg^\Pi(\overline{B}))=1\ \text{iff}\ m\in\overline{[(F_\mfkg\circ
r_M)^{-1}(\overline{B})]^\circ} = [(F_\mfkg\circ r_M)^{-1}(\overline{B})]^\circ.\enqu

If we extend the $Ad^*(G)$ to the whole $\dot{\mcl M}^G$ by the requirement of
$Ad^*(G)$-invariance of the points $m_\circ$ and $(\infty)$ we see, that $F_\mfkg$ is
$G$-equivariant:

\bequ\label{5.1.39(5)} F_\mfkg(\msg^*_gm)=Ad^*(g)F_\mfkg(m),\ \text{for\ all}\ m\in \mcl M^\Pi,\
g\in G.\enqu
\end{noti}
\pt\label{5.1.40}\rm The projection measure $E^\Pi_\mfkg$ on \fkgs\ together with the
$Ad^*$-action of $G$ determine a \emm macroscopic limit~ of the system $(\mfkA^\Pi,\msg_G).$ This
formulation together with the mapping $p_M$ of $\mS(\mfkA^\Pi)$ into the classical macroscopic
states of the system will enable us to generalize the notion of the macroscopic limit to much more
general situations. We shall investigate also the dynamics of the system $(\mfkA^\Pi,\msg_G)$
(resp. of its generalizations) if the time evolution were not included in the action $\msg_G$ as
the action of a one parameter subgroup  of $G$. The action $\msg_G$ of the `kinematical group' $G$
allows us, as we shall show, to in\-troduce rather wide class of 'mean-field-type' time evolutions
connected with noncompact groups $G$ - at least for a large $\msg_G$-invariant subset of states in
$\mS(\mfkA^\Pi)$. Also automorphic time evolutions $\tau:\ t\mapsto \tau_t \in\maut \mfkA$ of a
system $(\mfkA, \msg_G, \tau_\mbR)$ will be considered.

\section{Generalized macroscopic limits}\label{sec;5.2}

\pt\label{5.2.1}\rm We have considered, in the preceding section, a macroscopic limit  of the
system $(\mfkA^\Pi,\msg_G)$. This system was of a rather special type: the algebra $\mfkA^\Pi$ was
the infinite tensor product of identical copies $\mfkA_j\ (j\in\mbZ_+=:\Pi)$ of a \Ca\ $\mfkA_0$
and the automorphism group $\msg_G$ left each of the copies $\mfkA_j$ invariant:
$\msg_gx\in\mfkA_j$ for each $x\in\mfkA_j$, for all $g\in G$ and any $j\in\mbZ_+\equiv\Pi$.  We
shall now generalize the procedure of obtaining a macroscopic limit to much more general
situations. We shall ignore here possible quasilocal structures of the considered \Ca\ \fkA; the
usage of the term 'macroscopic limit' can be here understood in an analogy with the preceding
section.

The notion of the macroscopic limit introduced in this section is nonunique. A certain
arbitrariness is contained, however, also in the corresponding notion of Sec.\ref{sec;5.1}: The
generators $X_\xi^N$ of the restriction of $\msg_G$ to $\mfkA^N := \otimes_{j=1}^N \mfkA_j \subset
\mfkA^\Pi$ are determined up to additive constants $a_N(\xi),\ a_N\in\mfkgs,\
\xi\in\mfkg,$\footnote{$a_N$ forms a zero-dimensional orbit of $Ad^*(G):\ a_N([\xi,\eta])\equiv
0$.} hence also the choice of $p_G\in\mfk Z$ was arbitrary in a certain sense. We shall avoid
partly this kind of ambiguity in this section: we are dealing here just with the action of
$\msg_G$, and not with generators.

\pt\label{5.2.2}\rm Let $G$ be a connected Lie group, \fkg\ its Lie algebra, and \fkgs\ the dual
of \fkg. Let \fkA\ be an arbitrary \Ca, $\mfkA^{**}$ its double dual \Wa, and $\mfk Z$ is the
center of $\mfkA^{**}$. The algebra \fkA\ is naturally contained in $\mfkA^{**}$ as a
$\msg(\mfkA^{**},\mfkA^*)$-dense \Csa. Any state $\mome\in\mS(\mfkA)$:= the state space of \fkA,
has a {\bf natural extension}
\[\centerline{\emn$e_*\mome\in\mS_*(\mfkA^{**})$~ := the normal states of $\mfkA^{**}$.}\]

\noidt If \fk M\ is a \Csa\ of $\mfkA^{**}$, {\bf then} \emn$r_{\mfk M}:\ \mS(\mfkA^{**})\rarw
\mS(\mfk M)$~ is the \emm restriction mapping~; $r_{\mfk M}$ is
$\msg(\mfkA^{***},\mfkA^{**})-\msg(\mfk M^*,\mfk M)$ continuous and maps normal states onto normal
states.

{\bf Let \emn $\msg:\ G\rarw\maut\mfkA,\ g\mapsto\msg_g$~ be a given action of $G$}; by the same
symbol \emn$\msg_G$~ is denoted the canonical extension of $\msg_G\subset \maut\mfkA$ to the
action on $\mfkA^{**}$ - the double transpose of $\msg_G$. This system {\bf will be denoted by}
\emn$(\mfkA;\msg_G)$~. \emn$\mcl Z$~ {\bf will denote the spectrum space of} $\mfk Z=C(\mcl Z).$

 Let \fkgs\ be endowed with the structure of a Poisson manifold,
 \ref{5.1.37}, given by a {\bf tensor field} \emn$\mbs\mlam$~,\ usually
 $\mbs\mlam_F(\cdot,\cdot):= -F([\cdot,\cdot])-\theta_F(\cdot,\cdot)$,
 i.e.

 \bequ\label{5.2.2(1)} \{f,g\}(F):=
 -F([\rd_Ff,\rd_Fg])-\theta_F(\rd_Ff,\rd_Fg),\ F\in\mfkgs,\
 \theta_F\equiv\theta, \enqu

 \noidt where $f,g\in C^\infty(\mfkgs,\mbR)$ and \emn$\theta$ {\bf is a two
 form} on \fkg~\ satisfying

 \bequ\label{5.2.2.(2)}
 \theta(\xi_1,[\xi_2,\xi_3])+\theta(\xi_2,[\xi_3,\xi_1])+
 \theta(\xi_3,[\xi_1,\xi_2])=0,\enqu

 \noidt for all $\xi_j\in\mfkg,\ j=1,2,3.$ We assume that an
 action of $G$ on \fkgs\ is $\mphi:\ g\mapsto \mphi_g$, where
 $\mphi_G$ is a `maximal' group of Poisson morphisms, i.e.
 $\mphi_{gh}=\mphi_g\circ\mphi_h\ (g,h,\in G),\
 \mphi_e:=\id_{\mfkgs}\ (e:=$ the identity of $G$); each $\mphi_h$ is
 a diffeomorphism of \fkgs\ conserving the \emm Poisson structure~:

 \bequ\label{5.2.2(3)}\mphi^*_h\{f,g\}=\{\mphi_h^*f,\mphi^*_hg\},\quad
 f,g\in C^\infty(\mfkgs,\mbR),\ h\in G,\enqu

 \noidt and $\mphi_GF\ (\forall F\in\mfkgs)$ are the maximal
 integral submanifolds of \emn$\mbs\mlam$~,\ \cite[Def.3.1\&\,Thm.3.4]{marle}.
 Usually, one takes
\bequ\label{5.2.2(4)} \mphi_hF:=Ad^*(h)(F)+a_\theta(h),\
 h\in G,\ F\in\mfkgs,\enqu

 \noidt {\bf where \emn$a_\theta$~ is a unique differentiable mapping} from
 $G$ to \fkgs\ with the properties, \cite{marle}:
 \begin{enumerate}
\item[(i)] $a_\theta(gh)=Ad^*(g)(a_\theta(h))+a_\theta(g),\quad
\forall g,h,\in G$,
\item[(ii)] $T_ea_\theta(\xi)(\eta)=\theta(\xi,\eta),\quad\forall
\xi,\eta\in\mfkg,$ where $T_e a_\theta: \mfkg\rarw\mfkgs$\ is the tangent map of $a_\theta$ at
$e\in G$.
 \end{enumerate}

The system $(\mfkA;\msg_G)$ represents a quantal system {\bf and \emn$(\mfkgs,\mbs\mlam;\mphi_G)$~
is a (generalized) classical system} which will play the role of a macroscopic limit of the system
$(\mfkA;\msg_G)$. Let us introduce candidates for this micro-macro connection:

\begin{defs}\label{5.2.3}Let $\mcl B(\mfkgs)$ be the set of all complex-valued
uniformly bounded Borel functions on \fkgs\ and let $\Sigma_G$ be the Borel \sg-algebra of subsets
of \fkgs. Let the \emm $G$-measure~ $E$ (of the system $(\mfkgs,\mbs\mlam;\mphi_G)$, resp. of
$(\mfkA;\msg_G)$) be any projection-valued measure on \fkgs\ with values in \fk Z, which is
$G$-equivariant, i.e.
\begin{subequations}
\bequ\label{5.2.3(1)} E:\ \Sigma_G\rarw\mfk Z,\ B\mapsto E(B)=E(B)^*=E(B)^2\in\mfk Z\quad
(B\in\Sigma_G),\enqu

\bequ\label{5.2.3(2)} B_j\cap B_k=\emptyset\ (j\neq k,\ j,k\in\mbZ_+)\imply
E(\cup_jB_j)=\sum_jE(B_j),\enqu

\bequ\label{5.2.3(3)} E(\mphi_gB)=\msg_gE(B),\quad \text{for all}\ B\in\Sigma_G,\ \text{and for
all}\ g\in G.\enqu
\end{subequations}

\noidt {\bf Denote by \emn$E(f)\in\mfk Z$~ the integral of $f\in\mcl B(\mfkgs)$ over $E$}.\nl

\noidt  {\bf Let \emn$p_E:= E(\mfkgs)$~,\ $I:=$ the unit of $\mfkA^{**}$}.\nl

\noidt {\bf Denote by \emn$\mfk N(E)$~ the \Wsa\ of \fk Z\ generated by $E(f),\ f\in\mcl
B(\mfkgs).$}\nl

 {\bf Let} \emn$\mfk B(E)$~ denote the \emm Borel${}^*$-algebra~ \cite[4.5.5]{pedersen} in \fk Z\
 generated by all the $E(f),\
f\in\mcl B(\mfkgs)$; this means that $\mfk B(E)$ is the smallest \Csa\ of \fk Z\ containing all
the $E(B)\ (B\in\Sigma_G)$ and with each monotone (increasing or decreasing) {\em sequence}
$x_j\in\mfk B(E)_s$ it is also $s\=\lim x_j\in\mfk B(E).$ Clearly $\mfk B(E)\subset \mfk N(E).$
{\bf Let \emn$\mfk M_s$~ denote the set of all selfadjoint elements of a \Ca\ \fk M}. The
projector $p_E$ is the common unit of $\mfk B(E)$\ and $\mfk N(E)$. Any projector $q\in\mfk B(E)$
is of the form $q = E(B)$ for some $B\in\Sigma_G$, what need not be the case for $\mfk N(E)$.
Projections in $\mfk N(E)$ separate various kinds of spectra of $E$ (resp. of operators $E(f)$
etc.) what need not be the case of $\mfk B(E).$

Let $\supp\,E\subset\mfkgs$ be the minimal closed $B=\overline{B}\in\Sigma_G$ such that
$E(B)=p_E.$ On the other hand, $\supp\,E(B):=\{m\in \mcl Z=\mcl{ES}(\mfk Z):\ m(E(B))=1\}$ is a
clopen subset of $\mcl Z$. {\bf Let}
\[\centerline{\emn$\dim(F)$~:= dimension of the orbit $\mphi_GF\subset\mfkgs,\
\dim(F)=2k\leq\dim\mfkgs$,}\]

\noidt {\bf and}  \emn$\dim(E):=\max\{\dim(F): F\in\supp E\}$~. The \emn$G$-measure~ $E$ is \emm
trivial~ iff $\dim(E)=0.$ The quantal system $(\mfkA;\msg_G)$ has a \emm nontrivial macroscopic
limit~ {\bf in the classical system}\ $(\mfkgs,\mbs\mlam;\mphi_G)$ iff there is $E$ such that
$\dim(E)\geq 2.$ If there is an $E$ such that $\dim(E)= n_G,$ and for any other $G$-measure $E'$
it is $\dim(E')\leq n_G$, we say that the system $(\mfkA;\msg_G)$ has \emm $G$-macroscopic limit
of the dimension $n_G$~ (in the classical system $(\mfkgs,\mbs\mlam;\mphi_G)$). {\bf The number
\emm $n_G=:2k_G$~ is the} \emm $G$-macroscopic dimension~ of $(\mfkA;\msg_G)$ and $k_G$ is the
\emm $G$-macroscopic number~ {\bf of degrees of freedom} of the quantal system $(\mfkA;\msg_G).$
\end{defs}

\pt\label{5.2.4}\rm We shall assume in the following that $n_G\geq 2$ and we shall consider only
$G$-measures $E$ with $\dim(E)=n_G$. The projectors $p_E$ are, clearly, $G$-invariant:

\bequ\label{5.2.4(1)} \msg_g(p_E)= p_E\quad\ \text{for all}\ g\in G\ \text{and\ all\
$G$-measures}\ E.\enqu

Let $q\leq p_E$ be another $G$-invariant projector in \fk Z. Then {\bf we can define the
restriction of $E$ to $q$, the $G$-measure \emn$qE$~}, by

\bequ\label{5.2.4(2)} qE:\ \Sigma_G\rarw\mfk Z,\ B\mapsto qE(B);\ p_{qE}=qp_E.\enqu

If $p_Ep_{E'}=0$\ for two $G$-measures $E$ and $E'$ then the mapping

\bequ\label{5.2.4(3)} E+E':\ B\mapsto E(B)+E'(B)\quad (\forall B\in\Sigma_G)\enqu

\noidt is a $G$-measure with $\dim(E+E')=\max \{\dim(E),\dim(E')\}$,\ and\ $p_{E+E'}=p_E+p_{E'}$.
{\bf For any two $G$-measures $E$ and $E'$, there is a $G$-measure \emn$EsE'$~ given by}

\bequ\label{5.2.4(4)} EsE'(B):= E(B)+(I-p_E)E'(B)\quad \forall B\in\Sigma_G.\enqu

\noidt For the support projector $p_{EsE'}$ of the $G$-measure $EsE'$ we have

\bequ\label{5.2.4(5)} p_{EsE'}=p_E+p_{E'}-p_Ep_{E'}=p_{E'sE},\enqu

\noidt although, in general, $EsE'$ is different from $E'sE$. Now, one has
$\dim(EsE')\geq\dim(E).$ Since $p_{EsE'} = p_E\vee p_{E'} :=\lub[p_E;p_{E'}],$ we can endow the
set of classes $[E]$

\bequ\label{5.2.4(6)} [E]:=\{E':p_{E'}=p_E\} \enqu

\noidt with a partial ordering:\ind{$[E']\prec [E]$}

\bequ\label{5.2.4(7)} [E] \succ [E']\quad \Leftrightarrow\ p_E\geq p_{E'}.\enqu

\noidt {\bf This ordering makes the set $\{[E]\}$ of classes of $G$-measures a directed set.}

The same ordering will be considered for any set of subclasses $[E]'\subset [E]$ determined by
some further condition $C$, i.e. for classes

\bequ\label{5.2.4(8)} [E]':=\{E':p_{E'}=p_E,\ C(E')\}.\enqu

\noidt Here $C(E')$ means ``the $G$-measure $E'$ satisfies the condition $C$\,'', e.g.
$C(E'):=(\dim(E')=\dim(E_\circ))$, or $C(E):=(E(F)\neq 0\imply \dim(F)\neq 0)$, etc. The classes
\rref 5.2.4(8)~ could also be denoted by $[E]$.

\begin{lem}\label{5.2.5}The function $\dim: \mfkgs\rarw \mbR, F\mapsto
\dim(F)$ is \emn lower semicontinuous~. Hence, the sets $\{F\in\mfkgs: \dim(F)\geq n\}$ are open
and the sets $\{F\in\mfkgs : \dim(F)\leq n\}$ are closed in \fkgs\ for any $n\in\mbZ_+$.
Specifically, the set $\{F\in\mfkgs: \dim(F) = 0\}$ is closed, and the set $\{F\in\mfkgs : \dim(F)
= n\}$ is Borel.
\end{lem}
\begin{proof} It was assumed in \ref{5.2.2} that the action $\mphi_G$
 is a 'maximal' Poisson action, i.e. the orbits of $\mphi_G$ coincide
 with the maximal
integral manifolds of the Poisson structure $\mbs\mlam$\ on \fkgs, \cite{marle}. The dimension
$\dim(F)$ of $\mphi_GF$ is then given by the rank of the skew-symmetric 2-tensor $\mbs\mlam_F$ (:=
the value of $\mbs\mlam$\ in the point $F\in\mfkgs$), i.e. by the rank of the mapping
$\mbs\mlam_F: T_F\mfkgs\rarw T^*_F\mfkgs,\ v\mapsto \mbs\mlam_F(v,\cdot)$, {\bf denoted by}
\emn$\rank(\mbs\mlam_F)$~. Since $\mbs\mlam$\ depends smoothly on $F$, the function
$F\mapsto\dim(F)=\rank(\mbs\mlam_F)$ is lower semicontinuous. The remaining assertions then
follow.
\end{proof}

\pt\label{5.2.6}\rm Let $S_n := \{F\in\mfkgs: \dim(F)\leq n-1\}, 1\leq n\leq\dim G$. {\bf For such
a $G$-measure $E$ with $p_E\neq E(S_n)$\ let} \emn$r_nE$~ := $(p_E-E(S_n))E,$ see\rref 5.2.4(2)~.
Clearly, $\dim(E)=\dim(r_nE),$ if $r_nE\neq 0$. $E$ is a \emm purely nontrivial~ $G$-measure, if
$0\neq p_E$ and $r_1E = E$. If $0\neq E = r_nE$ and $r_{n+1}E = 0,\ E$ is called a \emm purely
n-dimensional~ $G$-measure. For $n := \dim(E)$ the measure $r_nE$ is purely n-dimensional. The
G-measures $E+E'$, $EsE'$ and $qE$ (with a G-invariant projector $q = qp_E\neq 0$) are purely
n-dimensional together with $E$ and $E'$. Let the ordering\rref 5.2.4(7)~ be given for the set of
classes $[E] := \{E': p_E'= p_E\ and\ E'=r_nE', r_{n+1}E'=0\}$. In any linearly ordered subnet of
such $[E]$'s there is a natural mapping

\bequ\label{5.2.6(1)} \pi_{EE'}:[E']\rarw [E],\ E'\mapsto \pi_{EE'}(E'):= p_EE'\in[E]\ \text{for}\
p_{E'}\geq p_E.\enqu

The {\bf mappings \emn$\pi_{EE'}$~ define} a \emm projective system~ \cite[Definition
20.1]{choquet} on the linearly ordered subset J of classes $[E]: \pi_{EE''}=\pi_{EE'}\circ
\pi_{E'E''}$ for $p_{E''}\geq p_{E'}\geq p_E$ and $\pi_{EE}=\id_{[E]}.$ If $p\leq p_{E'}$, and
$E:= pE'$, then $E'=EsE'$. We want to show that J has an upper bound in the set of classes $[E]$
of purely n-dimensional $G$-measures $E$. This would imply, by the Zorn's  lemma, the existence of
{\em the} maximal element in the set (uniqueness of the maximal element follows from the
directedness of the set).

\begin{lem}\label{5.2.7}Let L be a set of {\bf G-measures linearly
ordered} by \emn$E\leq E'\Leftrightarrow E'=EsE'$~. Then L has an upper bound.
\end{lem}
\begin{proof} For any $0\leq f\in \mcl B(\mfkgs)$ and $E'\geq E$
it is $E'(f)\geq E(f).$ Denote

\bequ\label{5.2.7(1)} E_L(f):= \lub\{E(f):\ E\in L\} = s \=\lim\{ E(f): E\in L\}.\enqu

The mapping $E_L$ can be extended by linearity to  \cl B(\fkgs):

\bequ\label{5.2.7(2)} E_L: \mcl B(\mfkgs)\rarw \mfk Z,\ f\mapsto E_L(f); \enqu

\noidt it is bounded: $\|E_L(f)\|\leq\|f\|:=\sup\{ |f(F)|:\ F\in\mfkgs\}.$ Due to continuity of
the product in the strong topology, the mapping $E_L$ is a $C^*$-homomorphism of the commutative
\Ca\ \cl B(\fkgs)\ into \fk Z. This implies the \sg-additivity of the set-function $E_L: B\mapsto
E_L(B):= E_L(\chi_B), B\in\Sigma_G$, hence, $E_L$ is a projection measure. Since
$\msg_g\in\maut\mfk Z$, and any automorphism of a \Wa\ is \sg-\sg-continuous, $E_L$ is a
$G$-measure. Clearly $E_L\geq E,\ \forall E\in L$.
\end{proof}
\begin{prop}\label{5.2.8}
The directed set of classes of purely n-dimensional $G$-measures has a maximal
element.\footnote{The present author was informed about some important set-theoretical concepts
connected with this Proposition by the late colleague Ivan Korec (1943 - 1998).}
\end{prop}
\begin{proof} Let $J$ be any linearly ordered subset of the directed set; cf.\rref 5.2.4(7)~.
We shall prove that it is possible to choose $E\in [E]$ in any $[E]\in J$ in such a way that $[E]
\prec [E']$ iff $E' = EsE'$. Then the result will follow from the Lemma \ref{5.2.7} and from the
Zorn lemma. It is clear that the choice $E\in [E]$ of the desired kind can be made in any finite
subset  $K_\circ\subset J,\ [E]\in K_\circ.$

The desired choice (it will be called a 'consistent choice') can be made in the subset $K_E :=
\{[E']\in J:\ [E']\prec [E]\}$ of $J$ by $E':= p_{E'}E$ for any $[E]\in J$, with any fixed $E\in
[E]$. We have to prove existence of a consistent choice on the whole $J$. Let $J_\circ$ be a \emm
well ordered~ \emm cofinal subset~ of $J$ (the well ordering of $J_\circ$ is that one induced by
the ordering of $J$ - it is possible by the axiom of choice, and cofinality means that for any
$[E]\in J$ there is an $[E_j]\in J_\circ : [E]\prec [E_j]$). Now we can choose $E_j\in [E_j]$ (for
all $[E_j]\in J_\circ$) in a consistent way: For the successor $[E_{j+1}]$\ of\ $[E_j]$ in
$J_\circ$ we shall choose $E_{j+1} := E_jsE_{j+1}'$ with any $E_{j+1}'\in [E_{j+1}]$, if $[E_j]$
has been defined before. If $[E_j]$ is not a successor in $J_\circ$, put $E_j^\circ :=\lub \{E_k:
[E_j]\succ [E_k]\in J_\circ,\ \text{all}\ E_k(\in [E_k])\ \text{are\ mutually\ consistent}\}$,
according to the Lemma \ref{5.2.7}, and choose $E_j := E_j^\circ s E'_j$ with any $E'_j\in [E_j]$.
Then we can 'to fill gaps' by setting $E := p_E E_j$ for all $[E]\prec [E_j]\in J_\circ$. This
provides a consistent choice $E\in [E]$ for all $[E]\in J$, if $J_\circ$ is considered as an
initial segment of the set of all ordinals.
\end{proof}
{\bf Note:} The same proof applies to purely n-dimensional measures $E$ of the form $E=qE$ for any
fixed $G-$invariant projector $q\in\mfk Z$.

\pt\label{5.2.9}\rm {\bf Let} \emn$[E]^\circ_G$~ be the {\bf maximal element of classes of purely
$n_G$-dimensional $G$-measures} and {\bf let} \emn$p_G^\circ := p_E$ for $E\in [E]^\circ_G$~. {\bf
Let} \emn$[E]^k_G$~ be {\bf the maximal element of classes of purely $(n_G-2k)$-dimensional
$G$-measures} of the form $E=(I-\sum_{j=0}^{k-1} p_G^j)E,$ and for $E\in [E]_G^k$ let $p_G^k:=
p_E,\ k=1,2,\dots, \frac{n_G}{2}$. Define now the {\bf class} \emn$[E]_G$~ of \emm maximal
$G$-measures~ by

\[ [E]_G := \sum_{k=0}^{\frac{n_G}{2}} [E]^k_G,\ \text{with}\ E\in [E]_G\
\text{iff}\ E = \sum_{k=0}^\frac{n_G}{2} E_k,\ E_k\in [E]_G^k, \]

\noidt and the sum of mutually \emn orthogonal $G$-measures~ is defined in\rref 5.2.4(3)~. The
choice of measures $E\in [E]_G$ for the realization of macroscopic limits corresponds to a
requirement of 'maximal sensitivity' of the corresponding macroscopic description of the system
$(\mfkA;\msg_G)$.

We shall not proceed further in an analysis of the set $[E]_G$ and we shall not try to specify
some 'most convenient' element $E\in [E]_G$ as a representative of the macroscopic limit. Let us
choose any fixed $E_\mfkg\in[E]_G$.

\begin{defs}\label{5.2.10}The projection-valued measure $E_\mfkg\in [E]_G$
 on the Poisson manifold \linebreak \emn$(\mfkgs,\mbs\mlam;\mphi_G)$~, the G-action on which
is 'maximal'\ (i.e. orbits $\mphi_GF$ are maximal symplectic immersed submanifolds of \fkgs\ the
Poisson bracket on which is given by $\mbs\mlam$, for any $F\in\mfkgs$), {\bf with values in}
\emn$\mfk Z$~\ ({\bf := the center of} $\mfkA^{**}$) is called the \emm $G$-macroscopic limit~
{\bf of the system $(\mfkA;\msg_G)$ in the classical system $(\mfkgs,\mbs\mlam;\mphi_G).$} The
{\bf projector} $p_G:= E_\mfkg(\mfkgs)$ is the \emm support projector~ of the macroscopic limit.
The dimension $n_G$ will be called also \emm the dimension of $E_\mfkg$~. The \emn Borel*- (resp.
the W*-)\,algebra~ \cite[4.5.5]{pedersen} generated by $E_\mfkg$\ (resp. by $E_\mfkg$ and
$I\in\mfk Z$) will be called the B*- (resp. W*-) \emm macroscopic algebra of $G$-definiteness~
(resp. the \emm $G$-macroscopic algebra~) of the system $(\mfkA;\msg_G)$ and will be denoted (in
the W*-cases) by $\mfk N_G$\ (resp. by $\mfk M_G$).

{\bf Denote by} \emn$p_M:\ \mS(\mfkA)\rarw \mSs(\mfk M_G)$~,\ $\mome\mapsto p_M\mome := r_{\mfk
M}\circ e_*(\mome)$, {\bf where} \emn$r_{\mfk M}$~ {\bf is the restriction} of $\mS(\mfkA^{**})$
to $\mS(\mfk M_G)$ and \emn$e_*$~ is {\bf the natural extension from \cS(\fkA)\ to}
$\mSs(\mfkA^{**})$. {\bf Let} \emn$\mu^\mome_\mfkg$~ be the probability measure on $\mcl M^G$ (:=
$\mfkgs\cup \{m_\circ\},\ m_\circ$ is an isolated point) given by

\bequ\label{5.2.10(1)} \mu^\mome_\mfkg(B):= \mome(E_\mfkg(B\setminus\{m_\circ\})) +
\mome(I-p_G)\delta_{m_\circ}(B),\ any\ Borel\ B\subset \mcl M^G,\enqu

\noidt compare\rref 5.1.36(1)~. Let us introduce the set

\bequ\label{5.2.10(2)} \mcl E_\mfkg :=\{\mome\in\mS(\mfkA): e_*\mome(p_G)=1,\
\mu^\mome_\mfkg(\xi^2)=[\mu^\mome_\mfkg(\xi)]^2<\infty,\ \forall\xi\in\mfkg\},\enqu

\noidt compare\rref 5.1.36(6)~, where $\xi\in\mfkg$ is considered as a linear function on
$\mfkgs$, since $\mfkg\subset\mfkg^{**}$.
\end{defs}

We can introduce also unbounded operators $X_\xi:= E_\mfkg(\xi)$ on the Hilbert space $\mH_u$ of
the universal representation of \fkA. Then we have

\begin{thm}\label{5.2.11}The Theorem \ref{5.1.38} as well as its proof are valid
also after the omission of the index $\Pi$ everywhere in its formulation and exchange of $Ad^*(G)$
by $\mphi_G$, with the interpretation of symbols according to \ref{5.2.10}.
\end{thm}

\begin{noti}\label{5.2.12} We could now, after the recognizing of
the Theorem, to continue in the choices of $E_\mfkg\in[E]_G$ according to the following idea:
Choose $E_\mfkg$ such that the sets $\mS_F$ of states with sharp values of the macroscopic
observables (cf. \ref{5.1.38}\ (iii)) are in a certain sense 'maximal'. We shall not make this
idea precise here. We believe, however, that continuing in this direction we could obtain
$E_\mfkg$ 'essentially uniquely' - up to natural coordinate transformations in the\break\ space
\fkgs.\end{noti}

\pt\rm {\bf A scheme of `macroscopic quantization'.}\label{5.2.13}\nl

Having once a classical limit in the form of the couple $\{(\mfkgs,\mbs\mlam;\mphi_G), (\mfk
M_G;\msg_G)\}$, where\break $\msg_G\subset\maut\mfk M_G$, we are interested in the question: Can
the original algebra \fkA\ be reconstructed from this classical limit ? Keeping in mind the model
of Sec.\ref{sec;5.1} we propose the following scheme for obtaining the algebra \fkA\ of a system
$(\mfkA;\msg_G)$, the macroscopic limit of which is $(\mfk M_G;\msg_G; E_\mfkg)$\ (here the
measure $E_\mfkg$ symbolizes the connection with the classical system
$(\mfkgs,\mbs\mlam;\mphi_G)$),\ (let us denote by MQ the following scheme):\nl

{\em{\bf(MQ)}\label{MQ}\ Find a faithful representation $\rho$ of $\mfk M_G$ in a Hilbert space
$\mH_\rho$ (necessarily nonseparable) with the properties:}
\begin{enumerate}

\item[(i)] There is a simple \Csa\ \fkA\ of $\mcl L(\mcl H_\rho)$\
such that the center of its commutant \fkA'\ contains $\rho(\mfk M_G)$; $\msg_G$ extends to an
automorphism group of \fkA.
\item[(ii)] \fkA\ is expressible as the norm-closure of union of a
net of von Neumann subalgebras $\mfkA_j\ (j\in J:=$ a directed set): $j\prec k\imply
\mfkA_j\subset\mfkA_k.$
\item[(iii)] Each $\mfkA_j$ is a $\msg_G$-invariant subset of
\fkA\ and the restriction of $\msg_G$ to any $\mfkA_j\ (j\in J)$ is unitarily implementable (i.e.
it exists a strongly continuous unitary representation $U^j$ of $G$ in $\mH_\rho$ such that
$\msg_g(x)=U^j(g)xU^j(g^{-1})$ for all $x\in\mfkA_j,\ g\in G$ and $j\in J$).
\item[(iv)] Each $\mfkA_k\ (k\in J)$ is generated by all $\mfkA_j$
with $j\prec k\ (j\neq k)$ as well as by the bounded Borel functions of the selfadjoint generators
$X_\xi^k\ (\xi\in\mfkg)$ of the one parameter groups $t\mapsto U^k(\exp(t\xi))$.

\end{enumerate}

Hence the proposed `quantization procedure' of the classical system $(\mfkgs,\mbs\mlam;\mphi_G)$
consists in finding an `\emn imprimitivity system~' $(\mfk M_G,\msg_G)$ (cf. \cite{varad})
determined by a choice of a $G$-measure $E_\mfkg$ (in some commutative \Ca\ $\mfk M_G$, where
$\msg_G\subset\maut\mfk M_G$ is determined by $\msg_gE_\mfkg(B) := E_\mfkg(\mphi_g B),\ g\in G,
B=$ Borel subsets in \fkgs), and afterwards applying the scheme (MQ) of `\emn macroscopic
quantization~' to $(\mfk M_G;\msg_G)$. We shall not investigate here conditions of existence and a
`degree of uniqueness' of this recipe. The scheme is nonempty, since it is fulfilled e.g. by the
models considered in Sec.\ref{sec;5.1} if $U(G)$ is irreducible, \ref{5.1.3}.

The question of obtaining a microscopic quantum dynamics of this `quantized macroscopic system'
corresponding to its given classical time evolution is posed and solved in the next Chapter
\ref{Ch6}.

\newpage

\chapter{Mathematical structure of QM mean-field theories}\label{Ch6}

\section{General considerations}\label{sec;6.1}

\pt\label{6.1.1}\rm The formalism developed in Chap. \ref{Ch5} will be used in this chapter for a
determination of a microscopic time evolution of an infinite quantum system from the macroscopic
(classical) evolution. It is clear that such an unusual determination of \emn microscopic
dynamics~ is possible for a very special type of interactions only. We shall show that this is the
case of a wide class of \emn quantum mean-field theories~,\footnote{For some history, general
meaning and technical construction of dynamics (given by full and correctly solved microscopic
evolutions - without any approximations) of ``Quantum mean-field theories'' see also \cite{bon1},
and for some of its applications look in \cite{bon2}.} at least in the time invariant subset
$\mS^\Pi_\mfkg$ of the set $\mS(\mfkA^\Pi)$ of all the microscopic states on the quasilocal
algebra $\mfkA^\Pi$, cf. Sec. \ref{sec;5.1}, esp. \ref{5.1.32}; cf. also '\emn classical states~'
in \cite{hp+lie1}. The systems of the considered type are determined by the couple
$(\mfkA;\msg_G)$ consisting of a \Ca\ \fkA\ (:= $\mfkA^\Pi$, e.g.; the upper indices $\Pi$ will be
usually omitted in this chapter) and of a representation $\msg(G) := \msg_G\subset\maut\mfkA$, cf.
\ref{5.2.2}, as well as by a $G$-measure $E_\mfkg$, \ref{5.2.3}, and by a classical Hamiltonian
function $Q\in C^\infty(\mfkgs,\mbR)$. A subclass of these systems consists of \emn thermodynamic
limits~ $N\rarw\infty$ of systems of the total number $N$ of quantal (mutually equal) subsystems
with dynamics described by \emn local Hamiltonians $Q^N$~. These local Hamiltonians are invariant
\wrt any permutations of $N$ subsystems and the $k$-body interaction constants (i.e. coefficients
at products of $k$ operators corresponding to $k$ different subsystems) are proportional to
$N^{1-k}$. We can construct such a sequence of the '\emn local time evolutions~'
$\tau^N\subset\maut\mfkA$ in the following way:

Let us keep the notation of Sec. \ref{sec;5.1}, and let a basis $\xi_j\ (j=1,\dots n)$ of \fkg\ be
fixed, the dual basis being $\{f_j: j=1,2,\dots n\}\subset \mfkgs$. Let $X_j\ (j=1,2,\dots n)$ be
the selfadjoint generators of the one parameter unitary groups $t\mapsto U(\exp(t\xi_j))$ on \H,
\ref{5.1.3}. Let $Q$ be a polynomial in $n$ variables and with a prescribed order of
multiplication of variables in such a way that the element $Q(\xi_1,\xi_2,\dots \xi_n)$ of the Lie
algebra envelope has the following property:

\nl \noidt {\bf (SA)}\quad \emph{Let \emn$\mathbf{Q \equiv \sum_{k=0}^q Q_k}$~, where $Q_k$ is a
homogeneous polynomial of degree $k$. In any continuous unitary representation $U$ of the group
$G$ on a separable Hilbert space \H, the operators $Q_k(X_1,X_2,\dots, X_n)$ defined on analytic
elements of $U$ are essentially selfadjoint, for all $k=1,2,\dots, q.$}\nl

This \emn property (SA)~ is fulfilled e.g. {\bf if all the \emn$Q_k(\xi_1,\xi_2,\dots, \xi_n)$~
are symmetric and elliptic}, cf. \cite[Chap.11]{bar&racz}. Then we define the \emn local
Hamiltonians $Q^N$~, with\rref 5.1.3(6)~, denoting by $N$ also an $N$-point subset of $\Pi$:

\bequ\label{6.1.1(1)} Q^N:= N\,Q(X_{1N},X_{2N},\dots, X_{nN}),\quad X_{jN}:=\frac{1}{N}X_j^N,\quad
N=1,2,\dots;\enqu

\noidt cf.\rref 5.1.29(1)~, i.e.
\[ X_j^N := \sum_{k=1}^{|N|} \pi_k(X_j),\quad j=1,2,\dots n;\ k\in N\subset\Pi, \]

\noidt which \ind{$X_j^N$} can be considered as (essentially) selfadjoint operators on $\mH_\Pi\ (=
\mH_N\otimes\mH_{\Pi\setminus N}\equiv \Pi$-tuple tensor product). For any $x\in\mfkA \
(:=\,\mfkA^\Pi)\subset \mcl L(\mH_\Pi)$ we set

\bequ\label{6.1.1(2)} \tau^N_t(x) := \exp(itQ^N)\,x\, \exp(-itQ^N),\  t\in\mbR,\enqu

\noidt and these mappings $\tau_t^N$ clearly form a one parameter group of \autm s of \fkA\ for
each finite $N$. Systems of this type were introduced in \cite{hp+lie1} for the case of spin
systems (i.e. $\dim\mH$\ was finite). It was shown in \cite{hp+lie1,bon1} that the sequence
$\{\tau^N : N = 1,2,\dots\}$ determines an evolution $\tau^Q$ of the observables of the form
$X_{\xi\Pi}$, cf. \ref{5.1.7} and \ref{5.1.9}, which is expressed in our notation by the formula

\bequ\label{6.1.1(3)} \tau_t^Q(X_{\xi\Pi}) := w^*_0\, \hbox{-}\lim_{N\rarw\infty} \tau^N_t(X_{\xi
N}) = \int f_\xi(\mphi_t^Q F)\, E_\mfkg(\rd F),\enqu

\noidt where $w^*_0$-topology on a von Neumann algebra containing \fkA\ and $X_{\xi\Pi}\
(\xi\in\mfkg)$ is determined by the set of the 'classical states'. The integral in\rref 6.1.1(3)~
corresponds to the integral in \cite[(2.29)]{hp+lie1}, which, specified to our case, reads:

\bequ\label{6.1.1(4)} \lim_{N\rarw \infty}\mome(\tau^N_t(X_{\xi N})) = \int f_\xi(\mphi_t^QF)\,
\mome(E_\mfkg(\rd F)),\quad \mome\in\mS_\mfkg.\enqu

\noidt We have used notation $f_\xi(F):= F(\xi)\ (\xi\in\mfkg,\,F\in\mfkgs)$, and $\mphi^Q$ is the
classical flow on \fkgs\ corresponding to the Hamiltonian function $Q\in C^\infty(\mfkgs, \mbR),\
Q(F):= Q(F_1,F_2,\dots F_n)$, with $F_j := f_{\xi_j}(F)=F(\xi_j),\ (F\in\mfkgs)$; the introduction
of the flow $\mphi^Q$ will be discussed later in this section. The natural question is, however.
whether the limits

\bequ\label{6.1.1(5)} \tau^Q_t(x):= (some\ topology)\ \hbox{-}\lim_{N\rarw\infty}\tau_t^N(x) \enqu

\noidt exist for some $t>0$ and for sufficiently many $x\in\mfkA$, so that $\tau^Q_t$ could be
extended to a one parameter group (resp. semigroup) of mappings of \fkA\ (or of some of its
completions) representing in a reasonable manner some time translations. We shall show that this
is indeed the case, and not only for the spin systems. The resulting family of transformations
$\tau^Q_t$ {\bf does not consist}, however, (for general $Q$) {\bf of automorphisms of the
original} (i.e. that one used at the determination of the infinite system) {\bf quasilocal \Ca\
\fkA}. The family of ${}^*$-isomorphisms of \fkA, $\tau^Q$, can be extended to a one parameter
group $\tau^Q$ of \autm\ of a \Csa\ of $\mfkA^{**}$ containing \fkA\ as a \Csa. The resulting
picture of the $\tau^Q$-time evolution has the properties of the \emn quantum mean-field
evolutions~ according to the usual understanding, cf. also \cite{bon1}. We shall write down
explicit formulas for the evolution of an arbitrary element of the extended algebra of observables
(including also an algebra of classical - intensive - observables) in terms of solutions of finite
dimensional differential equations.

In the present section, we shall introduce some basic concepts used in the general construction of
the automorphism group $\tau^Q$. We shall sketch here also a scheme of the general construction.
Details will be proved in the following sections of this chapter.

\pt\label{6.1.2}\rm {\bf Let} \emn$(\mfkgs,\mbs\mlam;\mphi_G)$~ be a \emm Poisson manifold~ with
the {\bf Poisson action} \emn$\mphi_G$~ of the Lie group $G$ the orbits of which coincide with the
maximal integral submanifolds of the Poisson structure $\mbs\mlam$, cf. \cite{marle}, and
\ref{5.1.37}, and \ref{5.2.2}. {\bf We shall assume, for simplicity, that \emn$\mphi_G$~$ :=
Ad^*(G)$ and \emn$\mbs\mlam_F(\rd f,\rd g)$~$ := - F([\rd_Ff,\rd_Fg])$ for all $f,g \in
C^\infty(\mfkgs,\mbR)$}. Let $Q\in C^\infty(\mfkgs,\mbR)$ be such a fixed function on \fkgs, that
the corresponding \emn Hamiltonian vector field $\msg_Q$~ on \fkgs,\rref 5.1.37(2)~, is \emn
complete~. This means that there is a one parameter group $t\mapsto\mphi^Q_t\ (\mphi^Q_{t+s} =
\mphi_t^Q\circ\mphi^Q_s$\ for all $t,s\in\mbR$) of Poisson morphisms of $(\mfkgs;\mbs\mlam)$ the
derivative of which is $\msg_Q$. Remember that $\msg_Q$ is complete for any $Q$ in the case of
compact groups $G$, in which case the $Ad^*(G)$-orbits are compact. The tangent spaces $T_F\mfkgs\
(F\in\mfkgs)$ will be identified with the linear manifold \fkgs\ in the canonical way. Then we
have also the canonical identification $T_F^*\mfkgs = \mfkg$ of the cotangent spaces in any point
$F\in\mfkgs$ with the Lie algebra \fkg\ of $G$. {\bf Let \emn$f_\xi\in C^\infty(\mfkgs,\mbR)$~\
(for any $\xi\in\mfkg$) be the linear function}
\[\centerline{\emn$f_\xi: F\mapsto f_\xi(F) := F(\xi)$~}\].

\noidt Any element $\xi$ of the Lie algebra \fkg\ determines also a covector field on \fkgs:

\bequ\label{6.1.2(1)} \rd f_\xi: F\mapsto \rd_Ff_\xi =\xi\in\mfkg= T^*_F\mfkgs.\enqu

The Hamiltonian (contravariant) vector field corresponding to the Hamiltonian function $f_\xi$
coincides with the vector field $\msg_\xi$ determined by the flow

\bequ\label{6.1.2(2)} \mphi^\xi: (t;F)\mapsto \mphi^\xi_t F:= Ad^*(\exp(t\xi))F \enqu

\noidt on \fkgs. We have the relations:

\bequ\label{6.1.2(3)} \{h,f_\xi\}(F) = - F([\rd h,\rd f_\xi])=\rd_Ff_\xi(\msg_h) =
 -\rd_Fh(\msg_\xi),\quad h\in C^\infty(\mfkgs,\mbR),\enqu

\noidt where $\msg_ h$ is the Hamiltonian vector field corresponding to the Hamiltonian function
$h$, cf.\rref 5.1.37(2)~ and\rref 5.1.37(3)~.

\pt\label{6.1.3}\rm {\bf Let \emn$g_Q : \mbR\times\mfkgs\rarw G,\ (t;F)\mapsto g_Q(t,F)$~ be a
function determining the Hamiltonian flow} $\mphi_t^Q$ with the help of the action $\mphi_G:=
Ad^*(G)$ in the following sense:

\bequ\label{6.1.3(1)} Ad^*(g_Q(t,F))F = \mphi_t^Q(F):=\mphi_t^QF,\ \text{for all}\ t\in\mbR, \
\text{and for all}\ F\in\mfkgs.\enqu

\noidt Such functions $g_Q$ exist due to $\mphi^Q$-invariance of the maximal integral submanifolds
of $\mphi_G$ (i.e. the orbits of $Ad^*(G)$) \wrt\ any Hamiltonian flow. Let us assume
differentiability of $g_Q$ {\bf and set} \ind{$\beta^Q_F$}

\bequ\label{6.1.3(2)} \beta^Q_F :=\left.\frac{\rd}{\rd t}\right|_{t=0}g_Q(t,F),\ \text{for all}\
F\in\mfkgs.\enqu \noidt A necessary condition for fulfilment of\rref 6.1.3(1)~ is the fulfilment
of
\bequ\label{6.1.3(3)}  F([\beta^Q_F, \eta])=\rd_FQ(\msg_\eta)\ (= -\Omega_F(\msg_Q,\msg_\eta) =
- \rd_F f_\eta(\msg_Q)),\quad \eta\in\mfkg,\ F\in\mfkgs,\enqu

\noidt (cf.\rref 6.1.2(3)~), where $\Omega$ is the standard \emn Kirillov-Kostant symplectic form~
on $\mfkgs$, since the following relation is valid:

\bequ\label{6.1.3(4)} \left.\frac{\rd}{\rd
t}\right|_{t=0}Ad^*(g_Q(t,F))F(\eta)=-F([\beta^Q_F,\eta]),\quad \eta\in\mfkg, F\in\mfkgs. \enqu

\noidt If we require, in addition to\rref 6.1.3(3)~, fulfilment of the following '\emm cocycle
identities~':

\bequ\label{6.1.3(5)} g_Q(s,\mphi_t^QF)g_Q(t,F)=g_Q(t+s,F),\ g_Q(0,F)\equiv e,\enqu

\noidt for all $t,s\in\mbR$ and all $F\in\mfkgs$ (with $e :=$ the identity of $G$), then the
condition\rref 6.1.3(3)~ will be also sufficient for the validity of\rref 6.1.3(1)~. {\bf Let
\emn$\beta^\circ: F\mapsto \beta^\circ_F\in\mfkg$~ be any differentiable function on \fkgs\
satisfying}

\bequ\label{6.1.3(6)} F([\beta^\circ_F,\eta])=0,\quad \text{for all}\ F\in\mfkgs,\
\eta\in\mfkg.\enqu

Elements $\beta^\circ_F\in\mfkg$ determine one parameter subgroups of the stability groups of
$F\in\mfkgs$ for the coadjoint action $Ad^*(G)$, cf. Lemma \ref{3.2.4}.  If a given $\beta^Q_F$
satisfies\rref 6.1.3(3)~, then also the substitution of \bequ\label{6.1.3(7)}
\beta^{'Q}_F:=\beta^Q_F+\beta^\circ_F \enqu

\noidt in place of $\beta^Q_F$  in\rref 6.1.3(3)~ will give a valid equality. Let $\beta^Q_F$ be
an infinitely differentiable function of $F\in\mfkgs$ with values in \fkg\ satisfying\rref
6.1.3(3)~. The equation\rref 6.1.3(5)~ with the condition\rref 6.1.3(2)~ can be rewritten in the
form of a differential equation on the group manifold $G$:

\bequ\label{6.1.3(8)} \frac{\rd}{\rd t}\, g_Q(t,F) = T_e(R_{g_Q(t,F)})\beta^Q_{F_t},\quad \forall
t\in\mbR,\ F\in\mfkgs,\enqu

\noidt where $F_t := \mphi_t^Q(F)$, and $R_G$ is the right action of the group $G$ on itself:
$R_g(h) := hg\ (g,h\in G);\ T_e$ is the tangent mapping restricted to the tangent space $T_eG
=\mfkg$ of the group $G$ at the identity $e\in G,\ T_e(f):\ T_eG\rarw T_{f(e)}G$,

\[ \xi\mapsto T_e(f)\xi := f_*\xi :=\left.\frac{\rd}{\rd
t}\right|_{t=0}f(\exp(t\xi))\]

\noidt for any differentiable function $f\!: G\rarw G$. According to the general theory of
ordinary differential equations, there is a unique solution of\rref 6.1.3(8)~ with the initial
condition $g_Q(0,F) = e$. The solution $g_Q$ depends, however, on the choice of the covector field
$\beta^Q$ which is, according to\rref 6.1.3(7)~, nonunique in the general case.

The \emm cocycle $g_Q$~ is, as we shall see later, the basic dynamical object determining fully
the microscopic time evolutions in the mean-field theories of the considered type. Various choices
of $\beta^Q$ corresponding to the various possible choices of $\beta^\circ$ according to\rref
6.1.3(7)~ will lead to the same classical evolution $\mphi^Q$ of the subalgebra of classical
(intensive) quantities of the extended algebra of quantal observables of the infinite system. The
time evolutions of local (microscopic) observables corresponding to various choices of
$\beta^\circ$ in\rref 6.1.3(7)~ are, however, mutually different. We shall see that the {\bf
thermodynamic limits described in \ref{6.1.1} correspond to the choice} \ind{$\beta^Q_F$}

\bequ\label{6.1.3(9)} \beta^Q_F := \rd_FQ,\quad F\in\mfkgs.\enqu

\noidt If we write $Q(F)$ in the terms of coordinate functions $F_j := F(\xi_j)$ as in
\ref{6.1.1}, then we have

\bequ\label{6.1.3(10)} \rd_FQ =\sum_{j=1}^n \frac{\partial Q(F)}{\partial F_j}\,\xi_j
\in\mfkg.\enqu

\noidt Let the structure constants of $\mfkg$ in the basis $\{\xi_j\}$ are $c^j_{kl}\in\mbR$, i.e.

\bequ\label{6.1.3(11)} [\xi_k,\xi_l] = c^j_{kl}\xi_j.\enqu

\noidt Then we have for the Poisson bracket of two classical Hamiltonians $Q_1$ and $Q_2$ the
expression (called also the \emn Berezin bracket~):

\bequ\label{6.1.3(12)} \{Q_1,Q_2\}(F) := - F([\rd_FQ_1,\rd_FQ_2]) = - c^j_{km}\,\frac{\partial
Q_1}{\partial F_k}\, \frac{\partial Q_2}{\partial F_m}\, F_j.\enqu

\pt\label{6.1.4}\rm Let us describe here, in a heuristic manner, the basic idea leading to the
definition of the time evolutions $\tau^Q$ mentioned in \ref{6.1.1} which will be described in
Sec.\ref{sec;6.3} in details. It will be also shown in Sec.\ref{sec;6.3} that the evolutions
obtained from the thermodynamic limits in the 'polynomial cases' (mentioned in \ref{6.1.1} and
investigated in Sec.\ref{sec;6.2}) are special cases of the general definition of $\tau^Q$ based
on the following general ideas.

The \emn cocycle $g_Q$~ reproduces an arbitrary classical Hamiltonian evolution on the Poisson
mani\-fold $(\mfkgs,\mbs\mlam;Ad^*(G))$ (since $Q$ is an arbitrary Hamiltonian function) via the
given (fixed!) action $Ad^*(G)$, cf.\rref 6.1.3(1)~. We have given an action
$\msg(G)\in\maut\mfkA$ and also the corresponding dual action $\msg^*(G)$ on the set $\mS(\mfkA)$\
of states on \fkA, cf.\rref 5.1.10(4)~. We have also a canonical decomposition of an arbitrary
state $\mome\in\mS(\mfkA)$ into the states $\mome_m$ corresponding to classical phase space points
$m\in \mcl M$, namely\rref 5.1.38(1)~, resp. the corresponding statement in \ref{5.2.11}. For
$\mome\in\mS_\mfkg := p_G\mS(\mfkA)$, the states $\mome_m$ lying in the support of the
corresponding measure $\hat{\mu}_\mome$ on $\mS(\mfkA^{**})$ can be indexed by $F_m\in\mfkgs$,
where the classical measure on \fkgs\ corresponding to the state $\mome_m\in\mcl E_\mfkg$ is
concentrated on the one point set $\{F_m\}$, cf. \ref{5.1.36} and \ref{5.1.39}. Hence we can use
the family of mappings

\bequ\label{6.1.4(1)} t\mapsto \msg^*(g_Q(t,F_m)),\ t\in\mbR,\ m\in \mcl M, \enqu

\noidt for a definition of time translations of the states $\mome_m$. Such a definition makes
sense since the projection measure $E_\mfkg$ (:= the G-macroscopic limit of the system
$(\mfkA;\msg(G))$ in $(\mfkgs,\mbs\mlam;Ad^*(G))$, \ref{5.2.10}\,) is $G$-equivariant,\rref
5.2.3(3)~, what implies that the classical point-measure corresponding to
$\msg^*(g_Q(t,F_m))\mome_m\in \mcl E_\mfkg$ is concentrated on $Ad^*(g_Q(t,F_m))F_m
=\mphi_t^Q(F_m)\in \mfkgs$; hence the cocycle identity\rref 6.1.3(5)~ can be used to prove the
group property of mappings\rref 6.1.4(1)~. A heuristic definition of the time evolution $\tau^Q$
is then given with the help of the decomposition\rref 5.1.38(1)~ by the formula:

\bequ\label{6.1.4(2)} \mome(\tau^Q_t(x)) :=\int_\mcl M\msg^*(g_Q(t,F_m))\mome_m(x)\,\mu_\mome(\rd
m),\quad \forall t\in\mbR,\ \mome\in\mS_\mfkg.\enqu

We shall see in Sec.\ref{sec;6.3} that this intuitive construction leads to a rigorously defined
group $\tau^Q$ of \autm s of a \Csa\ of the \Wa\ $p_G\mfkA^{**}$ containing the algebra \fkA\ as
well as an algebra $\mfk N^c$ of classical observables in a natural manner. The algebra $\mfk N^c$
is then $\tau^Q$-invariant: $\tau^Q_\mbR(\mfk N^c) = \mfk N^c,$ contrary to the algebra \fkA\ (in
general case).

\begin{rem}\label{6.1.5}\rm The general definition of mean-field time evolutions
$\tau^Q$ based on the formula\rref 6.1.4(2)~ depends on a topology determined by the subset
$\mS_\mfkg := p_G\mS(\mfkA)$ of states on \fkA\ (and their canonical normal extensions to
$\mfkA^{**}$), so called '\emn classical states~'. The reason why we cannot use the set of all
states \cS(\fkA)\ for the definition of $\tau^Q$ can be seen from the thermodynamic limits of
polynomial interactions described in Sec. \ref{sec;6.2}: In the representations of \fkA\
containing the GNS-representations of states $\{\mome:\ \mome(p_G)\neq 1\}$ as their
subrepresentations the thermodynamic limits of the local evolutions $\tau^N$ do not exist for a
general $Q$. This fact can be seen from the definition of the projector $p_G$ in\rref 5.1.29(3)~
as well as from considerations in Sec.\ref{sec;6.2}. Although the resulting (algebraic) concept of
$\tau^Q$ can be used in certain cases to a definition of time evolution of all states on \fkA,
such a definition scarcely can be considered as a physically correct consequence of the given
interaction $Q$. This interaction does not lead to any reasonable (from the point of view of
physics) time evolution of states \ome\ of the infinite system, the central supports $s_\mome$ of
which are orthogonal to $p_G: s_\mome p_G=0,$ i.e. $\mome(p_G)=0.$ Since the set $\mS_\mfkg =
p_G\mS(\mfkA)$ is $\tau^Q$-invariant (as will be clear later), the time evolution of states
$\mome\in(I-p_G)\mS(\mfkA)$, where $I$ is the identity of $\mfkA^{**}$, can be determined
arbitrarily with a help of some group $\tau_\mbR\subset\maut(I-p_G)\mfkA^{**}$. The group
$\tau_\mbR$ has nothing to do, in a general case, with the evolution $\tau^Q$. For special choices
of the function $Q$, however, the evolution $\tau^Q$ can be defined on a larger subalgebra of
$\mfkA^{**}$ than $p_G\mfkA^{**}$, hence also an evolution of a set of states larger than
$\mS_\mfkg$ can be defined in a natural way, cf. also \cite[Sec.II.C]{bon1}. This can be seen on
the following (seemingly trivial) example.\end{rem}

\begin{exm}\label{6.1.6} An important class of 'mean-field' evolutions is
obtained by choosing $Q := f_\eta \in\mfkg^{**}, f_\eta(F) := F(\eta), \eta\in\mfkg$. We have in
this case

\bequ\label{6.1.6(1)} g_Q(t,F)=g_\eta(t,F) := \exp(t\eta),\quad\forall F\in\mfkgs,\ t\in\mbR.\enqu

\noidt The corresponding time evolution is (due to the independence of $g_Q$ on $F\in\mfkgs$):

\bequ\label{6.1.6(2)} \tau_t^Q = \tau^\eta_t :=\msg(\exp(-t\eta))\in{}^*{\rm \=Aut}\,\mfkA,\
t\in\mbR.\enqu

\noidt This time evolution is 'representation independent' (contrary to the general case of an
arbitrary $Q$) and the definition of the evolution of an arbitrary state $\mome\in\mS(\mfkA)$ is
straightforward. Equally straightforward is the canonical extension of $\tau^\eta$ to the (equally
denoted) group $\tau^\eta\in\maut\mfkA^{**}.$ This evolution (for unbounded $X_\eta$, especially
that one obtained by the extension to $\mfkA^{**}$) is highly discontinuous, however, and some
appropriate continuity properties can be found in a restriction to a properly chosen subset of
states of \cS(\fkA)\ (this 'properly chosen set of states' will be possibly larger than
$p_G\mS(\mfkA)$).

The group $G$ in the cases of this example is a 'dynamical group' of the system $(\mfkA,\msg(G))$
containing the time-evolution one parameter group as the subgroup $\{\exp(-t\eta) : t\in\mbR
\}\subset G$.
\end{exm}

\section{Spin systems with polynomial local
Hamiltonians $Q^N$}\label{sec;6.2}

\pt\label{6.2.1}\rm Let us consider the system described in Sec. \ref{sec;5.1}: The \Ca\ of
quasilocal observables \fkA\ is the C${}^*$-inductive limit of the sequence of the von Neumann
algebras $\mfkA^N:=\mcl{L(H}_N),\ \mfkA:=\mfkA^\Pi.$ A compact Lie group $G$ acts on \fkA\ by the
subgroup $\msg(G) := \msg_G$ of \autm s of \fkA\ introduced in \ref{5.1.5}. It is assumed in this
section that the generators $X_\xi\ (\xi\in\mfkg)$ of the representation $U(G)$ in \H\ introduced
in \ref{5.1.3} are bounded operators. We shall use the notation of the subsections from
\ref{5.1.2} to \ref{5.1.28}; we shall write $\Pi$ for the set of all positive integers. It will be
convenient for definiteness and for some technical reasons to work in the subrepresentation
$s_G\pi_u$ of the universal representation $\pi_u$ of the algebra \fkA\ in the Hilbert space
$\mH_u$. The bidual $\mfkA^{**}$ is canonically identified with the bicommutant $\pi_u(\mfkA)''$
of $\pi_u(\mfkA)$  in $\mcl L(\mH_u)$, and $s_G\in\mfk Z:= \pi_u(\mfkA)'\cap \pi_u(\mfkA)''
\subset\mcl L(\mH_u)$ is defined in \ref{5.1.11}. The following considerations could be extended
to the larger representation $p_G\pi_u$, where $p_G\in\mfk Z$ is introduced in \ref{5.1.29}. Hence
we shall work in the framework of the von Neumann algebra $s_G\mfkA^{**}$ which is isomorphic with
the subalgebra $P_G\mfk B^\#$ of $\mcl L(\mH_\Pi)$ via the mapping $\rho_G$, cf. \ref{5.1.11}. The
quasilocal algebra \fkA\ will be identified with its representation $s_G\pi_u(\mfkA)$ in the
Hilbert space $s_G\mH_u$ or, equivalently, with the corresponding \Csa\  of the abstract \Wa\
$s_G\mfkA^{**}$. Remember that \fkA\ is simple, hence any of its nonzero representations as a \Ca\
is faithful.

Let us introduce notation for various elements and subsets of $s_G\mfkA^{**}$:
\begin{notat}
\label{6.2.2}

Let us denote:
\begin{itemize}

\item[(i)] \emn$E_\mfkg$~ {\bf denotes the projection measure} (G-measure, \ref{5.2.3})
on the linear space \fkgs\ generated by $E_\mfkg(F)\ (F\in\mfkgs)$ from \ref{5.1.13}; in the
notation of \ref{5.1.16} $E_\mfkg(B)= \mbs c(B)$ for any subset $B\equiv \mbs B$ of \fkgs.

\item[(ii)] $E_\mfkg(f):= \int f(F)\,E_\mfkg(\rd F)$ for any complex valued
function $f\in L^1(\mfkgs,\mu^\mome_\mfkg)$ for all $\mome\in\mS_*(s_G\mfkA^{**})$ := the normal
states on $s_G\mfkA^{**}$, i.e. the integral $E_\mfkg(f)$ is assumed to converge in the
$w^*$-sense.

\item[(iii)] \emn$\mfk B_0^N$~$:=\mfkA^N\cup\{X_{\xi K}:\ \xi\in\mfkg,\
K\in\Pi\}\subset s_G\mfkA^{**}$,\ if the generators $X_\xi\in\mLH$ are bounded, \ref{5.1.3}.

\item[(iv)] {\bf Let \emn$\mfk N^c$~ be the \Csa\ of $s_G\mfkA^{**}$ generated} by all
the elements $E_\mfkg(f)$ with uniformly bounded continuous $f\in C_b(\mfkgs,\mbC).$

\item[(v)] {\bf \emn$\mfk C^N$~:=\ the \Ca\ generated by $\mfkA^N$ and
$\mfk  N^c$;} $\mfk C^N$ is isomorphic to the $C^*$-tensor product $\mfk A^N\otimes\mfk N^c$, the
isomorphism being: $x\otimes z\mapsto xz\in\mfk C^N\ (x\in\mfkA^N,\ z\in\mfk N^c)$,\ cf
\cite[1.22]{sak1} and \cite[IV.4.7]{takesI}.

\item[(vi)] \emn\fk C~\ {\bf will denote the \Ca\ generated by} $\{\mfk C^N:\
N\in\Pi\}$; \fk C\ is isomorphic to the tensor product $\mfkA\otimes\mfk N^c$, cf. \ref{6.2.13}.
\end{itemize}
\end{notat}

\begin{notat}\label{6.2.3}\rm Let $\{\xi_j:j=1,2,\dots n\}$ be a
fixed basis of \fkg. Let

\bequ\label{6.2.3(1)} X_j^N:=|N|\,X_{jN} := \sum_{k=1}^{|N|} \pi_k(X(\xi_j)),\ X(\xi):=
X_\xi,\enqu

\noidt for any $N\in\Pi$, be the selfadjoint element of \fkA\ introduced in \ref{5.1.3} as a
selfadjoint operator on $\mH_\Pi$ and identified now with $s_G\pi_u(X_j^N)$. Let us denote

\bequ\label{6.2.3(2)} \textbf{b} := \max\{1+\|X(\xi_j)\|:j=1,2,\dots n:=\dim G\}.\enqu

We shall use the Einstein summation rule for the summation over repeated vector indices in \fkg\
and \fkgs. Let $c^m_{jk}$ be the structure constants of \fkg\ in the given basis:

\bequ\label{6.2.3(3)} [\xi_j,\xi_k]= c_{jk}^m\xi_m.\enqu

\noidt Then we have from\rref 5.1.2(1)~:

\bequ\label{6.2.3(4)} [X_j^K,X_k^K]= i\,c^m_{jk}X_m^K,\quad \text{for\ all}\ K\in\Pi.\enqu

Let $Q$ be a polynomial specified in \ref{6.1.1}, hence satisfying the property \ref{6.1.1}(SA).
Let $Q$ be written in the form of linear combination of {\bf $p$ monomials of the maximal degree
$q$} with the upper bound $M\geq 1$ of the absolute values of the coefficients. Let $Q^K$ be given
by\rref 6.1.1(1)~ for all $K\in\Pi$. Let us introduce the notation:
\begin{eqnarray}
\textbf{c} &:=& \max\{|c^m_{jk}|:\ j,k,m = 1,2,\dots n\};\label{6.2.3(5)}\\
a_N &:=& \max(n\textbf{c}; 2|N|\textbf{b}),\ N\in\Pi;\label{6.2.3(6)}\\
\textbf{b}(x)&:=& \max(\textbf{b}; \|x\|),\ x\in\mfkA.\label{6.2.3(7)}
\end{eqnarray}

\noidt We shall use the standard notation for the multiple commutators:

\bequ\label{6.2.3(8)} [y,x]^{(m+1)}:=[y,[y,x]^{(m)}],\ [y,x]^{(0)}:=x,\ [y,x]:= yx-xy,\enqu

\noidt for any $x,y\in\mfkA^{**}$. [We shall use also $|J|:=$ the number of elements of the set
$J$.]
\end{notat}

\begin{lem}\label{6.2.4}The following estimate is valid for any $x\in\mfk B_0^N$
and for all positive integers $N, K (\geq N), m$:

\bequ\label{6.2.4(1)}
\|[Q^K,x]^{(m)}\|<\frac{\textbf{b}(x)}{q}(m-1)!\,(Mpq^2\textbf{b}^{q-1}a_N)^m.\enqu
\end{lem}
\begin{proof} Each mu1tiple commutator in\rref 6.2.4(1)~ can be written in the
form of a finite linear combination of monomials $P^{(m)}$ in the variables $X_{jK}$ and $y_r$,
where $y_r\in\mfk B_0^N$ is of one of the forms of the multiple commutators occurring in the two
following formulas:

\bequ\label{6.2.4(2)} \|[X^N_{j_1},[X^N_{j_2},\dots [X^N_{j_r},x]\dots ]]\|\leq
(2\textbf{b}N)^r\|x\|,\ x\in\mfkA^N;\enqu

\bequ\label{6.2.4(3)} \|[X^K_{j_1},[X^K_{j_2},\dots [X^K_{j_r},X_{kL}]\dots]]\|\leq
(n\textbf{c})^r\textbf{b},\ L\in \Pi.\enqu

These estimates of $\|y_r\|$ are easy consequences of the definitions as well as of the
relations\rref 6.2.3(4)~. Let $r\in\mbZ_+$ be called the degree of any of the variables denoted by
$y_r$. Then the sum $\sum_j r_j$ of degrees of all the variables $y_{r_j}$ occurring in any of the
monomials $P^{(m)}$ is less or equal to $m$. The maximal degree of any of the monomials $P^{(m)}$
is $m(q-1)+1$, hence we have the estimate:
\bequ\label{6.2.4(4)} \|P^{(m)}\|\leq \textbf{b}(x)(a_N
\textbf{b}^{q-1})^m,\enqu

\noidt where we have used the fact that a variable $y_r$ of the form given in\rref 6.2.4(2)~
occurs in any of the monomials $P^{(m)}$ at most in the first power (what implies the first power
of \textbf{b}(x) in\rref 6.2.4(4)~), as well as the inequalities $a_N>\textbf{b}>1$ were used in
the derivation of\rref 6.2.4(4)~.

The maximal value of coefficients at the monomials $P^{(m)}$ is $<M^m$. The maximal number of
monomials $P^{(m)}$ occurring in the expression of $[Q^K,x]^{(m)}$ can be calculated recursively,
using the derivation property of the commutators. One has the identity
\bequ\label{6.2.4(5)}
[x_{j_1}\dots x_{j_q}, y_{k_1}\dots
y_{k_s}]=\\
 \sum_{i=1}^q \sum_{j=1}^s x_{j_1}\dots
x_{j_{i-1}}y_{k_1}\dots y_{k_{j-1}}[x_{j_i},y_{k_j}]y_{k_{j+1}}\dots y_{k_s}x_{j_{i+1}}\dots
x_{j_q},\enqu

\noidt in which the commutator of two monomials of degrees $q$ and $s$ is expressed as a sum of
$qs$ monomials of degree $q+s-1$  (some of the
monomials could be equal to zero). If $[Q^K,x]^{(m)}$ is a sum of $n_m$ monomials $const. P^{(m)}$
of the maximal degree $s_m := m(q-1) + 1$, then $[Q^K,x]^{(m+1)}$ is a sum of $n_{m+1}$ monomials,
where

\bequ\label{6.2.4(6)} n_{m+1}\leq n_mpqs_m\leq n_mmpq^2.\enqu

Since $n_1\leq pq,$ we obtain the estimate:

\bequ\label{6.2.4(7)} n_m\leq \frac{(m-1)!}{q}(pq^2)^m.\enqu

After the multiplication of the \rhs of\rref 6.2.4(4)~ by the \rhs of\rref 6.2.4(7)~ and by the
upper bound $M^m$ of the coefficients at $P^{(m)}$, we obtain the estimate\rref 6.2.4(1)~.
\end{proof}

\begin{lem}\label{6.2.5} Let us {\bf define}

\bequ\label{6.2.5(1)} \kappa_N\,:= (Mpq^2\textbf{b}^qa_N)^{-1},\ \text{for\ all}\ N\in \Pi.\enqu

\noidt Let $|t|\leq$ \emn$\kappa_N$~,\ $x\in\mfk B_0^N$\ for a given $N\in\Pi.$ Then:\nl

\noidt (i) The sums \bequ\label{6.2.5(2)} \tau^K_t(x):= e^{itQ^K}\,x\,e^{-itQ^K}=\sum_{m=0}^\infty
\frac{(it)^m}{m!}[Q^K,x]^{(m)},\ K\in\Pi,\enqu

\noidt are convergent in the norm-topology of \fkA, and this convergence is uniform on
$\{K:K\in\Pi\}\times\{t: |t|\leq \kappa_N\}\times\{x: x\in\mfk B_0^N,\|x\|\leq a\}$ for any
$a\in\mbR_+$. \nl

\noidt (ii) The following limits exist in $s_G\mfkA^{**}$:

\bequ\label{6.2.5(3)} \tau_t^Q(x) := s^*\hbox{-}\lim_{|K|\rarw\infty}\tau^K_t(x),\enqu

\noidt where the {\bf convergence is understood in the}
\emm$s^*(s_G\mfkA^{**},s_G\mfkA^*)$-topology~ generated by the \emm seminorms~
\emn$\hat{p}_\mome$~ and \emn$\hat{p}^*_\mome$~ for all $\mome\in\mS_*(s_G\mfkA^{**})$:

\bequ\label{6.2.5(4)} \hat{p}_\mome: x\mapsto \hat{p}_\mome(x):=\sqrt{\mome(x^*x)},\
\hat{p}_\mome^*: x\mapsto \hat{p}_\mome^*(x):=\sqrt{\mome(xx^*)}.\enqu
\end{lem}

\begin{proof} The estimates\rref 6.2.4(1)~ are independent of $K\in\Pi$
and the corresponding majorizing power series for\rref 6.2.5(2)~ is uniformly convergent on the
product of the disc $\{t: |t|\leq \kappa_N, t\in\mbC\}$ and the ball $\{x: x\in\mfk B_0^N,
\|x\|\leq a\}$ for any nonnegative $a$. This proves (i). The definition of $s_G$ in \ref{5.1.11}
implies the existence of the limits

\bequ\label{6.2.5(5)} X_{\xi\Pi}:= s^*\hbox{-}\lim_{|K|\rarw\infty} X_{\xi K} =E_\mfkg(f_\xi)\in
s_G\mfkA^{**},\ \xi\in\mfkg, \enqu

\noidt what implies, in turn, together with the uniform boundedness in $K\in\Pi$ of the multip1e
commutators in\rref 6.2.5(2)~, the existence of the limits

\bequ\label{6.2.5(6)} s^*\hbox{-}\lim_{|K|\rarw\infty}[Q^K,x]^{(m)}\in s_G\mfkA^{**}.\enqu

The statement (i) together with these facts imply (ii).
\end{proof}

\begin{lem}\label{6.2.6} {\bf Let \emn$\mfk B^N$~ be the \Csa\ of \fkA\ generated by
$\mfk B_0^N$.} Each of the mappings $\tau_t^Q: \mfk B_0^N\rarw s_G\mfkA^{**}\ (|t|\leq \kappa_N)$
can be extended to a unique ${}^*$-homomorphism of the \Ca\ $\mfk B^N$ into $s_G\mfkA^{**}$.
\end{lem}
\begin{proof} The mappings $\tau_t^K$ are inner automorphisms of \fkA, and
their canonical extensions to $\mfkA^{**}$ leave the center $\mfk Z$ elementwise invariant. Hence,
we can consider $\tau_t^K$ as (inner) automorphisms of $s_G\mfkA^{**}$:

\bequ\label{6.2.6(1)} \tau_t^K \in {}^*\text{-Aut}\ {s_G\mfkA^{**}},\quad \text{for all}\
t\in\mbR,\ K\subset\Pi.\enqu

The properties of the $s^*$-limit imply that $\tau_t^Q\ (|t|\leq \kappa_N,\ t\in\mbR)$ are
${}^*$-homomorphisms of the symmetric set $\mfk B_0^N$ into $s_G\mfkA^{**}$, as well as they are
${}^*$-homomorphisms of the minimal ${}^*$-algebra in \fkA\ containing $\mfk B_0^N$ into
$s_G\mfkA^{**}$. The obvious norm-boundedness of these homomorphisms gives by continuity the
wanted (equally denoted) extensions $\tau_t^Q$.
\end{proof}

\noidt {\bf Note:} The values $\tau_t^Q(x)$ can be calculated according to the formula\rref
6.2.5(3)~ for all $x\in\mfk B^N$. This is a consequence of the norm-continuity of
$C^*$-homomorphisms, and it is easily verified by an elementary calculation.

\begin{lem}\label{6.2.7} Let $|t|\leq \kappa_1,\ \xi\in\mfkg,\
E_\mfkg(f_\xi)=X_{\xi\Pi}\in s_G\mfkA^{**}$, cf. \ref{6.2.2}\ (ii). Then the limits

\bequ\label{6.2.7(1)} \tau_t^Q(E_\mfkg(f_\xi)) := s^*\hbox{-}\lim_{L\rarw\infty}\tau_t^Q(X_{\xi
L}) \enqu

\noidt exist.
\end{lem}
\begin{proof} One has

\bequ\label{6.2.7(2)} \tau_t^Q(X_{\xi L})=\sum_{m=0}^\infty\frac{(it)^m}{m!} \,
s^*\hbox{-}\lim_{K\rarw\infty} [Q^K,X_{\xi L}]^{(m)},\enqu

\noidt and the bounds\rref 6.2.4(1)~ give the estimates independent of $K$ and $L$. After the
substitution of $x := X_{\xi L}$ into the sum in\rref 6.2.5(2)~, this sum is norm-convergent
uniformly in $(K;L)\in \Pi\times \Pi.$  Hence we have

\bequ\label{6.2.7(3)} \tau_t^Q(E_\mfkg(f_\xi))=\sum_{m=0}^\infty s^*\hbox{-}\lim_L\,
s^*\hbox{-}\lim_K\, [Q^K,X_{\xi L}]^{(m)}\frac{(it)^m}{m!},\enqu

\noidt cf. also\rref 6.2.5(5)~ and\rref 6.2.5(6)~, and the limit\rref 6.2.7(1)~ exists (cf. also
\cite[Proposition 3.5]{bon1}).
\end{proof}

\pt\label{6.2.8}\rm It will be shown next that the elements $E_\mfkg(f_\xi)\ (\xi\in\mfkg)$ of the
algebra $\mfkA^{**}$ {\bf generate the abelian \Ca\ \emn$\mfk N^c$~ of (bounded continuous)
classical observables,} cf. \ref{6.2.2}(iv), given on the support of $E_\mfkg$ in \fkgs. We shall
show after this that the transformations $\tau_t^Q$ in\rref 6.2.7(1)~ leave this \Ca\ invariant,
and that their unique extension for all $t\in\mbR$ reproduces the classical flow $\mphi^Q$,
\ref{6.1.2}, restricted to the support $\supp E_\mfkg$, \ref{5.2.3}. These results will lead to a
natural definition of the {\bf unique extension of} \emn$\tau_t^Q:\mfkA^N\rarw s_G\mfkA^{**}$~ for
all $t\in\mbR$, such that these mappings together with the mappings\rref 6.2.7(1)~ leave the
tensor product $\mfk C^N=\mfkA^N\otimes\mfk N^c$, \ref{6.2.2}(v), invariant, and have a unique
extension to a (equally denoted) one parameter group of \autm s of this composite quantal
($\mfkA^N$) and classical ($\mfk N^c$) system.

Let $\mphi:\mfkgs\rarw \mfkgs$ be a Poisson automorphism,\rref 5.2.2(3)~, leaving all the
$Ad^*$-orbits invariant. Then, using the bicontinuity of $\mphi$ and the $G$-equivariance of the
$G$-measure $E_\mfkg$, one can prove that the {\bf $s_G\mfkA^{**}$-valued function
\emn$\hat{\mphi}E_\mfkg$~ of Borel subsets} $B\subset \mfkgs$,

\bequ\label{6.2.8(1)} \hat{\mphi}E_\mfkg: B\mapsto \hat{\mphi}E_\mfkg(B) :=
E_\mfkg(\mphi^{-1}B),\enqu

\noidt is again a projection-valued mesure with the same support:

\bequ\label{6.2.8(2)} \supp\hat{\mphi}E_\mfkg = \supp E_\mfkg.\enqu.

\begin{prop}\label{6.2.9}Let $E_\mfkg$ and $\mphi$ be as above. Then the mapping

\bequ\label{6.2.9(1)} E_\mfkg:f\mapsto E_\mfkg(f) :=\int f(F)\,E_\mfkg(\rd F),\ f\in C(\supp
E_\mfkg),\enqu

\noidt introduced in \ref{6.2.2}(ii) is a $C^*$-isomorphism of the commutative \Ca\ of continuous
complex valued functions $C(\supp E_\mfkg)$ on the compact subset $\supp E_\mfkg$ of \fkgs\
($X_\xi$'s are now bounded!) onto $\mfk N^c$.

The \Ca\ $\mfk N^c$ is generated by the finite set $E_\mfkg(f_{\xi_j}),\ j = 1,2,\dots n$ of its
elements ($\xi_j$'s form a basis of \fkg). The mapping

\bequ\label{6.2.9(2)} \mphi^*:f\mapsto \mphi^*f,\ with\ \mphi^*f(F):= f(\mphi F),\enqu

\noidt restricted to $f\in C(\supp E_\mfkg)$ is a ${}^*$-automorphism of $C(\supp E_\mfkg)$. One
has

\bequ\label{6.2.9(3)} \hat{\mphi}: E_\mfkg(f)\mapsto\hat{\mphi}(E_\mfkg(f)) :=\hat{\mphi}
E_\mfkg(f)=E_\mfkg(\mphi^*f),\ f\in C(\mfkgs),\enqu

\noidt and the mapping $\hat{\mphi}$ in\rref 6.2.9(3)~ is a ${}^*$-automorphism of $\mfk N^c$.
\end{prop}
\begin{proof} Since $\supp E_\mfkg$ is compact (due to the compactness of
spectra of all the $X_\xi$'s), the function set $C(\supp E_\mfkg)$ is a \Ca\ {\bf generated by
polynomials in the variables} \emn$F_j := F(\xi_j) = f_{\xi_j}(F)$~ according to the classical
Weierstrass theorem. The ${}^*$-morphism property of $E_\mfkg$ in\rref 6.2.9(1)~ is a consequence
of the standard functional calculus of normal operators determined by a projection measure. One
can show that if $f(F_0)\neq 0$ for some $F_0\in \supp E_\mfkg$ and a continuous $f$, then
$E_\mfkg(f)\neq 0$, and this implies that the mapping $E_\mfkg$ in\rref 6.2.9(1)~ is the
$C^*$-isomorphism of $C(\supp E_\mfkg)$ onto $\mfk N^c$.

The mapping $\mphi^*$ is a norm preserving ${}^*$-morphism of $C(\supp E_\mfkg)$ into itself,
hence, it is a ${}^*$-automorphism.

The automorphism property of $\hat{\mphi}$ in\rref 6.2.9(3)~ is then a consequence of the
relation\rref 6.2.8(2)~, since both the mappings $E_\mfkg^{-1}: \mfk N^c\rarw C(\supp E_\mfkg)$
and $\hat{\mphi}E_\mfkg: C(\supp E_\mfkg)\rarw \mfk N^c$ are ${}^*$-iso\-morphisms, and we have:

\bequ\label{6.2.9(4)} \hat{\mphi}(E_\mfkg(f))=\hat{\mphi}E_\mfkg\circ E_\mfkg^{-1}(E_\mfkg(f)),\
f\in C(\supp E_\mfkg).\enqu

\noidt The equality in\rref 6.2.9(3)~ can be obtained from\rref 6.2.8(1)~ and the integral
representation\rref 6.2.9(1)~. This concludes the proof.
\end{proof}
\begin{prop}\label{6.2.10} The mappings $\tau^Q_t$ introduced in\rref 6.2.7(1)~
leave the algebra $\mfk N^c$ invariant. The family $\tau^Q$ has a unique extension to a strongly
continuous one parameter group of ${}^*$-automorphisms of $\mfk N^c$. This group satisfies the
equality

\bequ\label{6.2.10(1)} \tau^Q_t(E_\mfkg(f))= E_\mfkg(\mphi_t^{Q*}f),\  f\in C(\supp E_\mfkg),\enqu

\noidt where $\mphi^Q$ is the classical flow corresponding to the Hamiltonian function $Q$,
\ref{6.1.2}.
\end{prop}

\begin{proof} The classical flow $\mphi^Q$ forms a group of
$Ad^*$-orbits-preserving Poisson automorphisms of \fkgs. According to \ref{6.2.9}, the right side
of\rref 6.2.10(1)~ defines a one parameter group of ${}^*$-automorphisms of $\mfk N^c$. The strong
continuity of this group (i.e. the continuity in the norm of all the functions $t\mapsto
E_\mfkg(\mphi^{Q*}_tf))$ follows from the differentiability (hence continuity) of

\bequ\label{6.2.10(2)} \mphi^Q: (F;t)\mapsto \mphi^Q_t(F), \enqu

\noidt what is uniformly continuous on compacts in $\mfkgs\times\mbR$\ (\fkgs\ is endowed by the
linear space topology), as well as from the norm-continuity of the isomorphism $E_\mfkg$. Hence,
it suffices to prove the validity of the equation\rref 6.2.10(1)~ for small $t$.

Let us calculate the limits in\rref 6.2.7(3)~. We intend to prove

\bequ\label{6.2.10(3)} s^*\hbox{-}\lim_L\, s^*\hbox{-}\lim_K i^m[Q^K,X_{\xi L}]^{(m)} = E_\mfkg
(\{Q,f_\xi\}^{(m)}),\ \xi\in\mfkg,\ m\in\mbZ_+. \enqu

\noidt Here $\{Q,f\}^{(0)}:= f, \{Q,f\}^{(m+1)} := \{Q,\{Q,f\}^{(m)}\}$, and $\{Q,f\}$ is the
classical Poisson bracket on the Poisson manifold \fkgs. The limits in\rref 6.2.10(3)~ do exist,
cf. \ref{6.2.5}. The local Hamiltonians $Q^K$ are polynomials of the form\rref 6.1.1(1)~ and the
commutators as well as the Poisson brackets are bilinear, antisymmetric, satisfying the Jacobi
identity and the derivation property: $[a,bc] = [a,b]c + b[a,c]$.

We have also

\bequ\label{6.2.10(4)} s\hbox{-}\lim_L\, s\hbox{-}\lim_K i[X^K_\xi,X_{\eta L}]=s\hbox{-}\lim_L
X_{[\eta,\xi]L}=E_\mfkg(\{f_\xi,f_\eta\}),\ \xi,\eta\in\mfkg.\enqu

\noidt what can be seen from\rref 5.1.2(5)~,\rref 6.2.5(5)~ and\rref 1.3.7(4)~. The morphism
properties of $E_\mfkg$ then lead to the formula\rref 6.2.10(3)~.

Inserting\rref 6.2.10(3)~ into\rref 6.2.7(3)~, we obtain

\bequ\label{6.2.10(5)} \tau^Q_t(E_\mfkg(f_\xi))=\sum_{m=0}^\infty \frac{t^m}{m!}\,
E_\mfkg(\{Q,f_\xi\}^{(m)}).\enqu

The estimates\rref 6.2.4(1)~ and the isometry of the mapping $E_\mfkg$ from\rref 6.2.9(1)~ give,
with the help of\rref 6.2.10(3)~, the norm-convergence (in the algebra $C(\supp E_\mfkg$)) of the
sum defining the element $f_{\xi t}\in C(\supp E_\mfkg)$:

\bequ\label{6.2.10(6)} f_{\xi t}(F):=\sum_{m=0}^\infty\frac{t^m}{m!}\{Q,f_\xi\}^{(m)}(F),\quad
F\in\supp E_\mfkg,\ |t|\leq \kappa_1.\enqu

\noidt The norm-continuity of the morphism $E_\mfkg$ then leads from\rref 6.2.10(5)~ to

\bequ\label{6.2.10(7)} \tau_t^Q(E_\mfkg(f_\xi)) = E_\mfkg(f_{\xi t}),\ \xi\in\mfkg,\ |t|\leq
\kappa_1:= (Mpq^2b^qa_1)^{-1}.\enqu

The derivative of the function $t\mapsto f_{\xi t}$ is, according to\rref 6.2.10(6)~:

\bequ\label{6.2.10(8)} \frac{\rd}{\rd t}f_{\xi t}(F)=\sum_{m=0}^\infty
\frac{t^m}{m!}\{Q,\{Q,f_\xi\}^{(m)}\}(F),\enqu

\noidt the series in\rref 6.2.10(8)~ being again absolutely and uniformly convergent in $F\in
\supp E_\mfkg$ and $|t|\leq \kappa_1$,\rref 6.2.4(1)~, i.e.

\bequ\label{6.2.10(9)} (t;F)\in \{u:\ u\in\mbR,\ |u|\leq \kappa_1 \}\times\supp E_\mfkg.\enqu

The classical Hamilton equations written in the form of Poisson brackets for the case of the
Hamiltonian function $Q$ with the flow $\mphi^Q$ have the form

\bequ\label{6.2.10(10)} \frac{\rd}{\rd u}f(\mphi^Q_uF)=\{Q,f\}(\mphi^Q_uF),\quad F\in\mfkgs,\
u\in\mbR.\enqu

\noidt Let us substitute $\mphi^Q_uF$ instead of $F$ into the formula\rref 6.2.10(8)~. From\rref
6.2.10(10)~ we obtain

\bequ\label{6.2.10(11)} \frac{\rd}{\rd t}f_{\xi t}(\mphi_u^QF)=\sum_{m=0}^\infty
\frac{t^m}{m!}\,\frac{\rd}{\rd u}\{Q,f_\xi\}^{(m)}(\mphi_u^QF).\enqu

The uniform convergence in $u\in\mbR$ for any given $(t;F)$ from\rref 6.2.10(9)~ and the known
theorem on the differentiation of series of functions lead to the equality:

\bequ\label{6.2.10(12)} \frac{\rd}{\rd t}f_{\xi t}(\mphi_u^QF)=\frac{\rd}{\rd u}f_{\xi
t}(\mphi_u^QF)=\{Q,f_{\xi t}\}(\mphi^Q_uF),\enqu

\noidt where the second equality was obtained by an application of\rref 6.2.10(10)~. Setting $u =
0$ in\rref 6.2.10(12)~ and comparing with\rref 6.2.10(10)~ we get:

\bequ\label{6.2.10(13)} f_{\xi t}(F)=f_{\xi 0}(\mphi_t^QF)\equiv\
f_\xi(\mphi_t^QF)=\mphi^{Q*}_tf_\xi(F),\enqu

\noidt since $f_{\xi 0} = f_\xi$ according to\rref 6.2.10(6)~. Insertion of $f_{\xi t}$ from\rref
6.2.10(13)~ into\rref 6.2.10(7)~ gives\rref 6.2.10(1)~ with $f := f_\xi\ (\xi\in\mfkg)$. The
algebra $C(\supp E_\mfkg)$ is generated by $f_\xi$'s, and $\mphi_t^{Q*}$ is a ${}^*$-isomorphism
of $C(\supp E_\mfkg)$,\rref 6.2.9(2)~. The norm-continuity of $C^*$-morphisms gives then the
validity of\rref 6.2.10(1)~ for the general $f\in C(\supp E_\mfkg)$.
\end{proof}

\begin{lem}\label{6.2.11}The mappings $\tau_t^Q\ (|t|\leq \kappa_N)$
defined in \ref{6.2.5}(ii) map the \Ca\ $\mfkA^N$ into the \Ca\ $\mfk C^N$ (which is generated in
$s_G\mfkA^{**}$ by $\mfkA^N$ and $\mfk N^c$).
\end{lem}

\begin{proof} We can write the definition of $\tau_t^Q\ (|t|\leq \kappa_N)$
on $\mfkA^N$,\rref 6.2.5(2)~ and\rref 6.2.5(3)~, in the form

\bequ\label{6.2.11(1)} \tau_t^Q(x):= \sum_{m=0}^\infty
\frac{t^m}{m!}\,s^*\hbox{-}\lim_K\,[i\,Q^K,x]^{(m)},\ x\in\mfkA^N.\enqu

\noidt Each multiple commutator in\rref 6.2.11(1)~ can be expressed in the form of a polynomial in
the variables $X_{\xi K}$
 and some of the variables $y_s$ of the form, cf. also\rref
 6.2.4(2)~ and\rref 6.2.4(3)~:

 \bequ\label{6.2.11(2)} y_s:=[X^K_{j_1},[X^K_{j_2},\dots
 [X^K_{j_s},x]\dots ]]\in\mfkA^N,\ K\in\Pi,\enqu

 \noidt with the coefficients independent of $K$. Due to\rref
 6.2.5(5)~ and the independence of any $y_s$ of $K$, the strong
 limits in\rref 6.2.11(1)~ are elements of $\mfk C^N$.
 The norm convergence of the sum
in \rhs\ of\rref 6.2.11(1)~ and the closeness of $\mfk C^N$  in the norm-topology give then the
result.
\end{proof}

\begin{lem}\label{6.2.12} For any $x\in\mfkA$ and any $z\in\mfk N^c$,
the equality $xz = 0$ implies the validity of $\|x\|\cdot\|z\|=0.$
\end{lem}

\begin{proof} For $z\neq 0$, we have $z=E_\mfkg(f)$ with $|f(F_0)|\neq 0$
 for some
$f\in C(\supp E_\mfkg)$ and some $F_0\in\supp E_\mfkg$. Let, for the definiteness, be $f(F_0)> 0$.
Then there is a subset $B_0\subset \mfkgs$ such that $E_\mfkg(B_0)\neq 0$ and
$f(F)>\frac{1}{2}f(F_0)$ for all $F\in B_0$.
 Since $\mfk N^c$ is in the commutant of \fkA\ in
$s_G\mfkA^{**}$, the product of the positive (i.e. nonnegative) operator $x^*x\in\mfkA$ with the
positive operator $(E_\mfkg(f)- \frac{1}{2}f(F_0))E_\mfkg(B_0)\in\mfk N^c$ is a nonnegative
operator in $\mfk C$. Then $xz = 0$ implies

\bequ\label{6.2.12(1)} 0\leq x^*x\,(E_\mfkg(f) - \frac{1}{2}f(F_0))E_\mfkg(B_0) = -
\frac{1}{2}\,f(F_0)\ x^*x\,E_\mfkg(B_0).\enqu

\noidt Hence we have $x\,E_\mfkg(B_0) = 0$. The mapping: $x\mapsto x\,E_\mfkg(B_0)$ is a nonzero
(nondegenerate) representation of the simple \Ca\ \fkA\ in $\mfkA^{**}$, hence $x=0$.
\end{proof}

\begin{lem}\label{6.2.13} {\bf Let \emn$\mfkA^N\otimes\mfk N^c$~ and
\emn$\mfkA\otimes\mfk N^c$~ be the} \emm$C^*$-products~ (uniquely defined, since $\mfk N^c$ is
abelian, \cite[1.22.5.]{sak1}), with the canonical inclusion $\mfkA^N\otimes\mfk
N^c\subset\mfkA\otimes\mfk N^c$. Let $\mlam_0^{-1}$ be the homomorphism of $\mfkA\otimes\mfk N^c$
into $\mfk C$, \ref{6.2.2}, determined by the association:

\bequ\label{6.2.13(1)} \mlam_0^{-1}:\ \sum_j x_j\otimes z_j\mapsto \sum_j x_jz_j\in\mfk C,\
x_j\in\mfkA,\ z_j\in\mfk N^c.\enqu

\noidt Then $\mlam_0^{-1}$ {\bf can be extended to a unique ${}^*$-isomorphism}
\emn$\mlam_0^{-1}=: (\mlam_0)^{-1}$~ of the \Ca\ $\mfkA\otimes\mfk N^c$ onto $\mfk C$, the
restrictions of which to the subalgebras $\mfkA^N\otimes\mfk N^c\ (N\in\Pi)$ are
${}^*$-isomorphisms onto $\mfk C^N\ (N\in\Pi)$, cf. \ref{6.2.2}.
\end{lem}

\begin{proof} The existence of an isomorphism onto \fk C\
extending $\mlam_0^{-1}$ is a direct consequence of \cite[Exercise IV.2]{takesI}, due to our Lemma
\ref{6.2.12}. The uniqueness is the trivial consequence of the norm-continuity of
$C^*$-homomorphisms, since the finite sums in\rref 6.2.13(1)~ form dense sets in the corresponding
\Ca s. The same considerations are applicable to the restrictions to $\mfkA^N\otimes\mfk N^c$,
hence we have the assertions of the Lemma.
\end{proof}

\begin{lem}\label{6.2.14} {\bf Let \emn$\tau_K\ (K\in\Pi)$~, resp. \emn$\tau_c$~,
be a ${}^*$-homomorphism of $\mfkA^K$, resp. of $\mfk N^c$}, into $\mfk C^K$. Assume that
$\tau_c(\mfk N^c)\subset\mfk N^c$. Then there is a unique ${}^*$-homomorphism $\tau: \mfk
C^K\rarw\mfk C^K$ such that:

\bequ\label{6.2.14(1)} \tau(xz)=\tau_K(x)\tau_c(z),\quad\text{for all}\ x\in\mfkA^K,\ z\in\mfk
N^c.\enqu

\end{lem}

\begin{proof} Let $\mlam_0: xz\mapsto x\otimes z$ be the isomorphism of
$\mfk C^K$ onto $\mfkA^K\otimes\mfk N^c$ determined in \ref{6.2.13}. According to
\cite[IV.4.7.]{takesI}, there is a unique homomorphism $\tau_0$ of $\mfkA^K\otimes\mfk N^c$ into
$\mfk C^K$ such that

\bequ\label{6.2.14(2)} \tau_0(x\otimes z)=\tau_K(x)\tau_c(z), \quad x\in\mfkA^K,\ z\in\mfk
N^c.\enqu

\noidt Since the $C^*$-norm on $\mfkA\otimes\mfk N^c$ is a \emn cross norm~ (see
\cite[IV.]{takesI}), the ${}^*$-property of $\tau_0$ follows from the norm continuity and from the
${}^*$-property of $\tau_K$ and $\tau_c$. We shall define $\tau$ as the composition

\bequ\label{6.2.14(3)} \tau:= \tau_0\circ\mlam_0.\enqu

\noidt The uniqueness of $\tau$ is then a consequence of linearity and continuity in the
norm-topology.
\end{proof}

\begin{prop}\label{6.2.15} There is a unique family
$\tau^Q:=\{\tau_t^Q\ ;|t|\leq \kappa_N,\ t\in\mbR\}$ of $C^*$-morphisms of $\mfk C^N$ into itself
such that their restriction  to $\mfkA^N\subset\mfk C^N$ is given by\rref 6.2.11(1)~, and their
restriction to $\mfk N^c\subset\mfk C^N$ is given by\rref 6.2.10(1)~. {\bf This family
\emn$\tau^Q$~ has a unique extension to an (equally denoted) one parameter group of
${}^*$-automorphisms of $\mfk C^N$},\ for any $N\in\Pi$.
\end{prop}
\begin{proof} After the identification of $\tau_K$ (resp. $\tau_c$)
from \ref{6.2.14} with $\tau_t^Q$ from\rref 6.2.11(1)~ (resp. with $\tau_t^Q$ from\rref
6.2.10(1)~) for any real $t:\ |t|\leq r_K\ (K\in\Pi)$, the wanted morphism $\tau_t^Q:\ \mfk
C^K\rarw \mfk C^K$ is obtained by its identification with $\tau$ from\rref 6.2.14(1)~. It suffices
to prove the group property of these morphisms $\tau_t^Q$ of $\mfk C^N$ into itself (with
$N\in\Pi$) for small $t\in\mbR$. Since the restrictions of $\tau^Q$ to $\mfk N^c$  form an
automorphism group of $\mfk N^c$, and the algebra $\mfk N^c$ is in the center of $\mfk C^N$, it
suffices to prove

\bequ\label{6.2.15(1)} \tau^Q_{t_1+t_2}(x)=\tau^Q_{t_1}(\tau^Q_{t_2}(x))\ \text{for\ all}\
x\in\mfkA^N,\enqu

\noidt and for all sufficiently small nonzero $t_j$\ (e.g., for all $t_j:\,
\max(|t_1|,|t_2|)<\frac{1}{2}\kappa_N$). For such $t_j$'s, we have according to \ref{6.2.5}(ii)
and\rref 6.2.11(1)~:

\bequ\label{6.2.15(2)} \tau^Q_{t_1}(\tau^Q_{t_2}(x))=
\tau^Q_{t_1}(s\hbox{-}\lim_{K\rarw\infty}\tau^K_{t_2}(x))= \sum_{m=0}^\infty\frac{(it_2)^m}{m!}\,
\tau^Q_{t_1}(s\hbox{-}\lim_{K\rarw\infty}\,[Q^K,x]^{(m)}),\enqu

\noidt where the norm continuity of $\tau_{t_1}^Q$ and the norm-convergence of the series were
used. (We write here $s\hbox{-}\lim$\ instead of $s^*\hbox{-}\lim$, where the
$s(s_G\mfkA^{**},s_G\mfkA^*)$-topology is generated by the seminorms $\hat{p}_\mome$ from\rref
6.2.5(4)~. This notation is used for brevity only; the existence and equality of both the limits
$s\hbox{-}\lim$ and $s^*\hbox{-}\lim$ is clear from the proof of \ref{6.2.5}.) Considering the
structure of the multiple commutators in\rref 6.2.15(2)~ according to the discussion in the proof
of \ref{6.2.11}, by the morphism property of $\tau^Q_{t_1}$ on $\mfk C^N$ as well as the
definition\rref 6.2.7(1)~ with\rref 6.2.10(1)~ we obtain:

\begin{eqnarray}
\label{6.2.15(3a)}\tau^Q_{t_1}(s\hbox{-}\lim_K[Q^K,x]^{(m)})&=&
s\hbox{-}\lim_K\tau^Q_{t_1}([Q^K,x]^{(m)})\\
\label{6.2.15(3b)}&=& s\hbox{-}\lim_K\,[\tau^Q_{t_1}(Q^K),\tau^Q_{t_1}(x)]^{(m)}.
\end{eqnarray}

Since any ${}^*$-morphism $\tau^Q$ is a \emn contraction~, the bounds from \ref{6.2.4} are valid
also for the multiple commutators in\rref 6.2.15(3b)~. From the norm-convergence of the sums we
obtain consequently:

\begin{eqnarray}\label{6.2.15(4)}
\tau^Q_{t_1}(\tau^Q_{t_2}(x)) &=& s\hbox{-}\lim_K\sum_{m=0}^\infty\frac{(it_2)^m}{m!}\,
[\tau^Q_{t_1}(Q^K),\tau^Q_{t_1}(x)]^{(m)}\nonumber \\
&=& s\hbox{-}\lim_K \tau^Q_{t_1}(\tau^K_{t_2}(x)).
\end{eqnarray}

\noidt One has also

\bequ\label{6.2.15(5)} \tau^K_{t_2}(x)\in\mfk B^N\ \text{for all}\ x\in\mfkA^N,\ \text{and for
all}\ K\in\Pi.\enqu

\noidt Then, according to the Lemma \ref{6.2.6} and the formula\rref 6.2.5(3)~, one obtains:

\begin{eqnarray}\label{6.2.15(6)}
\tau^Q_{t_1}(\tau^K_{t_2}(x)) &=& s\hbox{-}\lim_L
\tau^L_{t_1}(\tau^K_{t_2}(x)) \nonumber \\
&=& s\hbox{-}\lim_L\sum_{k,m=0}^\infty\frac{(it_1)^k}{k!}\,\frac{(it_2)^m}{m!}\,
[Q^L,[Q^K,x]^{(m)}]^{(k)}.
\end{eqnarray}

The norms of the multiple commutators in\rref 6.2.15(6)~ for $L\geq K\geq N$ are bounded from
above according to the estimate (cf. also \ref{6.2.3})

\bequ\label{6.2.15(7)} \|[Q^L,[Q^K,x]^{(m)}]^{(k)}\|<\frac{\textbf{b}(x)}{q}(m+k-1)!\, (Mpq^2{\mbs
b}^{q-1}a_N)^{m+k},\enqu

\noidt what can be obtained by the considerations analogous to those used in the proof of
\ref{6.2.4}. Hence the sum in\rref 6.2.15(6)~ converges in norm, uniformly in
$(K;L)\in\Pi\times\Pi$ with $L\geq K\geq 1$. Then the continuity of the product of elements of a
\Wa\ in the s-topology leads to:
\begin{eqnarray}\label{6.2.15(8)}
s\hbox{-}\lim_{K\rarw\infty} \tau^Q_{t_1}(\tau^K_{t_2}(x))&=& s\hbox{-}\lim_K\sum_{k,m=0}^\infty
\frac{(it_1)^k}{k!}\,\frac{(it_2)^m}{m!}\,
[Q^K,[Q^K,x]^{(m)}]^{(k)}\nonumber\\
&=& s\=\lim_K\sum_{p=0}^\infty
\frac{(t_1+t_2)^p}{p!}\,[iQ^K,x]^{(p)} \nonumber \\
&=& s\=\lim_K \tau^K_{t_1+t_2}(x)=\tau^Q_{t_1+t_2}(x).
\end{eqnarray}

The relations\rref 6.2.15(4)~ and\rref 6.2.15(8)~ give the desired group property\rref 6.2.15(1)~
for all sufficiently small nonzero $t_1,\ t_2$, hence $\tau^Q_t\in\maut\mfk C^N$ (due to the
consequent invertibility of $\tau^Q_t$ on $\mfk C^N$), and $\tau^Q$ is a one-parameter group of
automorphisms of $\mfk C^N$ (for any given $N\in\Pi$).
\end{proof}

\begin{noti}\label{6.2.16} We have worked in this section in the
framework of the subalgebra $s_G\mfkA^{**}$ of the von Neumann algebra $\mfkA^{**}$. The {\bf only
properties of the projector} \emn $s_G\in\mfk Z$~ we have used in the previous considerations was
the existence of the limits $X_{\xi\Pi} := s^*\=\lim_N s_GX_{\xi N}$ for all $\xi\in\mfkg$ (here
the elements $X_{\xi N}\in\mfkA$ are identified with $\pi_u(X_{\xi N})$, cf. \ref{6.2.1}) as well
as the $\msg(G)$-invariance: $\msg(g)(s_G)=s_G$ for all $g\in G$. {\bf Any projector \emm
$s_\pi\in\mfk Z$~ with these two properties}, i.e. $s_\pi$ such that:

\nl \noidt (i)\ the limits $s^*\=\lim_N X_{\xi N}s_\pi$ exist in
$s^*(\mfkA^{**},\mfkA^*)$-topology for all $\xi\in\mfkg$,

\nl \noidt (ii)\ $s_\pi$ is $\msg(G)$-invariant: $\msg(g)(s_\pi)=s_\pi$ for all $g\in G$, \nl

\noidt could be used instead of $s_G$ in the considerations of this section. Such projectors form
a lattice in \fk Z\ with the maximal element $p_G$ defined in \ref{5.1.29}. The $G$-measure
corresponding to $p_G$ was introduced in \ref{5.1.33} and denoted by $E_\mfkg^\Pi$. Then the
$G$-measure used up to now in this section was $E_\mfkg=s_G E_\mfkg^\Pi$, and the $G$-measure
$E_G^\pi$ corresponding to {\bf another projector} \emn$s_\pi\in\mfk Z$~ satisfying (i) and (ii)
{\bf equals to} \emn $s_\pi E_\mfkg^\Pi$~. The algebra $\mfk N^c_\pi :=
E^\pi_\mfkg(C_b(\mfkgs,\mbC))$ corresponding to the projector $s_\pi$, hence also the quasilocal
algebra $\mfk C_\pi:=\mfkA\otimes\mfk N^c_\pi,$ depend nontrivially on the choice of $s_\pi$. If,
however, $s_G\leq s_\pi\leq p_G$, then $\mfk N^c_\pi$ is isomorphic to $\mfk N^c$. This is an
immediate consequence of the Proposition \ref{6.2.9} as well as of the following Lemma
\ref{6.2.17}.
\end{noti}

\begin{lem}\label{6.2.17} Let the projector $s_\pi\in\mfk Z$ (:=
the center of $\mfkA^{**}$) satisfy \ref{6.2.16}\,(i)+(ii). Let $s_G\leq s_\pi\leq p_G,$ and let
$E_\mfkg^\pi := s_\pi E_\mfkg^\Pi$. Then $\supp E_\mfkg^\pi=\supp E_\mfkg^\Pi (= \supp E_\mfkg$,
consequently).
\end{lem}

\begin{proof}: Let $sp(X_\xi)\subset \mbR\ (\xi\in\mfkg)$ be spectrum
of the bounded selfadjoint operator $X_\xi \in\mLH$. {\bf Let \emn{\rm conv}$(B)$~ be the} \emm
convex hull~ of the subset $B$ of a linear space. We have $X_{\xi N}\in\mfkA\ (N\in\Pi)$, hence
the {\bf spectrum \emn$sp(\pi(X_{\xi N}))$~ does not depend of the representation} $\pi$ of \fkA\
(\fkA\ is simple). From the construction of $X_{\xi\Pi}$ in \ref{5.1.7} and \ref{5.1.8} we obtain
successive1y: \ind{$sp(X_{\xi N})$}

\bequ\label{6.2.17(1)} sp(X_{\xi N})\subset {\rm conv}\left(sp(X_\xi)\right),\quad \xi\in\mfkg,\ N\in\Pi,\enqu

\noidt what can be seen from \cite[Theorem VIII.33]{R&S}; from the spectral resolution of $X_\xi$
with a help of \ref{5.1.8} one has

\bequ\label{6.2.17(2)} \{\mlam\in\mbC:\ \mlam=(\mphi,X_\xi\,\mphi),\ \|\mphi\|=1,\
\mphi\in\mH\}={\rm conv}(sp(X_\xi))\subset sp(X_{\xi\Pi});\enqu

\noidt hence by \cite[Theorem VIII.24]{R&S}:

\bequ\label{6.2.17(3)} sp(X_{\xi\Pi})={\rm conv}(sp(X_\xi)).\enqu

\noidt The equality\rref 6.2.17(3)~ is independent of such representations $\pi$ of \fkA\ in
which\rref 6.2.17(2)~ is valid, i.e. for

\bequ\label{6.2.17(4)} X_{\xi\pi} := s^*\=\lim_N s_\pi X_{\xi N}\in\mfkA^{**} \enqu

\noidt we have the implication:

\bequ\label{6.2.17(5)} {\rm conv}(sp(X_\xi))\subset sp(X_{\xi \pi})\imply sp(X_{\xi\pi})= {\rm
conv}(sp(X_\xi)).\enqu

We have $X_{\xi\Pi}:= X_{\xi\pi}$ for $s_\pi := s_G$ and the spectrum of $X_{\xi\pi}$ cannot
decrease with increasing $s_\pi$. This proves the conclusion of\rref 6.2.17(5)~ for all $s_\pi\geq
s_G,\ \xi\in\mfkg$. Hence the spectra of $X_{\xi\pi}$ are independent of $s_\pi$ for $s_G\leq
s_\pi\leq p_G.$ The construction of the projection measure $E_\mfkg^\pi$ according to\rref
5.1.31(3)~ and \ref{5.1.33} shows that $F\in\supp E_\mfkg^\pi$ implies $F(\xi)\in sp(X_{\xi\pi})$:

\bequ\label{6.2.17(6)} X_{\xi\pi}=\int F(\xi)\,E^\pi_\mfkg(\rd F)= E^\pi_\mfkg(f_\xi).\enqu

\noidt This formula shows also that $\mlam\in sp(X_{\xi \pi})$ implies the existence of such an
$F\in \supp E_\mfkg^\pi$ that $F(\xi)=\mlam$. We shall show in the next Lemma that $\supp E_\mfkg$
is a convex subset of \fkgs. Let

\bequ\label{6.2.17(7)} B_\mfkg :=\{F\in\mfkgs:\ F(\xi)\in{\rm conv}(sp(X_\xi)),\ \forall
\xi\in\mfkg\}.\enqu

\noidt The set $B_\mfkg$ is convex and closed in \fkgs. We have

\bequ\label{6.2.17(8)} \supp E_\mfkg^\pi\subset B_\mfkg\  \text{for\ any}\ s_\pi\geq s_G\
(s_\pi\leq p_G).\enqu

\noidt Let $B=\overline{B}={\rm conv}(B)\subset B_\mfkg$ be such that for any $\xi\in\mfkg$ the
following implication is valid:

\bequ\label{6.2.17(9)} \mlam\in {\rm conv}(sp(X_\xi))\imply \exists\ F\in B:F(\xi)=\mlam.\enqu

\noidt The set $B:= \supp E_\mfkg$, and also $B:= B_\mfkg$ has the property\rref 6.2.17(9)~.

Let $F_0\in\mfkgs$ does not belong to $B:\ F_0\not\in B$. Then, according to
\cite[Lemma(B.26)]{hew&ross}, there is an element of $\mfkg^{**} = \mfkg,\ \xi_0\in\mfkg$, such
that

\bequ\label{6.2.17(10)} \inf \{F(\xi_0):\ F\in B\}> F_0(\xi_0).\enqu

But from\rref 6.2.17(9)~ and from $B\subset B_\mfkg$ we see that $\{F(\xi): F\in B\} = {\rm
conv}(sp(X_\xi))$ for all $\xi\in\mfkg$, hence $F_0(\xi_0)\not\in {\rm conv}(sp(X_{\xi_0}))$, and
this implies that $F_0\not\in B_\mfkg$. We have proved that $B=B_\mfkg$, hence $\supp E_\mfkg =
B_\mfkg$. But

\bequ\label{6.2.17(11)} s_G\leq s_\pi\imply E_\mfkg \leq E^\pi_\mfkg \imply \supp E_\mfkg\subset
\supp E^\pi_\mfkg,\enqu

\noidt what with the help of\rref 6.2.17(8)~ gives now the desired result.
\end{proof}

\begin{lem}\label{6.2.18}
$\supp E_\mfkg$ is convex.
\end{lem}
\begin{proof} The projection measure $E_\mfkg$ introduced in \ref{6.2.2}.(i) is
built of its values $E_\mfkg(F)$,\rref 5.1.13(1)~, calculated on one point sets
$\{F\}\subset\mfkgs$. The measure $E_\mfkg$ is isomorphically mapped onto the measure $E_\mfkg^\#
:= \rho_G\circ E_\mfkg$ acting in the Hilbert subspace $P_G\mH_\Pi$ of the infinite (complete)
tensor product space $\mH_\Pi$, cf. \ref{5.1.11}. According to the definitions in \ref{5.1.7},\
\ref{5.1.9} and \ref{5.1.11}, $F\in\supp E_\mfkg$ means that there is a product-vector
$\Psi\in\mH_\Pi$:

\bequ\label{6.2.18(1)} \Psi := \bigotimes_{k\in\Pi}\mphi_k,\ \mphi_k\in\mH_k:= u_k\mH,\
\|\mphi_k\|=1,\ \text{for all}\ k\in\Pi,\enqu

\noidt such that the following relations are valid:

\bequ\label{6.2.18(2)} \lim_{N\rarw\infty}\frac{1}{N}\sum_{k=1}^N (\mphi_k,\,\pi_k(X_\xi)\mphi_k)
= F(\xi),\ \text{for all}\  \xi\in\mfkg. \enqu

Let $F^{(j)}\in \supp E_\mfkg\ (j = 1,2)$ be determined according to\rref 6.2.18(2)~ by the
product vectors $\Psi^{(j)}:=\otimes_{k\in\Pi}\mphi_k^{(j)}\in\mcl D_\Pi(\mfkg)$. We shall
construct a product vector $\Psi\in\mcl D_\Pi(\mfkg)$, for any rational number $c:
0<c=\frac{r}{s}<1,$ such that the corresponding value of $F\in \mfkg$, cf.\rref 6.2.18(1)~
and\rref 6.2.18(2)~, is

\bequ\label{6.2.18(3)} F=cF^{(1)}+(1-c)F^{(2)}.\enqu

\noidt This will prove the convexity of $\supp E_\mfkg$, since $\supp E_\mfkg$ is a closed subset
of \fkgs.

We shall construct the sequence $\{\mphi_k: k\in\Pi\}$ defining $\Psi$ according to\rref
6.2.18(1)~ from the sequence $\{\mphi_k^{(j)}:\ k\in\Pi,\ j=1,2\}$ for any two natural numbers
$0<r<s$ as follows:

\begin{eqnarray}\label{6.2.18(4)}
\mphi_{ms+j} &:=& \mphi^{(1)}_{mr+j}\ ,\ \text{for}\ j=1,2,\dots r;\
m\in\mbZ_+;\label{6.2.18(4a)}\\
&:=& \mphi^{(2)}_{m(s-r)+j-r}\ ,\ \text{for}\ j=r+1,r+2,\dots s;\ m\in\mbZ_+.\nonumber
\end{eqnarray}

\noidt(Here we have identified $\mH_k$ with \H\ $(k\in\Pi)$. The formally correct rewriting of the
formula\rref 6.2.18(4a)~\ includes, e.g.,\
$\mphi_{ms+j}:=u_{ms+j}u^{-1}_{mr+j}\mphi^{(1)}_{mr+j}$.)

Let

\bequ\label{6.2.18(5)} \Psi^{(j)}_k(\xi) := (\mphi_k^{(j)},\pi_k(X_\xi)\mphi_k^{(j)}),\ j=1,2;\
\Psi_k(\xi):=(\mphi_k,\pi_k(X_\xi)\mphi_k).\enqu

\noidt Inserting from\rref 6.2.18(4)~ into the \lhs of\rref 6.2.18(2)~ we obtain:
\begin{eqnarray}\label{6.2.18(6)}
\frac{1}{ms+j}\sum_{k=1}^{ms+j}\Psi_k(\xi)&=&
\frac{1}{ms+j}\sum_{k=1}^j \Psi_{ms+k}(\xi)
\\ &+&
\frac{ms}{ms+j}\left(\frac{r}{s}\,\frac{1}{mr}\sum_{k=1}^{mr}
\Psi^{(1)}_k(\xi)+\frac{s-r}{s}\,\frac{1}{m(s-r)}\sum_{k=1}^{m(s-r)}
\Psi^{(2)}_k(\xi)\right).\nonumber
\end{eqnarray}

Taking the limit $m\rarw\infty$ on both sides of\rref 6.2.18(6)~ $(j\in\{1,2,\dots s\})$, we
obtain\rref 6.2.18(3)~.
\end{proof}

\begin{prop}\label{6.2.19} Let $s_\pi\leq p_G$ be a $\msg(G)$-invariant
projector in the center $\mfk Z$ of $\mfkA^{**}$. Let $E_\mfkg^\pi := s_\pi E_\mfkg^\Pi$ be the
corresponding $G$-measure. Then $\mfk N_\pi^c:= E_\mfkg^\pi(C_b(\mfkgs,\mbC)) \subset\mfk N^c$,
cf. \ref {6.2.2}(iv). Specifically, $\mfk N_\pi^c=\mfk N^c$ for $s_\pi\geq s_G$. (Here we have
identified ${}^*$-isomorphic \Ca s.)
\end{prop}
\begin{proof}If $s_{\pi j}\ (j=1,2)$ are two such projectors $s_\pi$
with $s_{\pi 1}\leq s_{\pi 2}$, then for the corresponding $G$-measures one has $\supp
E_\mfkg^{\pi 1}\subset \supp E_\mfkg^{\pi 2}\subset \supp E_\mfkg$, cf. \ref{6.2.17}. The
Proposition \ref{6.2.9} and its proof is applicable to any $G$-measure in the case of bounded
generators $X_\xi\ (\xi\in \mfkg)$. Since $C(\supp E_\mfkg^{\pi 1})\subset C(\supp E_\mfkg^{\pi
2})\subset C(\supp E_\mfkg)$, and $\mfk N^c_\pi = E_\mfkg^\pi(C(\supp E_\mfkg^\pi))$ is an
isomorphic image of $C(\supp E_\mfkg^\pi)$, the result follows.
\end{proof}

\noidt {\bf Note:} With a help of this proposition one can show that $s_G$ can be replaced by
$s_\pi$,\ with\ $s_G\leq s_\pi\leq p_G$, everywhere in this Section \ref{sec;6.2}.

\begin{thm}\label{6.2.20} Let $\mfkA:=\mfkA^\Pi$ be the quasilocal
algebra introduced in \ref{5.1.4}; $\msg(G)\subset \maut\mfkA$ is generated by the continuous
unitary representation $U(G)$ in \H\ of a Lie group G with bounded generators $X_\xi=X_\xi^*\
(\xi\in\mfkg)$, cf. \ref{5.1.5} and \ref{5.1.3}. Let $s_\pi\leq p_G$ be a $\msg(G)$-invariant
central projector in $\mfkA^{**}$, where $p_G$ is introduced in \ref{5.1.29}. Let $E_\mfkg^\pi :=
s_\pi E^\Pi_\mfkg,$ {\bf where \emn$E^\Pi_\mfkg$~ is defined in \ref{5.1.33}. Let \emn$\mfk
N^c_\pi$~ and \emn$\mfk C_\pi$~ be defined as in \ref{6.2.16} and} \emn$\mfk C_\pi^N :=
\mfkA^N\otimes\mfk N^c_\pi$~; the algebras $\mfk N^c_\pi,$\ \emn$\mfk C^N_\pi$~,\ and $\mfk C_\pi$
are considered as \Csa s\ of $s_\pi\mfkA^{**}$ in the canonical way, cf. \ref{6.2.1}, \ref{6.2.2}
and \ref{6.2.13}. Let Q be a polynomial with the property (SA) of \ref{6.1.1}. Then one has:

\nl \noidt (i)\quad The sequence $\{\tau^K:\ K\in\Pi\}$ of the one parameter \autm\ groups of
\fkA\ generated by $Q^K$ according to\rref 6.1.1(2)~ determines a unique one parameter group
\emn$\tau^Q\subset \maut\mfk C$~ (with $\mfk C := \mfk C_\pi$ for $s_\pi := p_G$) such that for
any $N\in\Pi$ and for all $|t|\leq \kappa_N$\ (cf.\rref 6.2.5(1)~)

\bequ\label{6.2.20(1)} \tau_t^Q(x)=s^*\=\lim_{K\rarw\infty}\tau_t^K(x),\quad\forall x\in\mfkA^N:=
p_G\pi_u(\mfkA^N).\enqu

\noidt {\bf The} \emm$s^*(p_G\mfkA^{**},p_G\mfkA^*)$-topology~ is determined by the seminorms
from\rref 6.2.5(4)~ with $\mome\in\mS_*(p_G\mfkA^{**})$.

\nl\noidt (ii) The \Csa s $\mfk N_\pi^c,\ \mfk C_\pi$\ and $\mfk C_\pi^N\ (N\subset\Pi,\ s_\pi\leq
p_G)$ of \fk C\ are invariant \wrt\ $\tau^Q$. {\bf Let the restriction of $\tau^Q$ to $\mfk C_\pi$
be denoted by} \emn$\tau^\pi$~. ({\em {\bf Note:} We have changed the notation here. It was
denoted by $\tau^Q$ the group $\tau^\pi$ with $s_\pi:= s_G$ in the preceding subsections.})

\nl\noidt (iii) The restriction of $\tau^\pi$ to $\mfk N^c_\pi$ reproduces the classical flow
$\mphi^Q$ corresponding to the Hamiltonian function $Q$ on the Poisson manifold \fkgs\ in the
sense that

\bequ\label{6.2.20(2)} \tau_t^\pi(E^\pi_\mfkg(f))= E^\pi_\mfkg(\mphi_t^{Q*}f),\quad f\in
C(\mfkgs).\enqu

\nl\noidt (iv) The group $\tau^Q$ is a strongly continuous subgroup of $\maut\mfk C$, i.e. the
functions

\bequ\label{6.2.20(3)} t\mapsto \tau_t^Q(y) \enqu

\noidt are norm-continuous for all $y\in\mfk C$: The triple $\{\mfk C,\mbR,\tau^Q\}$ is a
$C^*$-dynamical system, \cite[2.7.1]{bra&rob}.

\nl\noidt (v) $\tau^\pi$ (for any $s_\pi$ specified above) is a $\msg(\mfk
C_\pi,s_\pi\mfkA^*)$-continuous group of automorphis of $\mfk C_\pi$, i.e. the functions

\bequ\label{6.2.20(4)} t\mapsto \mome(\tau_t^\pi(y)) \enqu

\noidt are continuous for all states $\mome\in s_\pi\mfkA^*\ (:= \{f\in \mfkA^*: f(s_\pi x)=f(x),\
\forall x\in\mfkA^{**}\})$ and for all $y\in \mfk C_\pi$, and for all such \ome\ one has:

\bequ\label{6.2.20(5)} \mome\circ \tau_t^\pi\in s_\pi\mfkA^*,\quad \forall t\in\mbR.\enqu

\nl\noidt (vi) The infinitesimal generator of $\tau^\pi$ is the \emm derivation $\delta_\pi$~ on
$\mfk C_\pi$ such that
\begin{subequations}\label{6.2.20(6)}
\bequ\label{6.2.20(6a)} \delta_\pi(y)=i\,\sum_{j=1}^n E^\pi_\mfkg(\partial_jQ)\,[X_j^N,y],\
\text{for all}\ y\in\mfkA^N,\enqu

\bequ\label{6.2.20(6b)}\delta_\pi(E^\pi_\mfkg(f))=E^\pi_\mfkg(\{Q,f\})\ \text{for}\ f\in
C^1(\mfkgs),\enqu
\end{subequations}

\noidt where the square bracket in\rref 6.2.20(6a)~ is the commutator, and\,\  $\mfkA^N$ is
considered there as $s_\pi\pi_u(\mfkA^N)$ (\fkA\ is simple!), and the partial derivatives
$\partial_jQ$ denote the differentiation of Q \wrt the components $F_j := F(\xi_j)$ of
$F\in\mfkgs$ in the dual basis to the basis $\{\xi_j: j=1,2,\dots n\}$ of \fkg, $X_j:= X_{\xi_j}$.
The compound bracket in \rhs\ in\rref 6.2.20(6b)~ denotes the classical Poisson bracket on \fkgs.
The operator $\delta_\pi$ determined by\rref 6.2.20(6)~ determines the group $\tau^\pi\in\maut\mfk
C_\pi$ uniquely:

\bequ\label{6.2.20(7)} \tau_t^\pi(y)=\sum_{m=0}^\infty\, \frac{t^m}{m!}\,\delta^m_\pi(y),\
\text{for all}\ y\in \mfk B^\#,\ |t|\leq \kappa_N,\ N\in\Pi.\enqu
\end{thm}
\begin{proof} We shall use here the fact mentioned in the Note in \ref{6.2.19}
that in the assertions of this section we can replace $s_G$ by $p_G$; we shall refer to the
assertions and their proofs in Sec.\ref{sec;6.2} as if they were reformulated with this
replacement.\nl

\noidt(i) The restrictions of $\tau^Q$ to the subalgebras $\mfk C^N$ given in \ref{6.2.15}
determine a unique group $\tau^Q\subset \maut\mfk C,$ since each of the mappings

\bequ\label{6.2.20(8)} \tau^Q_t:\ y\mapsto \tau_t^Q(y),\ y\in\mfk C^N,\ N\in\Pi,\ t\in\mbR,\enqu

\noidt is norm-continuous and $\{y:\ y\in\mfk C^N,\ N\in\Pi\}$ is norm-dense in \fk C.\nl

\noidt (ii) After the replacement of $s_G$ by $s_\pi$ (hence also $E_\mfkg$ by $E_\mfkg^\pi$) in
\ref{6.2.10} and \ref{6.2.15} we obtain the invariance of $\mfk N_\pi^c$ and of $\mfk C_\pi$ due
to $\msg(G)$-equivariance of $E_G^\pi$. The $\tau^Q$-invariance of $\mfk C_\pi^N$ is clear.\nl

\noidt (iii) Immediately from \ref{6.2.10}, since
$\tau^\pi_t(E^\pi_\mfkg(f))=s_\pi\tau_t^Q(E_\mfkg^\Pi(f)).$\nl

\noidt (iv) It suffices to prove the continuity in\rref 6.2.20(3)~ for $t\rarw 0$. With $y :=
x\in\mfkA^N$ the continuity is given by the uniform convergence in\rref 6.2.11(1)~, and this
implies the continuity for all $x\in\mfkA$ (by an $\meps/3$- argument). For $y := E^\pi_\mfkg(f),\
f\in C(\supp E^\pi_\mfkg)$, it suffices to prove

\bequ\label{6.2.20(9)} \lim_{t\rarw 0}\|\mphi_t^{Q*}f-f\|=0,\enqu

\noidt since $f\mapsto E^\pi_\mfkg(f)$ is a $C^*$-morphism. The validity of\rref 6.2.20(9)~ is a
consequence of the joint continuity of the classical flow $\mphi^Q$,

\bequ\label{6.2.20(10)} \mphi^Q:\ (t;F)\mapsto \mphi^Q_t(F)\in\mfkgs,\enqu

\noidt as well as of the compactness of $\supp E^\pi_\mfkg$ and of the continuity of $f$.\nl

\noidt (v) The continuity in\rref{6.2.20(4)}~ is a consequence of (iv). Let us consider
$\tau^\pi_t\ (t\in\mbR)$ as a family of representations of $\mfkA :=
\pi_u(\mfkA)\subset\mfkA^{**}$ in the subalgebra $s_\pi\mfkA^{**}$ of $\mfkA^{**}$. The unique
$\msg(\mfkA^{**},\mfkA^*)-\msg(s_\pi\mfkA^{**},s_\pi\mfkA^*)$-continuous extensions of these
representations to $\mfkA^{**}$, \cite[1.21.13]{sak1}, {\bf will be denoted by
\emn$\overline{\tau}^Q_t$~ (resp. \emn$\overline{\tau}^\pi_t$~ for $s_\pi<p_G$)}. From
\ref{6.2.10} and from its proof one can see

\bequ\label{6.2.20(11)}\overline{\tau}^\pi_t(s_\pi)=
\overline{\tau}^\pi_t(E_\mfkg^\pi(\mfkgs))=s_\pi.\enqu

\noidt We have also $\overline{\tau}^\pi_t(\id_\mfkA-s_\pi) = 0$, and the restrictions of
$\overline{\tau}^\pi_t\ (t\in\mbR)$ to the $\overline{\tau}^\pi_t$-invariant subalgebra
$s_\pi\mfkA^{**}$ form a family of ${}^*$-automorphisms which are automatically \sg-continuous,
\cite[4.1.23]{sak1}. The definition $\tau^\pi$ with a he1p of strong limits, cf.\rref 6.2.7(1)~,
shows that the restriction of $\overline{\tau}^\pi_t$ to $\mfk C_\pi$ coincides with the above
defined $\tau_t^\pi\in\maut \mfk C_\pi$. This proves the normality of $\mome\circ\tau^\pi_t$ for
any normal \ome (i.e. $\mome\in s_\pi\mfkA^*$), hence\rref 6.2.20(5)~.\nl

\noidt (vi) The automorphism group $\tau^\pi$ of $\mfk C_\pi$ is determined uniquely by the
determination of $\tau^\pi_t(x)$ for all $x\in\mfk B^N$\ (cf. Lemma\ \ref{6.2.6} and
\ref{6.2.2}(iii) for notation), for $|t|\leq \kappa_N,\ N\in\Pi$; this is clear from \ref{6.2.5}
and from its consequences. The series in the formula

\bequ\label{6.2.20(12)} \frac{\rd}{\rd t}\tau_t^\pi(x)=i\,\sum_{m=0}^\infty
\frac{(it)^m}{m!}\,s^*\=\lim_K\,[Q^K,[Q^K,x]^{(m)}]\enqu

\noidt converges uniformly in the disc $|t|\leq \kappa_N\ (x\in\mfk B^N)$, hence the equality\rref
6.2.20(12)~ is valid. Considerations similar to those used in the dealing with\rref 6.2.15(6)~
lead to the equalities:

\begin{eqnarray}\label{6.2.20(13)}
s\=\lim_K\, [iQ^K,[iQ^K,x]^{(m)}]&=&s\=\lim_K\, s\=\lim_L\,
[iQ^L,[iQ^K,x]^{(m)}]\nonumber \\
&=&s\=\lim_K\,\delta_\pi([iQ^K,x]^{(m)}),
\end{eqnarray}

\noidt where for all $N\in\Pi$:

\bequ\label{6.2.20(14)}\delta_\pi(x):= s\=\lim_L\, i[Q^L,x],\ \text{for all}\ x\in \mfk
B^N:=s_\pi\pi_u(\mfk B^N).\enqu

\noidt The derivation property of commutators and the polynomial form of $Q$ together with\rref
6.2.5(5)~ lead to the expression\rref 6.2.20(6a)~ for $\delta_\pi$ in\rref 6.2.20(14)~. Setting $t
= 0$ in\rref 6.2.20(12)~, we see that so defined $\delta_\pi(x)$ is the value of the generator
$\delta_\pi$ of $\tau^\pi$ on $x\in\mfkA^N\ (N\in\Pi)$. By the differentiation of\rref 6.2.10(1)~
with $f\in C^1(\supp E^\pi_\mfkg)$ we obtain\rref 6.2.20(6b)~, cf.\rref 6.2.10(12)~ and notes in
\cite{bra&rob} above 3.2.29. From the continuity properties of $\tau^\pi$ and the corresponding
closedness of $\delta_\pi$, cf. \cite[3.1.6]{bra&rob}, we obtain by the repeated use of\rref
6.2.20(13)~:

\bequ\label{6.2.20(15)} s\=\lim_K\, \delta_\pi ([iQ^K,x]^{(m)}) = \delta_\pi(s\=\lim_K\,
[iQ^K,x]^{(m)}) = \delta_\pi^{m+1}(x).\enqu

\noidt Insertion from\rref 6.2.20(14)~ and\rref 6.2.20(15)~ into\rref 6.2.11(1)~, cf. the note
following\rref 6.2.15(2)~, gives for $x\in\mfk B^N,\ |t|\leq \kappa_N$ the norm-convergent series:

\bequ\label{6.2.20(16)} \tau_t^\pi(x)=\sum_{m=0}^\infty \frac{t^m}{m!}\,\delta^m_\pi(x).\enqu

 This proves that the operator $\delta_\pi$ from\rref 6.2.20(6)~
 determines $\tau^\pi$.
 \end{proof}

\section{Time evolution in generalized mean-field
theories}\label{sec;6.3}

\pt\label{6.3.1}\rm We shall construct in this section a general class of time evolutions $\tau^Q$
of the infinite quantum systems $(\mfkA;\msg(G))$ defined in Sec.\ref{sec;5.2}. The time evolution
$\tau^Q$ is determined in a canonical way by an arbitrary classical Hamiltonian function $Q$ on
the (generalized) homogeneous classical phase space \fkgs\ as well as by the automorphism group
$\msg(G)$ of \fkA. It will be shown later that the here presented construction leads to the same
evolution what was denoted by $\tau^Q$ in Sec.6.2. in the case of $\mfkA:=\mfkA^\Pi,\ \msg(G)$
being defined according to \ref{5.1.5}, and with $Q$ being a polynomial in a basis of \fkgs\ dual
to any fixed basis $\{\xi_j,\ j=1,2,\dots n\}$; the generators of the continuous representation
$U(G)$ in the 'one-spin space' \H, \ref{5.1.3}, are supposed to be bounded in this special case.

We shall start with the general case, the specifications to the cases considered in
Sec.\ref{sec;5.1}, and the further specification to the cases of Sec.\ref{sec;6.2} will be made
later on. Let us fix here some general assumptions valid throughout of this section.

Using the notation of Sec.\ref{sec;5.2}, let $E_\mfkg$ be a fixed nontrivial $G$-measure
associated with the system $(\mfkA;\msg(G))$ such that, with $p_G:= E_\mfkg(\mfkgs)$, the
following implication is valid:
 \bequ\label{6.3.1(1)} \mome\in p_G\mcl S(\mfkA) =: \mS_\mfkg\imply
g\ (\in G)\mapsto \mome(\msg(g)(x))\quad {\rm is\ continuous\ for\ all}\ x\in\mfkA.\enqu

It will be shown in \ref{6.3.10} that this assumption is fulfilled by $\msg(G)$ from \ref{5.1.5}
with $p_G$ from \ref{5.1.29}. We shall assume that \fkA\ is a unital \Ca\ which is simple (this
last assumption is made only for brevity of our expression). The nontriviality of $E_\mfkg$ means
a certain 'breaking of symmetries' occurring in the system, cf.\ our \ref{5.2.3} for basic
definitions, and for illustration of the phenomenon of ``spontaneous symmetry breaking'' see
 \cite{emch1,rieck},\ \cite[Sec. 4.3.4]{bra&rob},\ \cite[IV.A]{bon2},\ \ref{6.5.5}.

The time evolution $\tau^Q$ will be defined with a help of the group-valued function $g_Q(t,F)$ on
$\mbR\times\mfkgs$ defined in \ref{6.1.3} with\rref 6.1.3(9)~ (another possible choice of
$\beta^Q_F$ will not change the general construction of $\tau^Q$, so that the nonuniqueness of
$\beta^Q_F$ leads to various possibilities for the definition of the time evolutions $\tau^Q$).
The notation introduced in \ref{6.1.2} and \ref{6.1.3} will be used here. Let us note that the
equation\rref 6.1.3(8)~ for $g_Q$ can be written in any continuous unitary representation $U(G)$
in \H\ in the form

\bequ\label{6.3.1(2)} i\,\frac{\rd}{\rd t}U(g_Q(t,F))=X(\beta^Q_{F_t})U(g_Q(t,F)),\
F\in\mfkgs,\quad t\in\mbR,\enqu

\noidt where $F_t := \mphi^Q_t(F);\ X(\xi) := X_\xi\ (\xi\in\mfkg)$ are the selfadjoint generators
of $U(G)$, and $\beta^Q_{F}\in\mfkg$ was introduced in \ref{6.1.3}. The equation\rref 6.3.1(2)~ is
of the form of (linear) quantum-mechanical evolution equation with the time-dependent Hamiltonian
operator $X(\beta^Q_{F_t})$. The equation\rref 6.3.1(2)~ describes, in the setting of
Sec.\ref{sec;5.1}, the time evolution of any 'individual' quantum subsystem placed in any fixed
site $k\in\Pi$ in the surrounding 'mean field' $\mphi_t^Q(F)\in\mfkgs$ generated by the whole
collection of the quantal subsystems (for all the sites $k\in\Pi$) interacting by an 'infinitely
weak and of infinitely long-range' interaction with each other. The equation\rref 6.3.1(2)~ will
be useful in the analysis of thermodynamic properties of the considered systems.

\begin{defs}\label{6.3.2}\nl

\noidt (i)\  \emm $C_b:= C_b(\supp E_\mfkg,\mbC)$~ {\bf will denote} the set of all uniformly
bounded complex-valued continuous functions on $\supp E_\mfkg\subset \mfkgs$, see \ref{5.2.3} for
the definition of $\supp E_\mfkg$.\nl

\noidt (ii) The \emm $s^*$-topology on \fkA ~  is determined by seminorms $\hat{p}_\mome,\
\hat{p}^*_\mome$ (cf.\rref 6.2.5(4)~) for all $\mome\in p_g\mS(\mfkA)$.\nl

\noidt (iii) Let \emm $\mfk C_{bs}:=\mfk C_{bs}(\supp E_\mfkg,\mfkA)$~ be the set of all
\fkA-va1ued, uniformly bounded $s^*$-continuous functions on $\supp E_\mfkg$, i.e. \emm
$\mbs{f}\in\mfk C_{bs}$~ means that the function

\bequ\label{6.3.2(1)} \mbs{f}: F(\in\supp E_\mfkg)\mapsto\mbs{f}(F)\ (\in\mfkA)\enqu

\noidt is bounded in the sense

\bequ\label{6.3.2(2)} \|\mbs{f}\|:=\sup\{\|\mbs{f}(F)\|:\ F\in\supp E_\mfkg\}<\infty,\enqu

\noidt and all the functions

\begin{subequations}
\bequ\label{6.3.2(3a)} F\mapsto\mome\left((\mbs{f}(F)-\mbs{f}(F_0))^*
(\mbs{f}(F)-\mbs{f}(F_0))\right),\ \mome\in p_G\mS(\mfkA),\ F_0\in\supp E_\mfkg,\enqu

\bequ\label{6.3.2(3b)}F\mapsto\mome\left((\mbs{f}(F)-\mbs{f}(F_0))
(\mbs{f}(F)-\mbs{f}(F_0))^*\right),\ \mome\in p_G\mS(\mfkA),\ F_0\in\supp E_\mfkg,\enqu
\end{subequations}

\noidt converge to zero for $F$ converging to $F_0$ in the norm-topology of \fkgs.\nl

\noidt (iv) For any $\msg(G)$-invariant \Csa\ $\mfkA^J$ of \fkA:

\bequ\label{6.3.2(4)} \msg(g)(x):=\msg_g(x)\in\mfkA^J\ \text{for all}\ x\in\mfkA^J,\ g\in G,\enqu

\noidt let \emm $\mfk C^J_{bs}:=\mfk C_{bs}(\supp E_\mfkg,\mfkA^J)$~ be defined equally as it was
defined $\mfk C_{bs}$ in (iii) with the replacement of \fkA\ by $\mfkA^J$.\nl

\noidt (v) Let \emm $\mfk C^G_{bs}$~ (resp. \emn$\mfk C^{GJ}_{bs}$~) be the \Csa\ (cf.
\ref{6.3.4}) of $\mfk C_{bs}$ (resp. of $\mfk C^J_{bs}$) generated by all the functions
$\mbs{f}_0\in\mfk C_{bs}$ of the form

\bequ\label{6.3.2(5)} \mbs{f}_0:\ F\mapsto\msg_{g_0 (F)}(x)\,f(F),\ f\in C_b,\ g_0\in C(\supp
E_\mfkg,G), \enqu

\noidt with any $x\in\mfkA$\ (resp. any $x\in\mfkA^J$). {\bf The set} \emm $C(\supp E_\mfkg,G)$~
consists of all continuous $G$-valued functions on $\supp E_\mfkg$.\nl

\noidt (vi) We shall use also $K := \supp E_\mfkg$, resp. \emm $K\subset\mfkgs$~ {\bf will denote}
any $Ad^*(G)$-invariant closed subset of the generalized classical phase space in more general
cases. {\bf We shall identify} \emm $C_b := C_b(K,\mbC)$~ with the subset $\mfk C_{bs}(K,\mbC\,
\id_\mfkA)$ of $\mfk C_{bs}$ in the canonical way: $f\in C_b$ is identified with the function

\bequ\label{6.3.2(6)} \mbs{f}:\ F\mapsto \id_\mfkA\,f(F),\ F \in K ,\ \id_\mfkA\ {\rm is\ the\
identity\ of}\ \mfkA. \enqu
\end{defs}

\begin{prop}\label{6.3.3}The set $\mfk C_{bs}$ is a ${}^*$-algebra with
respect to the natural (pointwise) algebraic operations determined by the corresponding operations
in the range \fkA\ of the elements $\mbs{f}\in\mfk C_{bs}$:

\begin{eqnarray}\label{6.3.3(1)} (\mbs{f}_1+\mlam\mbs{f}_2)(F)&:=&
\mbs{f}_1(F)+\mlam\mbs{f}_2(F),\
(\mbs{f}_1\mbs{f}_2)(F):=\mbs{f}_1(F)\mbs{f}_2(F),\nonumber \\
\mbs{f}^*(F)&:=& [\mbs{f}(F)]^*,\forall F\in K,\mlam\in\mbC,\ \mbs{f}_j\ and\ \mbs{f}\in\mfk
C_{bs},\end{eqnarray}

\noidt and it is a normed algebra with the norm $\|\mbs{f}\|$ of $\mbs{f}\in\mfk C_{bs}$ given
by\rref 6.3.2(2)~. This normed ${}^*$-algebra $\mfk C_{bs}$ is a \Ca, and its subsets $\mfk
C^J_{bs}$ and $C_b$ endowed with the induced algebraic operations and the norm are \Csa s of $\mfk
C_{bs}$.
\end{prop}

\begin{proof}
 The continuity properties of the product in \fkA\ \wrt\ the
$s^*$-topology are given by Proposition 1.8.12. and Theorem 1.8.9. of \cite{sak1}. Then the
uniform boundedness of $\mbs{f}\in\mfk C_{bs}$ and the continuity of the ${}*$-operation in the
$s^*$-topology gives the invariance of $\mfk C_{bs}$ \wrt\ the algebraic operations\rref
6.3.3(1)~. The norm properties of the function given in\rref 6.3.2(2)~ are easily verified, and
the $C^*$-property of the norm:

\bequ\label{6.3.3(2)} \|\mbs{f}\|^2=[\sup_F \|\mbs{f}(F)\|]^2= \sup_F \|\mbs{f}(F)\|^2=\sup_F
\|\mbs{f}(F)^*\mbs{f}(F)\|=\|\mbs{f}^*\mbs{f}\|,\enqu

\noidt is valid too. We shall verify completness of $\mfk C_{bs}$ in this norm. For any Cauchy
sequence $\{\mbs{f}_n; n\in\mbZ_+\}$ in $\mfk C_{bs}$, the sequence $\{\mbs{f}_n(F),n\in\mbZ_+\}$
is Cauchy in \fkA\ for any $F\in K$. The completness of \fkA\ gives the existence of pointwise
limits

\bequ\label{6.3.3(3)} \mbs{f}(F):= n\=\lim_k \mbs{f}_k(F)\in\mfkA,\quad F\in K.\enqu

By defining the norm of any function $\mbs{f}: K \rarw\mfkA$\ ($\|\mbs{f}\|$ could be  infinite in
general) by\rref 6.3.2(2)~, we have the norm-convergence of $\mbs{f}_k$ to $\mbs{f}$ from\rref
6.3.3(3)~: If $\|\mbs{f}_n-\mbs{f}_m\| < \delta$ for all $n,m> n_\delta$, then $\|\mbs{f}_n -
\mbs{f}\| <\delta$ for all $n> n_\delta$, for any positive $\delta$, since $\lim_m \|\mbs{f}_n(F)-
\mbs{f}_m(F)\|=\|\mbs{f}_n(F)-\mbs{f}(F)\|$\ for all $F\in K$. Considering the cyclic
representation $(\pi_\mome,\mH_\mome,\Omega_\mome)$ corresponding to any $\mome\in p_G\mS(\mfkA)$
as a subrepresentation of the universal representation $\pi_u$ in $p_G\pi_u(\mfkA)$, we have with
the identification of \fkA\ with $p_G\pi_u(\mfkA)$ (cf.\rref 6.2.5(4)~):

\bequ\label{6.3.3(4)} \hat{p}_\mome(\mbs{f}(F)-\mbs{f}(F_0)) =
\|(\mbs{f}(F)-\mbs{f}(F_0))\Omega_\mome\|\leq
2\|\mbs{f}_m-\mbs{f}\|+\|(\mbs{f}_m(F)-\mbs{f}_m(F_0))\Omega_\mome\|, \enqu

\noidt and the s-continuity of $\mbs{f}_m$'s gives the s-continuity of $\mbs{f}$. A use of the
norm-continuity of the ${}^*$-operation gives us the $s^*$-continuity of $\mbs{f}$, i.e.
$\mbs{f}\in\mfk C_{bs}$. The remaining assertions of the proposition follow now easily.
\end{proof}

\begin{lem}\label{6.3.4} The functions $\mbs{f}_0$ from\rref 6.3.2(5)~
 belong to $\mfk C_{bs}$. Hence,
$\mfk C_{bs}^G$ and $\mfk C_{bs}^{GJ}$ are\nl \Csa s of $\mfk C_{bs}$.
\end{lem}

\begin{proof} Since $f\in C_b$ can be considered as an element of $\mfk C_{bs}$,
it suffices to prove $\mbs{f}_0\in \mfk C_{bs}$ for $\mbs{f}_0$ given by\rref 6.3.2(5)~ with $f
:=$ constant function. This will be proved by proving the $s^*$-continuity of $\msg(G)$. For any
$x\in \mfkA$ and any $\mome\in p_G\mS(\mfkA)$, we have

\begin{eqnarray}\label{6.3.4(1)}
&\hat{p}_\mome\left(\msg_g(x)-\msg_{g_0}(x)\right)^2=\mome\left((\msg_g(x^*)-\msg_{g_0}(x^*))
(\msg_g(x)-\msg_{g_0}(x))\right)=  \\
&=\mome\left(\msg_g(x^*x)-\msg_{g_0}(x^*x)\right) +
\mome\left((\msg_{g_0}(x^*)-\msg_g(x^*))\,\msg_{g_0}(x)\right)+\mome\left(\msg_{g_0}(x^*)
(\msg_{g_0}(x)-\msg_g(x))\right),\nonumber
\end{eqnarray}

\noidt and the s-continuity follows from the assumption\rref 6.3.1(1)~ by repeated use of the
polarization identity (expressing nondiagonal matrix elements of bounded operators in a Hilbert
space by a finite linear combination of the diagonal ones). The $s^*$-continuity is then obtained
by the replacement of $x$ by $x^*$ in the above considerations.
\end{proof}

\pt\label{6.3.5}\rm The quasilocal \Ca\ \fkA\ of quantum (microscopic) observables is naturally
embedded into $\mfk C_{bs}$ as a \Csa\ by the identification of any $x\in\mfkA$ with a constant
function $\mbs{f}\in\mfk C^G_{bs}$:

\bequ\label{6.3.5(1)} \mbs{f}(F):= x =\textbf{1}(F)\,\msg_e(x),\ F\in K,\enqu

\noidt where $\textbf{1}(F) := 1$ for all $F\in\mfkgs$. The classical (macroscopic) observables
are embedded into $\mfk C^G_{bs}$ according to the formula\rref 6.3.2(6)~, where the classical
observables are represented by functions belonging to $C_b(K,\mbC)$. We can (and we shall)
consider $\mfk C^G_{bs}$, or $\mfk C_{bs}$, as the (extended) \Ca\ of observables of the systems
with 'mean-field' dynamics. It might be useful, however, to embed this new algebra of observables
in a canonical way into the \Wa\ $\mfkA^{**}$, since there is a canonical bijection between the
set of all states $\mome\in\mS(\mfkA)$ and the set of all normal states $\mome\in\mSs(\mfkA^{**})$
on the double dual $\mfkA^{**}$ of \fkA: any $\mome\in\mS(\mfkA)$ corresponds to its (equally
denoted) canonical normal extension $\mome\in\mSs(\mfkA^{**})$. Hence, after obtaining an \emm
embedding of $\mfk C_{bs}$ into $\mfkA^{**}$~ such that $\mfkA\subset\mfk C_{bs}$ is mapped onto
$\pi_u(\mfkA)\subset \mfkA^{**}$ or onto its subrepresentation, we shall obtain a certain
canonical extension of any state $\mome\in\mS(\mfkA)$ (or of any state $\mome\in p_G\mS(\mfkA)$,
where $p_G\in\mfk Z$ is the projector onto the above mentioned subrepresentation of $\pi_u$) to a
state on $\mfkA^{**}$ (resp. on $p_G\mfkA^{**}$), and this in turn gives to us a certain canonical
extension of states on \fkA\ to states on $\mfk C_{bs}$. Such an embedding is given in the
following proposition.

\begin{prop}\label{6.3.6}Let us consider the integral decomposition of
any $\mome\in p_G\mSs(\mfkA^{**})$ [where \fkA\ is simple] given by the formula\rref 5.1.38(1)~
according to Theorem \ref{5.2.11}, and let $F_\mfkg: \mcl M\rarw \dot{\mfkg}^*$ be given as in
\ref{5.1.39}. There is a $C^*$-isomorphism of $\mfk C_{bs}$ into $p_G\mfkA^{**}$ formally written
in the form

\bequ\label{6.3.6(1)} E_\mfkg:\ \mbs{f}\,(\in \mfk C_{bs})\mapsto E_\mfkg(\mbs{f}) :=\int
\mbs{f}(F)\,E_\mfkg(\rd F),\enqu

\noidt where $E_\mfkg$ denotes the $G$-measure (as before) as well as the presently introduced
isomorphism. The isomorphism $E_\mfkg$ is uniquely determined by the formula\footnote{Note: For
noncompact $\supp E_\mfkg$, the integral is a limit of integrals over bounded subsets
$B\subset\mfkgs$: $\int\dots :=
\lim_{B\uparrow\mfkgs}\int\mome_m(E_\mfkg(B)\mbs{f}(F_m))\,\mu_\mome(\rd m).$}

\bequ\label{6.3.6(2)} \mome(E_\mfkg(\mbs{f})) := \int \mome_m(\mbs{f}(F_m))\,\mu_\mome(\rd
m),\quad\forall \mome\in p_G\mSs(\mfkA^{**}),\enqu

\noidt where the decomposition\rref 5.1.38(1)~ was used, and (cf. \ref{5.1.39})

\bequ\label{6.3.6(3)} F_\mfkg: m \mapsto F_m:= F_\mfkg(m),\quad m\in\mcl N\subset \mcl M,\enqu

\noidt is defined on the {\bf spectrum space} \emn$\mcl N$~ of the (commutative) subalgebra $\mfk
N(E_\mfkg)$ of $p_G\mfkA^{**}$, cf. \ref{5.2.3}.

The mapping $E_\mfkg$ leaves $\mfkA := p_G\pi_u(\mfkA)$ invariant and maps $C_b$ onto a \Csa\
$\mfk N^c$ of $\mfk N(E_\mfkg) =:\mfk N_G$, see also \ref{5.2.10}.\end{prop}

\begin{proof} Let $B\subset \mcl N$ be any Borel set and $\chi_B$ is its
characteristic function. The functions

\bequ\label{6.3.6(4)} m\mapsto \mome_m(x)\,\chi_B(m),\quad x\in\mfkA^{**},\enqu

\noidt are Borel functions on $\mcl N$ for any $\mome\in p_G\mSs(\mfkA^{**})$. Since the function
$F_\mfkg$ in\rref 6.3.6(3)~ is continuous, the measurability of the functions

\bequ\label{6.3.6(5)} m\mapsto \mome_m(\mbs{f}\circ F_\mfkg(m)),\ \mbs{f}\in\mfk C_{bs},\enqu

\noidt can be proved with a help of a sequence $F_\mfkg^{(n)}$ of functions from $\mcl N$ into the
one point compactification $\dot{\mfkg}^*$ of $\mfkgs$\ assuming each only a finite number of
values and pointwise converging to $F_\mfkg$ in the natural topology of $\dot{\mfkg}^*$. Then the
functions

\bequ\label{6.3.6(6)} m\mapsto \mome_m(\mbs{f}\circ F_\mfkg^{(n)}(m)),\ \mbs{f}\in\mfk C_{bs},\
\mome\in p_G\mSs(\mfkA^{**})\enqu

\noidt are finite sums of functions of the form\rref 6.3.6(4)~, hence the functions\rref 6.3.6(6)~
are measurable. The $s^*$-continuity of $\mbs{f}$ implies then the pointwise convergence of the
functions\rref 6.3.6(6)~ to the function\rref 6.3.6(5)~ for $n\rarw\infty$. According to a known
theorem in measure theory, cf. e.g. \cite[§6.l0.VII.]{najm}, the pointwise limit of uniformly
bounded measurable functions is measurable, hence\rref 6.3.6(5)~ are Borel functions. We have
proved the existence of the integrals in\rref 6.3.6(2)~ for any $\mbs{f}\in\mfk C_{bs}$. The
function

\bequ\label{6.3.6(7)} E_\mfkg(\mbs{f}):\ \mome\ (\in\mS(\mfkA))\mapsto
\mome(E_\mfkg(\mbs{f}))\in\mbC, \enqu

\noidt is affine: The extension mapping $e_*$ of \cS(\fkA)\ onto \Ss(\Ass)\ is affine, and the
association of subcentral (hence orthogonal, hence regular Borel) measures to the states
$\mome\in\mSs(\mAss)$ defined by (cf.\rref 5.1.38(2)~)

\bequ\label{6.3.6(8)} \hat{\mu}:\ \mome\mapsto \hat{\mu}_\mome\in \{{\rm probability\ measures\
on\ }\ \mS(\mAss)\},\enqu

\noidt where the measure $\hat{\mu}_\mome$ corresponds to the decomposition of
$\mome\in\mS(\mAss)$ given by the commutative subalgebra $\pi_\mome(\mfk M_G)''$ in $\mcl L(\mcl
H_\mome)$ (cf. \cite[4.1.25.]{bra&rob}, and for the definition of $\mfk M_G$ see \ref{5.2.10}), is
also affine. The affinity of\rref 6.3.6(8)~ can be proved on the basis of the fact that all the
measures in\rref 6.3.6(8)~ are obtained from the same algebra $\mfk M_G\subset\mfk Z$ by
considering $\hat{\mu}_\mome$ and $\mlam_1\hat{\mu}_{\mome_1}+\mlam_2\hat{\mu}_{\mome_2}$\ (with
$\mome:=\mlam_1\mome_1+\mlam_2\mome_2$) as limits of the nets of measures which correspond to the
net of finite dimensional subalgebras of $\mfk M_G$, compare Lemma 4.1.26. in \cite{bra&rob}:

\bequ\label{6.3.6(9)} \mome(x)=\
\sum_j\mome(p_jx)=\mlam_1\sum_j\mome_1(p_jx)+\mlam_2\sum_j\mome_2(p_jx),\
\sum_jp_j=\id_\mfkA,\enqu

\noidt for any finite set of mutually orthogonal projectors $p_j\in\mfk M_G$. Hence,\rref
6.3.6(8)~ is an affine mapping:

\bequ\label{6.3.6(10)} \hat{\mu}_\mome=\mlam_1\hat{\mu}_{\mome_1}+\mlam_2\hat{\mu}_{\mome_2},\
\text{for}\ \mome:=\mlam_1\mome_1+\mlam_2\mome_2.\enqu

The relation\rref 6.3.6(10)~ has a unique extension to all $\mome_j\in\mAs\ (\mlam_j\in\mbC)$.
Writing for $\mome\in p_G\mS(\mfkA)$:

\bequ\label{6.3.6(11)} \mome(E_\mfkg(\mbs{f}))=\int\mphi\left(\mbs{f}(F_\mfkg\circ
r_M(\mphi))\right)\hat{\mu}_\mome(\rd\mphi),\enqu

\noidt what is meaningful for $\mphi\in\supp\hat{\mu}_\mome$ (cf. the proof of \ref{5.1.38}), we
obtain now affinity of\rref 6.3.6(7)~ which can be uniquely extended to linearity on the whole \As
($\ni\mome$). The boundedness of the mapping\rref 6.3.6(7)~ is a direct consequence of\rref
6.3.6(11)~ as well as of the boundedness of the function $\mbs{f}$. This proves that
$E_\mfkg(\mbs{f})\in\mAss$, where the linear extension of\rref 6.3.6(7)~ is denoted by the same
symbol. We shall consider \Ass\ as a \Wa\ in the canonical way: $\mAss:=\pi_u(\mfkA)''\subset\mcl
L(\mcl H_u)$. We shall prove the morphism property of $E_\mfkg$ in\rref 6.3.6(1)~. The linearity
of\rref 6.3.6(1)~ is clear from\rref 6.3.6(2)~ and from the linearity of each of $\mome_m$. By a
'polarization procedure' one can prove

\bequ\label{6.3.6(12)} \mome(E_\mfkg(\mbs{f})y)=\int\mome_m(\mbs{f}(F_m)y)\,\mu_\mome(\rd m),\
y\in\mAss,\ \mome\in p_G\mSs(\mAss).\enqu

\noidt Since $\mome_m(yE_\mfkg(\mbs{f}))= \mome_m(y\mbs{f}(F_m))$ for all $\mome\in
p_G\mSs(\mAss),\ m\in\supp \mu_\mome,\ y\in\mAss$ and $\mbs{f}\in\mfk C_{bs}$, we have also

\begin{eqnarray}\label{6.3.6(13)}
\mome(E_\mfkg(\mbs{f}_1)E_\mfkg(\mbs{f}_2))&=&
\int\mome_m(\mbs{f}_1(F_m)E_\mfkg(\mbs{f}_2))\,\mu_\mome(\rd m)\\
&=&\int\mome_m(\mbs{f}_1(F_m)\mbs{f}_2(F_m))\,\mu_\mome(\rd m)=\mome(E_\mfkg(\mbs{f}_1\mbs{f}_2)),
\nonumber
\end{eqnarray}

\noidt which proves $E_\mfkg(\mbs{f}_1\mbs{f}_2)=E_\mfkg(\mbs{f}_1)E_\mfkg(\mbs{f}_2)$ for all
$\mbs{f}_j\in\mfk C_{bs}\ (j=1,2)$. The ${}^*$-property follows by the decomposition of
$\mbs{f}\in\mfk C_{bs}$ into the real and imaginary parts in\rref 6.3.6(2)~.

We shall show that the kernel of the morphism $E_\mfkg: \mfk C_{bs}\rarw p_G\mAss$ is trivial. We
shall use here the simplicity of the \Ca\ \fkA. Let $\mbs{f}>0$ be a positive element of $\mfk
C_{bs}, \|\mbs{f}\|>0$. If $\mbs{f}(F_0)\neq 0,\ F_0\in K$,  then there is a state
$\mome\in\mS(\mfkA)$ with $\mome(\mbs{f}(F_0))\neq 0$. The $s$-continuity of $\mbs{f}\in\mfk
C_{bs}$ implies that the set

\bequ\label{6.3.6(14)} B:=\{F\in K:\ \mome(\mbs{f}(F))>\frac{1}{2}\mome(\mbs{f}(F_0))\}\subset
\mfkgs \enqu

\noidt {\bf is open in} \emn$K := \supp E_\mfkg$~. Hence $E_\mfkg(B)\neq 0$, and

\bequ\label{6.3.6(15)} \|\mbs{f}(F)\|>\frac{1}{2}\,|\mome(\mbs{f}(F_0))|> 0,\ \text{for all}\ F\in
B.\enqu

Any state $\mome_0\in\mS(\mfkA)$ supported by $E_\mfkg(B): \mome_0(x) = \mome_0(E_\mfkg(B) x)\
(x\in\mfkA)$, is decomposed according to\rref 5.1.38(1)~ into the states $\mome_m$ with $F_m\in B$
for all $m\in\supp \mu_{\mome_0}$. Since \fkA\ is simple, there is an element $x_m\in\mfkA$ for
any such $\mome_m$ that

\bequ\label{6.3.6(16)} \mome_m(x^*_mx_m)=1,\ {\rm and}\ \mome_m(x^*_m\mbs{f}(F_m)x_m)\neq 0.\enqu

\noidt The state $\mphi_m\in\mS(\mfkA),\ \mphi_m(y):=\mome_m(x_m^*yx_m)$ is also supported by
$E_\mfkg(B_m)$ with any open $B_m\subset K$ containing $F_m$. Hence the decomposition\rref
5.1.38(1)~ of $\mome :=\mphi_m$ is concentrated on the one point set $\{m\}$. This means that

\bequ\label{6.3.6(17)} \mphi_m(E_\mfkg(\mbs{f})) := \mphi_m(\mbs{f}(F_m)) \neq 0,\enqu

\noidt hence $E_\mfkg(\mbs{f})\neq 0$ for any nonzero $\mbs{f}\in\mfk C_{bs}$. This proves the
isometry of $E_\mfkg$, hence $E_\mfkg$ is a $C^*$-isomorphism of $\mfk C_{bs}$ into
$E_\mfkg(\mfkgs)\mAss=p_G\mAss$. The remaining assertions are clearly valid.
\end{proof}

\begin{lem}\label{6.3.7} Let $\mbs{f}\in\mfk C_{bs},\ \mome\in p_G\mS(\mfkA)$.
Then the function

\bequ\label{6.3.7(1)} (g;F)\mapsto \mome\left(\msg_g^{-1}(\mbs{f}(F))\right)\in\mbC,\ (g;F)\in
G\times K,\enqu

\noidt is jointly continuous on the topological product $G\times\supp E_\mfkg$.\end{lem}

\begin{proof}
Let $\mbs{f} := \mbs{f}_0$, cf.\rref 6.3.2(5)~. Then

\bequ\label{6.3.7(2)} \msg^{-1}_g(\mbs{f}_0(F)) = \msg(g^{-1}g_0(F))(x)\,f(F),\enqu

\noidt and the joint continuity of the group operation

\bequ\label{6.3.7(3)} (g_1;g_2)\ (\in G\times G)\mapsto g^{-1}_1g_2 \in G \enqu

\noidt gives the joint continuity in\rref 6.3.7(1)~ with $\mbs{f} := \mbs{f}_0$. It can be
verified directly, cf. e.g.\rref 6.3.4(1)~, that the function in\rref 6.3.7(2)~ is even
$s^*$-continuous in the couple $(g;F)\in G\times K$. But the finite algebraic combinations as well
as the uniform limits of $s^*$-continuous bounded functions are $s^*$-continuous. Since $\mfk
C^G_{bs}$ is generated by functions of the form $\mbs{f}_0$, we have proved that the functions

\bequ\label{6.3.7(4)} (g;F)\mapsto \msg^{-1}_g(\mbs{f}(F))\in\mfkA,\ \text{for all}\
\mbs{f}\in\mfk C^G_{bs},\enqu

\noidt are even $s^*$-continuous.
\end{proof}

\begin{prop}\label{6.3.8} Let, with the notation of \ref{6.1.3},
be $\mbs{f}\in\mfk C^G_{bs}$, and for a fixed $Q\in C^\infty(\mfkgs,\mbR)$ and for any $t\in\mbR$,
$F\in K$, let

\bequ\label{6.3.8(1)} \mbs{f}_t(F):=\msg(g_Q^{-1}(t,F))(\mbs{f}(\mphi_t^QF)). \enqu

Then $\mbs{f}_t\in\mfk C^G_{bs}$ and the mappings $\mbs{f}\mapsto\mbs{f}_t$ form a one-parameter
group of ${}^*$-automorphisms of $\mfk C^G_{bs}:\ \mbs{f}_{t+s}=(\mbs{f}_t)_s$, for all
$t,s\in\mbR.$
\end{prop}

\begin{proof} From the continuity properties of $g_Q$ and $\mphi^Q$
($g_Q$ and $\mphi^Q$ depend smoothly on $t$ and $F$), and from the $s^*$-continuity of
functions\rref 6.3.7(4)~, we have $\mbs{f}_t\in\mfk C_{bs}$ for any $\mbs{f}\in \mfk C_{bs}$. The
${}^*$-morphism properties of the mapping $\mbs{f}\mapsto \mbs{f}_t$ are fulfilled due to the
morphism properties of the pull-back $\mphi^*$ by any diffeomorphism $\mphi$ of K,

\bequ\label{6.3.8(2)} \mphi^*:\ \mbs{f}\mapsto \mphi^*\mbs{f},\ \mphi^*\mbs{f}(F):= \mbs{f}(\mphi
F),\ F\in K\subset\mfkgs, \enqu

\noidt as well as of $\msg(g)\in \maut\mfkA$. The group property follows immediately from the
group property of the flow $\mphi^Q$ and from the cocyc1e property\rref 6.1.3(5)~ of $g_Q$. The
group property implies invertibility, hence isometry of the considered mappings.
\end{proof}

\pt\label{6.3.9}\rm We have just proved existence of a certain 'time evolution' in the \Ca\ $\mfk
C_{bs}$ containing \fkA\ and $\mfk N^c$. This evolution is determined by an arbitrary classical
Hamiltonian function $Q$ and by the representation $\msg(G)$ of the group $G$ of `macroscopic
symmetries' with the help of the formula\rref 6.3.8(1)~. To have possibility to see eonnections
with the 'mean-field evolutions' discussed in Sec.\ref{sec;6.2}, we shall transfer this evolution
into \Ass\ by a use of the isomorphism $E_\mfkg$ from\rref 6.3.6(1)~. We shall see that the time
evolutions defined by a limiting procedure in Sec.\ref{sec;6.2} can be defined directly by the
formula\rref 6.3.8(1)~ (transferred into $p_G\mAss$). The same possibility of a definition of
'mean-field evolutions' arises in all the systems considered in Sec.\ref{sec;5.1}. To make this
possibility clear, let us prove the property\rref 6.3.1(1)~ for those systems.

\begin{lem}\label{6.3.10} Let us consider the systems determined with a
help of infinite tensor product considered in Sec.\ref{sec;5.1}. Then the group
$\msg(G)\subset\maut\mfkA\ (\mfkA:=\mfkA^\Pi)$ has the property\rref 6.3.1(1)~: The functions
$g\mapsto \msg_g(x)$ on $G$ are $s^*$-continuous for all $x\in\mfkA$, the $s^*$-continuity being
determined by the seminorms $\hat{p}_\mome$ and $\hat{p}_\mome^*$ from\rref 6.2.5(4)~ with
$\mome\in p_G\mS(\mfkA)$, and $p_G$ was defined in \ref{5.1.29}.\end{lem}

\begin{proof} The implication ``\!\rref 6.3.1(1)~ $\imply s^*$-continuity'' was
proved in Lemma \ref{6.3.4}. Since the set of local elements $x\in\cup_{N\subset \Pi} \mfkA^N$ is
norm-dense in \fkA, it suffices to prove the continuity in\rref 6.3.1(1)~ for $x$ local. We have
assumed in \ref{5.1.29} the existence of the generators $X^N_\xi\ (\xi\in \mfkg, N\subset\Pi)$ of
all one parameter subgroups of the unitary group $V_N(G)$ acting in $\mH_N$, cf. \ref{4.3.8}\ and
\ref{5.1.2}, as well as the existence of (equally denoted) generators for the unitary groups
$p_G\pi_u(V_N(\exp(\xi t)))$ for all $\xi\in\mfkg$. For $\mome\in p_G\mS(\mfkA)$ and $x\in\mfkA^N$
we have \bequ\label{6.3.10(1)} \mome(\msg(\exp(\xi t))(x)) = (\Omega_\mome,
\exp(-itX_\xi^N)\pi_u(x)\exp(itX^N_\xi)\Omega_\mome),\enqu

\noidt what continuously depends on t. We have to prove the strong-continuity of the group $U(g)
:= p_G\pi_u(V_N(g))$ from the strong continuity of all one parameter subgroups $U(\exp \xi t)=:
\exp(-itX_\xi),\nl (\xi\in\mfkg)$; we write here $X_\xi$ instead of $X_\xi^N$. Let
$\xi_j\in\mfkg,\ j= 1,2,\dots n$ be a fixed basis in \fkg\ and set $X_j:=X_{\xi_j}$. Let us
parametrize $g\in G$ in a neighbourhood of the unity $e\in G$ by $\textbf{t}:= (t_1,t_2,\dots
t_n)\in\mbR^n$ in the following way, cf. \cite[Lemma II.2.4]{helgas}:

\bequ\label{6.3.10(2)} g\equiv g(\textbf{t}):= \exp(t_1\xi_1)\exp(t_2\xi_2)\dots
\exp(t_n\xi_n).\enqu

Now we can prove weak continuity of $U(g(\textbf{t}))$ in $\textbf{t}=0\in\mbR^n$ from the known
strong continuity of $U_j(t) := U(\exp \xi_j t) = exp(-itX_j)$, for all $j = 1,2,\dots n$. Since
$U$ is a representation of $G$, we can write

\bequ\label{6.3.10(3)} U(g(\textbf{t}))-I= \prod_{j=1}^n
U_j(t_j)-I=\sum_{k=1}^n\left[\prod_{j=1}^{k-1} U_j(t_j)\right]\,(U_k(t_k)-I),\enqu

\noidt where $I$ is the unit operator in the Hilbert space of the representation and the product
of zero number of factors equals to $I$. Since the unitary operators do not change the norm of
vectors, we have for any unit vectors $\Psi_1$ and $\Psi_2$ in the Hilbert space:

\bequ\label{6.3.10(4)} |(\Psi_1,(U(g(\textbf{t}))-I)\Psi_2)|\leq \sum_{k=1}^n
\|(U_k(t_k)-I)\Psi_2)\|.\enqu

This estimate gives weak, hence strong continuity of $U(g)$.
\end{proof}

\begin{defi}\label{6.3.11} Let $E_\mfkg$ be the ${}^*$-isomorphism of
$\mfk C_{bs}$ into $p_G\mAss$ described in\rref 6.3.6(1)~. Let $\tau_t^Q\in\maut p_G\mAss$
$(t\in\mbR)$ denote the one-parameter group determined by

\bequ\label{6.3.11(1)} \tau_t^Q(E_\mfkg(\mbs{f})) := E_\mfkg(\mbs{f}_t),\ t\in\mbR,\
\mbs{f}\in\mfk C^G_{bs},\enqu

\noidt where $\mbs{f}_t\in\mfk C^G_{bs}$ was introduced in\rref 6.3.8(1)~. The uniqueness of the
extension of\rref 6.3.11(1)~ to the whole $P_G\mAss$ is given by uniqueness of the normal
extension of the representations $\tau_t^Q:\ \mfkA\rarw p_G\mAss$ to the representations of \Ass\
in $p_G\mAss$, \cite[1.21.13]{sak1}, and the automorphism property of these extensions is given by
the $\tau^Q$-invariance of $p_G$ (hence, $\tau^Q_t(\id_{\mAss}-p_G)=0$ for all $t$ and $Q$). The
automorphism group $\tau^Q$ will be called the \emm mean-field time evolution~ of the system
$(\mfkA;\msg(G))$ {\bf determined by the classical Hamiltonian function $Q$}.
\end{defi}

\begin{thm}\label{6.3.12} Let $E_\mfkg$ be a nontrivial G-measure
associated with the system $(\mfkA;\msg(G))$, cf. \ref{5.2.3}, with $K := \supp E_\mfkg\subset
\mfkgs$ such that $\msg(G)\subset\maut \mfkA$ is $\msg(\mfkA,p_G\mAs)$-continuous $(p_G :=
E_\mfkg(K))$. Let $\tau^Q\subset\maut E_\mfkg(\mfk C^G_{bs})$ be the mean-field time evolution of
$(\mfkA;\msg(G))$ determined by any $Q\in C^\infty(\mfkgs,\mbR)$. Let $\mfkA^J$ be any
$\msg(G)$-invariant \Csa\ of \fkA. Then:\nl

\noidt (i)\quad $\mfk N^c := E_\mfkg(C_b)$ and $\mfk C^J := E_\mfkg(\mfk C^{GJ}_{bs})$ are
$\tau^Q$-invariant \Csa s  of the 'algebra of mean-field observables' $\mfk C := E_\mfkg(\mfk
C^G_{bs})\subset p_G\mAss$. \nl

\noidt (ii)  $\tau^Q$ is a $\msg(\mfk C,\mS_\mfkg)$-continuous group, i.e. for any $y\in\mfk C$
and for any $\mome\in p_G\mSs(\mAss)=: \mS_\mfkg$ the function $t\mapsto \mome(\tau^Q_t(y))$ is
continuous and the states $\mome\circ \tau_t^Q: y\mapsto\mome(\tau^Q_t(y))$ belong to $\mS_\mfkg,
\ \mome\circ\tau_t^Q\in p_G\mAs$. \nl

\noidt (iii) Let $\{\xi_j: j=1,\dots n\}$ be a fixed basis of \fkg\ and $F_j := F(\xi_j)$ be the
coordinates of $F\in\mfkgs$ in the dual basis. Let \emn$\delta_{\xi_j}:\mfkA\rarw\mfkA$~ be {\bf
the derivations} (defined on $\msg(\mfkA,p_G\mAs)$-dense domains in \fkA) of the one parameter
subgroups $\msg(\exp t\xi_j)$ of $\msg(G)$. Then the infinitesimal generator of the group $\tau^Q$
is the \emm derivation $\delta_Q$~ on $\mfk C$ expressed by:

\barr\label{6.3.12(1)} \delta_Q(E_\mfkg(\mbs{f})) &:=& \left.\frac{\rd}{\rd
t}\right|_{t=0}\tau^Q_t(E_\mfkg(\mbs{f}))=\\
&=&\sum_{j=1}^n\,\int\left(\partial_j\mbs{f}(F)\{Q,F_j\}(F)-\partial_jQ(F)\,
\delta_{\xi_j}(\mbs{f}(F))\right)\,E_\mfkg(\rd F),\nonumber \earr

\noidt where the derivation is taken in the $\msg(\mfk C,\mS_\mfkg)$-topology, the symbol
$\partial_jf(F)$ means the derivative of a function on \fkgs\ with respect to the variable $F_j$
in the point $F\in\mfkgs$, and the meaning of the integral is explained in \ref{6.3.6}.
$\{Q,F_j\}$ is here the Poisson bracket on \fkgs, \ref{6.1.2}.\nl

\noidt (iv) If the group $\msg(G)$  is strongly continuous (i.e. $g\mapsto \msg_g(x)$ is
continuous in norm for each $x\in\mfkA$), and if $K$ is compact, then the group $\tau_t^Q$ will be
strongly continuous.
\end{thm}

\begin{proof} The group $\tau^Q$ is considered here as an
automorphism group of the $\tau^Q$-invariant subalgebra $E_\mfkg(\mfk C^G_{bs})=:\mfk C$ of
$p_G\mAss$.\nl

\noidt (i) The invariance of $\mfk N^c$ is given by the invariance of $C_b$ with respect to the
transformations\rref 6.3.8(1)~, which is valid due to the invariance of scalars in \fkA\ \wrt\
$\msg(G):\ \msg_g(\mlam\,\id_{\mfkA})=\mlam\,\id_\mfkA,\ \mlam\in\mbC,\ \forall g\in G$.
Similarly, the relation $\msg(G)(\mfkA^J)=\mfkA^J$ gives the $\tau^Q$-invariance of $\mfk C^J$.\nl

\noidt (ii) The continuity of the functions $t\mapsto \mome(\tau^Q_t(y))\ (\mome\in\mS_\mfkg,\
y\in\mfk C)$ can be obtained from the definition of the evolution $\mbs{f}\mapsto \mbs{f}_t$ in
$\mfk C^G_{bs}$ as well as from the definition\rref 6.3.6(2)~ of $E_\mfkg(\mbs{f})$ as follows:

Due to the $s^*$-bicontinuity of the mappings\rref 6.3.7(4)~ and due to the (bi-)continuity of the
functions $g_Q$ and $\mphi^Q$, the functions

\bequ\label{6.3.12(2)} \Psi(m):\ t\mapsto\Psi_t(m):=\mome_m(\mbs{f}_t(F_m)),\
m\in\supp\mu_\mome,\enqu

\noidt are continuous for any fixed $\mome\in\mS_\mfkg$ and $\mbs{f}\in\mfk C^G_{bs}.$ We have
proved in\rref 6.3.6(5)~ the measurability of all the functions $\Psi_t:\ m\mapsto \Psi_t(m)$.
Since $|\Psi_t(m)|\leq \|\mbs{f}\|\ (t\in\mbR,\ m\in\supp\mu_\mome)$ and $\mu_\mome$ is finite, an
application of the Lebesgue dominated convergence theorem gives

\bequ\label{6.3.12(3)} \lim_{t\rarw 0}\mome(\tau^Q_t(E_\mfkg(\mbs{f}))=\lim_{t\rarw
0}\int\Psi_t(m)\,\mu_\mome(\rd m)=\int\Psi_0(m)\,\mu_\mome(\rd m)=\mome(E_\mfkg(\mbs{f})).\enqu

\noidt This gives the desired continuity.

Any $\tau_t^Q$ can be considered as a ${}^*$-automorphism of the \Wa\ $p_G\mAss$, and each such
automorphism is $\msg(p_G\mAss,\mS_\mfkg)-\msg(p_G\mAss,\mS_\mfkg)$-continuous, cf.
\cite[4.1.23]{sak1}. This implies that the state $\mome\circ\tau^Q_t$ is a normal state on
$p_G\mAss$  together with \ome, hence $\mome\in\mS_\mfkg$ implies that
$\mome\circ\tau^Q_t\in\mS_\mfkg$.\nl

\noidt (iii) We shall calculate the derivation $\delta_Q$ from\rref 6.3.12(1)~ by calculating the
derivatives of the functions $\Psi(m)$ in\rref 6.3.12(2)~. For 'sufficiently nice' elements
$E_\mfkg(\mbs{f})\in D(\delta_Q)$ (:= the domain of $\delta_Q$) we have:

\bequ\label{6.3.12(4)} \left.\frac{\rd}{\rd t}\right|_{t=0}
\mome(\mbs{f}_t(F))=\left.\frac{\rd}{\rd
t}\right|_{t=0}\mome(\mbs{f}(\mphi_t^QF))+\left.\frac{\rd}{\rd t}\right|_{t=0}
\mome\left(\msg(g^{-1}_Q(t,F))(\mbs{f}(F))\right).\enqu

\noidt For the calculation of the first term we shall use the classical evolution equation\rref
6.2.10(10)~, where we shall consider $f(F)$ as a function of coordinates $F_j:= F_j(0),\
F_j(t):=F_j(\mphi_t^QF):=\mphi_t^QF(\xi_j)$:

\bequ\label{6.3.12(5)} \frac{\rd}{\rd t}f(\mphi_t^QF)=\sum_{j=1}^n
\partial_jf(\mphi_t^QF)\,\frac{\rd}{\rd t}F_j(\mphi_t^QF)=\sum_{j=1}^n
\partial_jf(\mphi_t^QF)\{Q,F_j\}(\mphi_t^QF).\enqu

\noidt Insertion of $f(F):=\mome(\mbs{f}(F))$ into\rref 6.3.12(5)~ and setting $t=0$ we obtain

\bequ\label{6.3.12(6)} \left.\frac{\rd}{\rd t}\right|_{t=0}
\mome(\mbs{f}(\mphi_t^QF))=\sum_{j=1}^n
\partial_j\mome(\mbs{f}(F))\{Q,F_j\}(F).\enqu

\noidt The second term in\rref 6.3.12(4)~ can be calculated with a help of\rref 6.1.3(2)~ +\rref
6.1.3(9)~ +\rref 6.1.3(10)~, and by considering that for any $\xi\in\mfkg$ we have defined

\bequ\label{6.3.12(7)} \left.\frac{\rd}{\rd t}\right|_{t=0} \mome(\msg(\exp
t\xi)(x))=\mome(\delta_\xi(x)),\quad x\in D(\delta_\xi)\subset\mfkA.\enqu

\noidt One obtains

\bequ\label{6.3.12(8)} \left.\frac{\rd}{\rd t}\right|_{t=0}
\mome(\msg(g_Q(t,F))(x))=\sum_{j=1}^n\partial_j Q(F)\mome(\delta_{\xi_j}(x)),\ x\in\bigcap_{j=1}^n
D(\delta_{\xi_j}).\enqu

\noidt Combining\rref 6.3.12(6)~ and\rref 6.3.12(8)~, where we set $\mome:=\mome_m,\ F:=F_m$ and
$x := \mbs{f}(F_m)$, we obtain for the `sufficiently nice' $\mbs{f}\in\mfk C^G_{bs}$:

\bequ\label{6.3.12(9)} \left.\frac{\rd}{\rd
t}\right|_{t=0}\mome(\tau_t^QE_\mfkg(\mbs{f}))=\sum_{j=1}^n
\int\mome_m\left(\partial_j\mbs{f}(F_m)\{Q,F_j\}(F_m)-
\partial_jQ(F_m)\delta_{\xi_j}(\mbs{f}(F_m))\right)\,\mu_\mome(\rd
m).\enqu

\noidt The change of the sign is caused by the replacement of $g_Q$ by $g_Q^{-1}$ in\rref
6.3.12(8)~. The comparison of\rref 6.3.12(1)~ with\rref 6.3.12(9)~ gives the result.\nl

\noidt (iv) We have to prove that the functions

\bequ\label{6.3.12(10)} t\mapsto \|\mbs{f}_t-\mbs{f}\|\ \text{for\ all}\ \mbs{f}\in\mfk C_{bs}^Q
\enqu

\noidt are continuous at $t=0$. Let us write

\barr\label{6.3.12(11)} \|\mbs{f}_t(F)-\mbs{f}(F)\| &=&
\|\msg^{-1}(g_Q(t,F))(\mbs{f}(\mphi_t^QF))-\mbs{f}(F)\|\leq\nonumber \\
\leq \|\msg(g^{-1}_Q(t,F))(\mbs{f}(F)-\mbs{f}(F_0))\| &+&
\|\msg(g_Q^{-1}(t,F)(\mbs{f}(F_0))-\mbs{f}(F_0)\|+\|\mbs{f}(F_0)-\mbs{f}(F)\|\nonumber\\
= 2\|\mbs{f}(F_0)-\mbs{f}(F)\| &+& \|\msg(g^{-1}_Q(t,F))(\mbs{f}(F_0))-\mbs{f}(F_0)\|. \earr

\noidt The strong continuity of $\msg(G)$ and the joint continuity of $g_Q$ lead to existence of
an open interval $I(F_0,\mveps)\subset\mbR$ containing $t = 0$ as well as of an open neighbourhood
of $F_0,\ \mcl U(F_0,\mveps)\subset K$, corresponding to any $F_0\in K$ and to any $\mveps >0$,
such that

\bequ\label{6.3.12(12)} \|\msg(g^{-1}_Q(t,F))(\mbs{f}(F_0))-\mbs{f}(F_0)\|<\frac{\mveps}{3},\
\text{for all}\ (t;F)\in I(F_0,\mveps)\times \mcl U(F_0,\mveps).\enqu

\noidt The strong continuity of $\msg(G)$ leads also to norm continuity of the functions
$\mbs{f}_0$ in\rref 6.3.2(5)~ which generate $\mfk C^G_{bs}$, hence all $\mbs{f}\in\mfk C^G_{bs}$
are continuous in norm in the present case. This shows that we can choose the neighbourhoods $\mcl
U(F_0,\mveps)$ in such a way that

\bequ\label{6.3.12(13)} \|\mbs{f}(F)-\mbs{f}(F_0)\|<\frac{\mveps}{3},\ {\rm if}\ F\in \mcl
U(F_0,\mveps),\ \text{for any}\ F_0\in K.\enqu

\noidt Since $K$ is compact, we can find a finite set $\{F_p: p=1,2,\dots P\}\subset K$ such that
the union of $\{\mcl U(F_p,\mveps): p=1,2,\dots P\}$ covers $K$. Let $I(\mveps)$ be the
intersection of the intervals $\{I(F_p,\mveps): p = 1,2,\dots P\}$. Then

\bequ\label{6.3.12(14)} \|\mbs{f}_t(F)-\mbs{f}(F)\|<\mveps,\ \text{for all}\ (t;F)\in
I(\mveps)\times K.\enqu

\noidt Taking supremum in\rref 6.3.12(14)~ we obtain the desired continuity in\rref 6.3.12(10)~.
\end{proof}

\pt\label{6.3.13}\rm To compare the derivations $\delta_Q$ from\rref 6.3.12(1)~ with $\delta_\pi$
from the formulas\rref 6.2.20(6)~, it suffices to take $\mbs{f}\in\mfk C^{GJ}_{bs}$ where
$\mfkA^J:=\mfkA^N$ is a $\msg(G)$-invariant 'local algebra'. For such an $\mbs{f}$ we have
\bequ\label{6.3.13(1)} \msg(\exp t\xi)(\mbs{f}(F)) = \exp(-itX_\xi^N)\mbs{f}(F)\exp(itX_\xi^N),\
t\in\mbR,\ F\in\mfkgs,\enqu

\noidt for any $\xi\in\mfkg$; here we made the usual identifications, cf. notation in
\ref{6.3.10}. Then we have \bequ\label{6.3.13(2)}
\delta_\xi(\mbs{f}(F))=-i\,[X_\xi^N,\mbs{f}(F)],\enqu

\noidt where the commutator is taken between operators in the Hi1bert space $p_G\mH_u$. We can sea
easily now that the derivations $\delta_\pi$ and $\delta_Q$ are expressed by identical formulas.
This proves the identity of the time evo1utions determined in Sec.\ref{sec;6.2} with the
evolutions from the present section in the case of the \emm UHF-algebra~ $\mfkA := \mfkA^\Pi$
(cf. \cite[2.6.12]{bra&rob},\cite[6.4.1]{pedersen}; UHF:=``uniformly hyperfinite'') with the
polynomial $Q$. This shows also that the derivation $\delta_Q$ for the case of a nonseparable
$\mfkA^\Pi$ and unbounded $X_\xi$ is described by the same formulas as $\delta_\pi$ is.

\section{Equilibrium states}\label{sec;6.4}

\pt\label{6.4.1}\rm Let us consider in this section those states of physical systems which
describe the situations corresponding to the \emn thermodynamic equilibrium~ at a given
temperature $T\geq 0$. For quantal systems these states are specified usually by the \emm
KMS-condition~, cf. e.g. \cite{ruelle1,emch1,bra&rob2,pedersen}. We shall investgate here the \emn
KMS states~\footnote{KMS is for Kubo, Martin and Schwinger.} of systems considered in this
chapter, i.e. the systems specified by the triple $(\mfkA;\msg(G);\tau^Q)$, cf. also \cite{bon2}.
To avoid possible technical complications, we shall concentrate our attention here on the cases of
strongly continuous time evolutions $\tau^Q$ including, e.g. the cases described in
\ref{6.3.12}(iv). Let us use the notation of Theorem \ref{6.3.12}, hence $\mfk C := E_\mfkg(\mfk
C^G_{bs})$ be the \Ca\ of (generalized) observables describing the considered system with the
dynamics $\tau^Q$. Instead of the above mentioned triple, we shall use also the couple $(\mfk
C;\tau^Q)$ for denoting the system. In most of the analysis of this section an additional
structure of the system will be used. Let $\Pi$ be a locally compact noncompact group and
$\pi(\Pi)$ be its representation on $\mfk C$, i.e. $\pi(p)\in\maut\mfk C$ for all $p\in\Pi$. Let
$\pi(\Pi)$ commutes with $\tau^Q:$

\bequ\label{6.4.1(1)} \tau^Q_t\circ\pi(p)=\pi(p)\circ\tau^Q_t\quad\text{for all}\  t\in\mbR,\
p\in\Pi.\enqu

 We shall assume usually that $\pi(\Pi)$ has some \emn asymptotic abelianess~
properties. As an example of such a $\pi(\Pi)$ consider the situations described in
Sec.\ref{sec;5.1}. (i.e. $\mfkA := \mfkA^\Pi$ is a tensor product of the mutually commuting 'local
algebras' $\mfkA_p := \mcl L(\mH_p)$), where the set $\mbZ_+\setminus\{0\}$ is replaced by
$\Pi:=\mbZ^r$ (with easy modifications of the whole formalism). Let us write $\pi_p: \mcl
L(\mH)\rarw \mcl L(\mH_\Pi)$ for the isomorphism defined in\rref 5.1.3(4)~, $p\in\Pi$. Now we
define $\pi(p)\in\maut \mfkA^\Pi$ by

\bequ\label{6.4.1(2)} \pi(p)(\pi_j(A)):=\pi_{j+p}(A),\quad\text{for all}\  A\in\mLH,\quad p,\
j\in\Pi.\enqu

\noidt Since the elements $\pi_j(A)\ (j\in\Pi,\ A\in\mLH)$ generate $\mfkA^\Pi$,\rref 6.4.1(2)~
determines an automorphism $\pi(p)$ of $\mfkA^\Pi$ uniquely. This automorphism can be extended
naturally to an (equally denoted) automorphism group $\pi(\Pi)$ of $\mfk C:= E_\mfkg(\mfk
C^G_{bs})$ by the relation

\bequ\label{6.4.1(3)} \pi(p)(E_\mfkg(\mbs{f})):=\int\pi(p)(\mbs{f}(F))\,E_\mfkg(\rd F).\enqu

\noidt The group $\pi(\Pi)$ is \emn norm-asymptotically abelian~, i.e.

\bequ\label{6.4.1(4)} \lim_{p\rarw\infty}\|[\pi(p)(x),y]\|=0,\quad\text{for all}\  x,y\in\mfk
C.\enqu

In more general cases, the abelianess  properties of the action of $\Pi$ on $\mfk C$ can be
weaker. Systems with this structure {\bf will be denoted}
\[\centerline{\emn$(\mfk C;\tau^Q;\pi(\Pi))$~,
{\bf or} \emn$(\mfkA;\msg(G);\tau^Q;\pi(\Pi))$~.}\]

\noidt We shall use, as usual,  \emn$\beta := T^{-1}:= (kT)^{-1}$~ to denote the inverse \emn
temperature~ in convenient units. The following definitions are found e.g. in \cite[5.3.1, 5.3.18,
and 5.3.21]{bra&rob2}, and \cite[8.12]{pedersen}.

\begin{defi}\label{6.4.2} Let $(\mfk C,\tau)$ be a \emn$C^*$-dynamical
system~, i.e. the one parameter group \linebreak $\tau\subset\maut\mfk C$ is strongly continuous.
The state $\mome\in\mS(\mfk C)$ is defined to be a \emm $\tau$-KMS state at value $\beta\in\mbR$~,
or a \emn $(\tau,\beta)$-KMS state~, if
\bequ\label{6.4.2(1)}
\mome(x\,\tau_{i\beta}(y))=\mome(yx),\ \text{for all}\  x,y,\in\mfk C^\circ_\tau,\enqu

\noidt where $\mfk C^\circ_\tau$ is a norm-dense, $\tau$-invariant ${}^*$-subalgebra of the set
$\mfk C_\tau$ of the entire analytic elements of $\mfk C$:
\bequ\label{6.4.2(2)} y\in\mfk
C^\circ_\tau\eequiv\ {\rm the\ function}\ z\mapsto\tau_z(y)\ {\rm is\ analytic\ for\ all}\
z\in\mbC.\enqu

Let $\delta_\tau$ be the generator of $\tau$. Then $\mome\in\mS(\mfk C)$ is called a \emm
$\tau$-ground state~ if
\bequ\label{6.4.2(3)} -i\,\mome(y^*\delta_\tau(y))\geq 0,\quad \text{for
all}\  y\in D(\delta_\tau).\enqu

\noidt In this case, \ome is also called a \emm $\tau$-KMS state at value $\beta=\infty$~.
\end{defi}

\begin{defi}\label{6.4.3}Let $(\mfk C;\tau)$ be a $C^*$-dynamical
system with a unital \Ca\ \fk C, and let $\delta_\tau$ be the infinitesimal generator of $\tau$.
Then $\mome\in\mS(\mfk C)$ is said to be a \emm passive state~ if
\bequ\label{6.4.3(1)}
-i\,\mome(u^*\delta_\tau(u))\geq 0\enqu

\noidt for any $u\in D(\delta_\tau)$ belonging also to the connected component of the identity of
the unitary group of \fk C in the norm topology. \end{defi}

\pt\label{6.4.4}\rm {\bf Let us collect here some important properties of the sets \emn$\mcl
K_\beta$~ of $(\tau,\beta)$-KMS states:}

Proofs of the listed facts can be found in \cite[Chap.5]{bra&rob2}, or in \cite[4.3]{sak2}. We
shall consider $\beta\in (0,\infty]$, the set $\mcl K_\infty$ being the set of all ground states
$\mome\in\mS(\mfk C)$. Let $(\mfk C,\tau)$ be a $C^*$-dynamical system. Then: \nl

\noidt (0) Any state $\mome\in\mcl K_\beta$ is $\tau$-invariant: $\mome\circ\tau_t=\mome\
(t\in\mbR)$.\nl

\noidt (i) Any $\mcl K_\beta$ is a \emn convex $W^*$-compact~ subset of $\mS(\mfk C)$.\nl

\noidt (ii-a) For $\beta\neq\infty$, $\mcl K_\beta$ is a \emn simplex~ in $\mS(\mfk C)$.\nl

\noidt (ii-b) $\mcl K_\infty$ is a \emn face~ in $\mS(\mfk C)$.\nl

\noidt (iii-a) The set $\mcl{EK}_\beta$ of extrema1 points $\mome\in\mcl K_\beta$\
$(\beta\neq\infty)$ consists of \emn factor states~: The centers of $\pi_\mome(\mfk C)''$\  are
trivial.\nl

\noidt (iii-b) The extremal points $\mome\in\mcl K_\infty$, i.e. $\mome\in\mcl{EK}_\infty$, are
\emn pure states~: $\mome\in\mcl{ES}(\mfk C)$, i.e. $\pi_\mome(\mfk C)''=\mcl L(\mH_\mome)$.\nl

\noidt (iv) $\mome_j\in\mcl{EK}_\beta\ (\beta\neq\infty, j=1,2)$ implies either $\mome_1=\mome_2$,
or $\mome_1 \perp \mome_2$, i.e. $\mome_1$ and $\mome_2$ are mutually \emn disjoint~, i.e. the
\emn central cover~s $s_{\mome_1}$ and $s_{\mome_2}$ of the corresponding GNS-representations are
mutually orthogonal. \nl

\noidt (v) The \emn extremal decomposition~ of $\mome\in\mcl K_\beta\ (\beta\neq\infty)$ coincides
with its \emn central decomposition~, cf. \cite[Chap. 4]{bra&rob}, \cite[Chap. 4]{pedersen}. The
corresponding probability measure $\mu_\mome^c$ on $\mS(\mfk C)$ is \emn pseudosupported~ (cf.
\cite[Chap. 6]{bra&rob2})  by $\mcl{EK}_\beta$ and if the Hilbert space of the GNS-representation
$\mH_\mome$ is separable, then $\mu_\mome^c$ is supported by $\mcl{EK}_\beta:\
\mu^c_\mome(\mcl{EK}_\beta)=\mu_\mome^c(\mS(\mfk C))=1$.

\begin{lem}\label{6.4.5} Let $\mome\in\mS(\mfk C)$ be a $\tau$-ground state.
Let $(\pi_\mome,\mH_\mome,\Omega_\mome)$ be the corresponding GNS representation. Then for the
unique selfadjoint operator $Q_\mome$ on $\mH_\mome$ determined by the relation:

\bequ\label{6.4.5(1)} \exp(it\,Q_\mome)\,\pi_\mome(y)\Omega_\mome :=
\pi_\mome(\tau_t(y))\Omega_\mome, \quad\forall t\in\mbR,\enqu

\noidt the following is valid:

\bequ\label{6.4.5(2)} Q_\mome\geq 0,\ {\rm and\ for\ all}\ t\in\mbR\ {\rm one\ has}\
\exp(it\,Q_\mome)\in\pi_\mome(\mfk C)''.\enqu
\end{lem}

\begin{proof} See \cite[5.3.19]{bra&rob2}.\end{proof}

\pt\label{6.4.6}\rm Any $(\tau,\beta)$-KMS state, according to
\ref{6.4.4}(i), can be approximated in the $w^*$-topology by
convex combinations of extremal KMS states at the same temperature
$\beta^{-1}$. The set $\mcl K_\beta$ may be void for a genera1
dynamical system and for a given $\beta\in (0,\infty]$. Occurrence
of more than one points in $\mcl K_\beta$ means occurrence of
several mutually disjoint states in $\mcl{EK}_\beta$. Orthogonal
central projectors $s_1$ and $s_2$ (the central covers of the
corresponding GNS representations) are supporting such disjoint
states; these $s_j\in\mfk Z$\ (:= the center of $\pi_u(\mfk C)''$)
may be interpreted as corresponding to distinct values of a
macroscopic (global, classical) quantity for distinct $j = 1,2$.
We interpret this situation as possibility of existence of several
mutually different 'phases' of the considered system at the
temperature $T = \beta^{-1}$. This interpretation is especially
intuitive in cases of quasilocal algebras \fk C\ when the extremal
KMS (hence factor) states have \emn short range correlations~ (cf.
e.g. \cite{lanf&ruel}) - the necessary property of the states
representing pure phases of a spatially extended system
\cite[§6.5]{ruelle1}. We shall investigate general properties of
the extremal $(\tau^Q,\beta)$-KMS states of the systems $(\mfk
C;\tau^Q)$ and $(\mfk C;\tau^Q;\pi(\Pi))$ representing the
generalized mean-field theories.

\begin{prop}\label{6.4.7} Let $\mome\in\mcl K_\beta$ be an extremal
$\tau^Q$-KMS state of a generalized mean-field theory $(\mfkA;\msg(G);\tau^Q)$. Then there is an
element $F_\mome\in\supp E_\mfkg$ such that the central support $s_\mome\leq E_\mfkg(B)$ for any
open $B\subset\mfkgs$ containing $F_\mome:\ F_\mome\in B$. {\bf The point} \emn$F_\mome$~ is a
fixed point of the {\bf c1assical flow} \emn$\mphi^Q$~ on \fkgs. The state \ome\ is invariant
\wrt\ the one parameter {\bf subgroup of $\msg(G)$ generated by the element}
\emn$\beta^Q_{F_\mome}\in\mfkg$~,\rref 6.1.3(9)~, and the {\bf generator} \emn$Q_\mome$~ of
\emn$\tau^Q$~ in \emn$\pi_\mome(\mfkA)$~ {\bf implements} this subgroup in the sense that

\bequ\label{6.4.7(1)} \pi_\mome\left(\msg(\exp(-\beta^Q_{F_\mome}t))(x)\right)\Omega_\mome=
\exp(itQ_\mome)\pi_\mome(x)\Omega_\mome,\ t\in\mbR,\ x\in\mfkA.\enqu

The image $\pi_\mome(\mfk C)$ of $\mfk C:= E_\mfkg(\mfk C^G_{bs})$ coincides with
$\pi_\mome(\mfkA), \, \mfkA=E_\mfkg(\mfkA)\ (\mfkA\subset\mfk C^G_{bs}$ represents here
\fkA-valued constant functions).

Assume that the whole group $\msg(G)$ is unitarily implemented in the representation
$(\pi_\mome,\mH_\mome,\Omega_\mome)$. Then we can choose the generators $X_\mome(\xi)$ of the one
parameter subgroups $\exp(t\xi)$ in such a way that

\bequ\label{6.4.7(2)} Q_\mome=X_\mome(\beta^Q_{F_\mome})=
\sum_{j=1}^n\,\partial_jQ(F_\mome)\,X_\mome(\xi_j) \enqu

\noidt for any basis $\{\xi_j:\, j=1,2,\dots n\}$ in \fkg.
\end{prop}

\begin{proof} The factor state \ome  is projected by $p_M$ onto a pure state on
$\mfk N_G$, \ref{5.1.35}, hence the decomposition of \ome  in\rref 6.3.6(2)~ is concentrated on a
one point set $F_\mome\in\supp E_\mfkg$. Let $\mbs{f}_j\ (j = 1,2)$ be any such elements of $\mfk
C_{bs}^G$ that $\mbs{f}_1(F_\mome)=\mbs{f}_2(F_\mome)$. Then

\bequ\label{6.4.7(3)} \mome(E_\mfkg(\mbs{f}_1))=\mome(\mbs{f}_1(F_\mome)) =
\mome(\mbs{f}_2(F_\mome))=\mome(E_\mfkg(\mbs{f}_2)).\enqu

\noidt This proves that $\pi_\mome(\mfk C)=\pi_\mome(\mfkA)$. The state
$\mome\circ\tau_t^Q\equiv\mome$ is then concentrated (in the above described sense) on
$\mphi_t^Q(F_\mome)$, and states $\mome_1$ and $\mome_2$ concentrated on $F_1\neq F_2$ are
disjoint: $\mome_1\perp \mome_2$. Hence, $\mphi_t^Q(F_\mome)=F_\mome$ for all $t\in\mbR$. This
means, however, that the classical Poisson bracket $\{Q,f\}(F_\mome) = 0$ for any function $f$. It
follows that for the generator $\delta_Q$,\rref 6.3.12(1)~, in the representation $\pi_\mome$, one
has:

\bequ\label{6.4.7(4)} \mome(x\delta_Q(E_\mfkg(\mbs{f}))y)=
-\sum_{j=1}^n\,\partial_jQ(F_\mome)\,\mome(x\delta_{\xi_j}(\mbs{f}(F_\mome))y),\ x,y\in\mfkA.\enqu

 The definition of the time evolution in \ref{6.3.8} and the
$\mphi^Q$-invariance of $F_\mome$ shows the identity of the time evolution of $\pi_\mome(\mfkA)$
with the action of the one-parameter group $g_Q^{-1}(t,F_\mome),$ cf.\rref 6.1.3(5)~, with the
generator -$\beta^Q_{F_\mome}$, cf.\rref 6.1.3(2)~. According to\rref 6.1.3(9)~ and\rref
6.1.3(10)~, we obtain the remaining assertions of the proposition.
\end{proof}

 \begin{noti}\label{6.4.8} The
generator of the mean-field time evolution \emn$\tau^Q$~ of local perturbations of an extremal
equilibrium state \ome\ given in\rref 6.4.7(2)~ is usually called the \emn Bogoliubov-Haag
Hamiltonian~, cf. \cite{bogoliub,haag1,thir&wehrl}.
\end{noti}

\pt\label{6.4.9}\rm We shall assume in the following that \fkA\  is a quasilocal \Ca\ generated by
a net $\{\mfkA^J: J\subset \Pi, J\, {\rm finite}\}$ of local subalgebras $\mfkA^J$ commuting with
each other for disjoint $J$'s:

\bequ\label{6.4.9(1)} x\in\mfkA^J,\ y\in\mfkA^{J'},\ J\cap J'=\emptyset\imply [x,y]=0.\enqu

\noidt Here $\Pi$ is a countable infinite commutative group acting on \fkA\ by the
representation\break  $\pi: \pi_p\in\maut\mfkA$, in such a way that $\pi_p:\
\mfkA^J\rarw\mfkA^{J+p}$ is an isomorphism for any $J\subset\Pi$. This is the situation from\rref
6.4.1(2)~, where \LH\ is identified with $\mcl L(\mH_u)$, $\pi_0=\pi(0)=\id_{\mcl L(\mfkA)}$ ($0$
is here the identity of the group $\Pi$), hence $\pi(p)=\pi_p\ (p\in\Pi)$.

It will be assumed in the following that each $\mfkA^J\ (J\subset\Pi)$ is \sg(G)- invariant, and
that the action of \sg(G)\  commutes with $\pi(\Pi)$. Then also\rref 6.4.1(1)~ will be fulfilled
($\pi(\Pi)$ is naturally extended to the equally denoted automorphism groups of $\mfk C$ and of
\Ass).

In this situation, let $\mome\in\mS(\mfkA)$ be a factor state which is invariant \wrt\ the action
of $\pi(\Pi)$:
\bequ\label{6.4.9(2)} \mome(\pi_p(x))=\mome(x),\ \text{for all}\ x\in \mfkA,\
p\in\Pi.\enqu

 The locally normal factor states have \emn short range correlations~, \cite{lanf&ruel},
 \cite[Thm.2.6.10]{bra&rob},
  hence they are \emn weakly $\pi(\Pi)$-clustering~, and
\bequ\label{6.4.9(3)} \lim_{p\rarw\infty}\mome(\pi_p(x)y)=\mome(x)\mome(y),\ \text{for all}\
x,y,\in\mfkA.\enqu

 If $\mfkA^J$ are faithfully represented in Hilbert spaces $\mH_J$, as it was
the case of Sec.\ref{sec;5.1}, then $\pi_p$ will be used also for translations of unbounded
operators acting on $\mH_J$ to unitarily equivalent operators acting on $\mH_{J+p}$ (e.g. by
translating their spectral projectors belonging to $\mfkA^J$); this can be done if the
isomorphisms of $\mfkA^J\subset \mcl L(\mH_J)$ with $\mfkA^{J+p}\subset \mcl L(\mH_{J+p})\
(J\subset\Pi,\ p\in\Pi)$ are spatial. We shall write also $\mfkA_p := \mfkA^J$ with $J:=\{p\}$ :=
the one-point set, $p\in\Pi$. Let all the $\mfkA^J\ (J\subset\Pi)$ have common unit and let the
\Ca s $\mfkA_p$ with $p\in J$ generate $\mfkA^J\ (J\subset \Pi)$.

With the introduced notation and assumptions, we shall prove now the following:

 \begin{thm}\label{6.4.10} Let us consider a system
 $(\mfkA;\msg(G);\tau^Q;\pi(\Pi))$ with
 simple \Ca\ \fkA\ and `local' subalgebras $\mfkA^J\subset\mfkA$ being factors for all finite $J$.
Let $\mome\in\mS(\mfkA)$ and let $\mome^0$ be the restriction of \ome\ to the subalgebra $\mfkA_0\
(:=\mfkA^J$ with the one-point set $J$ containing the identity $0\in\Pi)$. Then the following two
statements are equivalent:\nl

\noidt (i) \ome\ is a locally normal extremal $\tau^Q$-KMS state at a positive temperature
$\beta^{-1}>0$.\nl

\noidt (ii) \ome = $\overline{\mome}$, where $\overline{\mome}$ is the $\pi(\Pi)$-invariant
product state determined by the relation

\bequ\label{6.4.10(1)} \overline{\mome}(\pi_{p_1}(x_1)\pi_{p_2}(x_2)\dots
\pi_{p_m}(x_m))=\prod_{j=1}^m\,\mome^0(x_j), \enqu

\noidt with $x_j\in\mfkA_0,\ p_j\in\Pi\ (p_j\neq p_k\ {\rm for}\ j\neq k)$, $j= 1,2,\dots m,\
\forall m\in\mbN$ and $\mome^0$ is the faithful normal KMS-state at $\beta$ on $\mfkA_0$
corresponding to the one-parameter subgroup $\{\msg(\exp(-t\beta^Q_{F_\mome})):\ t\in\mbR\}$ of
$\maut \mfkA_0$ with \bequ\label{6.4.10(2)} \mphi^Q_t(F_\mome)=F_\mome,\ \text{for all}\ t\in\mbR
\enqu

\noidt for some element $F_\mome\in\mfkgs$. Moreover, the `consistency condition'

\bequ\label{6.4.10(3)} \mome(E_\mfkg(f_\xi))=F_\mome(\xi),\quad (\xi\in\mfkg,\,  f_\xi(F):=
F(\xi)\ {\rm for}\  F\in\mfkgs)\enqu

\noidt is fulfilled.\footnote{The stationarity\rref 6.4.10(2)~ is a consequence of the
``consistency condition''\rref 6.4.11(1)~, i.e. of\rref 6.4.10(3)~; hence\rref 6.4.10(2)~\,
\&\!\rref 6.4.10(3)~ can be replaced by\rref 6.4.11(1)~.}
\end{thm}

\begin{proof} (i) implies
$\pi_\mome(\tau_t^Q(x))=\pi_\mome(\msg(\exp(-t\beta^Q_{F_\mome}))(x))$ according to\rref
6.4.7(1)~. Hence \ome\ satisfies the KMS-condition \wrt\ the group
$\msg(\exp(-t\beta^Q_{F_\mome}))$ at $T^{-1}$ and the same is true for $\mome^0$, since
$\msg(G)(\mfkA_0)=\mfkA_0$. Let $X(\beta^Q_{F_\mome})$ be the restriction of
$X_\mome(\beta^Q_{F_\mome})$ onto $\overline{\pi_\mome(\mfkA_0)\Omega_\mome}$. \ome\ is faithful
on \fkA\ (\fkA\ is simple) and the cyclic vector $\Omega_\mome$ is separating for
$\pi_\mome(\mfkA)''$, cf. \cite[5.3.9]{bra&rob2}. Hence $\mome(x^*x)\neq 0$\ for $x\neq 0$, and
$\mome^0$ is faithful on $\mfkA_0$. The local normality of \ome\ implies normality of $\mome^0$.
According to the \emn Takesaki's theorem~ \cite[5.3.10]{bra&rob2}, the one-parameter automorphism
group of $\pi_\mome(\mfkA_0)$:

\bequ\label{6.4.10(4)} t\mapsto\exp(itX(\beta^Q_{F_\mome}))\pi_\mome(x)
\exp(-itX(\beta^Q_{F_\mome})),\quad x\in\mfkA_0,\enqu

\noidt coincides with the corresponding modular automorphism group of $\pi_\mome(\mfkA_0)$
determined by the state $\mome^0$ (up to a resca1ing of time $t$). According to
\cite[5.3.29]{bra&rob2}, the KMS state at $\beta:= T^{-1}\in\mbR$ on the factor $\mfkA_0$
corresponding to its automorphism group $\msg(\exp(-t\beta^Q_{F_\mome}))$ is uniquely determined
faithful normal state on $\mfkA_0$.

We have to prove that \ome\ is a $\pi(\Pi)$-invariant product state on \fkA, i.e. that\rref
6.4.10(1)~ (with $\mome\hookrightarrow \overline{\mome}$) is satisfied. Let $y := \pi_p(x)$ for
some $x\in\mfkA_0,\ p\in\Pi$. From the commutativity of $\pi(\Pi)$ with $\msg(G)$ we have for
$y':= \pi_p(x'):$

\bequ\label{6.4.10(5)}\mome(\tau^Q_t(y)y')=\mome\circ \pi_p(\tau^Q_t(x)x'),\ \text{for all}\
x,x'\in\mfkA_0,\ t\in\mbR.\enqu

\noidt We can write here $\mome^p\in\mS(\mfkA_p)$ instead of \ome. The state $\mome^p$ is a
KMS-state, hence $\mome^p\circ\pi_p\in\mS(\mfkA_0)$ is the unique KMS state $\mome^0$:

\bequ\label{6.4.10(6)} \mome^p\circ\pi_p=\mome^0,\ \text{for all}\ p\in\Pi.\enqu

\noidt Since all the $\mfkA^J$ are factors ($J$ finite), we can repeat the above considerations
for the restrictions $\mome^J$ of \ome\ to $\mfkA^J$ (with $J$ replacing the one point set
$\{0\}\subset\Pi):\  \mome^J$ is the unique KMS state at $T^{-1}$ of $\mfkA^J$ corresponding to
the group $\msg(\exp(-t\beta^Q_{F_\mome}))\in\maut\mfkA^J$, and

\bequ\label{6.4.10(7)} \mome^{J+p}\circ\pi_p=\mome^J \ \text{for\ all\ finite}\  J\subset\Pi,\
p\in\Pi.\enqu

\noidt For an arbitrary local element $x\in\mfkA^J$ one obtains:

\bequ\label{6.4.10(8)}\mome\circ\pi_p(x)=\mome^{J+p}\circ\pi_p(x)= \mome^J(x)=\mome(x),\enqu

\noidt hence we have the translation invariance $\mome\circ\pi_p=\mome$ of the extremal
$\tau^Q$-KMS state \ome\ at positive temperature $T$.

The restriction to $\mfkA^J$ of the product state $\overline{\mome}$ on the \rhs\ of\rref
6.4.10(1)~ satisfies the KMS condition at $T^{-1}$ with respect to the one parameter group
$\{\msg(\exp(-t\beta^Q_{F_\mome})):\ t\in\mbR\}\subset\maut\mfkA^J$, since for all $x_j,\
y_j\in\mfkA_0,\ j = 1,2,\dots m$, one has the identity
\barr\label{6.4.10(9)} &
\overline{\mome}\left(\pi_{p_1}(x_1)\pi_{p_2}(x_2)\dots\pi_{p_m}(x_m)
\tau_t(\pi_{p_1}(y_1)\pi_{p_2}(y_2)\dots\pi_{p_m}(y_m))\right)=\nonumber\\
& \overline{\mome}\left(\pi_{p_1}(x_1\tau_t(y_1))\pi_{p_2}(x_2\tau_t(y_2))\dots
\pi_{p_m}(x_m\tau_t(y_m))\right)=\\
& \prod_{j=1}^m\mome^0(x_j\tau_t(y_j)),\ \text{for\ all\ $m\=$tuples}\ \{p_1,p_2,\dots
p_m\}\subset\Pi,\ m=1,2,\dots, \nonumber \earr

\noidt where $\tau_t\in\maut\mfkA$ leaves all $\mfkA^J$ invariant: $\tau_t(\mfkA^J)=\mfkA^J,\
J\subset\Pi$. Setting $\tau_t := \msg(\exp(-t\beta^Q_{F_\mome}))$, we obtain the KMS-property of
$\overline{\mome}$ from the proved KMS-property of the state $\mome^0$, since the finite linear
combinations of the products
\bequ\label{6.4.10(10)}\pi_{p_1}(x_1)\pi_{p_2}(x_2)\dots\pi_{p_m}(x_m),\ x_j\in\mfkA_0,\
p_j\in\Pi,\ m\in\mbZ_+\setminus\{0\},\enqu

\noidt form such a subset $\mfkA_L^0$\ of \fkA, that the values
\bequ\label{6.4.10(11)}
\overline{\mome}(y)\in\mbC,\ y\in\mfkA_L^0,\enqu

\noidt determine any locally normal state $\overline{\mome}\in\mS(\mfkA)$ uniquely. The uniqueness
of the KMS-states on $\mfkA^J$ ($J$ finite) gives the restrictions of \ome\  to all the $\mfkA^J$,
hence we have equality $\overline{\mome}=\mome$ of the states on \fkA, hence the relation\rref
6.4.10(1)~. [Warning: This does not imply uniqueness of the $\tau^Q$-KMS states on $\mfk C$, but
we have proved uniqueness of the KMS states on $\mfk C$ \wrt\ one parameter groups
$\msg(\exp(t\xi)) =: \msg_\xi(t)$. Different extremal $\tau^Q$-KMS states at the same temperature
$T$ give different values of $F_\mome$ and of $\beta^Q_{F_\mome}$, hence lead to different one
parameter groups $\msg_\xi\ (\xi := \=\beta^Q_{F_\mome})$.]

Let now $\mome^0\in\mS(\mfkA_0)$ be a given faithful normal KMS-state at the temperature $T > 0$
corresponding to the group $\msg_\xi$ with $\xi:=\=\beta^Q_{F_\mome}$, where $F_\mome\in\mfkgs$
satisfies\rref 6.4.10(2)~. Then the product state $\overline{\mome}$ from\rref 6.4.10(1)~ is
locally normal, since the finite product of normal states is a normal state on the tensor product
of \Wa s, \cite[Sec.IV.5]{takesI}. The factoriality is trivial for product states,
\cite[2.6.10]{bra&rob}. According to the Pusz-Woronowicz theorem, \cite[5.3.22]{bra&rob2},
$\mome^0$ satisfies the passivity condition\rref 6.4.3(1)~ with $\tau := \msg_\xi\
(\xi:=\=\beta^Q_{F_\mome})$. This implies the satisfaction of\rref 6.4.3(1)~ \wrt\  the same group
by the state $\overline{\mome}$. The cluster property of the product state gives now the
KMS-property of $\overline{\mome}$ \wrt\ the $\msg_\xi$. Since $\mome^0$ satisfies $\msg_\xi$-KMS
condition with $T\neq 0$ positive, the same is true for $\overline{\mome}$. Since $F_\mome$ is a
fixed point of $\mphi^Q$, the derivations of the $\msg_\xi$ and of $\tau^Q$ coincide in the
GNS-representations corresponding to the states supported by $E_\mfkg(F_\mome)$, cf.\rref
6.4.7(4)~. The assumption\rref 6.4.10(3)~ ensures, that the macroscopic limit of the product state
$\overline{\mome}$ from\rref 6.4.10(1)~ is concentrated on $F_\mome$, hence the evolutions
$\tau^Q$ and $\msg_\xi\ (\xi:=\=\beta^Q_{F_\mome})$ coincide in the representation $\pi_\mome$
corresponding to the state $\mome:=\overline{\mome}$ from\rref 6.4.10(1)~.
\end{proof}

\begin{corl}\label{6.4.11}Let $\mfkA:=\mfkA^\Pi$\ and the system
$(\mfkA;\msg(G);\pi(\Pi))$ be defined according to Sec.\ref{sec;5.1}, i.e. the G-measure $E_\mfkg$
is given by \ref{5.1.33} and $\msg(G)$ is locally implementable in states $\mome\in\mS_\mfkg$.
Let, with the assumptions of Theorem \ref{6.4.10}, \ome\ be locally normal extremal $\tau^Q$-KMS
state at $T>0$. Let $X_\xi\ (\xi\in\mfkg)$ be the generators of the ($\msg(G)$-defining)
representation $U(G)$ on $\mH_0:=\mH,\ \mfkA_0=\mLH,\
\msg(\exp(t\xi))(y):=\exp(-itX_\xi)\,y\,\exp(itX_\xi)$ for all $y\in\mfkA_0$. Then

\bequ\label{6.4.11(1)} \mome^0(exp(itX_\xi))=\exp(itF_\mome(\xi)),\ \forall \xi\in\mfkg,\enqu

\noidt where $F_\mome$ is given by the (trivially fulfilled) `consistency condition'

\bequ\label{6.4.11(2)} \mome(E_\mfkg(f_\xi))=F_\mome(\xi),\ \xi\in\mfkg.\enqu
\end{corl}

\begin{proof}

Since $\exp(itX_\xi)\in\mfkA_0$, the generators of the restriction of $\msg(G)$ onto
$\mfkA_p:=\pi_p(\mfkA_0)$ are $\pi_p(X_\xi)$, where

\bequ\label{6.4.11(3)} \exp(it\pi_p(X_\xi)) := \pi_p(\exp(itX_\xi)). \enqu

\noidt The generators of the restriction of $\msg(G)$ onto $\mfkA^J$\ (finite $J\subset \Pi$) are
$X_\xi^J:=\sum_{p\in J}\pi_p(X_\xi)$,

\bequ\label{6.4.11(4)} \exp(itX_\xi^J) := \prod_{p\in J}\exp(it\pi_p(X_\xi))\in\mfkA^J. \enqu

\noidt \ome\ is expressed by\rref 6.4.10(1)~, hence according to\rref 5.1.29(3)~:

\barr\label{6.4.11(5)} \exp(itF_\mome(\xi)) & = & \mome(\exp(itX_{\xi\Pi})) =\lim_J
\mome(\exp(it\frac{1}{|J|}\, X_\xi^J)) = \nonumber
 \\
&=& \lim_J \prod_{p\in J} \mome^0(\exp(\frac{it}{|J|}\, X_\xi)) =
\lim_{|J|\rarw\infty}[\mome^0(\exp(\frac{it}{|J|}\, X_\xi))]^{|J|}. \earr

The result\rref 6.4.11(1)~ is now obtained from\rref 6.4.11(5)~ by the `law of large numbers'
(\cite[II.Ch.XVII.§1.Thm.1]{feller}) applied to the arithmetic means of $|J|$ copies of
independent real-valued variables with equal distributions $\mu^0_\xi$. The probability measure
$\mu^0_\xi$ on \bR\ is given here by the projection-valued spectral measure $P_\xi$ of $X_\xi$:

\bequ\label{6.4.11(6)} X_\xi = \int_\mbR \mlam\,P_\xi(\rd\mlam). \enqu

\noidt Then we set

\bequ\label{6.4.11(7)} \mu^0_\xi(\rd\mlam) := \mome^0(P_\xi(\rd\mlam)), \enqu

\noidt and we can write:

\bequ\label{6.4.11(8)} [\mome^0(\exp(\frac{it}{|J|}\,X_\xi))]^{|J|}=\int_{\mbR^{|J|}}
\exp\left(\frac{it}{|J|}\,\sum_{p\in J}\mlam_p\right)\bigotimes_{m\in J}\mu^0_\xi(\rd\mlam_m),
\enqu

\noidt where $\otimes_{m\in J}\mu^0_\xi(\rd\mlam_m)$ is the tensor product of $|J|$ copies of the
measures\rref 6.4.11(7)~ describing the simultaneous probability distribution of the $|J|$
independent random variables. Combining\rref 6.4.11(5)~ and\rref 6.4.11(8)~ gives the wanted
result\rref 6.4.11(1)~. \end{proof}

\begin{prop}
Let us consider the system $(\mfkA;\msg(G);\tau^Q;\pi(\Pi))$ as in Theorem \ref{6.4.10}. Assume
that $\mome^p\ (p\in\Pi)$ are ground states for the restriction of the group
$\msg(\exp(-t\beta^Q_{F_\mome}))$ to the subalgebras $\mfkA_p$. Let the product-state

\bequ\label{6.4.12(1)} \mome:=\bigotimes_{p\in\Pi}\mome^p \enqu

\noidt satisfy the `consistency condition'

\bequ\label{6.4.12(2)} \mome(E_\mfkg(f_\xi))=F_\mome(\xi),\ for\ all\ \xi\in\mfkg.\enqu

\noidt Then \ome\ is a factor ground state of the evolution $\tau^Q$. If all the $\mome^p$ are
pure, then \ome\ is an extremal $\tau^Q$-ground state.
\end{prop}

\begin{proof}The factoria1ity of \ome\ is a consequence of cluster properties,
cf. e.g.\ \cite{bra&rob,bra&rob2}. The condition\rref 6.4.2(3)~ is fulfilled for $\tau_t :=
\msg(\exp(-\beta^Q_{F_\mome}t))$. An application of Proposition \ref{6.4.7} shows the fulfillment
of the ground state condition also for $\tau:=\tau^Q$. The validity of the remaining assertions is
clear.
\end{proof}

\begin{noti} A brief version of the  here presented theory
together with applications to models of \emn BCS theory~ and of Josephson junction was  published
in \cite{bon1,bon2}. Cf. also the next section.
\end{noti}

\section{An example: The B.C.S. model of superconductivity}\label{sec;6.5}

\pt\label{6.5.1}\rm We shall illustrate in this section the above developed theory by description
and analysis of a perhaps simplest nontrivial and physically interesting mathematical model: The
strong coupling version of the \emn Bardeen-Cooper-Schrieffer model~ of the phenomenon of
superconductivity in the \emn quasi spin formulation~; it was formulated and analyzed in
\cite{thir&wehrl,thirr5,jelinek}, in the framework of the traditional QM formalism. It can be
presented, completed, and solved in the framework of the constructions of the present work as
follows:

It is a tensor product type model of Sec.\ref{sec;5.1} with $G:=SU(2)$, $\mH:=\mH_0:=\mbC^2,\
\Pi:=\mbZ,$\ the generators of $U(G)$ in $\mbC^2$ are

\bequ\label{6.5.1(1)} X_{\xi_j}:= \left.i\,\frac{\rd}{\rd t}\right|_{t=0}
U(\exp(t\xi_j))=\frac{1}{2}\msg_j,\ j=1,2,3,\enqu

\noidt where $\msg_j$ are the Pauli matrices and the elements $\xi_j\in\mfkg$ of the chosen basis
satisfy the relations

\bequ\label{6.5.1(2)} [\xi_j,\xi_k]=\mveps_{jkm}\xi_m,\ j,k,(m) = 1,2,3.\enqu

Let $F_j:= F(\xi_j)$ be used for the functions $f_{\xi_j}$ on $\mfkgs\ni F$\ as well as for their
numerical values in the points $F\in\mfkgs$. The dynamics of the system is specified by the
function $Q$ on \fkgs:

\bequ\label{6.5.1(3)} Q(F)=-2\mveps F_3 - \mlam(F_1^2+F_2^2),\quad \mveps,\mlam\ \text{are some
positive numbers}.\enqu

\noidt This specifies the model completely.

\pt\label{6.5.2}\rm The Poisson structure on $\mfkgs=su(2)^*$ is determined by the Poisson
brackets

\bequ\label{6.5.2(1)} \{F_j,F_k\}=-\mveps_{jkm}F_m,\ j,k,(m) = 1,2,3,\enqu

\noidt which are obtained from\rref 6.5.1(2)~ according to\rref 5.1.37(3)~. The classical dynamics
corresponding to the given Hamiltonian function $Q\in C^\infty(su(2)^*,\mbR)$ is then described by
the flow $\mphi^Q$ on $su(2)^*$ which is determined by the Hamilton equations

\bequ\label{6.5.2(2)} \dot{F}_j(\mphi^Q_tF):=\frac{\rd}{\rd t}
F_j(\mphi^Q_tF)=\{Q,F_j\}(\mphi^Q_tF),\ t\in\mbR,\ j=1,2,3.\enqu

We see from\rref 6.5.2(1)~ that $\mphi^Q$ is nontrivial for a general $Q$, hence the symplectic
(even dimensional) $Ad^*$-orbits in $su(2)^*$ (which is 3-dimensional) are two-dimensional (with
the exception of a zero-dimensional orbit consisting of the point $F=0$). Since $SU(2)$ is a
compact group, orbits are compact orientable two-dimensional manifolds in $su(2)^*$. They are
submanifolds of the spheres $S^2_r$:

\bequ\label{6.5.2(3)} F^2:=F^2_1+F^2_2+F^2_3= r^2,\enqu

\noidt because

\bequ\label{6.5.2(4)} \{F^2,F_j\}=0\ \text{for}\ j=1,2,3.\enqu

\noidt Hence the $Ad^*(su(2))$-orbits are the spheres $S^2_r$. The equations of motion with $Q$
from\rref 6.5.1(3)~ are

\bequ\label{6.5.2(5)}
\dot{F}_j=\{Q,F_j\}=-2\mveps\{F_3,F_j\}-2\mlam(F_1\{F_1,F_j\}+F_2\{F_2,F_j\}),\enqu

\noidt that is

\begin{subequations}\label{6.5.2(6)}
\bequ\label{6.5.2(6a)} \dot{F}_1 = 2(\mveps-\mlam F_3)F_2,\enqu

\bequ\label{6.5.2(6b)} \dot{F}_2 = -2(\mveps-\mlam F_3)F_1,\enqu

\bequ\label{6.5.2(6c)} \dot{F}_3 = 0.\enqu
\end{subequations}

\noidt The solution is elementary: With

\bequ\label{6.5.2(7)} F_\pm :=F_1\pm iF_2,\enqu

\noidt one has the flow $\mphi^Q$ determined by the equations

\begin{subequations}\label{6.5.2(8)}
\bequ\label{6.5.2(8a)}F_3(t) = F_3 \equiv F_3(0),\ t\in\mbR, \enqu
\bequ\label{6.5.2(8b)}F_+(t)=F_+(0)\,\exp(-i2(\mveps-\mlam F_3)\,t).\enqu
 \end{subequations}

We shall assume $\mlam\neq 0$. The set of all \emn stationary points $F\in su(2)^*$~ of the flow
$\mphi^Q$ consists of points satisfying the conditions:

\noidt Either
\begin{subequations}\label{6.5.2(9)}
 \bequ\label{6.5.2(9a)} F_+=0,\ \text{and}\ F_3=\text{arbitrary real number},\enqu
 or
 \bequ\label{6.5.2(9b)} F_3=\frac{\mveps}{\mlam},\ \text{and}\ F_+=\text{arbitrary complex
 number}.\enqu
 \end{subequations}
The `physical region' for the values $F$ of the considered quantum mechanical system consists,
however, of the points $F\in\supp E_\mfkg\subset su(2)^*$.

\begin{lem}\label{6.5.3} $\supp E_\mfkg = \{F\in su(2)^*:F^2\leq \frac{1}{4}\}$.\end{lem}

\begin{proof} The spectra of the generators $X_{\xi_j}\ (j=1,2,3)$\ are the two-point sets
$\{\mlam=\pm\frac{1}{2}\}$. According to the proof of Lemma \ref{6.2.17}, $\supp
E_\mfkg=\{F\in\mfkgs: F(\xi)\in{\rm conv}(sp(X_\xi))\ \forall \xi\in\mfkg\}$. Since $\supp
E_\mfkg$ is $Ad^*$-invariant and the $Ad^*$-orbits are spheres $S^2_r$, the set $\supp E_\mfkg$ is
the ball $\{F:F\in S^2_r, 0\leq r\leq\frac{1}{2}\}$. \end{proof}

\pt\label{6.5.4}\rm The quantum evolution $\tau^Q$ is determined according to \ref{6.3.8} and
\ref{6.3.11} by $\mphi^Q$ as well as by the cocycle $\msg(g^{-1}_Q(t,F))\in\maut\mfkA$, where
\fkA\ is the quasilocal algebra of our spin system. The action of this cocycle on the local
algebra $\mfkA_0$ (:= the algebra of the $\frac{1}{2}$-spin sitting at the site $0\in\Pi$) is
given by the unitary family $U(g_Q(t,F))$ satisfying the Schr\"odinger-type evolution equation

\bequ\label{6.5.4(1)} i\frac{\rd}{\rd t}\,U(g_Q(t,F)) = X(\beta^Q_{F(t)})U(g_Q(t,F)),\
F(t):=\mphi_t^Q(F), \enqu

\noidt as can be seen from\rref 6.1.3(8)~. The elements $\beta^Q_F\in su(2)$ are defined by\rref
6.1.3(9)~, i.e.

\bequ\label{6.5.4(2)} \beta^Q_F := \rd_F Q = -2\mveps \xi_3-2\mlam(F_1\xi_1 +F_2\xi_2).\enqu

\noidt In the representation $g\mapsto U(g)$ one has

\bequ\label{6.5.4(3)} X(\beta_F^Q) = -\mveps\msg_3 - \mlam(F_1\msg_1
+F_2\msg_2)=-a(F)\,\textbf{n}(F)\cdot\mbs{\msg}, \enqu

\noidt where $\mbs{\msg}:=\{\msg_1,\msg_2,\msg_3\}$ is the 3-vector of \sg-matrices,

\bequ\label{6.5.4(4)} a(F):=\sqrt{\mveps^2+\mlam^2F_+F_-},\enqu

\noidt and $\textbf{n}(F):= \{n_1,n_2,n_3\}$ with

\bequ\label{6.5.4(5)} n_1:=\frac{\mlam F_1}{a(F)},\ n_2:=\frac{\mlam F_2}{a(F)},\
n_3:=\frac{\mveps}{a(F)},\enqu

\noidt and $\textbf{n}\cdot\mbs{\msg}:=n_j\msg_j$ is the scalar product.

If $F\in su(2)^*$ is one of the stationary points\rref 6.5.2(9)~, then the function $t\mapsto
g_Q(t,F)$ will be a one-parameter subgroup of $SU(2)$ with the generator $\beta^Q_F$. This
subgroup is the stability subgroup of $F$ \wrt\ the $Ad^*(SU(2))$-representation (for $F\neq 0$).
The time evolution $\tau^Q$ in those states \ome\ the classical projection of which is
concentrated on $F_\mome=F$ is now identical with the evolution according to the subgroup of
$\msg(SU(2))$ specified by the element $\beta^Q_F\in\mfkg$. The generator $Q_\mome$ of this
evolution in the representation $\pi_\mome$ can be expressed by its commutators with
$\pi_\mome(y), y\in\mfkA^J$ (finite $J\subset\Pi$):

\bequ\label{6.5.4(6)} [Q_\mome,\pi_\mome(y)]=[\pi_\mome(X^J(\beta^Q_F)),\pi_\mome(y)]\ \text{for}\
y\in\mfkA^J,\ J:=\{p_1,\dots p_m\},\enqu

\noidt where the usual notation $X^J(\xi):=\sum_{p\in J}\pi_p(X(\xi))$ was used, cf. also\rref
6.4.7(2)~. The generator $Q_\mome$ is a well defined selfadjoint operator on the space $\mH_\mome$
of the representation $\pi_\mome$ chosen so that $Q_\mome\mOme_\mome=0$ on the cyclic vector
$\mOme_\mome$. This is the meaning of the Bogoliubov-Haag Hamiltonian operator $Q_\mome$ in the
GNS-representations of \emn macroscopically pure~ and \emn macroscopically stationary~ states of
the system.\nl

\pt\label{6.5.5}\rm {\bf The KMS-states of} $\mbs{(\mfkA;\tau^Q)}$ {\bf at positive temperature}
$\mbs{T>0:}$\nl

The algebra \fkA\ is separable, hence the representation space $\mH_\mome$ of any cyclic
representation is separable and the \emn KMS-states~ \ome\ of this system are supported by the
\emn extremal KMS states~. This means, roughly speaking, that any KMS-state can be constructed as
an integral of the extremal KMS states at the same temperature  $T$. Hence, the evaluation of all
extremal KMS states is sufficient to characterization of all KMS states of the system. Let us
consider now the extremal KMS states.

Any \emn extremal $\tau^Q$-KMS state~ at $T>0$ (hence at $\beta:=T^{-1}\neq\infty$) is determined
uniquely by its restriction $\mome^0$ to $\mfkA_0$, cf. Theorem \ref{6.4.10} (remember that all
states on the UHF-algebra \fkA\ are locally normal). Let $F_\mome\in\mfkgs$ be the classical phase
point corresponding to a given extremal $\tau^Q$-KMS state on \fkA. Then the strong version\rref
6.4.11(1)~ of the `consistency condition' is valid, i.e.

\bequ\label{6.5.5(1)} \mome^0(X_\xi)=F_\mome(\xi)\ \text{for all}\ \xi\in\mfkg.\enqu

\noidt Here $\mome^0$ is the (unique, if it exists) KMS-state on $\mfkA_0$ at the same temperature
$T$ as the state $\mome\in\mcl S(\mfkA)$, corresponding to the evolution given by the generator
$-X(\beta^Q_{F_\mome})$. There is one-one  correspondence between the extremal $\tau^Q$-KMS states
of the infinite system and the states $\mome^0$ satisfying the above listed conditions for some
stationary point $F_\mome$ of the classical equations lying in the physical domain, $F_\mome\in
\supp E_\mfkg$.

Let a stationary point $F_\mome\in\supp E_\mfkg$ be given. Then any
$\msg(\exp(-t\beta^Q_{F_\mome}))$-KMS state $\mome^0$ on $\mfkA_0$ coincides with the Gibbs state
$\mome^0_T$ at some temperature $T$. The state $\mome^0_T$ is given by:

\bequ\label{6.5.5(2)} \mome^0_T(y):=
\left(Tr\,\exp\left(\frac{a(F_\mome)}{T}\,\mbs{n}(F_\mome)\cdot\mbs\msg\right)\right)^{-1}
Tr\,\left(\exp\left(\frac{a(F_\mome)}{T}\,\mbs{n}(F_\mome)\cdot\mbs\msg\right)y\right),\enqu

\noidt for all $y\in\mfkA_0$. It is sufficient to calculate\rref 6.5.5(2)~ for $y=\msg_j,\
j=1.2.3$. We obtain

\bequ\label{6.5.5(3)} \mome^0_T(\msg_j)= n_j(F_\mome)\, \tanh(T^{-1}a(F_\mome)),\ j=1,2,3,\enqu

\noidt and the consistency condition\rref 6.5.5(1)~ means:

\bequ\label{6.5.5(4)} n_j(F_\mome)\,\tanh(T^{-1}a(F_\mome))=2F_\mome (\xi_j),\ j=1,2,3,\enqu

\noidt which is equivalent to the following conditions:

\begin{subequations}\label{6.5.5(5)}
\bequ\label{6.5.5(5a)} \frac{\mlam
F_\mome(\xi_j)}{a(F_\mome)}\,\tanh(T^{-1}a(F_\mome))=2F_\mome(\xi_j),\ j=1,2;\enqu

\bequ\label{6.5.5(5b)} \frac{\mveps}{a(F_\mome)}\,\tanh(T^{-1}a(F_\mome))=2F_\mome(\xi_3).\enqu
\end{subequations}

\noidt These conditions are satisfied by $F_\mome = F$, where
\begin{itemize}
\item[(i)]\ either (in the cases of arbitrary positive \veps\ and \lam)

\bequ\label{6.5.5(6)} F_1=F_2=0,\ \text{and}\  F(\xi_3):= F_3 = \frac{1}{2}
\,\tanh\left(\frac{\mveps}{T}\right),\ T>0,\enqu

\item[(ii)]\ or (in the cases with $0<2\mveps<\mlam$)
\bequ\label{6.5.5(7)} F(\xi_3)=\frac{\mveps}{\mlam},\ 2a(F)=\mlam\,\tanh(T^{-1}a(F)),\ 0<T<T_c
:=\mveps\left(\tanh^{-1}\left(\frac{2\mveps}{\mlam}\right)\right)^{-1}.\enqu
\end{itemize}

\noidt Note that the condition\rref 6.5.5(7)~ can be fulfilled with $F_+\neq 0$ only, hence the
sets of values $F\in\mfkgs$ determined by the two conditions\rref 6.5.5(6)~ and\rref 6.5.5(7)~ are
mutually disjoint. These relations allow us to give the list of all $F_\mome$ corresponding to
extremal $\tau^Q$-KMS states at a given temperature $T>0$:

\begin{itemize}
\item[(i)] $T\geq T_c$; in this case $F_\mome(\xi_1)= F_\mome(\xi_2)=0$,\
$F_\mome(\xi_3)=\frac{1}{2}\,\tanh\left(\frac{\mveps}{T}\right).$

\item[(ii)] $0<T<T_c$; here one has a state with $F_\mome$ described in (i) above, and, if $0<2\mveps<\mlam$,
one has, moreover, a one-parameter family of possible $F_\mome\in su(2)^*$ such that:
$$ F(\xi_3)=\frac{\mveps}{\mlam},\ 2a(F_\mome)=\mlam\,\tanh(T^{-1}a(F_\mome)). $$

There is one-one correspondence between the elements $F_\mome$ corresponding to a given value of
$T>0$ in this list and the \emn extremal $(\tau^Q,\beta:=T^{-1})$-KMS states~ of the infinite
quantal system.
\end{itemize}

We see that, in the considered model, a KMS-state exists at any positive $T$, and for $T\geq T_c$
this state is unique. For $0<T<T_c$, except of the `trivial possibility'\rref 6.5.5(6)~, there is
a circle of points $F_\mome\in\mfkgs$ numbering the elements of pairwise \emn mutually disjoint~
extremal KMS states at the same temperature. If we call the subgroup $\exp(t\xi_3)$ the `gauge
group', then the \emn gauge-invariant KMS-states~ exist at all $T>0$ (the trivial
possibilities\rref 6.5.5(6)~ are gauge invariant); the extremal KMS states for temperatures
$0<T<T_c$ are not invariant with respect to the gauge group and they are transformed by the group
actions into one another: here appears the \emm spontaneous symmetry breaking~ phenomenon.  For
$0<T<T_c$, there is another gauge invariant state $\mome^s_T\in K_\beta\subset \mcl S(\mfkA),\
\beta:= T^{-1},$ given by the integral of the states $\mome^F_T\in\mcl{EK}_\beta$ corresponding to
the values $F$ from\rref 6.5.5(7)~:

\bequ\label{6.5.5(8)}
\mome^s_T(y)=\frac{1}{2\pi}\int_0^{2\pi}\mome^F_T(\msg(\exp(\iota\xi_3))(y))\,\rd\,\iota,\
0<T<T_c.\enqu

Let us denote by \emn$\mome^n_T$~ the (extremal) KMS-state at $\beta=T^{-1}$ corresponding to the
values\rref 6.5.5(6)~ of $F_\mome=F$. The states $\mome^n_T\ (T>0)$ are interpreted as describing
the `\emn normal conducting phase~', and the states \emn$\mome^s_T\ (T<T_c)$~ represent the `\emn
superconducting phase~'. The mixtures $\mome_T:=\mlam\mome^s_T + (1-\mlam)\mome^n_T$ are also
$(\tau^Q,\beta=T^{-1})$-KMS states at $0<T<T_c,\ 0\leq\mlam\leq 1$. The equilibrium states of the
considered system can be defined as the thermodynamic limits of the (unique) Gibbs states of local
systems $(\mfkA^J;\tau^J),\ |J|<\infty$, where $\tau^J_t\in\maut\mfkA^J$ is generated by the local
Hamiltonians $Q^J$ defined in\rref 6.1.1(1)~. According to \cite{jelinek}, these thermodynamic
limits coincide with $\mome^n_T$ for $T\geq T_c$, whereas for $0<T<T_c$ the limit $J\rarw \Pi$
leads to the state $\mome^s_T$.\nl

\pt\label{6.5.6}{\bf The ground states of} $\mbs{(\mfkA;\tau^Q):}$\nl

\rm Let us consider now an \emn extremal $\tau^Q$-ground state~ \ome\ of our system,
$\mome\in\mcl{EK}_\infty$. Let $F_\mome$ be the corresponding classical stationary point in $\supp
E_\mfkg$. The restriction $\mome^0$ of \ome\ to the subalgebra $\mfkA_0$ is the unique ground
state of the generator $X(\beta^Q_{F_\mome})$,\rref 6.5.4(3)~, corresponding to its eigenvector
$\chi(F_\mome)\in\mbC^2$ with the minimal eigenvalue: \bequ\label{6.5.6(1)} {\bf
n}(F)\mbs{\cdot\msg}\chi(F)=\chi(F),\ F\in su(2)^*.\enqu

 \noidt Due to the uniqueness of the ground state $\mome^0\in\mcl S(\mfkA_0)$ corresponding to a given
 $F_\mome\in su(2)^*$, any \emn extremal $\tau^Q$-ground state~ is  an $\pi(\Pi)$-invariant product
 state. Conversely, the $\pi(\Pi)$-invariant product state constructed from a vector $\chi(F)$
 defined in\rref 6.5.6(1)~ will be a pure ground state of $(\mfkA;\tau^Q)$ iff the `consistency
 condition' [$(\chi_1,\chi_2)$ is here the scalar product in $\mbC^2$]
\bequ\label{6.5.6(2)} (\chi(F),X(\xi)\chi(F))=F(\xi),\ \xi\in su(2),\enqu

 \noidt will be satisfied. This is a consequence of the considerations in Section
 \ref{sec;6.4}. Let us solve\rref 6.5.6(2)~ for $F$. For $\xi:=\xi_j\ (j=1,2,3)$ one has
\bequ\label{6.5.6(3)} (\chi(F), X(\xi_j)\chi(F))=\frac{1}{2}\,n_j(F),\ j=1,2,3,\enqu

 \noidt where $n_j(F)$ is defined in\rref 6.5.4(5)~. The obtained condition
 \bequ\label{6.5.6(4)} n_j(F)=2F(\xi_j),\ j= 1,2,3,\enqu

 \noidt leads to the following possibilities for $F=F_\mome,\ \mome\in\mcl{EK}_\infty$:

 \begin{itemize}
 \item[(i)] if \veps\ and \lam\ are arbitrary positive, then one can have:

 \bequ\label{6.5.6(5)} F_1=F_2=0,\ F_3=\frac{1}{2}; \enqu

 \item[(ii)] for $0<2\mveps<\mlam$, one has, moreover, the possibilities:

 \bequ\label{6.5.6(6)} F_1^2+F_2^2=\frac{1}{4}-\left(\frac{\mveps}{\mlam}\right)^2,\
 F_3=\frac{\mveps}{\mlam}.\enqu
\end{itemize}

Hence, in the case $0<2\mveps<\mlam$, the set of ground states has similar classical picture in
$su(2)^*$ as the set of $\tau^Q$-KMS states with temperatures lying under the `critical
temperature' $T_c$. Let $\mome^n_0\in \mcl K_\infty\subset\mcl S(\mfkA)$ corresponds to the value
$F_\mome$ from\rref 6.5.6(5)~, and let $\mome_0^s$ be given by\rref 6.5.5(8)~ with $T=0$ and with
$\mome_0^F\in\mcl{EK}_\infty$ corresponding to any value of $F$ given in\rref 6.5.6(6)~. According
to \cite{jelinek}, the thermodynamic limit of the (unique) local ground states on $\mfkA^J$
corresponding to the Hamiltonians $Q^J$ coincides with $\mome^s_0.$

\newpage

\chapter{Some models of ``quantum measurement''}\label{Ch7}

\section{Introductory notes}\label{sec;7.1}

\pt\label{7.1.1}\rm The interactions in the models of large quantal systems described in Chapter
\ref{Ch6} were of specific long-range type. All the elementary subsystems (``particles'' or
``spins'') mutually interacted with each other `in the same way' as if all the subsystems were not
distinguishable from each other, i.e. the multi-particle interaction was invariant \wrt\
permutations of the particles independent of their positions in the lattice $\Pi$, as specified
by\rref 6.1.1(1)~. Such interactions led in infinite limit of the number $N$ of the subsystems to
the dynamics of ``mean-field type'', i.e. to such a dynamics that each individual subsystem moved
as if it was immersed in an external (in general time dependent) field produced by the whole
collection of the infinite number of all the subsystems and independent of any changes of the
state of any of these subsystems. The resulting dynamics was such that macroscopic (classical)
parameters of the infinite system were varying in time according to the dynamics of some classical
mechanical Hamiltonian system.

In this chapter we will describe several specific models of large quantal systems whose elementary
subsystems interact by short range interactions. The macroscopic, or ``classical'', variables of
the infinite systems will change now just in the limit $t\rarw\infty$, because the short range
interaction results in finite velocity of spreading of local changes across the infinite system,
hence in finite times only local variables corresponding to changes of finite subsystems are
changed.

\pt\label{7.1.2}\rm We shall briefly describe here a few quantum-mechanical model systems
describing interactions of a `microscopic system' with a `macroscopic system' leading to a
`macroscopic change' in the second system. This means that such systems describe schemes modeling
dynamics of processes like `\emn quantum measurement~' as a process ascribing a classical
probability distribution of `measurement results' (given by macroscopically distinct states of the
`macroscopic system' which plays the role of the `\emn measuring apparatus~') to the corresponding
(according to the `measured observable') quantum-mechanical linear decomposition of the wave
function of the `microscopic system'. Construction of these models was inspired mainly  by the
classical paper by \emn Klaus Hepp~ \cite{hp-meas}, cf. also \cite{primas,primas1}. According to
the previous chapters, we are able to describe in QM in a mathematically clear way macroscopic
systems (with coordinates undergoing classical behaviour) by models of infinite quantal systems
only. Of course, the infinity of the number of degrees of freedom should be considered as a
convenient approximation to large but finite systems. Also infinite time duration of the processes
of changing macroscopic parameters corresponding to considered microscopic influences is connected
with this infinity. In this connection, it is relevant to be interested in the speed of the
corresponding macroscopic changes. In the `infinite models' presented here the convergence to a
macroscopic change is very slow.

A much larger speed of convergence is reached in the model of finite (arbitrary long) `Quantum
Domino' - spin chain (cf. \ref{7.1.3} and Sec \ref{sec;7.3}) interacting with fermion field in
such a way that after all the spins in the chain changed their orientations into the opposite ones
the chain emits a fermion. In this case the speed of convergence to final stationary state is
`almost exponential'. The model is described in \cite{bon-sir}. Its interpretation as ``a model of
quantum measurement'' is, however, questionable: Due to its finite dimension a definition of
``macroscopic difference'' is ambiguous and it would need probably a longer discussion. Cf. notes
on this problem in the original Hepp's work \cite{hp-meas}, and also in our Section \ref{sec;7.7}.

It should be stressed that we do not intend to present the described models of micro-macro
interaction as a definitive solution of the `measurement problem in QM', cf. Sec. \ref{sec;7.7}.
They could be considered rather as an illustration of possibilities of the standard quantum
mechanical formalism to include, by using this specific way of description of macroscopic
observables, some descriptions of possible responses of large systems (hence changes of their
`macroscopic variables') to some of their interactions with microsystems. It is shown how can
various states of a microsystem interacting with a macrosystem lead in QM to various
`corresponding changes' of values of their macroscopic (resp. `classical') observables.

 \pt\label{7.1.3}\rm We present here four models, the
second of which is based on the first one, the "\emn Quantum Domino~" (\emn QD~), published
originally in \cite{bon6}. The idea of the third model is similar to that of QD, but it is based
on the known X-Y model of the spin chain \cite{lieb-XY}. QD is a model of an infinite quantum
system - an infinite (or semiinfinite) spin chain with a short range interaction - in which any
local (microscopic) change of a specific stationary state leads to subsequent evolution (with time
$t\rarw\infty$) to a new, macroscopically different stationary state. The initial local changes of
these stationary states of this model are realized ``by hand'', i.e. a lo\-cal\-ly perturbed
stationary state is chosen as an initial condition for the forthcoming time-evolved states of that
system. This local perturbation can be realized by a change of quantum state of a single spin (say
the first one in the semiinfinite chain), and this spin can be considered, e.g. as an additional
microsystem (the `measured system') interacting with the infinite rest of the chain.\footnote{In
the case of some different choices of (locally perturbed stationary) initial states in this model,
the subsequent time evolutions of the chain could be different: e.g., an initial segment could
move quasiperiodically and the infinite rest of the chain will converge to a macroscopically
different state.}

The second model consists in the composition of two systems: of the previous (QD) one and of a
point particle scattered on it; the QD-spin chain occurs initially in its specific stationary
state. The scalar particle (moving in the configuration space $\mbR^3$) perturbs locally the
infinite system (by scattering on its `first two' spins) and the chain develops then (after
$t\rarw\infty$) with some probability to a macroscopically different state. This process can be
interpreted as modeling detection of the particle by a macroscopic detector. The model is
interesting by that that it does not correspond to an ``\emn ideal measurement~'' the results of
which are described usually by a \emn projector valued measure (PVM)~ realizing, e.g., the
spectral decomposition of some selfadjoint operator - the `measured observable'. In our case,
however, we obtain a \emn positive operator valued measure (POVM)~ describing the probabilities of
responses to incoming states of the particle; this expresses the technical characteristics of the
detector with less than $100 \%$ efficiency. This model is presented here in detail, since it is
presented here for the first time - it is a more complex and more complete version of an older
model. The original version of this model was published in \cite{bon7}.

The third presented model is based on an `X-Y modification' of the Heisenberg spin-chain models,
cf. e.g. \cite{robinson}. This ``model of quantum measurement'' consists of the 1/2-spin chain
with a nearest neighbourhood interaction, which is interrupted in one link, and in the point of
the interruption an additional 1/2-spin modeling a simplest possible ``measured microsystem'' is
included (together with its interaction with the rest of the chain).

The fourth model consists of a finite portion of QD of the length $N\gg 1$ coupled to Fermi field
and working so that in the initial state ``all the $N$ spins are pointing down'', but after
reversing the first spin the chain moves until all the spins are ``pointing up'' and, after
reversing the last $N$-th spin, the chain emits a Fermi particle. With the time $t\rarw\infty$ the
particle moves freely to infinity and the chain remains in a new stationary state with ``all spins
pointing up''. The finite length of the chain needs a different interpretation as a ``measuring
device'' in comparison with the preceding three infinite models.

\section{On `philosophy' of "models"}\label{sec;7.2} The term "\emn model~" is used
repeatedly in this Chapter, as well as in science in general. This word is generally used in
various connections and meanings. It is usually considered as denoting human constructs (material
or mental) approximating in some way an aspect of a considered `part of reality'. But, can we
determine where there is a borderline between `only approximation' and `full picture of truth'?
What is the `\emn reality~'? What is the \emn meaning of `the truth'~ (as it was asked also by \emn
Pontius Pilate~ very appropriately in Bible - New Testament: John 18:38)?

Let us consider (not only here) any human symbolic formulation of any \emn knowledge as a
``model''~. Hence, also our laws of nature including the whole physics are models - they are
provisional and waiting for further completions and/or reformulations.

It is motivating and orientating for researchers to believe in the existence of some `\emn final
truth~'. It is an important psychological aspect of scientific progress. The faith in our
`reliably verified knowledge' is perhaps necessary also for the success of our practical life.
But if a theory is completed (i.e. if it is in agreement with all available `trustworthy'
experimental results), it can be (and eventually should be) challenged in science.

Any theory, as well as \emn any concept~ appearing in our consciousness or/and used in our
communication {\bf is a} \emm human construction~. Hence it is dependent on human interests and
activities, and these activities are perpetually evolving--sometimes even substantially changing.
Hence, also our attention and interests are changing. This implies that the motives for our
intellectual activity are perpetually developing. The resulting our `\emn pictures of the world~',
either global, or various special, are correspondingly changing along with these other changes.
And, people also look then on `the same things' by different ways and from different points of
view than before.

The `models' presented in this chapter are just very simple abstractions imitating certain
features of mutual interactions of general classes of physical systems: microsystems described
adequately by QM, and macroscopic systems (usually described by CM) consisting of a large number
of microsystems. We tried to be mathematically rigorous in proceeding from basic axioms of QM to
definitions of introduced concepts and constructions of the mathematical models, as well as to
description and obtaining the consequences of the used dynamics. This emphasis on mathematical
rigor was motivated by our desire to show clearly that the obtained results are exact consequences
of the currently generally accepted formal theory of QM.

\section{Quantum Domino}\label{sec;7.3}

\pt\label{7.3.1}\rm We shall describe here briefly (for more details we refer to
\cite{bon-disert,bon6}) the model of infinite spin chain which we call, due to the character of
its time evolution, \emm Quantum Domino~ (\emn QD~). The 1/2 spins are ordered by the values of
the index $i\in\mathbb{Z}$ and the Hamiltonian produces a local nearest three body interaction.
This interaction can be described easily as follows: If the hamiltonian acts on the state with the
$i$-th spin "pointing up" and the $(i+2)$-nd spin "pointing down", then the $(i+1)$-st spin
changes its orientation to the opposite one. The dynamics of the two sided infinite spin-1/2
quantum chain has spin configurations ``all spins pointing up'', and ``all spins pointing down''
as stationary states, which are unstable: If we reverse the direction of one of the spins in these
states, the new state will develop in the limit $t\rightarrow\infty$ into another stationary (and
`macroscopically' stable) state, in which all the spins lying on one side of the reversed spin are
also reversed, and all the spins lying on the other side of that spin stay unchanged. Since this
evolution leads to the change of the value of a macroscopic observable of the chain, it can be
used as a model for `quantum measurement' of microscopic observables of a single spin of the
chain.  We shall show in this section how such model works.

\pt\label{7.3.2}\rm Let the $C^*$-algebra of observables  $\frak{A}$ be the $C^*$-tensor product
of countably infinite set of copies of the algebra of complex $2\times 2$ matrices generated by
the spin creation and annihilation operators $a_j^*, a_j,\ j\in\mathbb{Z}$ satisfying the
following (anti)commutation relations
\begin{equation}\label{7.3.2(1)}
a_ia_j-a_ja_i =:[a_i,a_j]=[a^*_i,a_j]=0, \ \  i\neq j
\end{equation}
\begin{equation*}
a_ia_i=0,\ \ a^*_ia_i+a_ia^*_i=1,
\end{equation*}
for all $i, j\in\mathbb{Z}$. The algebra $\mathfrak{A}$ is simple, hence each its nonzero
representation is faithful. We shall describe the dynamics in $\frak{A}$ in the ``vacuum''
representation, i.e. in the GNS representation corresponding to the ``vacuum state'' $\omega_0\in
\frak{A}^*_{+1}\equiv\mathcal{S}(\frak{A})$  that is given by the relation

\begin{equation}\label{7.3.2(2)}
\omega_0(a_j^*a_j)= 0,\  \text{for all}\  j\in\mathbb Z.
\end{equation}

This state is pure, hence the GNS representation is irreducible. We shall call the spins in this
state to be ``pointing down'', to be specific in verbal expression. Let the {\bf cyclic vector
(``vacuum'' in the lattice gas terminology) of this representation be denoted by} \emn$\Omega_0$~,
i.e. for all elements $x\in\frak{A}$ it is

\bequ\label{7.3.2(3)} \omega_0(x)=\lb\Omega_0|x|\Omega_0\rb,\  \text{for all}\  x\in \mathfrak A.
\enqu

Here and in the following we shall denote the elements of $\frak A$ and their operator
representatives in the considered irreducible Hilbert space representation by the same symbols.
Let us {\bf denote this Hilbert space by} \emn$\mH_{vac}$~.

Let us define a ``finite-subchain Hamiltonian'' \emn$H_{(j,k)}$~:
\begin{equation}\label{7.3.2(4)}
H_{(j,k)}:=\sum_{n=j+1}^{k-2}a_n^*a_n(a^*_{n+1}+a_{n+1})a_{n+2}a^*_{n+2}.
\end{equation}
Local time evolution automorphisms of $\frak{A}$ are given by

\bequ\label{7.3.2(5)} \tau^n_t(x):= \exp(itH_{(-n,n)})\,x\, \exp(-itH_{(-n,n)}), \enqu

\noidt and the norm limits

\bequ\label{7.3.2(6)}\tau_t(x):= \text{norm}\=\!\lim_{n\rightarrow\infty}\tau^n_t(x) \enqu

\noidt determine the time evolution in $\frak{A}$ (in the ``Heisenberg picture'').

In our vacuum representation, this evolution is determined by a selfadjoint Hamiltonian $H$,

\bequ\label{7.3.2(7)}\tau_t(x)= e^{itH}x\ e^{-itH}. \enqu

\noidt Here, the (unbounded) operator $H$ can be written in the evident form (its obvious
definition and a proof of selfadjointness is given in \cite[Prop.II.1]{bon6})

\begin{equation}\label{7.3.2(8)}
H:=\sum_{n\in\mathbb Z}a_n^*a_n(a^*_{n+1}+a_{n+1})a_{n+2}a^*_{n+2}.
\end{equation}
This evolution is \emn time-reflection invariant~, but it is not invariant with respect to the
\emn space reflection~ $n\mapsto -n$. Let us introduce the operators
\[
g_j:=a_ja_j^*a^*_{j+1}a_{j+1}. \]

\noidt These quantities \ind{g_j:=a_ja_j^*a^*_{j+1}a_{j+1}} are integrals of motion. One can also prove that the Hilbert space
$\mH_{vac}$ can be decomposed into $H$-invariant orthogonal subspaces and on each of them the
restriction of the Hamiltonian $H$ is a bounded operator.

Let $X\subset\mathbb Z$ be of finite cardinality, and {\bf let} $\Omega_X:= \prod_{j\in X}a_j^*\
\Omega_0$. The vectors \emn$\Omega_X$~ with all mutually distinct finite $X\subset\mathbb Z$, {\bf
with} \emn$\Omega_\emptyset:=\Omega_0$~, form an orthonormal basis in $\mH_{vac}$. Each finite
$X\subset\mathbb Z$ is of the form $Y_1\bigcup Y_2\bigcup\dots\bigcup Y_r$, where all
$Y_k\subset\mathbb Z$ are nonempty finite, mutually disjoint and of the form
$\{j_k+1,j_k+2,\dots,j_k+m_k\}$, with $j_{k+1}>j_k+m_k$, $|Y_k|\equiv m_k$, i.e the sets
$Y_k\subset X\ (k=1,2,\dots, r)$ form mutually separated ``connected islands'' consisting of
``pointing up'' spins. All the vectors $\Omega_X$ are eigenvectors of all the operators $g_j$. For
the set $X$ of the just described structure we have

\bequ\label{7.3.2(9)} g_j\Omega_X=\begin{cases} \Omega_X & \text{for}\ j=j_k, \ \ k=1,2,\dots, r \\
0 & \text{otherwise}.  \end{cases}
\end{equation}

This implies that the time evolution of the vectors $\Omega_X$ conserves the number  of islands,
leaving the initial (``left'') points $j_k+1$ of each $Y_k\ (k=1,2,\dots, r)$ unchanged
(``occupied'', or ``pointing up''), and the places $j_k,\ k=2,3,\dots, r$ as well as $j_1-n\
(n\in\mathbb Z_+)$ remain all the time ``unoccupied'' (i.e. spins are there ``pointing down'').
Hence, the subspaces $\mH_{\{{\bf j}\}}$ spanned by all such vectors with a fixed set $\{{\bf
j}\}:= \{j_1,j_2,\dots, j_r\}$ are left invariant \wrt the action of the Hamiltonian $H$. Then
the space $\mH_{vac}$ decomposes as

\bequ\label{7.3.2(10)} \mH_{vac} = \bigoplus_{\{\bf j\}}\mH_{\{\bf j\}}, \enqu

\noidt where the orthogonal sum is taken over all mutually different $\{\bf j\}$; note that {\bf
the stationary subspace \emn$\mH_{\{\mbs\emptyset\}}:=\{\mlam\mOme_0:\mlam\in\mathbb{C}\}$~ is one
dimensional}.

 The structure of the Hamiltonian $H$ shows, moreover, that each  $\mH_{\{{\bf j}\}}$ can be
 written as (i.e. it is isomorphic to) the tensor product of a vector (resp. of a one-dimensional
 subspace) and a finite number of Hilbert spaces corresponding to restricted subchains  of
 spins:

\bequ\label{7.3.2(11)}
 \mH_{\{{\bf j}\}}=\Omega^0_{(-\infty,j_1]}\otimes \mH_{(j_1,j_2)}\otimes \mH_{(j_2,j_3)}\otimes
 \dots\otimes\mH_{(j_r,+\infty)}, \enqu

\noidt where $\Omega^0_{(-\infty,j_1]}$ is one-dimensional space containing the vector with all
spins numbered by $j\leq j_1$ ``pointing down'', and the spaces $\mH_{(j_k,j_{k+1})}$ are spanned
by $j_{k+1}-j_k-1$ vectors corresponding to the ``islands'' $Y_k$ of all permitted lengths $1\leq
|Y_k|< j_{k+1}-j_k$. Here we understand that $j_{r+1}\equiv +\infty$.  We see from the form of the
Hamiltonian that the time evolution of vectors in the subspaces $\mH_{\{{\bf j}\}}$ is described
by (``mutually independent'') evolutions in each $\mH_{(j_k,j_{k+1})}$ determined by the
Hamiltonians $H_{(j_k,j_{k+1})}$, cf.\rref 7.3.2(4)~; for more details see \cite{bon6,bon-disert}.

\pt\label{7.3.3}\rm The result of these considerations is that the evolution of general vectors of
our representation (hence also the evolution of any states from $\mathcal{S}(\frak{A})$) can be
described by two simpler kinds of evolution, namely, the evolutions in finite chains described by
Hilbert spaces $\mH_{(j_k,j_{k+1})}$, as well as in the Hilbert spaces $\mH_{(j_r,+\infty)}$
spanned by vectors
 of arbitrary one-sidedly unrestricted lengths. Because the interaction in our infinite
chain is translation invariant, we can describe these two possibilities as\footnote{We shall use
here the Dirac bra - ket notation for convenience.}

(1) the evolution in the finite-dimensional Hilbert space $\mH_{(0,N+1)}$ spanned by the vectors

\begin{subequations}\label{7.3.3(1)}
\bequ |m\rb := a^*_1a^*_2\dots a^*_m\Omega_0\ (m=1,2,\dots, N)\enqu

\noidt by the unitary evolution group $U_N(t):=e^{-itH_N}$ with the Hamiltonian $H_N:=H_{(0,N+1)}$
from\nl \rref 7.3.2(4)~, and

(2) the evolution in the infinite-dimensional Hilbert space $\mH_{(0,\infty)}$ spanned by the
vectors

\bequ |m\rb := a^*_1a^*_2\dots a^*_m\Omega_0\ (m\in\mbZ, m\geq 1)\enqu
\end{subequations}

 \noindent by the unitary evolution operators
$U_\infty(t):=e^{-itH}$ with the Hamiltonian $H:=H_{(0,+\infty)}$.

Let us express these two instances of dynamics by the matrix elements $\lb n|U(t)|m\rb$.
 The result can be obtained by explicitly solving the eigenvalue problem for $H_N$. The action of $H_N$ is:
\begin{subequations}\label{7.3.3(2)}
\begin{eqnarray}
H_N|1\rb &=& |2\rb,\\
H_N|m\rb &=& |m-1\rb +|m+1\rb,\ m=2,3,\dots,N-1, \\
H_N|N\rb &=& |N-1\rb, \\
H_N|k\rb &=& 0\ \text{for}\ k>N.
\end{eqnarray}
\end{subequations}

\noidt For the eigenvectors \emn$\psi_E:\ H_N\psi_E=E\psi_E$~ written in the basis of vectors
$|m\rangle$: \bequ\label{7.3.3(3)} \psi_E=\sum_{m=1}^N c_m(E) |m\rb\enqu

\noidt we obtain the eigenvalue problem in the form:

\begin{subequations}\label{7.3.3(4)}
\begin{eqnarray}
Ec_1(E) &=& c_2(E),\\
Ec_m(E) &=& c_{m-1}(E)+c_{m+1}(E),\ m=2,3,\dots N-1,\\
Ec_N(E) &=& c_{N-1}(E).\label{7.3.3(4a)}
\end{eqnarray}
\end{subequations}

The equations\rref 7.3.3(4)~ lead to

\bequ\label{7.3.3(5)} c_m(E) = \mcl U_{m-1}(E/2)c_1(E),\enqu where

\bequ \mcl U_{m-1}(z):= \frac{\sin(m\arccos z)}{\sin(\arccos z)}\enqu

\noidt are the \emm Tshebyshev polynomials~ of the second kind \cite[8.940]{RG}. This is seen from
the recurrent relations for \emn$\mcl U_n$~ following from\rref 7.3.3(4)~, cf.
\cite[III.(27)]{bon6}:

$$ \mcl U_{n+1}(z) = 2z\,\mcl U_n(z) - \mcl U_{n-1}(z),\ \mcl U_0(z) = 1,\ \mcl U_1(z) = 2z. $$

 The equation\rref 7.3.3(4a)~ has now the form

\bequ\label{7.3.3(6)} \mcl U_N(E/2)=0,\enqu

\noidt which is the secular equation corresponding to our eigenvalue problem. Its solutions are

\bequ\label{7.3.3(7)} E_j=2\cos\left(\frac{j\pi}{N+1}\right),\ j=1.2.\dots N,\enqu hence we have
the expressions

\bequ\label{7.3.3(8)}
c_m(E_j)=\left[\frac{2}{N+1}\right]^{1/2}\sin\left[\frac{jm\pi}{N+1}\right].\enqu

We shall need also the following definition:

\bequ\label{7.3.3(9)} J^{(N)}_n(z):=\frac{i^n}{N+1}\sum_{j=1}^N\exp\left[-i
z\cos\left(\frac{j\pi}{N+1}\right)\right] \cos\left(n\frac{j\pi}{N+1}\right). \enqu

\noidt This is an integral sum of \emn Sommerfeld integral representation~ of the \emn Bessel
function $J_n(z)$~, see also \cite[8.41]{RG}:

\bequ\label{7.3.3(10)} J_n(z)=\frac{i^n}{\pi}\int_0^\pi e^{-iz \cos \alpha}\cos(n\alpha)
\textrm{d}\alpha. \enqu

We can now write the desired expression for the Green function of a finite chain:

\bequ\label{7.3.3(11)} \lb n|U_N(t)|m\rb = (-i)^{n-m} J^{(N)}_{n-m}(2t) -
(-i)^{n+m}J^{(N)}_{n+m}(2t), \enqu

\noidt what can be obtained by a standard way using the completeness of the orthonormal system of
vectors \rref 7.3.3(3)~ in $\mH_{(0,N+1)}$.

\noidt This, for an infinite chain with $N\rarw\infty$, gives:

\bequ\label{7.3.3(12)} \lb n|U_\infty(t)|m\rb = (-i)^{n-m} J_{n-m}(2t) - (-i)^{n+m}J_{n+m}(2t).
\enqu

\pt\label{7.3.4}\rm Let us now consider the local perturbation $\omega_1(x):=\omega_0(a_1xa^*_1)\
(x\in\mathfrak{A})$ of the time-invariant vacuum state $\omega_0$. The state $\omega_1$ describes
the infinite spin-chain in the state where all the spins except of the one sitting in the site
$j=1$ are pointing down. Its time evolution $\omega_1(\tau_t(x))\equiv \omega^t_1(x)$ can be
expressed in terms of the results given above. Let us, for example, calculate the expectation of
``flipping up'' of the spin placed in the $j$-th place at the time $t$. We have

\bequ\label{7.3.4(1)} \omega_1^t(a^*_ja_j)=\sum_{m=1}^\infty \lb 1|e^{itH}a_j^*a_j|m\rb \lb
m|e^{-itH}|1\rb = \sum_{m=j}^\infty \lb 1|e^{itH}|m\rb \lb m|e^{-itH}|1\rb =
1-\sum_{m=1}^{j-1}|\lb m|e^{-itH}|1\rb|^2, \enqu

\noidt since

\bequ a_j^*a_j|m\rb =
\begin{cases} 0 & (m<j),\\ |m\rb & (m\geq j),     \end{cases} \nonumber \enqu

\noidt and the set of vectors $\{ |m\rb : m\in\mathbb Z\}$ forms an orthonormal basis in the
relevant Hilbert space. From\rref 7.3.3(12)~ and from the recurrent formula for Bessel functions
\begin{subequations}
\bequ\label{7.3.4(2a)} J_{p+1}(z)+J_{p-1}(z)=\frac{2p}{z}J_p(z), \enqu
 we obtain

\bequ\label{7.3.4(2)} \omega_1^t(a^*_ja_j)=1-\sum_{m=1}^{j-1}\left[\frac{m}{t} J_m(2t)\right]^2.
\enqu
\end{subequations}
Because of the asymptotic behaviour of the Bessel function for large real arguments
$|\xi|\rarw\infty$, given by $J_p(\xi)=O(|\xi|^{-\frac{1}{2}})$, we obtain asymptotic behaviour of
our expectation:

\bequ\label{7.3.4(3)} \omega_1^t(a^*_ja_j) \asymp 1-\frac{\text{const.}}{|t^3|},\ \ (\forall
j\in\mathbb N) \ \text{for}\ t\rightarrow\infty. \enqu

Hence the local perturbation of the state ``all spins are pointing down'' converges according to
\rref 7.3.4(3)~ to the state ``all spins sitting in sites with $j>0$ are pointing up''. For more
details see also \cite{bon-m,bon-disert,bon6}.

\pt\label{7.3.5}\rm This can be used for construction of models imitating the `quantum measurement
process'. For instance, let the infinite chain without the spin sitting in the site $j=0$ model an
``apparatus'' and the spin at $j=0$ serve as a
 ``measured microsystem''. If the apparatus is initially in the state $\omega_\downarrow$ with
all its spins pointing down, and the measured spin in a superposition $\mphi:= c_\downarrow
|\downarrow\rb + c_\uparrow |\uparrow\rb$, then the compound system ``measured microsystem +
apparatus'' is in the time $t=0$ in the state described by the state-vector
$c_\downarrow\Omega_0+c_\uparrow a^*_0\Omega_0$, which is a coherent superposition of vectors in
the `vacuum representation' of the algebra of observables of the compound system. Then the final
state of the chain (at $t=\infty$) will be (as a state on the algebra $\mathfrak A$ of the
 compound system ``measured system + apparatus'')\footnote{\label{7.3.5-f}We consider here, for
 the
sake of simplicity, the measured system after the measurement as a part of the apparatus, what
makes no difference for observing results of measurements via various macrostates - the
macroobservables of the compound system are identical with those of the measuring apparatus alone.
See however the subsection \ref{7.3.7} below.} in an incoherent genuine mixture $\omega_f$
according to the above described dynamics: $\omega_f=|c_\downarrow|^2 \omega_0 + |c_\uparrow|^2
\omega_\uparrow$, where the state $\omega_\uparrow$ means that all spins of the compound system
lying in sites $j\geq 0$ are pointing up, whereas the spins lying in sites $j<0$ remain pointing
down. The states $\omega_0$ and $\omega_\uparrow$ on $\mathfrak A$ are mutually \emn disjoint~;
this is interpreted here as ``macroscopic difference'' of these states. Also, the states
$\omega_0$ and $\omega_\uparrow$ define two representations of the algebra of quasi-local
observables (see also \cite{bra&rob,bra&rob2,sak1,sak2}  for further details) which are not
unitary equivalent, and can be distinguished by a measurement of a macroscopic observable.

As the macroscopic observable distinguishing these states could be chosen, e.g., the weak limit
$\gamma\in\mfkA^{**}$ for $n\rarw\infty$ of the sequence

\bequ\label{7.3.5(1)} \gamma_n := \frac{1}{2n+1}\sum_{j=-n}^n a^*_ja_j,\enqu

\noindent and for the states $\mome_0, \mome_\uparrow$ (now considered as being extended to normal
states on the von Neumann algebra $\mfkA^{**}$) we obtain: $\mome_0(\gamma)=0,\
\mome_\uparrow(\gamma)=\frac{1}{2}.$ This is an example in the spirit of the models proposed in
the classical paper by Hepp \cite{hp-meas} for modeling the ``quantum measurement process''.

\pt\label{7.3.6}\rm Observable quantities in QM, or ``observables'', are described usually by
selfadjoint operators $A$ acting on the Hilbert space where the ``observed'' states of a
considered physical system appear. In another setting, we can speak instead of a selfadjoint
operator $A$ about its \emm projection-valued measure~ ($\equiv$ \emn projector-valued measure~)
(PM) $\Lambda\mapsto E_A(\Lambda)$ for $\Lambda\subseteqq\Gamma\equiv$ the set (with a given
\sg-algebra structure) of possible values of the observable (specifying the operator uniquely);
here $E_A(\Lambda)$ are mutually commuting orthogonal projectors satisfying \sg-additivity \wrt\
set unions of various disjoint arguments $\Lambda\subset\Gamma$, with $E_A(\Gamma)=I_\mH$.

 More general concept of ``\emm observable~'' in QM is again \sg-additive
 \emm positive operator valued measure~ (\emn
POVM~)\quad $\Lambda\mapsto\mathbf{A}(\Lambda)$, with $\mathbf{A}(\Lambda)\in\mLH,\,
0\leq\mathbf{A}(\Lambda)\leq\mathbf{A}(\Gamma)=I_\mH,\, \Lambda_i\cap\Lambda_j=\emptyset\ (\forall
i,j)\imply\mathbf{A}(\cup_k\Lambda_k)=\sum_k\mathbf{A}(\Lambda_k)$, which also specifies a
selfadjoint operator $A$, but is not specified by it uniquely. The different
$\mathbf{A}(\Lambda),\ \Lambda\subset\Gamma$, need not be now mutually commutative. According
to a general `philosophy' of QM, to each observable corresponds a measuring apparatus (better: a
class of equivalent apparatuses) characterized abstractly by the observable, by which it can be
measured. Conversely, if we perform a measurement on some quantum-mechanical system, some
observable is measured. The results of the measurement of $A$ on the state \rh\ is found
in the set $\Lambda\subseteqq\Gamma$ with the probability
$pr_{\mathbf{A}}(\mrh,\Lambda)=Tr(\mrh\mathbf{A}(\Lambda))$. If $\Lambda\mapsto pr(\mrh,\Lambda)\
(\Lambda\subseteqq\Gamma)$ is a probability measure for any \rh\ and this mapping depends on \rh\
affinely: $pr(\mlam\mrh_1+(1-\mlam)\mrh_2,\Lambda)\equiv\mlam\, pr(\mrh_1,\Lambda)
+(1-\mlam)pr(\mrh_2,\Lambda)$, then there is a unique observable $A$ of the measured
system such that $pr(\mrh,\Lambda)\equiv Tr(\mrh\mathbf{A}(\Lambda))$. If the distribution of the
results of a measurement is expressed in this way by some POVM $\mathbf{A}\neq E_A$, the
measurement is often called a \emm nonideal measurement~. For more complete formulations cf.
\cite{davies,heino&zim}.

We are dealing in this work with infinite quantal systems described by \Ca s having many mutually
inequivalent representations. Hence, we cannot restrict the concept of observables to operators
acting e.g. on a Hilbert space $\mH_\mome$ of a specific cyclic representation. If we want stay in
a framework of the above presented scheme, we can, and we presently shall, use the universal
representation of \Ca\ \fkA\ in $\mH_u$, resp. of its weak closure, which is a \Wa\ isomorphic to
the double dual \Ass\ of \fkA. For some comments on this reformulation see e.g. \cite[Sec.
2.5]{davies}.

\pt\label{7.3.7}\rm We can now ask, which observable (in the sense of \ref{7.3.6}) was measured by
the `measuring apparatus' modeled by our QD, as it was sketched in \ref{7.3.5}. The `microsystem'
being measured consists in the spin sitting in the point $j=0$ of the infinite spin-chain and the
rest of the chain is the `measuring apparatus'. Let us consider as the apparatus the half-infinite
chain with spins sitting in the points numbered by $j=1,2,\dots\infty$ only, because the spins
sitting in the points with $j<0$ do not take part in these measurements.\footnote{\label{7.3.7-f}
In accordance with that, the notation in this subsection will be
 changed slightly \wrt\ the notation in the subsection \ref{7.3.5}.}
 An integral part of the characterization of the apparatus
is, however, also its initial state `with all spins pointing down', as well as its dynamics
including the interaction with the measured spin. The results of these measurements are read by
looking at the final states of the apparatus.\footnote{\label{7.3.7-ff}We are speaking here about
the states on the algebra generated by $a_j, a^*_j$ with $j>0$ only.} There are just two
possibilities in this process: The state $\mome_\downarrow$ with all spins pointing down, i.e.
$\mome_\downarrow(a_ja^*_j)\equiv 1$, and the state $\mome_\uparrow$ with all spins pointing up,
i.e. $\mome_\uparrow(a_j^*a_j)\equiv 1$, which is \emn disjoint~ from the state
$\mome_\downarrow.$ If these states are (uniquely) extended to normal states on the double dual of
the algebra of measuring apparatus, their values can be calculated on the `macroscopic observable'
$\gamma$ defined now as the weak limit of the sums

\bequ\label{7.3.7(1)} \gamma_n:=\frac{1}{n}\sum_{j=1}^n a_j^*a_j. \enqu

Then it is $\mome_\downarrow(\gamma)=0,\ \mome_\uparrow(\gamma)=1.$ The ``spectral set'' $\Gamma$
from \ref{7.3.6} consists now of only two points, let us denote them (arbitrarily, but taking into
account the actual measurement process) $\pm\frac{1}{2}$, hence
$\Gamma:=\{\frac{1}{2},-\frac{1}{2}\}.$

The initial (=measured) state of the `microsystem' in the example of \ref{7.3.5} was given by the
normalized vector $|\mphi\rb := c_\downarrow |\downarrow\rb + c_\uparrow |\uparrow\rb$
corresponding to the density matrix $\mrh = |\mphi\rb \lb\mphi|$  being just the one-dimensional
projector on the pure state $|\mphi\rb$ of the measured system. The final state of the apparatus
was in this case (according to \ref{7.3.5}) $\mome_f:=
|c_\downarrow|^2\mome_\downarrow+|c_\uparrow|^2\mome_\uparrow,$ where $|c_\downarrow|^2,\
|c_\uparrow|^2$ are the desired probabilities $pr(\mrh,\mp\frac{1}{2}).$ From the linearity of the
tensor products, as well as of time evolution, we can see that the extension of the previously
introduced function $pr(\mrh,\mp\frac{1}{2})$ to general density matrices \rh\ is an affine
function of \rh. Hence, e.g. for convex combination of two `pure' density matrices,

\bequ\label{7.3.7(2)} \mrh:=\mlam_1|\mphi_1\rb\lb\mphi_1|+\mlam_2|\mphi_2\rb\lb\mphi_2|,\
\text{with}\ |\mphi_j\rb:=c_{j\downarrow}|\downarrow\rb + c_{j\uparrow} |\uparrow\rb,\ j=1,2,
\enqu

\noidt we obtain

\barr\label{7.3.7(3)} pr(\mrh,-\frac{1}{2})&=&\mlam_1\,
pr(|\mphi_1\rb\lb\mphi_1|,-\frac{1}{2})+\mlam_2\,
pr(|\mphi_2\rb\lb\mphi_2|,-\frac{1}{2})=\mlam_1|c_{1\downarrow}|^2+\mlam_2|c_{2\downarrow}|^2,\\
 pr(\mrh,\frac{1}{2})&=&\mlam_1\, pr(|\mphi_1\rb\lb\mphi_1|,\frac{1}{2})+\mlam_2\,
pr(|\mphi_2\rb\lb\mphi_2|,\frac{1}{2})=\mlam_1|c_{1\uparrow}|^2+\mlam_2|c_{2\uparrow}|^2.\nonumber
\earr

Let us define the operator $A:=
\frac{1}{2}|\uparrow\rb\lb\uparrow|-\frac{1}{2}|\downarrow\rb\lb\downarrow|$ on the Hilbert state
space of the measured system. Its spectral projections are $P_\uparrow:=|\uparrow\rb\lb\uparrow|$
and $P_\downarrow:=|\downarrow\rb\lb\downarrow|$ and the corresponding mutually distinct
eigenvalues are chosen to be $\pm\frac{1}{2}$. Then, for our density matrix there holds

\bequ\label{7.3.7(4)} pr(\mrh,-\frac{1}{2})= Tr(P_\downarrow \mrh),\quad pr(\mrh,\frac{1}{2})=
Tr(P_\uparrow \mrh). \enqu

Hence, our measuring process corresponds to measurement of operators with PM given by the
one-dimensional orthogonal projectors $P_{\uparrow,\downarrow}$. Our choice of the values of
elements in the set $\Gamma$ corresponds to the observable describing a component of the
$\frac{1}{2}\hbar$-spin, which is usually described in this way. We did not need here a
generalized observable determined by a POVM, which will be, however, the case of the following
example.

\section{Particle detection - a ``nonideal'' measurement}\label{sec;7.4}

\pt\label{7.4.1}\rm This model describes a compound system of a spin chain $A$ with a particle
$B$; it is a completed version of the model presented originally in \cite{bon7}. The model of the
spin chain is the half-infinite chain of the form described in the section \ref{sec;7.3}, and the
particle is a nonrelativistic scalar particle.

Let us use (essentially) the notation of section \ref{sec;7.3}. Hence, the algebra \fkA\ of the
observables of the spin chain is now generated by the elements $a_n^*, a_n,\ n\geq 1.$ Let the
Hamiltonian of the chain be the operator (cf.\rref 7.3.2(8)~)

\bequ\label{7.4.1(1)} H_A:=\sum_{n\geq 1}a_n^*a_n(a^*_{n+1}+a_{n+1})a_{n+2}a^*_{n+2}\enqu

\noidt acting in the Hilbert space $\mH_{vac}$ of the GNS-representation of \fkA\ with {\bf the
cyclic vector} \emn$\Omega_0$~ corresponding to the state

\bequ\label{7.4.1(2)} \omega_\downarrow^A(a_j^*a_j)= 0,\  \text{for all}\ j\geq 1.
\end{equation}

The particle B is moving in the 3-dimensional Euclidean space and is described as in elementary QM
by operators acting in the space $\mH_B:=L^2(\mbR^3,\rd^3 x)$, so that its states are described by
vectors (resp. the corresponding unit rays) $\psi\in\mH_B$. The free particle's Hamiltonian will
be just the kinetic energy (in conveniently chosen units and in the ``\textbf{x}-representation'')

\bequ\label{7.4.1(3)} H_B:= \hat{\mathbf{p}}^2=-\sum_{j=1}^3\frac{\partial^2}{\partial
x_j^2}.\enqu

\noindent The \emm interaction Hamiltonian~ will be \emn$V_\mphi$~, with

\bequ\label{7.4.1(4)} V_\mphi := (a_1^*+a_1)a_2a^*_2\otimes |\mphi\rb\lb\mphi|\ \in\mcl
L(\mH_{vac}\otimes\mH_B), \enqu

\noidt where $\mphi\equiv |\mphi\rb\in\mH_B$ is a conveniently chosen normalized vector, hence
$|\mphi\rb\lb\mphi|\equiv P_\mphi$ is a one-dimensional projector in $\mH_B$.

The \emm total Hamiltonian $H$~ of the compound system \{spin chain \& particle\} will be

\bequ\label{7.4.1(5)} H:= H_A + H_B + \mgam V_\mphi,\ \mgam\in\mbR.\enqu

\noidt Some restrictions on the interaction constant \gam\ and on the unit vector $\mphi$\ will be
specified later.

\pt\rm\label{7.4.2} We want to prove, for conveniently chosen parameters \gam\ and $\mphi$ of
interaction and for suitable initial states $\psi\in\mH_B$ of the particle as well as for given
initial state of the spin chain with all spins ``pointing down'', that the
 compound system will evolve for $t\rarw\infty$ with positive probability into a convex combination
of two mutually \emn disjoint~ (hence `macroscopically different') states, one of which
corresponds to the unchanged initial state of the apparatus and in the other the apparatus has all
its spins reversed to the ``pointing up'' direction. If we denote by $\mfk B:=\mcl L(\mH_B)$ the
algebra of all bounded operators on $\mH_B$, which is the \Ca\ of the observables of the particle,
and {\bf by \emn$\mfk C:= \mfkA\otimes\mfk B$~ the \Ca\ of the compound system}, then \emn$\mS(\mfk C)$~ {\bf
will be the state-space} $\mfk C^*_{+1}$ (i.e. positive normalized elements of the topological
dual of \fk C) of the compound system.

We will prove that {\bf the initial state} \emn$\mome_0^{A\&
B}$~$\equiv\mome_\downarrow^A\otimes\mome_{\psi}^B\in \mS(\mfk C)$, {\bf where}
$a\mapsto\mome_{\psi}^B(a):=\lb\psi|a|\psi\rb$ for $a\in\mfk B$, will evolve to the state
$\overline{\mome}\in\mS(\mfk C)$, $\overline{\mome}= (w(\psi)\,\mome^A_\uparrow +
(1-w(\psi))\,\mome^A_\downarrow)\otimes\mome^B_0$, and where $\mome^B_0\in\mS(\mfk B)$ is the
state without particles, cf. \ref{Notation}, and $0<w(\psi)<1$ for any of the considered initial
state-vectors $\psi$.

If we ask ``which observable is measured by this process'', the relevant answer is -- if we
consider only the mathematical expression of the ``observable'' appearing in the question --  in
the expression of the \emm probability $w(\psi)$ as a diagonal matrix element~ of a {\bf positive
operator} \emn$W\equiv W_\mgam$~ between the state vectors of the particle's initial state $\psi$:
$w(\psi)=\lb\psi|{\rm W}|\psi\rb$. The operator $W$, $0<W<1,\ W\neq W^2$, replaces here the usual
appearance of a projector from the PM of measured selfadjoint operator in the cases of `\emm ideal
measurements~', cf.also \cite{heino&zim}. Our simple specific model represents more general
instances of measurements: The `\emm nonideal measurement~' is described by a \emn POVM~ (=\emm
positive operator valued measure~). Hence, our model illustrates the concept of ``generalized
observables'' introduced in \cite[Sec.3.1]{davies}, cf. also our \ref{7.3.6} and \ref{7.3.7}, and
its usefulness. The quantity $w(\psi)=\lb\psi|{W}|\psi\rb$ has to be interpreted as the measured
probability of one of two possible results of a two-valued observable of the particles prepared at
$t=0$ in the state $\psi$. A verbal expression of the intuitive physical meaning of ``the
particle's observable $W$'' might be here just something like ``what can be registered by this
specific measuring apparatus'', with two different pointer values: to be or not to be registered
by this specific apparatus.

\begin{notat}\label{7.4.3}
We shall use the following symbols: \label{Notation}
\begin{enumerate}
\item The state without particles could be defined in a standard way, e.g. as the vacuum state
in the Fock representation, where the algebra of observables of particles is constructed by
creation-annihilation operators, cf. \cite{bon7}. To avoid this (here unnecessary) complication,
we shall define the no-particle state as the normal linear functional $\mome_0^B\in\mS(\mfk B)$ on
$\mfk B= \mcl L(\mH_B)$,\ $\mome_0^B: b\mapsto\mome_0^B(b)\ (\text{remember that}\
\dim\mH_B=\infty)$ such that
$$ \mome_0^B(b)= 1,\ \text{if}\  b=I_{\mH_B}; \ \mome_0^B(b)= 0,\ \text{if}\
 b=|\psi_1\rb\lb\psi_2|,\ \psi_j\in\mH_B.
 $$

 This will give equivalent results of our considerations to those obtained from the
 considerations using the formalism of nonrelativistic quantum field theory.

\item Let us introduce also the symbol \emn$\mH_A$~ for the Hilbert (sub-)space of the chain generated by
the vectors $\{ |m\rb\ |\ m=1,2,\dots \}$ introduced in\rref 7.3.3(1)~. We shall use also:
\emn$U_t:=\exp(-itH)$~ with $H$ from\rref 7.4.1(5)~, and \emn$\tau_t{c}:= e^{itH}c\, e^{-itH}$~
for $c\in \mfk C.$ The vector \emn$\Omega_0$~ $= |0\rangle$ is defined in\rref 7.3.2(2)~ and\rref
7.3.2(3)~. We shall also use \emn$\Omega_0^\chi$~ $:= \Omega_0\otimes\chi,\ \chi\in\mH_B$.

\item \label{7.4.3(1)} Let $\mphi\in\mH_B,\ \|\mphi\|_2=1,$ be the vector appearing in the interaction
Hamiltonian $V_\mphi$ in\rref 7.4.1(4)~, and let $\psi\in\mH_B$ be the (also normalized) initial
state-vector of the particle. We shall {\bf introduce the symbols} \emn$F^0(t)$~, \emn$g(t)$~, and
\emn$F(t)$~ as:
\begin{subequations}
\bequ\label{7.4.3(1a)} F^0(t)\equiv F^0_\mphi(\psi)(t):= \lb\mphi|e^{-itH_B}|\psi\rb,\
g(t):=F^0_\mphi(\mphi)(t)\equiv\lb\mphi|e^{-itH_B}|\mphi\rb. \enqu \bequ\label{7.4.3(1b)}
F(t)\equiv F_\mphi(\psi)(t):= \lb \mphi|\otimes\lb 0|e^{-itH}|0\rb\otimes |\psi\rb\equiv
\lb\Omega_0^\mphi|e^{-itH}|\Omega_0^\psi\rb, \enqu
\end{subequations}
\noidt where $H:= H_A+H_B+\gamma V_\mphi$ is the total Hamiltonian of the compound system\rref
7.4.1(5)~.\nl \noindent The symbols \emm$f_m(t)$~, \emm$f(t)$~ will be also useful abbreviations
(cf.\rref 7.3.3(12)~):
\begin{eqnarray}\label{7.4.3(2)} f_m(t)&:=& \lb m|e^{-it H_A}|1\rb =
(-i)^{m-1}\,\frac{m}{t}\,J_m(2t),\ m=1,2,\dots \\
\label{7.4.3(3)} f(t)&:=& g(t)f_1(t)= \lb\mphi|\otimes\lb1|e^{-it(H_A+H_B)}|1\rb\otimes |\mphi\rb.
 \end{eqnarray}

\item To restrict a function $t\mapsto h(t)$ defined on the whole real line $t\in\mbR$\ to the
positive (resp. negative) values of its argument $t\in\mbR_+$ (resp. $\mbR_-$), we shall use the
(Heaviside) \emn$\theta(t)$-function~ equal to zero for $t<0$ and equal to one for $t\geq 0$. We
shall denote these restrictions as \emn $h_\pm(t)$~:
 \bequ\label{7.4.3(4)} h_+(t) := \theta(t) h(t),\text{resp.}\
h_-(t):= \theta(-t) h(t),\quad t\in\mbR. \enqu

\noidt Such restrictions \emn$f\mapsto f_+$~ will be useful here, e.g., for rewriting certain
equations in the convolution form.
\item The \emm convolution $f\ast h(t)$~ of two complex-valued integrable functions is defined by

\bequ\label{7.4.3(5)} f\ast h(t)=\int_{-\infty}^{+\infty}\rd \tau f(t-\tau)h(\tau)= h\ast
f(t).\enqu

\noindent For more details on existence conditions of convolutions see e.g. \cite[�IX.4]{R&S}.
The operation $\ast$ is not only commutative, but also associative. It can be trivially extended
to functions $t\mapsto h(t)$ defined for $t\in\mbR^n$, as well as to some other classes of
functions and of distributions, see e.g. \cite{R&S,vladimirov}.

\item Let us define and denote, for purposes of the present section, to any integrable function
$h\in L^1(\mbR)$, its Fourier transformed \ind{Fourier transform $\mcl F$} function \emn$\mcl
F(h)\equiv \hat{h}$~:
\begin{subequations}
\bequ\label{7.4.3(6)} \hat{h}(u) \equiv \mcl F(h)(u)
:=\frac{1}{\sqrt{2\pi}}\int_{-\infty}^{+\infty} e^{-itu} h(t)\,\rd t,\ u\in\mbR. \enqu

\noidt In the case of higher dimensional arguments of the \bC-valued functions $h\in L^1(\mbR^n)$
the analogous formula applies:

\bequ\label{7.4.3(7)}\hat{h}(u) \equiv \mcl F(h)(u)\equiv \mcl F(h(t))(u)
:=(2\pi)^{-\frac{n}{2}}\int_{\mbR^n} e^{-it\cdot u} h(t)\,\rd^n t,\ u\in \mbR^n.\enqu

\noidt The inverse \emn$\mcl F^{-1}$~ of  \cl F\ defined on the image \emn$\hat{h}=\mcl F(h)$~ has
the similarly looking form:

\bequ\label{7.4.3(8)} h(t)=\mcl F^{-1}(\hat{h})(t)=\mcl F(\hat{h}(-u))(t)=
(2\pi)^{-\frac{n}{2}}\int_{\mbR^n} e^{it\cdot u} \hat{h}(u)\,\rd^n u,\ t\in \mbR^n.\enqu

\end{subequations}

  Generalizations to various classes of functions $h$ and also to
tempered distributions is very useful in process of solution of various equations. Many important
properties of the Fourier transformation can be found, e.g. in \cite{R&S,vladimirov}. One of the
most useful properties of \cl F\ is the possibility to extend it from $L^1(\mbR^n)$ to a unitary
transformation in the Hilbert space $L^2(\mbR^n)$ - the Plancherel theorem: The scalar product
$\lb\cdot|\cdot\rb$ is invariant \wrt\ the transformation \cl F; for $\mphi,\psi\in L^2$ it means:
$\lb\mphi|\psi\rb=\lb\hat{\mphi}|\hat{\psi}\rb.$ Moreover, the following important property
concerning the interconnection between the convolution and the Fourier transformation is valid:
\bequ\label{7.4.3(9)} \mcl F(h_1\ast h_2)=(2\pi)^{\frac{n}{2}}\mcl F(h_1)\mcl F(h_2)\equiv
(2\pi)^{\frac{n}{2}}\hat{h_1}\hat{h_2},\enqu

\noidt with the pointwise multiplication of functions.
\end{enumerate}

\end{notat}

For the proof of our main result formulated in Theorem \ref {7.4.8}, we shall also need several
following lemmas. The first one together with its proof can be deduced from \cite{hegerfeldt}:

\begin{lem}\label{7.4.4} Let $H$ be a lower-bounded selfadjoint operator on a Hilbert space \H\ with its
spectrum $sp(H)\geq a$. Then, for any two of nonzero vectors $\mphi,\psi\in\mH$, it is:
\begin{description}
\item[(a)] either $\lb\mphi|e^{-itH}|\psi\rb\equiv 0,\ \forall t\in\mbR,$
\item[(b)] or $\lb\mphi|e^{-itH}|\psi\rb\neq 0$ for $t$ in an open dense subset of \bR\ of total Lebesgue
measure.
\end{description}

If the above chosen  $\mphi$ is fixed, then the set of all $\psi\in\mH$ satisfying {\bf (a)} forms
a closed linear subspace of \H, hence the open complement in \H\ of this set contains those
$\psi\in\mH$ which satisfy the point {\bf (b)}.

\end{lem}
\begin{proof} Let $\mlam\mapsto E_H(\mlam)$ be the projection measure of $H$. According to the
functional calculus (cf. e.g. \cite{R&S}) it is

\bequ\label{7.4.4(1)} \lb\mphi|e^{-itH}|\psi\rb=\int_a^\infty\rd\mlam\, e^{-it\mlam}\lb\mphi|
E_H(\mlam)|\psi\rb. \enqu

\noidt This function of time $t\in\mbR$ can be analytically continued to the lower complex
half-plain of $t$, i.e. extended to $t\mapsto t-i\mveps =:z,\ \mveps\geq 0$:

\bequ\label{7.4.4(2)}\lb\mphi|e^{-i(t-i\mveps)H}|\psi\rb =\int_a^\infty\rd\mlam
e^{-i(t-i\mveps)\mlam}\lb\mphi|E_H(\mlam)|\psi\rb \equiv \lb\mphi|e^{-izH}|\psi\rb,\ \Im\,z\leq 0,
\enqu

\noidt which is analytic in the open lower complex half-plain of $z$ and continuous in the closed
lower half-plain, hence also on the real line $z=t-i\mveps\rarw t-i 0+$. Assume that
$\lb\mphi|e^{-itH}|\psi\rb\equiv 0,\, \forall t\in I\subset\mbR$, where $I$ is an interval of
positive length. Then, according to the Schwarz reflection principle, the analytic function
$z\mapsto \lb\mphi|e^{-izH}|\psi\rb$ is complex-analytic also on this interval $I$, hence it is
identically zero also in lower complex half-plain. Due to its continuity on \bR, the function
$t\mapsto \lb\mphi|e^{-itH}|\psi\rb\equiv 0\ (\forall t\in\mbR).$

In the other cases, there is no interval of  nonzero length  $I\subset\mbR$ on which the function
$t\mapsto \lb\mphi|e^{-itH}|\psi\rb$ identically vanishes. Since it is continuous, it is $\neq 0$
on open intervals composing an open dense subset of \bR. But union of all these intervals is a set
of total Lebesgue measure on \bR, as is shown in \cite{hegerfeldt}. Hence the function $t\mapsto
\lb\mphi|e^{-itH}|\psi\rb\neq 0$ a.e. \wrt the Lebesgue measure.

Linearity of the set of the $\psi$'s satisfying (a) is clear. That this subspace is closed in \H\
follows from the norm-continuity of the matrix elements $\psi\mapsto\lb\mphi|e^{-itH}|\psi\rb$;
the last assertion follows from the other proved assertions of this Lemma.
\end{proof}

\begin{lem}\label{7.4.5} The condition $\mphi\in\mcl D(\mbR^3)$ for the choice of the vector $\mphi$\ occurring
 in the definition of the interaction Hamiltonian in\rref 7.4.1(4)~, as well as the condition
 $\psi\in\mH_B\cap L^1(\mbR^3)$ for the choice of the
particle's initial vector $\psi$, both imposed in the Theorem \ref{7.4.8}, guarantee the following
properties of the functions  $t\mapsto F^0_\mphi(\psi)(t)$\rref 7.4.3(1a)~ of the time variable
$t\in\mbR$:

\bequ\label{7.4.5(0)}  F^0_\mphi(\psi)\in L^2(\mbR)\cap L^1(\mbR),\ \quad g\equiv
F^0_\mphi(\mphi)\in L^2(\mbR)\cap L^1(\mbR).\enqu

The set $L^1(\mbR^3)\cap L^2(\mbR^3)\supset \mcl D(\mbR^3)$ is
dense in $\mH_B$ together with $\mcl D(\mbR^3)$.

\end{lem}
\begin{proof}According to the Theorem IX.30 of \cite{R&S}, there is for $\psi\in L^1(\mbR^3)\cap
L^2(\mbR^3)$:

\bequ\label{7.4.5(1)}\text{ess}\sup_{\textbf{x}\in\mbR^3}|e^{-itH_B}\psi(\textbf{x})|\equiv
\|e^{-itH_B}\psi\|_\infty \leq |t^{-\frac{3}{2}}|\,\|\psi\|_1.\enqu

\noidt The function $\mphi$ has finite support, say $\mphi(\textbf{x})\neq 0\imply
|\textbf{x}|<R<\infty.$ Let us denote by $B_R\subset\mbR^3$ the ball of radius $R$ containing the
support of $\mphi$. Due to the implication $\chi\in L^2(\mbR^3)\imply |\chi|\in L^2(\mbR^3)$, we
have

\barr\label{7.4.5(2)} |\lb\mphi|e^{-itH_B}\psi\rb|\leq \int_{B_R}\rd^3
\textbf{x}\,|\mphi(\textbf{x})|\cdot|e^{-itH_B}\psi(\textbf{x})|\leq \nonumber \\
\int_{B_R}\rd^3
\textbf{x} \,|\mphi(\textbf{x})|\cdot \frac{\|\psi\|_1}{|t^{\frac{3}{2}}|}
=\frac{\|\mphi\|_1\|\psi\|_1}{|t^{\frac{3}{2}}|}.\earr

\noidt But the matrix element of a unitary operator between two normalized vectors in $\mH_B$ is
bounded by unity: $|F^0_\mphi(\psi)(t)|\leq 1,$ hence we have

\bequ\label{7.4.5(3)} |\lb\mphi|e^{-itH_B}\psi\rb|\leq \min{\{\ 1;\
\frac{\|\mphi\|_1\|\psi\|_1}{|t^{\frac{3}{2}}|}\}},\ \text{for all}\  t\in\mbR, \enqu

\noidt and the obtained estimate is

\bequ\label{7.4.5(4)} |F^0_\mphi(\psi)(t)|\leq
\theta\left((\|\mphi\|_1\|\psi\|_1)^{\frac{2}{3}}-|t|\right)+ \theta\left(|t|-
(\|\mphi\|_1\|\psi\|_1)^{\frac{2}{3}}\right)\,\frac{\|\mphi\|_1\|\psi\|_1}{|t^{\frac{3}{2}}|}.\enqu

\noidt The function $t\mapsto |t^{-\frac{3}{2}}|\,\theta(|t|- k),\ k>0,$ belongs to $L^1(\mbR)\cap
L^2(\mbR)$, hence $F^0_\mphi(\psi)$ also belongs there $\forall\psi\in L^1(\mbR^3)\cap
L^2(\mbR^3)$. Since also our $\mphi\in\mcl D(\mbR^3)\subset L^1(\mbR^3)\cap L^2(\mbR^3)$, the both
relations in\rref 7.4.5(0)~ are proved. The density of $\mcl D(\mbR^3)$ in $L^2(\mbR^3)$ is easily
seen, cf. e.g. \cite[I.1.7]{vladimirov}.

\end{proof}

\begin{lem}\label{7.4.6} Let $G\in L^1(\mbR^n)\cap L^\infty_0(\mbR^n)$ and $G'\in L^p(\mbR^n)\cap
L^\infty_0(\mbR^n)$ ($1\leq p\leq\infty$), where $L^\infty_0$ is the space of (essentially)
uniformly bounded functions converging to zero at infinity. Then the convolution $G\ast G'\in
L^p\cap L^\infty_0$.
\end{lem}
\begin{proof}According to the Theorem 1.3. in \cite{stein&weiss}, $\|G\ast G'\|_p\leq \|G\|_1\cdot
\|G'\|_p$, and also $\|G\ast G'\|_\infty\leq \|G\|_1\cdot\|G'\|_\infty$, hence $G\ast G'\in
L^p\cap L^\infty$. It remains to prove the convergence to zero at infinity.

Let us choose $\delta>0$. For any such $\delta$ there is a $T_\delta >0$ such, that $\forall\
|\mbbs \tau|> T_\delta\,\imply\ |G'(\mbbs \tau)|<\delta.$ Then for $|\mbbs t|>T_\delta$ it is

\begin{eqnarray}\label{7.4.6(1)} |G\ast G'(\mbbs t)|& \leq& \int_{|\mbbs\tau|<
T_\delta}\rd^n\mbbs\tau\, |G(\mbbs t-\mbbs\tau) G'(\mbbs \tau)|\ +\
\delta\int_{|\mbbs\tau|>T_\delta}\rd^n\mbbs\tau\, |G(\mbbs
t-\mbbs\tau)| \nonumber \\
& \leq & \|G'\|_\infty\, \Omega_n(T_\delta)\sup_{|\mbbs \eta|\geq||\mbbs t|-T_\delta|}\, |G(\mbbs
\eta)|\ +\ \delta\,\|G\|_1,
\end{eqnarray}
\noidt where $\Omega_n(T)$ is the Euclidean volume of the n-dimensional ball of radius $T$. With
any fixed $\delta$, the supremum converges to zero with $|\mbbs t|\rarw\infty$. Hence, by a
convenient choice of $\delta>0$ and for sufficiently large $|\mbbs t|$, the right hand side of
\rref 7.4.6(1)~ can be made arbitrarily small, hence the left hand side converges with $|\mbbs
t|\rarw\infty$ to zero.
\end{proof}

A similar useful Lemma for functions of $t\in\mbR$ restricted to $\mbR_+$\ claims:

\begin{lem}\label{7.4.7} For $h\in L^1(\mbR)\cap L^\infty_0(\mbR)$ and $k\in L^1(\mbR)$ it is:
\bequ\label{7.4.7(1)}
 h_+\ast k_+ \in L^1(\mbR)\cap L_0^\infty(\mbR).\enqu
\end{lem}
\begin{proof}Again from the known $L^p$-estimate \cite{stein&weiss} there is $h_+\ast k_+ \in L^1(\mbR)\cap
L^\infty(\mbR),$ and also $\|h_+\ast k_+\|_p\leq \|h_+\|_p \|k_+\|_1$ for $p=1,\infty.$ Let us
prove the convergence to zero. It is

\bequ\label{7.4.7(2)} h_+\ast k_+(t) = \theta(t) \int_0^{\frac{t}{2}}\rd\tau\,[\,h(t-\tau)k(\tau)
+ h(\tau) k(t-\tau)\,],\enqu

\noidt and the needed estimate is:

\bequ\label{7.4.7(3)} |h_+\ast k_+(t)|\leq \theta(t)\, \left[\,\|k\|_1
\sup_{\tau>\frac{t}{2}}|h(\tau)| +\|h\|_\infty \int_{\frac{t}{2}}^{+\infty}\rd\tau\,
|k(\tau)|\,\right].\enqu

\noidt The first term on the \rhs\ converges for $t\rarw+\infty$ to zero because the function $h$
converges to zero. The second term converges to zero due to integrability of $k\in L^1(\mbR)$.
This shows that $h_+\ast k_+\in L^\infty_0$. The assertion is proved.
\end{proof}

\noindent We shall give here a proof of the main result of this section:

\begin{thm} \label{7.4.8} Let the dynamics of the compound system: nonrelativistic point particle B
(as a ``detected microsystem'') and the one-dimensional spin chain A, described in Sec.
\ref{sec;7.3} (as a ``detector''), be given by the Hamiltonian\rref 7.4.1(5)~ defined in the
ground-state representation (corresponding to the state $\mome^A_\downarrow$ of the chain with
``all spins pointing down'').

Let the particle's initial normalized state-vector be $\psi\in\mH_B\cap L^1(\mbR^3)\equiv
L^2(\mbR^3)\cap L^1(\mbR^3)$, and the initial state of our half-infinite chain be
$\mome_\downarrow^A$ from\rref 7.4.1(2)~. The normalized vector $\mphi\in L^2(\mbR^3)$ occurring
in the Hamiltonian $H$ in\rref 7.4.1(4)~ will be chosen as a rapidly decreasing $C^\infty(\mbR^3)$
function with compact support: $\mphi\in\mcl D(\mbR^3)\subset \mH_B\cap L^1(\mbR^3)$. To ensure a
nontrivial interaction of the particle with the chain, let us assume that\ (cf. Lemma \ref{7.4.4})

\bequ\label{7.4.8(1)} F^0_\mphi(\psi)(t)\equiv \lb\mphi|\exp(-itH_B)|\psi\rb \not\equiv 0,\quad
t\in\mbR.\enqu

\noidt  We require, moreover, a condition on the upper bound of the interaction constant $\gamma$
to be fulfilled:

\bequ\label{7.4.8(2)} 0<\|\gamma g\|_1 < 2,  \enqu

\noidt with $g\equiv F^0_\mphi(\mphi)$.

If these conditions are satisfied, then there exist, for all $a\in\mfkA,\ b\in\mcl L(\mH_B)$, the
limits

\begin{subequations}\label{7.4.8(3)}
\bequ \lim_{t\rarw\infty}\momeAB t(ab) = \left(w(\psi)\,\mome^A_\uparrow(a) +
(1-w(\psi))\,\mome^A_\downarrow(a)\right)\,\mome^B_0(b),\enqu
 \noidt with  $\momeAB t(ab):=\lb\psi\otimes\Omega_0|e^{itH}a\otimes b\,
e^{-itH}|\Omega_0\otimes\psi\rb, $

\noidt i.e. \bequ w^*\hbox{-}\lim_{t\rarw\infty}\,\momeAB0\circ\tau_t\equiv
w^*\hbox{-}\lim_{t\rarw\infty} \momeAB t=\left(w(\psi)\,\mome^A_\uparrow +
(1-w(\psi))\,\mome^A_\downarrow\right)\otimes\mome^B_0.\enqu
\end{subequations}

 \noidt The \emm probability of the detection~ \emn$w(\psi)$~ is here positive: $w(\psi)> 0$,  and, moreover,
 it depends on the initial state $\psi$ of the particle as:

\bequ\label{7.4.8(4)} \psi \mapsto \lb\psi|W|\psi\rb\equiv\lb\psi|W_\gamma|\psi\rb\equiv w(\psi),
\enqu

\noidt where $W\equiv W_\gamma\in\mcl L(\mH_B)$ is a positive operator $0<W_\gamma< I_{\mH_B},$
independent of $\psi$. Moreover, for sufficiently small nonzero interaction constants
$\gamma\in[-\gamma_0,\gamma_0]\subset\mbR$ it is $W^2_\gamma\neq W_\gamma,$ hence $W_\gamma$ is
not a projector.
\end{thm}

\begin{proof}Let us use the notation introduced in \ref{Notation}.
We want to prove the existence of the limit\rref 7.4.8(3)~  first. Let the state-vectors of the
chain $|m\rb,\ m=0,1,2,\dots\infty$ be defined as in\rref 7.3.3(1)~ {\bf with} $|0\rb :=\Omega_0$.
The {\bf Hilbert subspace} \emm$\mcl K\equiv (\mH_A\oplus\mH_{\{\mbs\emptyset\}})\otimes\mH_B$~\
{\bf of the state-space of the (initial-state representation of the) compound system} ``the spin
half-chain \& the particle'' generated by vectors $|m\rb\otimes|\psi\rb\ (m=0,1,\dots),
\psi\in\mH_B,$ is $H$-invariant, hence also invariant \wrt\ the time evolution $U_t\equiv
\exp[-it(H_A + H_B + \mgam V_\mphi)]$. Let $P_{\mcl K}$ be the orthogonal projector onto \cl K.
Let us define the partial isometries $P_{nm}$ in \cl K\ by

\bequ\label{7.4.8(5)} P_{nm}|k\rb\otimes |\psi\rb = \delta_{mk} |n\rb\otimes |\psi\rb,\ \text{for
all}\  \psi\in\mH_B,\ n,m,k=0,1,2,\dots\enqu

\noindent Let $P_n:= P_{nn},\ \forall n$. Denote also by $P_\psi,\ \psi\in \mcl H_B\
(\|\psi\|_2=1),$ the one dimensional projector $|\psi\rb\lb\psi|$ in $\mcl H_B$. Clearly
$P_0\Omega_0^\psi=\Omega_0^\psi\equiv |0\rb\otimes |\psi\rb,$ and for all $k,l,m,n
\in\mathbb{Z}_+$ it is

\bequ\label{7.4.8(6)} P^*_{nm}=P_{mn},\ P_{nk}P_{lm}=\delta_{kl} P_{nm},\ \  \sum_{m=0}^\infty
P_m=P_{\mcl K}.\enqu

We shall write elements $x=a\otimes b\in \mfk A\otimes\mfk B$ as $x=ab$\ (hence also $a\equiv
a\otimes I_{\mH_B},\  b\equiv I_{\mfkA}\otimes b$), if a confusion would be improbable. So, we are
looking for limits

\bequ\label{7.4.8(7)} \overline{\mome}(x) := \lim_{t\rarw\infty}\momeAB t(x),\  x\in \mfk
A\otimes\mfk B=\mfk C.\enqu

We shall see that the limits\rref 7.4.8(7)~ for $x\in\mfk A\subset\mfk C$ are expressible in terms
of $\overline{\mome}(P_{mn})$. The very well known Dyson equation\rref 7.4.8(8)~ expressing the
unitary evolution group $U_t=\exp[-it(H_A + H_B + \mgam V_\mphi)]$ of a system with the
interaction $\gamma V_\mphi$ in terms of this interaction and of the free system (without
interaction) evolution group $U_t^0\ (t\in\mbR)$:

\bequ\label{7.4.8(8)} U_t = U_t^0 - i\gamma\int_0^t\rd \tau U^0_{t-\tau}V_\mphi U_\tau, \enqu

\noindent with $U_t^0:=\exp[-it(H_A+H_B)]$, will be used repeatedly in our work here. We shall
work in the Hilbert space \cl K\ (for $t<\infty$). The restriction of the interaction Hamiltonian
$V_\mphi$ to the subspace \cl K\ has the form

\bequ\label{7.4.8(9)} P_{\mcl K}V_\mphi = (P_{01}+P_{10})P_\mphi.\enqu

Due to the commutativity of $U^0_t$ with $P_0$, we obtain for $m\neq 0$ after the insertion from
\rref 7.4.8(9)~ into\rref 7.4.8(8)~ :

\bequ\label{7.4.8(10)}P_mU_tP_0=-i\gamma\int_0^t\rd\tau\ P_mU^0_{t-\tau} P_{10}P_\mphi P_0U_\tau
P_0.\enqu

\noindent For $m=0$, we obtain similarly:

\bequ\label{7.4.8(11)} P_0U_tP_0=P_0U^0_t - i\gamma\int_0^t\rd\tau\ U^0_{t-\tau} P_{01}P_\mphi
P_1U_\tau P_0.\enqu

\noindent Substitution of\rref 7.4.8(10)~ with $m=1$ to this equation leads, after a linear change
of integration variables, to an integral equation for $P_0U_tP_0$:

\bequ\label{7.4.8(12)} P_0U_tP_0= P_0U_t^0 - \gamma^2\int_0^t\rd t'\int_0^{t-t'}\rd\tau \
U^0_{t-t'-\tau}P_{01}P_\mphi U^0_\tau P_\mphi P_{10}P_0U_{t'}P_0.\enqu

\noindent Also the commutativity of $P_\mphi$ with $P_{mn}$ was used here. Since $U_t^A:=
e^{-itH_A}$ leaves the vector $\Omega_0$ invariant, it is also $P_0U^0_t=P_0\exp(-itH_B)$, and with \rref 7.4.3(3)~ we have:

\bequ\label{7.4.8(13)} P_{01}P_\mphi U^0_\tau P_\mphi P_{10} = \lb\mphi|e^{-i\tau H_B}|\mphi\rb\lb
1|U_\tau^A|1\rb P_\mphi P_0\equiv f(\tau) P_\mphi P_0.\enqu

The integral equation\rref 7.4.8(12)~ can be rewritten now in the form:

\bequ\label{7.4.8(14)} P_0U_tP_0=P_0 e^{-itH_B} -\gamma^2\int_0^t\rd t'\int_0^{t-t'}\rd\tau\,
e^{-i(t-t'-\tau)H_B}P_\mphi f(\tau)P_0U_{t'}P_0.\enqu

With the symbols from\rref 7.4.3(1a)~ and\rref 7.4.3(1b)~, by taking the matrix elements of both
sides of this equation  as in\rref 7.4.3(1b)~, we can write the equation for $F(t)$, cf. Notation
\ref{Notation}:

\begin{subequations}\label{7.4.8(15)}
\bequ\label{7.4.8(15a)} F(t)=F^0(t)-\gamma^2 \int_0^t\rd t'\int_0^{t-t'}\rd\tau\, g(t-t'-\tau)
f(\tau) F(t'). \enqu

 If we take the restrictions of these functions to the values of the argument $t\geq 0$
 according to\rref 7.4.3(4)~, we can rewrite\rref 7.4.8(15a)~ as a convolution equation,
  cf. also\rref 7.4.3(5)~:\footnote{\label{7.4.8(15f)}Note that, due to time-reflection
  symmetry of all the systems considered here, quite analogical equations and the
  corresponding results could be obtained also for the function $t\mapsto F(-t),\ t\geq 0$.}

\bequ\label{7.4.8(15b)} F_+=F_+^0 -\gamma^2 g_+\ast f_+\ast F_+. \enqu
\end{subequations}

We shall express now the quantities $\mome^{A\& B}_t(P_m)$ in terms of the subsection
\ref{Notation}, with a help of\rref 7.4.8(10)~:
\bequ\label{7.4.8(16)} \mome^{A\& B}_t(P_m)=\gamma
^2 \int_0^t\rd t'\int_0^t\rd t'' \bar{F}(t') F(t'') g(t'-t'') \bar{f}_m(t-t') f_m(t-t''),\
m=1,2,\dots .\enqu

To obtain a similar expression for $\mome^{A\& B}_t(P_0)$ we shall use completeness of the set of
projections $\{P_m: m\in\mbZ_+\}$ in the subspace \cl K, cf.\rref 7.4.8(6)~. We can sum over $m$
in the argument of $\mome^{A\& B}_t(\cdot)$ in\rref 7.4.8(16)~ because of normality of the state
$\mome^{A\& B}_t\in \mcl S(\mfk C)$ for finite $t$. After the summation we can perform also
$\lim_{t\rarw\infty}$. Summation over $m$ in\rref 7.4.8(16)~ can be performed under the integral
signs due to Lebesgue dominated convergence theorem, cf. the definition of $f_m$. The completeness
of the orthonormal basis $\{\ |m\rb |\ m=1,2,\dots\}$ in $\mH_A$ gives also:

\bequ\label{7.4.8(17)} \sum_{m=1}^\infty \bar{f}_m(t-t')f_m(t-t'') = f_1(t'-t'').\enqu

\noidt We then obtain:

\bequ\label{7.4.8(18)} \mome^{A\& B}_t(P_0)= 1-\gamma^2 \int_0^t\rd t'\int_0^t\rd t''\,
\bar{F}(t')F(t'')f(t'-t'').\enqu

To see the asymptotic properties of $\mome^{A\& B}_t(P_m)\ (t\rarw+\infty)$, we shall need some
properties of the solution $F(t)$ of\rref 7.4.8(15)~. We shall obtain them by expressing the
solution of the Volterra equation\rref 7.4.8(15b)~ in the form of (Carl) Neumann series

\bequ\label{7.4.8(19)} F_+= \sum_{n=0}^\infty (-\gamma^2 g_+\ast f_+\ast)^n F_+^0, \enqu

\noidt converging uniformly on any bounded interval for any $\gamma$ and any continuous $f,\,
g$.\footnote{To see this, calculate $\sum_{n=0}^\infty (h_+\ast)^n(t)$ for $h\equiv const.$} Since
the free particle Hamiltonian $H_B:=\hat{\mathbf{p}}^2$ has an absolute continuous spectrum, the
functions $F^0(t)$ and $g(t)$ from\rref 7.4.3(1a)~ are continuous converging to $0$ for
$t\rarw\infty$. With our assumptions it is  (see also  \cite[Sec. IX.4]{R&S})\quad $|f(t)|\leq 1
\imply \|f\|_1\leq\|g\|_1=2\|g_+\|_1$. This implies

\bequ\label{7.4.8(20)} \|\gamma^2 g_+\ast f_+\|_1 < \gamma^2 \|g_+\|_1\cdot\|f_+\|_1 < 1, \enqu

\noindent which
is a sufficient condition for also the $L^1$-norm convergence of the series in
\rref 7.4.8(19)~. In this way we obtained (cf. also Footnote \ref{7.4.8(15f)})

\bequ\label{7.4.8(21)}  F\in L^2(\mbR)\cap C_0(\mbR).\enqu

\noidt We conclude from the preceding  that

\begin{subequations}\label{7.4.8(22)}
 \bequ\label{7.4.8(22a)}
\lim_{t\rarw\infty} \mome_t^{A\& B}(P_m) = 0,\quad \text{for all}\  m\geq 1. \enqu

The corresponding limit for $m=0$ is obtained from\rref 7.4.8(18)~. Written in the form of the
scalar product $(\bullet,\circ)\in\mbC$ in $L^2(\mbR)$, it has the form:

\bequ\label{7.4.8(22b)} \lim_{t\rarw\infty} \mome_t^{A\& B}(P_0) = 1 - \gamma^2(F_+,F_+\ast
f).\enqu
\end{subequations}

We can prove the assertion\rref 7.4.8(3)~ of the Theorem now. Since the space \cl K\ of the used
representation of \fk C\ is time invariant \wrt\ our dynamics of the interacting systems, we shall
restrict our work to investigation of the limits $\lim_{t\rarw\infty}\mome^{A\& B}_t(a b)$ for
$a=P_{mn},\ m,n\in\mbZ_+,$ resp. $ a=I_{\mH_{vac}}$\, , and $ b=|\psi_1\rb \lb\psi_2|,\
\psi_j\in\mH_B$, resp. $b=I_{\mH_B}$; for possibly more details cf. \cite{bon7}.

Let $|\Omega_t^\psi\rb := \exp(-itH)|0\rb\otimes |\psi\rb\equiv |\Omega_t(\psi)\rb,\quad
|\Omega_0(\psi_j)\rb := |0\rb\otimes|\psi_j\rb.$ On the basis of the following elementary
estimates:
 \bequ\label{7.4.8(23)} |\mome_t^{A\& B}(P_{mn} b)|\equiv | \lb P_m\Omega_t^\psi| P_{mn}
b P_n\Omega_t^\psi\rb | \leq \|b\| \sqrt{\mome_t^{A\& B}(P_{m})\mome_t^{A\& B}(P_{n})} \enqu

\noidt we obtain from\rref 7.4.8(22)~
\bequ\label{7.4.8(24)} \lim_{t\rarw\infty}\mome_t^{A\&
B}(P_{mn} b) = 0,\ \text{for}\ m+n>0.\enqu

Let us calculate now for arbitrary $\psi_{1,2}\in\mH_B$

\bequ\label{7.4.8(25)} \mome_t^{A\& B}(P_0|\psi_1\rb\lb\psi_2|)=
\lb\Omega_t^\psi|\Omega_0(\psi_1)\rb\lb\Omega_0(\psi_2)|\Omega_t^\psi\rb.\enqu

\noidt We find, according to the notation from \ref{Notation}\ (\ref{7.4.3(1)}.) (used now for
arbitrary $\psi',\ \psi\in\mH_B$), and according to the equation\rref 7.4.8(15)~, that
\bequ\label{7.4.8(26)} \lb\Omega_0(\psi')|\Omega_t^\psi\rb_+ \equiv
\lb\Omega_0(\psi')|P_0U_tP_0|\Omega_0(\psi)\rb_+ = F^0_{\psi'}(\psi)_+(t) - \gamma^2
F^0_{\psi'}(\mphi)_+\ast f_+\ast F_+(t).\enqu

It follows from\rref 7.4.8(21)~ that the \rhs\ of\rref 7.4.8(26)~ converges with $t\rarw +\infty$
to zero, hence also the \rhs in\rref 7.4.8(25)~ converges  to zero (for all $\psi_j$). Hence

\bequ\label{7.4.8(27)} \lim_{t\rarw\infty}\mome_t^{A\& B}(P_0 |\psi_1\rb\lb\psi_2|)=0,\ \text{for
all}\  \psi_{1,2}\in L^2(\mbR^3)=\mH_B.\enqu

 \noidt Let us note that a different situation appeared in the case $b:= I_{\mH_B}$ in which case
  the equation\rref 7.4.8(22b)~ is valid.

It remained to find the limit of the expressions $\mome_t^{A\& B}(|\psi_1\rb\lb\psi_2|)\equiv
\mome_t^{A\& B}(I_{\mH_{vac}}\otimes |\psi_1\rb\lb\psi_2|)$. Because we are working in the
time-invariant subspace $\mcl K\subset \mH_{vac}\otimes\mH_B$, and the projection onto it is
$P_{\mcl K}=\sum_{m=0}^{\infty}P_m$, we shall write this sum instead of  $I_{\mH_{vac}}$ in
$\mome_t^{A\& B}$. The summation over $m$ in its argument  should be done, however,  before
performing the limit $\lim_{t\rarw\infty}\mome_t^{A\& B}(P_m |\psi_1\rb\lb\psi_2|)$.

With the help of\rref 7.4.8(10)~, we can obtain \bequ\label{7.4.8(28)} \mome_t^{A\& B}(P_m
|\psi_1\rb\lb\psi_2|)=\gamma^2\,\int_0^t\rd t' \int_0^t\rd t'' \bar{F}(t')
F(t'')F^0_\mphi(\psi_1)(t'-t)\bar{F}^0_\mphi(\psi_2)(t''-t)\, \bar{f}_m(t-t') f_m(t-t''). \enqu

\noidt {\bf Let us introduce the  functions} \emn$G(t',t'')$~ and \emn$g(\psi_1,\psi_2)(t',t'')$~
of $\{t',t''\}\in\mbR^2$:
$$ G(t',t''):= \bar{F}_+(t')F_+(t'');\ g(\psi_1,\psi_2)(t',t''):=f_1(t''-t')
F^0_\mphi(\psi_1)_{-}(-t')\bar{F}^0_\mphi(\psi_2)_{-}(-t'')$$

\noidt where e.g. $F^0_\mphi(\psi)_{-}(-t):= \theta(-t)F^0_\mphi(\psi)(-t).$

 \noidt A use of\rref 7.4.8(17)~ leads us to:

\begin{eqnarray}
\label{7.4.8(29)}  \mome_t^{A\& B}((P_{\mcl K}-P_0)|\psi_1\rb\lb\psi_2|)&=& \gamma^2 \int_0^t\rd
t' \int_0^t\rd t'' \bar{F}(t')
F(t'')F^0_\mphi(\psi_1)(t'-t)\bar{F}^0_\mphi(\psi_2)(t''-t)\,f_1(t'-t'')\nonumber \\ &=&
\gamma^2\,G\ast g(\psi_1,\psi_2)(t,t),\end{eqnarray}

\noidt where $\ast$ denotes the 2-dimensional convolution. From the given properties of the
entering functions (cf. also our Lemma \ref{7.4.6}, and the $L^p$-estimates in
\cite{R&S,stein&weiss}), and with the use of\rref 7.4.8(27)~, we obtain the desired result:

\bequ\label{7.4.8(30)} \lim_{t\rarw\infty}\mome_t^{A\& B}(|\psi_1\rb\lb\psi_2|)=0,\
\psi_j\in\mH_B.\enqu

The existence of a limit state $\overline{\mome}:=w^*\hbox{-}\lim_{t\rarw +\infty}\mome^{A\& B}_t$
according to\rref 7.4.8(3)~ is proved; its form as a product state\rref 7.4.8(3)~ in $\mS(\mfk
A\otimes\mfk B)$ can be seen by checking its values on elements of $\mfk A\otimes\mfk B$, cf. also
\cite{bon7} and \cite[I.4.5.Proposition 2]{dix1}. By comparing the definition in \ref{Notation} of
the no-particle state $\mome_0^B$ on \fk B\ with our results, and considering the results\rref
7.4.8(22)~,\rref 7.4.8(24)~, and\rref 7.4.8(27)~ (together with\rref 7.4.8(30)~) we finally
obtain:
 \bequ\label{7.4.8(131)}
\overline{\mome}:=w^*\hbox{-}\lim_{t\rarw\infty}\mome_t^{A\& B}= (w\,\mome^A_\uparrow +
(1-w)\,\mome^A_\downarrow)\otimes\mome^B_0,\ \text{with} \ w:= \gamma^2(F_+,\,F_+\ast f).\enqu

\noidt Let us show next that the probability $w$ in\rref 7.4.8(131)~ is positive and has the form

\bequ\label{7.4.8(132)} w=\lb\psi|W_\gamma|\psi\rb,\ \text{where}\  W_\gamma\in\mcl
L(\mH_B),\,\quad 0<W_\gamma\neq W_\gamma^2, \enqu

\noidt where $\psi\in\mH_B$ is the initial state-vector of the scattered particle.

Remember that the function $f$ does not depend on the initial state $\psi$ of the scattered
particle,\rref 7.4.3(3)~ . The function $\psi\mapsto F(t)\equiv F_\mphi(\psi)(t),\ \psi\in\mH_B$\
is, according to its definition\rref 7.4.3(1b)~, a bounded linear functional of the initial
state-vector $\psi$, and the same is valid for $F_+(t)$. Hence, the probability $w =: w(\psi)$\
in\rref 7.4.8(131)~ is a quadratic function of $\psi\in\mH_B$. We can rewrite it, by applying to
it the \emm polarization identity~, into a sesquilinear form dependent on two vectors
$\psi_1,\,\psi_2\in\mH_B$ being ''occasionally'' chosen in the expression of $w(\psi)$ to be
equal: $\psi_1=\psi_2\equiv \psi.$ So, let us write $w(\psi)=:\mcl W(\psi,\psi)$, and define:

\begin{subequations}\label{7.4.8(31)}
\bequ\label{7.4.8(31a)} \mcl W(\psi_1,\psi_2):=\frac{1}{4}\sum_{\alpha=\pm i,\pm  1}\alpha\,
w(\alpha\psi_1+\psi_2)\enqu

\noidt which is the wanted bounded sesquilinear form on $\mH_B$ depending on $\psi_1$
antilinearly; hence, it can be written as a matrix element of a bounded linear operator on
$\mH_B$. Let us denote this operator as $W_\gamma$:

\bequ\label{7.4.8(31b)}  \lb\psi_1|W_\gamma|\psi_2\rb:= \mcl
W(\psi_1,\psi_2)=\frac{1}{4}\sum_{\alpha=\pm i,\pm 1}\alpha\, \mcl
W(\alpha\psi_1+\psi_2,\,\alpha\psi_1+\psi_2),\ W_\gamma\in\mcl L(\mH_B),\enqu,

\noidt and we can write the probability $w$ in the form of a diagonal element of $W\equiv
W_\gamma$:

\bequ\label{7.4.8(31c)} w\equiv w(\psi):= \gamma^2(F_\mphi(\psi)_+,\,F_\mphi(\psi)_+\ast f) =
\lb\psi|W_\gamma|\psi\rb,\ \psi\in\mH_B,\enqu

\noidt where the first bracket $(\cdot,\cdot)$ denotes the scalar product in $L^2(\mbR)$, and the
second one: $\lb\cdot|\bullet|\cdot\rb$ is a matrix element in $\mH_B= L^2(\mbR^3)$.
\end{subequations}

If we notice that the function $f$ from\rref 7.4.3(3)~ entering\rref 7.4.8(31c)~ is of positive
type (because it is a diagonal matrix element of $\exp[-it(H_A+H_B)]$), cf. \cite[Thm. IX.9]{R&S},
and if we reconsider the (commutative) convolution operation $f\ast$ in\rref 7.4.8(31c)~ as a
linear operator $f\ast \in\mcl L(L^2(\mbR))$, we can immediately see that the operator $W_\gamma$
is a positive operator on $\mH_B,\ W_\gamma\geq 0$. It remains to check that the matrix element
$\lb\psi|W_\gamma |\psi\rb$ in\rref 7.4.8(31c)~ is different from zero, if the assumptions of our
Theorem are fulfilled.

To proceed further, let us rewrite the expression\rref 7.4.8(31c)~ of $w$  in terms of Fourier
transforms.

Let us take Fourier transform of the equation\rref 7.4.8(15b)~ for $F(t)=F_\mphi(\psi)(t)$. We
shall use the notation:\footnote{This notation should not be confused with $\mcl
F(F)_+:=\theta\cdot\mcl F(F)\equiv(\hat{F})_+ $, differing by the place where the sign ``+''
occurs.}

\begin{subequations}\label{7.4.8(34)}
\bequ\label{7.4.8(34a)} \hat{F_+}(u)\equiv \mcl F(F_\mphi(\psi)_+(\bullet))(u)\equiv\mcl
F(\theta\cdot F_\mphi(\psi))(u),\enqu

\noidt and similarly for other functions $g_+\mapsto\hat{g}_+,\ f_+\mapsto\hat{f}_+$, or also

\bequ\label{7.4.8(34b)} \mcl F(F^0_+)\equiv ({F^0_+})\,\hat{}\equiv \hat{F^0_+}\equiv \mcl
F(F^0_\mphi(\psi)_+(\bullet))\equiv \mcl F(\theta\cdot F_\mphi^0(\psi)).\enqu
\end{subequations}

\noidt We obtain then from\rref 7.4.8(15b)~ the transformed equation:

\bequ\label{7.4.8(35)} \hat{F_+}= \hat{F^0_+}-2\pi\,\gamma^2\,\hat{g}_+\hat{f_+}\hat{F_+},\enqu

\noidt which can be solved immediately:

\begin{subequations}\label{7.4.8(36)}
\bequ\label{7.4.8(36a)} \hat{F_+}(u)=\frac{\hat{F^0_+}(u)}{1+2\pi\,
\gamma^2\,\hat{g}_+(u)\hat{f_+}(u)},\quad\  u\in\mbR,\enqu

\noidt or in another form

\bequ\label{7.4.8(36b)} \mcl F(F_+)(u)=\frac{\mcl F(F^0_+)(u)}{1+2\pi\, \gamma^2\,\mcl
F(g_+)(u)\mcl F(f_+)(u)}.\enqu
\end{subequations}

\noidt This is the Fourier transform of the explicit expression\rref 7.4.8(19)~ of the solution
of\rref 7.4.8(15)~ obtained with the help of Carl Neumann series.

Let us rewrite the expression\rref 7.4.8(31c)~ for the probability
$w(\psi)\equiv\lb\psi|W_\gamma|\psi\rb$ with the help of\rref 7.4.8(36)~ (remember the
notation\rref 7.4.3(1b)~): \bequ\label{7.4.8(37)}
\lb\psi|W_\gamma|\psi\rb=\gamma^2\,(\hat{F_+},\hat{F_+}\cdot\hat{f})=\gamma^2\sqrt{2\pi}
\int_{\mbR}\rd u\, \hat{f}(u)\,
\frac{|\hat{F^0_+}(u)|^2}{|1+2\pi\,\gamma^2\,\hat{g}_+(u)\hat{f_+}(u)|^2}.\enqu

\noidt Let us investigate properties of the above integrand in some details. Let us express first
the function $\hat{f_+}(u)= \mcl F(f_1\cdot
g\cdot\theta)(u)=\sqrt{2\pi}\hat{f_1}\ast\hat{g_+}(u)=\sqrt{2\pi}\mcl
F(f_1\cdot\theta)\ast\hat{g}(u)$. The Fourier image $\hat{f_1}(u)$ of $f_1(t)\equiv
\frac{1}{t}\,J_1(2t)$ can be obtained with a help of its integral representation taken from
\cite[3.752-2]{RG}: \bequ\label{7.4.8(38)} f_1(t) = \frac{1}{t}\,J_1(2t)= \frac{4}{\pi}\int_0^1
\cos(2tx)\,\sqrt{1-x^2}\,\rd x. \enqu

\noidt We can rewrite this expression to the forms
\begin{eqnarray}\label{7.4.8(39)}
 f_1(t) &=&
 \frac{1}{2\pi}\int_{-2}^{2}e^{itu}\sqrt{4-u^2}\,\rd u \nonumber \\
 &=&
\frac{1}{2\pi}\int_{-\infty}^{+\infty}e^{itu} \,\theta(2-|u|)\,\sqrt{4-u^2}\,\rd u \nonumber \\
&=& \left[\mcl F^{-1}\left(\frac{1}{\sqrt{2\pi}}\,\theta(2-|u|)\,\sqrt{4-u^2}\right)\right](t),
\end{eqnarray}

\noidt hence, we obtain from\rref 7.4.8(39)~ the wanted Fourier image immediately:

\bequ\label{7.4.8(40)} \hat{f_1}(u)\equiv\mcl F(f_1)(u)
=\frac{1}{\sqrt{2\pi}}\,\theta(2-|u|)\,\sqrt{4-u^2}. \enqu

\noidt The expression\rref 7.4.8(38)~ of $f_1$ leads, in agreement with its definition\rref
7.4.3(2)~, to the estimates \bequ\label{7.4.8(41)} |f_1(t)|\leq \frac{4}{\pi}\int_0^1 \rd x
|\cos(2tx)|\sqrt{1-x^2}\leq \frac{4}{\pi}\int_0^1 \rd
x\,\sqrt{1-x^2}=\frac{4}{\pi}\int_0^{\frac{\pi}{2}}\rd\malp\cos^2\malp =1, \enqu

\noidt where we used the change of the integration variable $x := \sin\malp$, the identity
$\sin^2\malp +\cos^2\malp\equiv 1$, and the symmetry properties of the goniometric functions.
Since both functions $f_1,\, g$ are continuous, $g(t)=(\mphi,\exp(-itH_B)\mphi)$,\,\ $\mphi\in\mcl
D(\mbR^3)\imply$ the Fourier image $\hat{\mphi}\in\mcl S(\mbR^3)$ is an entire analytic function
of three complex variables \cite[Thm. IX.12]{R&S}, the function $t\mapsto g(t)\neq 0\ (\text{a.e.
for}\ t\in\mbR)$ according to Lemma \ref{7.4.4}, and the continuous function $f_1(t)$ is not
constant, hence the function $|f_1(t)|<1$ on certain intervals of \bR, the estimate for
$L^1$-norms gives:

\bequ\label{7.4.8(42)} \|f\|_1\equiv \|f_1\cdot g\|_1< \|g\|_1,\enqu

\noidt hence we have here obtained the sharp inequality. From the definition of the Fourier
transformation it is seen that the following trivial inequality is valid for any function $h\in
L^1(\mbR)$:

\bequ\label{7.4.8(43)} \|\hat{h}\|_\infty \leq\frac{1}{\sqrt{2\pi}} \|h\|_1.\enqu

These considerations give an estimate for the denominator in\rref 7.4.8(37)~ by
\bequ\label{7.4.8(44)} \|2\pi\gamma^2 \hat{g_+}\hat{f_+}\|_\infty \leq
2\pi\gamma^2\|\hat{g_+}\|_\infty \|\hat{f_+}\|_\infty \leq  \gamma^2 \|g_+\|_1 \|f_+\|_1=
 \frac{\gamma^2}{4}\|g\|_1 \|f\|_1< \frac{\gamma^2}{4} \|g\|_1^2. \enqu

\noidt This proves, also due to the condition $\|\gamma\, g\|_1<2$\ in\rref 7.4.8(2)~, that the
denominator of the integrand in\rref 7.4.8(37)~ is everywhere different from zero and finite.

Another part of the integrand in\rref 7.4.8(37)~ is the function $\hat{f}=\mcl F(f_1\cdot g)=
\frac{1}{2\pi}\,\hat{f_1}\ast \hat{g}.$ The Fourier image of $g(t)\equiv \lb\mphi|\exp(-it
H_B)\mphi\rb$ is
\begin{subequations}\label{7.4.8(45)}
\bequ\label{7.4.8(45a)}
\hat{g}(u)=\frac{1}{\sqrt{2\pi}}\int_{-\infty}^{+\infty}e^{-itu}\lb\mphi|e^{-itH_B}\mphi\rb\,\rd t
= \frac{1}{\sqrt{2\pi}}\int_{-\infty}^{+\infty}\rd t\,e^{-itu}\int_0^{+\infty}\rd \lambda\,
e^{-it\lambda}\lb \mphi|\, E_{H_B}(\lambda)|\mphi\rb,\enqu

\noidt where $E_{H_B}(\lambda) := E_{H_B}((-\infty,\lambda])$ is the projection-measure of the
selfadjoint operator $H_B$. Because the spectrum of $H_B$ is positive (and absolutely continuous
\wrt\ Lebesgue measure on \bR), and the function $g(t)$ is proportional to the Fourier image of
$\mlam\mapsto\lb\mphi|E_{H_B}(\mlam)|\mphi\rb$, one has \bequ\label{7.4.8(45b)}
\hat{g}(u)=\theta(-u)\mcl F(g)(u)=\sqrt{2\pi}\lb\mphi| E_{H_B}(-u)|\mphi\rb.\enqu

\noidt This can be rewritten in the ``p-representation'', which allows us to see better the
dependence on the specific functions $\mphi$. We shall write the element of the solid angle $\phi$
in terms of the Euler angles $\theta,\mphi$ in $\mbR^3$ as
$\rd\phi:=\sin\theta\,\rd\theta\,\rd\mphi$, and the function
$\hat{\mphi}(\vec{p})\equiv\hat{\mphi}(p,\phi)\ (p:=|\vec{p}|)$. It is
\bequ\label{7.4.8(45c)}
g(t)=\lb\mphi|e^{-itH_B}\mphi\rb\ = \int_{\mbR^3}\rd^3\vec{p}\ \overline{\hat{\mphi}(\vec{p})}e^{-itp^2}
\hat{\mphi}(\vec{p})= \int_0^{+\infty}\rd p\, p^2
e^{-itp^2}\int_{4\pi}\rd\phi\,|\hat{\mphi}(p,\phi)|^2,\enqu

\noidt which, after the change of variables $\lambda := p^2$, leads to

\bequ\label{7.4.8(45d)}g(t)=\frac{1}{2}\int_0^{+\infty}\rd \lambda\, \sqrt{\lambda}e^{-it\lambda}
\int_{4\pi}\rd\phi\,|\hat{\mphi}(\sqrt{\lambda},\phi)|^2;\enqu

\noidt this has the form of the Fourier image of

\bequ\label{7.4.8(45e)} \mcl
F^{-1}(g)(\lambda):=\theta(\lambda)\sqrt{\frac{\pi}{2}}\,\sqrt{\lambda}\,
\int_{4\pi}\rd\phi\,|\hat{\mphi}(\sqrt{\lambda},\phi)|^2,\enqu

\noidt and the Fourier image $\hat{g}$ has now the form \bequ\label{7.4.8(45f)} \hat{g}(u) =
\hat{g}(u)\,\theta(-u) =\mcl F^{-1}(g)(-u) =
\theta(-u)\sqrt{\frac{\pi}{2}}\,\sqrt{-u}\,\int_{4\pi}\rd\phi\,
|\hat{\mphi}(\sqrt{-u},\phi)|^2.\enqu
\end{subequations}

Similar considerations could be applied also to
$f(t)\equiv\lb\Omega_1\otimes\mphi|\exp(-it(H_A+H_B))|\Omega_1\otimes\mphi\rb$; the spectrum of
$H_A$ from\rref 7.4.1(1)~ acting on the Hilbert space $\mH_{vac}$ of the used representation
consists of a single eigenvalue $\{0\}$, and of absolutely continuous part consisting of the
interval $[-2,+2]\subset\mbR,$ which can be seen from the Section \ref{sec;7.3}, and from
\cite{bon6}. So the function $f(t)=\lb\Omega_1|\exp(-itH_A)|\Omega_1\rb\cdot
\lb\mphi|\exp(-itH_B)|\mphi\rb=f_1(t)g(t)$ has the Fourier image
$\hat{f}(u)=(2\pi)^{-\frac{1}{2}}\hat{f_1}\ast\hat{g}(u)$, which with the help of\rref 7.4.8(40)~
and\rref 7.4.8(45f)~ gives
\begin{eqnarray} \label{7.4.8(46)}
\hat{f}(u)&=& \frac{1}{\sqrt{2\pi}}\,\hat{f_1}\ast\hat{g}(u)= \frac{1}{\sqrt{2\pi}}\int\rd\tau
\hat{f_1}(\tau)\hat{g}(u-\tau) \\
&=& \frac{1}{2\pi}\int_{-2}^2 \rd\tau\,
\sqrt{4-\tau^2}\,\sqrt{\frac{\tau-u}{2}}\,\theta(\tau-u)\int_{4\pi}\rd\phi\,
|\hat{\mphi}(\sqrt{\tau-u},\phi)|^2.\nonumber
\end{eqnarray}

\noidt Remember that $\mphi\in\mcl D(\mbR^3)$, hence its Fourier image $\hat{\mphi}\in\mS(\mbR^3)$
is an entire analytic function of three complex variables, so that the function $p\mapsto
\int_{4\pi}\rd\phi\, |\hat{\mphi}(p,\phi)|^2 > 0,\, \text{a.e. for}\  p>0.$ Then\rref 7.4.8(46)~
implies that $\hat{f}(u)=0$ for $u>2$, and $\hat{f}(u)> 0$ for almost all $u<2$.

For checking finally the conditions of the positivity of $w(\psi)$ from its expression\rref
7.4.8(37)~, we have to check under which conditions it is $|\hat{F^0_+}(u)|\theta(2-u) > 0,\,\
u\in S\subset\mbR$, for some $S$ of positive Le\-bes\-gue measure.

Let us assume that $\hat{F^0_+}(u)\equiv 0$ in some nonzero interval: $u\in I\subset\mbR$. The
function $\hat{F_+^0}(u)\equiv\frac{1}{\sqrt{2\pi}}\int_0^{+\infty}\,e^{-itu}
\lb\mphi|\exp(-itH_B)|\psi\rb\,\rd t$, cf.\rref 7.4.3(1a)~, can be continued to a function
analytic in the lower complex half plane $\Im\,u<0$ and continuous on the real axis \bR. The
identical vanishing of this function on an interval $I\subset\mbR$  would imply (with the help of
the Schwarz Reflection Principle) its analyticity on $I$, and consequent vanishing everywhere in
the analyticity domain, hence also on the whole real axis (i.e. vanishing also on the boundary of
the analyticity domain). The identical vanishing $\hat{F^0_+}(u)\equiv 0,\ \forall u\in\mbR$,\
would imply, however, the identical vanishing $\lb\mphi|\exp(-itH_B)|\psi\rb\equiv 0,$ which
contradicts\rref 7.4.8(1)~. This proves that, for $\gamma^2>0$ satisfying\rref 7.4.8(2)~, it is
$\lb\psi|W_\gamma|\psi\rb >0$, iff $\psi$ satisfies\rref 7.4.8(1)~. Since the condition\rref
7.4.8(1)~ does not depend on the parameter $\gamma$, the subspace of $\mH_B$ consisting of those
vectors $\psi$ for which it is $\lb\psi|W_\gamma|\psi\rb =0$ does not depend on $\gamma$, hence
also {\bf its orthogonal complement \emn$\mH_W\subset\mH_B$~ is independent of} $\gamma$, cf.
Lemma \ref{7.4.4}.

 It remains to show that, at least for some values of $\gamma\in\mbR$, {\bf it is}
 \emn$W_\gamma^2\neq W_\gamma$~, i.e. that {\bf the positive operator $W_\gamma$ is not  a projector}.
 For any nonzero orthogonal projector
$P\in\mLH$ there exists a subspace $P\mH\equiv\mH_P\subset\mH$ such that for any normalized vector
$\psi\in\mH_P$ it is $\lb\psi|P|\psi\rb=1$, and for all vectors $\psi$ from its orthogonal
complement: $\psi\in\mH_P^\perp:=\mH\ominus\mH_P$, it  is $\lb\psi|P|\psi\rb =0$. If an operator
$W_{\gamma}$ would be a nonzero projector, for all the normalized  vectors $\psi\in\mH_W$ it would
be $\lb\psi|W_{\gamma}|\psi\rb =1$. Such a $\psi$ would necessarily satisfy\rref 7.4.8(1)~, and
then $\lb\psi|W_\gamma|\psi\rb >0$ for any $\gamma$ satisfying\rref 7.4.8(2)~.

For any given normalized $\psi$ satisfying\rref 7.4.8(1)~, the numerical function
$\gamma^2\mapsto\lb\psi|W_\gamma|\psi\rb$ expressed in\rref 7.4.8(37)~ is continuous and
monotonically increasing  in a nonzero interval $\gamma^2\in [0,\gamma_0^2]\subset\mbR$. For an
arbitrary normalized $\psi\in\mH_B$, it is $\lb\psi|W_{\gamma=0}|\psi\rb=0$, and it is $0<
\lb\psi|W_\gamma|\psi\rb<1$ for all sufficiently small $|\gamma|>0$ and all normalized
$\psi\in\mH_B$.  Hence, at least for sufficiently small nonzero $\gamma\in\mbR$, it is
$\lb\psi|W_\gamma|\psi\rb\neq 1$ for normalized $\psi\in\mH_W$, so that $W_\gamma^2\neq W_\gamma$,
i.e. the positive operator $W_\gamma$ is not a projector. The theorem is proved.

\end{proof}

\section{The X-Y chain as a measuring device}\label{sec;7.5}

\pt\label{7.5.1}\rm {\bf The X-Y chain}\nl

Let us formulate first what we understand here under the ``X-Y chain'' (cf. \cite{robinson}, and
also \cite{ruelle1}, \cite{bon-disert}, \cite{emch2}) - a special case of the Heisenberg spin
chains:

It is again a model of one-dimensional spin chain with \Ca\ of observables \fkA\ generated by spin
creation-annihilation operators $a_j^*,\ a_j\ (j\in\mbZ)$, as it was introduced in \ref{7.3.2}.
The algebra \fkA\ is the $C^*$-inductive limit of the sequence of its local subalgebras $\mfkA_n\
(n\in\mathbb{N})$, each generated by $a_j^*,\ a_j\ (|j|\leq n)$. The dynamics in any subalgebra
$\mfkA_n$ is given by the {\bf local Hamiltonian} \emn$H_n$~ (without interaction with external
magnetic field):

\bequ\label{7.5.1(1)} H_n:=\frac{\kappa}{2}\sum_{j=-n}^{n-1}(a^*_ja_{j+1}+a^*_{j+1}a_j),\enqu

\noidt where $\kappa\in\mbR$. These local Hamiltonians define the time-evolution
$(t;x)\mapsto\tau_t^{(n)}(x)$ of local elements $x\in\mfkA_n$:

\bequ\label{7.5.1(2)} \tau^{(n)}_t(x):=e^{itH_n}x e^{-itH_n},\ x\in\mfkA_n,\ n\in\mbN,\
t\in\mbR.\enqu

The evolution in the whole algebra \fkA\ is obtained by taking first the limit $n\rarw\infty$ in
norm of \fkA\ for any fixed $t\in\mbR$ and any local $x\in\mfkA$, and afterwards obtaining the
result by the norm-continuity, extending it to all $x\in\mfkA$:

\bequ\label{7.5.1(3)} \tau_t(x):= \rn\=\lim_{n\rarw\infty}\tau^{(n)}_t(x).\enqu

Note that the term ``X-Y model'' comes from the form of the hamiltonian if it is rewritten in the
terms of \emm Pauli \sg-matrices~: $\msg^x_j:= a_j^*+a_j,\ \msg^y_j:=ia_j-ia^*_j,\ \msg_j^z:=
2a^*_ja_j-1$, i.e.

\bequ\label{7.5.1(4)} H=\frac{\kappa}{4}\sum_j (\msg_j^x\msg_{j+1}^x+\msg_j^y\msg_{j+1}^y).\enqu

\noidt We shall write often $H$ instead of $H_n$, also without specifying the local characters of
the entering algebraic elements $x$, or $A\in\mfkA$, \dots, to simplify the notation and the
corresponding comments; the reader could easily add the necessary specifications on his own.

We shall use the known formula to express the automorphism\rref 7.5.1(3)~:

\bequ\label{7.5.1(5)} e^{itH} A e^{-itH}= \sum_{m=0}^\infty \frac{(it)^m}{m!} [H,A]^{(m)},\enqu

\noidt where $[H,A]^{(0)}:= A$, and higher elements are recurrently defined with a help of the
commutator $[H,A]^{(1)}:= [H,A]\equiv HA-AH$:

\bequ\label{7.5.1(6)} [H,A]^{(m+1)}:= [H,[H,A]^{(m)}]. \enqu

\noidt The application of\rref 7.5.1(5)~ to norm-bounded elements $A$ (with also $H\hookrightarrow
H_n$) makes no principal problems, but calculations of time evolved elements in\rref 7.5.1(5)~ of
e.g. $A\hookrightarrow a_j$ is technically complicated and it is much easier to work, instead with
the spin operators $a_j$, with elements \emn$b_j\in\mfkA$~ satisfying the Fermi {\bf canonical
anticommutation relations} (\emn CAR~). This can be reached by the \emm Jordan-Wigner
transformation~ (\cite{jordan&wigner}, and also \cite[Ch.3,§ 2]{emch1}):

\bequ\label{7.5.1(7)} b_j:= a_j\prod_{k=-n-1}^{j-1} (1-2a^*_ka_k),\ b_j^*:=(b_j)^*,\enqu

\noidt for $|j|\leq n$. Although these elements become to be nonlocal with $n\rarw\infty$, their
bilinear combinations remain local, and this is sufficient for our calculations. Note also that
there is the inverse transformation expressing $a_j$ in terms of $b_j$, which has the same form
as\rref 7.5.1(7)~ after the exchange $a_{j,k}\leftrightarrow b_{j,k}$.

The elements $b_j, b_k,\ j,k\in[-n,n]$ satisfy CAR:

\bequ\label{7.5.1(8)} [b_j,b_k]_+\equiv 0,\ b_jb^*_k+b^*_kb_j=:[b_j,b^*_k]_+=\delta_{jk}.\enqu

The local Hamiltonians \emn$H_n$~ from\rref 7.5.1(1)~ can be written now as

\bequ\label{7.5.1(9)} H_n = \frac{\kappa}{2}\sum_{j=-n}^{n-1} (b^*_jb_{j+1}+b_{j+1}^*b_j).\enqu

We can calculate now the time evolution of the elements $b_j\in\mfkA$. We shall need later the
estimates for $\tau_t(a_j^*a_j)$, and due to equality $a_j^*a_j = b_j^*b_j$ the explicit
expressions for $\tau_t(b_j)$ will be sufficient for us. We can use\rref 7.5.1(5)~ to calculate
$\tau_t(b_j)$. One easily checks that the multiple commutators have the form:

\bequ\label{7.5.1(10)} [H,b_j]^{(m)}=\sum_p c_j^{(m)}(p) b_p, \enqu

\noidt where the c-number coefficients $c_j^{(m)}(p)$ ($m\in\mbZ_+,\ j,p\in\mbZ$) satisfy
following recurrent relations:

\begin{eqnarray}
\label{7.5.1(11)}  c^{(m+1)}(p)&=&-\frac{\kappa}{2} (c^{(m)}(p-1) + c^{(m)}(p+1)), \\
\text{where}\quad c^{(0)}(p)&=& \delta_{0p},\quad c^{(m)}(j-p)\equiv c_j^{(m)}(p). \nonumber
\end{eqnarray}

\noidt It is seen that the coefficients $c_j^{(m)}(p)$ depend on $p-j$ only:  they are expressible
as linear combinations of the Kronecker deltas $\delta_{j,p+c}$. Notice also that
$c^{(m)}(-p)=c^{(m)}(p),\ \forall p\in\mbZ$. Note moreover that for each $m\geq 0$ only finite
number of the coefficients $c^{(m)}(j-p)$ is nonzero.

From\rref 7.5.1(5)~ and\rref 7.5.1(10)~ we have:

\begin{subequations}
\bequ \label{7.5.1(12a)} \tau_t(b_j)= \sum_{k\in\mbZ} C_t(j-k)\, b_k, \enqu

\noidt {\bf where} \bequ\label{7.5.1(12b)} C_t(r) := \sum_{m=0}^\infty \frac{(it)^m}{m!}\,
c^{(m)}(r).\enqu

\end{subequations}

The Bessel functions of the first kind $J_r(t),\ r\in\mbZ_+,\ t\in\mbR$, can be expressed by the
power series:

\bequ \label{7.5.1(13)} J_r(t) = \sum_{k=0}^\infty (-1)^k\, \left(\frac{t}{2}\right)^{2k+r}
\frac{1}{k!(r+k)!}. \enqu

By calculation of coefficients $c^{(m)}(r)$ in\rref 7.5.1(12b)~ with the help of\rref 7.5.1(11)~
and by comparison of coefficients at equal powers $t^m$ of the variable $t\in\mbR$ in the
expressions\rref 7.5.1(12b)~ for \emn$C_t(r)$~ and in\rref 7.5.1(13)~ for $J_r(t)$, we can see
that for $r\in\mbZ_+$ it is

\bequ \label{7.5.1(14)} C_t(r) \equiv (-i)^r J_r(\kappa t).\enqu

After inserting this into\rref 7.5.1(12a)~ (keep in mind that $C_t(-r)=C_t(r)=(-i)^{|r|}
J_{|r|}(\kappa t)$) we obtain explicit expression for time evolution of elements $b_j\in\mfkA$,
hence the time-automorphism group $\tau_t,\ t\in\mbR$, of \fkA\  in terms of standard special
functions $J_r,\ r\in\mbZ_+$.

\pt\label{7.5.2}\rm {\bf Interaction with a small system.} Let us use the just described X-Y spin
chain to construction of an alternative ``model of quantum measurement'' now.

Let us represent the algebra \fkA\ in a subspace of the CTPS = $\otimes_{j\in\mbZ}\mbC^2_j$ (cf.
\ref{5.1.3}) corresponding to the product-vector
 $\Psi_0$ defined as follows: Let the spins on our chain be well ordered and
numbered by $j\in\mbZ$. Let $|\pm j\rb$ be the states of the $j$-th spin being eigenvectors  of
the Pauli matrix $\msg^z_j$ corresponding to the up-, resp. down-orientations: $\msg^z_j |\pm j\rb
= \pm |\pm j\rb$.  Let then

\bequ\label{7.5.2(1)} \Psi_0 := \bigotimes_{j\leq -1}|+ j\rb\otimes\bigotimes_{k\geq 0}|-
k\rb.\enqu

Let the Hamiltonian of this chain be

\bequ\label{7.5.2(2)} H_0:= \frac{\kappa}{2}\sum_{j\leq -2}(a^*_ja_{j+1}+a_{j+1}^*a_j)
+\frac{\kappa}{2}\sum_{k\geq 0}(a^*_ka_{k+1}+a_{k+1}^*a_k),\enqu

\noidt which is the Hamiltonian of the X-Y model without the term $(a_{-1}^*a_0+a_0^*a_{-1})$.
This chain with the Hamiltonian $H_0$ will play for us the role of the ``macroscopic (measuring)
system''. The state described by the vector $\Psi_0$ is stationary for this Hamiltonian:

\bequ\label{7.5.2(3)}  H_0\Psi_0 = 0. \enqu

The ``measured microsystem'' will be an additional $1/2$-spin (i.e. it does not belong to the
chain) with the interaction Hamiltonian

\bequ\label{7.5.2(4)}  V:= P_+\,\otimes\, \frac{\kappa}{2}\, (a_{-1}^*a_0+a_0^*a_{-1}), \enqu

\noidt where $P_+$ is the projector in the state space $\mbC^2$ of the added spin-microsystem
projecting onto the state $|+\rb$ in which the spin ``is pointing up'': $\msg^z|+\rb
=|+\rb$.\footnote{We shall omit usually in the following the tensor-product symbol $\otimes$,
according our preceding conventions.} If we write (in microsystem's state space $\mbC^2$)\
$P_-:=I-P_+$, the total Hamiltonian $\widetilde{H}$ of our compound system ``micro \& macro''
reads: \bequ\label{7.5.2(5)} \widetilde{H}= H_0 +V= HP_+ +H_0P_-,\enqu

\noidt where $H$ is the total Hamiltonian of the X-Y model\rref 7.5.1(4)~. Let the initial state
of the compound system be \bequ\label{7.5.2(6)} \Phi_0 := \mphi_0\otimes\Psi_0,\ \mphi_0:=
c_+|+\rb +c_-|-\rb,\enqu

\noidt where $\mphi_0$ is normalized: $|c_+|^2+|c_-|^2=1$, and $|\pm\rb$ are also normalized
eigenvectors of $\msg^z\in\mcl L(\mbC^2)$: \bequ\label{7.5.2(7)} P_\pm|\pm\rb =|\pm\rb,\quad
P_+P_-=0.\enqu

\noidt Since, in accordance with\rref 7.5.2(3)~,
 \bequ\label{7.5.2(8)} \widetilde{H}(|+\rb\otimes
\Psi_0)=|+\rb\otimes H\Psi_0,\ \widetilde{H}(|-\rb\otimes \Psi_0)=0,\enqu

\noidt the time evolution looks like:

\bequ\label{7.5.2(9)} \Phi_t:=e^{-it\widetilde{H}}\Phi_0 = c_+|+\rb\otimes e^{-itH}\Psi_0 +
c_-|-\rb\otimes \Psi_0.\enqu

We shall show, similarly as in \ref{sec;7.3}, that the pure state state vector $\Phi_t$ of the
compound system converges in the limit $t\rarw\infty$ to the incoherent linear combination of two
vectors, corresponding to two disjoint states of the compound system (as well as of the
macrosystem-chain); hence this limit is a vector which describes a mixture of two macroscopically
distinct states of the system. It is sufficient to check this assertion by calculation of the
quantities

\bequ\label{7.5.2(10)} \widetilde{\mome}_t(a^*_ja_j) := \lb\Phi_t|\, a^*_ja_j\, |\Phi_t\rb\quad
\text{for}\ j\in\mbZ, \enqu

\noidt i.e. of \bequ \label{7.5.2(11)} \widetilde{\mome}_t(a^*_ja_j) =
|c_+|^2\,\lb\Psi_0|\,\tau_t(a_j^*a_j)\,|\Psi_0\rb + |c_-|^2\,\lb\Psi_0|\,a_j^*a_j\,|\Psi_0\rb;
\enqu

\noidt here, the automorphisms $\tau_t$ are expressed in\rref 7.5.1(12a)~.

It can be proved now that the limit
$\overline{\mome}(A):=\lim_{t\rarw\infty}\lb\Psi_0|\tau_t(A)|\Psi_0\rb,\ A\in\mfkA,$ of a state
from\rref 7.5.2(11)~ exists, and the states $\overline{\mome},\ \mome_0\in\mS(\mfkA)$:
\bequ\label{7.5.2(12)}
\overline{\mome}(A):=\lim_{t\rarw\infty}\mome_t(A)\equiv\lim_{t\rarw\infty}\lb\Psi_0|\tau_t(A)|\Psi_0\rb,
\quad \mome_0(A):= \lb\Psi_0|A|\Psi_0\rb,\quad A\in\mfkA,\enqu

\noidt are mutually disjoint and macroscopically distinct. We shall prove now existence of the
limits\rref 7.5.2(12)~ in\rref 7.5.2(11)~ for $A=a_j^*a_j$. It is

\bequ\label{7.5.2(13)} \mome_0(a_j^*a_j) =
\begin{cases}1\quad\text{for}\ j\leq-1,\\
0\quad\text{for}\ j\geq 0. \end{cases}\enqu

Since according to\rref 7.5.1(7)~ it is $a_j^*a_j = b_j^*b_j$, we can use\rref 7.5.1(12a)~ to
obtain:

\label{7.5.2(14)}
\begin{eqnarray}
\tau_t(a^*_ja_j)&=& \tau_t(b^*_j)\tau_t(b_j) =\sum_{r,s}\overline{C_t}(j-r)C_t(j-s)\, b^*_rb_s=\nonumber \\
&=& \sum_{r,s}\overline{C_t}(j-r)C_t(j-s)\, a_r^*
\left\{\prod_{q=\min[r,s]}^{\max[r-1,s-1]}(1-2a^*_qa_q)\right\}a_s,
\end{eqnarray}

\noidt where the products $\prod_n^m B_q:=1$ if $m<n$. Hence \bequ\label{7.5.2(15)}
\mome_0(\tau_t(a^*_ja_j))=\sum_{r,s}\overline{C_t}(j-r)C_t(j-s)\, \mome_0\left(a_r^*
\left\{\prod_{q=\min[r,s]}^{\max[r-1,s-1]}(1-2a^*_qa_q)\right\}a_s\right),\enqu

\noidt and due to the properties\rref 7.5.2(1)~\,\&\!\rref 7.5.2(13)~ of $\mome_0$ and due to
commutation properties of the $a_j,\ a_k^*$ we see that the terms with $r\neq s$ are zeros.
According to\rref 7.5.1(14)~ we have:

\barr \label{7.5.2(15a)} \mome_t(a_j^*a_j)&=& \sum_{r=-\infty}^{+\infty} |C_t(j-r)|^2\,
\mome_0(a^*_ra_r) = \sum_{r=1}^{+\infty}|C_t(j+r)|^2 \nonumber \\
&=& \sum_{r=1}^{+\infty} J^2_{j+r}(\kappa t)\equiv \sum_{r=1}^{+\infty} J^2_{|j+r|}(\kappa t).
\earr

According to the known formula \cite[(21.8-26)]{2Korn}:

\bequ\label{7.5.2(16)} 1= J_0^2(z) + 2\sum_{k=1}^{+\infty} J_k^2(z), \enqu

\noidt and due to the asymptotic behaviour of Bessel functions

\bequ\label{7.5.2(17)} J_m(t) \asymp O(t^{-\frac{1}{2}}),\ m\in\mbZ, \enqu

\noidt we have finally

\bequ\label{7.5.2(18)}  \overline{\mome}(a_j^*a_j):=\lim_{t\rarw+\infty} \mome_t(a_j^*a_j) =
\frac{1}{2},\quad \text{for all}\quad j\in\mbZ.\enqu

Returning to the formulas\rref 7.5.2(10)~ $\&$\rref 7.5.2(11)~ of our main interest, we have
obtained:

\bequ\label{7.5.2(19)}
\widetilde{\mome}_\infty(a_j^*a_j):=\lim_{t\rarw +\infty}\widetilde{\mome}_t(a^*_ja_j)=|c_+|^2\
\overline{\mome}(a_j^*a_j)+|c_-|^2\  \mome_0(a_j^*a_j). \enqu

The last formula describes an (incoherent) mixture of two mutually macroscopically distinct, hence
disjoint states $\mome_0,\ \overline{\mome}$  on the \Ca\ \fkA\ of the infinite spin chain. This
can be checked in the explicit way by calculating values of a macroscopic observable in the states
$\mome_0$, resp. $\overline{\mome}$, e.g. of the observable constructed from\rref 7.3.7(1)~

\bequ\label{7.5.2(20)} \gamma := w\=\lim_{N\rarw\infty} \frac{1}{N}\sum_{n=1}^N a_n^*a_n\,\in\mfk
Z(\mfkA^{**})\subset\mfkA^{**}. \enqu

According to\rref 7.5.2(18)~ and\rref 7.5.2(13)~, it is:

\bequ\label{7.5.2(21)} \overline{\mome}(\gamma)= \frac{1}{2}\ \neq\ \mome_0(\gamma)=0.\enqu

\noidt Hence, again here, a microscopic system interacting with the macroscopic X-Y chain changed
the chain's initial state $\mome_0$ into a new, macroscopically distinct state
$\widetilde{\mome}_\infty=|c_+|^2\,\overline\mome\ +\ |c_-|^2\,\mome_0$. Here the probabilities
$|c_\pm|^2$ of occurrence of the mutually \emn disjoint states~ $\mome_0,\,\overline{\mome}$ in
the proper (resp. `genuine', cf. \ref{1.1.4}) mixture $\widetilde{\mome}_\infty$ are exactly the
probabilities of appearing of the states $|\pm\rb$ of the microsystem in its initial state
$\mphi_0$, cf.\rref 7.5.2(6)~. This corresponds again to the ``\emn ideal measurement~'', as it
was discussed in \ref{7.1.3}, \ref{7.3.6} and \ref{7.4.2}.

\section{Radiating finite spin chain}\label{sec;7.6}

\pt\label{7.6.1}\rm We shall present very briefly in this section, without proofs, the dynamics of
a model of a large but finite system  interacting with a Fermi field.\footnote{The formulation and
main features of the dynamics of this model were presented first time in \cite{bon-cologne}. The
technical details are described in \cite{bon-sir}.} The system's initial state is stationary but
unstable, as it was also the case of the models presented in the preceding  sections. After an
initial perturbation, the model evolves quickly into a new stationary state by simultaneous
radiation of a Fermi particle, which escapes into infinity. The process is very quick in contrast
to the time evolutions in the case of the models described in the previous sections \ref{sec;7.3},
\ref{sec;7.4} and \ref{sec;7.5}. The three preceding models might, however, serve as clear
mathematical pictures of ``quantum measurement'' in the sense that the time evolution of a large
system led with the time growing to infinity to the state ``macroscopically different'' from its
initial state. The ``macroscopic difference'' between states of the system is mathematically
expressed there as disjointness of the states on the \Ca\ of observable quantities of the large
system. The disjointness implies that if those states are represented as vectors in a Hilbert
space, their mutual linear combinations do not lead to any interference (the C*-algebra of
observables representing all possible observations on the model system is fixed!) and such a
linear combination is physically equivalent to a ``proper mixture'', or ``genuine mixture'' (cf.\
\ref{1.1.4}), i.e. to a classical statistical description of an ensemble in which the individual
copies of the large system are distributed between the uniquely determined `classical' states
under consideration. This unique decomposability to pure states on the algebra of classical
(macroscopic) observables is a consequence of the fact that the states of a classical system form
a simplex. This differs from ``mixed'' quantum states described by density matrices of standard QM
of finite-size systems having multiple convex decompositions to extremal (pure) states.

Since the model of a ``large'' system described in this section is finite (corresponding by
physical intuition to that consisting of finite number of some ``elementary'' or ``small''
subsystems, each of them described by elementary QM in separable Hilbert space ${\cal H_\nu}$\ with the
algebra of its observables coinciding with the whole ${\cal L(H_\nu)}$), there is no possibility of emergence of
any disjoint states, hence there is no unambiguously defined ``macroscopic difference'' between
some of its states.\footnote{An exception consists in possible introduction `by hand' by a
theoretician some `superselection rules' representing a model of `macroscopic difference' and
forbidding interference between vectors from specific subspaces of a Hilbert space \H, cf. e.g. \cite{jauch}.}
 Of course, the infinite size of the previous models is a mathematical idealization, and there should
be some empirical possibility of distinction between ``microscopic'' and ``macroscopic'', resp.
between ``quantum'' and ``classical'', also in `large but finite systems', as it is perceived in
our everyday life.\footnote{\label{R-penr} Another possibility is some, up to now not clearly
specified basic change of QM, as it was most urgently proposed by Penrose in several his
publications, e.g. in \cite{penrose1,penrose2,penrose3}; the main motivation for these
reformulations of QM was some inclusion of the usually postulated ``reduction of wave packet''
\cite{neum1}, called by Penrose the ``\emm process {\bf R}~'', into the dynamics of general QM
systems.}

 This distinction does not need to be, however, mathematically
sharp. Such a possibility was sketched in \cite{hp-meas}: In a verbal transcription it could be,
perhaps, formulated so that it would be very improbable to construct such an observation device on
states of large (however finite) system, which could ``see'' simultaneously sufficiently many
atoms of the system to be able to detect some interference phenomenon. This could be considered as
a rough `definition' of the notion that some set of states of the (now finite) apparatus consists
of elements being pairwise `almost macroscopically different' (cf. also
\cite{hp-meas}).\footnote{\label{emp-disj}Let us illustrate briefly this idea on a long but finite
spin\hbox{-}1/2 chain of the length $N$ with the \Ca\ \fkA\ of its observables generated by the
spin creation-annihilation operators $a_j, a^*_j\ (j=1,2,\dots N)$ acting on the finite
dimensional Hilbert space $\mH_N:=(C^2)^N$: If we are able to use apparatuses detecting the
observables of this chain occurring in an arbitrary of the \Csas\ $\mfk B\subset\mfk A$\ generated
by any of the fixed restricted set of operators $a_{j_m}, a^*_{j_m}\ (m=1,2,\dots K\ll N,\ 0\leq
j_m\leq N)$ {\bf only}, then the states $|\Psi\rb, |\Phi\rb$ from $\mH_N$ for which it holds
$\lb\Psi|B|\Phi\rb\equiv 0\ \forall B\in\mfk B$ could be considered as `almost macroscopically
different', resp. `empirically disjoint'. This happens, e.g., if in the state $|\Psi\rb$ all the
spins are `pointing up', and in the state $|\Phi\rb$ all the spins are `pointing down'.} To
proceed in these considerations, one would need to build some (more) general theory of
observational devices. E.g., as far as the present author knows, there were no published works
paying attention to the fact that human observers come into contact with measuring apparatuses by
 electromagnetic interactions, and probably only by them. Shortly, according to the
point of view proposed here: The formalized set of ``observables'' of any physical system should
depend on the existing possibilities of the construction of measuring devices in accordance with
physical laws and environmental conditions.

We have not stressed up to now, however, that the spin chain of our present model is also coupled
to a Fermi particle (resp. to the Fermi field) representing a sort of `\emm environment~'. The
particle occurs in the initial state of the system in its vacuum state, and afterwards it is radiated
by the chain and subsequently escapes into infinity; the state of the Fermi field containing the
radiated particle is in each finite time orthogonal to its vacuum state. This facilitates, in the
intuitive sense of some sort of a `decoherence program', cf. e.g.
\cite{zeh1,zurek,decoherence,schlosshauer}, the possibility of interpretation of the effective
absence of interference between the initial and final states of the spin chain in our model, as
representing the two different `macroscopically' distinguished `pointer positions'.

We shall keep in mind such an idea to be able to believe that also our finite system described in
this section can be considered as a model of ``quantum measurement'' process.

\pt\label{7.6.2}\rm Let us look at the Quantum Domino from Section \ref{sec;7.3}. We shall
restrict here that model to finite number of degrees of freedom, hence the spin chain will be of
finite length and its algebra of observables \fk A\ (with unity $I_{\mfk A}$) is generated by the
spin-1/2 creation and annihilation operators $a^*_j,\ a_j\ (j=0,1,\dots, N)$ satisfying\rref
7.3.2(1)~. This system will interact with the (nonrelativistic scalar) Fermi field, the {\bf
algebra} \emn\fk F~\ (with unity $I_{\mfk F}$)\ of which  is generated by the {\bf particle
creation-annihilation operators} \emn$b^*(\mphi),\ b(\mphi)$~ satisfying the relations
\bequ\label{7.6.2(1)}
 b(\varphi)^2=0, \
  b(\varphi)b^*(\psi)+b^*(\psi)b(\varphi)=(\varphi,\psi)I_{\mfk F},\
   (\text{for all}\  \varphi,\psi\in L^2(\mathbb{R}^3,d^3x)),\enqu

   \noidt with the linear dependence $\psi\mapsto b^*(\psi)$.

The dynamics is given by the Hamiltonian $H:=H_0 + V$, where
\begin{subequations}
\bequ\label{7.6.2(2a)}
H_0:=\left(\sum_{n=0}^{N-2}a^*_na_n(a^*_{n+1}+a_{n+1})a_{n+2}a^*_{n+2}-\varepsilon_0a^*_Na_N\right)
\otimes I_{\mfk F}+I_{\mfk A}\otimes d\Gamma(\mbs h), \enqu

\bequ\label{7.6.2(2b)} V:=v^2\left(a^*_{N-1}a_{N-1}a^*_N\otimes
b^*(\sigma)+a^*_{N-1}a_{N-1}a_N\otimes b(\sigma)\right). \enqu
\end{subequations}\nl

\noidt We can consider these algebras $\mfk A$ and $\mfk F$ as algebras of operators acting on the
Hilbert space $\mH_S:=(\mathbb{C}^2)^{N+1}$, and on the Fermi Fock space $\mH_F$ respectively, resp.
on their tensor product
 $\mH:= \mH_S\otimes\mH_F$. In the above written formulas, the {\bf
symbol} \emn$\rd\Gamma(\mbs h)$~ {\bf means} the {\bf ``second quantization''} (cf. \cite[Sec.
5.2.1]{bra&rob2}\footnote{The ``second quantization'' $d\Gamma(\mbs h)$ of the
`one-Fermi-particle-operator' $\mbs h$ is the linear operator acting in the Fermi Fock space
$\mH_F:=\oplus_{n=0}^\infty P_-\otimes_1^n\mfk h$, where $P_-$ is the antisymmetrization operator,
such that \linebreak $d\Gamma(\mbs h)P_-\otimes_{k=1}^n\psi_k:=P_-\sum_{j=1}^n
\psi_1\otimes\psi_2\otimes\dots\otimes\mbs h\psi_j\otimes\dots\otimes\psi_n$ for all $n\in\mbZ_+$.} )
of the operator \emn$\mbs h\in\mcl L(\mfk h):=\mcl L(L^2(\mbR^3,\rd^3x))$~ given by the function
\emn$\mathbf{p}\mapsto \mveps(\mathbf{p})$~ of one-particle momentum $\mathbf{p}$, hence acting on
the vectors of $\mfk h:= L^2(\mbR^3,\rd^3x)$ ``in the p-representation'' as multiplication by
$\mveps(\mathbf{p}): (\mbs h\psi)(\mathbf{p})\equiv \mveps(\mathbf{p})\psi(\mathbf{p})$. The
nonnegative function \emn$\mveps(\mathbf{p})$~, as well as the parameters $\varepsilon_0 >0,\
v\in\mbR,\ \msg\in L^2(\mbR^3,\rd^3x)$, will be specified later. In our expressions of action of
elements of \fk A, resp. \fk F,\ on vectors of $\mH_S\otimes\mH_F$, the unity operators of the
other algebra will be usually omitted, e.g. for $a\in\mfk A,\ |s\rb\otimes
|\mphi\rb\in\mH_S\otimes\mH_F$, we shall write $a\otimes I_{\mfk F} (|s\rb\otimes|\mphi\rb)\equiv
a(|s\rb\otimes|\mphi\rb)\equiv a|s\rb\otimes|\mphi\rb$.

Let \emn$\Omega_0^F$~ be the {\bf Fermi vacuum} in $\mH_F$, and \emn$\Omega_0^S\in\mH_S$~ be the
state of the spin chain ``with all spins pointing down'': $a_n\Omega_0^S=0,\, \forall n$. Notice
also that  {\bf here} $|n\rb :=a^*_0a^*_1\dots a^*_n\Omega_0^S,\ n=0,1,\dots N$. Let the {\bf
Hilbert subspace} \emn$\mH_{min}\subset\mH$~ be {\bf generated by the vectors}\ind{\beta_n}\ind{\beta_N(\psi)}
\[\{ \Omega_0 :=\Omega_0^S\otimes\Omega_0^F,\
 \beta_n :=|n\rb\otimes\Omega_0^F,\ \beta_N(\psi):= |N\rb\otimes b^*(\psi)\Omega_0^F;\ n=0,1,\dots
 N-1,\ \psi\in L^2(\mbR^3,\rd^3x)\}.\]

\noidt Then \ind{\Omega_0 :=\Omega_0^S\otimes\Omega_0^F}it is valid:

\begin{lem}\label{7.6.3} The space $\mH_{min}$ defined above is $H$-invariant:
$H\mH_{min}\subset\mH_{min}$.\end{lem}

A proof of this Lemma is presented in \cite{bon-sir}. Hence the description of our process can be
restricted to time evolution in the subspace $\mH_{min}\subset\mH$. We shall choose the parameters of
the model, namely the operator $\mbs h$ acting on $L^2(\mbR^3)$, and the quantities $\varepsilon_0
>0,\ v\in\mbR,\ \msg\in L^2(\mbR^3,\rd^3x)$, so that with our Hamiltonian given by\rref 7.6.2(2a)~
the relation

\begin{subequations}\label{7.6.3(1)}
\bequ\label{7.6.3(1a)} \lim_{t\rarw\infty}\lb\beta_n| e^{itH}a^*_Na_N e^{-itH}|\beta_n\rb =1,\
n=0,1,\dots N-1, \enqu

\noidt or more specifically: \bequ \label{7.6.3(1b)} \lb\beta_n| e^{itH}a^*_Na_N
e^{-itH}|\beta_n\rb =1-o(t^{-m}),\ n=0,1,\dots N-1,\ \text{for}\ t\rarw+\infty,\ \forall
m\in\mathbb{N},\enqu

\end{subequations}

\noidt will be satisfied. The meaning of\rref 7.6.3(1)~ is that the probability of emission of the
Fermi particle and simultaneous transition of the spin chain to the stationary state $\beta_N$
(i.e. all the spins in the chain ``are pointing up'' and the Fermi field is again in its vacuum
state) approaches certainty `almost exponentially quickly' if the time is growing to infinity.

The dynamics is investigated by a repeated use of {\bf Fourier transform $\mcl F$}, e.g. in
\cite[Lemma 2]{bon-sir}: \ind{Fourier transform $\mcl F$}

\begin{lem}\label{7.6.4}
\label{fourier} Let $e^{-itH}$ be any (unitary) time evolution group. Then the Fourier transform
of its (truncated) matrix elements for given $\phi, \psi \in \mathcal{H}$ is
\begin{equation}\label{7.6.4(1)}
\mathcal{F}[\theta(t)\lb\phi,e^{itH}\psi\rb](\xi)=\frac{i}{\sqrt{2\pi}}\lb\phi,R_{H}(\xi)\psi\rb,
\end{equation}
for  $\xi\in\mbC:\,\text{Im}\ \xi < 0$.
\end{lem}

\noidt The function $\theta$ is here the Heaviside function, and \emn$R_H(\xi)\equiv (H-\xi
I)^{-1}$~\ $(\xi\in\mbC,\, \xi\notin sp(H)\equiv$\ \emn spectrum of\ $H$~) is the \emm resolvent
of the operator $H$~.

 Another useful result is that we obtain the resolvent $R_H(\lambda)$ as a solution of an
operator equation, \cite[Lemma 3]{bon-sir}.

\begin{lem}\label{7.6.5} Suppose $H=H_0+V\in \mathcal{L}(\mathcal{H})$ and $\xi \notin
sp(H)\cup sp(H_0)$. Then the resolvent $R_H(\xi)$ is the solution of the operator equation

\bequ R_H(\xi)=R_{H_0}(\xi)(I-VR_{H}(\xi)). \enqu

\noidt Hence, the Fourier transform of the (truncated) matrix elements of the time evolution
operator for Im $\xi < 0$ is given by:
\begin{equation}\label{7.6.5(1)}
\mathcal{F}[\theta(t)\lb\phi,e^{itH}\psi\rb](\xi)=\frac{i}{\sqrt{2\pi}}\lb\phi,R_{H_0}(\xi)\psi\rb
-\frac{i}{\sqrt{2\pi}}\lb\phi,R_{H_0}(\xi)VR_{H}(\xi)\psi\rb.
\end{equation}
\end{lem}

\noidt Important for the following analysis are the matrix elements

 \bequ\label{7.6.5(2)}
F_{mn}:=\lb\beta_m,R_H(\xi),\beta_n\rb, \enqu

\noidt since e.g.:
\begin{equation}\label{7.6.5(3)}
\mathcal{F}[\theta(t)\lb\beta_m,e^{itH}\beta_n\rb](\xi)=\frac{i}{\sqrt{2\pi}}F_{mn}(\xi). \enqu

Now, the proper choice of the parameters of the model is, according to \cite{bon-sir}:

\begin{subequations}\label{7.6.5(4)}
\bequ\label{7.6.5(4a)} \mveps(\mathbf{p}):= a|\mathbf{p}|^2,\ a>0,\enqu

\bequ\label{7.6.5(4b)}  \mcl F(\msg)(\mathbf{p})=0\ (|\mathbf{p}|<b),\quad \mcl
F(\msg)(\mathbf{p})>0\  \text{for all}\ |\mathbf{p}|>b>0,\quad \msg\in\mcl S(\mbR^3),\enqu

\bequ\label{7.6.5(4c)} \varepsilon_0> ab^2+2,\quad \enqu
\end{subequations}

\noidt where $\mcl S(\mbR^3)$ is the set of all rapidly decreasing Schwartz complex valued
functions on $\mbR^3$, the symbol $\mcl F(\msg)$ again means the Fourier transform (i.e. the
transition to ``p-representation''), and the constants $a,\ b$ occurring in\rref 7.6.5(4c)~ are
the same as the ones occurring  in\rref 7.6.5(4a)~ and\rref 7.6.5(4b)~.

After making this choice it is possible, after a series of considerations and calculations
\cite{bon-sir}, to show that (cf. \cite[(4.33)]{bon-sir})

\bequ \mathcal{F}[\lb\beta_m,e^{itH}\beta_n\rb](p)=-\sqrt\frac{2}{\pi}\ \lim_{\nu \to
0^+}\text{Im}\ F_{mn}(p-i\nu)\in \mathcal{S}(\mathbb{R}). \enqu

But the Schwartz set $\mathcal{S}(\mathbb{R})$ of rapidly decreasing smooth functions is invariant
with respect to the Fourier transform, hence the function $t\mapsto\lb\beta_m,e^{itH}\beta_n\rb$
also belongs to $\mathcal{S}(\mathbb{R})$, what proves the `almost exponential decay' in time of
this matrix element. This result is crucial for the proof of Theorem \ref{7.6.6}.

 To formulate the main result as a
theorem, let us introduce also the notation:

\begin{subequations}
\bequ\label{7.6.5(5a)} \mu((-\infty,\mlam]):=\int_{\mveps(\mathbf{p})<\mlam}|\mcl
F(\msg)(\mathbf{p})|^2\,\rd^3\mathbf{p},\enqu

\bequ\label{7.6.5(5b)} \rho_\mu(\mlam):=\frac{\rd\mu((-\infty,\mlam])}{\rd\mlam}.\enqu
\end{subequations}

\begin{thm}\label{7.6.6}
In the above described model of finite spin chain QD interacting with nonrelativistic scalar Fermi
field, with the parameters specified in\rref 7.6.5(4)~, for either all such $\mveps_0$ with
possibly one exception, or for all

\[ \mveps_0>2+ab^2 + 2v^2\ \int_{ab^2}^\infty \frac{\rho_\mu(\mlam)}{\mlam-ab^2}\,\rd\mlam, \]

\noidt the time evolution of the probability of all the $ N + 1$ spins being turned up (realizing
the wanted final state of the spin chain), if initially the Fermi field was in the vacuum state
and the first $ n$  spins $(N-1\geq n \geq 0)$ were turned up, approaches unity almost
exponentially fast, i.e. the relation:

\bequ \label{7.6.6(1)} \lb \beta_n|e^{itH}a^*_Na_Ne^{-itH}|\beta_n\rb = 1 - o(t^{-m}),\ \text{for
all}\ 0\leq n\leq N-1,\ \text{for any}\  m\in\mbN,\enqu

\noidt is satisfied.
\end{thm}

A detailed proof of this theorem can be found in \cite{bon-sir}.

\pt\label{7.6.7}\rm Let us look at the result\rref 7.6.6(1)~ from the point of view of the Section
\ref{sec;7.7}, to make it more intuitive as a relevant assertion \wrt\ the ``measurement
problem'', cf.\rref 7.7(2)~.

As the ``measured system'' in this model can be considered the single spin lying at the
`beginning' of the spin chain. Let its \Ca\ of observables be generated by $\{a^*_0;\ a_0\}$
satisfying \rref 7.3.2(1)~, and let $\mphi_\downarrow, \mphi_\uparrow$ be its normalized state
vectors corresponding to the two opposite orientations of the spin. Let its initial normalized
state vector be $\mphi_0:=c_\downarrow\mphi_\downarrow + c_\uparrow\mphi_\uparrow$, with
$a^*_0\mphi_\downarrow=\mphi_\uparrow,\ a_0\mphi_\uparrow=\mphi_\downarrow$.

 The initial state of the whole composite system \{\text{\sl measured system}\ \&\ \text{\sl
 rest of the spin chain}\ \&\ \text{\sl Fermi field}\} is then $\widetilde{\Psi}_0:=
c_\downarrow\mOme_0+ c_\uparrow\beta_0= (c_\downarrow I_\mfkA+c_\uparrow
a^*_0)\mOme^S_0\otimes\mOme^F_0$. The time evolved states
$\widetilde{\Psi}_t:=\exp(-itH)\widetilde{\Psi}_0$ can be written, due to the Lemma \ref{7.6.3} as
well as the stationarity of $\mOme_0$, in the form

\bequ\label{7.6.7(1)} \widetilde{\Psi}_t=c_\downarrow\mOme_0 + c_\uparrow e^{-itH}\beta_0.\enqu

The second term in\rref 7.6.7(1)~ can be written, again due to the $H$-invariance of $\mH_{min}$, cf.
Lemma \ref{7.6.3}, in the form

\bequ\label{7.6.7(2)} e^{-itH}\beta_0=\sum_{n=0}^{N-1} d_n(t)\beta_n + \beta_N(\psi(t)).\enqu

\noidt Since $a_N\beta_n =0,\ (n=0,1,\dots N-1)$, and
$a_N^*a_N\beta_N(\psi)=(1-a_Na^*_N)\beta_N(\psi)=\beta_N(\psi)$, the expression from \rref
7.6.6(1)~ with our $n=0$ is

\bequ\label{7.6.7(3)} \lb\beta_0|e^{itH}a_N^*a_N e^{-itH}|\beta_0\rb=\|a_Ne^{-itH}\beta_0\|^2=  \|
\beta_N(\psi(t))\|^2, \enqu

\noidt and this converges very quickly, according to\rref 7.6.6(1)~, to unity. The vectors on the
\rhs\ of\rref 7.6.7(2)~ are mutually orthogonal and the whole \rhs has the constant norm equal to
1. Hence the norm of the sum on the \rhs\ of\rref 7.6.7(2)~ quickly converges to zero. All the
vectors $\beta_N(\psi)\ (\psi\in L^2(\mbR^3,\rd^3x))$ describe the states of the \emn composite
system~:\nl \[\centerline{$\{\text{\sl the measured system}\ \&\ \text{\sl the rest of the spin
chain}\ \&\ \text{\sl the Fermi field}\}$}\]

\noidt in which all the $N+1$ spins ``are pointing up'', which has to mimic the macroscopically
different state from the initial state $(c_\downarrow I_\mfkA+c_\uparrow
a^*_0)\mOme^S_0\otimes\mOme^F_0\equiv (c_\downarrow I_\mfkA+c_\uparrow a^*_0)\mOme_0$, as well as
from $\mOme_0$, of the compound system. For the wave function\rref 7.6.7(1)~ of the compound
system we obtain asymptotically for large times $t\rarw\infty$:

\bequ\label{7.6.7(4)} \widetilde{\Psi}_t=c_\downarrow\mOme_0 + c_\uparrow e^{-itH}\beta_0\,
\asymp\, c_\downarrow\mOme_0 + c_\uparrow \beta_N(\psi(t)), \enqu

\noidt which has the form of the formula\rref 7.7(2)~ for the (approximate expression of the)
``measurement dynamics'' in the conventional QM framework of considering of only finite systems
(as measuring apparatuses). The probabilities of the two different ``measurement results''
corresponding to the states $\mphi_\downarrow$, resp. $\mphi_\uparrow$, occurring in the
orthogonal decomposition of the initial state $\mphi_0$ of the measured system are, as it was
expected, the numbers $|c_\downarrow|^2$, resp. $|c_\uparrow|^2$. By `tracing out' the states of
the \emn environment $\equiv$ the Fermi field~ we obtain the density matrix for the spin chain,
and by tracing out the both \{Fermi field\ \&\ the spins\ 1,2,\dots N\} we obtain the \emn density
matrix~ $\mrh:= |c_\downarrow|^2 P_{\mphi_\downarrow}+|c_\uparrow|^2 P_{\mphi_\uparrow}$, with
$P_{\mphi_\uparrow}\equiv a_0^*a_0$ and $P_{\mphi_\downarrow}\equiv a_0a_0^*$, in the state space
of the measured system (i.e. of the spin placed in the point $0$ of the chain), corresponding
formally to the `\emn collapse of its wave packet~' $\mphi_0:=c_\downarrow\mphi_\downarrow +
c_\uparrow\mphi_\uparrow$, i.e. of its initial state of the just described process. Neither of
these density matrices can be, however, interpreted as describing a `proper', or `genuine'
probability distribution of quantal states in the sense of classical statistics. To interpret them
in that sense, and distinguish one decomposition of a density matrix as `more relevant' (i.e.
reflecting the classical-type statistics), some another additional assumption is needed. We have
had in our interpretations of the infinite models in previous Sections the requirement of \emm
disjointness~ of mutually noninterfering states, and this was ensured by existence of a \emm
macroscopic quantity~ obtaining mutually different values in these states. For some alternative
approaches, we could go back again to the attempts in the `decoherence programs',
\cite{zeh1,zurek,decoherence,schlosshauer}. More detailed mathematical and interpretational
considerations on decompositions of states of a \Ca\ can be found in our \ref{1.2.3}, \ref{1.3.3},
\ref{1.3.4}, \ref{1.4.3}, and citations therein, e.g.\cite[Chap.4]{bra&rob}.

\pt\label{7.6.8}\rm{\bf Notes on irreversibility.} This model of a radiating multispin system can
be also considered as a caricature reflecting one of the usual mechanisms of \emn irreversible
behaviour~ of large physical systems: Large systems usually (resp. `almost always') are not
isolated from their environment, and their interaction with (a `relative stable', and a `relative
stationary') environment leads to their motion to more stable stationary, e.g. thermodynamic
equilibrium, states. Some kind of radiation, as it was built in into our model, is a usual form of
interactions of large systems with their environment.

This approach reflects just one `aspect' of irreversible behaviour of physical systems. Another
often discussed `aspect' of theoretical descriptions of irreversible behaviour of finite
many-particle systems is their complicated mechanical motion even if they are isolated from any
environment. Then we are dealing with such phenomena as various types of ``chaos'', and with
``recurrences'' in their (deterministic and time-reversible) mechanical motion. We shall not
consider here such mechanical explanations of irreversibility, initiated by J. C. Maxwell and L.
Boltzmann. As concerns some study on these topics in the case of classical systems, it might be
interesting to look to nice conference or journal papers like, e.g. \cite{wergeland}, but more
elementary and also more complex information could be found in some books on the ``theory of
dynamical systems'' listed in our Bibliography, e.g.
\cite{abr&mars,arn1,arn3,arn&avez,K&S&F,walters,poston&stewart}. However long are durations of the
\emn Poincar\'e cycles~ corresponding to the above mentioned recurrences in mechanical motions of
isolated systems with several degrees of freedom (they are comparable with the lifetime of
Universe \cite{wergeland}), an evolution during which the system approaches some stable stationary
state cannot be reached in theoretical description of finite isolated mechanical systems. This
does not exclude, however, effectiveness of the statistical physics, which does not deal with a
unique phase-space trajectory of the considered system; here we have a certain physical
reinterpretation of the mechanics of motions in the system's phase space. But full effectiveness
of the statistical approach to description of behaviour of multiparticle systems, e.g.
mathematically clear description of thermal equilibria and phase transitions, is again possible in
the `thermodynamical limit' of infinitely large systems only, e.g.\cite{ruelle1}.

 It is seen that after making the finite quantum spin chain of our model to become an ``open system''
 by adding to
 the Hamiltonian of the restricted QD the term corresponding to the radiation of a fermion,
  the speed of the motion to the limiting  state was
enormously increased in comparison with the infinite, but isolated, QD-chain, cf.\rref 7.3.3(12)~
and\rref 7.3.4(2)~, i.e. \wrt\rref 7.3.4(3)~. The finite-sized version of the isolated QD would
behave, however, almost-periodically, cf.\rref 7.3.3(11)~. The addition of interaction of the
finite QD with Fermi field enabled us to obtain a system's state converging for $t\rarw\infty$ to
a new stationary state. But a clear and unambiguous interpretation of some states of a finite
system, e.g. the two states appearing in the sum on the \rhs\ of\rref 7.6.7(4)~, as being
approximately `mutually macroscopically different' (hence their quantum interference being `almost
impossible'), is still open to discussion. We shall not further investigate here some other
connections of these phenomena and questions.

\section{On the ``measurement problem'' in QM}\label{sec;7.7}

 Let us add here several notes
to the above mentioned ``measurement problem'', considered for a long time to be a fundamental
problem of the conceptual structure of QM, cf. e.g. \cite{QTM}, \cite{penrose1,penrose2} and
\cite[Ch. 29]{penrose3}. These notes should be also supplemented by the notes in \ref{7.6.1}, esp.
by the footnotes \ref{R-penr} and \ref{emp-disj}.

States of the physical systems are described in the mathematical theory of QM by mathematical
objects like ``wave functions'', ``density matrices'', or ``linear functionals \ome on algebras of
observable quantities'' (which generalize the former two classes of objects). The ``observable
quantities'' (represented by operators, resp. elements of an algebra) correspond to experimental,
or observational, arrangements of empirical situations, in which the observer is able, after
``installing'' a specific state \ome of the observed system, to perceive and appreciate by his
human senses some well determined, in advance expected feelings (optical, auditory, acquired by
touch or in another way) of some specific perceptions that are clearly distinguishable from others
(e.g. when reading positions of a pointer, or hearing a characteristic sound from a
counter,\dots), so that they can be formalized into a form suitable for further communication. A
single observable $A$ of a specific physical system appears in such an empirical situation through
a specific instance of a set of such clearly distinguishable phenomena, each of which can be (and,
as a rule, is) denoted by a number $\malp_j$\ ($\in\mbR$) called the {\bf result of single
measuring act in the state \ome\ of a value of $A$} (not to be confused with ``the value of $A$ in
\ome'' -- different single measuring acts of the same observable on \ome\ could lead to different
results!). Many experiments on microscopic systems performed in the history of microphysics have
shown that we are not able to prepare states of any microsystem in such a way that in a many times
repeated measurement of an observable on the same (prepared each time anew) state \ome one obtains
the same measured value {\bf for each observable} which can be chosen for these repeated
measurements. To state it briefly: For any state of any microsystem there is some observable {\bf
which does not have any specific value} in that state. This is reflected mathematically in, e.g.,
Heisenberg uncertainty relations. On the other hand, to each value $\malp_j$ of the given
observable $A$ there exists (for observables with discrete spectra) at least one state $\mome_j$
such that the repeated measurements of $A$ on it give with certainty the same value $\malp_j$. The
problem arises because there is (with certainty) some other observable $B$ such that the repeated
measurements of it on the same state $\mome_j$ give mutually different values $\beta_k\neq
\beta_l\dots \in\mbR$, i.e. the statistical dispersion of the measured values of $B$ in that
$\mome_j$ is nonzero. Sharp values (obtained consistently in the identical, many times repeated
measurements) $\beta_k$ of $B$ can be obtained in other states $\mome'_k$, for which, however, the
measurements of some other observables $A, C, \dots$ would have nonzero dispersions.

The existing very successful  mathematical model of QM provides solution of this problem which
consists in describing an arbitrarily chosen (but, by assumption,``pure'') state $\mome_j$ as a
\emm linear superposition~ of some (again pure) states $\mome'_k, \mome'_l\dots$, i.e., if we
express all the states in the form of vectors in a Hilbert space \H, in writing the state in
question as $\psi_j=\sum_k c_k\mphi_k$ , where the correspondence with the values of the
observables $A, B,\dots$ described now as linear operators on \H, is such that the
``state-vector'' $\psi_j$, corresponding to the state $\mome_j$, is an ``eigenvector'' of the
 operator $A$ (a common practice
is to use the same symbol for the operator as for the physical quantity represented by it):
$A\psi_j=\malp_j \psi_j$, and similarly the vectors $\mphi_k$ corresponding to the states
$\mome'_k$ are the eigenvectors of the operator $B:\ B\mphi_k=\beta_k\mphi_k$.

 All this is, of course, very well known, and we have also briefly described it in our
 Sec.\ref{sec;1.2}.
  We recall it here to stress the unusual intuition required when dealing with
the phenomena described by the mathematical model of QM,
  in comparison with the intuition provided by the `everyday life', whose formal reflection is
  contained in the mathematical models of classical physics.

   One of the prominent results of the history of observations and measurements mentioned above
   is that QM
   is considered an \emn irreducibly statistical theory~; i.e., that the probabilistic results
   of the measurements with nonzero dispersions are not necessarily due to the presence of some
\emn statistical ensembles~ of systems in various states, as they are in the \emn classical
statistical physics~,
   but that it is impossible to find any fully dispersionfree states even when considering
   individual (micro)systems.
   This is now (starting from 1920's) acceptable and included in a logically consistent manner
   into the description of our world.
   The resulting picture of the world is, however, not without problems, since
   its integral part is a class of counterintuitive phenomena encountered in QM. These are,
   pictorially
   expressed, the problems of the type of the well known ``Schr\"odinger's cat paradox'',
   which is just a popular representation
   of the ``measurement problem'' to be discussed further
   (the cat can be regarded here also as a measuring device).

    We are measuring with some macroscopic apparatuses which belong to the
   same world as microsystems do, but seem to be correctly described by a  theory
   that is very different from QM. Is QM a universal theory, or is there some borderline
   between  the two differently behaving parts of the world? If so, it should be explained
   in the theory {\bf where that borderline is located}.
   But the apparatuses are composite of many microsystems and
  (as far as the present author knows) no new aspect of microsystems was discovered which could
   effectively distinguish between them and  macrosystems. Thus, let us regard
   the apparatuses as some quantum-mechanical systems. Then any measuring process should look
   as follows:\footnote{We will work here with pure states (resp. vector states) only. In fact, it
   is not necessary to use density matrices in an analysis of the process of measurement in QM, as
   shown, e.g. by Wigner in \cite{wigner4}.}

   If the initial state of the measured microsystem is described by the normalized vector
   $\mphi_k$ corresponding to the value $\beta_k$
of the observable $B$, and the initial state of the apparatus capable to measure the quantity $B$
is described by the normalized vector $\Psi_0$ in its Hilbert space, installed independently of
the measured state, then the unitary process $U(t)$ corresponding to the time evolution of the
mutually interacting measured microsystem and apparatus will lead, after the `time of the
measurement' $t_m$, to the state

\bequ\label{7.7(1)} U(t_m)\left[\mphi_k\otimes\Psi_0\right]=\widetilde{\Psi}_k.\enqu

\noidt Here, in the `post-measurement state' $\widetilde{\Psi}_k$ of the compound system
microsystem\,$\&$\,apparatus, the ``pointer position'' of the apparatus corresponds to the value
$\beta_k$ of $B$. This is assumed to be valid for all $\beta_k$, hence for $\beta_k\neq\beta_j$
the pointer positions (i.e. certain macroscopic parameters) in the states $\widetilde{\Psi}_k$ and
$\widetilde{\Psi}_j$ are different from each other. The same unitary evolution should lead, after
the measurement by the same apparatus on the state $\psi:=\sum_k\,c_k\,\mphi_k$, due to its
linearity, to the state of the compound system

\bequ\label{7.7(2)} U(t_m)\left[(\sum_{k\in
J}\,c_k\,\mphi_k)\otimes\Psi_0\right]\equiv\widetilde{\Psi}:=\sum_k\,c_k\,\widetilde{\Psi}_k,\quad
\sum_{k\in J}|c_k|^2=1. \enqu

 The `macroscopic part of the world' appears here in the state $\widetilde{\Psi}$, expressed as
  a nontrivial linear
superposition $\widetilde{\Psi}$ of the states $\widetilde{\Psi}_k$ corresponding to different
values of some macroscopic parameter (different ``pointer positions'', distinguished here by the
index $k$). Such superpositions in QM do not mean only a probability distribution with nonzero
dispersion of the values of a macro-parameter corresponding to various $\beta_k$, but they should
also allow (according to the principles of QM) a realization of  measurements of some new
observable having a sharp value in the state $\widetilde{\Psi}$ (on the statistical ensemble of
equally prepared compound systems obtained in the process of the measurement of this new
observable on the microsystem). The states $\widetilde{\Psi}$ are representing in such a way an
{\bf interference of different values of a macro-parameter} ('the cat is simultaneously dead and
alive'). Thus, the apparent conceptual problem of QM does not consist in its probabilistic nature,
it rather consists in the unanswered question of the existence of the very counterintuitive ``\emm
macroscopic interference~'' we have just described, or/and in a dynamical explanation why they do
not occur.

The widely accepted `solution' of this ``measurement paradox'' (as termed by Penrose
\cite{penrose3}) consists in accepting of so called ``\emm reduction postulate~'', consisting in
the claim that there supposedly exists the phenomenon colloquially termed the ``reduction (or also
\emn collapse~) of the wave packet''. This can be rephrased, in terms of our preceding
considerations, in such a way that within some final phase of the process of measurement, either
during or just after the measurement (e.g. such as is sketched in\rref 7.7(2)~) performed on the
system, the system (i.e. either the measured system alone -- this is the traditional point of
view, or the apparatus, or -- which seems to the present author as the most acceptable possibility
-- the compound system microsystem\,$\&$\,apparatus) ends after each single run of the measurement
in a specific state corresponding to the obtained value of the measured observable, and after many
times repeated `identical' measurements on such a state we arrive at a statistical mixture (in the
sense of classical statistical physics, i.e. the ``proper'' or ``genuine'' mixture, cf. in
\ref{1.1.4}) of the set of (systems occurring in the) states which, in the case of compound
system, consists of

\bequ\label{7.7(3)} \{\widetilde{\Psi}_k:\ k\in J\}\ \text{with probabilities}\
|(\widetilde{\Psi},\widetilde{\Psi}_k)|^2 = |c_k|^2,\ k\in J.\enqu

This transition from superpositions to classical mixtures of states with different ``pointer
positions'' takes place, according to the reduction postulate, instantaneously, or in some
``negligibly short time''.

Many existing \emn theories of quantum measurements~ which have appeared up to the present day
analyze systematically possible results of various measurements (of corresponding observables) as
well as their mutual relations like their mutual consistency or `complementarity', see e.g.
\cite{davies,QTM,busch,lahti}. These theories, called by their authors
 ``operational'', are purely phenomenological,  built on the formal
structure of quantum kinematics and usually manifested no interest in the description of specific
dynamics of the considered processes. They are often mathematically highly elaborated, very
elegant and probably also useful from the point of view of applications of QM. We were not
concentrating ourselves here on these approaches and on the questions motivating them. The
avoidance  of the problems with the dynamics of the interaction of the measured microsystems with
the measuring macroscopic apparatuses indicates that in these phenomenological works one assumes,
at least implicitly, the existence of some unknown mechanism of the ``\emn wave packet
reduction~'', or equivalently ``wave packet collapse''. This is acceptable from the `practical
point of view', because in the usual praxis of manipulations with microsystems (e.g. measurements
on them) it is possible to deal with the results (e.g. the outcomes of the measurements) as if the
``wave packet reduction'' really happened. We are here, however, interested in the problem how
this process can be included into a noncontradictory quantum theory. An extensive discussion of
these problems by the  leading physicists up to 1980'ties contains \cite{w+z}.

The last decades, on the other hand, have seen experiments whose results  indicate that {\em the
interference of macroscopically different states is possible in suitable conditions}, cf. e.g.
\cite{leggett1,leggett2}. These `suitable conditions' consist, first of all, in sufficient
isolation of the considered quantum macro-system from any interactions with surrounding
environment, then, of course, in the ability of experimenters to discover some suitable
`macrointerference detecting' observable quantity, and finally in the inventiveness of
experimenters when constructing the desired measuring apparatus.

Our models described in the sections \ref{sec;7.3},\,\ref{sec;7.4} and \ref{sec;7.5} of this
chapter, mainly inspired by the ideas published in \cite{hp-meas}, show that in the limit
$t_m\rarw\infty$ the classical-like probability distributions of the measurement results (i.e.
probability without mutual interferences of results) can be reached. In these models, apparatuses
are treated as quantum collections of infinitely many ``small'' subsystems, and the time necessary
for reaching the ``reduction of the wave packet'' is infinitely long; also, the convergence to the
final states of the apparatuses of ``proper mixtures''-type is in these simple models -- contrary
to the ideal requirements -- very slow.

The last of our models described in Sec.\ref{sec;7.6} shows, however, that if we construct an
``apparatus'' as a large but finite collection of microsystems, interacting, moreover, with the
environment by radiating a particle, the convergence proceeds fast enough -- in the sense `almost
exponentially'. The problem here is nonvanishing possibility of interference of states with
different pointer positions, although such a possibility would be for `sufficiently large'
apparatuses very improbable. Again, an opened question is the existence and location of a possible
borderline for the validity of QM. A mathematically clear formulation of the dependence of
possible interferences between macroscopic states of a ``large system'' on its size will be,
probably, a subject of future investigations in theoretical physics. One cannot exclude, however,
that there is no sharp borderline between QM and CM, and instead, there is a continuous transition
from QM to CM dependent on more parameters than just the size of the measuring apparatus. Or, that
there is no borderline at all, QM is a universal theory, but our understanding of its possible
applications requires some completions.

\newpage

\def\autor{Pavel B\'ona: Classical Systems in Quantum Mechanics}


\newpage


\label{index}\hypertarget{index}{} \printindex



\end{document}